%% file: principal.tex
\documentclass[a4paper,11pt,openright]{thesis}  
\usepackage{superfig,morefloats}
\usepackage{graphicx}
\usepackage{dropping}         
\usepackage{setspace}
\usepackage{section,relsize}
\usepackage{caption2,topcapt}
\usepackage[Lenny]{fncychap}  
\usepackage{chicago}          
\usepackage{xspace}           
\usepackage{sidecap}          
\usepackage{rotating}
\usepackage{lscape}
\usepackage{fancyheadings}
\usepackage{overrulehere}

\usepackage{psfig}
\usepackage{amssymb}
\usepackage{amsmath}
\usepackage{amstext}
\usepackage{overrulehere}
\usepackage[Lenny]{fncychap}
\usepackage{rotate}
\usepackage{floatfig}
\usepackage{graphicx}
\usepackage{dropping}
\usepackage{rotating}
\usepackage{supertabular}
\usepackage{epsfig}
\usepackage{subfigure}
\usepackage{multirow}
\usepackage{longtable}
\usepackage{colortbl}
\usepackage{lscape}

\usepackage{natbib}
\usepackage{multicol}
\usepackage{chicago}
\usepackage[english]{babel}
\psfull
\newcommand{\clearemptydoublepage}{\newpage{\pagestyle{empty}\cleardoublepage}}

\def\farcs{\hbox{$.\!\!^{\prime\prime}$}}
\def\arcsec{\hbox{$^{\prime\prime}$}}
\definecolor{color1}{cmyk}{0.8, 0.4, 0.2, 0.2}
\definecolor{color2}{cmyk}{0.8, 0.4, 0.2, 0.2}
\definecolor{color3}{rgb}{0.7, 0.2, 0.2}

\renewcommand{\captionlabeldelim}{---}
\renewcommand{\captionfont}{\footnotesize}
\renewcommand{\captionlabelfont}{\small \sc}
\newcounter{savesection}
  {\setcounter{savesection}{\value{section}}%
   \setcounter{section}{0}%
   \renewcommand\thesection{\sffamily\thechapter.\Alph{section}}}%
  {\setcdropping.styounter{section}{\value{savesection}}}
\captionstyle{normal} \columnseprule=0pt \columnsep=2.0em
\def\lesssim{\mathrel{\hbox{\rlap{\hbox{\lower4pt\hbox{$\sim$}}}\hbox{$<$}}}}

\def\gtrsim{\mathrel{\hbox{\rlap{\hbox{\lower4pt\hbox{$\sim$}}}\hbox{$>$}}}}

\def\farcs{\hbox{$.\!\!^{\prime\prime}$}}
\def\arcsec{\hbox{$^{\prime\prime}$}}

\begin{document}
\sffamily
\frontmatter
\newpage
\thispagestyle{empty}

\noindent
{\Large \textbf{Santiago Vargas Dom\'inguez}}\\

\vspace{3cm}

\noindent
{\LARGE \textbf{Study of horizontal flows}}\\
{\LARGE \textbf{in solar active regions based on}}\\
{\LARGE \textbf{high-resolution image reconstruction techniques}}\\

\vspace{2cm}

\noindent
{\Large Doctoral thesis (\emph{Doctor Europeus})}\\
{\Large With 106 Figures, including 42 Color Figures,}\\
{\Large and 20 Tables.}\\

\vspace{6cm}

\noindent{\LARGE \textbf{2008}}

\clearemptydoublepage

\thispagestyle{empty}

\begin{center}
         \begin{large}
		{DEPARTAMENTO DE ASTROFISICA\\
		\vspace{3mm}
		Universidad de La Laguna}
	\vspace{2.6cm}
	\end{large}	
	
	{\Large{\sf {\expandafter{Estudio de flujos horizontales en regiones solares}}}}\\
	{\Large{\sf {\expandafter{activas basado en t\'ecnicas de alta resoluci\'on}}}}\\
	 {\Large{\sf {\expandafter{para reconstrucci\'on de im\'agenes}}}}
\vspace{1cm}	
	
{\Large{\sf {\expandafter{Study of horizontal flows in solar active regions}}}}  \\
{\Large{\sf {\expandafter{based on high-resolution image}}}} \\
{\Large{\sf {\expandafter{reconstruction techniques}}}}
			
\vspace{2.cm}	
	
		\sf Memoria que presenta\\
		{\large {\sf Santiago Vargas Dom\' {\i}nguez}}\\
                para optar al grado de \\
		Doctor en Ciencias F\'isicas\\
                Menci\'on \emph{Doctor Europeus}\\
	
\vspace{1.4cm}		

\begin{figure}[h]
\centering
\includegraphics[width=.2\linewidth]{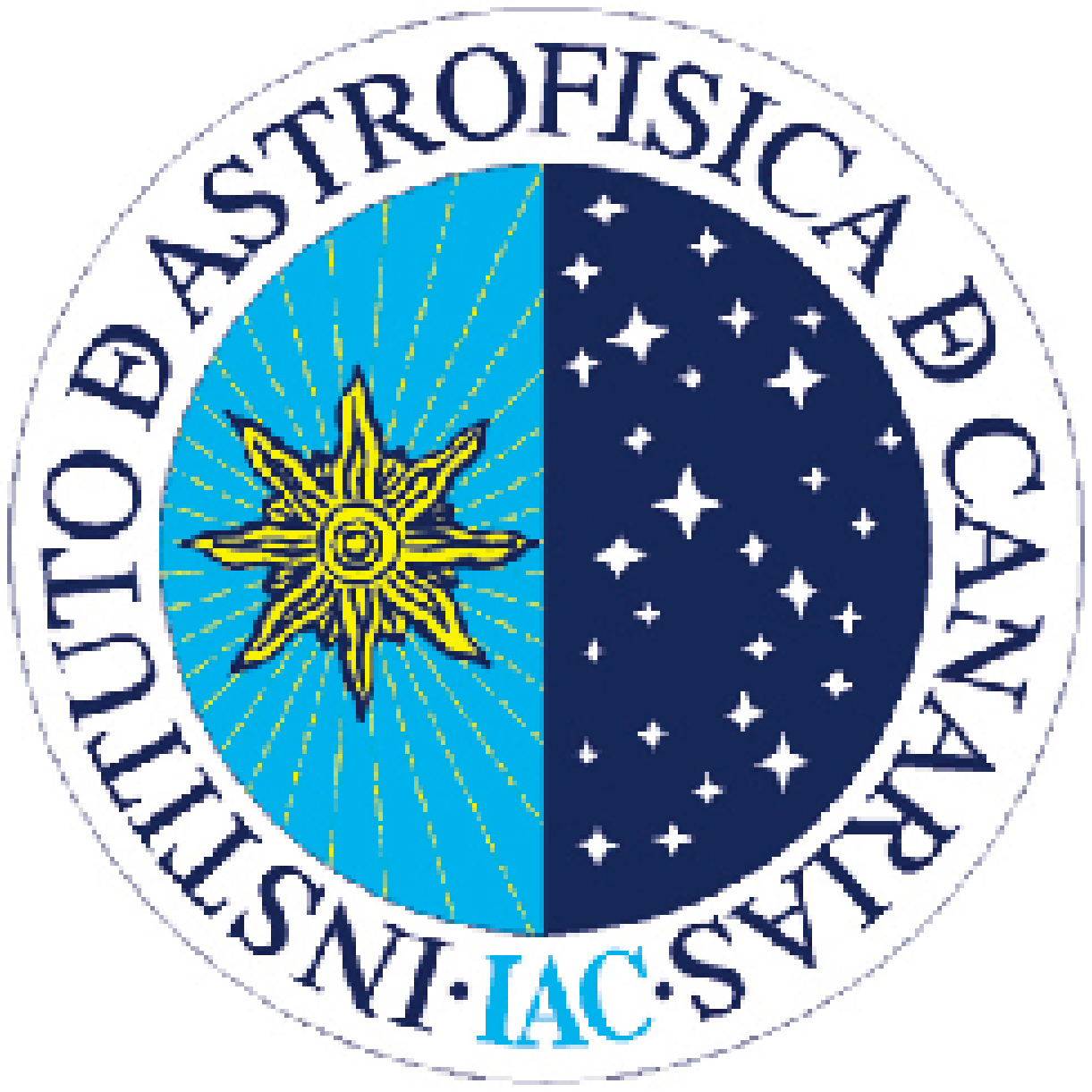}\\
\end{figure}
\vspace{.3cm}	
		INSTITUTO \de\ ASTROFISICA \de\ CANARIAS\\
				junio de 2008
	\end{center}

\newpage

\thispagestyle{empty}
\noindent	
\begin{minipage}[b]{1.\textwidth}
\vspace{14cm}
Examination date: December, 2008  \\
Thesis supervisor: Jos\'e Antonio Bonet Navarro\\
Thesis supervisor: Valent\'in Mart\'inez Pillet\\
 \\
\copyright  Santiago Vargas Dom\' {\i}nguez 2008\\
ISBN: xx-xxx-xxxx-x\\
Dep\'osito legal: TF-xxxx/2008\\
Some of the material included in this document has been already
published in \emph{The Astrophysical Journal}, \emph{The Astrophysical Journal Letters} and \emph{Astronomy \& Astrophysics}.\\
\end{minipage}

\clearemptydoublepage

\thispagestyle{empty}
\vspace{2cm}
\begin{verse}
\flushright 
\textit{A mis padres, \\mi hermano \\  y Laura}
\end{verse}
\clearemptydoublepage
\thispagestyle{empty}
\noindent	
 \hspace{5cm} \begin{minipage}[b]{.6\textwidth}
  \vspace{9cm}
\textsf{El cacique de Iraca y su sobrino Ramiquir\'{\i} gobernaban sobre la tierra en una noche absoluta. Para resolver la situaci\'on, el cacique de Iraca decidi\'o que su sobrino ascendiera sobre los cielos y trajera la luz. Este se dirigi\'o vertiginosamente hacia las alturas y de pronto se transform\'o en un astro incandescente y luminoso: Ramiquir\'{\i} se hab\'{\i}a convertido en el Sol.
Pero su t\'{\i}o no estaba satisfecho del todo pues una parte del d\'{\i}a se hallaba aun en tinieblas y esto le recordaba a la humanidad, con miedo y tristeza, la \'epoca en que todo era tinieblas. Fue entonces cuando el cacique de Iraca resolvi\'o hacer lo mismo que su sobrino, perdi\'endose en la b\'oveda celestial. Y se convirti\'o en un astro de luz m\'as tenue: la Luna. Su luz serv\'{\i}a para alegrar a la gente durante la ausencia del Sol.} \\

\textit{Leyenda muisca o chibcha. Colombia.}

\end{minipage}

\clearemptydoublepage \thispagestyle{empty}
%
%
\clearemptydoublepage
\include{resumen}

\newpage
\thispagestyle{empty}
\clearpage
\include{abstract}

\clearemptydoublepage
%
%
\include{introduction}

\tableofcontents
\clearemptydoublepage
\listoffigures
\listoftables

\mainmatter
\part{\sf Defining a method for in-flight calibration of IMaX aberrations}
\clearemptydoublepage
\include{chap1}
\clearemptydoublepage
\include{chap2}

\clearemptydoublepage
\part{\sf Study of proper motions in solar active regions}
\include{chap3}
\clearemptydoublepage
\include{chap4}
\clearemptydoublepage
\include{chap5}

\clearemptydoublepage
\include{chap6}

\clearemptydoublepage
\include{chap7}

\clearemptydoublepage
\include{conclusiones}

\clearemptydoublepage

%

\clearemptydoublepage
\appendix{
\include{appenA}

\include{appenB}

\include{appenC}}
\thispagestyle{empty}
\include{agra}

\clearpage

\end{document}

%% file: resumen.tex
\chapter*{\sf Resumen}
\dropping[0pt]{2}{E}ste trabajo de tesis se enmarca en un concepto mas general denominado "Alta resoluci\'on en f\'isica solar". El trabajo consiste de dos partes claramente definidas. La primera parte trata sobre el desarrollo instrumental para observaciones solares y la segunda parte est\'a dedicada a la explotaci\'on cient\'ifica de datos solares obtenidos con intrumentaci\'on solar puntera.\\

En la primera parte de la tesis se trabaja el tema de la alta resoluci\'on y la restauraci\'on de im\'agenes para la obtenci\'on de una alta calidad de imagen. Se comienza con una revisi\'on te\'orica del problema que representa la turbulencia atmosf\' erica y las aberraciones instrumentales en la calidad de las im\' agenes. Esto plantea la necesidad de implantar t\'ecnicas de restauracion post-facto de las im\'agenes, que sumadas a las correcciones en tiempo real de la Optica Adaptativa, nos den im\'agenes cada vez mas cercanas a la realidad, es decir al \emph{objeto verdadero}, que en nuestro caso es la regi\'on del Sol que queremos estudiar.\\

La forma evidente, aunque no por ello la mas sencilla, de evitar el efecto \mbox{negativo} de la turbulencia atmosf\'erica sobre la calidad de nuestras im\'agenes  es el uso de \mbox{telescopios} espaciales. Fuera de la atm\'osfera terrestre, las observaciones no estar\'ian afectadas por aberraciones atmosf\'ericas. Sin embargo, aun seguir\'ian \mbox{existiendo} aberraciones instrumentales degradando las im\'agenes, aunque con menos intensidad. El problema principal de una misi\'on espacial es su elevado costo
de puesta en \'orbita, mantenimiento y actualizaci\'on.\\

Un esfuerzo por tener una observaci\'on solar, sin el efecto contraproducente de la atm\'osfera terrestre, es el desarrollo de la misi\'on SUNRISE, una colaboraci\'on entre la Agencia Espacial Alemana, DLR, la Estadounidense NASA y el Programa Nacional Espa\~nol del Espacio. Este proyecto lanzar\'a un globo aerost\'atico con un telescopio de 1 metro de apertura que tendr\'a durante 15 d\'ias la posibilidad de observar ininterrumpidamente al Sol con alta resoluci\'on espacial, temporal y espectral sin precedentes. El objetivo principal de SUNRISE es el estudio de la formaci\'on de estructuras magn\'eticas en la atm\'osfera solar y su interacci\'on con los flujos convectivos de plasma. Para cumplir este objetivo se cuenta con el instrumento \emph{Imaging Magnetograph eXperiment} (IMaX), un magnet\'ografo desarrollado enteramente por instituciones espa\~nolas, lideradas por el Instituto de Astrof\'isica de Canarias en el cual he realizado el presente trabajo. Este instrumento ser\'a capaz de producir mapas del campo magn\' etico de regiones extensas de la superficie solar midiendo la polarizaci\' on de la luz en determinadas l\'ineas espectrales. Como miembro del equipo de IMaX, he desarrollado un m\'etodo de calibraci\'on en vuelo para caracterizar las aberraciones que afectar\'an las im\'agenes en IMaX. La descripci\'on del m\'etodo de calibraci\'on, as\'i como tambi\'en las pruebas de su robustez, constituyen el n\'ucleo de la primera parte de esta tesis doctoral.\\ 

En la segunda parte de la tesis nos centramos en el tema del estudio de flujos horizontales en regiones solares activas. Se utilizan datos de observaciones solares desde Tierra y desde el espacio y se aplica el m\'etodo de reconstrucci\'on de im\'agenes expuesto en la primera parte para restaurar el material observado. Estudiamos los movimientos propios de estructuras dentro y fuera de regiones solares activas. A trav\'es de t\'ecnicas de correlaci\'on local y los subsiguientes mapas de flujo que \mbox{generamos}, podemos cuantificar los flujos horizontales en las regiones observadas.\\

La primera regi\'on activa estudiada corresponde a un complejo grupo de manchas solares de configuraci\'on $\delta$. Se infiere el campo de velocidades horizontales sobre de una serie temporal de alta resoluci\'on, y a partir de este mapa de flujos se encuentra una correlaci\'on entre la presencia de flujos de gran velocidad alrededor de las manchas solares hacia afuera y la existencia de penumbra. La zona afectada por estos flujos se denomina \emph{foso} (en ingl\'es \emph{moat}). Se sugiere una relaci\'on entre flujos radiales hacia afuera a lo largo de los filamentos penumbrales (flujo Evershed) y los flujos fotosf\'ericos, tambi\'en radiales, en la granulaci\'on circundante a las manchas solares. Para confirmar este resultado, se estudia una muestra mas amplia de manchas solares con gran variedad de configuraciones penumbrales, y nuevamente se encuentra la misma dependencia foso-penumbra.  En las \'areas donde las umbras son adjacentes a la granulaci\'on circundante, no hay evidencias de presencia de estos flujos a gran escala.\\

Finalmente se estudia el campo de velocidades horizontales alrededor de poros (manchas solares sin penumbra). Se trabaja con im\'agenes restauradas de alta \mbox{resoluci\'on}, que conforman una serie temporal estable y de larga duraci\'on, sobre la cual se estudian nuevamente los movimientos propios en todo el campo de observaci\' on y sus propiedades alrededor de poros solares. Como resultado relevante, no se encuentra ninguna evidencia de \emph{flujo de foso} alrededor de los poros.

%% file: abstract.tex
\chapter*{\sf Abstract}
\dropping[0pt]{2}{T}he present thesis work can be framed in a more general concept designated as "High resolution in solar physics". The work consists of two clearly defined parts. The first part concerning instrumental development for solar observations and the second one devoted to the scientific exploitation of solar data acquired with cutting edge solar instrumentation.\\

The first part of this thesis is dedicated to the topic of high-resolution observations and image restoration to obtain high-quality images. It begins with a theoretical reviewing of the problem that represents the atmospheric turbulence and the instrumental aberrations on the image quality. This problem force us to implement post-facto image restoration techniques that, added to the real-time corrections performed by the Adaptive Optics, gives us images closer to reality, i.e. to the \emph{true object}, that in our case is the region of the Sun we want to study.\\

The most evident solution, although not the easiest one, to get rid of the negative effect of the atmospheric turbulence on the image quality, is the use of space telescopes. Out from the Earth's atmosphere, the observations would not be affected by atmospheric aberrations. Nevertheless, the instrumental errors would still be present degrading the images although with less strength. The counter part of this solution is the elevated costs of launching, maintenance and updating of a space instrument.\\

To have good solar observations overcoming the negative influence of the Earth' s atmosphere, one effort is being made with the development of the so-called SUNRISE mission, a collaboration between the German Space Agency DLR, the north American NASA and the National Spanish Space Program. This project consists in a balloon-borne mission that will launch a 1-meter telescope to the stratosphere and will record data uninterruptedly during 15 days, with unprecedented temporal, spatial and spectral resolution. The main aim of SUNRISE is to study the formation of magnetic structures in the solar atmosphere and their interaction with the convective plasma flows. In order to do so, the on-board instrument called \emph{Imaging Magnetograph eXperiment} (IMaX) has been designed and implemented entirely by Spanish institutions leaded by the Instituto de Astrof\'isica de Canarias in which I have developed the present work. This instrument will be able to produce magnetic field maps of extensive solar regions by measuring the light polarization in certain spectral lines. As a member of the IMaX team, I have been in charge of performing numerical simulations to identify and evaluate possible optical error sources. My main contribution to the project has been the development of an in-flight calibration method to characterize the aberrations affecting the images in IMaX. The description of this calibration method as well as the test to prove its robustness make up the core of the first part of this thesis.\\

The second part of the thesis is centered on the study of horizontal flows in solar active regions. Data from ground-based and space observations are used and the reconstruction techniques explained in the first part are successfully employed to restore the images. We focus on the proper motions of structures in and around solar active regions. The way to quantify the horizontal flows in the field-of-view consist of using local correlation tracking techniques that generate flow maps.\\

The first active region studied corresponds to a very complex sunspot group with $\delta$-configuration. The horizontal velocity field is inferred from a high-quality time series and, from these flow maps, it is found a correlation between the presence of strong velocity flows (moats) surrounding sunspots and the existence of penumbra. A relation between radial outflows along the penumbral filaments (Evershed flow) and the photospheric outflows in the granulation surrounding the sunspot is then suggested. To confirm this result, a larger sample of sunspots with a variety of penumbral configurations is studied and once again the same moat-penumbra dependence is found. In the areas where umbrae are adjacent to the surrounding photospheric granulation there are not evidence of these large-scale outflows. \\

Finally, the horizontal velocity field is studied around pores (sunspots lacking penumbrae). Working with restored high-resolution images conforming a stable and long-duration time series, the analysis of proper motions is performed again within the field-of-view and the properties of the motions around solar pores are studied and quantified in detail. As a relevant result, there are no evidences of moat-like flows around the pores.

%% file: introduction.tex
\chapter*{\sf Introduction}\label{introduction}
\dropping[0pt]{2}{T}he Sun is undoubtedly the most important astronomical object for the human kind. Although its proximity and the fact that it has been largely studied, our knowledge is yet quite poor in understanding how it really works: which are the mechanisms that rule its behaviour, with the eleven years solar cycle, and further more facts concerning the magnetic nature and its implications in the structure of our closest star.\\

The study of the Sun embraces nowadays multiple branches: from the study of the solar interior through helioseismology techniques up to the study of the solar wind that impacts the Earth after traveling a long distance across the emptied space. The majority of advances in solar physics have come thanks to the new telescopes and observational techniques. The quality of the images taken by telescopes has improved considerably in the last decades and nowadays it has almost reached the diffraction limit imposed by the instrument. Computational tools have also taken part in the development of solar physics by improving modelling and simulations which are in their turn corroborated by observations, hereby the importance of reaching high-resolution observing levels.\\

The Sun evolves and changes in different time and spatial scales. Added to the 11 years cycle that marks the biggest changes in the solar magnetic field structure, it also presents more rapid changes that can be seen from Earth. Appearance of sunspots, for instance, was first observed in the XVII century with the use of the telescope in astronomy though there were many records of sunspots observations since antiquity.\\

The aim of this thesis associated to the \emph{Imaging Magnetograph eXperiment (IMaX)} project developed by several Spanish institutions \footnote{\sf Instituto de Astrof\'isica de Canarias (IAC), Instituto de Astrof\'isica de Andaluc\'ia (IAA), Instituto Nacional de T\'ecnica Aeroespacial (INTA), Grupo Astronom\'ia y Ciencias del Espacio (GACE)} has two main branches. Firstly, the initial part of the thesis deals with instrumental aspects, reviewing the latest restoration techniques and concerning the design of a method to calibrate the in-flight instrumental aberrations of IMaX. The second part of the thesis is dedicated to the study of the dynamics of structures in the solar photosphere, in particular in solar active regions. A more detailed information of the contents for each chapter is described below:\\

In {\bf \sf Chapter~\S\ref{cap1}} we start reviewing the theory concerning the image formation which is crucial to understand many concepts developed through all the thesis, such as the degradation of images and the restoration techniques applied to reconstruct them.\\

{\bf \sf Chapter~\S\ref{cap2}} is devoted to the instrument IMaX, describing first the project it is associated to, the objectives of the mission and some important characteristics about it. Moreover, the main concern of this section is the definition of the method for the IMaX in-flight calibration and the subsequent test of the robustness of the method taking into account all possible error sources.\\

{\bf \sf Chapter~\S\ref{cap3}} is the first chapter within the second part of the thesis. It introduces the theory on solar active regions and the current knowledge in the field, i.e. the description of the structures conforming a solar active region, ranging from the widely studied umbra and penumbra of sunspots to the more recently discovered fine structures.\\

In {\bf \sf Chapter~\S\ref{cap4}} we employ the restoration techniques to correct solar data and show the first results yield by the study of horizontal proper motions in and around sunspots. Maps of horizontal velocities are presented for a complex solar active region with a $\delta$-configuration. Some of the results in this chapter have already been published in the \emph{The Astrophysical Journal Letters (ApJL)}.\\

{\bf \sf Chapter~\S\ref{cap5}} extends the study of proper motions to a larger sample of sunspots displaying various penumbral configurations. Most of the results presented in this chapter have already been published in the \emph{The Astrophysical Journal (ApJ)}.\\

{\bf \sf Chapter~\S\ref{cap6}} extends the study of solar active regions done in the previous two chapters by analyzing a sample of solar pores from ground-based and space observations. This chapter is a extended version of the work to be published in the \emph{The Astrophysical Journal (ApJ)}.\\

{\bf \sf Chapter~\S\ref{cap7}} summarizes the conclusions of the work and the final discussion.

\clearemptydoublepage

%% file: chap1.tex
\chapter{\sf Foundations on image restoration}\label{cap1}
\dropping[0pt]{2}{I}n the first part of this thesis I concentrate on the techniques we employ to achieve high-resolution solar images that enable us to study the Sun at very small spatial scales (fine details).

\section{Introduction}
Current problems trying to explain the physics of the Sun require to resolve very tiny structures as the first step to be able to model what is actually happening in different layers and regions in the Sun.\\ 

The Earth's atmosphere can be considered as an isotropic turbulent medium. The quality of the solar images captured by ground-based telescopes is severely affected by the atmospheric turbulence. The image degradation is generally described as the combination of three main contributions, as follows:

\begin{itemize}
\item Structures smearing (\emph{blurring}).
\item Global displacements of the image (\emph{image motion}).
\item Distortion of the structures caused by the differential image motion of different patches in the FOV (\emph{stretching}).
\end{itemize}

All these degradation effects considered together are usually referred to with the term \emph{seeing} and represent the first problem to face if we are interested in high resolution data. \\

There have been recent efforts trying to circumvent the atmospheric influence on the observed images. Space-based telescopes \citep[SOHO]{domingo1995} are undoubtedly the obvious best choice to get rid of the atmospheric effects. We all have got fascinated with the images obtained by space facilities and nowadays we are still getting astonishing data from them \citep[HINODE;][]{kosugi2007}. But the elevated cost of construction and operation of these space solar telescopes has made the scientific community to think about an alternative. \\ 

In recent years and in order to improve the ground-based observations, the Adaptive Optics \citep[AO][]{rimmele2000, scharmer2000}  has made possible to partially correct both, the instrumental and the atmospheric aberrations. The idea is conceptually simple and uses optical elements deforming in real-time to compensate the wavefront aberration induced by the atmosphere and the telescope.\\

Due to the temporal scale in the evolution of the seeing, the AO only pursues low-order corrections that moreover are limited to an isoplanatic patch of a few arc seconds, so that it is compulsory to implement post-facto computational techniques to complement the real-time corrections. Powerful numerical codes for image restoration have been thus developed in the last decade, every single one requiring an especially designed observing strategy.\\

In order to quantify how the aberrations affect the images, it is worth to dedicate the next section to the mathematical formalism describing the image formation in the telescope.

\section{Image formation}
\begin{figure}
\centering
\includegraphics[width=1.0\linewidth]{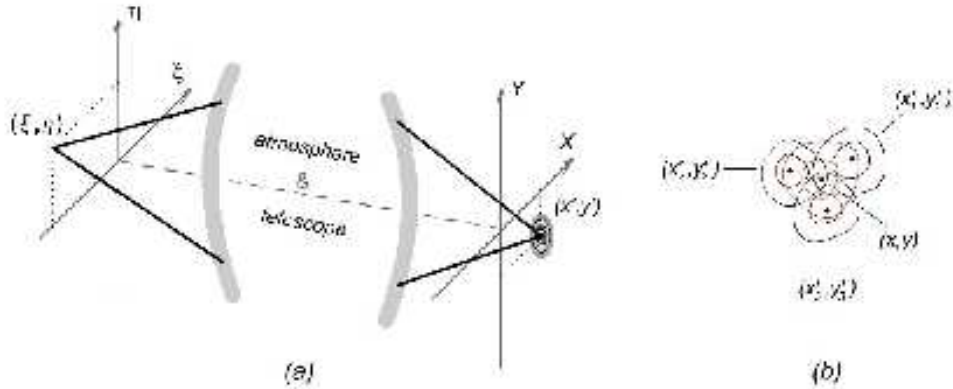} 
\caption[\sf Image formation process]{\sf a) Skecth representing the image formation process, and b) Detail of the image plane showing that the intensity at each point has a contribution from the distribution of intensities centered at points in the neighbourhood.}
\label{formacion}
\end{figure}

The basic process describing the formation of images is shown in Figure~\ref{formacion}a. The object of study is placed on the so-called object plane represented by the coordinates system ($\xi , \eta$), and another coordinate system ($x,y$) called the image plane is situated right in the focal plane of the telescope. The wave coming from a point source in the object plane passes through the atmosphere and part of it enters the telescope and forms the corresponding image (not a point anymore but a spot) centered at position  ($x',y'$)  in the image plane. \\

If the object source is the \emph{impulse unit function} its image corresponds to the \emph{impulse response function} of the optical transmission system (atmosphere + telescope). This response is also named Point Spread Function (PSF) of the system. The PSF can be viewed as the normalized distribution of intensity in the image of a point source and can be expressed as $s(x,y;x',y';t)$, where $t$ represent the variability of the transmission system (e.g.\ atmospheric turbulence evolves at a fast rate) and ($x',y'$) reflects that the PSF is, in general, space variant, i.e.\ the distribution of intensity in the image of a point source changes with its location ($\xi,\eta$) in the object plane. Note that ($x',y'$) is the image conjugated point of ($\xi,\eta$).\\

Considering an extended object like the Sun as composed by multitude of incoherent point sources, with individual intensities $i_{o}(\xi ,\eta)$, we can estimate the resulting intensity distribution in the focal plane of the telescope. Assuming a linear optical system and incoherent illumination, this intensity can be expressed as

\begin{equation}
  i(x,y,t) = \int\hspace{-1.5mm}\int\,i_{o}(x',y')\,s(x,y;x',y';t)\,\mbox{d}x'\,\mbox{d}y',
\label{intensidad}
\end{equation}
\\
\noindent where $i_{o}(x',y')$ represents the distribution of intensity in the ideal image (i.e.\ also in the object) that would produce a perfect system free from aberrations and with infinite aperture. The intensity at each point $(x,y)$ of the image has a contribution from the images centered at points in the neighbourhood, e.g.\ $(x_1^{\prime},y_1^{\prime}), (x_2^{\prime},y_2^{\prime}), (x_3^{\prime},y_3^{\prime})$ in Figure ~\ref{formacion}b~.

\vspace{3mm}

 For an isoplanatic  system having a spatially invariant PSF, i.e.\\
  $s(x,y;x',y';t) = s(x-x',y-y',t)$, the equation  \ref{intensidad} can be written as

\begin{eqnarray}
i(x,y,t) & = & \int\hspace{-1.5mm}\int\,i_{o}(x',y')\,s(x-x',y-y',t)\,\mbox{dx}'\,\mbox{dy}', \nonumber\\ & 
= & \int\hspace{-1.5mm}\int\, i_o(\mbox{\boldmath $q$}',t)\,s(\mbox{\boldmath $q$}-\mbox{\boldmath $q$}',t)\,\mbox{d}\mbox{\boldmath $q$}=i_{o}(\mbox{\boldmath $q$})\ast s(\mbox{\boldmath 
$q$},t),
\label{intensidadiso}
\end{eqnarray}

\noindent where $\ast$ stands for convolution and $\mbox{\boldmath $q$}$ is the vectorial notation for the coordinates in the image points. Using the convolution theorem \footnote{\sf The convolution theorem states that the Fourier transform of the convolution of two functions is the product of their respective Fourier transforms.}, the intensity can be expressed in the Fourier domain as the product

\begin{equation}
  I(\mbox{\boldmath $u$},t) = I_{o}(\mbox{\boldmath $u$})\,S(\mbox{\boldmath 
$u$},t),
\label{intensidadfourier}
\end{equation}

\noindent where capital letters stand for the Fourier transforms of the functions in lowercase,  $\mbox{\boldmath $u$}$ is the frequency vector in the Fourier domain and, $S$ is the so-called Optical Transfer Function (OTF) of the system. The amplitude of the OTF is defined as the Modulation Transfer Function (MTF): $\mbox{MTF}(\mbox{\boldmath $u$},t)=|\mbox{OTF}(\mbox{\boldmath $u$},t)|$. According to equation~\ref{intensidadfourier} the MTF is a filter that attenuates the amplitudes of the Fourier components of the ideal image to form the observed image.

\section{Image restoration as a particular case of the Inverse \mbox{Problem} in Physics}

Image restoration fits into the format of the \emph{Inverse Problem} in Physics\footnote{\sf The inverse problem frequently occurs in different branches of science where the values of some model parameters must be obtained from the observed data.} which, in general, can be considered to as the solution $f(\mbox{\boldmath $q$}')$ of the \emph{Fredholm inhomogeneous integral equation of the $1^{st}$ kind}

\begin{equation}
g(\mbox{\boldmath $q$}) = \int\hspace{-1.5mm}\int {\cal K}(\mbox{\boldmath $q$},\mbox{\boldmath $q$}') \, f(\mbox{\boldmath $q$}') \, \mbox{d}(\mbox{\boldmath $q$}'),
\label{fredholm}
\end{equation}

\noindent where  ${\cal K}(\mbox{\boldmath $q$},\mbox{\boldmath $q$}')$ is known as the kernel of the integral equation. Equation~\ref{intensidadiso} is a particular case of equation~\ref{fredholm}, where the kernel is the PSF.\\

The inverse problem in our particular case is usually called \emph{image reconstruction} or \emph{image restoration} since the target is to achieve an estimate of the ideal image or equivalently of the \emph{true object} starting from a \emph{degraded image}. Because in our particular case the \emph{inverse problem} can be formulated as a convolution equation (eq.~\ref{intensidadiso}), image restoration can also be referred to as a \emph{deconvolution problem}, the formal solution of which can be expressed from equation~\ref{intensidadfourier} as follows,

\begin{equation}
I_{o}(\mbox{\boldmath $u$})= \frac{I(\mbox{\boldmath $u$},t)}{S(\mbox{\boldmath $u$},t)}, 
\label{io_fourier}
\end{equation}

\noindent or in the measuring domain as

\begin{equation}
\hat{i}_o(\mbox{\boldmath $q$})= {\cal F}^{-1} \left [ I(\mbox{\boldmath $u$},t)\over S(\mbox{\boldmath $u$},t) \right ].
\label{i_measure}
\end{equation}

The symbol ~$\hat{}$~ over $i_0$ means that the restoration will not be complete in any case since the transmission system operates as a low-pass spatial frequency filter with a given cut-off. In other words, the restoration problem will render an estimate of the \emph{true object}.\\
 
In order to compute $\hat{i}_o(\mbox{\boldmath $q$})$ it is mandatory to characterize the PSF (or its Fourier transform the OTF) describing the optical system. Several approaches to the PSF determination lead to different numerical methods for the restoration of solar images \citep{bonet1999}.\\

\subsection{Noise contribution}

An additional difficulty in the inversion problem arises from the fact that in the real case, the observed image is affected by noise caused by different sources being the readout and the photon noise the more relevant components. Though the latter is proportional to the square root of the number of photons in the incoming signal, in most solar physics applications the assumption of uncorrelated signal and noise gives good results. In this manner, equation~\ref{intensidadiso} can be completed including the noise as and additive contribution, and reformulated as\footnote{\sf Note that the variable t is dropped from the formulae describing the image formation in order to shorten the notation. However, one has to keep in mind that the formulae describe instantaneous events.}

\begin{equation}
  i_{n}(\mbox{\boldmath $q$}) = i_{o}(\mbox{\boldmath $q$})\ast 
s(\mbox{\boldmath $q$}) + n(\mbox{\boldmath $q$}),
\label{intensidadcorta}
\end{equation}

\noindent and in the Fourier transformed domain

\begin{equation}
  I_{N}(\mbox{\boldmath $u$}) = I_{o}(\mbox{\boldmath $u$})\, S(\mbox{\boldmath 
$u$}) + N(\mbox{\boldmath $u$}) = I(\mbox{\boldmath $u$}) + N(\mbox{\boldmath 
$u$}).
\label{intcortafourier}
\end{equation}

The restoration based in formula~\ref{io_fourier} leads to

\begin{equation}
  {I_{N}(\mbox{\boldmath $u$})\over S(\mbox{\boldmath $u$})} = 
{I(\mbox{\boldmath $u$})\over S(\mbox{\boldmath $u$})} + {N(\mbox{\boldmath 
$u$})\over S(\mbox{\boldmath $u$})},
\label{intensidadrest}
\end{equation}

\noindent where the term $\displaystyle{N(\mbox{\boldmath $u$})/S(\mbox{\boldmath $u$})}$ represents a noise amplification, making compulsory to pursue a noise filtering, previous to the restoration process. 

\subsection{Noise filtering}

Filtering of the noisy signal is standardly done by using the so-called \emph{optimum filter}  \mbox{$\Phi_N(\boldmath u)$}, described for 1D problems by \cite{brault1971}, which is a real function that weights the diverse Fourier spectral components  according to the noise level at each frequency.

\begin{equation}
I_{F}(\mbox{\boldmath $u$}) = I_{N}(\mbox{\boldmath $u$}).\Phi_N(\mbox{\boldmath $u$}) = \bigl[ I(\mbox{\boldmath $u$}) + 
N(\mbox{\boldmath $u$})\bigr] \Phi_N(\mbox{\boldmath $u$}).
\label{filtrado}
\end{equation}

The \emph{optimum filter} is formulated as

\begin{equation}
  \Phi_N(\mbox{\boldmath $u$}) = {\vert I(\mbox{\boldmath $u$})\vert^{2}\over
\vert I(\mbox{\boldmath $u$})\vert^{2}+\vert N(\mbox{\boldmath $u$})\vert^{2} }.
\label{filtro}
\end{equation}

Combining the noise filtering together with the deconvolution, we eventually find the optimum filter for restoration $\Phi_R(\mbox{\boldmath $u$})$, better known as the \emph{Wiener-Helstrom} filter \footnote{\sf Named after the optimal estimation theory of Norman Wiener, this filter simply acts separating signals based on their frequency spectra. The gain of the filter at each frequency is determined by the OTF of the system and the relative amount of signal and noise at that frequency.} 

\begin{equation}
  \Phi_{\rm R}(\mbox{\boldmath $u$}) = {\vert I(\mbox{\boldmath 
$u$})\vert^{2}\over \vert I(\mbox{\boldmath $u$})\vert^{2}+\vert 
N(\mbox{\boldmath $u$})\vert^{2} }\cdot {1\over S(\mbox{\boldmath $u$})}={S^*(\mbox{\boldmath $u$})\over 
\mbox{MTF}^2(\mbox{\boldmath $u$})+1/\mbox{SNR}(\mbox{\boldmath $u$})},
\label{wiener}
\end{equation}

\noindent where SNR$(\mbox{\boldmath $u$})=\vert I_{o}(\mbox{\boldmath 
$u$})\vert^{2}/\vert N(\mbox{\boldmath $u$})\vert^{2}$ is the signal-to-noise ratio (SNR).  Since we do not know \emph{a priori} the function $ \displaystyle \vert I_{o}(\mbox{\boldmath 
$u$})\vert^{2}$, some models for SNR$(\mbox{\boldmath $u$})$ are commonly assumed \citep{collados1986}. Hereafter the superscript $^*$ stands for complex conjugate.

\section{The Phase Diversity Technique for image restoration}
\label{PD}

\begin{figure}
\centering
\includegraphics[width=0.7\linewidth]{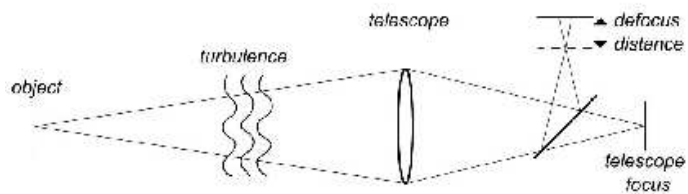} 
\caption[\sf Sketch of the phase diversity technique implementation]{\sf Sketch of a possible setup for the implementation of the phase diversity technique.}
\label{pd}
\end{figure}

The Phase Diversity (PD) technique was first proposed by \citet{gonsalves1979} as a new method to infer phase aberrations working with images of extended incoherent objects formed through an optical system. This technique of image reconstruction \citep[see also ][]{gonsalves1982, paxman1992} requires to use at least  two images of the object we want to reconstruct. One of the images is the conventional one degraded by an unknown aberration (atmosphere and telescope) and the second one is a strictly simultaneous image of the same object affected by the same unknown aberration plus a known intentionally induced aberration. Figure~\ref{pd} shows an optical setup where we induce a known amount of defocus by displacing backwards the camera with respect to the nominal focus.

\vspace{3mm}

The system of equations that mathematically describes the formation of both images can be expressed as

\begin{equation}
\left. \begin{array}{c}
   i_{1}(\mbox{\boldmath $q$}) = i_{o}(\mbox{\boldmath $q$})\ast 
s_{1}(\mbox{\boldmath $q$}) + n_{1}(\mbox{\boldmath $q$}), \\ \\
   i_{2}(\mbox{\boldmath $q$}) = i_{o}(\mbox{\boldmath $q$})\ast
s_{2}(\mbox{\boldmath $q$}) + n_{2}(\mbox{\boldmath $q$}), \\
   \end{array} \right\}
\label{pdformulacion}
\end{equation}

\noindent where :\\\\
\begin{tabular}{ll}
$i_{1}(\mbox{\boldmath $q$}),i_{2}(\mbox{\boldmath $q$})$ & is the focus-defocus image pair.\\
$s_{1}(\mbox{\boldmath $q$}), s_{2}(\mbox{\boldmath $q$})$ & are the corresponding PSF's for both channels.\\
$n_{1}(\mbox{\boldmath $q$}),n_{2}(\mbox{\boldmath $q$})$ & are the noise additive terms.\\
$i_{o}(\mbox{\boldmath $q$})$ & is the so-called \emph{true object}.\\
\end{tabular}

\vspace{4mm}
If we do not consider the noise terms the system above is determined since it has two equations and two unknowns: $i_{o}(\mbox{\boldmath $q$})$ and  $s_{1}(\mbox{\boldmath $q$})$. Note that $s_{1}(\mbox{\boldmath $q$})$ and $s_{2}(\mbox{\boldmath $q$})$ differ in a well-known  defocus value and consequently are analytically related.\\

Nevertheless, the noise terms force a statistical solution of the problem. ~\cite{paxman1992} propose a least squares solution for the system of equations. For Gaussian noise, they derived the following error metric in the Fourier domain to be minimized

\begin{equation}
L(I_{o},S_1) = \sum_{\mbox{\boldmath $u$}} \Bigl\{ 
\big\vert I_{1}(\mbox{\boldmath $u$}) - {I}_{o}(\mbox{\boldmath 
$u$}){S}_{1}(\mbox{\boldmath $u$}) \big\vert ^{2} + \gamma \, \big\vert 
I_{2}(\mbox{\boldmath $u$}) - {I}_{o}(\mbox{\boldmath $u$}) 
{S}_{2}(\mbox{\boldmath $u$}) \big\vert ^2 \Bigr\},
\label{minfourier}
\end{equation}

\noindent where capital letters stand for the Fourier transforms of the functions in lowercase in the system of equations~\ref{pdformulacion},  $\gamma = {\sigma}_{1}^{2} /{\sigma}_{2}^{2}$  is introduced by \cite{lofdahl1994} as a weighting factor to equalize the noise contributions for the case when the noise variances  ${\sigma}_{1}^{2}$ and ${\sigma}_{2}^{2}$  are not the same in both images. $S$ represents the OTF of the system which can be derived as the auto-correlation of the so-called \emph{generalized pupil function} $H$, 
 
\begin{equation}
S(\mbox{\boldmath $u$}) \propto \int\hspace{-1.5mm}\int H(\mbox{\boldmath $r$}) . H^{*}(\mbox{\boldmath $r$} -\lambda f \mbox{\boldmath $u$}) \mbox{d} \mbox{\boldmath $r$},
\label{OTFS}
\end{equation}

\noindent where $\lambda$ and $f$ stand for the working wavelength and the effective focal length of the optical system, respectively, and  \mbox{\boldmath $u$} is a vector with dimensions of spatial frequency. $H$, in turn, can be expressed in terms of the joint phase aberration $ \phi(\mbox{\boldmath $r$})$ caused by the telescope and the turbulence as

\begin{equation}
H(\mbox{\boldmath $r$}) = |H(\mbox{\boldmath $r$})| e^{i \phi(\mbox{\boldmath $r$})}.
\label{OTF_H}
\end{equation}

The phase aberration can be parametrized by using a Zernike polynomials expansion (see Appendix A)
 
 \begin{equation}
 \phi(\mbox{\boldmath $r$})=\phi(R\mbox{\boldmath $\rho$})= \sum \alpha_j Z_j(\mbox{\boldmath $\rho$}),
 \end{equation}
 
\noindent where $R$ is the pupil radius, $\mbox{\boldmath $\rho$}$ is the vectorial notation for $(\rho,\theta)$ with $\rho$ defined in a unit circle $(0 \le \rho \le1)$ and $\theta$ being the azimuthal angle.\\

Thus, equation~\ref{OTF_H} can be then written as

\begin{equation}
H(R\mbox{\boldmath $\rho$}) = |H(R\mbox{\boldmath $\rho$})| e^{i \left [ \delta(R\mbox{\boldmath $\rho$}) + \sum_{j=1}^J \alpha_j Z_j (\mbox{\boldmath $\rho$}) \right ]},
\label{OTFS2}
\end{equation}

\noindent where $\mbox{\boldmath $\alpha$} \equiv \left\{ \alpha_{j}, j=1,2,...,J \right\}$ are the coefficients of the terms in the Zernike expansion and $\delta(R\mbox{\boldmath $\rho$})$ is the diverse phase, i.e.\ the induced defocus in our case. The final result is that the OTF (equation~\ref{OTFS}) and therefore the error metric $L$ (equation~\ref{minfourier}) are parametrized by the expansion in Zernike polynomials, so that one can write: $L(I_o,\mbox{\boldmath $\alpha$})$. The diverse phase $\delta(R\mbox{\boldmath $\rho$})$ will be zero in the case of $S_1$ that corresponds to the focus image.\\

Part of the minimization of the equation \ref{minfourier} can be performed analytically. The solution of the equation  $\displaystyle {\partial L/\partial {I_o}} = 0$ yields an object estimate $\hat{I}_{o}(\mbox{\boldmath $u$})$ that minimizes the equation \ref{minfourier}, 

\begin{equation}
   \hat{I}_{o}(\mbox{\boldmath $u$}) = {I_{1}(\mbox{\boldmath 
$u$})\,{S}_{1}^{*}(\mbox{\boldmath $u$},\mbox{\boldmath $\alpha$}) + \gamma \, 
I_{2}(\mbox{\boldmath $u$})\,{S}_{2}^{*}(\mbox{\boldmath $u$},\mbox{\boldmath 
$\alpha$}) \over \vert {S}_{1}(\mbox{\boldmath $u$},\mbox{\boldmath $\alpha$}) 
\vert ^2 + \gamma \, \vert {S}_{2}(\mbox{\boldmath $u$},\mbox{\boldmath 
$\alpha$}) \vert ^2}.
\label{io}
\end{equation}

Replacing $\hat{I}_{o}(\mbox{\boldmath $u$})$ in the equation  \ref{minfourier}, we obtain a so-called \emph{modified error metric}

\begin{equation}
   L_{M}(\mbox{\boldmath $\alpha$}) = \sum_{\mbox{\boldmath $u$}} {\big\vert 
I_{1}(\mbox{\boldmath $u$})\,{S}_{2}(\mbox{\boldmath $u$},\mbox{\boldmath 
$\alpha$}) - I_{2}(\mbox{\boldmath $u$})\,{S}_{1}(\mbox{\boldmath 
$u$},\mbox{\boldmath $\alpha$})\big\vert^2 \over \vert {S}_{1}(\mbox{\boldmath 
$u$},\mbox{\boldmath $\alpha$}) \vert ^2 + \gamma \, \vert 
{S}_{2}(\mbox{\boldmath $u$},\mbox{\boldmath $\alpha$}) \vert ^2}.
\label{lm}
\end{equation}

Note that this modified metric is not explicitly depending on the Fourier transform of the object $\hat{I}_{o}(\mbox{\boldmath $u$})$ 
but only includes the unknown  \mbox{\boldmath $\alpha$} vector. By means of non-linear optimization techniques, we find the \mbox{\boldmath $\alpha$} vector, characterizing the aberration components, that minimizes the equation \ref{lm}. Once these components are determined, $\hat{S}_{1}$ and $\hat{S}_{2}$ can be calculated from equations~\ref{OTFS2} and \ref{OTFS} and the object estimate can be eventually derived from equation \ref{io}, completing in this way the restoration process.\\

In conditions of poor seeing, the restored images are sometimes contaminated by some artifacts like periodic strips or other regular patterns. This is a consequence of having zeros or quasi-zeros in the OTFs at some specific spatial frequencies which produce a poor SNR at these frequencies. Thus, the division in equation~\ref{io} by the squares of these nearly zero values cause excessive amplification of some spectral components. To circumvent this drawback, a sort of speckle summation of various realizations close in time has been successfully used. This summation carries out compensation for spectral information gaps in the Fourier domain. From equation~\ref{io} and summing up $K$ realizations, one can easily derive

\begin{equation}
\sum_{k=1}^K \biggl ( I_{k1}(\mbox{\boldmath $u$})S_{k1}^{\ast}(\mbox{\boldmath $u$},\mbox{\boldmath 
$\alpha$})+\gamma I_{k2}(\mbox{\boldmath $u$})S_{k2}^{\ast}(\mbox{\boldmath $u$},\mbox{\boldmath 
$\alpha$}) \biggr ) = \hat{I}_{o}(\mbox{\boldmath $u$}) \sum_{k=1}^K \biggl ( |S_{k1}(\mbox{\boldmath $u$},\mbox{\boldmath 
$\alpha$})|^2+\gamma |S_{k2}(\mbox{\boldmath $u$},\mbox{\boldmath 
$\alpha$})|^2 \biggr ),
\end{equation}

\noindent where $\hat{I}_{o}(\mbox{\boldmath $u$})$ is not affected by the $k$-subscript because it must be unique, i.e.\ the estimate of the true signal that we are seeking\footnote{\sf For more detailed information on the PD-method see e.g.\ \cite{bonet2003}; \cite{criscuoli2005} and references therein.}.

\section{Multi-Object Multi-Frame Blind Deconvolution}
\label{S:MOMFBD}

The MFBD (Multi-Frame Blind Deconvolution) is a restoration method that uses multiple frames, bringing in such a way complementary information to recover the aberrations affecting the images \citep[MFBD;][]{lofdahl02MFBD, lofdahl1996}. The method generally works in a better way when the contrast is high, the exposure time is short and the noise is low.\\

An extension of the MFBD, named Multi-Object Multi-Frame Blind Deconvolution  \citep[MOMFBD;][]{momfbd2005}, has made possible to use, apart from multiple frames, also multiple objects simultaneously observed, to restore the images in a more efficient way. By different objects we mean the same field-of-view in the Sun but observed in different wavelengths (within a rather narrow spectral range). For each object, PD focus-defocus pairs can also be included. The observations in all channels must be simultaneous so that we can assure a common atmospheric aberration. The joint restoration of several objects has also the advantage that almost perfect alignment can be achieved between all of them.\\

\begin{figure}
\centering
\includegraphics[width=.8\linewidth]{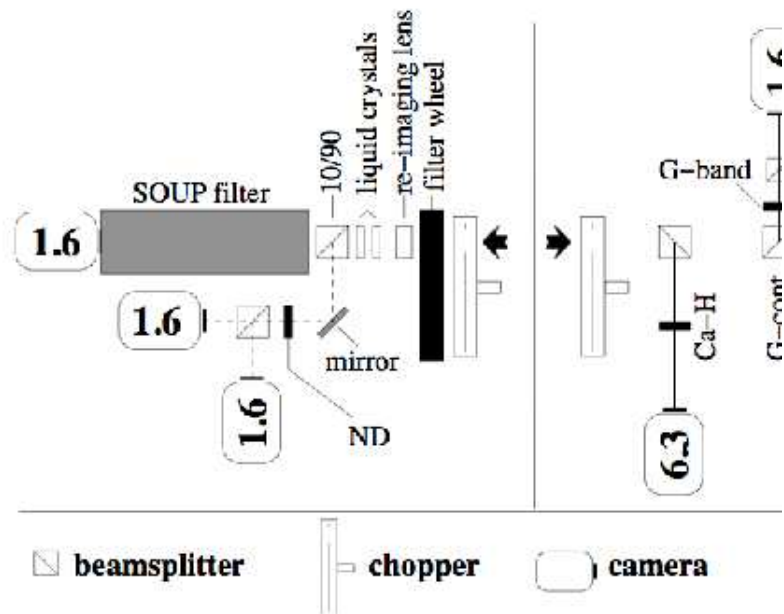} 
\caption[\sf Optical setup for the red and blue beam at the SST]{\sf Optical setup for the red (\emph{left}) and the blue (\emph{right}) beam at the SST showing the configuration of the optical elements and cameras in the proper disposition to pursue corrections with MOMFBD. Arrows mark the direction of the incoming light. Cameras marked as 1.6, 4.2, and 6,3 indicate the number of mega-pixels.  ND stands for a neutral density filter placed along the red beam to balance the exposure level. The SOUP filter is a tunable Lyot filter with 72 m\AA ~ bandpass for the Fe I 6302 \AA ~ line and 128 m\AA ~ bandpass for H$\alpha$. \emph{Taken from \citet{momfbd2005}}.}
\label{MFBDsetup}
\end{figure}

Figure~\ref{MFBDsetup} shows a sketch of the optical setup at the Swedish Solar Telescope (Observatorio Roque de los Muchachos). The MOMFBD technique is applied independently to the \emph{blue} and the \emph{red} channels. The G-band and G-cont beams in the figure are splitted into two channels (focus-defocus) that produce PD image-pairs.  Red channel has also its own PD-pair \footnote{\sf The detailed description of the setup will be commented in the next chapter when dealing with the observations we pursued in different campaigns.}.\\

Throughout this thesis work we employ all the restoration methods described above (i.e.\ PD, MFBD and MOMFBD) depending on the particular circumstances for each situation. The high-spatial-resolution achieved by using these methods is evident from the results shown in all the different chapters.

.

\clearemptydoublepage

%% file: chap2.tex
\chapter{\sf In-flight calibration of IMaX aberrations}
\label{cap2}

\dropping[0pt]{2}{O}ne of the main concerns of this thesis has an instrumental nature. This chapter is devoted to present the assignments I have developed in this field as a member of the \emph{Imaging Magnetograph eXperiment} (IMaX) project at the IAC in Tenerife. Initially, a brief preliminary part is presented to introduce the balloon-borne SUNRISE project conceived to pursue high-resolution solar observations and from which IMaX is part of. Afterwards, I will introduce the IMaX concept. Finally, I will explain with great detail the specific part I have performed within the IMaX project that consists in designing a robust method to calibrate the in-flight instrumental aberrations of IMaX.\\

\section{SUNRISE}
\begin{figure}
\begin{center}
\includegraphics[width=0.38\linewidth]{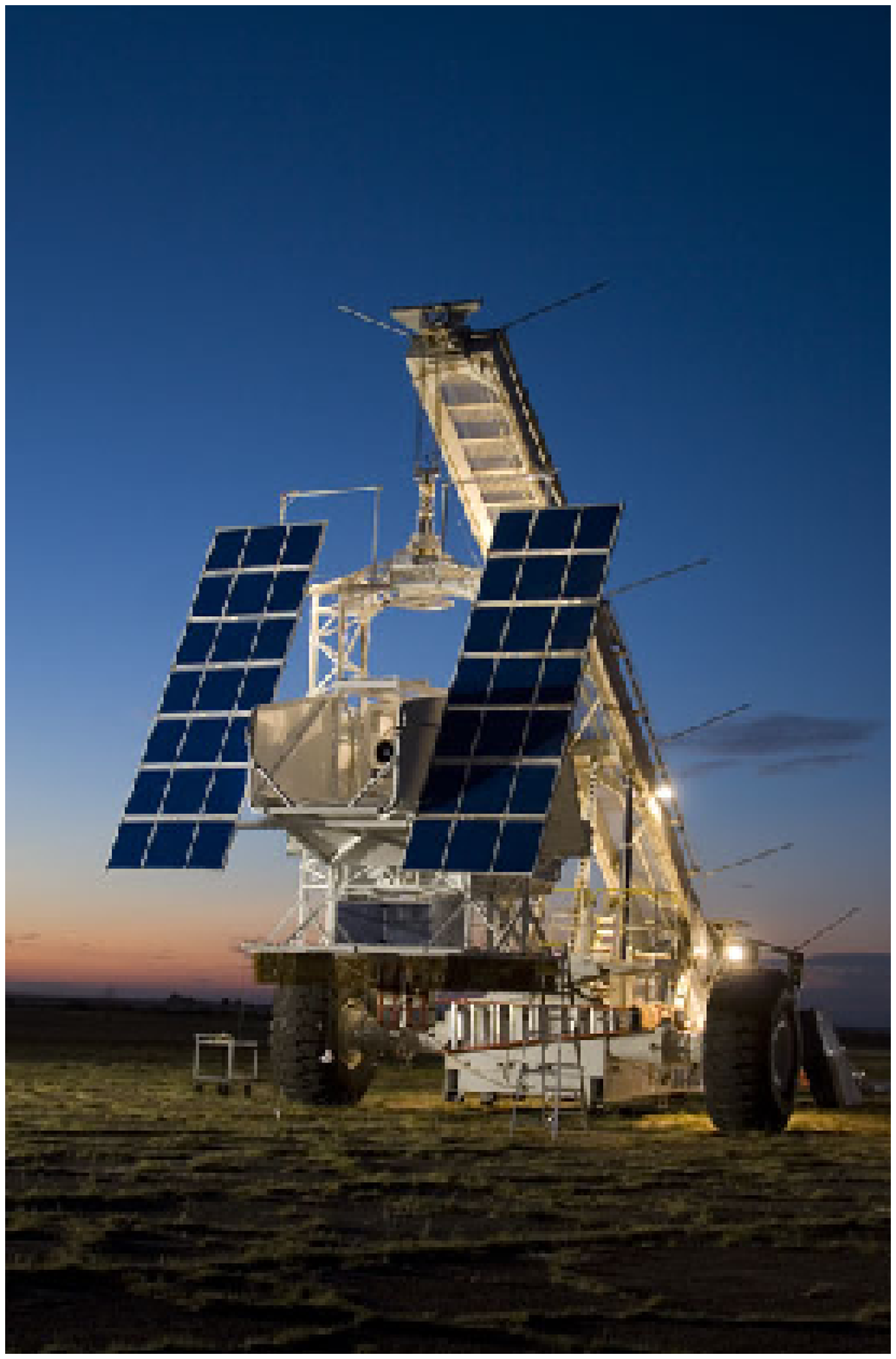}~~\includegraphics[width=0.387\linewidth]{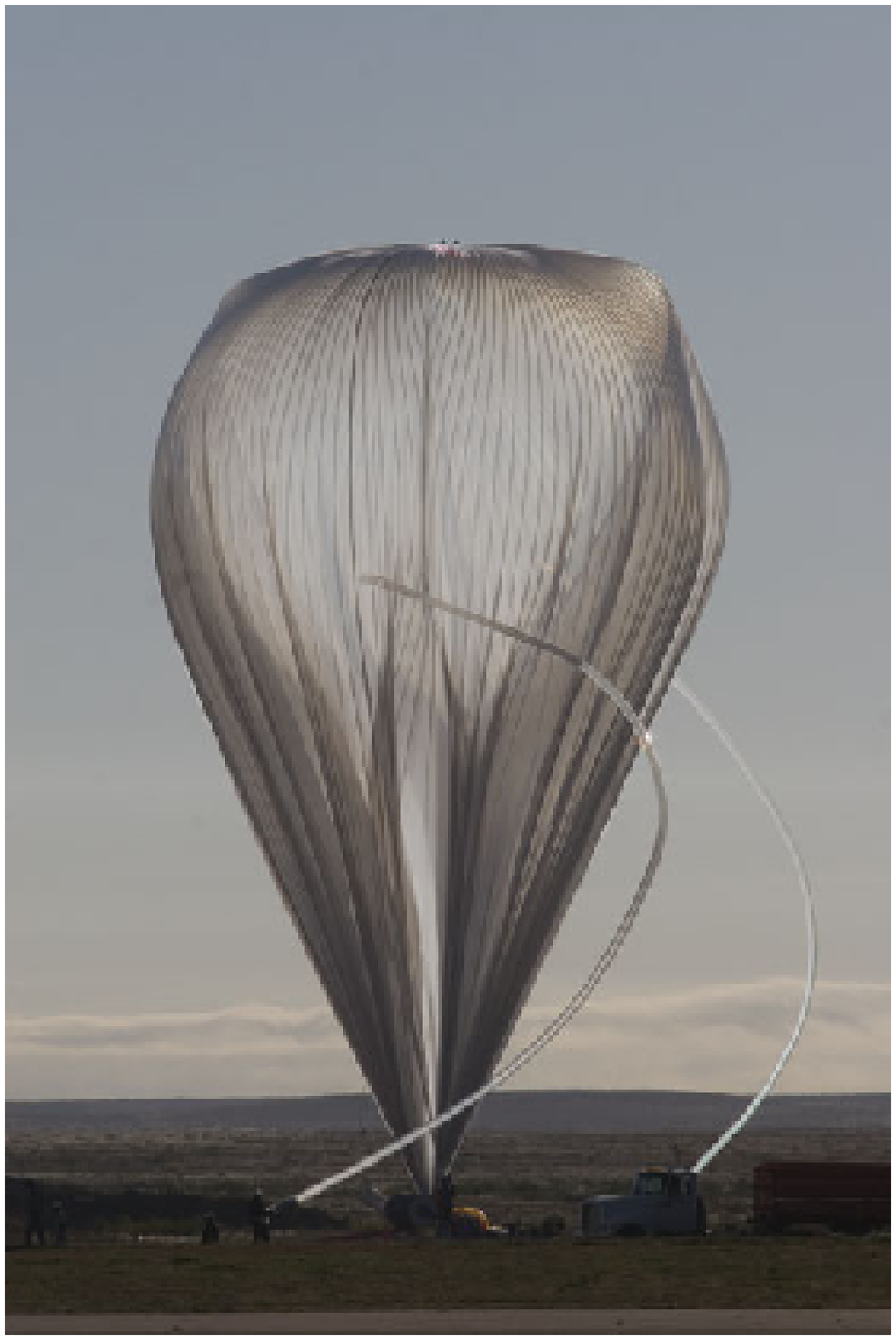}
\caption[\sf SUNRISE testing flight]{\sf SUNRISE instrument in full flight configuration. Photo taken by Carlye Calvin at the Columbia Scientific Balloon Facility in Fort Summer, New Mexico during the test flight on October 3, 2007. \emph{Courtesy SUNRISE team.}}
\label{sunrise}
\end{center}
\end{figure}

\begin{figure}
\begin{center}
\includegraphics[width=0.63\linewidth]{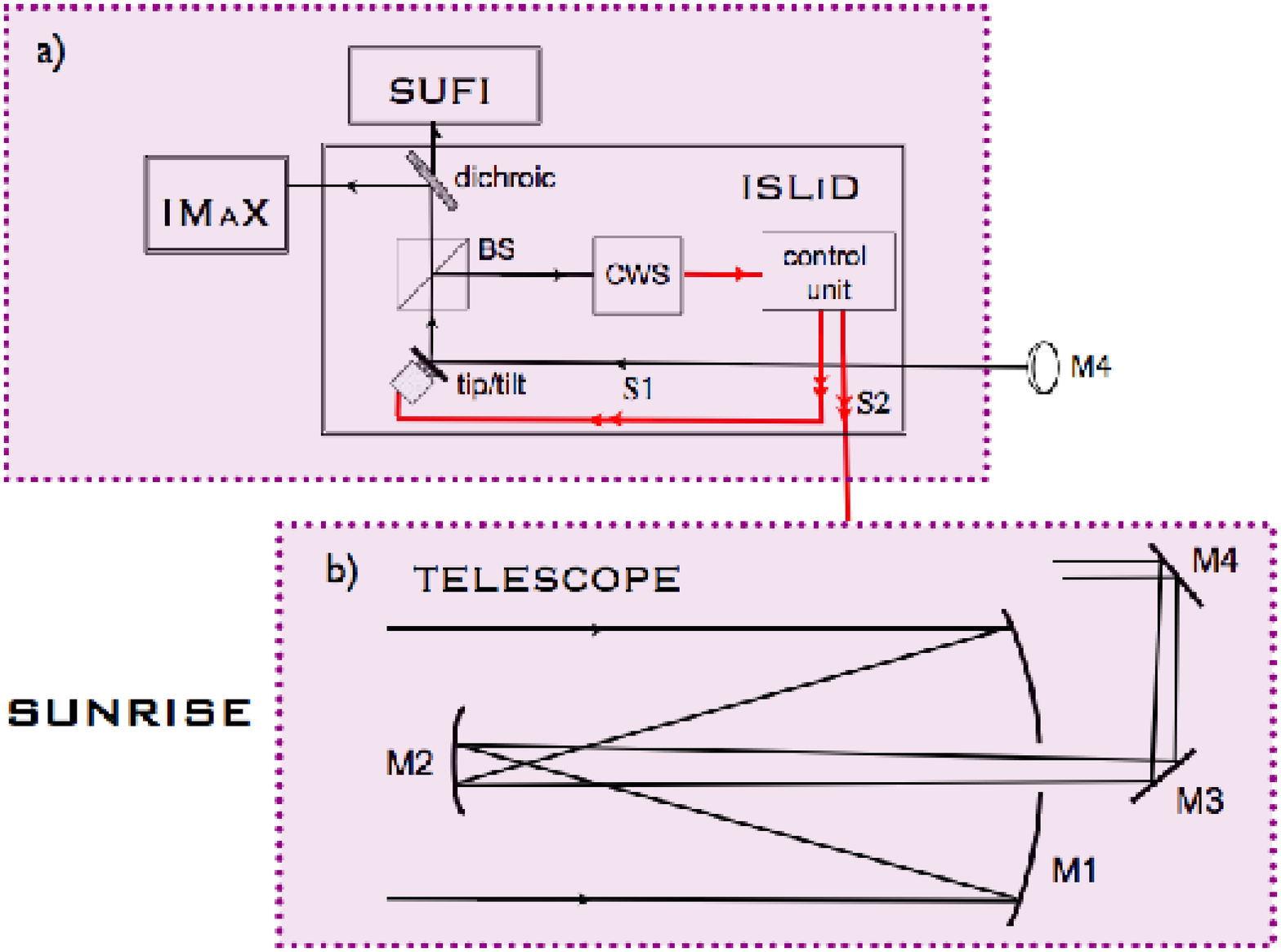}
\caption[\sf Box diagram of SUNRISE project]{\sf Box diagram showing the SUNRISE configuration. a) Light distribution system (ISLiD) and the two focal instruments (IMaX and SUFI). b) Telescope and folding mirrors. }
\label{sunrisebox}
\end{center}
\end{figure}

The SUNRISE project consists of a balloon-borne 1-m aperture solar telescope (Figure~\ref{sunrise}) that aims at high-resolution spectro-polarimetric observations of the solar atmosphere, to be flown in the framework of NASA's LDB (Long Duration Balloon) program in 2009, in a series of flights on circumpolar trajectories at a float altitude of 35~-~40 km. The main goal of the project is understanding the formation  of magnetic structures in the solar atmosphere and their interaction with the plasma convective flows. The SUNRISE equipment consist of the main telescope feeding two focal-plane instruments through a light distribution system. These instruments have been developed by several PI institutions \footnote{\sf Max Planck Institute for Solar System Research (MPS), Instituto de Astrof\'isica de Canarias (IAC), Kiepenheuer-Institut f\"ur Sonnenphysik (KIS), High Altitude Observatory (HAO), Lockheed-Martin Solar and Astrophysics Laboratory (LMSAL).}:  
\begin{itemize}
\item Telescope: Gregorian design. Carbon fiber based telescope structure with 1m Schott Zerodur lightweighted primary mirror.
\item ISLiD: Image Stabilisation and Light Distribution System that ensures capability of simultaneous observations with all science instruments, based on all-dielectric dichroic beam splitters (see Figure~\ref{sunrisebox}). An important part of the ISLID is the so-called Correllator and Wavefront Sensor (CWS) system in charge of image stabilisation and correction of optics misalignments and defocus. 
\item SUFI: The Sunrise Filter Imager is a filtergraph for high-resolution images in the visible and UV spectral ranges.
\item IMaX: Imaging Magnetograph eXperiment, is a magnetograph providing fast-cadence two-dimensional maps of the complete magnetic field vector and the line-of-sight velocity as well as white-light images with high spatial resolution.
\end{itemize}
With all this equipment, SUNRISE will provide spectra and images resolving spatial scales down to 35 and 70 km on the Sun in the ultra-violet and visible ranges, respectively. Figure~\ref{sunrisebox} sketches the concept of the SUNRISE project. The main telescope in panel (b) puts the light beam into the optical bench in panel (a) being M4 the interface between both subsystems. The light beam is folded by the tip/tilt mirror and afterwards splitted into two channels, one feeding the CWS for wavefront sensing and the other reaching a dichroic plate that separates the specific wavelength for SUFI and IMaX, respectively. The CWS generates electric signals: S1 for steering the tip/tilt mirror and S2 to control M2 in order to correct for defocus and optical misalignments.

\section{The Imaging Magnetograph eXperiment}
\label{sectionimax}

As briefly explained in the last section IMaX is one of the instruments part of the payload of the SUNRISE balloon. It will allow to study the dynamics and evolution of the solar magnetic field as well as its interaction with the plasma, with high temporal, spatial and spectral resolutions and unprecedented polarimetric sensitivity. This instrument will then provide magnetograms of extended solar regions by combining high temporal cadence and polarimetric precision while preserving the bidimensional integrity of the images. To meet this goal IMaX has to work as a:

\begin{itemize}
\item High-efficient image acquisition system.
\item Near diffraction limited imager.
\item High resolving power spectrograph.
\item High sensitivity polarimeter.
\end{itemize}

\begin{figure}
\begin{center}
\hspace{-0.3cm}\includegraphics[angle=-90,width=0.48\linewidth]{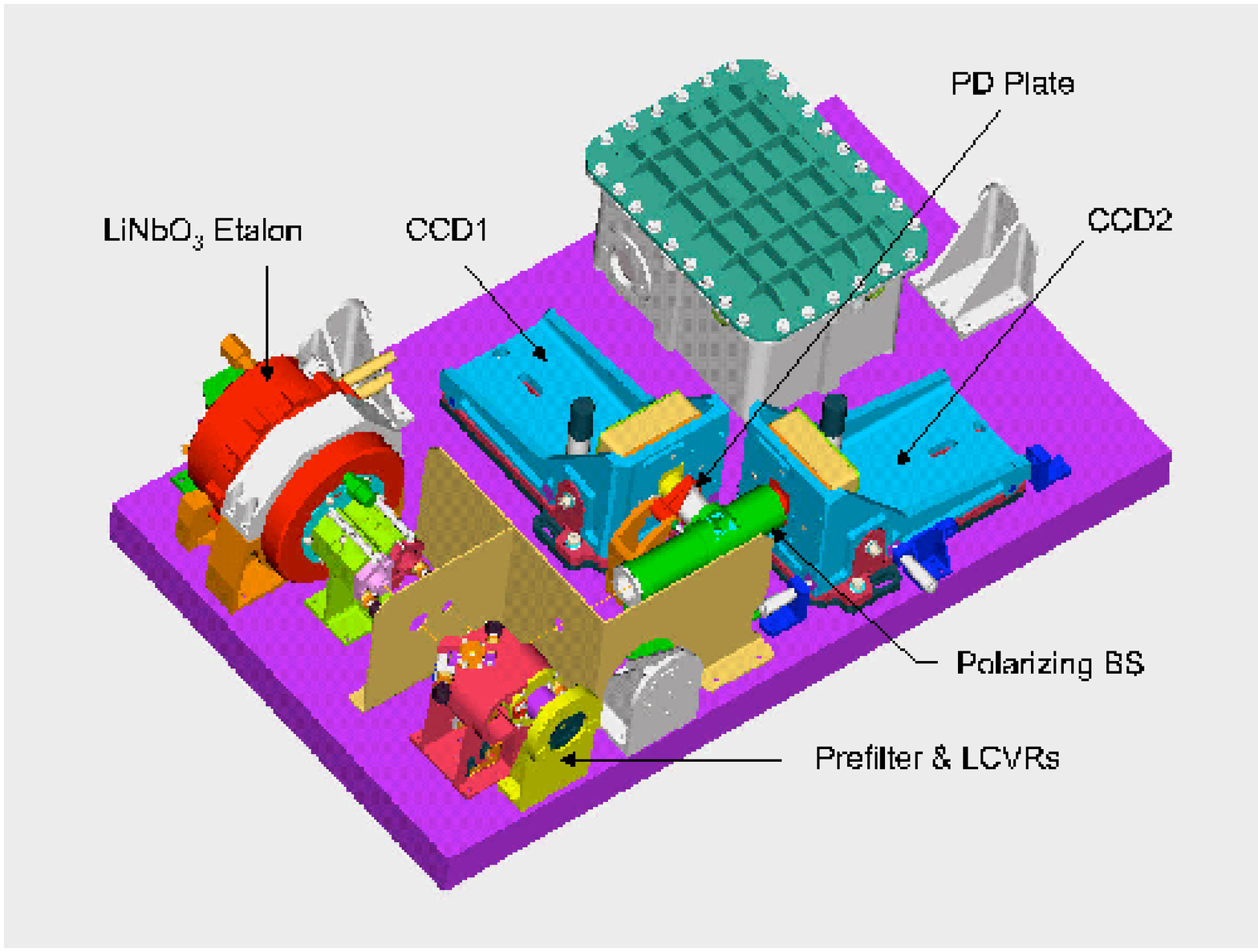}~\includegraphics[angle=-90,width=0.487\linewidth]{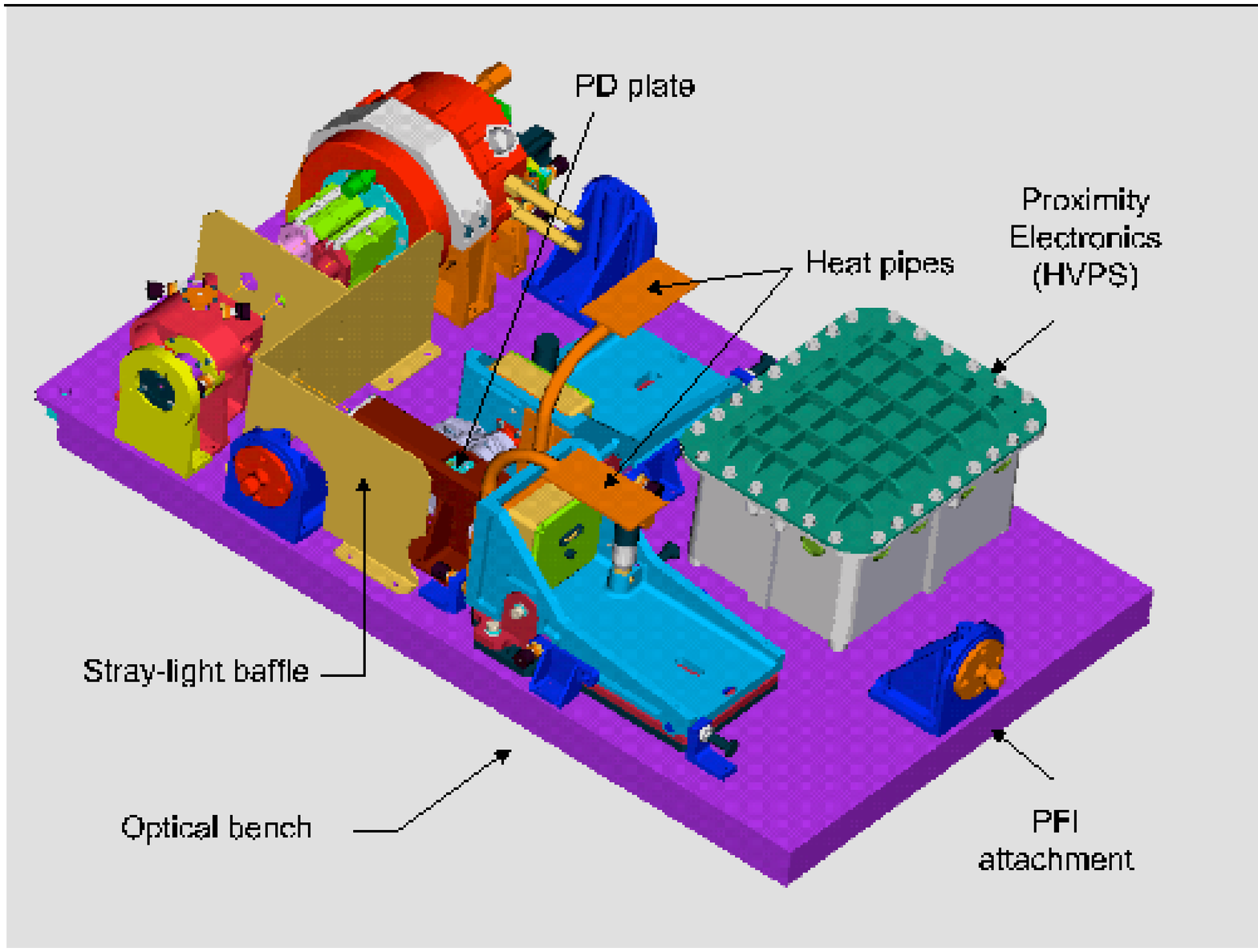}
\caption[\sf IMaX concept]{\sf Sketch showing the IMaX design with all its components, including among others the etalon working in double-pass, the two science cameras and the box devoted to the proximity electronics. The two CCD cameras will also be employed in combination with a plane-parallel glass plate to obtain bursts of PD image-pairs to calibrate the in-flight optical aberrations. All the information about the project can be found in the website \emph{http://www.iac.es/proyect/IMaX/}.}
\label{imax}
\end{center}
\end{figure}

Figure~\ref{imax} shows the final optical and mechanical design of the instrument in a 3D representation; the main optical components will be described in detail in the next section. Table~\ref{tablaimax} lists the main parameters of the instrument. 

\begin{table}
\centering
\sffamily
\caption[\sf Main IMaX parameters]{\sf Main IMaX parameters.}
\begin{tabular}{ll}\\
\footnotesize{PARAMETER} & \footnotesize{VALUE}\\ \hline
Aperture & 1 m\\
Effective focal length & 45.00 m \\
Pixel size  & 12 $\mu$m\\
Spatial sampling  & 0.055 arcsec/pix \\
Working wavelength $\lambda$ & 525.02 nm (FeI) \\
Pixels & 1024 $\times$ 1024 \\
FOV & 50 $\times$ 50 arcsec \\
Image exposure time & $\le$ 200 ms\\
IMaX weight & $\le$ 80 Kg\\
IMaX data rate & $\le$ 850 KB/s\\\hline
{\bf \sf {\scriptsize SPECTROPOLARIMETRIC MODE}} & \\
Number of wavelengths & 4 + 1 continuum\\
Time for I,Q,U,V & $<$ 30 sec SNR 1000\\
Time for I,V & $<$ 15 sec SNR 1000\\\hline
{\bf \sf {\scriptsize MAGNETOGRAPH MODE}} & \\
Number of wavelengths & 4 + 1 continuum\\
Time for I,Q,U,V & $<$ 20 sec SNR 1000\\
Time for I,V & $<$ 10 sec SNR 1000\\\hline
{\bf \sf {\scriptsize DEEP MAGNETOGRAPH MODE}} & \\
Number of wavelengths & 1\\
Time for I,V & $<$ 160 sec for SNR 1000\\\hline
\end{tabular}
\label{tablaimax} 
\end{table}
 
\subsection{Optical description}
\begin{figure}
\begin{center}
\vspace{-0.6cm}
\begin{tabular}{c}
\vspace{-1cm}\includegraphics[angle=-90,width=0.7\linewidth]{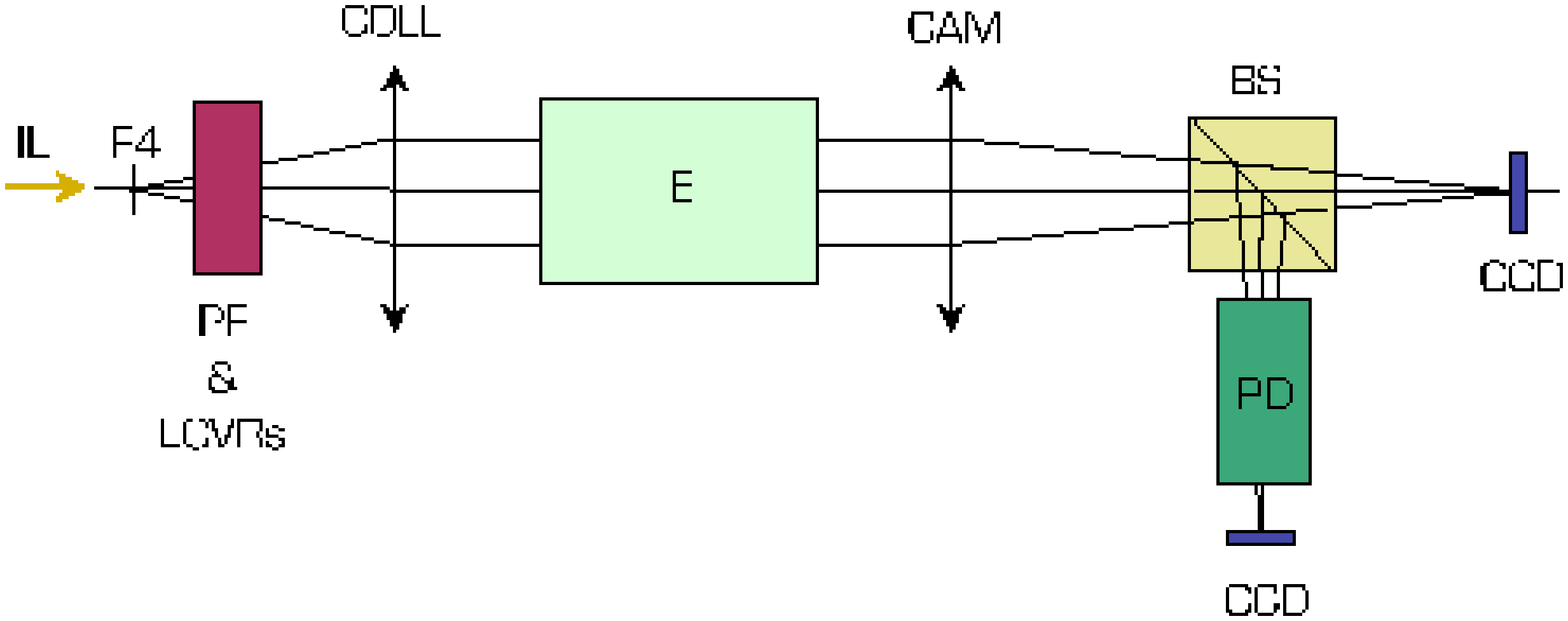}\\
\includegraphics[angle=-90,width=0.9\linewidth]{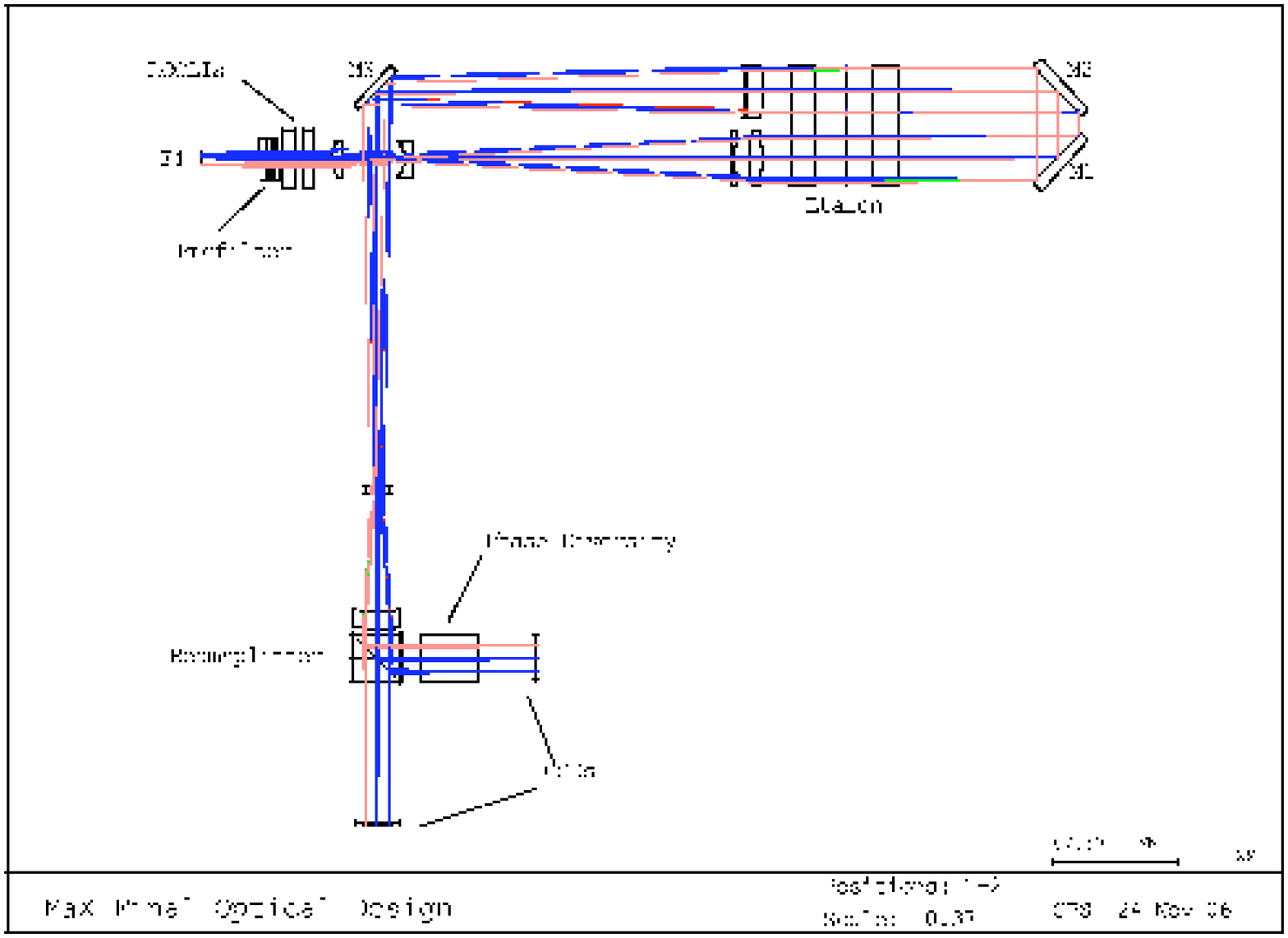}\\
\end{tabular}
\vspace{5mm}
\caption[\sf IMaX box diagram and optical layout]{\sf {\bf \emph{Upper panel}}: IMaX box diagram showing the direction of the incoming light (IL) from the Telescope and ISLiD and the focus position (F4) just before entering IMaX. The main IMaX optical components are: the prefilter and the Liquid Cristal Variable Retarders (PF \& LCVRs), the collimator (COLL), the etalon (E) located in a conjugate pupil and working in double pass, the camera optics (CAM), the cubic polarizing beamsplitter (BS) that divides the light beam into two branches, to form in the CCD respective images with orthogonal polarization states. In one of the branches, a phase diversity (PD) device can be optionally inserted for calibration. {\bf \emph{Lower panel}}: IMaX optical layout. The light goes from focus F4 to the CCD cameras after passing through all the optical elements drawn in the re-scaled diagram. ROCLIs are the same as LCVRs. \emph{Based in the document IMaX Final Optical Design, SUN-IMaX-RP-IX200-023}.}
\label{imaxdesign}
\end{center}
\end{figure}

The box diagram in Figure~\ref{imaxdesign} shows a concise description of the IMaX optical configuration\footnote{\sf From the \emph{IMaX Final Optical Design} document SUN-IMaX-RP-IX200-023.} (see figure caption).\\

Figure~\ref{imaxdesign} (\emph{lower panel}) shows a more detailed view of the IMaX optical layout that will be briefly commented in the next lines. The optical interface with SUNRISE is the ISLiD Focus F4. Next to F4 we locate the prefilter (PF) set (bandwidth 1\AA) and the modulator based on Liquid Cristal Variable Retarders (LCVRs or ROCLIs). After suffering the selected polarization, the beam passes through a collimator system consisting of 2 lenses and a doublet. In the collimated space we locate one solid Fabry-Perot Interferometer (which cavity is LiNbO$_3$\footnote{\sf Due to its unique electro-optical, photoelastic, piezoelectric and non-linear properties Lithium Niobate (LiNbO$_3$) is widely used in a variety of integrated and active acousto-optical devices.}) in the position of a pupil image of the SUNRISE system. The etalon operates in double pass which means that the light passes once through it, then reflects back by means of two folding mirrors and crosses for the second time the etalon. The beam is then focused by an imaging optical system (camera optics) that consist of a doublet and two lenses. A cubic polarizing beamsplitter divides the light beam into two branches, to form in the CCDs respective images with orthogonal polarization states. A parallel glass plate can be optionally inserted in the light path between the BS and one of the CCDs for calibration purposes by applying the post-facto phase diversity technique. We will extensively detail the phase diversity mechanism in section~\S\ref{pdmechanism}.

\section{The Phase Diversity plate}
\label{pdmechanism}

As commented in the last section, the IMaX design includes a mechanism to apply the Phase Diversity technique for calibration purposes (wavefront sensing). According to section~\S1.4 where the PD method is described, we need to combine information included in at least two simultaneous images of the object, one being the conventional focal-plane image that is degraded by the unknown system aberrations we are interested in, and the other one affected by the same unknown aberrations plus an extra known aberration intentionally induced. \\

For the sake of simplicity this extra aberration is commonly chosen as a defocus (see Appendix A) induced by simply displacing one of the cameras out of focus by a certain known distance that we will also term \emph{diversity}. For IMaX another alternative was implemented since the displacement of the camera could generate inertial problems caused by redistribution of masses or vibrations when moving a heavy CCD. One alternative option to induce defocus in one of the images without moving the CCD camera to avoid the above mentioned problems, consist on intercalating a light plane-parallel glass plate in front of one the CCDs to induce a known displacement of the image focus as will be described below.\\

\begin{figure}
\centering
\hspace{-12mm}\includegraphics[width=.8\linewidth]{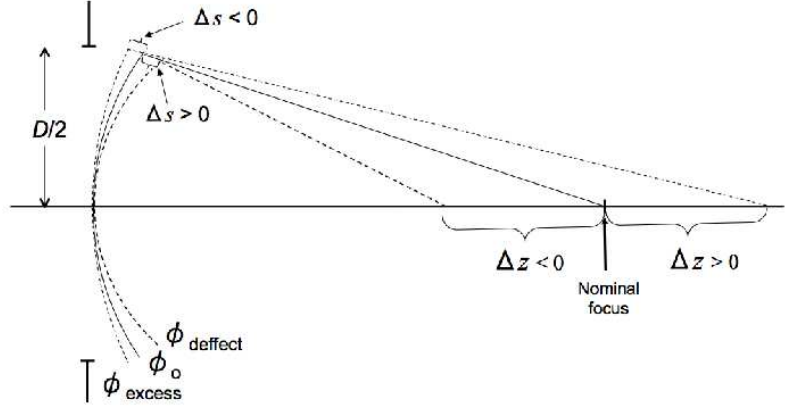}
\caption[\sf Geometrical configuration for a defocus system]{\sf Geometrical configuration for a defocus system. The sketch shows the correspondence between the PV defocus aberration ds and the longitudinal displacement of the image plane, $\Delta z$, along the optical axis.}
\label{defocussketch}
\end{figure}

The defocus aberration can be described as an excess/defect in the radius of the spherical wavefront with respect to the value that would be required to form the image of a point source at the nominal focus $F$ (see Figure~\ref{defocussketch}). The subtraction between both wavefronts give us the defocus aberration function $\phi$ that turns out to be a paraboloid function.\\

Geometric considerations lead to the following relationship between the axial displacement of the image plane ($\Delta z$) and the defocus aberration function

\begin{equation}
\phi(\rho)=-\frac{\pi \Delta z}{4 \lambda} \left (\frac{D}{f} \right )^2 \rho^2 \mbox{ rad},
\label{def}
\end{equation}

\noindent where $D$ is the aperture, $0 \le \rho \le 1$, $\rho D/2$ is the radial distance from the pupil center, $\lambda$ is the working wavelength and $f$ the effective focal length of the system.\\

On the other hand, the defocus aberration function expressed as a Zernike polynomial in the Noll's basis (see Appendix~A) is
 
 \begin{equation}
 \alpha_4\sqrt3 (2\rho^2-1) = 2\alpha_4\sqrt3\rho^2-\alpha_4\sqrt3,
 \label{c4}
\end{equation}

\noindent where $\alpha_4$ is the defocus weighting coefficient (in radian). The last term $\alpha_4\sqrt3$ in equation~\ref{c4} represents an offset that is not affecting the shape of the wavefront but only an axial shift (piston). Thus we can relate expressions in equations~\ref{def} and \ref{c4} as

\begin{equation}
 -\frac{\pi \Delta z}{4 \lambda} \left (\frac{D}{f} \right )^2 \rho^2 = 2\alpha_4\sqrt3\rho^2,
\end{equation}

and finally obtain the displacement in terms of the coefficient $\alpha_4$ (in radianº):
\begin{equation}
\Delta z = -\frac{8\sqrt3\lambda}{\pi} \left ( \frac{f}{D} \right )^2 \alpha_4.
\end{equation}

By replacing the IMaX working parameters (Table~\ref{tablaimax}) in the last expression we obtain $\Delta z$ as proportional to $\alpha_4$ as follows: $\Delta z$ = -~4.68923 $\alpha_4$ mm.\\

This relation will be used later on to have an intuitive and also quantitative feeling about how significant is a defocus aberration expressed as a certain $\alpha_4$ value.\\

Apart from the axial displacement $\Delta z$, another way to characterize the amount of defocus is by means of the \emph{Peak-to-Valley} (PV) optical path difference (OPD) which obviously corresponds to the value of the defocus aberration function at the pupil edge. According to expression~\ref{def} this value is
  
\begin{equation}
\displaystyle \mbox{PV} = \frac{2\pi \Delta s}{\lambda} = -\frac{\pi \Delta z}{4 \lambda} \left (\frac{D}{f} \right )^2, 
\label{ds}
\end{equation}

\noindent or in wavelength units

\begin{equation}
\displaystyle \frac{\Delta s}{\lambda} = -\frac{\Delta z}{8 \lambda} \left (\frac{D}{f} \right )^2.
\label{ds_lenght}
\end{equation}

Table~\ref{OPDtabla} presents the defocus aberration in terms of the PV optical path difference corresponding to the focus displacement along the optical axis for a telescope with $f_{\mbox{\scriptsize{Number}}} = f/D$ = 45.0 and $\lambda$525.02 nm, which are the adopted values in IMaX.\\

\begin{table}
\sffamily
\centering
\begin{tabular}{c|c}\hline\hline
{\footnotesize Optical Path Difference (OPD)} & {\footnotesize Focus displacement along the optical axis}\\
{\footnotesize at the pupil edge ($\Delta s/\lambda$) [number of waves]} & {\footnotesize $\Delta z$ [mm]  for  $f_{\mbox{\scriptsize{Number}}}$=45 and $\lambda$525.02 nm} \\\hline
1/18 & 0.472518 \\
1/15 & 0.567022 \\
1/14 & 0.607523 \\
1/13 & 0.654256 \\
1/12 & 0.708777\\
1/11 & 0.773211 \\
1/10 & 0.850532\\
1/9 & 0.945036\\
1/8 & 1.06317\\
1/7 & 1.21505\\
1/6 & 1.41755 \\
1/5 & 1.70106\\
1/4 & 2.12633\\
1/3 & 2.83511\\
1/2 & 4.25266 \\
1 & 8.50532\\\hline\hline
\end{tabular}
\caption[\sf Relation between the optical path difference and the focus displacement along the optical axis]{\sf Relation between the optical path difference (OPD) at the pupil edge in wavelength units and the focus displacement ($\Delta z$) along the optical axis for a particular telescope and wavelength  $\lambda$.
\label{OPDtabla}}
\end{table}

As a rule of thumb, an amount of defocus corresponding to $\Delta s$ = 1~$\lambda$ at the pupil edge has proved to be satisfactory for the phase diversity inversions. Using the IMaX parameters (see Table~\ref{tablaimax}), and by setting $\Delta s$ = 1~$\lambda$, we obtain a value for the focus displacement along the axis $\Delta z$ = 8.51 mm.\\

\begin{figure}[h]
 \hfill
\begin{minipage}[h]{.35\textwidth}    
 \caption[\sf Displacement on axis produced by a plane parallel plate]{\sf Displacement on axis produced by a plane parallel plate. An incident ray forming an angle $\alpha$ with the normal to the surface is deflected when entering the plate and eventually leaves the plate with the same angle but inducing an OPD as shown in the figure.}
 \label{plate-index}
    \end{minipage}
    \begin{minipage}[h]{.60\textwidth}
      \epsfig{angle=-90,file=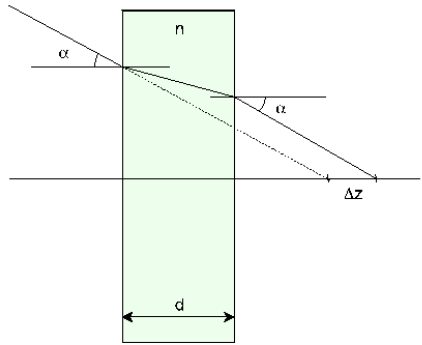, scale=0.27}
  \end{minipage}
  \hfill
\end{figure}

As mentioned above, we will induce the image plane displacement by employing a glass plate of thickness d, as shown in Figure~\ref{plate-index}. This displacement can be mathematically formulated by the expression 
\begin{equation}
\Delta z = d \left ( 1- \frac{1}{n} \right ),
\end{equation}

\noindent where $n$ is the refractive index of the material. The thickness of the plate required to produce a desired displacement depends on the refractive index of the optical glass commercially available, and for IMaX the selected material was \emph{Fused Silica}\footnote{\sf Fused Silica is a high purity synthetic amorphous silicon dioxide. This noncrystalline, colorless, silica glass combines a very low thermal expansion coefficient with excellent optical qualities and exceptional transmittance over a wide spectral range and is also resistant to scratching and thermal shock.} with $n$=1.461. Assuming a PV aberration of 1~$\lambda$ or equivalently $\Delta z$=8.51 mm as reported above, the required thickness for our PD plate would be $d$=27 mm.\\

\vspace{3mm}

\subsubsection{\bf Defocus tolerance}
The image blurring permitted for an instrument can be specified by the diameter of the blur spot or as the angle ($\beta$) subtended by such a diameter. For a given $\beta$  of acceptable blurring, the focus deepness is lower towards the lens that in the backward direction.\\

A criterion for the defocus tolerance can be made by fixing the blur angle $\beta$ permitted for our purposes. For instance, we can select $\beta$ as the value of the diffraction cutoff wavelength, that is slightly greater than the Full-Width at Half-Maximum (FWHM) of the Airy spot. According to this criterion the defocus tolerance $\Delta z_t$ can be expressed as

\begin{equation}
\Delta z_t=\pm ~\lambda \left(\frac{f}{D}\right)^2.
\label{epsilon4}\
\end{equation}

Nevertheless, this criterion is quite severe and there is another less-strict and more commonly used to establish the limit of the defocus tolerance in an optical system. This criterion is based on the loss of intensity in the central part of the PSF.\\

Rayleigh  \footnote{\sf Lord Rayleigh (John William Strutt), 1842-1919, Nobel prize in Physics in 1904.} first (for the spherical aberration) and other authors afterwards (for other aberrations), demonstrated that when the PV value of the error in the wavefront is $< \lambda/4$, the central intensity of the PSF of the system reduces by less than 20$\%$ with respect to the central intensity of the Airy function.\\

Marechal demonstrated that when the rms of the wavefront aberration in an optical system is $\le \lambda/14$, the central intensity of the PSF also reduces by less than 20$\%$, or in other words, the Strehl ratio\footnote{\sf The Strehl ratio of an optical system is defined as the quotient between the central intensities of the real PSF and the theoretical one assuming a diffraction limited system, the latter being the Airy function.   $SR=$PSF$(0,0)/Airy(0,0)$. This ratio is closely related to the sharpness criteria for optics defined by Karl Strehl.} is $\ge$ 0.8.\\

Both limits described above, $\lambda/4$ for the PV value or $\lambda/14$ for the rms, are commonly accepted as tolerable when establishing whether or not a system achieves the resolution limit for practical purposes. These limits are referred to as the $\lambda/4$-Rayleigh and the Marechal image quality criteria, respectively.\\

~\cite{bornwolf} study the distribution of intensity in a volume around the focus in a diffraction limited optical system and, in particular, the intensity in the central part of the PSF along the optical axis. In this study it is demonstrated that to reduce the central intensity of the PSF by a 20$\%$, the displacement as measured from the nominal focus is given by the equation 

\begin{equation}
\Delta z_t=\pm  ~2 \lambda \left(\frac{f}{D}\right)^2.
\label{epsilon5}
\end{equation}

This is the classical limit adopted as the tolerance for the defocus, an it is based on the above described Rayleigh and Marechal criteria. Note that the difference in both defocus tolerances in equations \ref{epsilon4} and \ref{epsilon5} is a factor of 2, meaning that the last one is more permissive as a quality criterion.\\

With the IMaX parameters (in Table~\ref{tablaimax}) the defocus tolerances $\Delta z_t$ for the instrument according to both defocus criteria are, respectively,

\begin{equation}
\Delta z_t= \pm ~5250.2\mbox{x}10^{-10} \left( \frac{45.0}{1.0} \right)^2=\pm ~1.06 \mbox{ mm},
\label{deltaz1}
\end{equation}

\begin{equation}
\Delta z_t = \pm ~2 \times 5250.2\mbox{x}10^{-10} \left( \frac{45.0}{1.0} \right)^2=\pm ~2.12 \mbox{ mm}.
\label{deltaz2}
\end{equation}

Nevertheless, the modelled defocus error in IMaX, stemming from optical tolerancing and thermal effects, remains within the tolerance range even in the more restrictive case ($\pm$~1.06 mm).


\section{Defining the method for the in-flight calibration of the IMaX image aberrations}
\label{modeloIMaX}
The goal of this section is to define the procedure that will be employed in IMaX to calibrate the image degradation during the flight. The proposed calibration procedure is based on the PD-speckle technique for image reconstruction and wavefront sensing (see section~\S1.4). To that aim IMaX has been provided by a PD-plate to intentionally induce a controlled defocus. \\

\begin{figure}[h]
 \hfill
\begin{minipage}[h]{.35\textwidth}    
\caption[\sf Sketch of the IMaX phase diversity mechanism.]{\sf Sketch of the IMaX phase diversity mechanism. The plate is meant to be put in front of one of the cameras during the calibrations to get defocus images.}
 \label{imaxpdsketch}
    \end{minipage}
    \begin{minipage}[h]{.60\textwidth}
      \epsfig{angle=-90,file=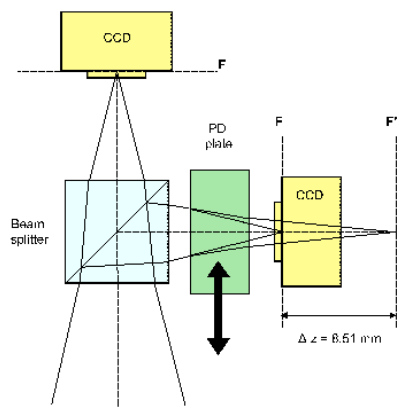, scale=0.3}
  \end{minipage}
  \hfill
\end{figure}

The PD-plate can be optionally intercalated in one of the IMaX imaging channels (see Figure~\ref{imaxpdsketch} and section~\ref{pdmechanism}) so that a simultaneous focus-defocus image-pair, i.e.\ a PD image-pair, can be recorded.  From this pair an estimate of the aberrations will be possible in post-processing by means of a PD inversion code. PD-inversions throughout this chapter have been performed with an IDL code developed by \cite{bonet2003}.\\

\vspace{3mm}

Assuming a long-term variation in the instrumental aberrations, the image acquisition for calibration could be performed with a cadence of one hour. A burst of 25-30 PD-pairs taken in a short time interval in the spectral continuum would be enough each time. This way we get a large amount of information \footnote{\sf Improving the SNR.} collected within a sufficiently short time period so that it can be assumed negligible evolution of the solar structures in the FOV. The averaged results from the PD-inversions of these calibration image-pairs in a post-facto process will provide a description of the aberrations also affecting our science spectral images and, in turn, the maps of the full-Stokes vector. Subsequently, the science images will be reconstructed by using a standard deconvolution code to nearly reach the diffraction limit of SUNRISE.\\

\begin{figure}
\begin{center}
\hspace{0cm}\includegraphics[angle=90,width=1.\linewidth]{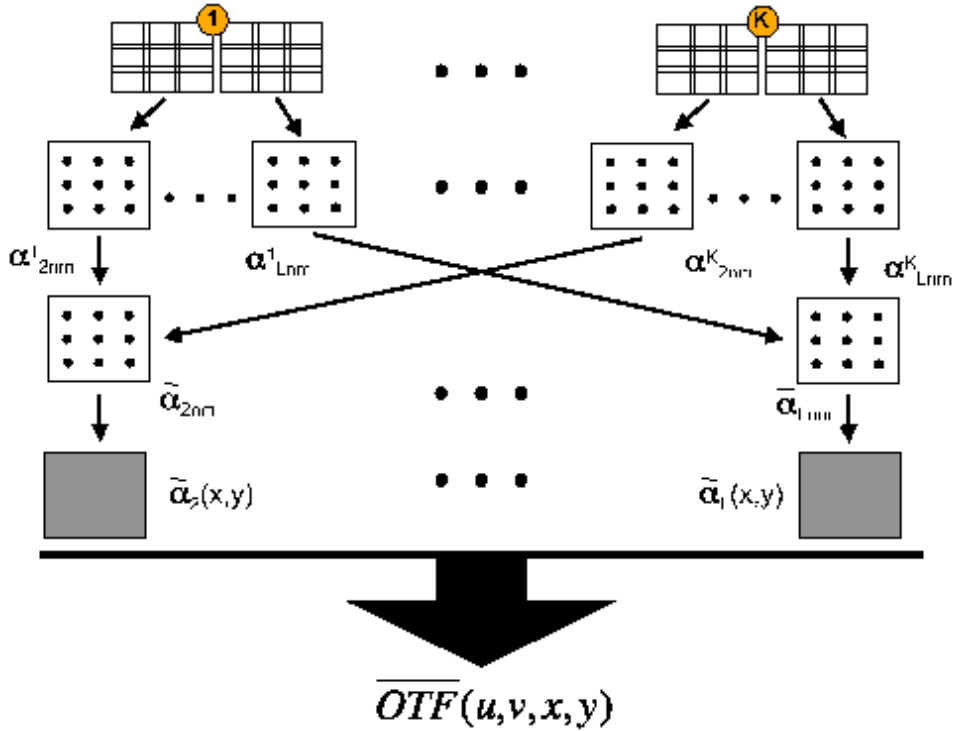}
\vspace{0.5cm}
\caption[\sf Organogram explaining the strategy calibration procedure that will be applied to the IMaX data]{\sf Organogram explaining the strategy for the calibration procedure that will be applied to the IMaX data. Within a time interval of $\sim$~10 s a burst of focus-defocus image-pairs are recorded and labeled from 1 to K. Every single image is divided in a total of N $\times$ M square isoplanatic patches. Note in the figure that the patches are overlapped since their borders will be corrupted by mathematical manipulations in the Fourier domain during the PD-inversion process. For all the focus-defocus image-pairs the PD inversions are independently performed for every single patch-pair (corresponding to a common focus-defocus portion in the FOV) to obtain a set of $L-1$ Zernike coefficients (from $\alpha_2$ to $\alpha_L$) describing the wavefront aberration affecting such a portion of the FOV. Thus, for instance, the nomenclature $\alpha_{4\mbox{{\tiny nm}}}^K$ stands for the Zernike coefficient of index 4 (defocus) computed from the patch centered at coordinates (n,m) in the last image-pair recorded which is labeled by K. The coefficients extracted from every patch are then averaged for all the K realizations and hereby we are able to describe the wavefront aberration in any given patch of the FOV. Interpolation over the whole FOV will result in continuous \emph{maps} for every aberration coefficient $\overline \alpha_l(x,y)$, with 2$ \le l \le L$. The OTF of the system for every location on the image is straightforward computed from these $\overline \alpha_l(x,y)$ functions (eqs.\ref{OTFS2} and \ref{OTFS}).}
\label{imaxcalibrate}
\end{center}
\end{figure}

The organogram of Figure~\ref{imaxcalibrate} shows a schematic representation of the sequential steps in the post-processing calibration procedure.\\

\section{Testing the robustness of the calibration method}

The purpose of this section is to evaluate by means of numerical simulations, the robustness of the method we have implemented for the wavefront error calibration in IMaX, versus a variety of aberration assumptions. To pursue the numerical experiments we select an isoplanatic patch consisting of a portion of one of the synthesized images of solar granulation, from the MHD model developed by \cite{vogler2005,schussler2003}. The spatial resolution of this synthetic image is much higher than that corresponding to the \emph{cutoff} frequency in SUNRISE\footnote{\sf In order to be as realistic as possible constructing the theoretical images one has to consider the spectral contamination by the wings and lobes of the filter transmission curve $T(\lambda)$. For this purpose a sequence of images are constructed from the theoretical model for different $\lambda$'s around our working $\lambda$: $i(x,y,\lambda)$. This sequence has been convolved with $T(\lambda)$: $i^{\prime}(x,y,\lambda)$=$i(x,y,\lambda) \ast T(\lambda)$ and from the resulting sequence $i^{\prime}(x,y,\lambda)$ the central image in our working $\lambda$ is extracted. These calculations have been done by D. Orozco Su\'arez.}, and represents what will be termed hereafter the \emph{true object}.\\

The testing procedure will consist in simulating the formation of PD image-pairs as produced by a telescope with the SUNRISE aperture and a given set of aberrations (model of aberrations). A total of 30 image-pairs for different photon noise realizations is hereby obtained. The resulting degraded image-pairs are re-sampled to the IMaX pixel size and inverted with the PD-code. This set of images will be processed according to the diagram sketched in Figure~\ref{imaxcalibrate} applied to the particular case of only one isoplanatic patch, and the set of averaged aberrations retrieved from the inversions will be compared to the input model of aberrations.\\ 

In our simulations we intend to reproduce the global aberration as close as possible to the real one affecting the image quality when the instrument is acquiring data in real-time. For this purpose we need to, firstly, identify the potential sources of error and, secondly, make realistic assumptions about the quantitative contribution from each of them.\\

\subsection{Identifying the error sources}
\label{identify} 
The contribution from the different error sources can be mathematically represented through the \emph{generalized pupil function} (see equation~1.18) as follows,

\begin{equation}
H(\rho,\theta) = |H(\rho,\theta)|~ \exp \biggl \{ i\bigl [{\delta(\rho,\theta)+\phi_{\emph{ripple}}(\rho,\theta)+\phi_e(\rho,\theta)+\phi(\rho,\theta)+\phi_{\emph{atm}}(\rho,\theta)} \bigr ] \biggr \},
\label{OTF_S_2}
\end{equation}
\\
\noindent
where the terms in the exponential stand for:\\

\hspace{-8mm}
\begin{tabular}{ll}
\noindent$\delta(\rho,\theta)$ & $\equiv $ \footnotesize{Differential aberration between focus/defocus channels $\equiv$ phase diverse.}\\
\noindent$\phi_{\emph{ripple}}(\rho,\theta) $ & $\equiv$ \footnotesize{Main mirror polishing error.}\\
\noindent$\phi_e(\rho,\theta) $ & $\equiv $ \footnotesize{Phase error from the etalon.}\\
\noindent$\phi(\rho,\theta)$ & $\equiv $ \footnotesize{Low-order optical system aberrations common in focus/defocus channels.}\\
\noindent$\phi_{\emph{atm}}(\rho,\theta)$ & $\equiv$ \footnotesize{Atmospheric aberration which is negligible in IMaX.}\\
\noindent$|H(\rho,\theta)|$ & $\equiv$ \footnotesize{Transmission function over the pupil.}\\\\
\end{tabular}

In addition we have to include in the error budget the pixel integration effect in the CCD and the noise.\\

$\delta(\rho,\theta)$ and $\phi(\rho,\theta)$ will be expressed as a polynomial expansion in a basis of Zernike as explained in section~\S1.4. The rest of the contributions will be introduced as amplitude- or phase-screens constructed according to a model or obtained from direct laboratory measurements. \\

We emphasize that the formulation of the Zernike's basis we use corresponds in all cases to that introduced by \cite{noll1976} as stated in section~\S1.4.\\

\subsection{Quantifying the contribution of the error sources}
\label{quantifyingerrors}
The opto-mechanical elements in SUNRISE which are affecting the formation of our images are: {\bf Telescope}, {\bf ISLiD}, {\bf mechanical interface} with IMaX and {\bf IMaX} itself (see Figure~\ref{sunrisebox} for a schematic view of the optical design including all these components). The first step is, consequently, the compilation of data from the design and specifications of all the different components\footnote{\sf Most of these data have already been published in the IMaX technical document: SUN-IMaX-RP-IX200-023 \emph{IMaX Final Optical Design (2007)}.}. \\

Table~\ref{tablesystems} lists a compendium of some nominal values for the rms wavefront error (\emph{rms-WFE}) derived from the optical design including the statistical modelling for the optical tolerancing and the thermal behaviour of the systems \footnote{\sf Thermal effects are simulated with an optical design software (\emph{CODE V}) by varying the refractive index and the geometry of the opto-mechanical components as a function of the temperature and the expansion coefficients. Temperature variations, for instance, are taken into account in the expected range except in those components having their own temperature control in a lower range (i.e Liquid Cristal Variable Retarders - LCVRs -, etalon, etc). Simulations are meant to study the aberration for three possible cases: {\bf operative}, {\bf hot} and {\bf cold}. It is mandatory to know the thermal variation of the refractive index ($dn/dT$) for every optical material and the thermal expansion coefficient for every opto-mechanical element. The thermal effects have been studied for IMaX as a whole and also for its main components independently (e.g.\ the PD-plate, etc). The calculations were made under the assumption of a temperature variation range of 15$\pm$7.5$^\circ$C. However, further improvements in the thermal system (May 2008) have provided a more restrictive range for the temperature excursions: 24.5$\pm$0.5$^\circ$C. This implies that the \emph{rms-WFE} values in the table, derived from the former thermal model, are rather pessimistic estimations. Nevertheless, we preserved them because new estimations are not available for all cases yet. Moreover, testing the robustness of our calibration method with over-degraded images will reinforce the conclusions about the reliability of the calibration method.
Optical tolerancing is a critical step in the design of an optical system. The objective is to define a fabrication and assembly tolerance budget and to accurately predict the resulting as-built performance, including the effects of compensation. Also part of the study is determining the best set of compensators.}.\\

The rms values shown hereafter are computed excluding piston and image displacement (tip/tilt) from the error budget.\\

\begin{table}
\sffamily
\centering
\caption[\sf Aberrations budget (\emph{rms-WFE}) from the modelling of different optical systems in SUNRISE]{\sf Aberrations budget from the modelling of different optical systems in SUNRISE. Polishing errors in the main mirror are not included.}
\begin{tabular}{clc}\\\hline\hline
 &  &  \footnotesize{{\bf rms-WFE}}\\
\footnotesize{~~~{\bf ITEM}}  & \footnotesize{~~~~~~~~~~~~~~~~~~~~~~~~~~~~~~~~~~~~~~~~~~~{\bf SYSTEM$^\dag$}} & \scriptsize{(no tip/tilt)}\\
& & \scriptsize{[waves]}\\\hline\\
1 & {\bf Telescope+ISLiD} \footnotesize{(nom. design) averaging 9 points in FOV} & 1/26 \\
2 & {\bf Telescope+ISLiD} \footnotesize{(nom. design+op.tolerancing+therm.eff.)$^\ddag$}\ & 1/8.25 \\
3 & {\bf Interface ISLiD-IMaX} \footnotesize{(misalignment)} & 1/42 \\
4 & {\bf IMaX} \footnotesize{without etalon (nominal design)} & 1/140 \\
5 & {\bf IMaX} \footnotesize{without etalon (nominal design + optical tolerancing)} & 1/22 \\
6 & {\bf IMaX} \footnotesize{without etalon (nominal design + thermal effects)} & 1/38 \\
7 & {\bf IMaX} \footnotesize{without etalon (nom. design+ op.tolerancing+therm. eff.)$^\triangle$} & 1/19\\
8 & {\bf Etalon} \footnotesize{(lab. calibration CSIRO$^\Diamond$)} & 1/23 \\
9 & {\bf IMaX} \footnotesize{(nom. design + op.tolerancing + thermal eff. + etalon)}$^\Box$ & 1/14.6 \\
10 & {\bf Total: Teles.+ISLiD+Interface+IMaX$^\oplus$} & 1/7.1 \\\\\hline\\
\end{tabular}
\begin{tabular}{ll}
\scriptsize{$\dag$ Symbol "+" is employed to include another element.} & \scriptsize{$\Diamond$ Measured by the etalon's manufacturer.} \\
\scriptsize{$\ddag$ RSS $\equiv$ Root Sum Square:}  & \scriptsize{$\Box$ Equivalent to the RSS of items 7 and 8.}\\
~~\scriptsize{$\mbox{rms-WFE}=\sqrt{(\mbox{rms-WFE}_1)^2+(\mbox{rms-WFE}_2)^2}$.} & \scriptsize{$\oplus$ Equivalent to RSS of items 2, 3 and 9.}\\
\scriptsize{$\triangle$ Equivalent to RSS of items 5 and 6.} & \\
\end{tabular}
\label{tablesystems}
\end{table}

\subsubsection{\bf Low-order optical system aberrations}
In addition to the above mentioned mean parameters describing the rms aberration in Table~\ref{tablesystems}, we also have available the sets of  Zernike coefficients corresponding to the nominal design for both, {\bf Telescope+ISLiD} and {\bf IMaX}, up to a total of 36 modes (7$^{\mbox{\footnotesize{th}}}$ order) starting from the term of index 4, i.e.\ defocus (see columns 2 and 3 in Table~\ref{tablebudget}). Respective \emph{rms-WFE} are included in Table~\ref{tablesystems} (1/26 and 1/140 waves).\\

\begin{table}
\sffamily
\centering
\caption[\sf Error budget]{\sf Error budget for different sub-systems in SUNRISE. Zernike coefficients in Noll's basis.}
\hspace{-3mm}\begin{tabular}{@{}c@{}@{}@{}c@{}@{}c@{}@{}c@{}@{}c@{}@{}c@{}@{}c@{}@{}c@{}@{}c@{}}\\\hline\hline
& \tiny {\bf  IMaX}~~~ & \tiny ~~{\bf Tel+ISLiD}~~~& \tiny ~{\bf \textcolor{color1}{T+IS+IM$^\triangle$}} ~& ~~~\tiny {\bf IMaX} ~~~& ~~~\tiny {\bf Tel.+ISLiD} ~~~& \tiny {\bf \textcolor{color2}{Sum prev.}}  & \tiny ~~ {\bf Etalon aberr.}~~ & \tiny ~~~{\bf \textcolor{color3}{TOTAL}} ~~~\\
{\bf Z} & \tiny no etalon & & \tiny \textcolor{color1}{Sum of re-scaled} & ~\tiny no etalon~ & \tiny re-scaled & \tiny {\bf \textcolor{color2}{two columns}} & \tiny Low-order modes& {\bf \tiny \textcolor{color3}{Sum prev.}} \\
& ~~~\tiny nom.aberr.~~~ &  \tiny nom.aberr. & \tiny \textcolor{color1}{nom.aberr.} & \tiny Zygo & \tiny nom.aberr. & & \tiny Zygo & {\bf \tiny \textcolor{color3}{two col.}}\\
 & \tiny [rad] & \tiny [rad] & \tiny \textcolor{color1}{[rad]} & \tiny [rad] &\tiny [rad] & \tiny \textcolor{color2}{[rad]} & \tiny [rad] &  \tiny \textcolor{color3}{[rad]}\\\hline
\tiny4 & \tiny 0.0445282 & \tiny -0.0714395 & \tiny \textcolor{color1}{0.0493187} & \tiny 0.0411549 & \tiny -0.278658 & \tiny \textcolor{color2}{-0.237503} & {\tiny 0.647168$^\clubsuit$}& \tiny \textcolor{color3}{-0.237503}\\
\tiny5 & \tiny -1.24691e-09 & \tiny 2.51327e-07 & \tiny \textcolor{color1}{9.71148e-07} & \tiny 0.0157080 & \tiny 9.80332e-07 & \tiny \textcolor{color2}{0.0157089} & \tiny -0.0565487& \tiny \textcolor{color3}{-0.0408397}\\
\tiny6 & \tiny -0.00246226 & \tiny 0.180514 & \tiny \textcolor{color1}{0.685980} & \tiny 0.279602 & \tiny 0.704116 & \tiny \textcolor{color2}{0.983718} & \tiny 0.175929& \tiny \textcolor{color3}{1.15965}\\
\tiny7 & \tiny -1.11203e-09 & \tiny -0.000384531 & \tiny \textcolor{color1}{-0.00149992} & \tiny 0.00314159 & \tiny -0.00149991 & \tiny \textcolor{color2}{0.00164168} & \tiny 0.0314159 & \tiny \textcolor{color3}{0.033057}\\
\tiny8 & \tiny 0.000775724 &  \tiny 0.0186024 &  \tiny \textcolor{color1}{0.0782744} &  \tiny -0.0408407 &  \tiny 0.0725608 &  \tiny \textcolor{color2}{0.0317201} &  \tiny -0.0753982 & \tiny \textcolor{color3}{-0.0436782}\\
\tiny9 & \tiny 1.52930e-09 &  \tiny0.000275518 &  \tiny\textcolor{color1}{0.00107470} &  \tiny0.00314159 &  \tiny0.00107469 & \tiny\textcolor{color2}{0.00421628} & \tiny-0.0502655 & \tiny \textcolor{color3}{-0.0460492}\\
\tiny10 & \tiny 3.28740e-07 & \tiny-0.000634476 & \tiny\textcolor{color1}{-0.00247243} & \tiny0.0219912 & \tiny-0.00247485 & \tiny\textcolor{color2}{0.0195163} & \tiny-0.0188496 & \tiny \textcolor{color3}{6.66745e-4}\\
\tiny11 & \tiny -0.00423551 & \tiny0.00933983 & \tiny\textcolor{color1}{0.00523408} & \tiny0.0219912 & \tiny0.0364311 & \tiny\textcolor{color2}{0.0584223} & \tiny-0.0188496 & \tiny \textcolor{color3}{0.0395727}\\
\tiny12 & \tiny -6.63747e-08 & \tiny-0.000165059 & \tiny\textcolor{color1}{-0.000644322} & \tiny0.0251327 & \tiny-0.000643833 & \tiny\textcolor{color2}{0.0244889} &\tiny-0.0125664 & \tiny \textcolor{color3}{0.0119225}\\
\tiny13 & \tiny 6.54083e-10 & \tiny6.28319e-08 & \tiny\textcolor{color1}{2.49901e-07} & \tiny0.00000 & \tiny2.45083e-07 & \tiny\textcolor{color2}{2.45083e-07} & \tiny0.0314159 & \tiny \textcolor{color3}{0.0314162}\\
\tiny14 &  \tiny -6.65045e-07 & \tiny2.51327e-06 &   \tiny\textcolor{color1}{4.90487e-06} & \tiny-0.00628319 & \tiny9.80332e-06 & \tiny\textcolor{color2}{-0.00627338} & \tiny0.0125664 & \tiny \textcolor{color3}{0.00629299}\\
\tiny15 & \tiny1.23446e-09 & \tiny0.00000 &	   \tiny\textcolor{color1}{9.09256e-09} & \tiny-0.0125664 & \tiny0.00000 & \tiny\textcolor{color2}{-0.0125664} & \tiny0.00628319 & \tiny \textcolor{color3}{-0.00628319}\\
\tiny16 & \tiny2.13090e-09 & \tiny-1.54566e-05 &	\tiny  \textcolor{color1}{-6.02747e-05} &		\tiny 0.0219912 &	\tiny-6.02904e-05 &	\tiny \textcolor{color2}{0.0219309} &	 \tiny0.0125664 & \tiny \textcolor{color3}{0.0344972}\\
\tiny17 & \tiny4.98823e-10 & \tiny-4.39823e-07 &	\tiny  \textcolor{color1}{-1.71191e-06} &		\tiny 0.0785398 &	\tiny-1.71558e-06 &	\tiny \textcolor{color2}{0.0785381} &		  \tiny-0.00628319 & \tiny \textcolor{color3}{0.0722549}\\
\tiny18 & \tiny7.93558e-10 & \tiny1.88496e-07 &	\tiny   \textcolor{color1}{7.41094e-07} &	      \tiny   -0.0219912 &	\tiny 7.35249e-07 &	\tiny\textcolor{color2}{-0.0219904} &		  \tiny 0.00628319 & \tiny \textcolor{color3}{-0.0157072}\\
\tiny19 &\tiny-7.84298e-10 &	 \tiny1.25664e-07 &	\tiny   \textcolor{color1}{4.84389e-07} &		\tiny 0.00628319 &	\tiny 4.90166e-07 &	\tiny \textcolor{color2}{0.00628368} &		\tiny  -0.0125664 & \tiny \textcolor{color3}{-0.00628270}\\
\tiny20 & \tiny5.43063e-09 &	\tiny 0.00000 &	   \tiny\textcolor{color1}{3.99998e-08} &		\tiny 0.0282743 &	\tiny 0.00000 &	\tiny \textcolor{color2}{0.0282743} &	 	 \tiny  0.00628319 & \tiny \textcolor{color3}{0.0345575}\\
\tiny21 & \tiny1.04887e-09 &\tiny	 0.00000 &	\tiny   \textcolor{color1}{7.72552e-09} &	 	\tiny-0.0125664 &	\tiny 0.00000 &\tiny	\textcolor{color2}{-0.0125664} &		  \tiny 0.0314159 & \tiny \textcolor{color3}{0.0188496}\\
\tiny22 & \tiny0.000234998 &	 \tiny5.34071e-06 &	\tiny   \textcolor{color1}{0.00175173} &		\tiny         0.185354 &	\tiny 2.08321e-05 &	\tiny \textcolor{color2}{0.185375} &\tiny	   0.0314159 & \tiny \textcolor{color3}{0.216791}\\
\tiny23 & \tiny-6.92688e-11 &	\tiny 0.00000 &	 \tiny \textcolor{color1}{-5.10206e-10} &	\tiny	 0.0691150 &\tiny	 0.00000 &\tiny	 \textcolor{color2}{0.0691150} &		  \tiny-0.00628319 & \tiny \textcolor{color3}{0.0628319}\\
\tiny24 & \tiny6.84204e-09 &	\tiny 0.00000 &	 \tiny  \textcolor{color1}{5.03957e-08} &		\tiny 0.0282743 &\tiny	 0.00000 &	\tiny \textcolor{color2}{0.0282743} &		  \tiny 0.0188496 & \tiny \textcolor{color3}{0.0471239}\\
\tiny25 & \tiny1.05472e-09 &	 \tiny0.00000 &	 \tiny  \textcolor{color1}{7.76868e-09} &		\tiny 0.0188496 &\tiny	 0.00000 &	\tiny \textcolor{color2}{0.0188496} &		  \tiny-0.00628319 & \tiny \textcolor{color3}{0.0125664}\\
\tiny26 &\tiny-7.45863e-07 &	 \tiny0.00000 &	  \tiny\textcolor{color1}{-5.49373e-06} &		\tiny 0.00314159 &\tiny	 0.00000 &	\tiny \textcolor{color2}{0.00314159} &		  \tiny 0.00000 & \tiny \textcolor{color3}{0.00314159}\\
\tiny27 &\tiny-1.08106e-09 &	 \tiny0.00000 &	  \tiny\textcolor{color1}{-7.96267e-09} &		\tiny 0.00000 &	 \tiny0.00000 &	\tiny \textcolor{color2}{0.00000} &	 	  \tiny 0.00000 & \tiny \textcolor{color3}{0.00000}\\
\tiny28 &\tiny 5.19823e-09 &	\tiny 0.00000 &	  \tiny \textcolor{color1}{3.82881e-08} &		\tiny 0.00000 &	 \tiny0.00000 &	\tiny \textcolor{color2}{0.00000} &		 \tiny  0.00000 & \tiny \textcolor{color3}{0.00000}\\
\tiny29 & \tiny-6.84537e-10 &\tiny 0.00000 &	\tiny  \textcolor{color1}{-5.04203e-09} &	\tiny 0.0691150 &	\tiny 0.00000 &	\tiny \textcolor{color2}{0.0691150} &		  \tiny 0.0188496 & \tiny \textcolor{color3}{0.0879646}\\
\tiny30 &\tiny-5.44505e-09 &	 \tiny0.00000 &	 \tiny \textcolor{color1}{-4.01060e-08} &		\tiny 0.0471239 &	\tiny 0.00000 &	\tiny \textcolor{color2}{0.0471239} &		  \tiny 0.00628319 & \tiny \textcolor{color3}{0.0534071}\\
\tiny31 &\tiny-3.69741e-10 &	 \tiny0.00000 &	 \tiny \textcolor{color1}{-2.72336e-09} &		\tiny 0.0188496 &	\tiny 0.00000 &	\tiny \textcolor{color2}{0.0188496} &		  \tiny 0.00628319 & \tiny \textcolor{color3}{0.0251327}\\
\tiny32 &\tiny-1.26343e-08 &	 \tiny0.00000 &	 \tiny \textcolor{color1}{-9.30591e-08} &		\tiny 0.0125664 &	\tiny 0.00000 &	\tiny \textcolor{color2}{0.0125664} &		   \tiny0.00628319 & \tiny \textcolor{color3}{0.0188496}\\
\tiny33 &\tiny 1.23635e-09 &	 \tiny0.00000 &	 \tiny  \textcolor{color1}{9.10646e-09} &		\tiny 0.00000 &	\tiny 0.00000 &	\tiny \textcolor{color2}{0.00000} &	\tiny	   0.00000 & \tiny \textcolor{color3}{0.00000}\\
\tiny34 & \tiny5.01911e-09 &	\tiny0.00000 &	  \tiny \textcolor{color1}{3.69687e-08} &		\tiny 0.00000 &	\tiny 0.00000 &	\tiny \textcolor{color2}{0.00000} &	\tiny	   0.00000 & \tiny \textcolor{color3}{0.00000}\\
\tiny35 & \tiny1.78661e-09 &	\tiny 0.00000 &	  \tiny \textcolor{color1}{1.31595e-08} &		\tiny 0.00000 &	\tiny 0.00000 &	 \tiny\textcolor{color2}{0.00000} &	\tiny	   0.00000 & \tiny \textcolor{color3}{0.00000}\\
\tiny36 &\tiny-8.78520e-09 &	\tiny 0.00000 &	 \tiny \textcolor{color1}{-6.47082e-08} &		\tiny 0.00000 &	\tiny 0.00000 &	\tiny \textcolor{color2}{0.00000} &	\tiny	   0.00000 & \tiny \textcolor{color3}{0.00000}\\\hline
\end{tabular}
\begin{tabular}{l}
\noindent
\hspace{-5mm}\scriptsize{ $\triangle$ Telescope+ISLiD+IMaX.}\\
\hspace{-5mm}\scriptsize{ $\clubsuit$ This contribution is set to zero ($\to$ 0.00000) for all our calculations hereafter. See text for detailed information.}
\end{tabular}
\label{tablebudget}
\end{table}

\begin{table}
\sffamily
\centering
\caption[\sf Aberrations budget from laboratory calibrations]{\sf Aberrations budget (\emph{rms-WFE}) from laboratory calibrations.}
\begin{tabular}{clc}\\\hline\hline
 &  &  \footnotesize{{\bf rms-WFE}}\\
\footnotesize{~~~{\bf ITEM}}  & \footnotesize{~~~~~~~~~~~~~~~~~~~~~~~~~~~~~~~~~~~{\bf SYSTEM$^\dag$}} & \scriptsize{(no tip/tilt)}\\
& & \scriptsize{[waves]}\\\hline\\
1 & {\bf IMaX} \footnotesize{excluding etalon and PD-plate$^\Box$. From Zernike coeffs.} & 1/16.8\\
2 & {\bf IMaX} \footnotesize{(item 1)} + ({\bf Telesc.+ISLiD})$^\triangle$. \footnotesize{From Zernike coeffs.} & 1/6\\
3 & {\bf Etalon \footnotesize{in double-pass:}} \footnotesize{From the CSIRO phase-screen$^\clubsuit$} & 1/26 \\
4 & {\bf Etalon} \footnotesize{in double-pass within the oven$^\Diamond$. From Z coeffs.} & 1/9.2 \\
5 & {\bf Etalon} \footnotesize{contribution in item 4 after removing defocus} & 1/28.5 \\
6 & {\bf Total:} \footnotesize{Low-order aberration terms$^\otimes$. From Zernike coeffs.} & 1/5.2 \\\hline\\
\end{tabular}
\begin{tabular}{l}
\scriptsize{$\dag$ Symbol "+" is employed to include another element.} \\
\scriptsize{$\Box$ Laboratory calibration with Zygo at INTA.}\\
\scriptsize{$\triangle$ Nominal coefficients of Teles.+ISLiD, re-scaled with the factor in item 2 of Table~\ref{tablesystems}.}\\
\scriptsize{$\clubsuit$ Optimal selection of two sub-apertures in the phase-screen provided by the manufacturer.} \\
~~~\scriptsize{Tip and tilt have been removed.}\\
\scriptsize{$\Diamond$ At operative temp. (23$^\circ$C) and voltage (2000 v). Lab. calib. Zygo, INTA.}\\
\scriptsize{$\otimes$ (re-scaled Teles.+ISLiD)+(IMaX excluding PD-plate and considering in the etalon} \\ \scriptsize{~~~ only the low-order aberration terms except for the defocus; Zygo).}
\end{tabular}
\label{tablesystems_lab}
\end{table}

Initially, at a first stage we represented the model for low-order aberrations in our numerical experiments by re-scaling the coefficients of the nominal design by a factor accounting for the thermal effects and optical tolerancing\footnote{\sf Let $\left\{ \alpha_{j}, j=1,2,...,J \right\}$ be the set of Zernike coefficients of the expantion in the Noll's basis, approaching a wavefront aberration. The orthogonality properties of these basis functions facilitate the calculation of the \emph{rms-WFE} as 
$\sqrt{\sum{\alpha_j^2}}$. Re-scaling the coefficients to provide a wavefront aberration with a given (\emph{rms-WFE})$^{\prime}$ consists of simply computing a new set of coefficients $\alpha_i^{\prime}$ such that $\alpha_i^{\prime}=\alpha_i$(\emph{rms-WFE})$^{\prime}/$(\emph{rms-WFE}).}, excluding the etalon (the contribution of which is explicitly introduced in the simulations by means of amplitude- and phase-screens supplied by the manufacturer) and the ISLiD-IMaX interface contribution. In Table~\ref{tablesystems} the re-scaling factors correspond to: {\bf Telescope+ISLiD} (with optical tolerancing and thermal effects, item 2) and {\bf IMaX} (with optical tolerancing and thermal effects, item 7), so that their values are 1/8.25 and 1/19, respectively. The resulting sets of re-scaled  Zernike coefficients were summed up to give the desired low-order aberrations model (column 4 in Table~\ref{tablebudget}).\\

A criticism to this procedure is that the simple re-scaling preserves the distribution curve of the nominal Zernike coefficients, which might be a rather arbitrary statement because the contribution from optical tolerancing and thermal effects could modify in a non-proportional way the different aberration terms. Having this in mind, we changed, according to our possibilities, the criterion to construct the model of aberrations affecting the images in IMaX.\\

\begin{figure}
\begin{center}
\vspace{-1.2cm}\includegraphics[angle=90,width=1.\linewidth]{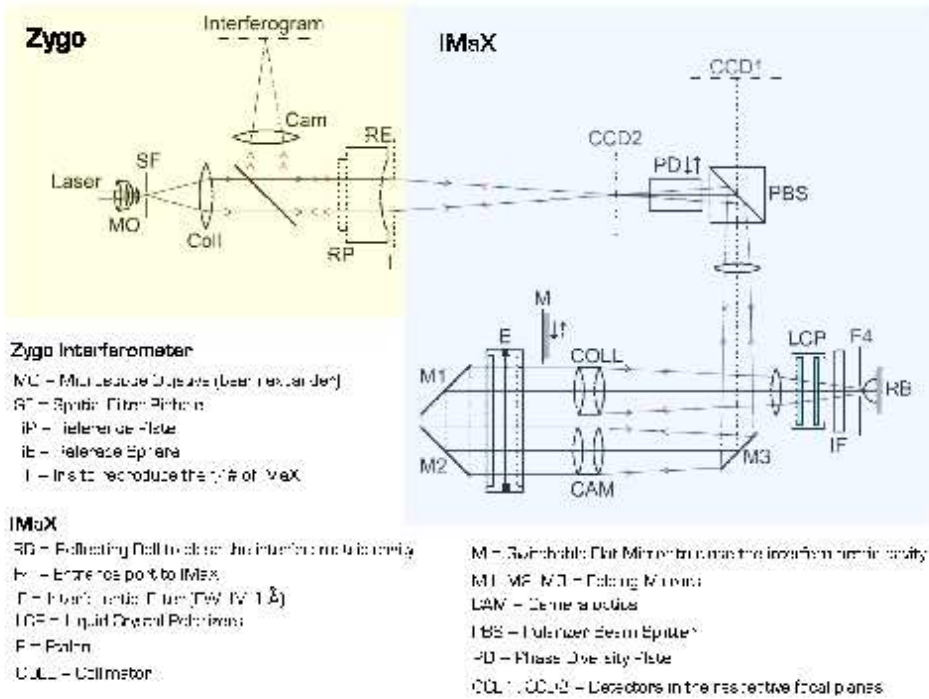}\\
\includegraphics[width=.9\linewidth]{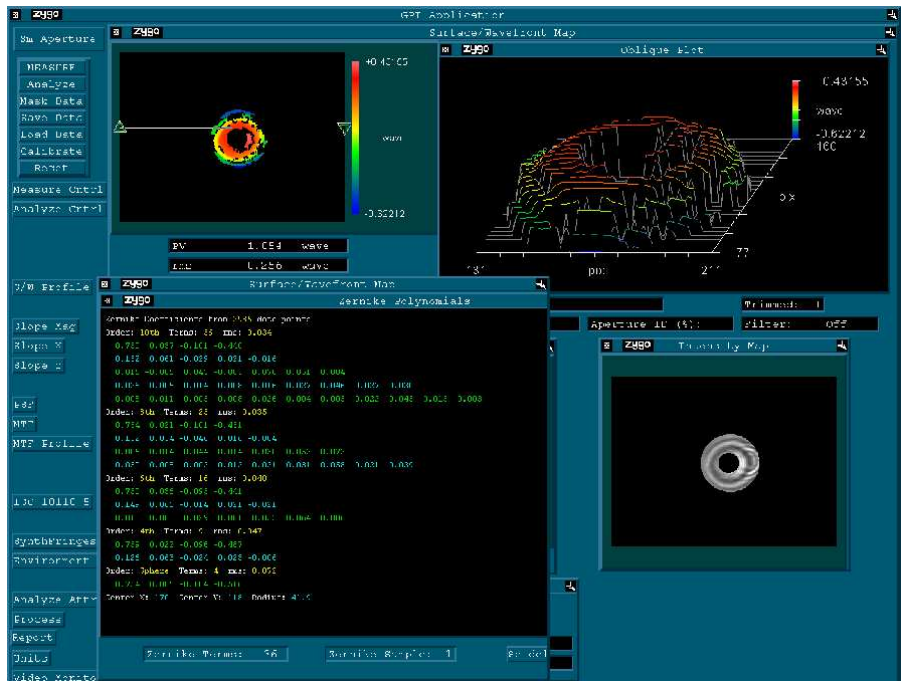}
\caption[\sf Optical configuration of the system including IMaX and the Zygo interferometer when calibrating the optical aberrations.]{\sf Optical configuration of the system including IMaX and the Zygo interferometer when calibrating the optical aberrations of the IMaX instrument (\emph{upper figure}) and a snapshot of the Zygo software interface (\emph{lower figure}).}
\label{Zygo}
\end{center}
\end{figure}

Table~\ref{tablesystems_lab} complements Table~\ref{tablesystems} by adding new empirical measurements and other considerations regarding the wavefront errors. Once IMaX has been manufactured and assembled (excluding etalon and PD-plate), the aberrations have been calibrated in the laboratory by using a Zygo Interferometer at INTA. Figure~\ref{Zygo} shows the arrangement in the laboratory including IMaX and Zygo, sketching the optical elements and the light paths along the whole system when calibrating the optical aberrations of IMaX. The etalon and PD-plate drawn in the sketch have been actually removed for this calibration as mentioned above.\\

From these measurements we obtained the Zernike coefficients approaching the real IMaX aberrations (excluding etalon and PD-plate) at laboratory conditions\footnote{\sf Atmospheric pressure and room temperature about 20$^\circ$C. At a first stage, laboratory experiments for checking the thermal behaviour of the system have not been performed.} (Column 5 in Table~\ref{tablebudget}). The corresponding value for the \emph{rms-WFE} is 1/16.8 waves (item 1 in Table~\ref{tablesystems_lab}).\\

\begin{figure}
\begin{center}
\includegraphics[width=1.\linewidth]{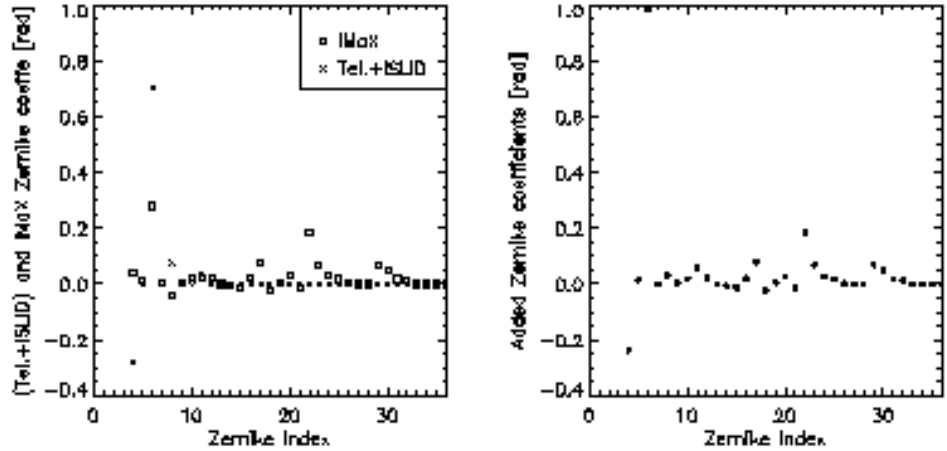}
\caption[\sf Plot of the Zernike coefficients representing the IMaX and Telescope+ISLiD degradation contributions]{\sf Plot of the Zernike coefficients representing the IMaX (excluding etalon and PD-plate) and Telescope+ISLiD degradation independently (\emph{left panel}), and their added contribution (\emph{right panel}). The coefficients correspond to the Noll's basis, starting from the term of index 4 (defocus).}
\label{coeffs_Tel-ISLiD}
\end{center}
\end{figure}

Concerning the Telescope+ISLiD contribution, we do not have the laboratory measurements and therefore we have preserved the procedure of using the coefficients from the nominal design after rescaling by a factor accounting for the average sources of error (optical tolerancing and thermal effects; item 2 in Table~\ref{tablesystems}) numerically modelled (column 6 in Table~\ref{tablebudget}). Figure~\ref{coeffs_Tel-ISLiD} (\emph{left panel}) plots the Zernike coefficients for IMaX as measured at the laboratory and the Telescope+ISLiD coefficients as re-scaled from the nominal ones. The total added contribution of both (Column 7 in Table~\ref{tablebudget}) is plotted in Figure~\ref{coeffs_Tel-ISLiD} (\emph{right panel}) and represents the set of coefficients we are going to use in the simulations. In all cases we present the Zernike coefficients corresponding to the Noll's basis and starting from coefficient $\alpha_4$ (defocus). The terms for piston, tip and tilt are excluded from our model since they do not represent figure errors.\\

The calculation of the \emph{rms-WFE} from the quadratic sum of coefficients in our model produce a joint aberration of 1/6 waves (item 2 in Table~\ref{tablesystems_lab}) to be compared with the value RSS (=1/7.6 waves) derived from items 2 and 7 in Table~\ref{tablesystems} that represent the global error effect (excluding the ISLiD-IMaX interface and the etalon) as modelled with \emph{CODE V\,\footnote{\sf Software by Optical Research Associates used at INTA for designing and analysing the IMaX optical configuration.}}. The difference might be partially ascribed to the approximate behavior of the RSS parameter representing the joint aberration of various optical systems.\\

\subsubsection{\bf Phase and transmission errors from the etalon}
\begin{figure}
\centering
\includegraphics[angle=-90,width=.7\linewidth]{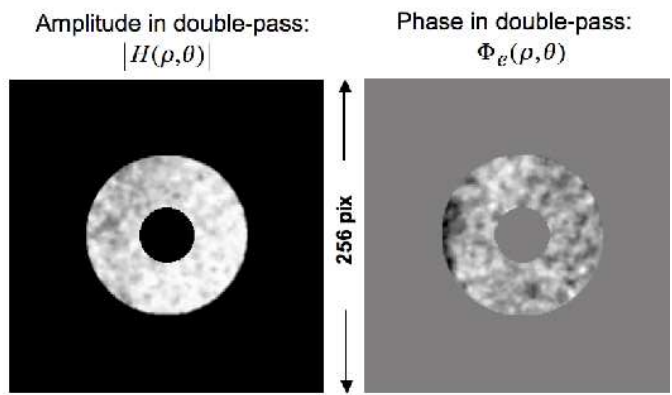}
\caption[\sf Amplitude and phase screens for the etalon.]{\sf Amplitude and phase aberration screens for the etalon (rms $\phi_e(\rho,\theta)$=1/26 waves).}
\label{etalonCSIRO}
\end{figure}

As described before, the etalon is placed onto a conjugate pupil of the SUNRISE system. Nevertheless, up to this step, we are not including in the model of aberrations the etalon contribution. The effects induced by the etalon in amplitude (transmission) $|H(\rho,\theta$)$|$ and phase $\phi_e(\rho,\theta)$ will be explicitly included in our simulations. The manufacturer, CSIRO\footnote{\sf Australian Commonwealth Scientific and Industrial Research Organisation,\\website: \emph{http://www.csiro.au/}}, provided computer files mapping the errors in amplitude and phase all over the etalon circular surface of 70 mm of diameter (this information is what we call amplitude- and phase-screens). Based on these maps we selected two optimal circular areas of 25 mm of diameter each. The criterion for this selection was to achieve the best compromise to minimize both the amplitude and phase errors. The etalon surface was oriented so that the selected circular areas fitted the position of the pupil image on the etalon operating in double pass mode. Considering the double pass through the selected areas we derived from the error maps supplied by the manufacturer, the amplitude- and phase-screens displayed in Figure~\ref{etalonCSIRO} that exhibit predominately high-spatial frequency structures (high-order Zernike modes in the phase-screen). The resulting \emph{rms-WFE} value is as small as 1/26 waves (Table~\ref{tablesystems_lab}, item 3). The residual piston and tip/tilt were removed from the phase-screen.\\

\begin{figure}
\centering
\includegraphics[width=1.\linewidth]{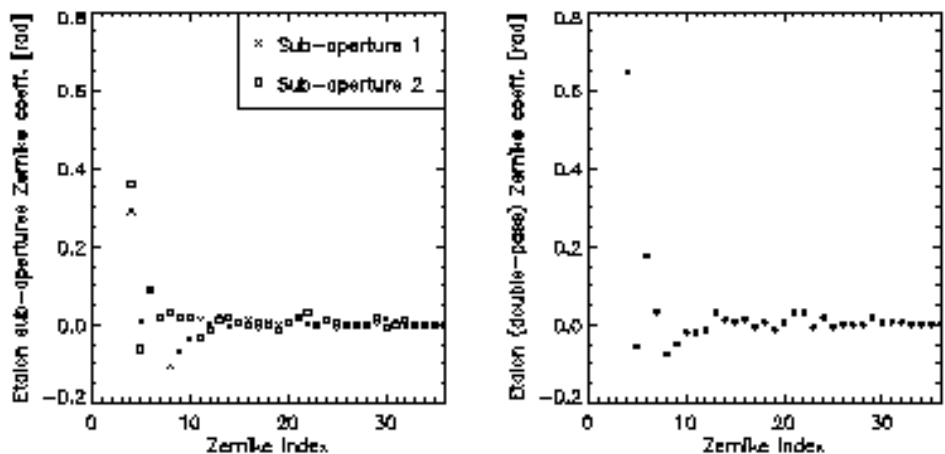}
\caption[\sf Plot of the Zernike coefficients representing the etalon degradation contribution]{\sf Plot of the Zernike coefficients representing the etalon degradation contribution from every sub-aperture (\emph{left panel}) and for the double-pass configuration (\emph{right panel}). The coefficients correspond to the Noll's basis, starting from the term of index 4 (defocus). Values for the rms of every single sub-aperture are 1/18.8 and 1/16.4 waves, respectively, and the one for the complete double-pass configuration 1/9.2 waves.}
\label{coeffs_etalon}
\end{figure}

Nevertheless, the problem gets complicated when enclosing the etalon in the pressurized and thermalized oven (the \emph{thermally controlled enclosure}). Interferometric measurements (with Zygo at INTA) of the wavefront error in double-pass and parallel-beam configuration, unveiled in the joint etalon+oven system a significant extra contribution of low-order Zernike modes (column 8 in Table~\ref{tablebudget} and Figure~\ref{coeffs_etalon}) with a \emph{rms-WFE} of 1/9.2 waves (item 4 in Table~\ref{tablesystems_lab}). Since this value comes from the measured Zernike coefficients, that do not fundamentally describe the high-spatial frequencies detected in the calibration maps from CSIRO, we will assume that both contributions are complementary and additive and so we will treat them in our simulations. The origin of low modes is not fully-characterized yet  though we speculate they are originated  by deformations of the oven windows caused by mechanical stresses or by the bulge of the etalon itself when applying high voltages (ranging from 0 to 2000 volts) while scanning our working spectral line.\\ 

\begin{figure}
\centering
\includegraphics[width=1.\linewidth]{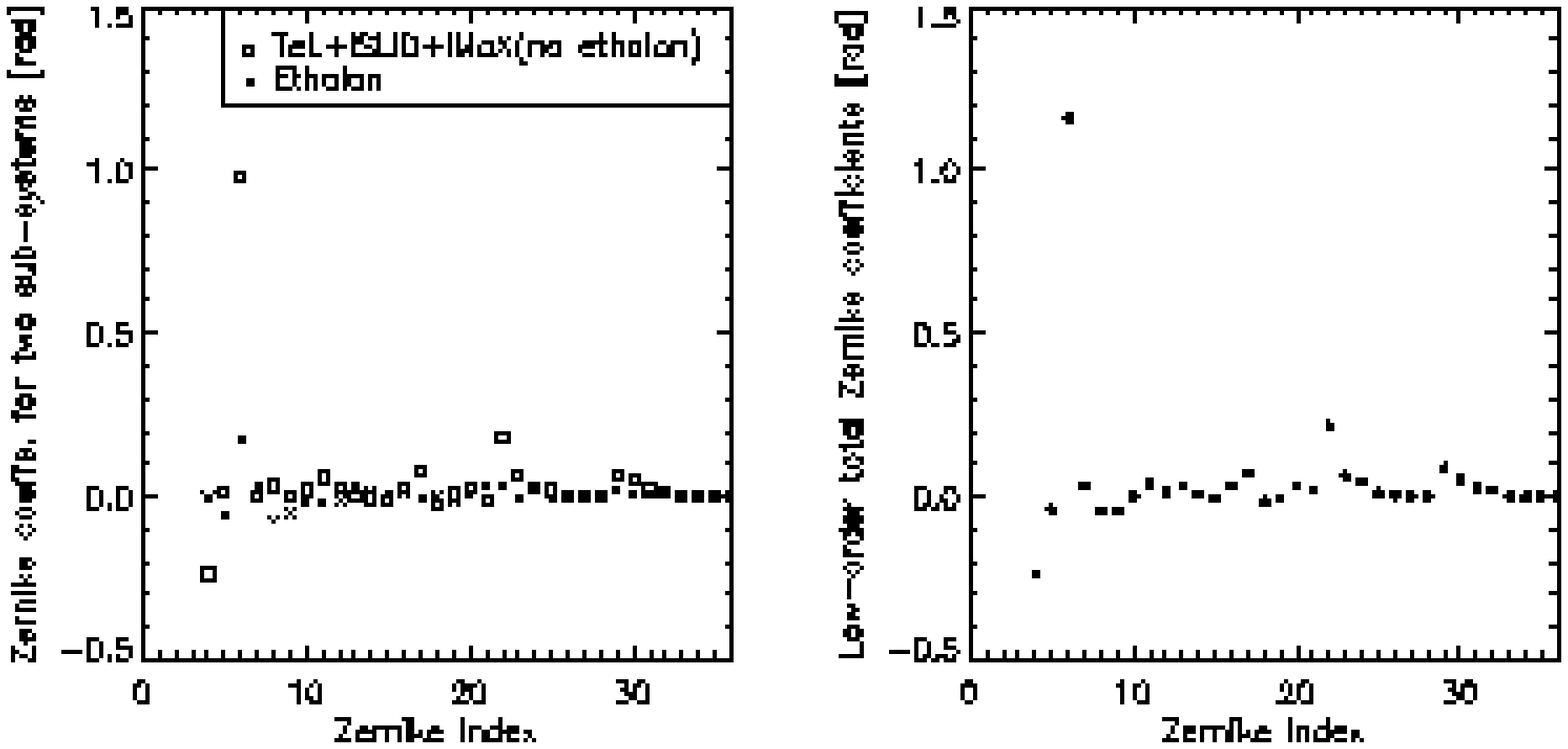}
\caption[\sf Plot of the Zernike coefficients representing the low-order total contribution]{\sf Plot of the Zernike coefficients representing the low-order contribution from the two sub-systems: 1) Telescope+ISLiD+IMaX {\scriptsize excluding etalon and PD-plate} (\emph{also shown in the right panel of Figure~\ref{coeffs_Tel-ISLiD}}), and 2) etalon with the defocus coefficient set to zero(\emph{compared with right panel of Figure~\ref{coeffs_etalon}}), independently (\emph{left panel}). The total added contribution is shown in the \emph{right panel}.}
\label{coeffs_total}
\end{figure}

Note that the major contribution to low-order modes in the etalon corresponds to the defocus coefficient ($\sim$0.65 rad or 3 mm of displacement), and we are confident it can be compensated when coupling the etalon+oven system into IMaX by optimizing the position of the image focal plane\footnote{\sf By means of an MTFs optical bench.}. For this reason this defocus contribution will be considered as null in our simulations (first value in column 8 in Table~\ref{tablebudget}), so that the \emph{rms-WFE} ascribed to the etalon+oven in low modes shrinks from 1/9.2 to 1/28.5 (item 5 in Table~\ref{tablesystems_lab}). The rest of the Zernike coefficients will be added to the low-order Zernike coefficients characterizing the system Tel+ISLiD+IMaX (column 7 in Table~\ref{tablebudget}) thus resulting the total budget of low-order aberration terms affecting the image formation in IMaX (column 9 in Table~\ref{tablebudget} and Figure~\ref{coeffs_total}) with a \emph{rms-WFE}=1/5.2 waves (item 6 in Table~\ref{tablesystems_lab}). The total phase error induced by low-order aberrations will be denoted hereafter as $\phi_Z(\rho,\theta)$.\\

In the final array of Zernike coefficients remains a non-zero value for defocus stemming mainly from ISLiD. This coefficient can not be set to zero in the simulations as we did in the case of the etalon defocus since the ISLiD system has been designed and integrated with total independence with respect to IMaX. Thus, we have not any chance to compensate the defocus contribution from ISLiD when fixing the optimum position of the image focal plane during the IMaX integration.\\

\subsubsection{\bf Main mirror polishing errors}
Another effect we are meant to include in our numerical experiment is the so-called \emph{ripple}. Commented briefly in section~\S\ref{identify} this error is caused by the polishing tool on the main mirror surface of the SUNRISE telescope. The wavefront error induced by this effect is quantified by a certain \emph{rms-WFE} that will be, from now on, represented as \emph{rms-ripple} and modelled by a phase screen arbitrarily chosen out of a sample of realizations which average power spectrum matches a von Karman\footnote{\sf Theodore von Karman (1881-1963), originally from Hungary, is responsible for many advances in aerodynamics, supersonic and hypersonic airflow characterization, among others. The von Karman spectrum is a power spectrum of the refractive index fluctuations describing the atmospheric turbulence. Thus, phase errors caused by atmospheric turbulence are also standardly modelled to match, in average, a von Karman power spectrum for given values of the \emph{inner-} and \emph{outer-scale} of turbulence. \emph{Inner-} and \emph{outer-scales} mean the smallest/largest spatial scales of the fluctuations also referred to as smallest/largest eddies in the turbulent medium.} power spectrum, for an \emph{outer-scale} equal to the size of the polishing tool (hereafter  \emph{ripple-scale}). The size assigned to the \emph{ripple-scale} in our simulations is 30 cm. Note that the amount of \emph{rms-ripple} we are considering is derived from the wavefront and consequently it is twofold the \emph{ripple} error in an optical surface working by reflection. The total phase error induced by ripple will be denoted hereafter as $\phi_{\emph{ripple}}(\rho,\theta)$.\\

\begin{figure}[h]
 \hfill
\begin{minipage}[h]{.3\textwidth}    
\caption[\sf Ripple screen]{\sf Ripple screen randomly chosen out of a sample of many realizations (von Karman power spectrum). The figure displays an example illustrating the contribution of high-order aberrations induced by the polishing tool.}
\label{ripple_screen}
    \end{minipage}
    \begin{minipage}[h]{.60\textwidth}
      \hspace{1.5cm}\epsfig{,file=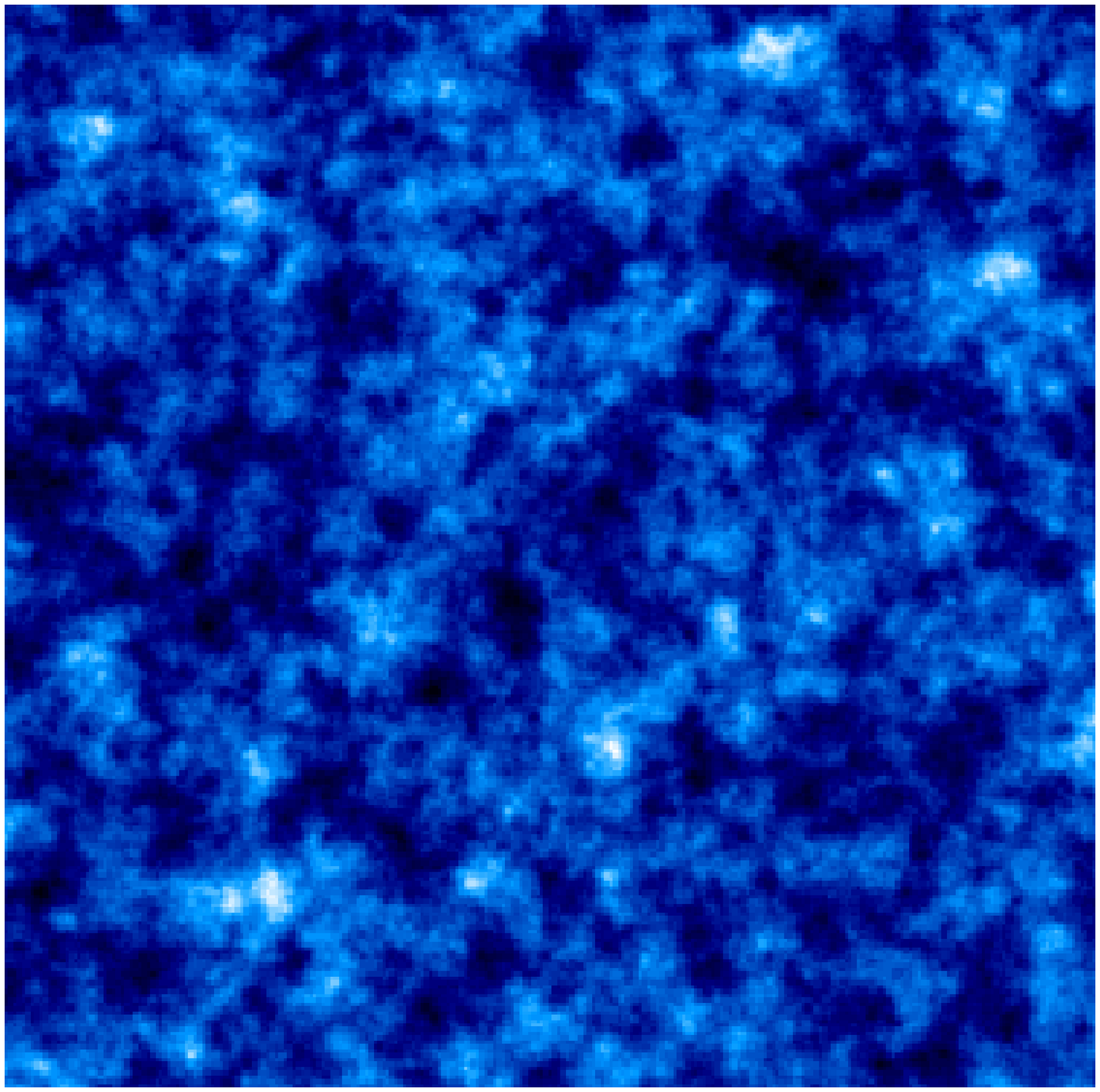, scale=0.25}
  \end{minipage}
  \hfill
\end{figure}

The \emph{ripple} can be classified as a high-order modes contribution to the global WFE. Figure~\ref{ripple_screen} shows a \emph{ripple screen} realization.

\vspace{3mm}

\subsubsection{\bf Phase Diversity plate}
\begin{figure}
\centering
\begin{tabular}{cc}
\includegraphics[angle=-90,width=.48\linewidth]{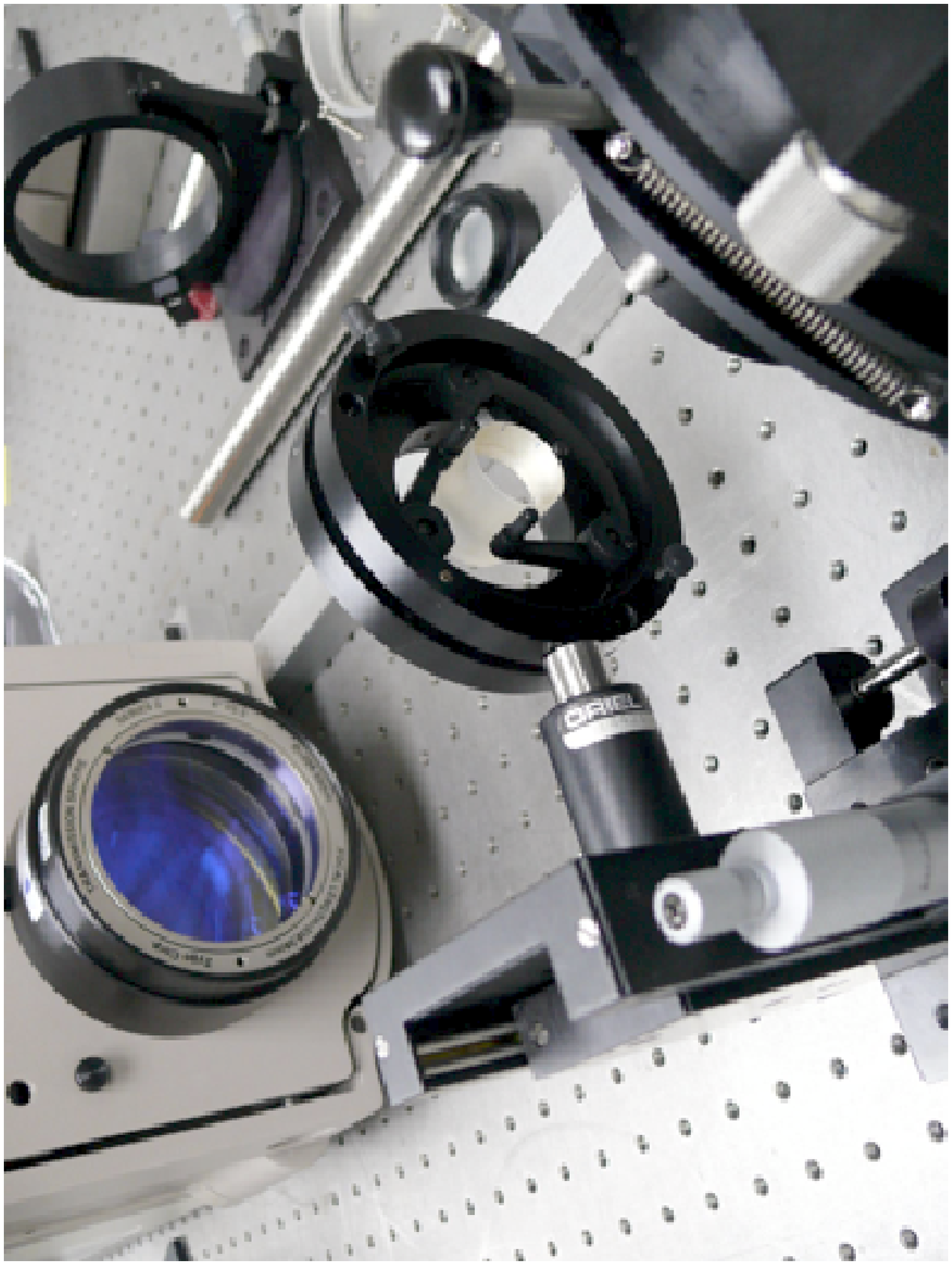} &
\includegraphics[angle=-90,width=.48\linewidth]{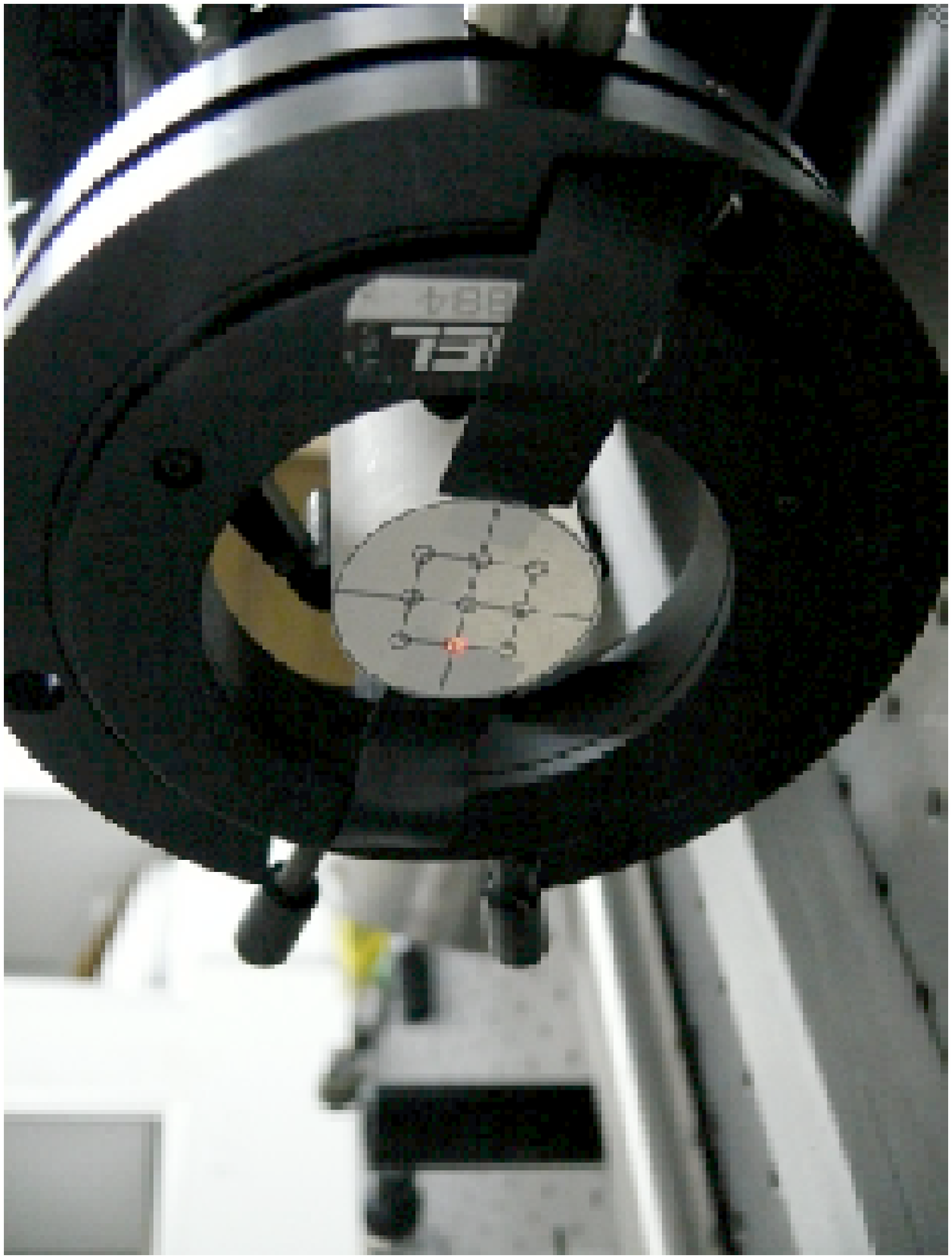}
\end{tabular}
\caption[\sf PD-plate during interferometric calibration]{\sf PD-plate during the interferometric calibration at the IAC. \emph{Left}: Zygo interferometer and PD-plate optical configuration. \emph{Right}: Cross-section of the PD-plate showing the positions and areas (\emph{footprints}) where the WFEs are evaluated. \emph{Courtesy of the IAC optical engineer F\'elix Gracia T\'emich}.}
\label{PD_plate}
\end{figure}

As commented in section~\S2.3 the PD-plate will introduce a defocus in one of the cameras. Apart from that, we were interested in finding out other possible contributions from the PD-plate to the error budget of the global system. To that aim an interferometric calibration of the isolated glass plate was made at the IAC optical laboratory\footnote{\sf The laboratory conditions were: temperature 23$^\circ$C and humidity 38$\%$.} by means of a Zygo interferometer. The aberrations were evaluated in 9 footprints of the IMaX telecentric beam (1.88 mm of diameter each) over the plate, arranged as shown in Figure~\ref{PD_plate} (\emph{right panel}). A view of the optical setup for the calibration is shown in Figure~\ref{PD_plate} (\emph{left panel}).\\
 
\begin{figure}
\centering
\includegraphics[width=.9\linewidth]{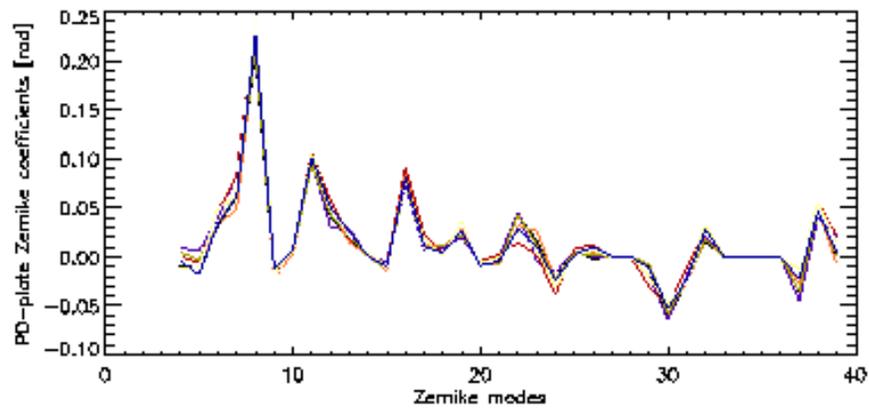} 
\caption[\sf Zernike coefficients of the PD-plate WFE for 9 footprints over plate]{\sf Zernike coefficients of the PD-plate WFE for 9 footprints over the plate (\emph{different colors}).}
\label{coeffs_PD_plate}
\end{figure}

The final results expressed by the Zernike coefficients for each single \emph{footprint} are plotted in Figure~\ref{coeffs_PD_plate}, and the obtained \emph{rms-WFE} is, in average, 1/21.3 waves. Although this is a small contribution, we suspect that it has been over-valuated (i.e.\ the aberration must be still smaller) because of some parasitic structures detected in the interferograms probably caused by reflections at the front and back faces of the plate. This reflections are the consequence of an inappropriate light source wavelength differing from the IMaX working wavelength for which the surface coating was designed and manufactured. In consequence, apart from its inherent task of displacing the image plane, we will assume a null contribution from the PD-plate to the error budget.\\

\begin{table}
\sffamily
\centering
\caption[\sf Thermal model for the PD-plate]{\sf Thermal model for the PD-plate (May 2008).}
\begin{tabular}{lccc}\\
&& \footnotesize{PD INDUCED DEFOCUS}\\
\footnotesize{THERMAL CASE} & \footnotesize{T [$^\circ$C]} & \footnotesize{Focus displacement [mm]} & \footnotesize{PV [waves]}\\\hline
Hot ............... & 25.0 & 8.46 & 1.008 \\
Operative ...... & 24.5 & 8.51 & 1.015\\
Cold .............. & 24.0 & 8.56 & 1.019 \\\hline
\end{tabular}
\label{temperatures}
\end{table}

Based on the IMaX thermal study\footnote{\sf Performed in May 2008 by Carmen Pastor Santos at INTA on the basis of the new thermal model calculations.}, Table~\ref{temperatures} lists for three different cases within the expected temperature variations range, the amount of defocus (in mm) induced by the PD-plate.

\subsubsection{\bf Detector contribution}
The CCD mainly produces three effects that degrade the image quality \citep{boreman2001}, namely:

\begin{itemize}
\item Detector footprint effect: It consist of the integration of the image information over the surface of the detector elements since they have a physical size.\\

A detector element with dimensions $a_x$ and $a_y$ performs a spatial averaging of the irradiance falling onto its surface, that in the frequency domain is equivalent to a spatial filtering represented by a transfer function OTF$_{\emph{footprint}}$, mathematically expressed as

\begin{equation}
\mbox{OTF}_{\emph{footprint}}(\nu_x,\nu_y)=\frac{\sin(\pi \nu_x a_x)}{\pi \nu_x a_x} \cdot \frac{\sin(\pi \nu_y a_y)}{\pi \nu_y a_y},
\label{ccdfootprint}
\end{equation}

where $\nu_x$, $\nu_y$  represent the components of the spatial frequency.

The smaller the pixel size the broader the resulting transfer function and, therefore, the lesser the smoothing in the image when recorded by the CCD.

\item Sampling effect: A sampled-imaging system is not shift-invariant and the position of the light reaching the detector, with respect to the pixels, will affect the final image. This effect is commonly treated as a statistical average of the relative image locations respect to the CCD. 

\item Crosstalk effect: Charge-carrier diffusion\footnote{\sf The absorption of photons in a semiconductor material is wavelength dependent: high for short-wavelength photons and decreases for longer-wavelength photons. With less absorption the long-wave photons penetrate deeper into the material and thus the charges generated should travel longer paths to be collected.} and charge-transfer inefficiency\footnote{\sf It is caused by incomplete transfer of charge packets along the CCD delay line.} over the detector induce a spurious signal on the neighbourhood.
\end{itemize}

In what follows, we only model in our simulations the first effect, which is the most significant.

\subsubsection{\bf Noise}
Typically, the noise in the image recorded by a CCD has two components: the photon noise and the readout noise. Due to the high performance of the CCDs in IMaX we can neglect, for the sake of simplicity, the contribution from the readout noise. We will simulate the photon noise as having a Gaussian distribution with a given rms. As an IMaX requirement and in order to reach a spectral sensitivity of 10$^{-3}$ decisive to perform precise polarimetric measurements, the value we adopt in our simulations is \emph{rms-noise}=10$^{-3}$ times the signal in the spectral continuum, i.e.\ a SNR=10$^3$ in the continuum.

\subsection{Numerical simulations and results}
\label{numsim}
In order to do a systematical presentation of the numerical experiments we are going to describe hereafter for various aberration assumptions, and taking into account all the different error sources described above, we consider appropriate to classify these error contributions in three main groups, as follows:

\begin{enumerate}
\item {\bf Low-order aberrations (LOA)} caused by the following optical sub-systems: SUNRISE telescope (excluding polishing errors), ISLiD, Interface ISLiD-IMaX, and IMaX excluding high-spatial frequency inhomogeneities in the etalon but preserving figure errors caused by deformations in the etalon and in the windows of its thermally controlled enclosure.\\

These aberrations, except for those from the ISLiD-IMaX Interface which will not be explicitly considered, are modelled by the set of Zernike coefficients listed in column 9 of Table~\ref{tablebudget} with a rms wavefront error \emph{rms-LOA} = 1/5.2 waves (Table~\ref{tablesystems_lab}). This will be our \emph{reference value} for LOA in our simulations. So as to avoid the danger of going to one extreme or the other, whereby either excessively pessimistic considerations when evaluating the error budget, or the possibility that unexpected error sources could arise (e.g
uncontrolled spikes in the temperature or pressure variations), we will include other cases of aberration in our simulations, with \emph{rms-LOA} values that vary from that of the reference. To do that as simple as possible, we will re-scale by a certain factor the list of coefficients in column 9 of Table~\ref{tablebudget}. Thus, different cases of LOA will be simulated by assuming the following values for \emph{rms-LOA}: 0, 1/12, 1/7, 1/5, 1/4, 1/3 and 1/2 waves.\\

The efficiency of the PD-inversion is strongly dependent on the particular set of LOA terms included in the simulation. For instance, the problem becomes simpler when the main contribution to the aberration comes from very low order terms. Then, the inversion code gives good results by assuming a small number of unknowns (e.g.\ 15 or 21 Zernike coefficients or equivalently 15 or 21 equations). This simplifies the iterative solution of the equations system and speed up the convergence.\\

\item {\bf High-order aberrations (HOA)} caused by ripple in the telescope main mirror and by high-spatial frequency inhomogeneities in the etalon surface as described by laboratory calibration maps. Figure errors in the etalon, induced by high voltages are not included in this budget but in the LOA.\\

This HOA enter into the simulation code as respective phase-screens. The \emph{rms-WFE} measured for the etalon in double-pass configuration (\emph{rms-etalon}) is 1/26 waves (Table~\ref{tablesystems_lab}) and will be considered as a fixed value in all numerical experiments in which this effect is included. For the polishing errors three different cases will be considered: \emph{rms-ripple}=0, 2/60 and 2/28 waves. The way we write down the precedent values as fractions with a nominator 2 is meant to denote the effect of reflection of the wavefront in the main mirror (e.g.\ in the case 2/60 we are adding two contributions: 1/60 + 1/60, each of them corresponding to the \emph{rms-ripple} characterizing the mirror surface). Apart from the etalon phase-screen, we also have available an amplitude-screen mapping its transmission inhomogeneities.

\item {\bf Pixel integration in the CCD and Noise}. The first effect is simulated by an analytical formula (see equation~\ref{ccdfootprint}) as a function of the pixel size (fixed value equals 12 $\mu$m). The noise contribution is introduced into the code as noise-screens generated by an IDL\footnote{\sf The Interactive Data Language is a widely known programming language that is a popular data analysis language among scientists.} simulator of pseudo-random signals with Gaussian distribution for a given rms value. As justified above, we take for all considered cases a fixed value \emph{rms-noise}=10$^{-3} \times$ (the signal in the spectral continuum).\\

\end{enumerate}

In what follows, a variety of aberration models will be made up by combining weighted contributions from different error sources. In an attempt to approach the real observations when recording bursts of 30 image-pairs for calibrations (see section~\S2.4), 30 realizations of noise will be systematically performed for every aberration model adopted.

Table~\ref{num-experiments} reports on the different ingredients composing the aberration models considered in every numerical experiment. The results of these experiments are displayed in Figures from \ref{tiras-exp1} to \ref{caso4_2}.

\subsubsection{\bf Key notes for the interpretation of Figures \ref{tiras-exp1} to \ref{caso4_2}}

Figures~\ref{tiras-exp1}, \ref{tiras-exp2}, \ref{tiras-exp3}, \ref{tiras-exp4} display PD-speckle reconstructed images from 30 realizations for every aberration assumption. Lying below every restored scene we present one out of the 30 realizations of the in-focus degraded images. To facilitate the visual comparison between different images, all of them are displayed by employing a common gray-scale ranging from the minimum to the maximum intensity values of the \emph{true object} shown at the upper right corner for reference. This way, real differences in contrast and resolution can easily be detected by eye. \\

Apart from the visual comparison, we also calculate the values of different parameters to quantify the degree of success achieved in our numerical experiments basically consisting of: \emph{degradation followed by PD-inversion}. Figures~\ref{caso1_1}, \ref{caso1_2}, \ref{caso2_1}, \ref{caso2_2}, \ref{caso3_1}, \ref{caso3_2}, \ref{caso4_1} and \ref{caso4_2} are arranged in three columns. The panels in left, middle and right columns of the figures correspond to different assumptions for the \emph{rms-ripple}: 0, 2/60 and 2/28 waves, respectively. The abscissae in all panels stand for different cases of \emph{rms-LOA}. Bullets linked by \emph{dotted lines} show the mean values resulting from PD-inversions. Vertical bars represent the dispersion of these results caused by 30 different realizations of photon noise. \emph{Dashed lines} represent the values of the different parameters in the degraded (smeared) images simulated.

To evaluate the reliability of the results from the PD-process we need some expected reference values for comparison, e.g.\ the contrast of the \emph{true object} that should ideally by reached after restoration, or the zero value that should ideally result from the subtraction of certain wavefronts (see \emph{thick solid lines} in Figures~\ref{caso1_1}, \ref{caso2_1}, \ref{caso3_1} and \ref{caso4_1}). We also use as a reference the results of the restoration $i_Z$ performed with expression~1.21, by only employing the input Zernike terms simulating the LOA of the system, $\phi_Z$ (see \emph{thin solid lines} in Figures~\ref{caso1_1}, \ref{caso2_1}, \ref{caso3_1} and \ref{caso4_1}).\\

Subscripts \emph{t,s,r} and \emph{Z} in Figures~\ref{caso1_1}, \ref{caso2_1}, \ref{caso3_1} and \ref{caso4_1} stand for \emph{true object}, \emph{smearing} (simulated degradation), \emph{reconstruction} with PD, and restoration performed with only the \emph{Zernikes} representing the LOA in the input, respectively.\\

\subsubsection{\bf Experiment 1}
Combined effects considered (see also Table~\ref{num-experiments}).\\
\hspace{6mm}\fbox{
\hspace{0cm}\parbox[c]{11cm}{{
\footnotesize $\bullet$ \emph{rms-ripple} = 0, 2/60 and 2/28 waves.}\\
{\footnotesize $\bullet$ \emph{rms-LOA} = 0, 1/12, 1/7, 1/5, 1/4, 1/3 and 1/2 waves.}\\
{\footnotesize $\bullet$ \emph{rms-noise} = 10$^{-3} \times$ continuum signal (30 realizations per aberration model.) }\\
{\footnotesize $\bullet$ No CCD effects.}\\
{\footnotesize $\bullet$ No etalon effects:  $|H(\rho,\theta)|=$1 , $\phi_e(\rho,\theta)=$0.}\\
{\footnotesize $\bullet$ PD-defocus in mm (degradation/inversion):  8.51 / 8.51.}\\
{\footnotesize $\bullet$ Inversion with 25 coefficients.}}\\
}\\
\vspace{4mm}

\begin{table}
\sffamily
\centering
\caption[\sf Parameters for the numerical experiments]{\sf \footnotesize{Error contributions for diverse numerical simulations}}
\begin{tabular}{llcccc}\\
\footnotesize{ERROR} & \footnotesize{~~~~CONTRIBUTION}& \multicolumn{4}{c}{{\footnotesize EXPERIMENT}}\\
\footnotesize{SOURCE} & \footnotesize{~~~~~~~~~~~VALUE} & 1 & 2 & 3 & 4\\\hline\\
\footnotesize{\emph{rms-ripple} ...............} & $0$, $\frac{2}{60}$, $\frac{2}{28}$ \footnotesize{waves} & $\surd$ & $\surd$ & $\surd$ & $\surd$\\
\\
\footnotesize{\emph{rms-LOA} .................} & $0$, $\frac{1}{12}$, $\frac{1}{7}$, $\frac{1}{5}$, $\frac{1}{4}$, $\frac{1}{3}$, $\frac{1}{2}$ 
\footnotesize{waves} & $\surd$ & $\surd$ & $\surd$ & $\surd$\\
\\
\footnotesize{\emph{rms-noise} .................} & \footnotesize{10$^{-3}\times$} \footnotesize{continuum signal} & $\surd$ & $\surd$ & $\surd$ & $\surd$\\
& \footnotesize{(30 realizations/aberr.model)} &  &  &  \\
\footnotesize{\emph{rms-etalon} .................} & $\frac{1}{26}$ \footnotesize{waves} & -& $\surd$ & $\surd$ & $\surd$\\
\\
\footnotesize{Etalon amplitude ......} & \footnotesize{$|H(\rho,\theta)| \ne 1$} & - & $\surd$ & $\surd$ & $\surd$\\\\
\footnotesize{CCD .......................} & \footnotesize{12 $\mu$m/pix $\to$ 0.055 "/pix}  &- & - & $\surd$ & $\surd$\\\\
\footnotesize{PD-defocus ................} & \scriptsize{DEGRADATION / INVERSION} & & & &\\
& \footnotesize{8.51 mm} \scriptsize{(PV 1.00$\lambda$)} / \footnotesize{8.51 mm} & $\surd$ & $\surd$ & $\surd$ & -\\
& \footnotesize{9.00 mm} \scriptsize{(PV 1.06$\lambda$)} / \footnotesize{8.51 mm} & - & - & - & $\surd$\\ 
\\\hline
\end{tabular}
\label{num-experiments}
\end{table}

A glance at Figure~\ref{caso1_1}, rows (a), (b), (c), reveals that the results derived from the application of PD to the simulated observations and the expected values for different parameters (thick solid line) are in good agreement when \emph{rms-ripple}=0 or 2/60 waves (\emph{left and middle columns}) and \emph{rms-LOA}=0, 1/12, 1/7, 1/5 or 1/4 waves. Thus the retrieved rms contrast for $i_r$ ranges from 16$\%$ to 17.4$\%$ versus the reference rms($i_t$)=17.5$\%$; the mean absolute difference mean$(|i_t-i_r|)$ remains always  below 2.5$\%$ and rms($i_t-i_r$) below 3.5$\%$. In many cases the results are also reasonably good for \emph{rms-LOA}=1/3 waves but always noticeably worse for the case \emph{rms-LOA}=1/2 waves.\\
 
The figure shows that the results from the PD-inversions deviate progressively from the reference values as long as the amount of \emph{ripple} and LOA augment in the simulated aberrations. This behavior reflects the limitations of the method stemming from: 1) the unsensed aberrations due to the limited number of Zernikes used in the inversion (25 coefficients versus 36 employed to simulate LOA in the total system -see Table~\ref{tablebudget}; 2) the influence of the also non-sensed HOA terms present in the ripple\footnote{\sf In further experiments also the inhomogeneities in the etalon phase will bring additional HOA contribution.}; 3) the noise in the data; and 4) the instabilities in the PD-method itself.\\

The comparison of the \emph{bullet-dotted} with the \emph{dashed lines} evidences the enormous improvement obtained from PD-inversions with respect to the smeared images. It is worth mentioning that Figure~\ref{caso1_1}, rows (a), (b), (c), (d), show the good coincidence (within the above mentioned range \emph{rms-ripple} $\le$ 2/60 and \emph{rms-LOA} $\le$ 1/4) between the \emph{bullet-dotted} and the \emph{thin solid lines}, the latter representing the results of the deconvolution performed by only employing the Zernike coefficients assumed as the input to simulate LOA in the total system. This coincidence, even extensive to the third column, reveals that the PD-algorithm compensate reasonably well at least for the degradation caused in the simulation by LOA terms. Note also that, for a given parameter and for \emph{rms-LOA} $\le$ 1/4, the variation trend in the \emph{bullet-dotted} and \emph{thin solid lines} is quite similar in the three columns, being the main difference between these variations a vertical offset which increases from the left to the right columns. We ascribe these offsets to a loss of contrast caused by the ripple.\\

Ripple errors are high-frequency phase errors and can be viewed as noise in the wavefront since they can hardly be modelled even with very high order Zernike terms in a polynomial expansion. The consequence of these errors is a transfer of energy from the peak of the PSF to its far wings thus producing stray-light and consequently an unrecoverable loss of contrast as evidenced in Figure~\ref{caso1_1}, row (a) --~this degradation increases as the \emph{ripple-scale} decreases \citep{bello2001}. Thus, large values of \emph{rms-ripple} have more impact on the image contrast than on the image resolution since the latter is related to the FWHM of the PSF that, for well corrected systems, changes rather slowly\footnote{\sf According to equations~1.15, 1.16 and the auto-correlation theorem, the PSF of an optical system is proportional to the power spectrum of the generalized pupil function. Consequently, the presence of high-spatial frequencies in $\phi(\rho,\theta)$ will cause a long tail in the power spectrum or equivalently long wings in the PSF of the system. Furthermore, the energy conservation forces the normalization of the volume enclosed by the PSF; thus the extension of its wings will be done at the expenses of an intensity decrease at shorter distances, in particular at its central point. If, apart from the ripple, the optical system is well corrected, this transfer of energy from short to long distances will provoke a clear decrease in the central intensity of the PSF as long as the ripple increases, but because of the very steep slopes in the PSF core, the FWHM of the PSF will scarcely be affected.}. Note that the Strehl ratio is strongly dependent on the amount of ripple (see panels in row (f) of Figure~\ref{caso1_1}). The case \emph{rms-ripple}=2/28 has only an academic interest in our simulations because we expect a better polishing quality in the SUNRISE main mirror surface\footnote{\sf In fact the expected rms values for polishing errors as measured on the mirror surface range from 5 to 10 nm. For our working wavelength, $\lambda$525.02 nm, this range corresponds to 1/100 to 1/50 waves, which includes the value 1/60 assumed as the second case for \emph{rms-ripple} in our simulations.}.\\

Stray-light in solar photometry represents a problem that classically has been faced from various approaches, all of them based on modelling the far wings of the PSF by linear combinations of Gaussian and Lorentzian functions, the parameters of which are determined from solar aureole measurements \citep[][and references therein]{pillet1992}. In spectro-polarimetry, being the case of IMaX, stray-light is a particularly delicate issue because it represents a contamination of the magnetic signal at a given feature by the signal originated in other structures located all around the feature even at quite large distances (e.g.\ $\sim$ arc minutes).\\

\begin{figure}
\centering
\includegraphics[width=.9\linewidth]{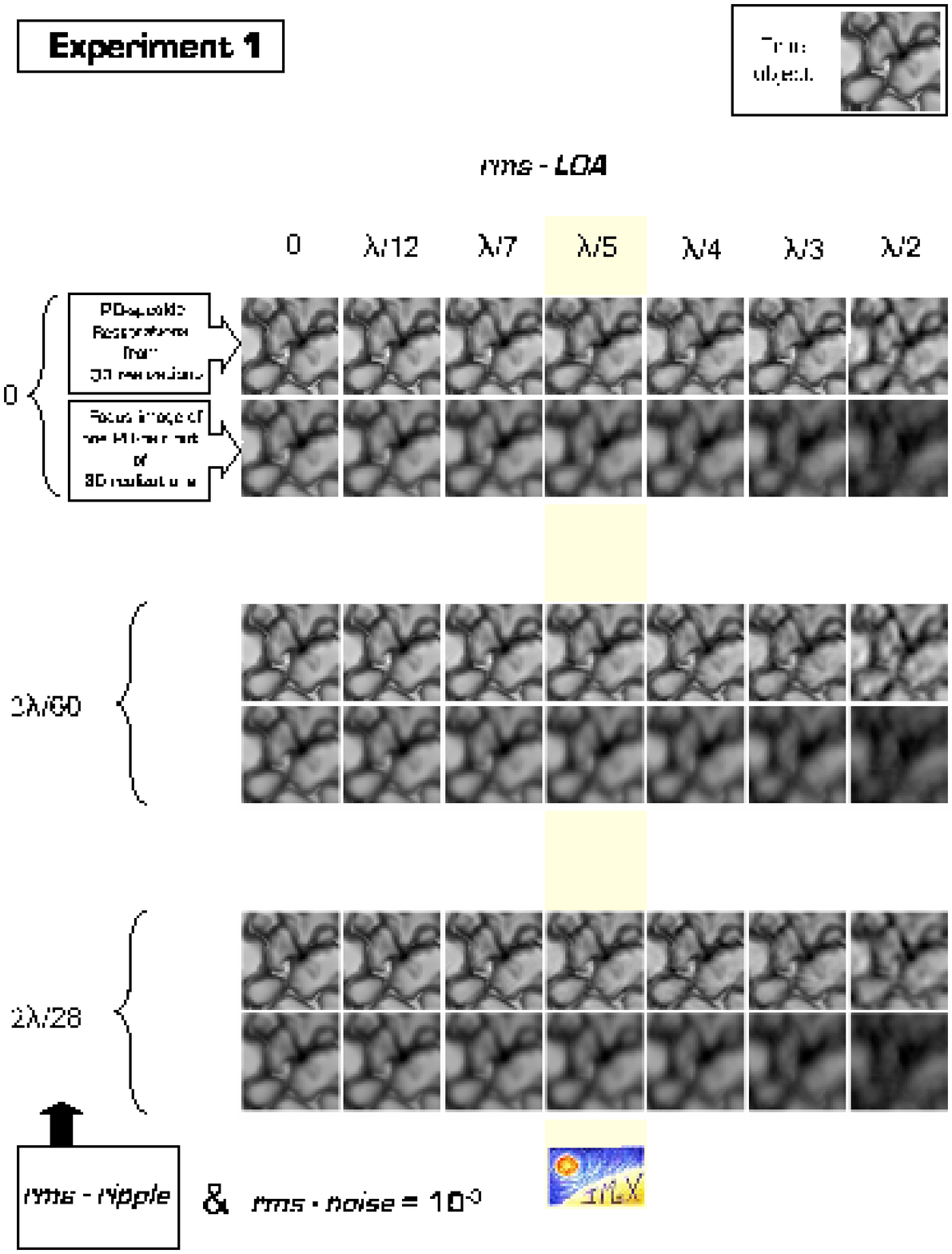} 
\caption[\sf PD reconstruction for different amounts of aberration: Experiment 1]{\sf Experiment 1: PD reconstruction for different amounts of aberration. Seven different contributions from low-order aberrations are adopted: \emph{rms-LOA} = 0, $\lambda/12$, $\lambda/7$, $\lambda/5$, $\lambda/4$, $\lambda/3$, $\lambda/2$, and for each of them three possible cases of polishing errors are considered: \emph{rms-ripple} = 0, $2\lambda/60$, $2\lambda/28$. The exam of the PD-speckle reconstructed images from 30 realizations allows us to check the robustness of the IMaX WFE calibration. The IMaX case is emphasized in \emph{yellow} and the so-called \emph{true object} representing the free-from-aberrations and -diffraction image is located in the upper right corner. All the images are displayed by employing a common gray scale ranging from the minimum to the maximum intensity values of the true object. Specific differences characterizing every Experiment are summarized at the bottom of the figure.}
\label{tiras-exp1}
\end{figure}

\begin{figure}
\centering
\begin{tabular}{c}
\footnotesize{{\bf \sf EXPERIMENT 1 ~/~ Inversions evaluation}}\\
\\
\hspace{-0.7cm}\includegraphics[width=1.05\linewidth]{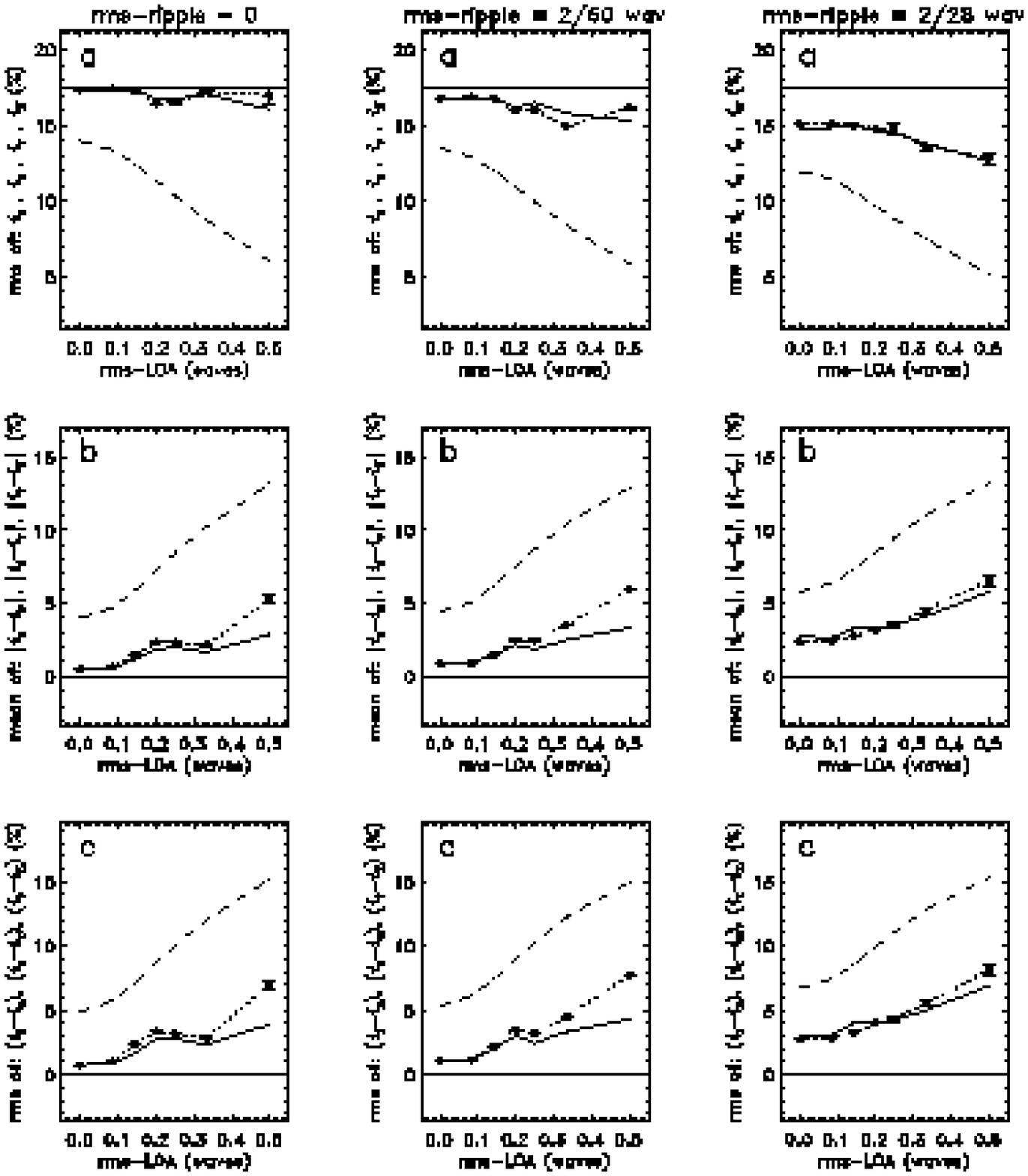} 
\end{tabular}
\caption[\sf Comparative plot for the numerical Experiment 1]{\sf Comparative plot for the numerical Experiment 1. To understand the meaning of the different symbols and labels see section~\S\ref{numsim}: \emph{Key notes for the interpretation of Figures~\ref{tiras-exp1} to \ref{caso4_2}.}}
\label{caso1_1}
\end{figure}

The loss of contrast and spatial resolution, as long as the \emph{rms-ripple} and \emph{rms-LOA} increase, can be visually appreciated in the strips of restored images in Figure~\ref{tiras-exp1}. Image resolution is particularly affected by \emph{rms-LOA} $>$ 1/4 waves: medium-high spatial frequency structures, grains-like or fringes-like patterns start arising. These artifacts originate because the OTF in these cases drops to close-to-zero values in a range at medium-high spatial frequencies\footnote{\sf This effect is favoured by the central obscuration in the SUNRISE's pupil inherent to the particular design adopted for this telescope. The central obscuration depresses the central region of the MTF in comparison with its typical shape for diffraction-limited telescopes with clear aperture. See Figure~\ref{imaxcaseMTF} illustrating this depression.} and are more evident in single-restored images before doing the speckle summation of equation~ 1.21 (see discussion at the end of section~\S1.4). Thus, the final product for single restorations is often arbitrarily affected by over restoration of some particular spectral components.\\

\begin{figure}
\centering
\begin{tabular}{c}
\footnotesize{{\bf \sf EXPERIMENT 1 ~/~ Zernike coefficients evaluation}}\\
\\
\hspace{-0.7cm}\includegraphics[width=1.05\linewidth]{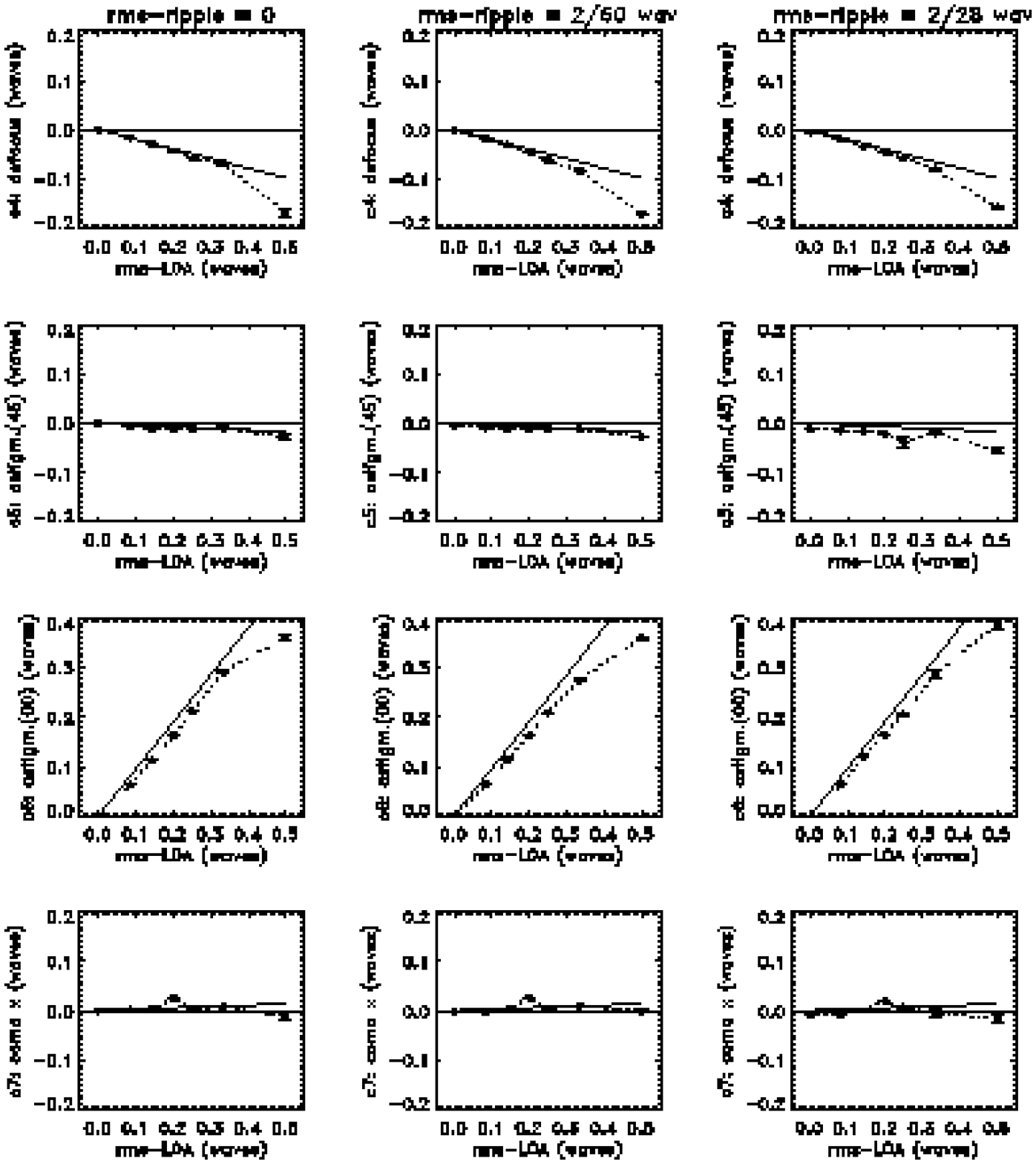} 
\end{tabular}
\caption[\sf Comparative plot of Zernike coefficients for the numerical Experiment 1]{\sf Comparative plot of Zernike coefficients for the numerical Experiment 1. \emph{Bullets} linked by \emph{dotted lines} represent the variation of the mean Zernike coefficients retrieved from the PD-inversions of 30 realizations. \emph{Thick solid lines} represent the variation of the original coefficients used as code inputs to simulate the degradation caused by low-order aberration (LOA) terms. Only the coefficients indexed from 4 to 11 are included in the figure.}
\label{caso1_2}
\end{figure}

In Figure~\ref{caso1_2}, \emph{thick lines} represent the variation of the Zernike coefficients (up to index 11) used as the input to simulate the LOA (i.e.\ the reference for comparison), whereas \emph{thin lines} mark simply the zero level. \emph{Bullet-dotted lines} represent again values retrieved from the PD-inversion process. In most cases the coincidence of the retrieved coefficients with the expected values is quite good even in the right column of the figure (we exclude from this comparison the case \emph{rms-LOA}=1/2 waves). Nevertheless, the deviations encountered from the reference values could probably be ascribed to cross-talk between the limited number of Zernike terms employed in the PD-inversions (25 terms) in an attempt to compensate for the contribution of higher-order terms which are in fact present in the simulated wavefront: we have used 36 terms to simulate the system LOA and, furthermore, we can not disregard the contribution from polishing errors and noise to the signal. In other words, as argued by \cite{lofdahl1994} \emph{the inversion program compensates for the missing high-order terms by introducing errors in the lower Zernike coefficients. To understand such compensations, we note that the inversion algorithm does not attempt to derive the best fit to the wavefront. Because of the chosen error metric (see equation~1.19), the inversion attempts to find two internally consistent (i.e.\ derived from the same Zernike coefficients) transfer functions for the focus and defocus images such that the error metric is minimized. The value of the transfer function at a specific frequency is given by an average over the aperture of all differences in phase taken at a fixed separation (see equation~1.15). Thus, mainly at low and intermediate frequencies, different wavefronts can give similar transfer functions. This gives the possibilities for compensations and allows the inversion procedure to find the wavefront that gives the best fit to the transfer functions given the observed data and the number of Zernike coefficients used to represent the wavefront. Nevertheless the improvement given by such compensations is limited because of the absence of high-spatial frequencies in the lower aberrations.}\\

The panels in rows (d) and (e) in Figure~\ref{caso1_1} deserve a particular comment. The inversion process yields 
rms($\phi_r$) values that fit rms($\phi_s$) very well, at least up to \emph{rms-LOA}=1/4 waves. In spite of this coincidence, the deviation of the parameter rms($\phi_s-\phi_r$) from zero increases as the \emph{ripple} and the LOA augment. This variation preserves a similar trend in the three columns of row (e) although it is affected by a vertical offset in the middle and right columns. Note that the parameter rms($\phi_s-\phi_Z$) is also affected by similar offsets. In fact these offsets coincide, as expected, with the \emph{rms-ripple} value assigned to each case. The slope in the variation of rms($\phi_s-\phi_r$) could be justified in terms of the compensations mentioned in the previous paragraph (cross-talk between low order modes), which reproduce good transfer functions but not so good wavefront shapes. Apparently, the difference in the wavefront shape does not imply a substantial difference between rms($\phi_s$) and rms($\phi_r$) as shown in the panels of row (d).

\begin{figure}
\centering
\begin{tabular}{c}
\footnotesize{{\bf \sf EXPERIMENT 1 ~/~ Inversions evaluation (\sf \emph{cont})}}\\
\\
\hspace{-0.7cm}\includegraphics[width=1.05\linewidth]{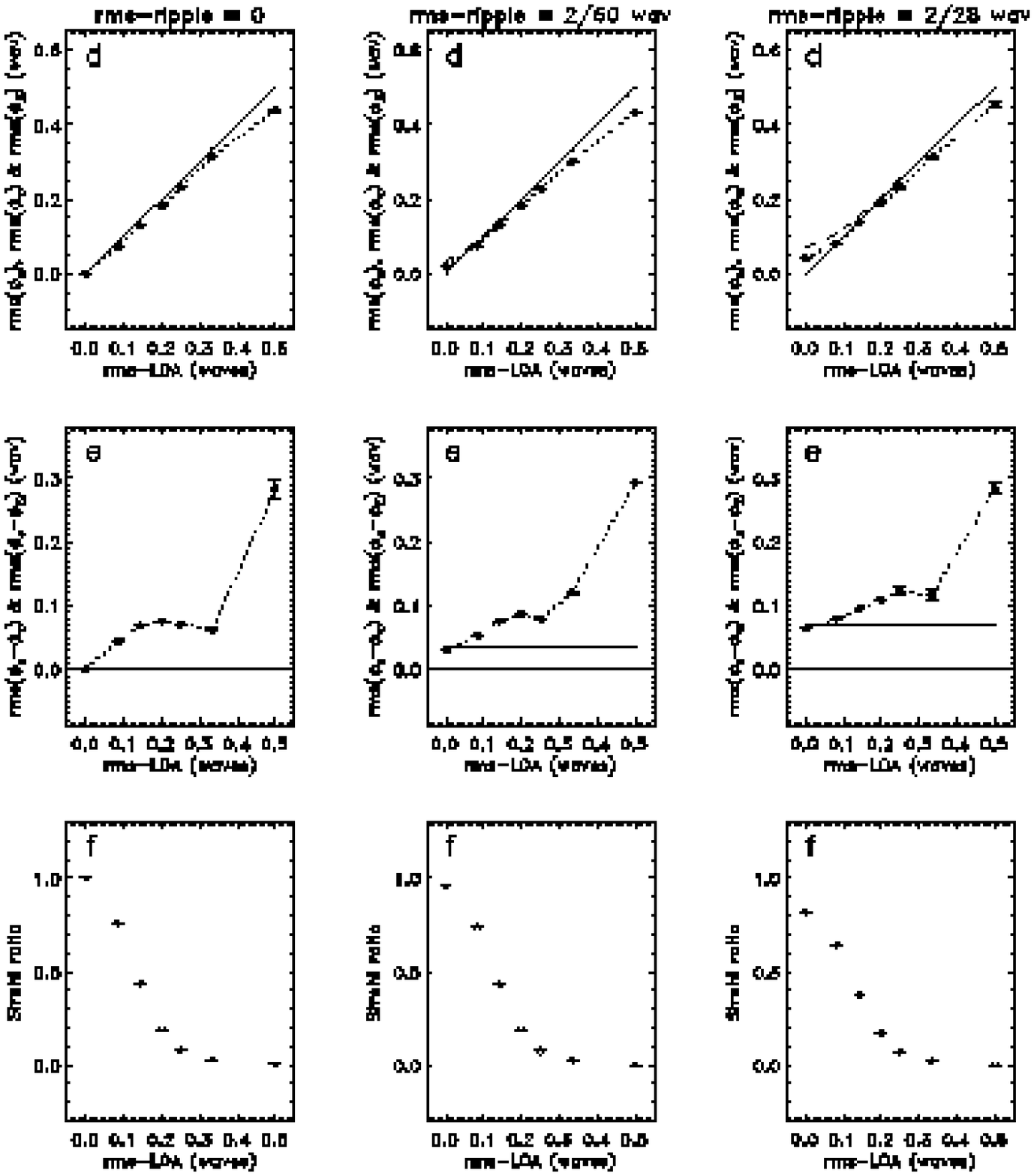} 
\end{tabular}
\vspace{6mm}
\flushleft {\footnotesize FIGURE \ref{caso1_1} --- \sf \emph{(cont).}}
\end{figure}

\begin{figure}
\centering
\begin{tabular}{c}
\footnotesize{{\bf \sf EXPERIMENT 1 ~/~ Zernike coefficients evaluation (\sf \emph{cont})}}\\
\\
\hspace{-0.7cm}\includegraphics[width=1.05\linewidth]{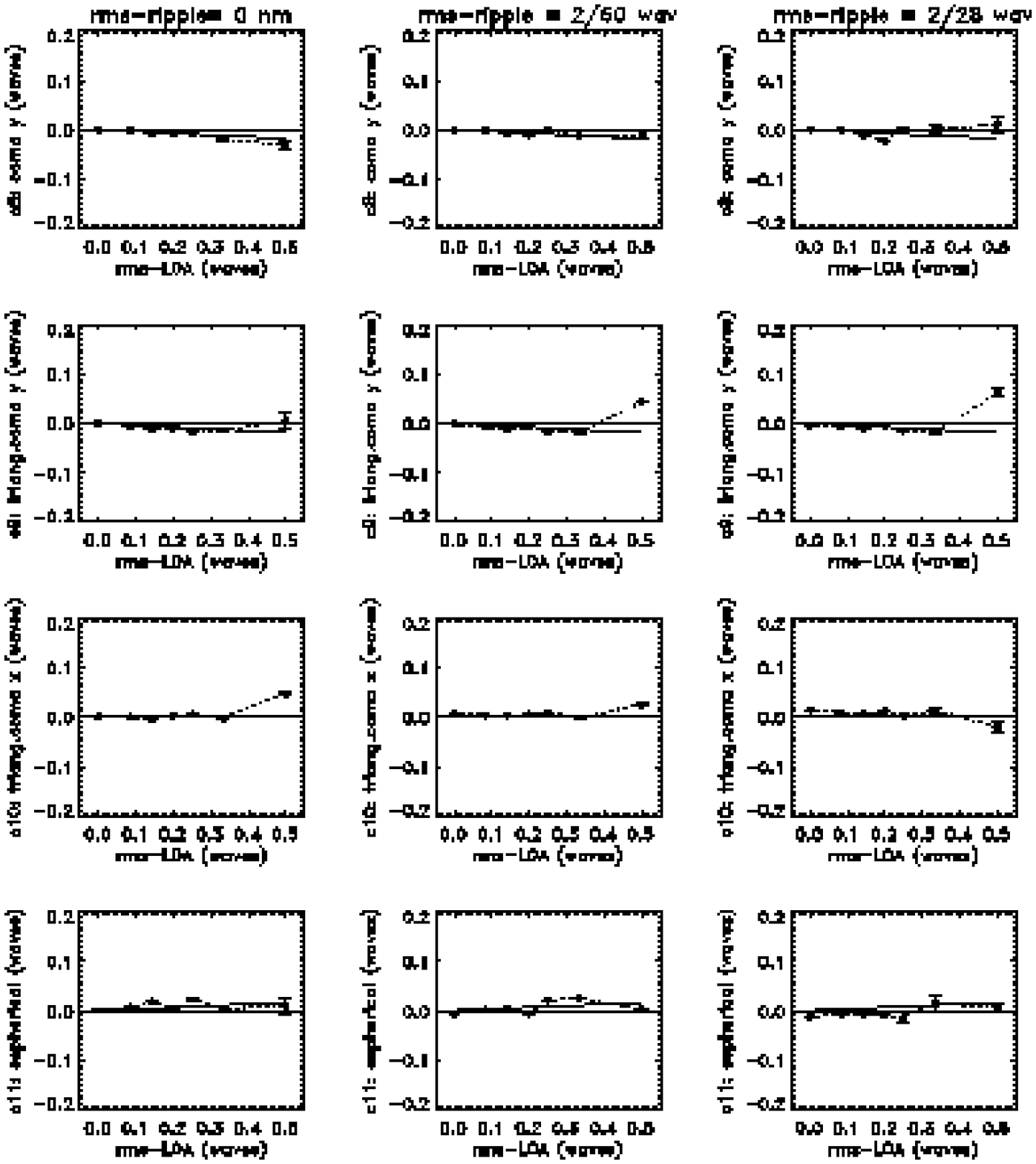} 
\end{tabular}
\vspace{6mm}
\flushleft {\footnotesize FIGURE \ref{caso1_2} --- \sf \emph{(cont).}}
\end{figure}

\subsubsection{\bf Experiment 2}
Combined effects considered (see also Table~\ref{num-experiments}).\\
\hspace{6mm}\fbox{
\hspace{0cm}\parbox[c]{11cm}{{
\footnotesize $\bullet$ \emph{rms-ripple} = 0, 2/60 and 2/28 waves.}\\
{\footnotesize $\bullet$  \emph{rms-LOA} = 0, 1/12, 1/7, 1/5, 1/4, 1/3 and 1/2 waves.}\\
{\footnotesize $\bullet$ \emph{rms-noise} = 10$^{-3} \times$ continuum signal (30 realizations per aberration model.) }\\
{\footnotesize $\bullet$ No CCD effects.}\\
{\footnotesize $\bullet$ Etalon effects:  $|H(\rho,\theta)|$ and $\phi_e(\rho,\theta)$ derived from lab. calibration maps.}\\
{\footnotesize $\bullet$ PD-defocus in mm (degradation/inversion):  8.51 / 8.51.}\\
{\footnotesize $\bullet$ Inversion with 25 coefficients.}}\\
}\\
\vspace{4mm}

The only difference of this Experiment with respect to Experiment 1 is that now we include in the simulated degradation the contribution from $|H(\rho,\theta)|$ and $\phi_r(\rho,\theta)$ mapping the transmission inhomogeneities and the HOA of the etalon, respectively. Note that the PD-code is only designed for the retrieval of the phase-error function in the pupil but not of the pupil transmission function. Thus, in the inversion process the function $|H(\rho,\theta)|$ cannot be considered as an unknown in any case but one has to presume a certain model for it. In a first attempt we used for the inversion the same amplitude-screen as the one applied to simulate the degradation but this way of proceeding generated a disastrous behaviour in the PD-code with strong fluctuations during the iterative process that minimizes the error metric, and eventually collapsed the calculations. So, we were compelled to use in the inversion the simple ideal model $|H(\rho,\theta)|$=1 all over the pupil.\\

Now we center our attention in the impact of these new complications in the efficiency of the PD-code to recover the global phase-error introduced as the input to simulate the degradation. Firstly, we concentrate in Figure~\ref{caso2_1}, rows (a), (b), (c), for \emph{rms-ripple}=0, 2/60 waves and for \emph{rms-LOA} $\le$ 1/4 waves. As in Experiment 1, a good agreement between the results of the inversion and the reference values is found. In most cases, the contrast rms$(i_r)$ ranges from 16.5\% to 18.4\%, that is, close to the reference value 17.5\%. The trends and ranges in the variations of the mean absolute differences mean$(|i_t-i_r|)$ and of the rms$(i_t-i_r)$ remain quite similar to those in Figure~\ref{caso1_1}. However, a significant difference with Experiment 1 deserves special discussion. Now, for the cases with small \emph{rms-LOA} $\le$ 1/7 waves, the inversion yields image contrasts rms$(i_r)$ overvalued with respect to the reference one.

\begin{figure}
\centering
\includegraphics[width=.9\linewidth]{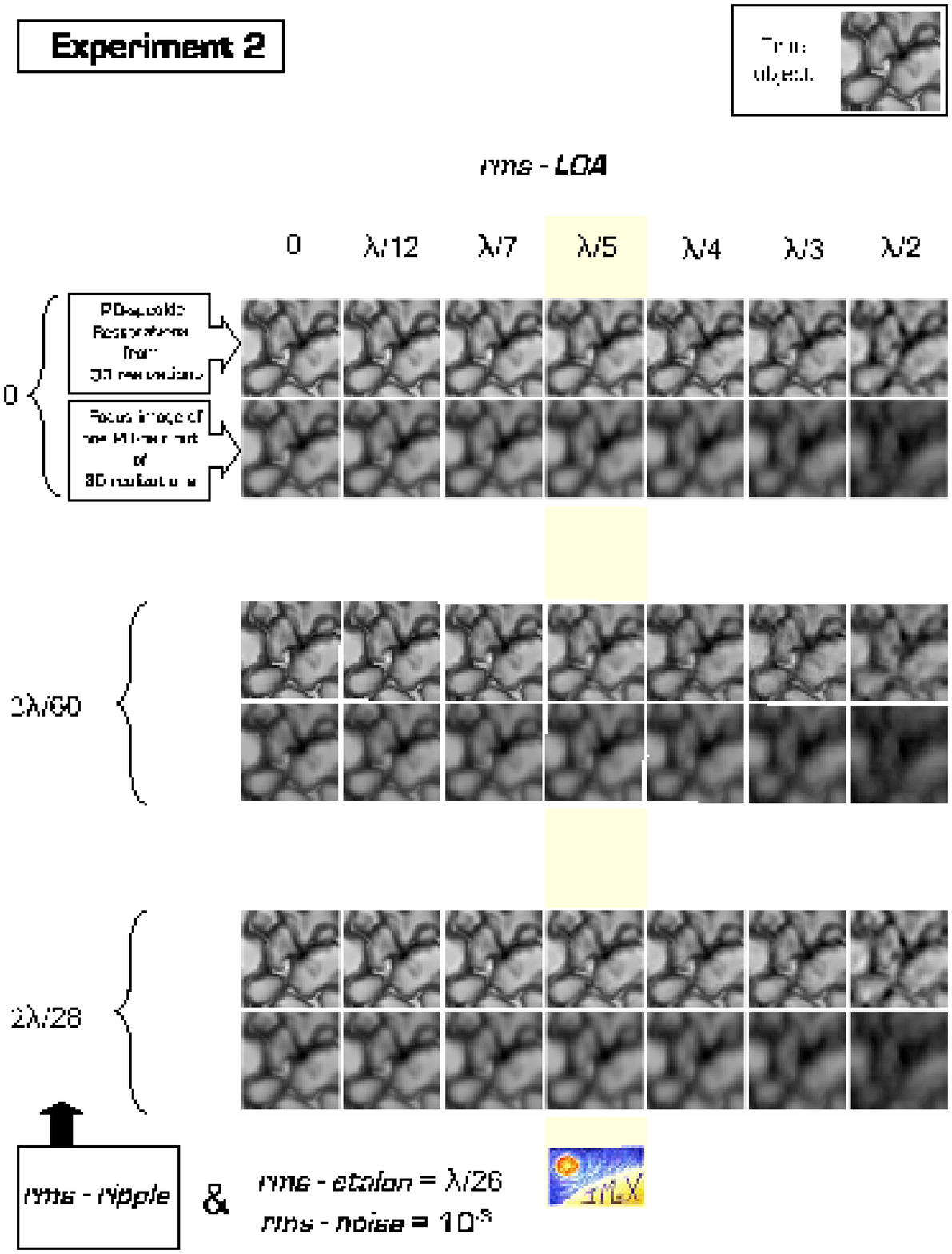} 
\caption[\sf PD reconstruction for different amounts of aberration: Experiment 2]{\sf Experiment 2: PD reconstruction for different amounts of aberration. Seven different contributions from low-order aberrations are adopted: \emph{rms-LOA} = 0, $\lambda/12$, $\lambda/7$, $\lambda/5$, $\lambda/4$, $\lambda/3$, $\lambda/2$, and for each of them three possible cases of polishing errors are considered: \emph{rms-ripple} = 0, $2\lambda/60$, $2\lambda/28$. The exam of the PD-speckle reconstructed images from 30 realizations allows us to check the robustness of the IMaX WFE calibration. The IMaX case is emphasized in \emph{yellow} and the so-called \emph{true object} representing the free-from-aberrations and -diffraction image is located in the upper right corner. All the images are displayed by employing a common gray scale ranging from the minimum to the maximum intensity values of the true object. Specific differences characterizing every Experiment are summarized at the bottom of the figure.}
\label{tiras-exp2}
\end{figure}

\begin{figure}
\centering
\begin{tabular}{c}
\footnotesize{{\bf \sf EXPERIMENT 2 ~/~ Inversions evaluation}}\\
\\
\hspace{-0.7cm}\includegraphics[width=1.05\linewidth]{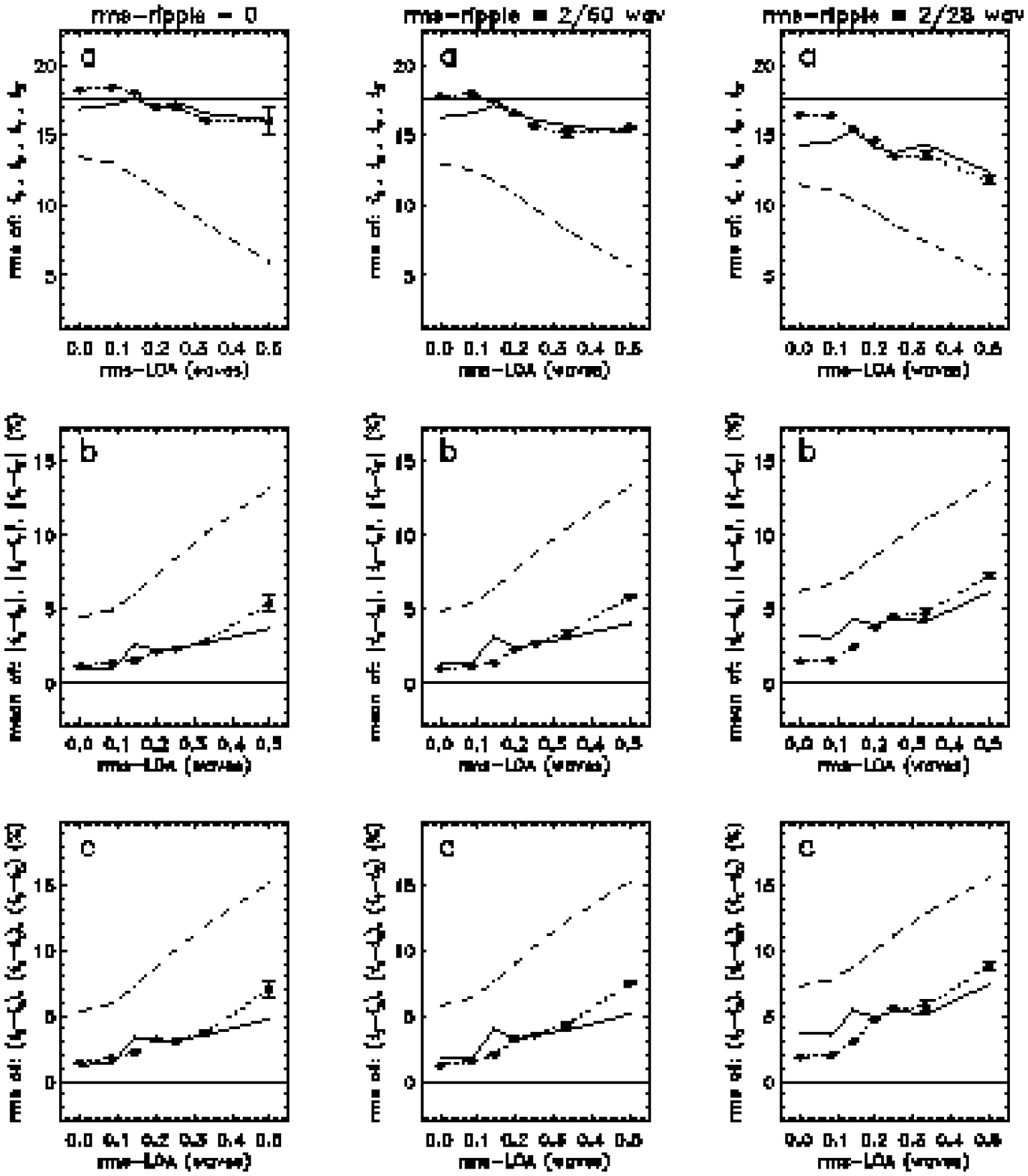} 
\end{tabular}
\caption[\sf Comparative plot for the numerical Experiment 2]{\sf Comparative plot for the numerical Experiment 2. To understand the meaning of the different symbols and labels see section~\S\ref{numsim}: \emph{Key notes for the interpretation of Figures~\ref{tiras-exp1} to \ref{caso4_2}.}}
\label{caso2_1}
\end{figure}

\begin{figure}
\centering
\begin{tabular}{c}
\footnotesize{{\bf \sf EXPERIMENT 2 ~/~ Inversions evaluation (\sf \emph{cont})}}\\
\\
\hspace{-0.7cm}\includegraphics[width=1.05\linewidth]{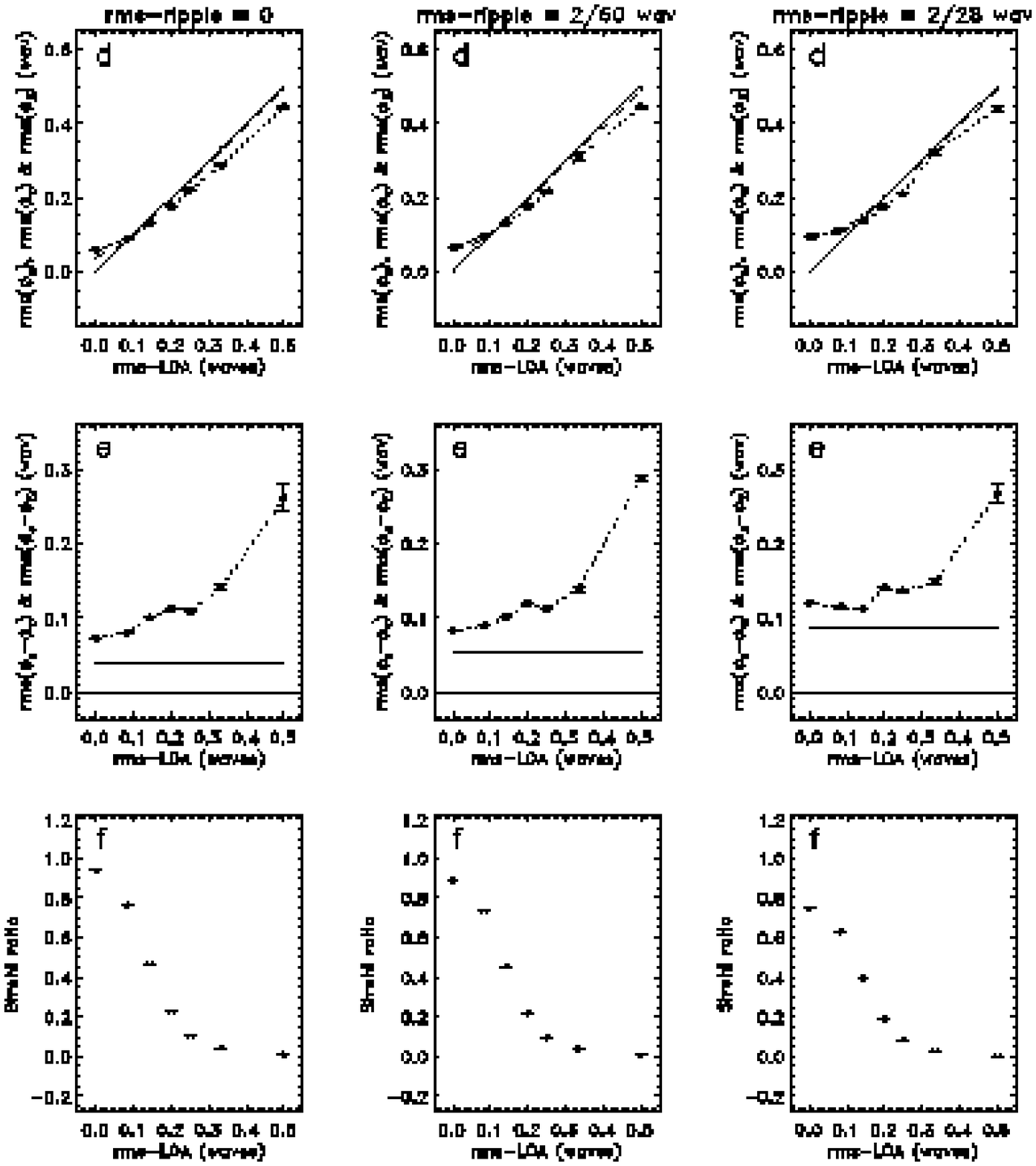}\\
\end{tabular}
\vspace{6mm}
\flushleft {\footnotesize F}{\scriptsize IGURE} {\footnotesize \ref{caso2_1} --- \sf \emph{(cont).}}
\end{figure}

\begin{figure}
\centering
\begin{tabular}{c}
\footnotesize{{\bf \sf EXPERIMENT 2 ~/~ Zernike coefficients evaluation}}\\
\\
\hspace{-0.7cm}\includegraphics[width=1.05\linewidth]{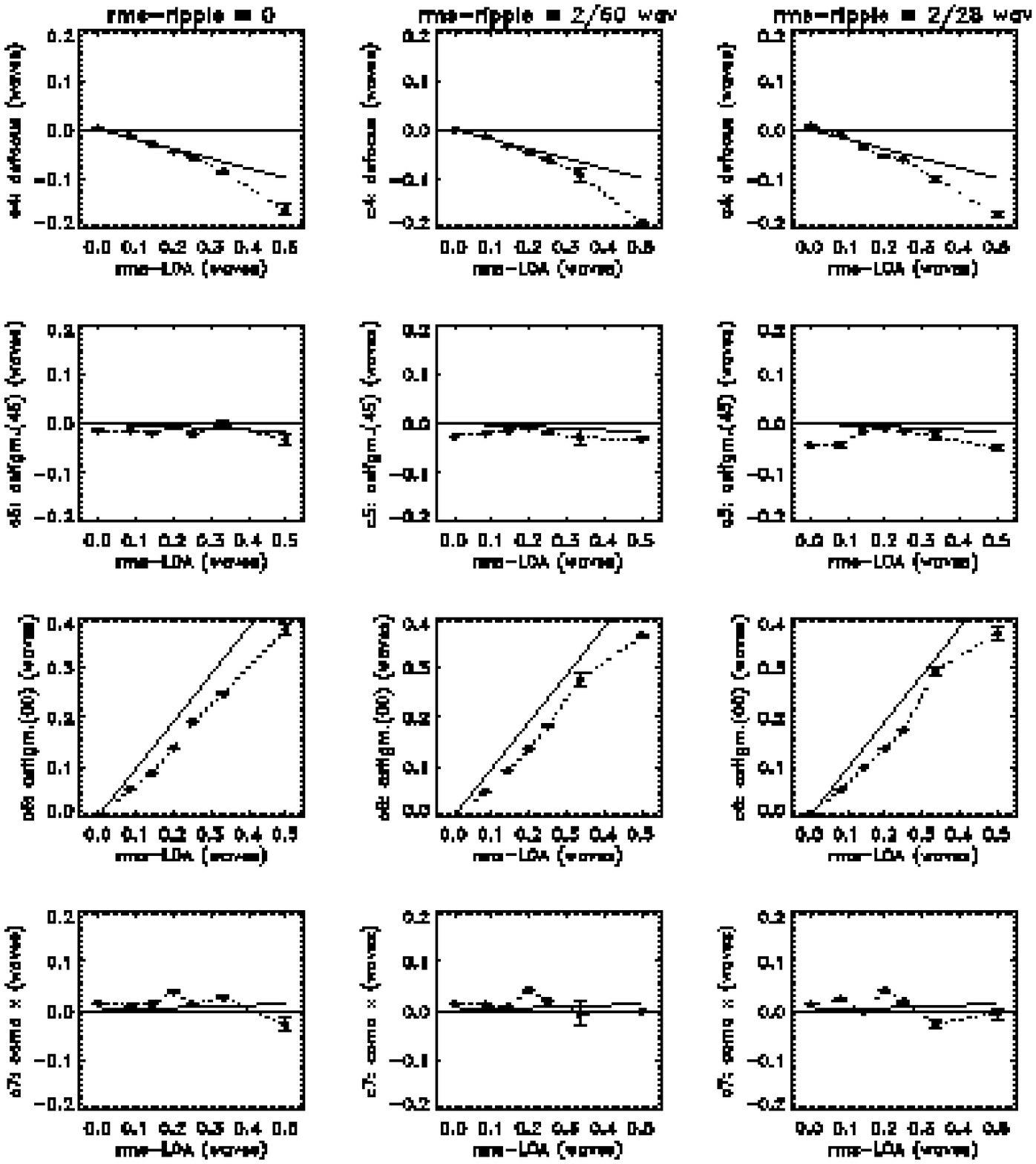} 
\end{tabular}
\caption[\sf Comparative plot of Zernike coefficients for the numerical Experiment 2]{\sf Comparative plot of Zernike coefficients for the numerical Experiment 2. \emph{Bullets} linked by \emph{dotted lines} represent the variation of the mean Zernike coefficients retrieved from the PD-inversions of 30 realizations. \emph{Thick solid lines} represent the variation of the original coefficients used as code inputs to simulate the degradation caused by low-order aberration (LOA) terms. Only the coefficients indexed from 4 to 11 are included in the figure.}
\label{caso2_2}
\end{figure}

\begin{figure}
\centering
\begin{tabular}{c}
\footnotesize{{\bf \sf EXPERIMENT 2 ~/~ Zernike coefficients evaluation (\sf \emph{cont})}}\\
\\
\hspace{-0.7cm}\includegraphics[width=1.05\linewidth]{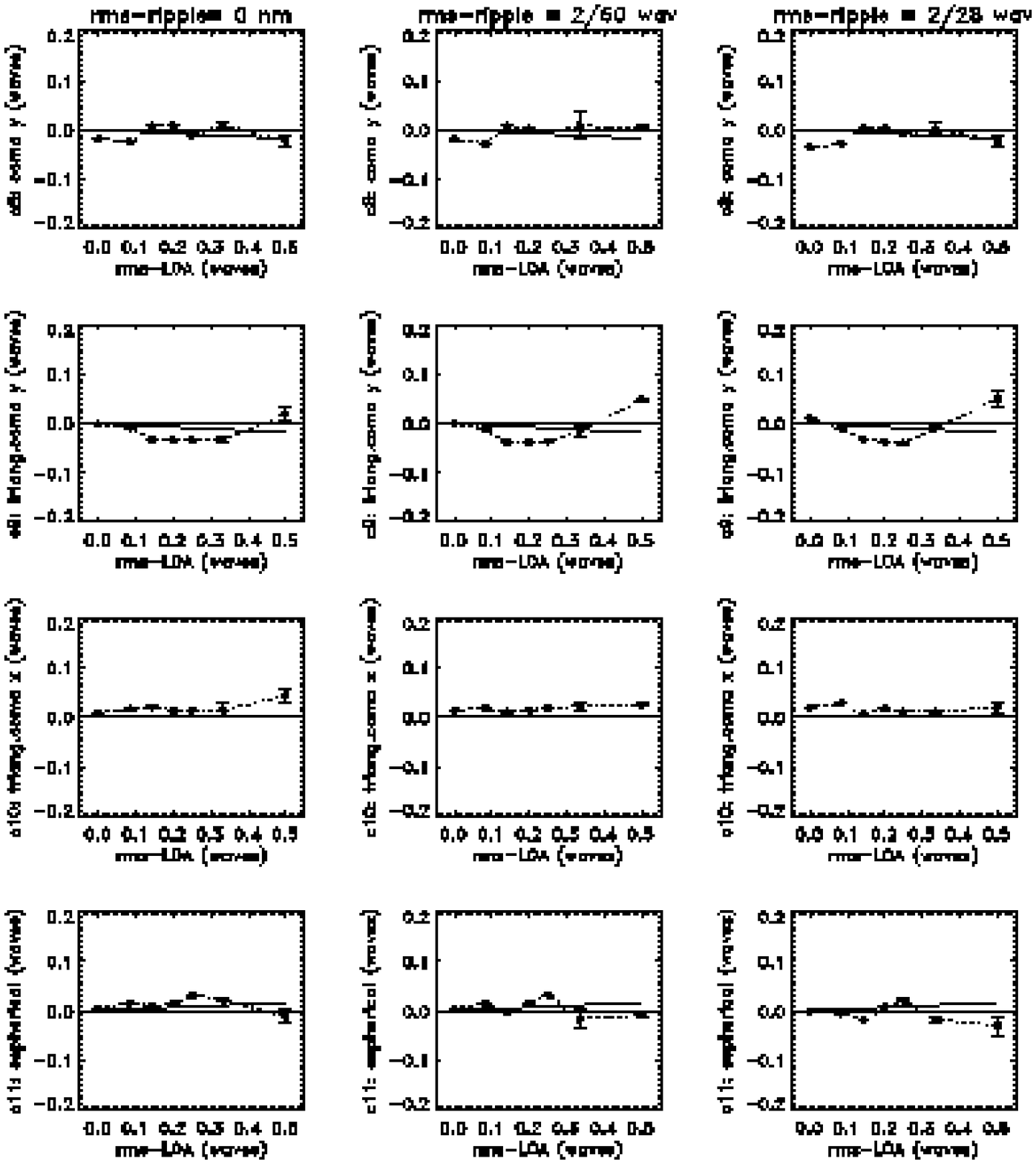} 
\end{tabular}
\vspace{6mm}
\flushleft {\footnotesize FIGURE \ref{caso2_2} --- \sf \emph{(cont).}}
\end{figure}

We ascribe this result to the non-realistic assumption made when setting $|H(\rho,\theta)|$=1 in the inversions. The error metric to be minimized in the PD-inversions does not explicitly contain neither the phase, $\phi$, nor the amplitude, $|H|$, that define the \emph{generalized pupil function} (expression~1.16), but the OTFs that can be derived from the latter.  Thus, the defective information assigned to the amplitude $|H|$ can be compensated by some extra errors --~in comparison with Experiment 1~-- in the phase functions provided that the resulting generalized pupil functions still yield appropriate OTFs that minimize the error metric. These extra errors are evident from the comparison of Figures~\ref{caso1_1} and \ref{caso2_1}, rows (d) and (e), and can be the cause of the over-restoration mentioned above. Figure~\ref{caso2_1}, row (e), also shows the deviations of rms($\phi_s-\phi_Z$) from zero. These vertical offsets are related to the HOA contribution from polishing errors in the main mirror and inhomogeneities in the etalon surface and, as expected, in the three panels they very closely amount to RSS=$\sqrt{\emph{rms-ripple}^2+\emph{rms-etalon}^2}$.

\subsubsection{\bf Experiment 3}
Combined effects considered (see also Table~\ref{num-experiments}).\\
\hspace{6mm}\fbox{
\hspace{0cm}\parbox[c]{11cm}{{
\footnotesize $\bullet$ \emph{rms-ripple} = 0, 2/60 and 2/28 waves.}\\
{\footnotesize $\bullet$ \emph{rms-LOA} = 0, 1/12, 1/7, 1/5, 1/4, 1/3 and 1/2 waves.}\\
{\footnotesize $\bullet$ \emph{rms-noise} = 10$^{-3} \times$ continuum signal (30 realizations per aberration model.) }\\
{\footnotesize $\bullet$ CCD effects.}\\
{\footnotesize $\bullet$ Etalon effects:  $|H(\rho,\theta)|$ and $\phi_e(\rho,\theta)$ derived from lab. calibration maps.}\\
{\footnotesize $\bullet$ PD-defocus in mm (degradation/inversion):  8.51 / 8.51.}\\
{\footnotesize $\bullet$ Inversion with 25 coefficients.}}\\
}\\
\vspace{4mm}

In this Experiment we add to the image degradation one more contribution with respect to Experiment 2. It consist of the integration over the pixel area, of the information reaching the image plane as commented in section~\S\ref{quantifyingerrors}. Table~\ref{num-experiments} ticks off the error contributions taken into account for this simulation. \emph{Dotted line} in Figure~\ref{imaxcaseMTF} shows the OTF$_{\emph{footprint}}$ representing the degradation effect which is incorporated to the total error budget by simple multiplication by the OTF describing the global aberration in Experiment 2. The new ingredient in the simulated aberration has no significant impact in the results of the PD-inversion. Thus, for \emph{rms-ripple} $\le$ 2/60 and \emph{rms-LOA} $\le$ 1/4 waves the differential impact in the parameters rms$(i_r)$, mean$(|i_t-i_r|)$ and rms$(i_t-i_r)$ --~Figure~\ref{caso3_1}, rows (a), (b), (c)~--, with respect to Experiment 2, is smaller than 2 tenths \%. Similar coincidences between the results from Experiments 2 y 3 are achieved for the parameter rms$(\phi_r)$. However, the influence of the CCD pixel integration is remarkable in the Strehl ratio as evidenced by comparing Figures~\ref{caso2_1} and \ref{caso3_1}, rows (f).

\begin{figure}
\centering
\includegraphics[width=.9\linewidth]{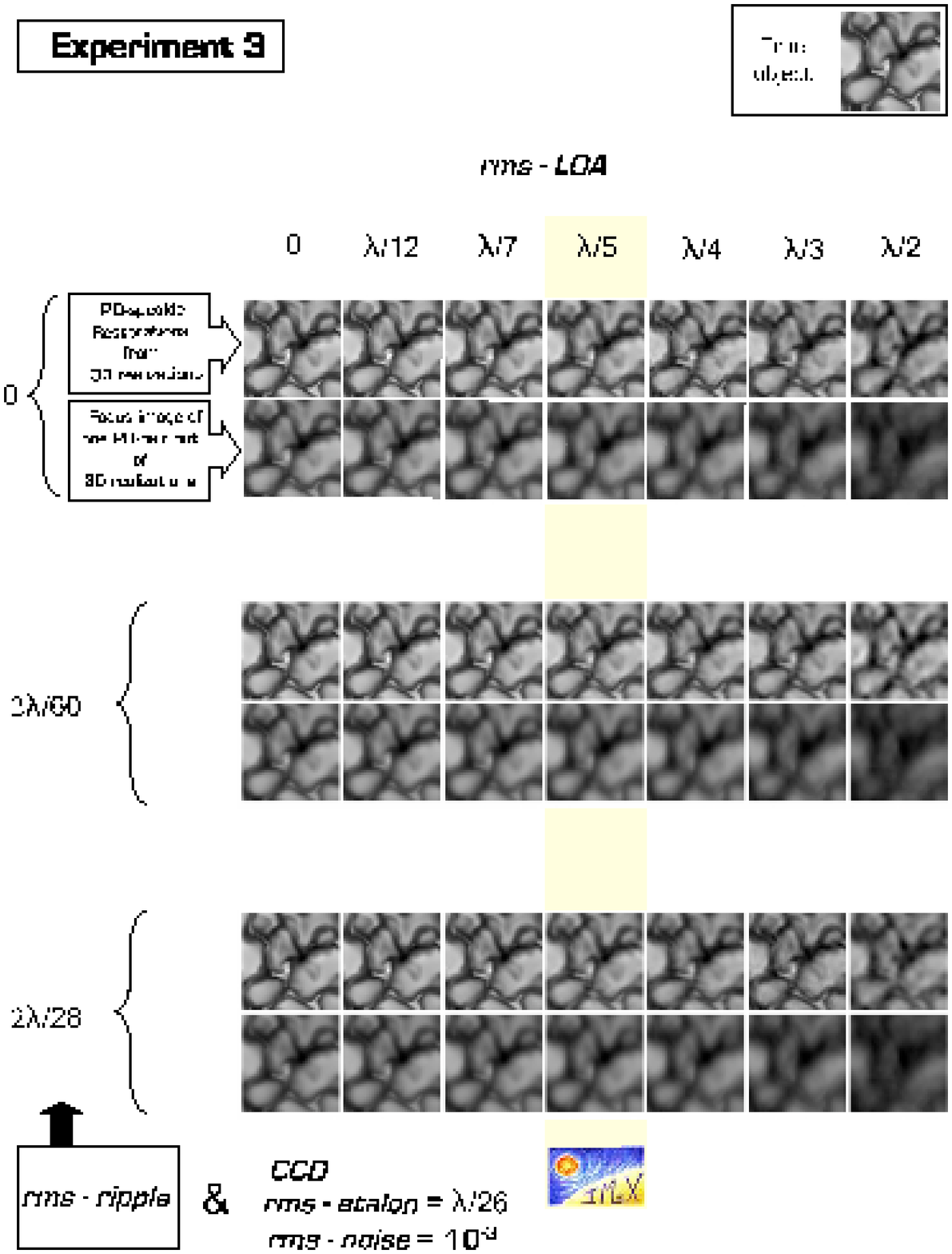} 
\caption[\sf PD reconstruction for different amounts of aberration: Experiment 3]{\sf Experiment 3: PD reconstruction for different amounts of aberration. Seven different contributions from low-order aberrations are adopted: \emph{rms-LOA} = 0, $\lambda/12$, $\lambda/7$, $\lambda/5$, $\lambda/4$, $\lambda/3$, $\lambda/2$, and for each of them three possible cases of polishing errors are considered: \emph{rms-ripple} = 0, $2\lambda/60$, $2\lambda/28$. The exam of the PD-speckle reconstructed images from 30 realizations allows us to check the robustness of the IMaX WFE calibration. The IMaX case is emphasized in \emph{yellow} and the so-called \emph{true object} representing the free-from-aberrations and -diffraction image is located in the upper right corner. All the images are displayed by employing a common gray scale ranging from the minimum to the maximum intensity values of the true object. Specific differences characterizing every Experiment are summarized at the bottom of the figure.}
\label{tiras-exp3}
\end{figure}

\begin{figure}
\centering
\begin{tabular}{c}
\footnotesize{{\bf \sf EXPERIMENT 3 ~/~ Inversions evaluation}}\\
\\
\hspace{-0.7cm}\includegraphics[width=1.05\linewidth]{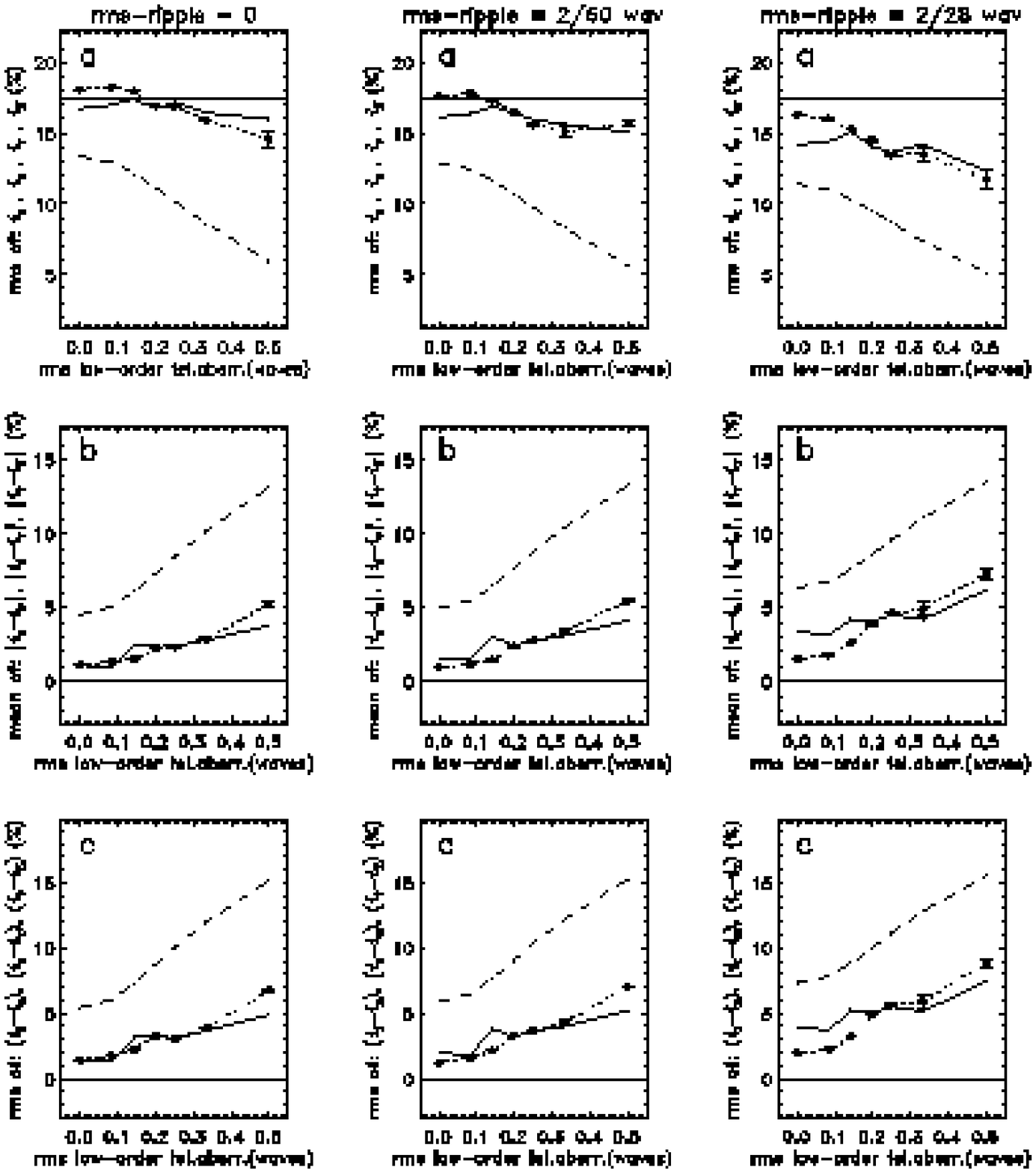} 
\end{tabular}
\caption[\sf Comparative plot for the numerical Experiment 3]{\sf Comparative plot for the numerical Experiment 3. To understand the meaning of the different symbols and labels see section~\S\ref{numsim}: \emph{Key notes for the interpretation of Figures~\ref{tiras-exp1} to \ref{caso4_2}.}}
\label{caso3_1}
\end{figure}

\begin{figure}
\centering
\begin{tabular}{c}
\footnotesize{{\bf \sf EXPERIMENT 3 ~/~ Inversions evaluation (\sf \emph{cont})}}\\
\\
\hspace{-0.7cm}\includegraphics[width=1.05\linewidth]{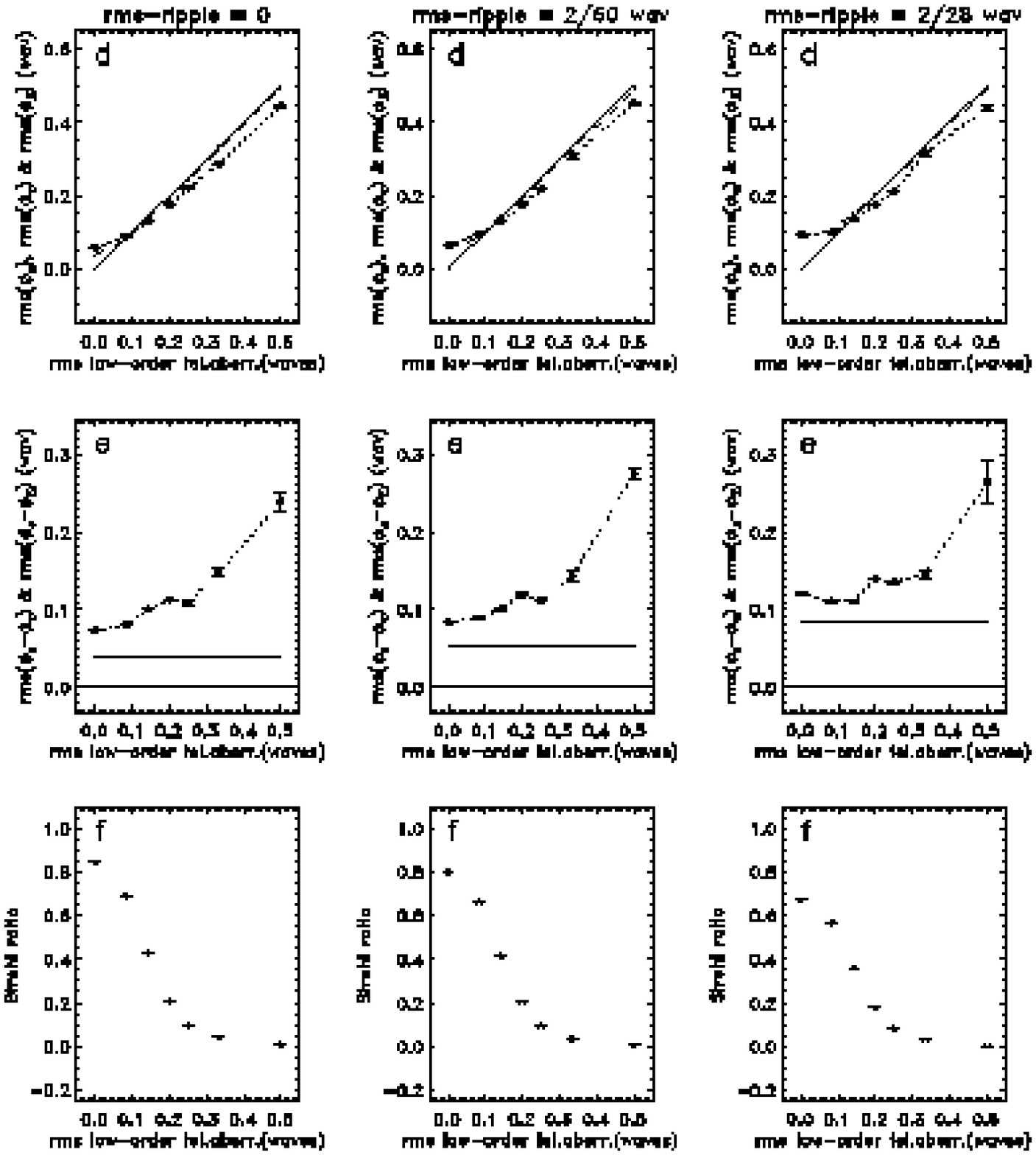}\\
\end{tabular}
\vspace{6mm}
\flushleft {\footnotesize F}{\scriptsize IGURE} {\footnotesize \ref{caso3_1} --- \sf \emph{(cont).}}
\end{figure}

\begin{figure}
\centering
\begin{tabular}{c}
\footnotesize{{\bf \sf EXPERIMENT 3 ~/~ Zernike coefficients evaluation}}\\
\\
\hspace{-0.7cm}\includegraphics[width=1.05\linewidth]{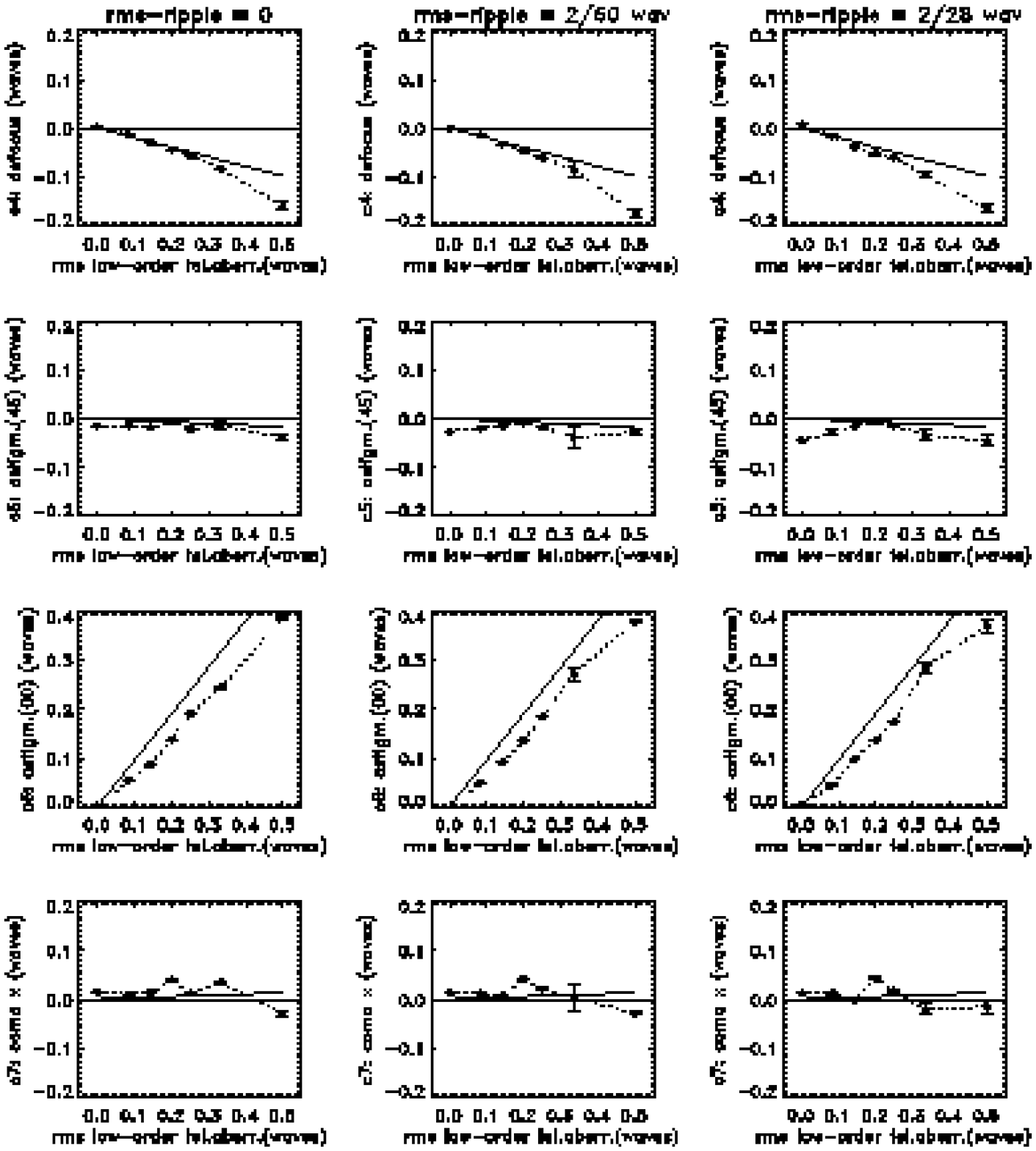} 
\end{tabular}
\caption[\sf Comparative plot of Zernike coefficients for the numerical Experiment 3]{\sf Comparative plot of Zernike coefficients for the numerical Experiment 3. \emph{Bullets} linked by \emph{dotted lines} represent the variation of the mean Zernike coefficients retrieved from the PD-inversions of 30 realizations. \emph{Thick solid lines} represent the variation of the original coefficients used as code inputs to simulate the degradation caused by low-order aberration (LOA) terms. Only the coefficients indexed from 4 to 11 are included in the figure.}
\label{caso3_2}
\end{figure}

\begin{figure}
\centering
\begin{tabular}{c}
\footnotesize{{\bf \sf EXPERIMENT 3 ~/~ Zernike coefficients evaluation (\sf \emph{cont})}}\\
\\
\hspace{-0.7cm}\includegraphics[width=1.05\linewidth]{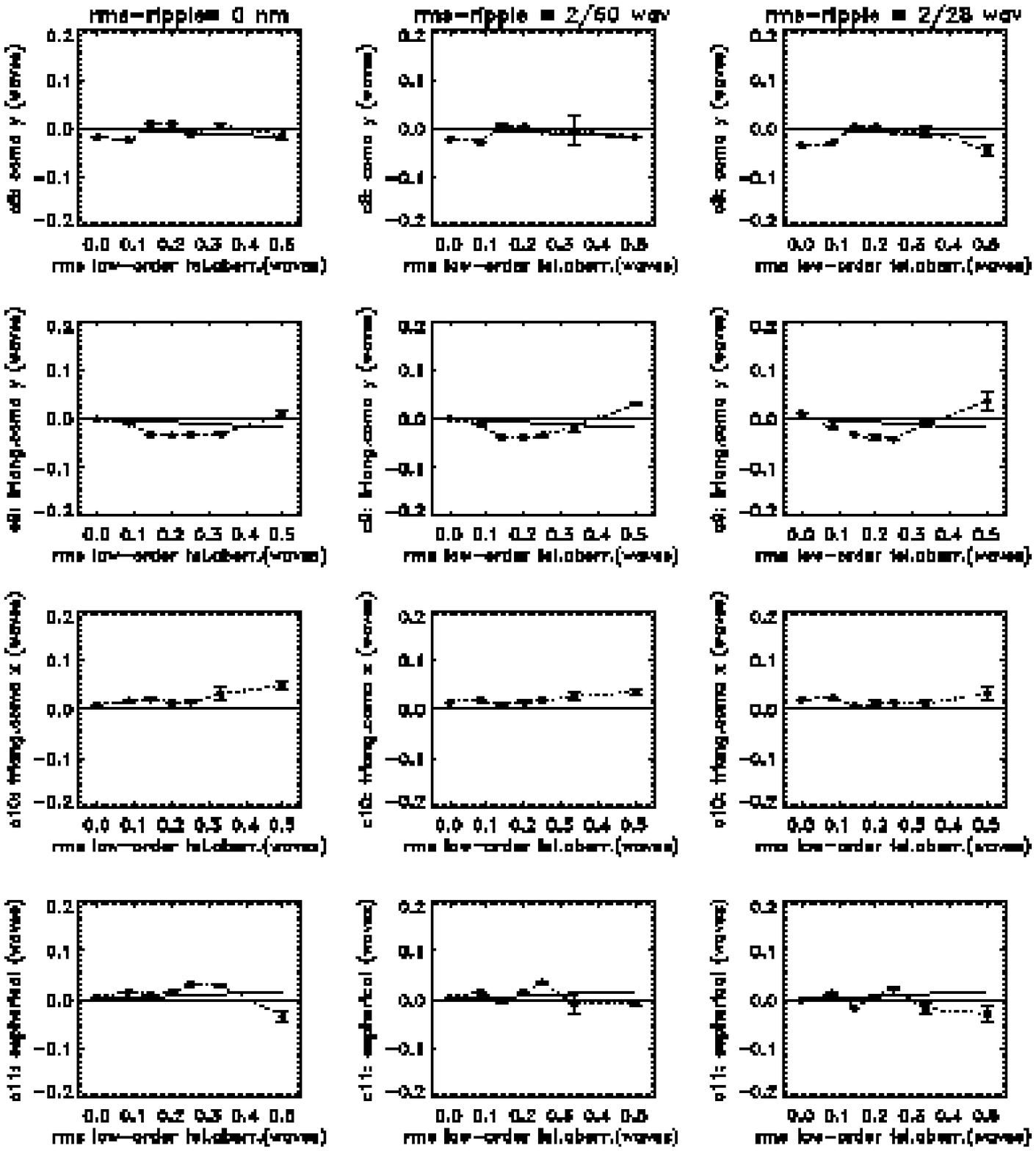} 
\end{tabular}
\vspace{6mm}
\flushleft {\footnotesize FIGURE \ref{caso3_2} --- \sf \emph{(cont).}}
\end{figure}

\vspace{5mm}
Concerning the retrieved Zernike coefficients from the PD-process, they look very similar in Figures~\ref{caso2_2} and \ref{caso3_2}, within the range of aberrations mentioned above (\emph{rms-ripple} $\le$ 2/60 and \emph{rms-LOA} $\le$ 1/4 waves). Nevertheless, as discussed in previous paragraphs, small differences in the Zernike coefficients of both Experiments would not be crucial provided that the resulting \emph{generalized pupil functions} yield appropriate OTFs to obtain a good minimization of the error metric. This seems to be the case in the present Experiment to judge by the good image restorations achieved. Therefore, we will consider the results from Experiment 3 as the reference for the next numerical test.

\subsubsection{\bf Experiment 4}
Combined effects considered (see also Table~\ref{num-experiments}).\\
\hspace{6mm}\fbox{
\hspace{0cm}\parbox[c]{11cm}{{
\footnotesize $\bullet$ \emph{rms-ripple} = 0, 2/60 and 2/28 waves.}\\
{\footnotesize $\bullet$ \emph{rms-LOA} = 0, 1/12, 1/7, 1/5, 1/4, 1/3 and 1/2 waves.}\\
{\footnotesize $\bullet$ \emph{rms-noise} = 10$^{-3} \times$ continuum signal (30 realizations per aberration model.) }\\
{\footnotesize $\bullet$ CCD effects.}\\
{\footnotesize $\bullet$ Etalon effects:  $|H(\rho,\theta)|$ and $\phi_e(\rho,\theta)$ derived from lab. calibration maps.}\\
{\footnotesize $\bullet$ PD-defocus in mm (degradation/inversion):  8.51 / 9.00.}\\
{\footnotesize $\bullet$ Inversion with 25 coefficients.}}\\
}\\
\vspace{4mm}

The recent IMaX thermal design (May 2008) predicts for the PD-plate, during the flight, a temperature variation range of 24.5 $\pm$ 0.5 $^o$C. From this thermal model, simulations with \emph{CODE V} estimate the defocus displacement induced by the PD-plate ranging from 8.46 mm (PV-WFE=1.008 waves; hot case) to 8.56 mm (PV-WFE=1.019 waves; cold case), being 8.51 mm (PV-WFE=1.015 waves) the displacement corresponding to the operative case (24.5 $^o$C).\\

The amount of displacement of the image plane produced by the PD-plate, also called \emph{diversity}, is an input parameter in the PD-code. One of our concerns has always been how critical the value assigned to the \emph{diversity} is, or in other words, how much robust is the PD-code in rendering good results versus possible errors in the value of the \emph{diversity}. This concern arises from the possibility of having during the flight unexpected peaks in the temperature fluctuation, out of the range predicted by the thermal model. \\

\begin{figure}
\centering
\includegraphics[width=.86\linewidth]{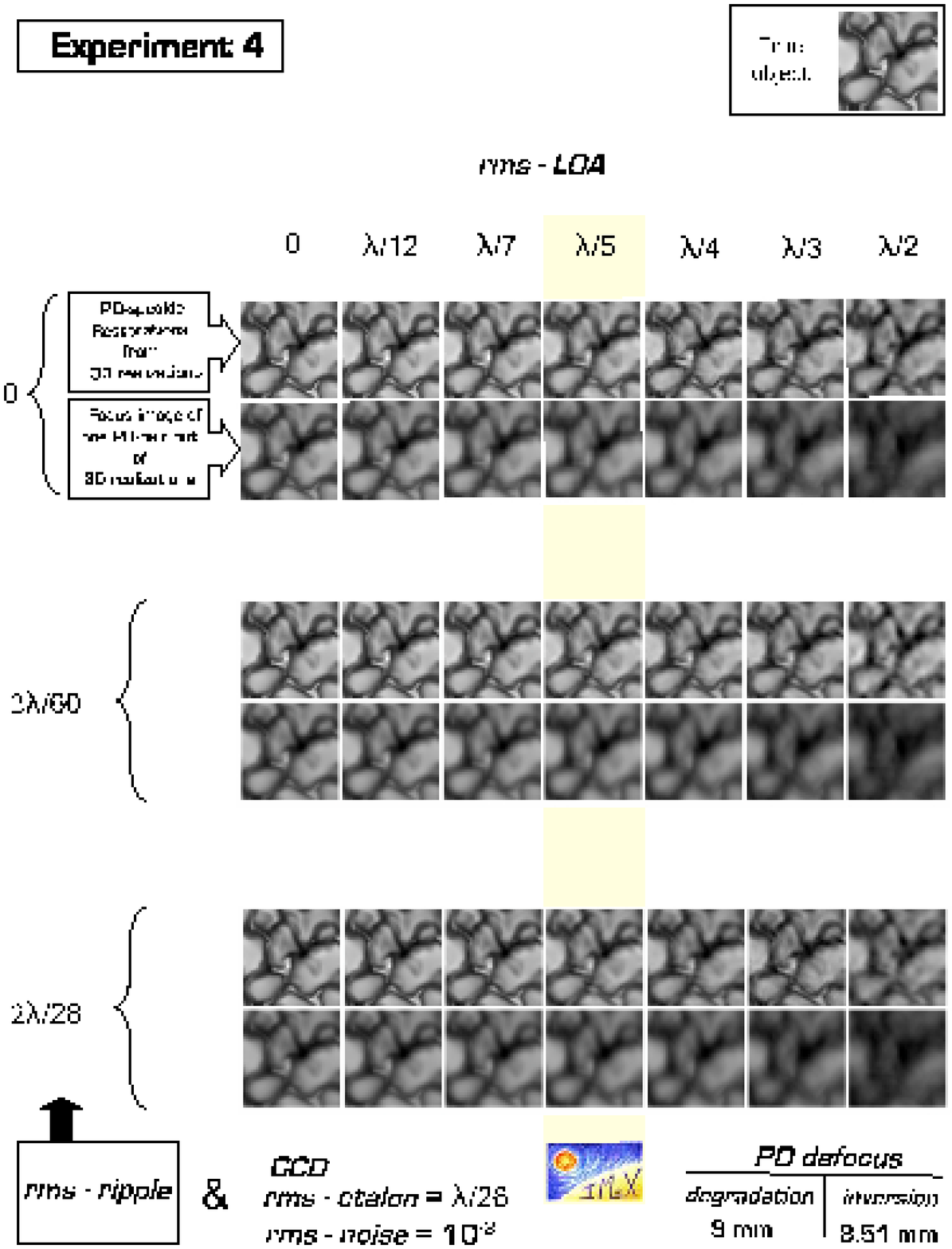} 
\caption[\sf PD reconstruction for different amounts of aberration: Experiment 4]{\sf Experiment 4: PD reconstruction for different amounts of aberration. Seven different contributions from low-order aberrations are adopted: \emph{rms-LOA} = 0, $\lambda/12$, $\lambda/7$, $\lambda/5$, $\lambda/4$, $\lambda/3$, $\lambda/2$, and for each of them three possible cases of polishing errors are considered: \emph{rms-ripple} = 0, $2\lambda/60$, $2\lambda/28$. The exam of the PD-speckle reconstructed images from 30 realizations allows us to check the robustness of the IMaX WFE calibration. The IMaX case is emphasized in \emph{yellow} and the so-called \emph{true object} representing the free-from-aberrations and -diffraction image is located in the upper right corner. All the images are displayed by employing a common gray scale ranging from the minimum to the maximum intensity values of the true object. Specific differences characterizing every Experiment are summarized at the bottom of the figure.}
\label{tiras-exp4}
\end{figure}

\begin{figure}
\centering
\begin{tabular}{c}
\footnotesize{{\bf \sf EXPERIMENT 4 ~/~ Inversions evaluation}}\\
\\
\hspace{-0.7cm}\includegraphics[width=1.05\linewidth]{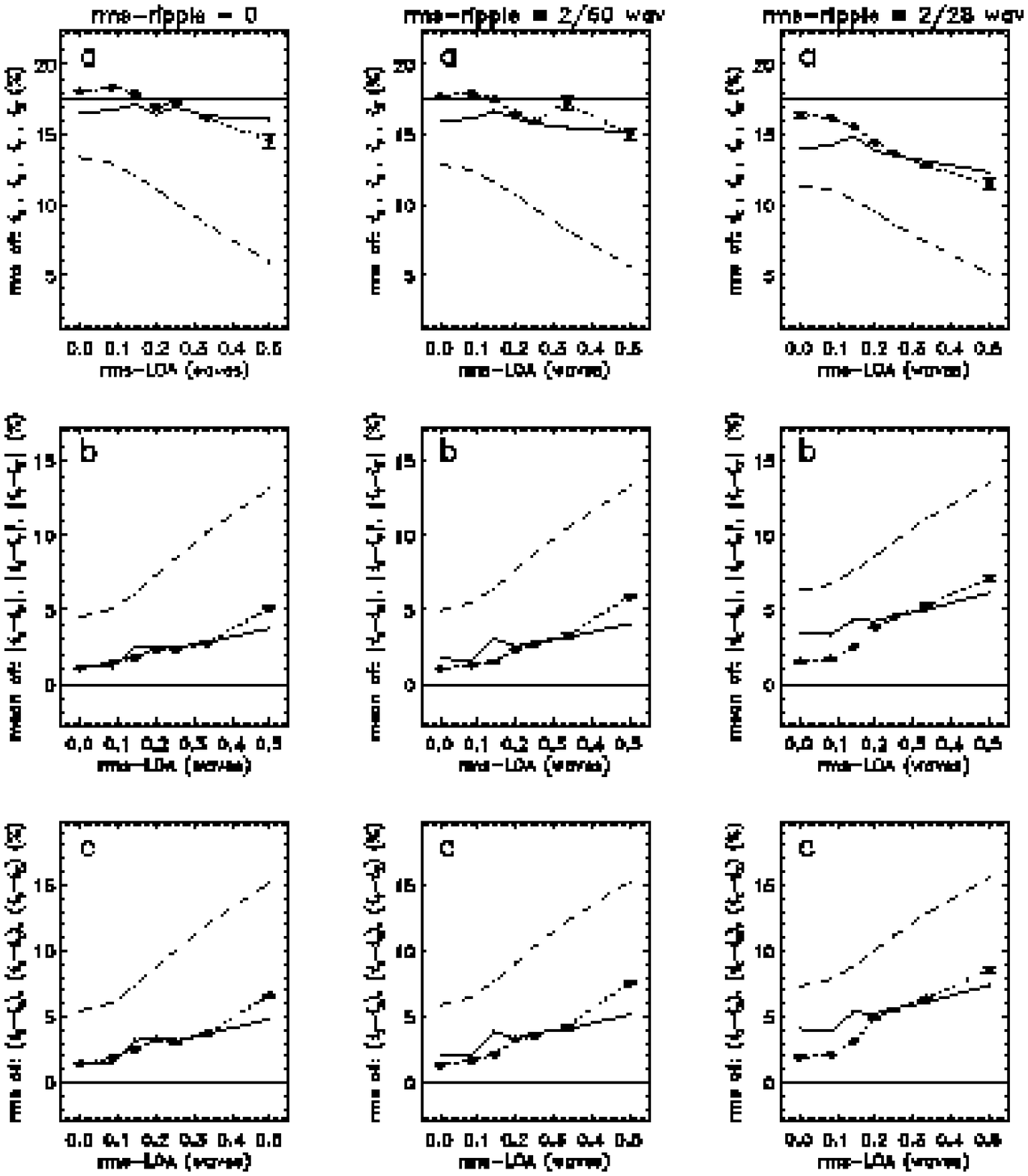} 
\end{tabular}
\caption[\sf Comparative plot for the numerical Experiment 4]{\sf Comparative plot for the numerical Experiment 4. To understand the meaning of the different symbols and labels see section~\S\ref{numsim}: \emph{Key notes for the interpretation of Figures~\ref{tiras-exp1} to \ref{caso4_2}.}}
\label{caso4_1}
\end{figure}

\begin{figure}
\centering
\begin{tabular}{c}
\footnotesize{{\bf \sf EXPERIMENT 4 ~/~ Inversions evaluation (\emph{cont})}}\\
\\
\hspace{-0.7cm}\includegraphics[width=1.05\linewidth]{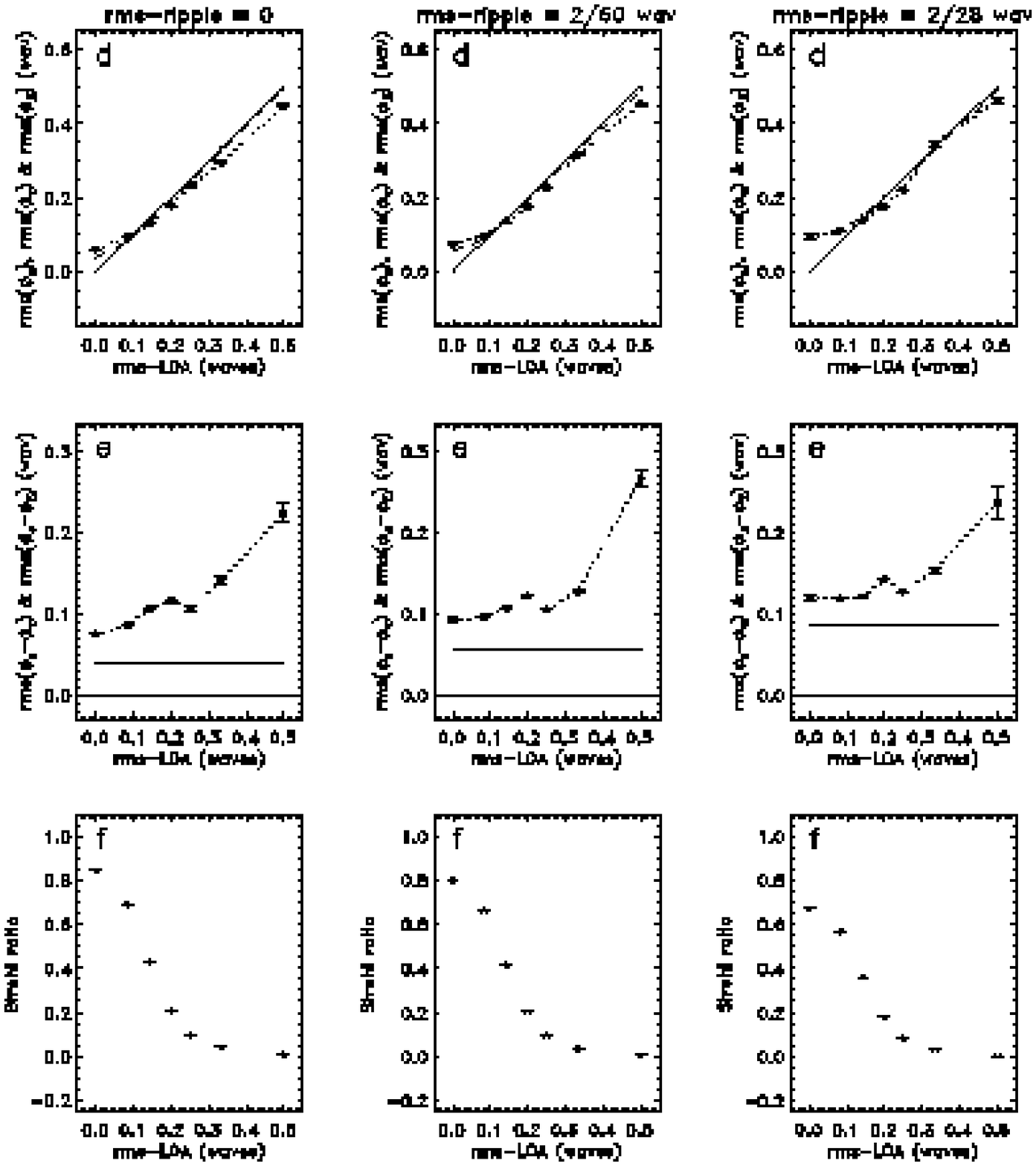} 
\end{tabular}
\vspace{6mm}
\flushleft {\footnotesize FIGURE \ref{caso4_1} --- \sf \emph{(cont).}}
\end{figure}

\begin{figure}
\centering
\begin{tabular}{c}
\footnotesize{{\bf \sf EXPERIMENT 4 ~/~ Zernike coefficients evaluation}}\\
\\
\hspace{-0.7cm}\includegraphics[width=1.05\linewidth]{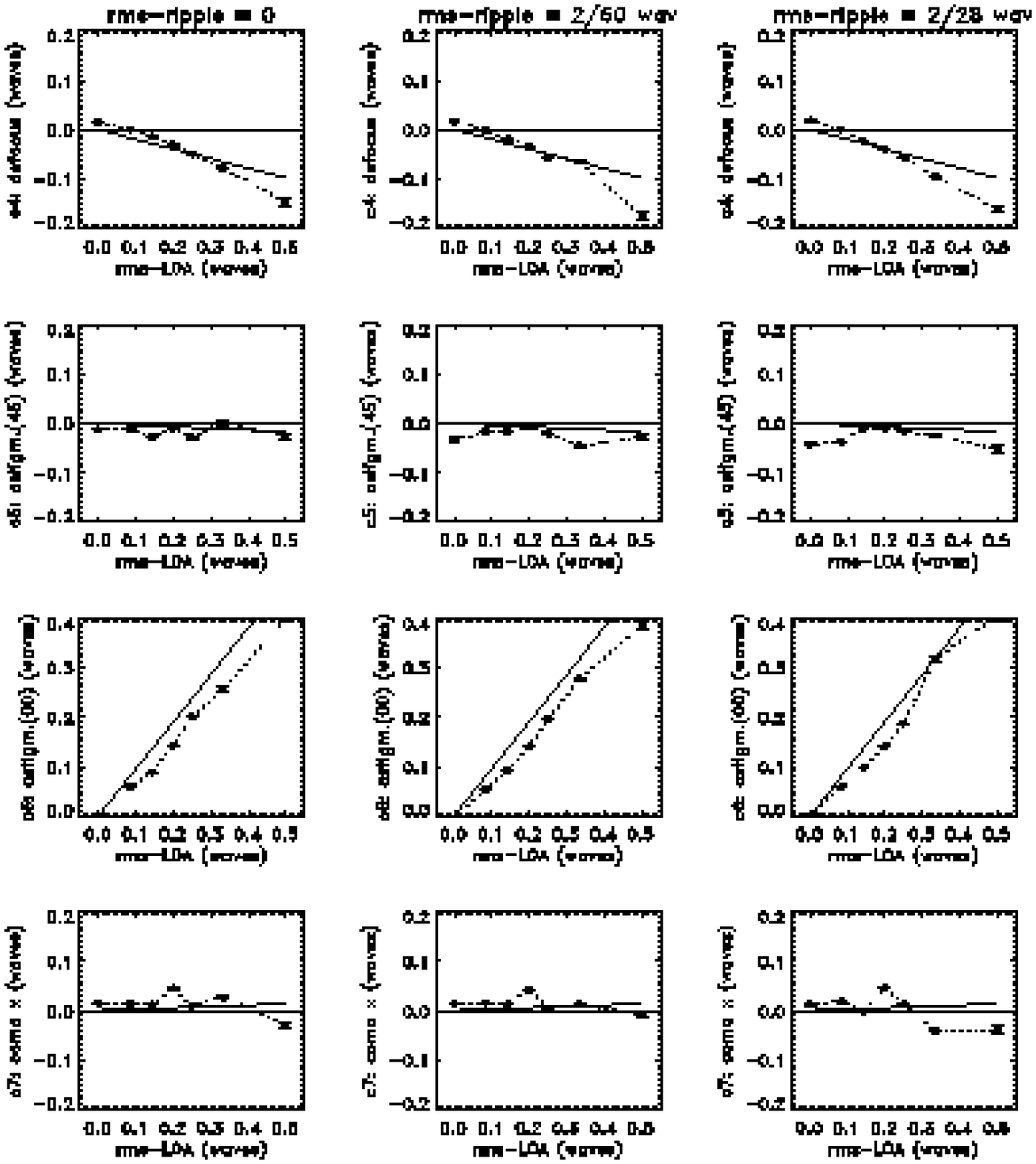} 
\end{tabular}
\caption[\sf Comparative plot of Zernike coefficients for the numerical Experiment 4]{\sf Comparative plot of Zernike coefficients for the numerical Experiment 4. \emph{Bullets} linked by \emph{dotted lines} represent the variation of the mean Zernike coefficients retrieved from the PD-inversions of 30 realizations. \emph{Thick solid lines} represent the variation of the original coefficients used as code inputs to simulate the degradation caused by low-order aberration (LOA) terms. Only the coefficients indexed from 4 to 11 are included in the figure.}
\label{caso4_2}
\end{figure}

\begin{figure}
\centering
\begin{tabular}{c}
\footnotesize{{\bf \sf EXPERIMENT 4 ~/~ Zernike coefficients evaluation (\emph{cont})}}\\
\\
\hspace{-0.7cm}\includegraphics[width=1.05\linewidth]{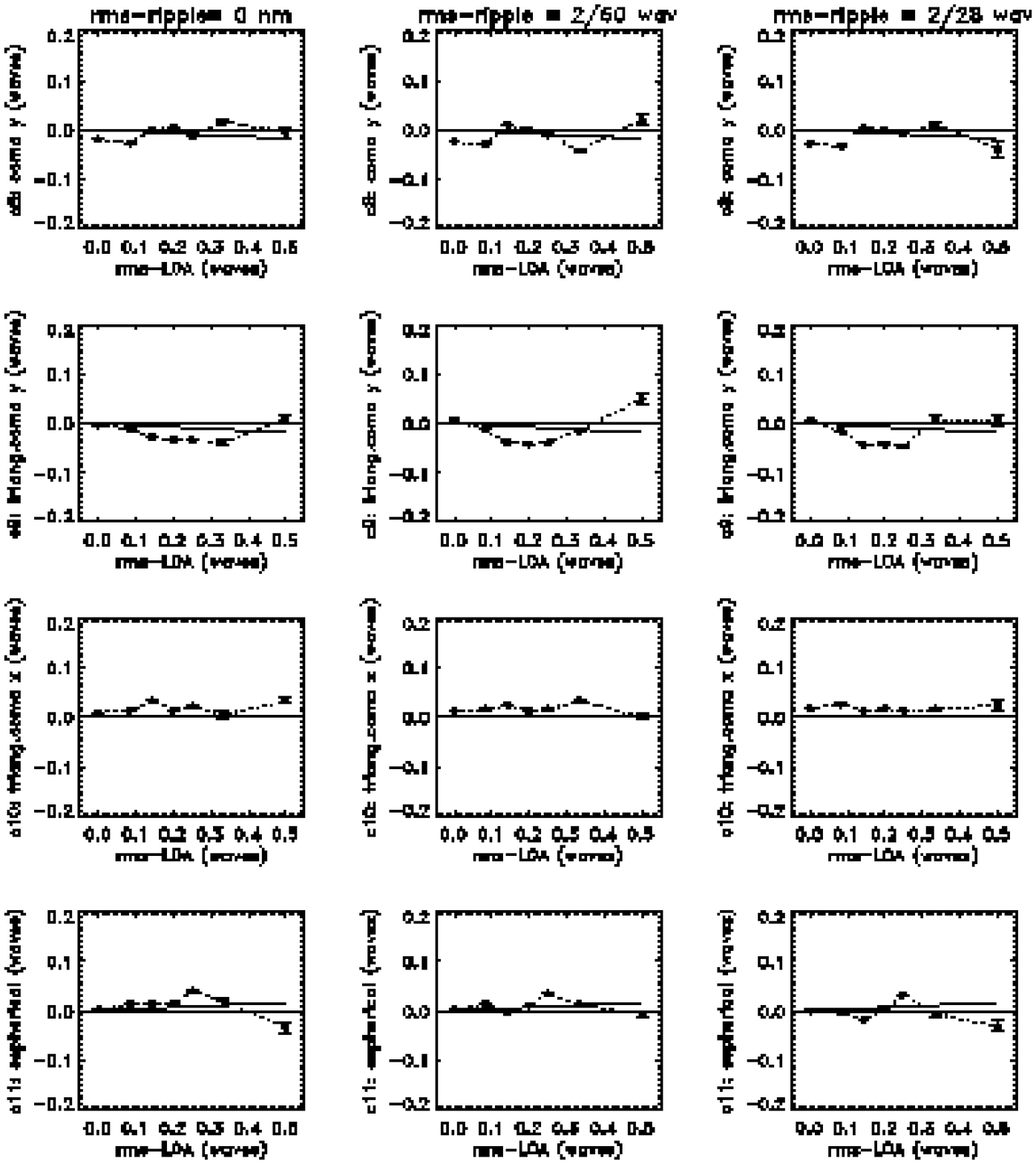} 
\end{tabular}
\vspace{6mm}
\flushleft {\footnotesize FIGURE \ref{caso4_2} --- \sf \emph{(cont).}}
\end{figure}

To clarify this question, we have designed the following numerical experiment: in the inversion process we will take for the \emph{diversity} the expected nominal operative value 8.51 mm, whereas in the degradation process, we will assign to the \emph{diversity} a different value thus simulating an unexpected temperature peak which we are not aware of.\\

In the first realization of the Experiment we have selected for the degradation process a rather large \emph{diversity} value (12 mm) as compared with the nominal one used in the inversion (8.51 mm). Always within our standard range of analysis (i.e.\ for \emph{rms-ripple} $\le$ 2/60 and \emph{rms-LOA} $\le$ 1/4 waves), the inversion produces over-restoration: rms$(i_r)$ ranging from 17.5 to 20.7 \%, mean$(|i_t-i_r|)$ from 2.6 to 5.5 \% and rms$(i_t-i_r)$ from 3.6 to 7.1 \%. Also the retrieved Zernike coefficients, particularly $\alpha_4$, differ substantially from the reference values. A curious result is that the values obtained for $\alpha_4$ are larger than those corresponding to the Experiment 3 by a roughly constant amount of about 0.1 waves, corresponding to a defocus distance of $\sim$ 3 mm, similar to the \emph{diversity} error (3.49 mm). It seems like if the error in the \emph{diversity} would have misled the PD-algorithm in such a way that this error is now ascribed to the image in nominal focus position, in the form of defocus aberration. This could explain the resulting over-restoration mentioned above. Since this is an extreme experiment considering \emph{diversity} values far away from the IMaX case, we do not display the results.\\

In a second and more realistic numerical experiment we have selected a \emph{diversity} of 9 mm for the degradation process whereas we have kept the nominal value for the inversion. Thus, the error introduced, 9.00 - 8.51 = 0.49 mm, is greater by a factor of 10 than the expected peak variation of the \emph{diversity} ($\pm$ ~0.05 mm) according to the IMaX thermal model (Table~\ref{temperatures}). Even so, the PD-inversion gives very similar values to those in Experiment 3 (compare rows (a) to (f) in Figures~\ref{caso3_1} and \ref{caso4_1}). However, small discrepancies can be detected in the retrieved Zernike coefficients, from the comparison of Figures~\ref{caso3_2} and \ref{caso4_2}. In particular, the \emph{bullet-dotted line} for $\alpha_4$ in the latter figure is shifted upwards by $\sim$ 0.015 waves ($\sim$ 0.44 mm of defocus distance) with respect to the corresponding line in the former figure. Again we find that the error in the knowledge of the real \emph{diversity} is transferred to the budget of real aberrations in the system, in the form of a defocus error. Nevertheless, in this case, the impact on the restorations can be consider as negligible since we get very similar results to those obtained in Experiment 3. \\

The repercussion of the discrepancies between the results of Experiments 1, 2, 3 and 4,are very subtle as to be easily detected by comparing the Figures~\ref{tiras-exp1}, \ref{tiras-exp2}, \ref{tiras-exp3} and \ref{tiras-exp4}.

\subsubsection{\bf Case of IMaX}
\begin{figure}
\centering
\begin{tabular}{cc}
\hspace{-1cm}\includegraphics[width=.55\linewidth]{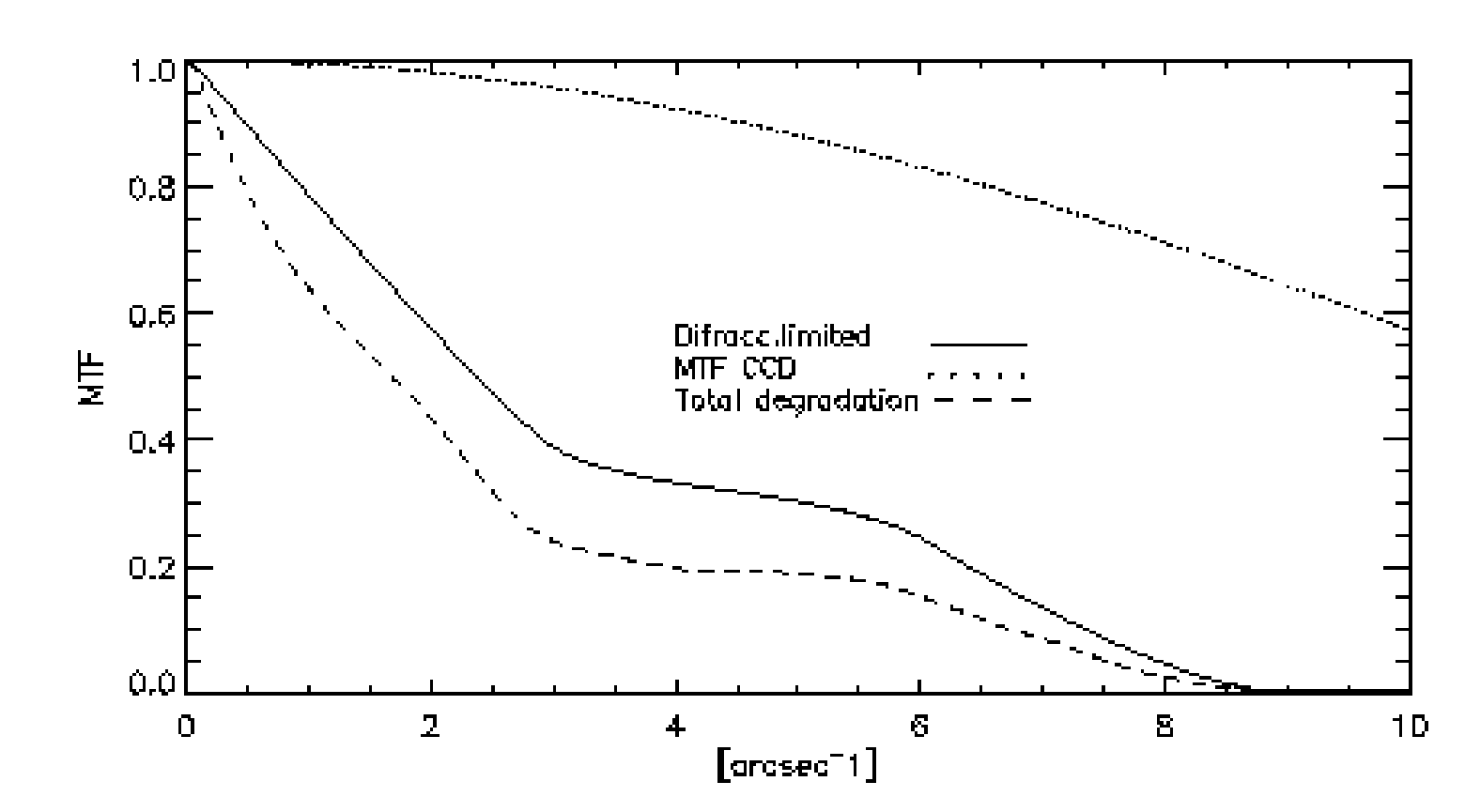} &
\includegraphics[width=.55\linewidth]{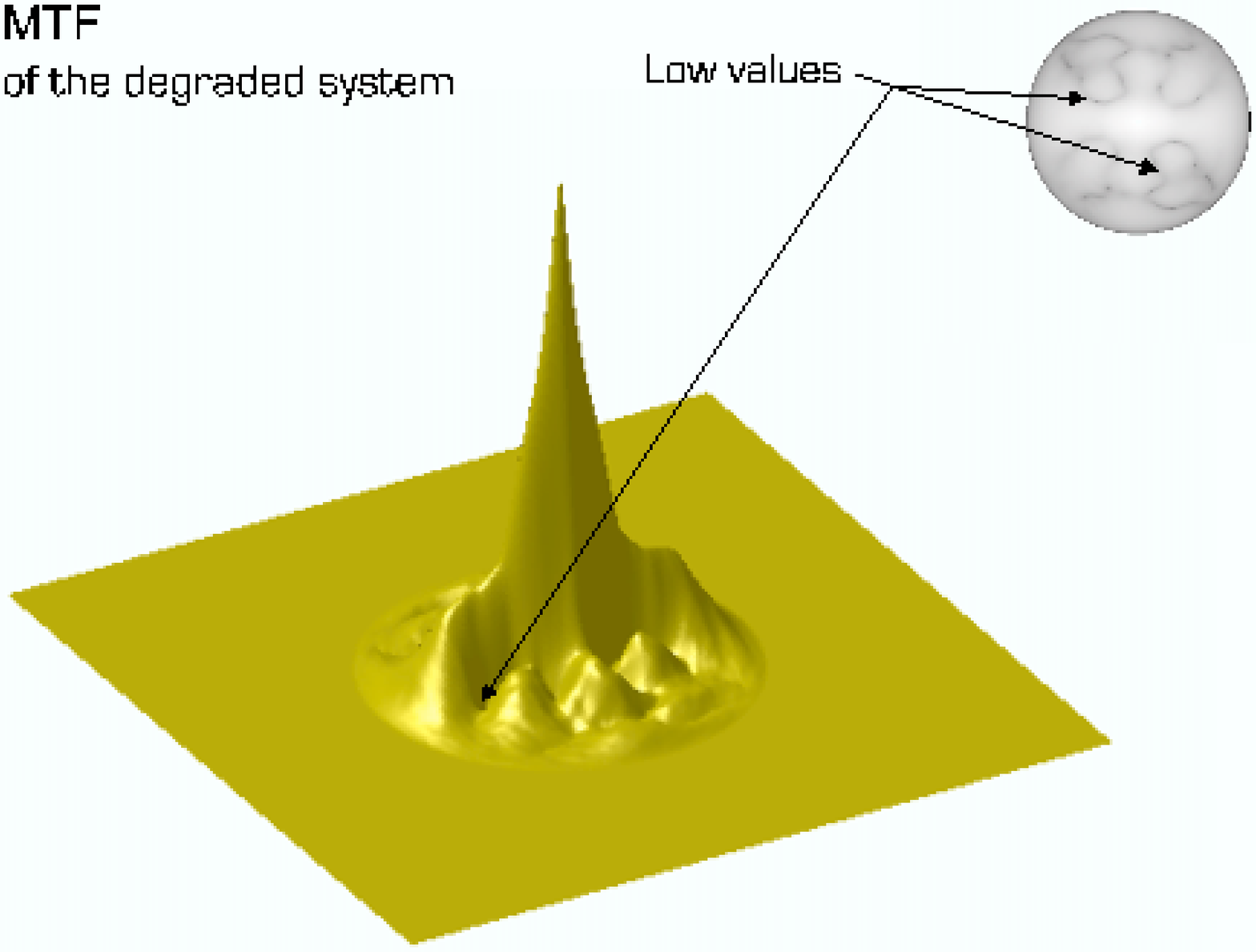} 
\end{tabular}
\caption[\sf MTF shape for the case of IMaX]{\sf MTF shape for the case of IMaX. \emph{Left panel}: Diffraction limited MFT (\emph{solid line}), MTF$_{\emph{footprint}}$  (\emph{dotted line}) and MTF resulting from the total error contribution (\emph{dashed line}). \emph{Right panel}: 3D representation of the latter.}
\label{imaxcaseMTF}
\end{figure}

According to the aberrations estimated from laboratory measurements and realistic models, the case of IMaX fits into Experiment 3 for \emph{rms-ripple} $\le$ 2/60 and \emph{rms-LOA} = 1/5 waves. This implies a loss of contrast of $\sim$ 5 \% of the real value and rms$(i_t-i_r)$ $\sim$ 3 \% (see Figure~\ref{caso3_1}). Figure~\ref{imaxcaseMTF} (\emph{left panel}) shows the diffraction limited MTF (\emph{solid line}) for SUNRISE. The \emph{dotted line} is the MTF$_{\emph{footprint}}$ representing the pixel integration effect in the CCD. The \emph{dashed line} is a section of the MTF resulting from the total error contribution estimated for the IMaX images including also the CCD pixel integration effect. The central depression in the MTF profile is very sensitive to the significance of LOA in such a way that it can almost reach a zero level at some particular spatial frequencies (see \emph{right panel} in Figure~\ref{imaxcaseMTF}). These singularities complicate the construction of the noise filter (see section~\S1.3.2) to be applied in the restoration process.\\

\begin{figure}
\centering
\includegraphics[width=.7\linewidth]{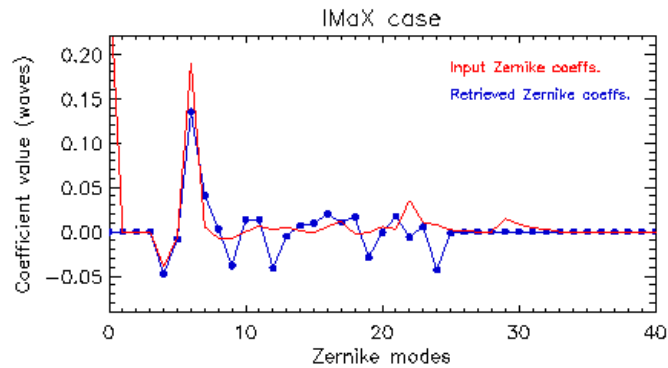}
\caption[\sf Comparative plot of Zernike coefficients for the IMaX case]{\sf Zernike coefficients for the IMaX case. The figure shows for comparison the coefficients used to degrade the image (\emph{input}) and those retrieved from the PD-inversion (\emph{output}).}
\label{imaxcaseZernikes}
\end{figure}

Figure~\ref{imaxcaseZernikes} is a comparative plot of the Zernike coefficients used for degrading the images (from index 4 to 36) and those retrieved from the inversion process (from index 4 to 25).
 
\section{Conclusions}
In this chapter, a method for in-flight calibration of aberrations in IMaX has been proposed. The method is based on the Phase Diversity (PD) inversion procedure.\\

The robustness of the method has been tested by numerical experiments simulating different aberration components. The sources of aberration have been modelled and added in subsequent experiments. The repercussion of every new added ingredient in the final result from the inversion has been evaluated.\\

The PD-code does not accurately reproduce the shape of the wavefront errors but provides reliable OTFs for subsequent satisfactory restorations. \\

Small-scale irregularities in the wavefront error are not detectable by the PD-code where a finite (rather low) number of Zernike terms is used. This limitation mainly produces stray-light all over the restored image and consequently a loss of contrast. This effect is tolerable within certain margins, and fixes constraints to the polishing quality in the main mirror of the SUNRISE's telescope and to the inhomogeneities in the etalon of IMaX.  \\

The inhomogeneities in the transmission function of the etalon are converted by the inversion process into some extra errors in the resulting wavefront which partially compensate for the loss of contrast caused by the high-order aberrations terms non-sensed by the PD-code.\\

Experiment 3 (see Table~\ref{num-experiments}) includes all the error sources assumed as significant for IMaX, and the results of its inversions can be considered reliable enough as to validate the method proposed for the calibration of errors in the images of IMaX, at least for aberrations lying in the range: \emph{rms-ripple} $\le$ 2/60 and \emph{rms-LOA} $\le$ 1/4 waves. The case of IMaX, with the expected values of \emph{rms-ripple} $<$ 2/60 and \emph{rms-LOA} $\sim$ 1/5 waves, is included in this range of aberrations and, consequently, the calibration method is applicable to IMaX.\\

For the case of IMaX, the residual errors in the proposed calibration method induce, in turn, errors in the subsequent restoration: mean$(|i_t-i_r|)$ $\le$ 2.5 \%, rms$(i_t-i_r)$ $\le$ 3.3 \%, and a loss of rms contrast $\le$ 5 \%. \\

The amount of defocus (also named \emph{diversity}) produced by the PD-plate is a critical parameter in order for the PD-algorithm to work optimally. We have got acceptable results for errors in the diversity of $\sim$ 5 \%.

%% file: chap3.tex
\chapter{\sf Solar active regions}\label{cap3}
\dropping[0pt]{2}{T}he second part of this thesis, devoted to the study of solar active regions, is introduced in this chapter based on the revision of literature published on this topic. Information about solar active regions is firstly presented starting from the historical point of view, secondly going into the theoretical and observational framework and finally reviewing the state-of-the-art in the research on the dynamics and evolution of photospheric structures around sunspots.

\section{Introduction}
\begin{figure}[h]
 \hfill
\begin{minipage}[h]{.35\textwidth}    
\caption[\sf Cartoon representing different layers in the Sun and their parameters]{\sf Cartoon representing different layers in the Sun:  Photosphere (Ph, \emph{yellow}) , Chromosphere (Ch, \emph{orange})  and Corona (Cr, \emph{blue}). The plot shows the stratification of temperature (\emph{black line}) and mass density (\emph{dashed line}) when moving radially outwards the Sun (height). The temperature minimum occurs at a height of 500 km above the deepest layer observable with optical telescopes. The temperature minimum marks the upper limit of the photosphere. The Transition Region (\emph{red}) is characterized by a huge increment of temperature in just a few hundredth kilometers. }
    \end{minipage}
    \begin{minipage}[h]{.60\textwidth}
      \epsfig{file=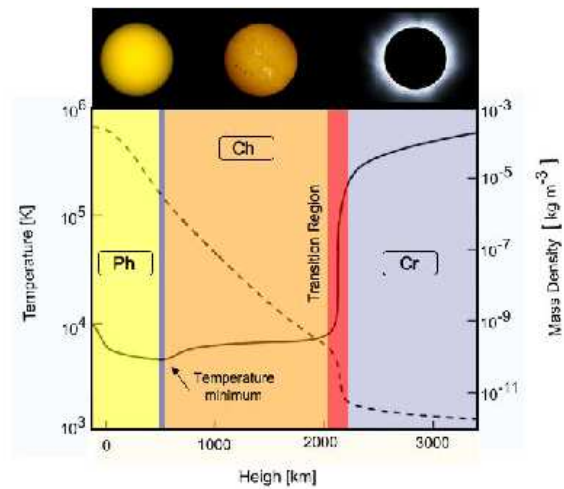, scale=0.38}
  \end{minipage}
  \hfill
\label{sunlayers}
\end{figure}

The importance of the Sun for humankind is more than justified as well as the research on solar physics 
in which the Sun appears to be a fabulous laboratory to study multitude of phenomena and fundamental processes in astrophysics. By the analysis of the radiation (\emph{photons}) coming from the different layers of the Sun (see Figure~\ref{sunlayers}) scientists have been able to figure out a large amount of information about the underlying physics behind our closest star. Telescopes and solar instrumentation have played an important role since the very first ground-based observations up to the more recent space telescopes and future missions trying to untangle the behavior of the Sun. 
The Sun as a variable star experiences a changing magnetic activity framed in what is called the solar cycle. The main cycle is composed by two consecutive 11 years cycles and represents a reversal of the global magnetic field. The solar cycle is evidenced by the variation of the number of sunspots, number of solar flares and solar irradiance among other parameters. Magnetism then dominates the Sun's behavior and can be studied in more detail than in any other stellar body. 
The vast majority of light leaving the Sun and reaching us (approx. 8 minutes later) comes from the photosphere, a thin layer of about 500 km of thickness (sort of 0.04$\%$ of the solar radius) that is actually the densest part of the solar atmosphere, composed by convective cells known as solar granulation. Underneath the photosphere, convection is taking place carrying the energy released by thermonuclear fusion in the solar core. Plasma motions in the underlying convection zone are therefore driving the dynamical behavior of the photosphere and reveal through the granulation pattern.

\vspace{3mm}

\section{Sunspots}
Sunspots are the most evident manifestation of the solar activity and of the solar cycle. They are related to the majority of violent phenomena occurring in the solar external layers and moreover a clear sign of the magnetism effects in the dynamics and structure of our closest star. 
Solar active regions (sunspots and plages) are labeled for identification by consecutive numbers as they appear on the visible solar disk. One of the most widely used identifiers is the NOAA (National Oceanic and Atmospheric Administration) catalogue. Many important facts about the sunspots have been studied for many years and we have learned things about its nature, origin and evolution, but there are still fundamental questions we are willing to clarify. One of the important issues about sunspots was to discover that they were the first astronomical object known to harbor a magnetic field \citep{hale1908}. \\

Sunspots are nowadays interpreted as complex structures having strong magnetic fields that inhibit the plasma convection reducing therefore the efficiency in the convective energy transport. The temperature of sunspots is for this reason lower than that in the surrounding photosphere. They are seen as black entities when comparing it with the outside brightness but actually the temperature inside sunspots goes up to 5000 K.\\

Study of sunspots is important because they are connected to intense and explosive phenomena in the chromosphere and solar corona. Coronal loops, for instance, are magnetic lines anchored to sunspots that after mutual interaction and reconnection liberate strong amounts of  energy  in the form of the so-called flares and coronal mass ejections (CME). The released energy travels millions of kilometers before interacting with the Earth's atmosphere and affecting the terrestrial communications and orbiting satellites. Clarifying the sunspots internal structure and their physical processes is demanding to understand and predict the solar activity and its influence on Earth.\\

\section{Historical overview}
The first observations of sunspots on the solar surface brought another view regarding the incorruptibility of heavens so that at the beginning many astronomers rejected to believe that sunspots were actually attached to the Sun. J. Fabricius was the first to write a book on sunspots in 1611 \citep{fabriciusMDCXI} and independently C. Scheiner claimed to observe sunspots the same year associating them to orbiting objects \citep{scheiner1623-30}. Scheiner is also regarded as the first who evidences velocity fields in the solar photosphere noting that sunspots near the equator traverse the solar disc faster than those at higher latitudes.
Galileo associated the sunspots to a sort of thick clouds sharing the solar rotation, as dense that they were able to block the Sun's light. He was the first that used the terminology of solar latitude and longitude to locate sunspots on the Sun. Galileo's accurate observations using a rather good telescope for the time being, allowed him to look at many details on the solar surface. He was also the first to discover brighter zones which he called \emph{faculae}.
But it was not until Keppler that sunspots were indeed put right on the solar surface \citep{keppler1955}. Keppler had first postulated that the Sun should rotate in less than three months but later on reduced the rotating period to just one day.
After proving by means of geometrical arguments that Scheiner was wrong when considering the behavior of sunspots at the solar limb, Keppler settled the question, being convinced that spots are on the Sun as we all know now.\\

Sunspots observations continued in the XVII century although there is a lapse from 1645 to 1715 where a very few sunspots were observed. This fact is known as the \emph{Maunder minimum} and corresponds to a time period in which the Earth suffered from very low temperatures and cold winters that are associated to the reduction of the solar activity.  Nowadays, the Sun is still affecting us directly and indirectly and the Sun`s changing magnetic behavior has an influence on the solar neighborhood climate.\\

More solar properties and parameters, as the solar axis inclination and differential rotation rate were inferred by \citet{carrington1859} and confirmed by several studies afterwards. The dynamics of the plasma in the solar photosphere was found to show several convective patterns at different spatial scales. In addition to the granular pattern (with typical diameters of 150-2500 km), the mesogranular one at intermediate scales (5-10 Mm) and the supergranular one at large scales  (over 20 Mm) organize the plasma in a sort of hierarchical convective structure. There is still an open debate concerning the nature, origin and structure of these convective patterns.\\

For a more extensive and detailed historical review on observations of sunspots see  \citet{casanovas1997} and references therein.

\section{Structure of sunspots}
The responsible for the origin and structure of sunspots is believed to be the deep-seated toroidal (azimuthally oriented) magnetic flux in the solar interior ~\citep{schussler2002}. This magnetic flux is thought to be situated between the convective zone and the radiative interior and increases by means of a dynamo action up to the formation of a kind of instability. The eruption of flux leads to the formation of $\Omega$-shaped magnetic loops coming up to higher layers, therefore reaching the convective zone and emerging through the solar surface as active regions ~\citep{zwaan1992}. Summarizing this idea, what we have is that the magnetic field generated underneath the convective zone in the Sun, emerges to the solar surface forming sunspot groups and active regions and spreads around its surface through convective mechanisms. \\

Diverse models for sunspots have been proposed through the years such as the cluster model and the monolithic model. On the one hand the cluster (\emph{spaghetti}) model developed by ~\citet{parker1979} in which the magnetic field divides into many separate flux tubes in the first few Mm below the surface. On the other hand the monolithic model by ~\citet{cowling1957} establishes that the magnetic field underneath the solar surface is confined to a single flux tube and the plasma flows crossing it are prohibited.\\
 
Sunspots, as complex magnetic structures embedded in a convective plasma,
show many active and changing features on multiple scales. Convective flows and
large-scale plasma circulation plays an important role in the dynamics and
evolution of solar active regions \citep[see, e.g.,][]{schrijver2000}\\

Important physical processes are present in sunspots, some of them at spatial-scales beyond the resolution limit of the current solar telescopes. In the last years and using the latest restoration techniques and advances in adaptive optics, the complex magnetic and dynamic structure of sunspots have been unveiled in some detail. Nevertheless, there are many important questions concerning them that remain unanswered yet.\\

\subsection{Umbra}
The umbra is the central dark area that corresponds to the coolest part of a sunspot. The temperature inside the umbra is low ($\sim$ 3500-5000  K) due to the presence of a strong magnetic field that inhibits the convective energy transport underneath ~\citep{biermann1941}. The magnetic field orientation inside the umbra is vertical respect to the solar surface and becomes slightly inclined when approaching the umbra-penumbral boundary. Larger sunspots with bigger umbrae exhibit a more intense magnetic field strength and the range inside umbrae goes from 2000 to 3500 Gauss in average. The energy radiation in the umbra is roughly 20$\%$ of the photospheric value.

Although the umbra seems to be completely dark and with non internal structure, when observed in high-resolution, a lot of fine features in its interior can be detected, for instance the \emph{umbral dots} which are bright dots with sizes of about 0.2-0.5 arc seconds and hotter than the umbral background. These features are still under study to clarify and confirm whether they are related to convective phenomena transporting some sort of flow. Light bridges are other structures inside the umbra that look like luminous streaks or narrow bands with different ages and sizes. 

\subsection{Penumbra}
\label{penumbra}

\begin{figure}
\centering
\begin{tabular}{lr}
\includegraphics[width=0.56\linewidth]{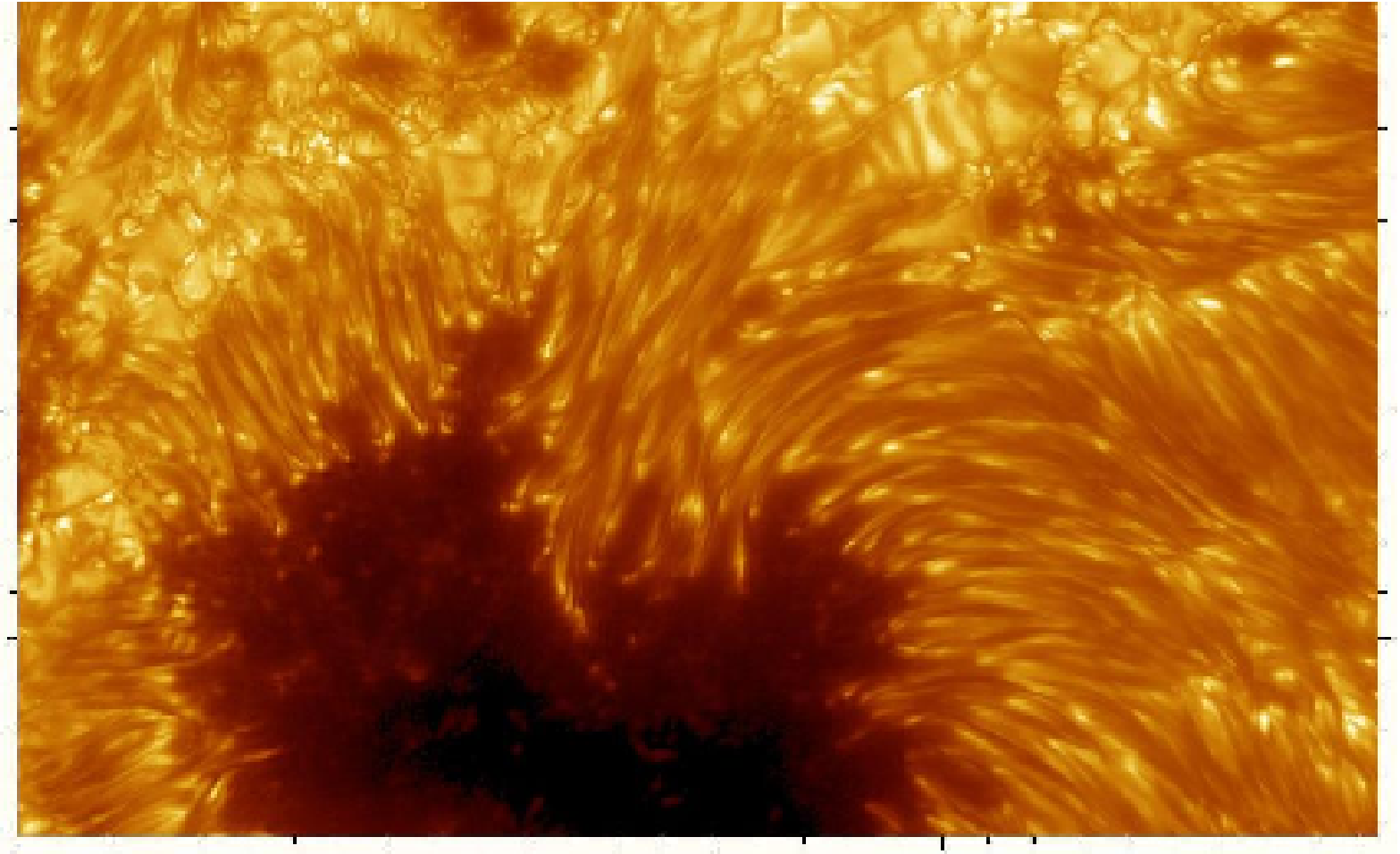} &
\includegraphics[width=0.35\linewidth]{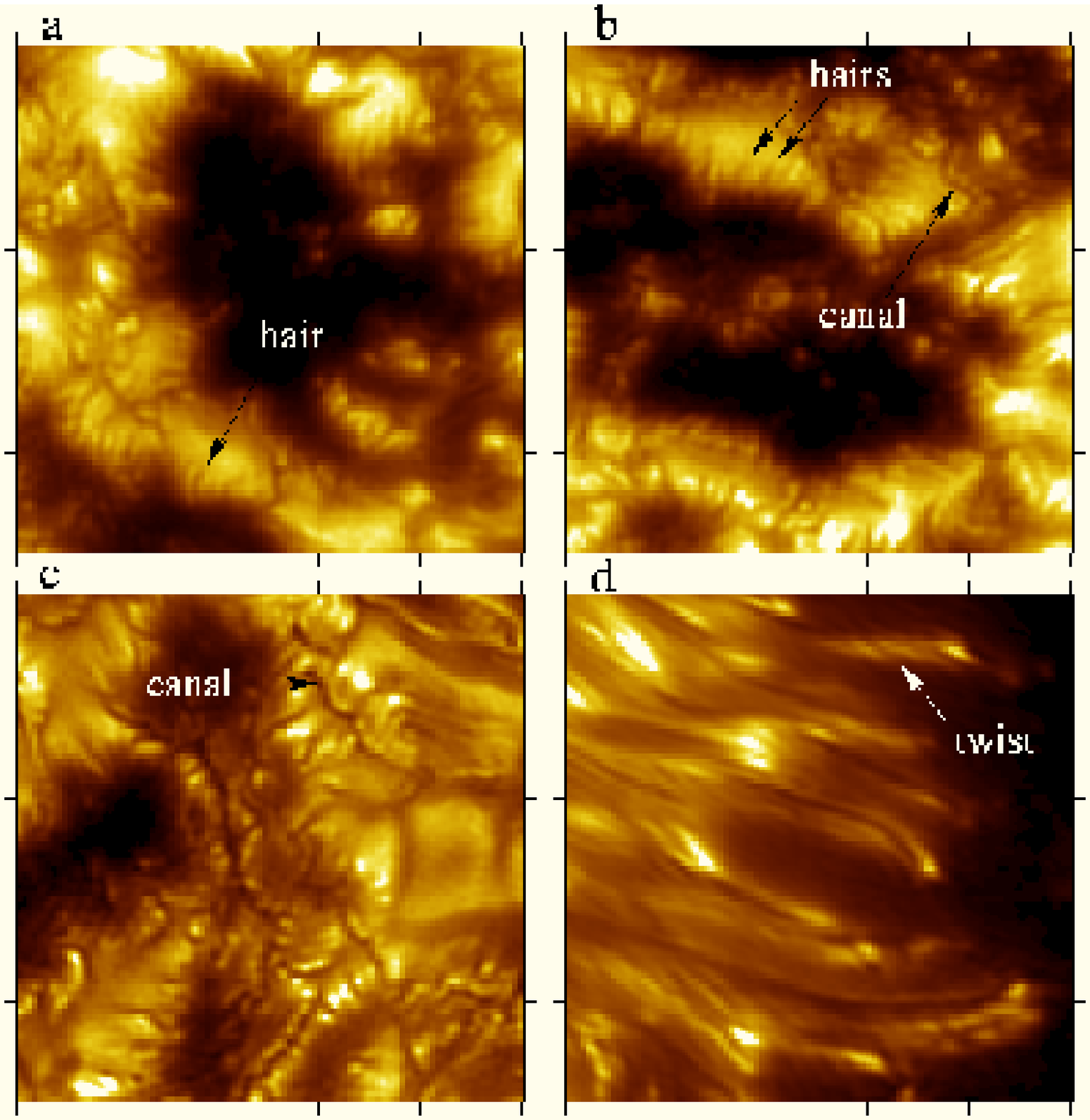}
\end{tabular}
\caption[\sf Images of sunspot penumbra and fine-scale solar structures]{\sf Image showing a highly-resolved sunspot penumbra in a sunspot:  AR 10030,  observed the 15 July 2002 with the 1-meter SST in la Palma (\emph{left}) and some fine-scale structures (\emph{right}) placed inside and around solar pores. \emph{Courtesy Institute of Solar Physics, Stockholm}.}
\label{sstimages}
\end{figure}

The penumbra is the filamentary and deeply-rooted (thick) structure in a sunspot that forms the ring-shaped part surrounding the umbra, where a significant part of the total magnetic flux is carried  ~\citep{schmidt1991, thomas1992}. The penumbra is seen in continuum images as radial alternating bright and dark filaments \citep[e.g.][]{collados1988} and exhibits an intermediate brightness between the umbra and the quiet photosphere \citep[see also the reviews by][]{deltoro2001,solanki2003}. Figure~\ref{sstimages} shows some high resolution images taken with the SST in La Palma, where we can see the intricate filamentary structure of the penumbra and the fine features inside and around pores.\\

\begin{figure}
\centering
\includegraphics[width=0.7\linewidth]{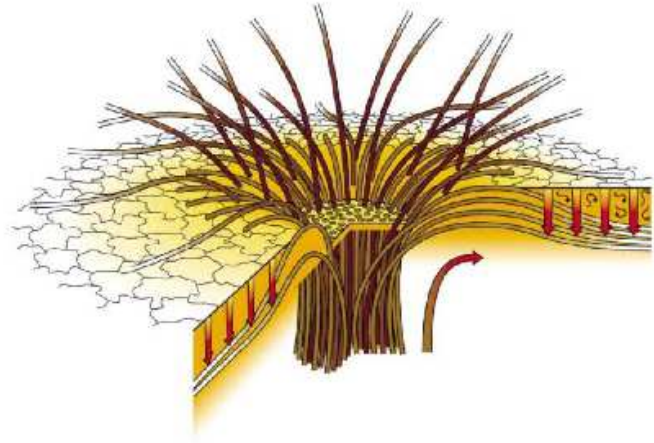} 
\caption[\sf Cartoon showing the structure of the magnetic field in a sunspot penumbra]{\sf Cartoon showing the interlocking structure of the magnetic field in the filamentary penumbra of a sunspot. Radial bright filaments where the magnetic field is inclined ($\sim 40\,^{\circ}$ respect to the horizontal direction in the external penumbra) coexist with the dark filaments having an almost horizontal magnetic field. \emph{Courtesy John H. Thomas}.}
\label{thomastubes}
\end{figure}

Magnetic field is in general stronger in the inner part closest to the umbra and decreases in the outer frontier and its  energy radiation is approximately 75$\%$ the photospheric value. The picture of the fine structure of the penumbra has not always been the same but evolved during the last years. Nowadays it is commonly accepted that the penumbral magnetic field is \emph{uncombed} so that at least two different inclinations of the magnetic field coexist on a small scale in a sort of interlocking structure (see Figure~\ref{thomastubes}).\\

The main two distinct inclinations correspond to the component where the magnetic field is tilted by $\sim$ 40-50 degrees (with respect to the vertical to the solar surface) with a strength of $\sim$2000 G, and the component in which the magnetic field is almost horizontal and weaker \citep{lites1993}. Vertical magnetic inclinations were first associated to bright filaments whereas the more horizontal ones to dark filaments though nowadays this picture has changed significantly since the discovery of two narrow lateral brightenings located at each side of the central obscuration, as will be commented below.\\

Recent works using high-resolution images have improved the measurements of the field inclination in the penumbral filaments \citep{bello2005,langhans2005}. Results from full-Stokes profiles have confirmed and firmly established that variations of inclination and field strength are indeed present in sunspot penumbrae, and moreover they are anti-correlated: the less inclined components (known as \emph{spines}) have stronger magnetic field than the more  horizontal ones \citep[know as \emph{intraspines},][]{lites1993}. In the intraspines, some flux tubes (collection of multiple magnetic field lines) overtake the external penumbral border \citep{sainz2005} whereas some other immerse below the surface due to the turbulent effect of the granular convection pushing them downwards ~\citep{westendorp1997,thomas2002}. We will come back again to this topic.\\

Although many unknowns have been unveiled, there are still many fundamental and unanswered questions such as why sunspots have a penumbra and how it is maintained for a few days before it disappears. A lot of internal structure has been resolved in penumbral filaments and they seem to be composed by a dark core accompanied by a bright halo at both sides. This new features named dark-cored penumbral filaments, recently discovered by \cite{scharmer2002}, have brought great interest as they can unveil the very fundamental influence of magnetoconvection at small scales occurring in sunspot penumbrae. At the beginning of the filament, the dark core is not present and the lateral brightenings merge into a larger bright structure known as \emph{penumbral grains} (PG) that seems to move radially in the penumbra.\\

Several authors have worked on this new observational evidence to investigate on the fine structure of the penumbra and nowadays there some models of the penumbral fine structure trying to explain all diverse phenomena. Firsly, the most mentioned and confronted one, the \emph{uncombed} model proposed by \cite{solanki1993} in which the penumbra is seen as a collection of horizontal flux tubes embedded in an almost vertical and strong background magnetic field. Numerical simulations of the temporal evolution of flux tubes \citep[the moving-tube model of][]{schlichenmaier1998a,schlichenmaier1998b} have modelled them starting at the magnetopause, then rising adiabatically as a result of radiative heating, by magnetic buoyancy, and eventually being bent into the horizontal where they cool down due to radiative losses while continue outward from the center of the spot. Recently, some new observational and theoretical works support the uncombed model\footnote{\sf Also known as embedded flux tube model.} \citep[e.g.][]{muller2002,bellot2003,borrero2005,bellot2007b,jurcak2008,ruizcobo2008}. A second model has been proposed by \cite{spruit2006}, the so-called \emph{gappy} penumbral model \citep[see also][]{scharmer2006}. These authors interpret the penumbral filaments as due to convection in field-free, radially aligned gaps below the surface of the penumbra and argue that this solves the discrepancy between the large heat flux and the low velocities present in penumbra or, in other words, that their model can explain the heating of the penumbra. Apart from these two models, alternative ones have come out on the scene based on MIcro-Structured Magnetic Atmosphere \citep[MISMA,][]{salmeida1998,salmeida2005}. These models assume that the penumbra is formed by optically thin magnetic fibrils a few km in diameter and moreover that every resolution element contains a bunch of unorganized field lines with random strengths and inclinations. In general, all models have some problems \citep[see][]{bellot2007a} and the debate is not sealed yet.\\

New features and fine-scale structures have been recently observed thanks to the improvements in the real-time and post-facto correction of the images.  It is likely that inside dark cores there is more unseen fine structure.  This new structures will add more ingredients to this already complicated scenario. \\

The penumbra not only shows an intricate magnetic structure but also a complex dynamics. The PGs for instance, are moving in two privileged directions, radially towards the umbra in the inner penumbra with velocities of the order of 500 m s$^{-1}$ ~\citep{muller1976} and towards the surrounding photosphere in the outer penumbra  ~\citep{sobotka1999,sobotka2001}. The line dividing both distinct directions of movement is placed at a distance from the outer penumbral border of about one third of the total penumbra longitude. Some of the PGs cross the penumbra-photosphere boundary and continue to move in the granulation as small bright features \citep{bonet2004}.\\
 
Apart from the radial proper motions of PGs, \cite{marquez2006} find, by using local correlation tracking techniques, that radial motions diverge away from bright filaments to converge toward dark filaments, leading to an association between downflows and dark features that could be a sign of convection in penumbra. This result is reinforced by \cite{salmeida2007} on the basis of spectroscopic measurements, with high-angular resolution, of the Evershed effect (see next section).\\

\vspace{3mm}

\subsection{Evershed flow}
The Evershed Flow (hereafter EF) is associated to an observational effect in the penumbra registered as a global wavelength shift for spectral lines forming in the penumbra of sunspots. There are actually two kinds of EF registered: 1) the normal EF at photospheric level where blueshifts are observed in the penumbra on the sunspot side close to the solar disc center whereas redshifts are detected on the sunspot side towards the solar limb. It corresponds to mainly radial and horizontal flows traveling in the penumbra of the sunspots towards the external photosphere; 2) at chromospheric level, the so-called inverse EF, where the material flows towards the umbra, contrary to the normal EF and much less studied so far. From here forth, we will focus on the normal EF.\\

This effect was first observed by Evershed in 1909 \citep{evershed1909} at Kodaikanal Observatory in India and is interpreted as a Doppler shifting caused by the radial and predominantly horizontal flow along the penumbra. The EF has been interpreted as a siphon flow along magnetic flux tubes \citep{meyer1968,thomas1988}.\\

There have been several studies trying to explain the properties and nature of the EF in the last forty years \citep[e.g.][]{haugen1969,maltby1975,moore1981,dialetis1985,dere1990}. For a very recent review the reader is referred to the work by \cite{bellot2007a}.\\

In more recent studies and helped by high-resolution observations, it has been confirmed that there is a strong correlation between the EF and the horizontal magnetic fields in the penumbra. It is the expected behavior as resulted from magnetohydrodynamical considerations that require for the flow to be aligned respect to the magnetic field lines ~\citep{bellot2003}.  ~\cite{rimmele1995} found that the EF is confined in thin channels which elevate upon the continuum level over the vast majority of the penumbra. The last result has been afterwards confirmed in different observations. It has also been found that some of these channels and their associated field lines re-enter the solar surface in the outermost part of the penumbra and beyond ~\citep{westendorp1997,schlichenmaier2000}. Some recent results indicate that the EF does not end abruptly at the white-light sunspot boundary, but that at least part of it continues beyond the visible boundary and forms an extensive magnetic canopy \citep[e.g.][]{solanki1994,rezaei2006}. The existence of radial outflow in the sunspot canopy is still an ongoing debate.\\

\vspace{3mm}

\section{Photosphere surrounding sunspots}
\begin{figure}[h]
 \hfill
\begin{minipage}[h]{.5\textwidth}
      \epsfig{file=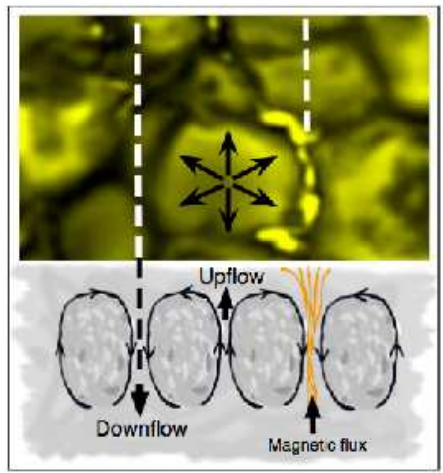, scale=0.43}
  \end{minipage}
  \hfill
\begin{minipage}[h]{.40\textwidth}    
\caption[\sf Sketch showing the convective pattern mechanism in a granulation region]{\sf Sketch showing the underneath convective pattern mechanism in a solar granulation region. The black arrows plotted over the granulation image stand for the divergence directions of the exploding granule whereas the short black arrows below show the position of the global up and downflows.  Magnetic lines (\emph{in orange}) are also delineated and correspond to small flux tubes seen as bright structures in the inter-granular areas when reaching the photospheric level.}
\label{undergranule}
 \end{minipage}
  \hfill
\end{figure}

Sunspots are immersed in the convective plasma as mentioned above, and this plasma is observed mainly as granulation. Granulation is the first pattern that can be identified when looking at the solar photosphere and covers the vast majority of its surface. Figure~\ref{undergranule} shows the underneath convective pattern mechanism forming the granulation that we see on the photospheric layer. In the inter-granular lanes we can detect the presence of magnetic flux coming from below, sometimes visible as bright and elongated structures as shown in the high-spatial resolution image in Figure~\ref{undergranule}.\\

The granulation convective pattern surrounding sunspots is perturbed by the presence of magnetic elements that move radially outwards through an annular cell called the "moat". Next sections comment on the physical properties of these magnetic structures and on the plasma flows around sunspots.

\vspace{3mm}

\subsection{Moving Magnetic Features}
The Moving Magnetic Features (MMFs) are bright features that correspond to small magnetic elements  of mixed polarity traveling radially outwards while immersed in the granulation surrounding sunspots \citep[][for a recent review]{sheeley1972,harvey1973,hagenaar2005}. These features move with velocities of some hundredths of meters per second up to 3 km s$^{-1}$ and can be generally classified in three main groups, as follows:

\begin{figure}
\centering
\hspace{18mm}\includegraphics[width=0.8\linewidth]{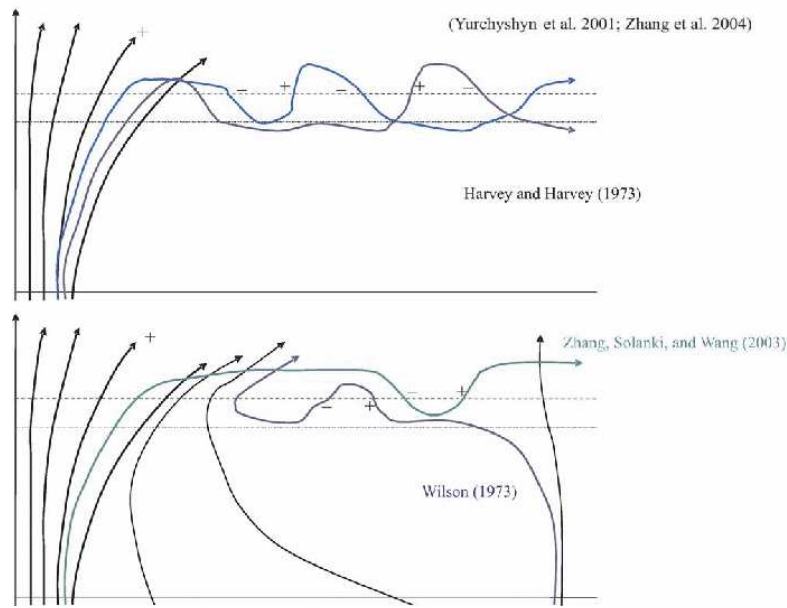} 
\caption[\sf Diverse interpretations for the Type I  MMFs]{\sf Diverse interpretations for the bipolar Type I  MMFs}
\label{MMF}
\end{figure}
\vspace{3mm}

\begin{itemize}
\item Type I:  Bipolar pairs of magnetic elements moving together outwards the sunspot inside the moat, with velocities in a range between 0.5 a 1  km s$^{-1}$. They appear just outside the penumbra along directions that are extensions of the dark penumbral filaments. These bipolar pairs are interpreted as being footpoints of a loop emerging to the solar surface belonging to a penumbral flux tube in which there is a strong vertical convection. The loop is then swept away by the moat flow. Several loops can emerge in different positions and different times along the same flux tube and this mechanism is seen when successive MMF Type I tend to follow similar trajectories across the moat region. As long as the flux tube is attached to the sunspot, this type of MMFs associated to the tube are not responsible for the sunspot decay. Figure~\ref{MMF} shows the diverse interpretations of bipolar type I MMFs ~\citep{yur2001,wilson1973, harvey1973}.

\item Type II: Consist of unipolar magnetic elements with the same polarity as the parent sunspot that are shed from the sunspot border, moving across the moat with velocities of 2 and 3 km s$^{-1}$. They are interpreted to be flux tubes that have been separated from the main flux and are dragged by the moat. The loss of flux represented by this features is a possible mechanism involved in the sunspot decay ~\citep{hagenaar2005}.  Nevertheless, further investigations should be done to know whether or not this type of MMFs is more common in rapidly decaying sunspots.

\item Type III: The last type of MMFs are also unipolar features but in this case with magnetic polarity opposite to that of the parent sunspot, having high velocities in a range similar to that of the Type II. They are more transitory than the first two types and travel outward a few thousand km before disappearing, and are thought to be produced when a flux tube section emerges to the surface generating a rapid horizontal displacement of the footpoints in the moat.
\end{itemize}

\subsection{Moat flow}
The existence of an annular region around sunspots was first reported by \citet{sheeley1969} who termed it \emph{moat}. Radially oriented outflows in moats were found by \citet{sheeley1971}. \\

The sunspot moat defines an organized horizontal flow pattern that ends quite abruptly at a distance that can be comparable to supergranular sizes or even larger. \cite{sheeley1972} reported Doppler speed of these outflows 0.5~-~1.0 km s$^{-1}$ for distances 10-20 Mm beyond the outer boundary of the sunspot. \cite{brickhouse1988} performed a study of the properties of moats and concluded that the moat radius is roughly proportional to the penumbral radius. \\

\begin{figure}
\centering
\includegraphics[width=0.8\linewidth]{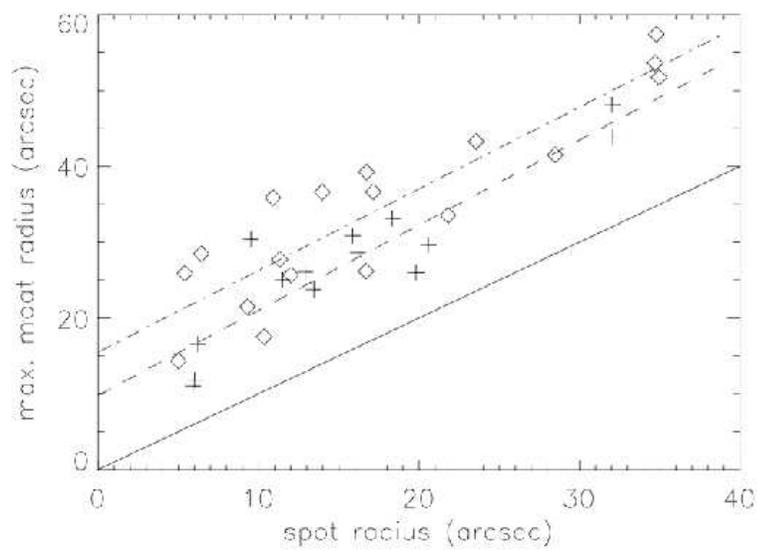} 
\caption[\sf Plot of the maximum moat radius versus sunspot radius.]{\sf Plot of the maximum moat radius versus sunspot radius. Young spots are marked by "$+$" and old spots by "$\Diamond$". Linear fits are shown for young spots (dashed line) and old ones (dash-dot). Solid line correspond to values $R_{max} = r_s$. \emph{Taken from \citet{sobotka2007}}.}
\label{fig2sobotka}
\end{figure}

\begin{figure}
\centering
\includegraphics[width=0.9\linewidth]{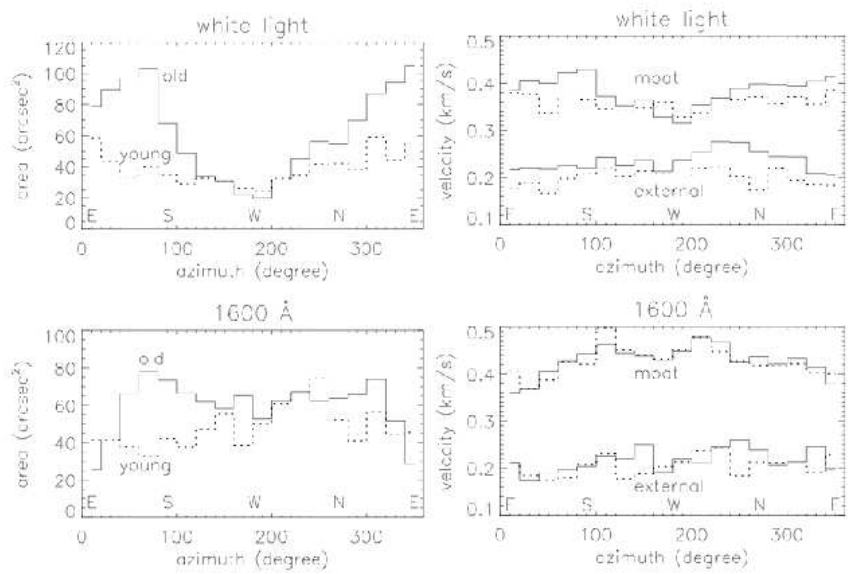} 
\caption[\sf Average azimuthal distribution of the moat area and of the velocities inside and outside moats.]{\sf Plot of the average azimuthal distribution of the moat area (\emph{left}) and  average azimuthal distribution of velocities inside moats and outside (\emph{right}) for old spots (\emph{solid line}) and young sunspots (\emph{dotted line}). \emph{Taken from \citet{sobotka2007}}.}
\label{fig3sobotka}
\end{figure}

The characteristics and temporal evolution of moats derived at two heights in the atmosphere for a large sample of sunspots with different sizes, shapes and evolutionary stages have been recently studied by \cite{sobotka2007}. These authors establish a relation between sunspot types and their occurrence, areas, and horizontal velocities of the moats in the photosphere and in the transition region and found that moats do not show substantial changes for periods of about 12 hours. Table~\ref{tablemoat} lists the averaged horizontal velocities inside and outside the moat for young and old spots and Figures~\ref{fig2sobotka} and \ref{fig3sobotka} show plots (for young and old spots) of the maximum moat radius versus the sunspot radius, and the average azimuthal distribution of the the moat area and velocities respectively.\\

\cite{bonet2005} have found that MMFs in the moats are mostly located in radially oriented \emph{channels} between the local divergent motions of mesogranular scale. This motions confine the MMFs within the channels. They also found that most MMFs move passively with the same average speeds as the neighbouring granules, that is, they are carried along by the same large-scale moat flows as the granules. However, a small fraction of them (6\%) move away from the spot more rapidly (they reach velocities grater than 1.4 km s$^{-1}$) than the neighbouring granules which can be interpreted as the consequence of interactions with fast moving small granular fronts or fragments.\\

\begin{figure}
\centering
\includegraphics[width=0.95\linewidth]{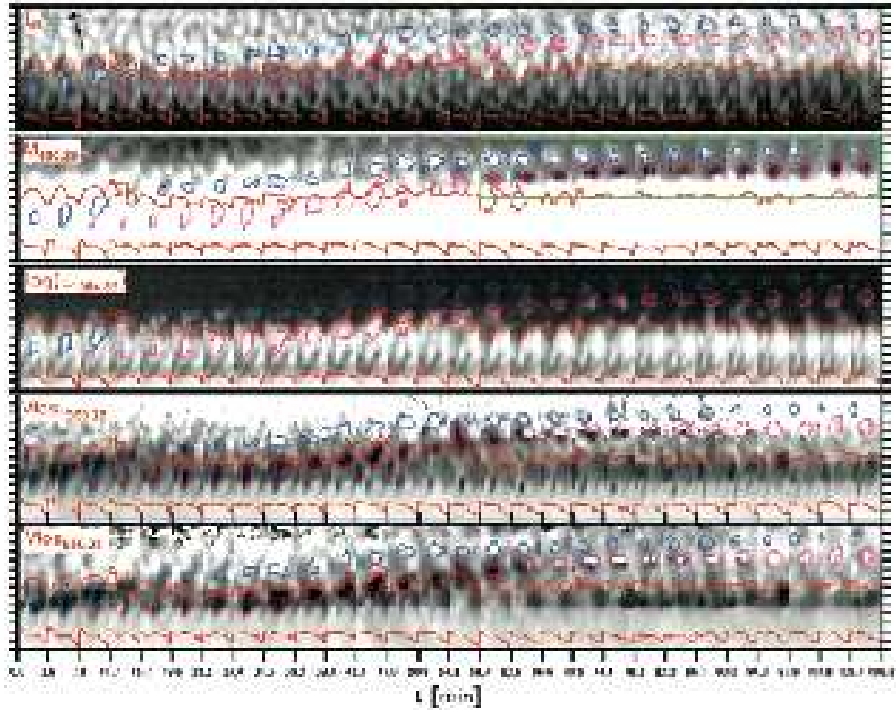} 
\caption[\sf Temporal evolution of the center-side penumbra and moat region]{\sf Temporal evolution of the center-side penumbra and moat region as studied by \cite{cabrera2006}. The figure shows (\emph{from top to bottom}): Continuum intensity at 1565 nm, magnetogram signal from Fe I 630.25 nm, logarithm of the total linear polarization of Fe I 1564.85 nm, Stokes V zero-crossing velocity of Fe 1565.28 nm, and Stokes V zero-crossing velocity of Fe I 630.25 nm. Negative velocities are blue shifts. Pink and blue contours delimit features A and B respectively. Red lines indicate the inner and outer penumbral boundaries. Tick marks in x-axis represent 1\arcsec. The black arrow points to the disc center and the time scale starts at June 30, 9:34 UT.}
\label{figcabrera}
\end{figure}

\begin{figure}
\centering
\includegraphics[width=0.95\linewidth]{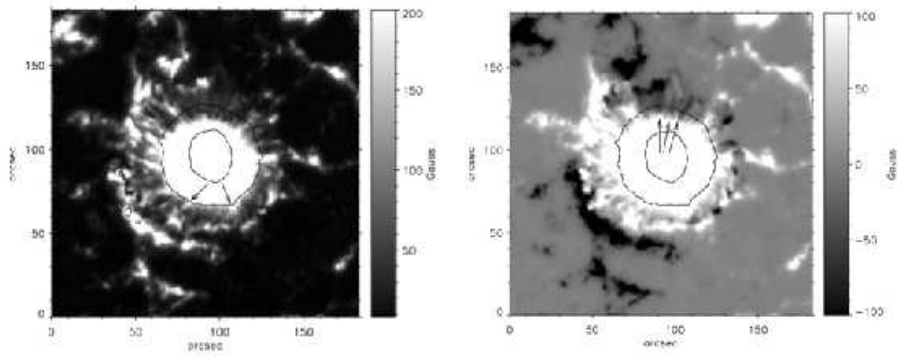} 
\caption[\sf Magnetograms showing the presence of moat filaments around the spots]{\sf Average of the absolute value (\emph{left}) and true sign magnetogram  (\emph{right}) showing the presence of moat filaments all around the sunspot. Black arrows point out some examples. The mean umbral and penumbral boundaries are overlaid. \emph{Taken from \citet{sainz2005}}.}
\label{figsainz}
\end{figure}

New findings \citep{sainz2005} have shown that the penumbral filaments extend beyond the sunspot boundary entering the region dominated by the moat flow where the MMF activity is detected. Thus, the temporal average of magnetograms has unveiled the existence of moat filaments: horizontal, filamentary structures coming from the penumbra and reaching the photospheric network as an extension of the penumbral filaments. Moreover, some MMF have been found \citep{sainz2005,ravindra2006,cabrera2006} starting just inside the outer penumbral boundary in its way out from the sunspot (see Figures~\ref{figcabrera} and \ref{figsainz}). \cite{kubo2007} have recently established a relationship between the vertical component (spines) of the magnetic field in the so-called uncombed structure of the penumbra (section~\S\ref{penumbra}) and MMFs (with the same polarity as the sunspot) observed in the moat regions.

\begin{table}[]
\sffamily
\caption[\sf Averaged horizontal velocities inside and outside moat ]{\sf Averaged horizontal velocities [m\,s$^{-1}$] inside and outside moat. \emph{Taken from \citet{sobotka2007}}.}
\centering
\vspace{4mm}
\begin{tabular}{llcc}
\hline
Position & Passband & Young spots & Old spots \\
\hline
 \multirow{2}{*}{Inside moat    ....} & white light & 380 $\pm$ 30  & 410  $\pm$ 30\\
                      & 1600 \AA & 450  $\pm$ 70 & 450  $\pm$ 40  \\
 \multirow{2}{*}{Outside moat ....} & white light & 200 $\pm$ 40 & 240 $\pm$ 40  \\
                       & 1600 \AA & 200 $\pm$  40 & 210 $\pm$  50 \\
\hline
\end{tabular}
\label{tablemoat}
\end{table}
 
\vspace{3mm}

\vspace{3mm}

\vspace{3mm}

\vspace{3mm}

\clearemptydoublepage

%% file: chap4.tex
\chapter{\sf Proper motions in a complex active region}\label{cap4}
\dropping[0pt]{2}{T}his chapter is the first one devoted to the analysis of solar observations, showing the results from the application of the image restoration process commented in the former part of the thesis. The study is based on ground-based observations of a complex $\delta$-configuration solar active region in which we performed a detailed analysis of proper motions at photospheric level describing the dynamics of the plasma in the FOV. Some results presented in this chapter are an extensive version of \cite{vargas2007}.\\

\section{Introduction}
\label{S:intro}

Sunspots that contain umbrae of opposite magnetic polarity within a single penumbrae are known as $\delta$-configuration sunspots. They are complex active regions that usually evolve rapidly and are likely to produce energetic solar flares ~\citep{zirin1989}. The present chapter is based on an excellent $\sim$79 minutes time series of a sunspot group with a $\delta$-configuration. The long duration of the series, besides the high and stable quality throughout the entire period, substantially improved after image reconstruction, make of this material an excellent data set to study morphology and dynamical behavior of sunspots and their surroundings. We are interested here in the relation between the existence of penumbra in this active region and the dynamics of the convective plasma surrounding it. The proper motions inside penumbrae are also analyzed and a statistical study between the different regions over the FOV is performed.

\section{Observations}
\label{S:obs}

The first observing campaign I was involved in at the Swedish 1 m Solar Telescope ~\citep[SST\footnote{\sf The SST is operated by the Institute for Solar Physics (ISP) of the Royal Swedish Academy of Sciences at the Roque de los Muchachos Observatory in La Palma, Canary Islands.},][]{scharmer2003a} corresponding to the so-called Spanish Time\footnote{\sf The Spanish Time corresponds to the 20\% of the total observing time at the Observatories of the Canary Islands, assigned to Spain by the institutions the telescopes belongs to, by virtue of the \emph{International Agreements on Cooperation in Astrophysics}.}, was held in April 2004. During this campaign I was particularly interested on becoming familiar with the instrumentation, software and observing techniques that have been in constant evolution and development in recent years. Moreover, the observed images were used to test the restoration algorithms (see chapter~\S1).\\

The second observing campaign was held at the SST in July 2005 and corresponded to the so-called International Time Program (ITP)\footnote{\sf Period corresponding to the 5\% of the total observing time (see previous footnote), aimed to promote astrophysical research in cooperation with various telescopes and institutions.}. In this framework, several Japanese scientists came to the Canary  Observatories \footnote{\sf Observatorio del Teide (OT) in Tenerife and Observatorio Roque de los Muchachos (ORM) in La Palma.} for a joint programme of simultaneous observations using various solar telescopes: the Vacuum Tower Telescope (VTT), the Solar Swedish Telescope (SST), the Dutch Open Telescope (DOT), and the satellites Transition Region and Coronal Explorer (TRACE) and Solar and Heliospheric Observatory (SOHO). The observations were focused on targets displaying solar active regions and, in particular, regions inside and around sunspots where various magnetic features (e.g.\ MMFs, see section~\S3.4) are widely observed. The SST and the DOT were devoted to imaging, pursuing observations of the photosphere and chromosphere with high -spatial and -temporal resolutions.
The main objective of these coordinated observations was to figure out what kind of magnetic features in the photosphere/chromosphere are responsible for quasi-steady heating and enhanced activities well observed in the active corona.\\

\begin{figure}
\begin{center}
\begin{tabular}{cc}
\includegraphics[width=0.49\linewidth]{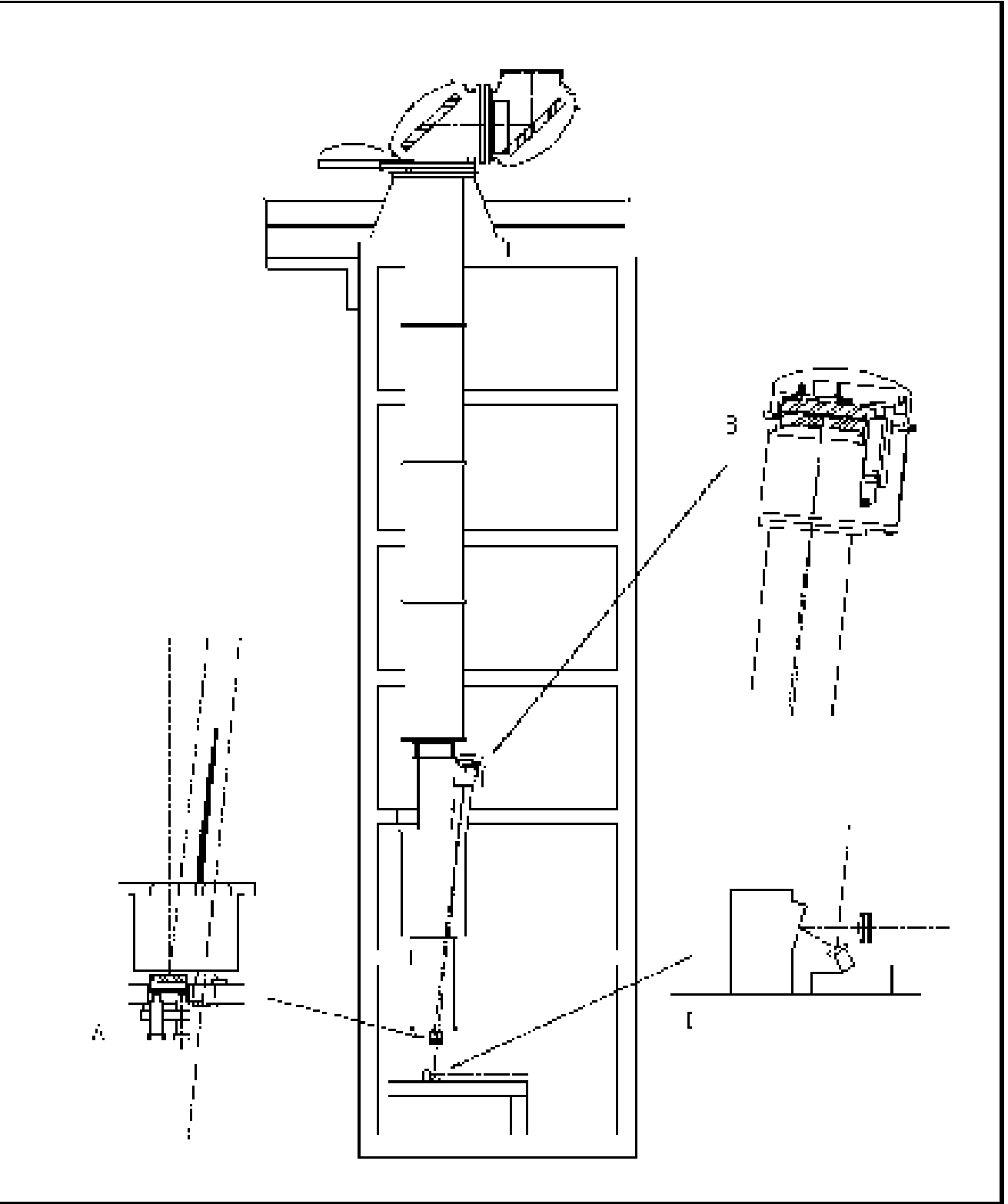} &
\includegraphics[width=0.47\linewidth]{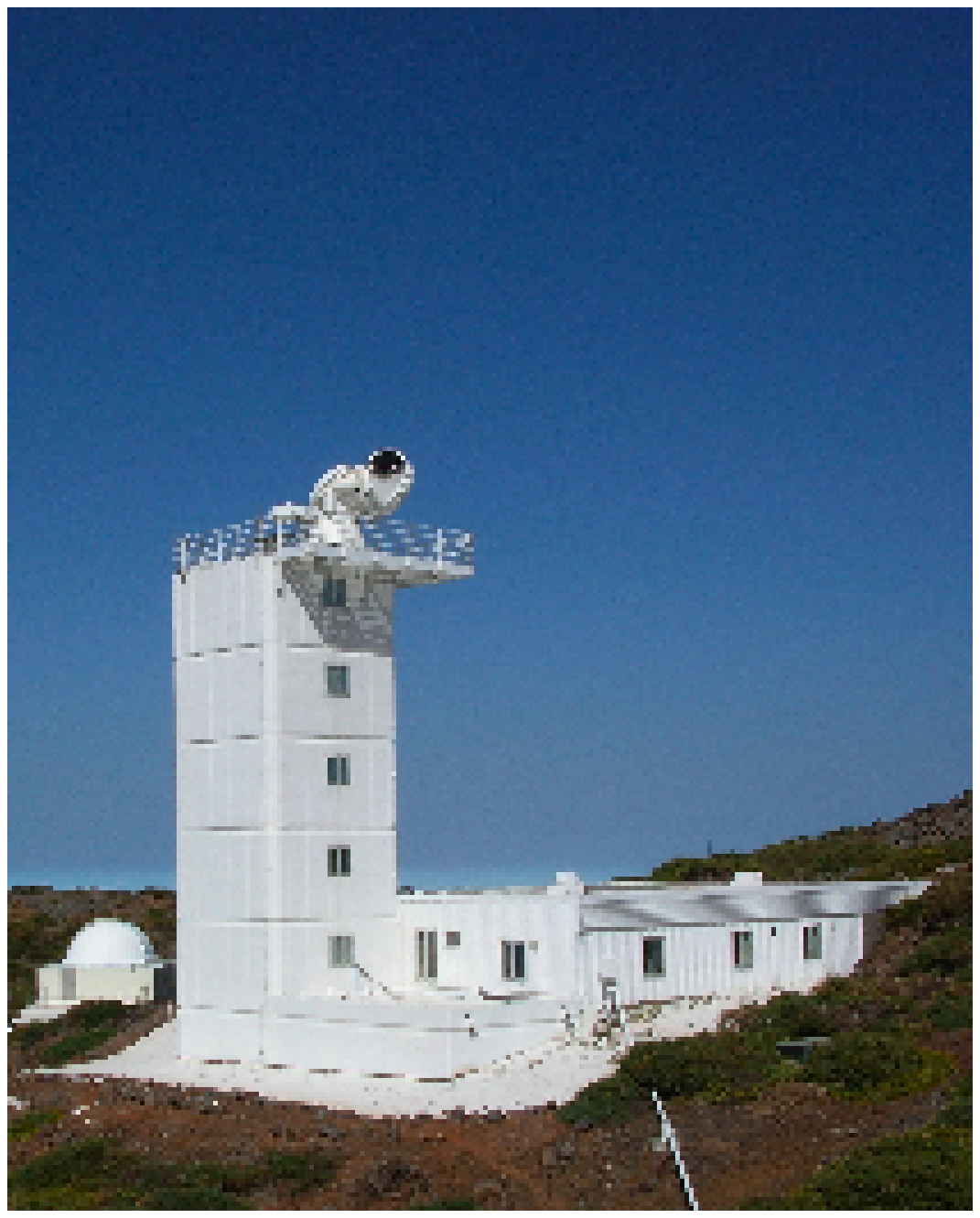}
\end{tabular}
\caption[\sf Sketch of the Swedish 1 m Solar Telescope (SST) at the Roque de los Muchachos Observatory]{\sf  \emph{Left figure}: Sketch showing the design of the Swedish 1-m Solar Telescope (SST) installed at the Roque de los Muchachos Observatory (more information available at the website \emph{http://www.solarphysics.kva.se/}). The diagram plots the primary focus (\emph{A}) where the rays coming from the lens converge, the Schupmann system (\emph{B}) to correct for chromatic aberration, and the deformable mirror of the adaptive optics (\emph{C}). \emph{Courtesy Institute of Solar Physics, Stockholm}. \emph{Right figure}: Photography of the SST taken by the author during the 2005 campaign.}
\label{telescope}
\end{center}
\end{figure}
\vspace{3mm}

\begin{table}
\sffamily
\centering
\begin{tabular}{|ll|}\hline
~~~~~~~~~~~~~~~~SST & CHARACTERISTICS \\\hline\hline
Type & Evacuated refractor\\
Mount & Turret design\\
Aperture & 97 cm\\
Primary focus length  ($f_1$) & 20.35 m\\
Secondary focus length ($f_2$) & 45.8 m\\
Wavelength ($\lambda$) range & 0.35 $\mu$m - 0.8 $\mu$m\\
Image scale in $f_2$ & 4.5"/mm \\\hline
Post-focus instruments & Correlation Tracker (CT)\\
   & Adaptive Optics (AO)\\
   & High resolution CCDs\\
   & SOUP Tunable Filter\\
   & Spectrograph \\\hline
Capabilities & Low-order AO \\
             &(up to 35-modes\\
	    & Karhunen-Lo\'eve)\\
            &  Real-time frame selection\\\hline	    
\end{tabular}
\caption[\sf Main parameters and characteristics of the SST]{\sf Main parameters and characteristics of the SST.}
\label{SSTtable}
\end{table}

\subsection{Instrumentation and setup}
\label{S:instru}

The work and results presented in all this chapter are based on the observations done at the SST (during the ITP collaboration, see Figure~\ref{telescope}), using the adaptive optics (AO) system. The AO system has a bi-morph mirror equipped with 37 electrodes and a Shack-Hartmann wavefront sensor with 37 microlenses. A thin-film filter (dichroic filter) separates the light in two beams,the blue and the red beams. The blue beam for imaging and the red one to feed a longitudinal magnetograph 
(the SOUP Tunable Filter that will be described in detail later on). Table~\ref{SSTtable} enumerates the main characteristics of the telescope.\\

In order to apply the restoration techniques described in chapter~\S1 we recorded two \emph{objects}: images in G-band\footnote{\sf Spectral band of molecular lines at $\lambda$430.5 nm.} and G-cont\footnote{\sf Relatively clear continuum, close to the G-band, at $\lambda$436.3 nm.}. The setup included then, for the blue beam, two channels taking simultaneous images. In the first channel, G-cont ($\lambda$436.3 nm, FWHM=1.15 nm), two Kodak Megaplus 1.6 CCD cameras with a 10-bits dynamical range and 1536 $\times$ 1024 pixels were employed, as corresponding to the PD focus-defocus image-pair. The defocus image was generated by moving backwards the camera a certain distance to generate a larger optical path in comparison to the focus one, hence inducing a known amount of defocus. The second channel was devoted to G-band ($\lambda$430.5 nm, FWHM=1.08 nm). The pixel size of the cameras was 0$\farcs$041 square.\\

Apart from the two above described, a third independent channel (asynchronous respect to G-band and G-cont) was acquiring reference images in the Ca~II~H line ($\lambda$396.9 nm, FWHM=0.3 nm). Table \ref{data} groups the main characteristics for the blue channel.\\

\begin{table}
\sffamily
\centering
\begin{tabular}{|l|ccc|}\hline
FILTER & G-band & G-cont & Ca II H \\\hline\hline
$\lambda$ [\AA] & 4305.6 & 4363 & 3969  \\
$\Delta\lambda_{\mbox{\scriptsize{fwhm}}}$ [\AA] & 10.8 & 11.5 & 3  \\
Exposure time [ms] & $\sim$13 & $\sim$13 & $\sim$80 \\
CCD & 1536 $\times$ 1024 & 1536 $\times$ 1024 & 2048 $\times$ 2048\\
Field of view ["] & 63 $\times$ 42 & 63 $\times$ 42 & 84 $\times$ 84 \\
Pixel size ["] & 0.041 & 0.041 & 0.041\\\hline 
\end{tabular}
\caption[\sf Parameters of the observation for the 2005 campaign]{\sf Parameters for the blue channel during the 2005 campaign.}
\label{data}
\end{table}

\begin{figure}
\begin{center}
\includegraphics[width=1.\linewidth]{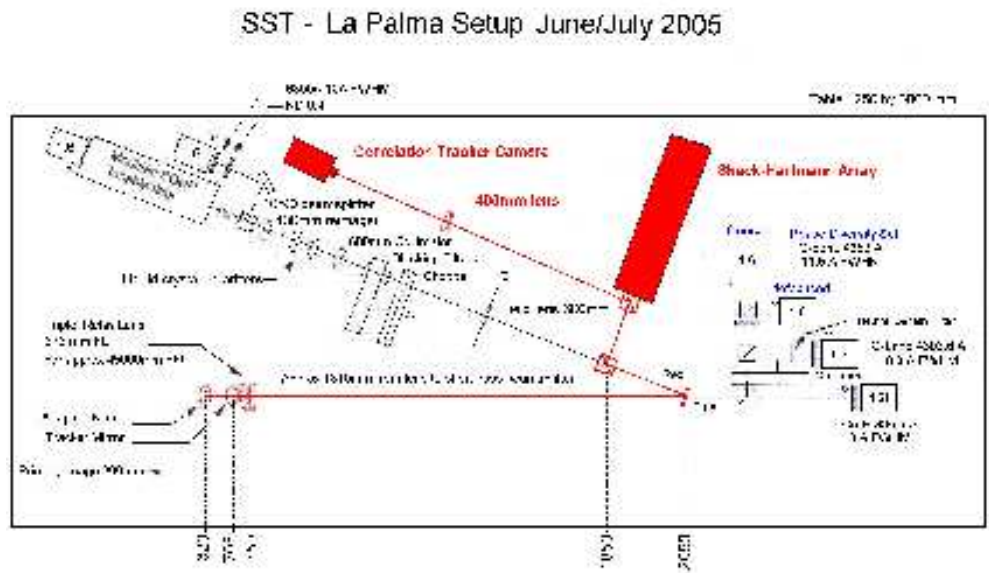}
\caption[\sf Setup mounted at the SST for the 2005 observing campaign]{\sf Setup mounted at the SST for the July 2005 observing campaign. The optical paths and instruments employed by the adaptive optics system are drawn in red. On the right-hand side the blue beam includes the PD image-pair (G-cont), the G-band channel (synchronized with G-cont by means of a \emph{chopper}) and the independent Ca H channel.}
\label{SST_set}
\end{center}
\end{figure}

Figure~\ref{SST_set} shows the setup with all the optical elements used in the configuration we mounted for the campaign. The optical paths and instruments employed by the adaptive optics system are drawn in \emph{red} in the figure. On the right-hand side, the blue beam includes the PD image-pair (G-cont), the G-band channel (synchronized with G-cont by means of a \emph{chopper}) and the independent Ca~II~H channel.

\vspace{3mm}

The Solar Optical Universal Polarimeter \citep[SOUP,][]{title1986} tunable filter, owned by the Lockheed Martin Solar and Astrophysics Laboratory (LMSAL), was placed along the red beam. This instrument is essentially  a temperature compensated Lyot filter. It operates by rotating a series of linear polarizers and 1/2-wave retarders placed between calcite crystal optics. The selected wavelength is adjusted by tuning the servo motors for each rotating element to a previously calibrated position. Two bandpass modes, wide ($>$100 m\AA) and narrow ($<$ 100 m\AA), are selectable for any given line tuning. The complete configuration of the SOUP also includes two Liquid Crystal Variable Retarders (LCVRs) for measuring the polarization state of the incoming light. 
The SOUP filter is used for observations in the H$\alpha$ spectral line that forms in the solar chromosphere and to obtain magnetograms using the magnetically sensitive line Fe I 6302~\AA.\\ 

A beam splitter in front of the SOUP divides the light beam to obtain simultaneous images in the continuum near the selected bandwidth. These images, understood as additional \emph{objects}, will help us to restore the narrow band images (by MOMFBD) and to obtain high-resolution magnetograms.\\

\begin{figure}
\begin{center}
\begin{tabular}{cc}
\includegraphics[width=0.47\linewidth]{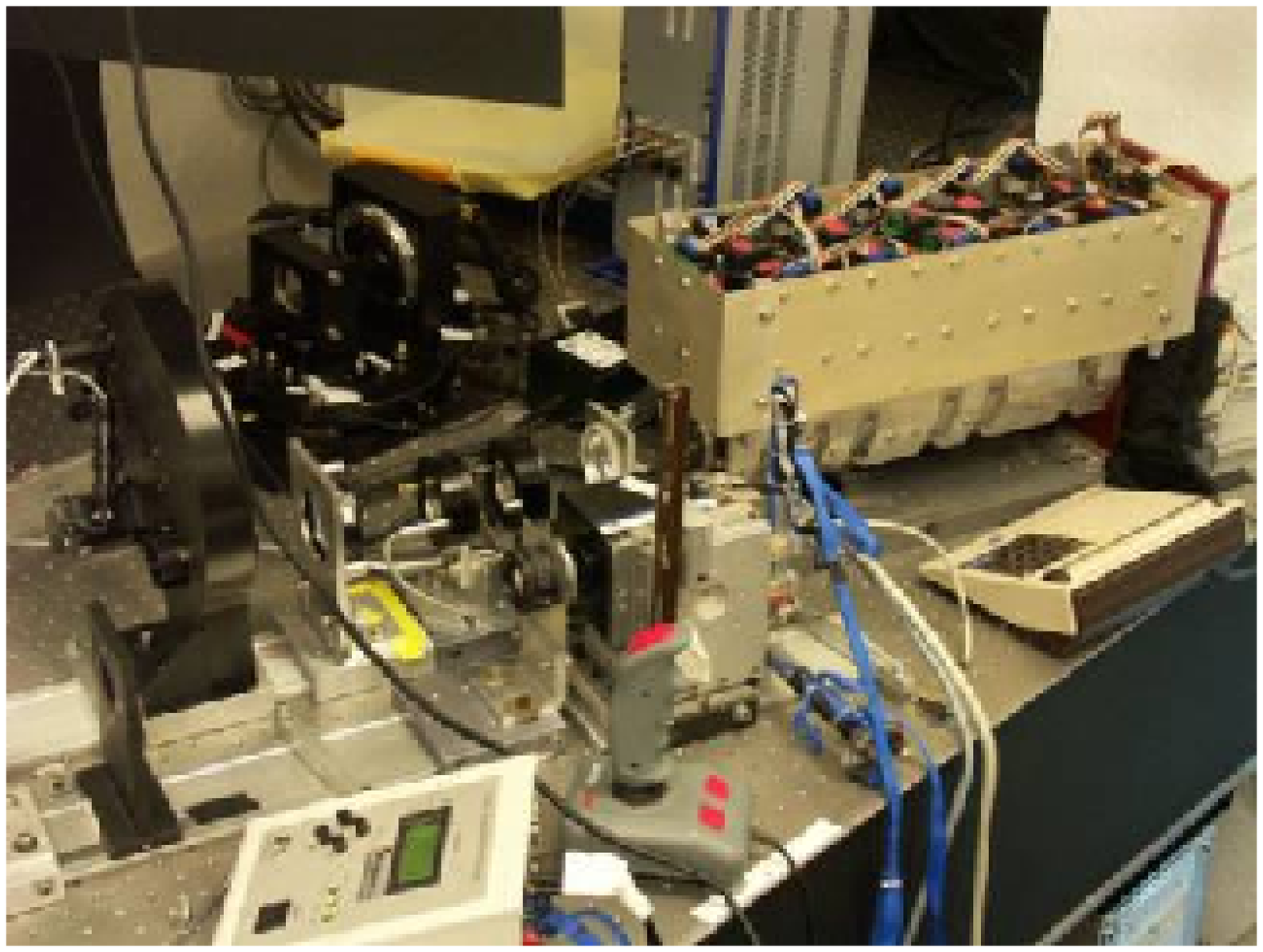} &
\includegraphics[width=0.5\linewidth]{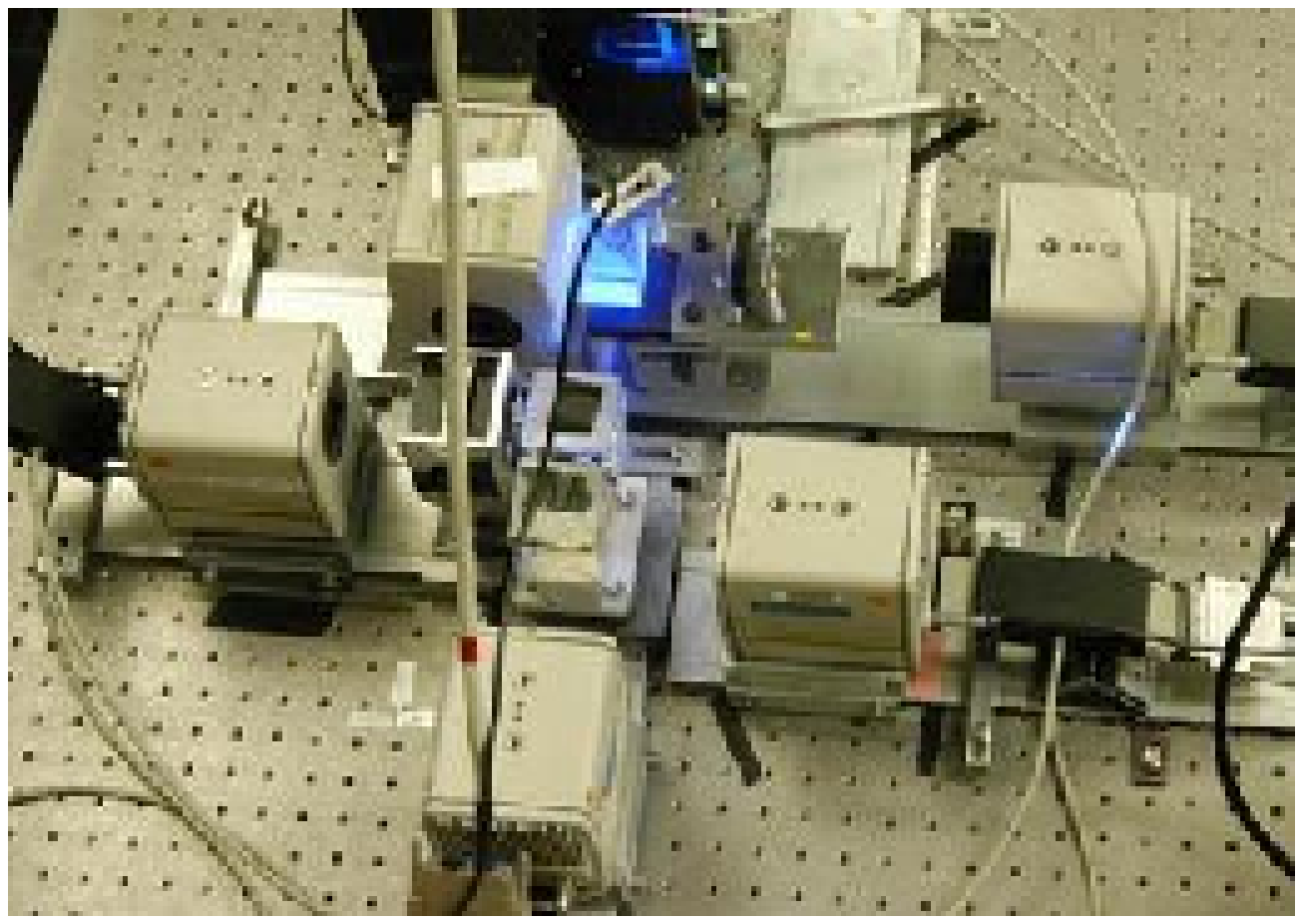}
\end{tabular}
\caption[\sf The Solar Optical Universal Polarimeter tunable filter and the science cameras  at the SST]{\sf The Solar Optical Universal Polarimeter (SOUP) tunable filter (\emph{left}) and the science high-resolution cameras (\emph{right}) on the SST optical bench.}
\label{SOUP}
\end{center}
\end{figure}

Figure~\ref{SOUP} shows the SOUP (\emph{left}) and the science high-resolution cameras of the blue beam (\emph{right}) as placed on the optical bench at the SST \footnote{\sf The complete description of the instrumentation and the manuals of operation of the SST can be found in the website \emph{http://www.solarphysics.kva.se/LaPalma/}}.

\vspace{4mm}

\begin{table}
\sffamily
\centering
\begin{tabular}{|cllc|}\hline
\bf{DATE} & \bf{TIME, UT} & \bf{ACTIVE REGION} & \bf{QUALITY}\\\hline \hline
03.07.05 & 8:38-11:47  & 10781 & \footnotesize{medium}\\
04.07.05 & 16:45-17:31 & 10781 \footnotesize{(\emph{follower})} & \footnotesize{poor}\\\hline
05.07.05 & 7:58-9:15   &  10781 \footnotesize{(\emph{follower})} & \footnotesize{medium}\\
05.07.05 & 14:16-14:44 &  10781 \footnotesize{(\emph{follower})} & \footnotesize{good}\\
05.07.05 & 16:23-17:33 &  10781 \footnotesize{(\emph{leader})} & \footnotesize{medium}\\\hline
06.07.05 & 7:50-8:57 &  10786 \footnotesize{(\emph{leader})} & \footnotesize{good} \\\hline
07.07.05 & 8:55-10:12  & 10786 \footnotesize{(\emph{leader})} & \footnotesize{medium} \\
07.07.05 & 10:36-12:58  & 10786 \footnotesize{(east of the \emph{leader})} & \footnotesize{good} \\
07.07.05 & 13:40-14:25  & 10786 \footnotesize{(\emph{leader})} & \footnotesize{medium} \\\hline
08.07.05 & 8:00-8:45 & 10786 & \footnotesize{medium} \\\hline
09.07.05 & 7:44-9:30  & 10786 & \footnotesize{excellent}\\\hline
10.07.05 & 8:18-9:07 & 10786 & \footnotesize{medium}\\
10.07.05 & 16:18-16:47 & 10786 \footnotesize{(north region)} & \footnotesize{good} \\\hline
11.07.05 & 7:44-10:01  & 10789 \footnotesize{(\emph{follower})} & \footnotesize{medium}  \\
12.07.05 & 8:30-9:53 & 10789 & \footnotesize{medium}  \\
13.07.05 & 8:36-9:42  & 10789 \footnotesize{(\emph{follower})} & \footnotesize{good} \\\hline\hline
\end{tabular}
\caption[\sf Log of observations for the 2005 ITP campaign]{\sf Log of observations for the 2005 International Time Program (ITP) campaign. The quality of the images is quantified by averaging the absolute value of the local gradients inside a window located in a granulation region within the FOV. The quality values have been translated into a more qualitative scale varying from poor quality images up to excellent quality ones. The quality was measured using the G-band images as reference.}
\label{campana}
\end{table}
 
\subsection{Data acquisition}
\label{S:data}
 
During the period 3-13 July 2005, images were acquired as listed in Table~\ref{campana}. The weather conditions at the observatory through all the campaign were quite variable obtaining rather medium quality images though for the observing run on July 9 we were able to record very good images (close to the diffraction limit after restoration) due to an excellent seeing.\\

\begin{figure}
\begin{center}
\includegraphics[width=1.\linewidth]{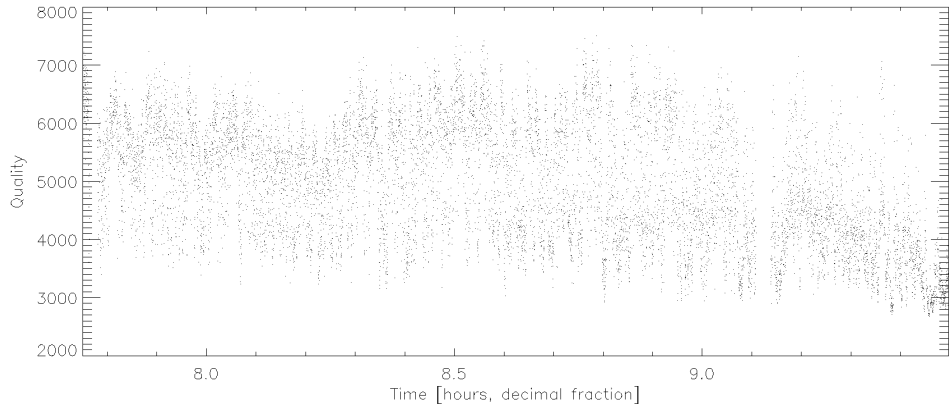}
\caption[\sf Quality of the G-band images]{\sf Plot of the quality (in arbitrary units) of the G-band images during the entire observation on 9 July 2005. The quality scale is such that 4000 indicates good quality images and there are astonished good quality images having values over 7000. At the end of the recorded images (at $\sim$ 9:30 UT)  the quality went down to values around 3000  which forced us to stop the observations due to the bad seeing conditions after that.}
\label{gbquality}
\end{center}
\end{figure}

Our targets of interest were sunspots in different evolutionary stages close to the solar disc center both, leaders and followers. Within all the images recorded, a post-facto pre-selection was made in order to get rid of the data below a certain quality threshold. The quality of the images is quantified by averaging the local gradients in absolute value inside a granulation window within the FOV. The quality was measured by using the G-band images as reference. Figure~\ref{gbquality} shows a plot of the quality for the G-band images through the entire observation on July 9, 2005.\\ 

\begin{figure}[h]
  \hfill
 \hspace{7mm} \begin{minipage}[t]{.53\textwidth}
      \epsfig{file=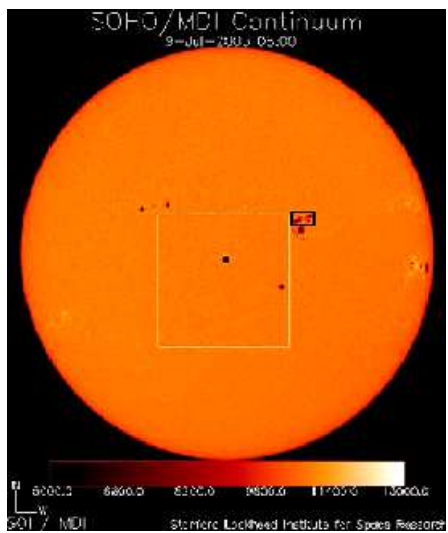, scale=0.42}
  \end{minipage}
  \hfill
  \begin{minipage}[b]{.38\textwidth} 
\caption[\sf Solar active region NOAA 10786 observed on July 9 2005 taken by the SOHO satellite]{\sf Solar active region NOAA 10786 observed on July 9 2005 (\emph{small black box}). Image taken by the SOHO satellite. The \emph{white box} indicates the SOHO high-resolution FOV and the disc center is marked with a black dot.}  
\label{solfig}
  \end{minipage}
  \hfill
\end{figure}

Among all the acquired data, we decided to work with the best quality images that corresponded to the NOAA Active Region 10786 observed on July 9, 2005. This complex region, displaying a $\delta$-configuration, was placed at heliocentric position $\mu = $0.9 as shown in the SOHO full-disk continuum image in Figure~\ref{solfig}. The sequences span for more than one hour, from 7:47 UT to 9:06 UT, following the evolution of the sunspots.

\section{Data processing}
\label{S:dataproc}

Real-time corrections with the AO of the telescope and further post-processing techniques rendered up images near the diffraction limit. The main steps performed in the processing of the images are sequentially detailed in the next sections. 

\subsection{Flatfield and darkfield corrections}
\label{S:flatdark}

Standard procedures for flatfielding and dark-current subtraction were firstly applied. During the observations, apart from the science images, we also acquired the corresponding \emph{flat} and \emph{dark} images\footnote{\sf To improve the SNR, every \emph{flat} and \emph{dark} image resulted from real-time averages of 100 single exposures. The \emph{flats} were taken while the telescope, out-of-focus, scanned the central area of the solar disc at fast speed.}. The corrections were made by applying the following expression

\begin{equation}
i_{corr}(i,j)=\frac{i_{raw}(i,j)-d(i,j)}{f(i,j)-d(i,j)},
\label{flatcor}
\end{equation}

\noindent where $i_{corr}$ is the final image after flatfield and dark current corrections, $i_{raw}$ is the observed raw image, $d$ and $f$ are the mean dark and flat images, respectively, and indexes $(i,j)$ stand for the pixel position in the image matrix. Special care has been taken during these corrections to eliminate dark and hot pixels, and also the black borders in the flat images, as a crucial step to ensure the images will not be affected by spurious artifacts when applying the restoration code.

\subsection{Restorations}
\label{S:restore}

\begin{figure}
\centering
\begin{tabular}{cc}
\textsc{G-band} & \textsc{G-cont } \\
\includegraphics[width=.48\linewidth]{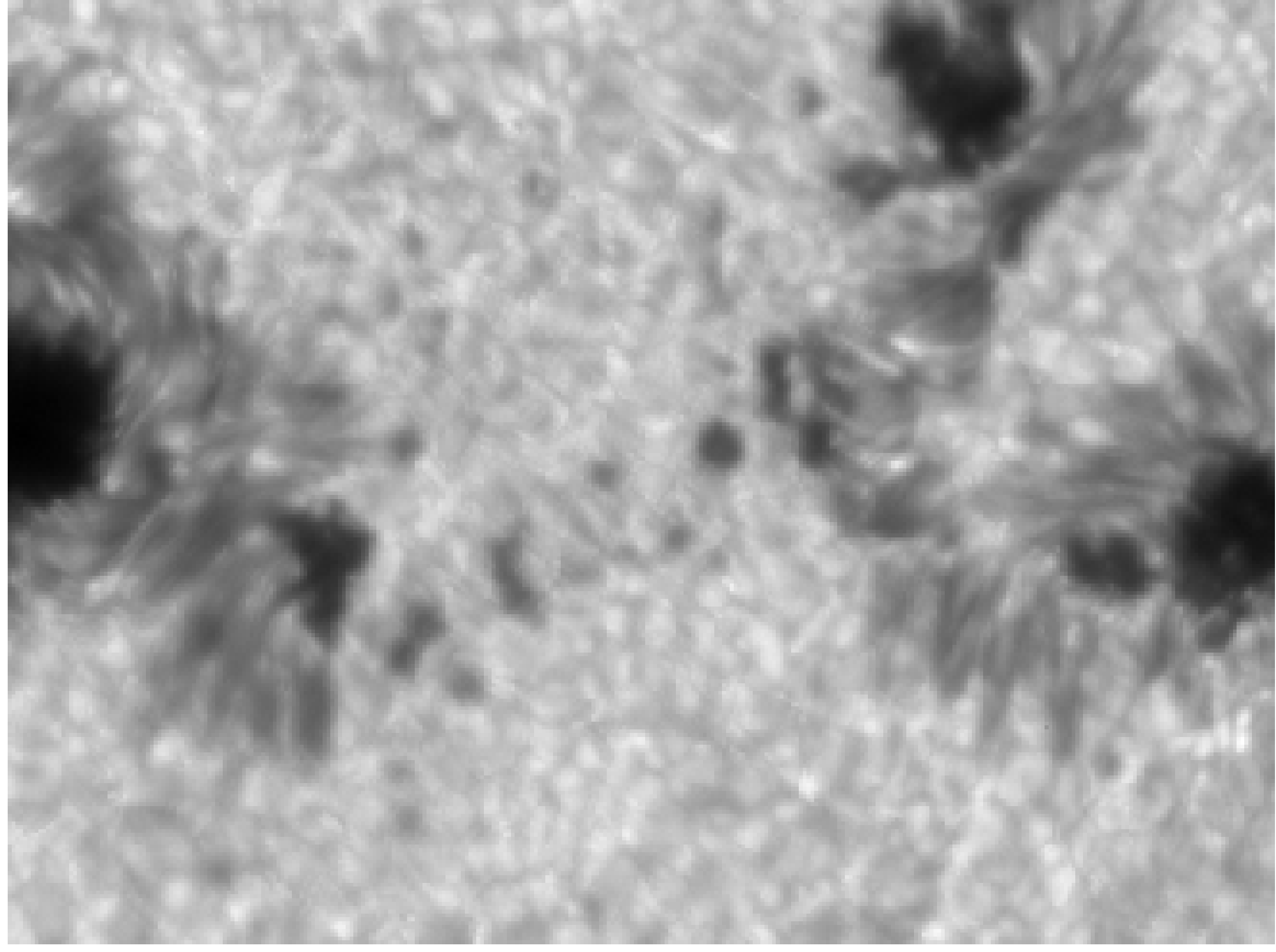} &
\includegraphics[width=.48\linewidth]{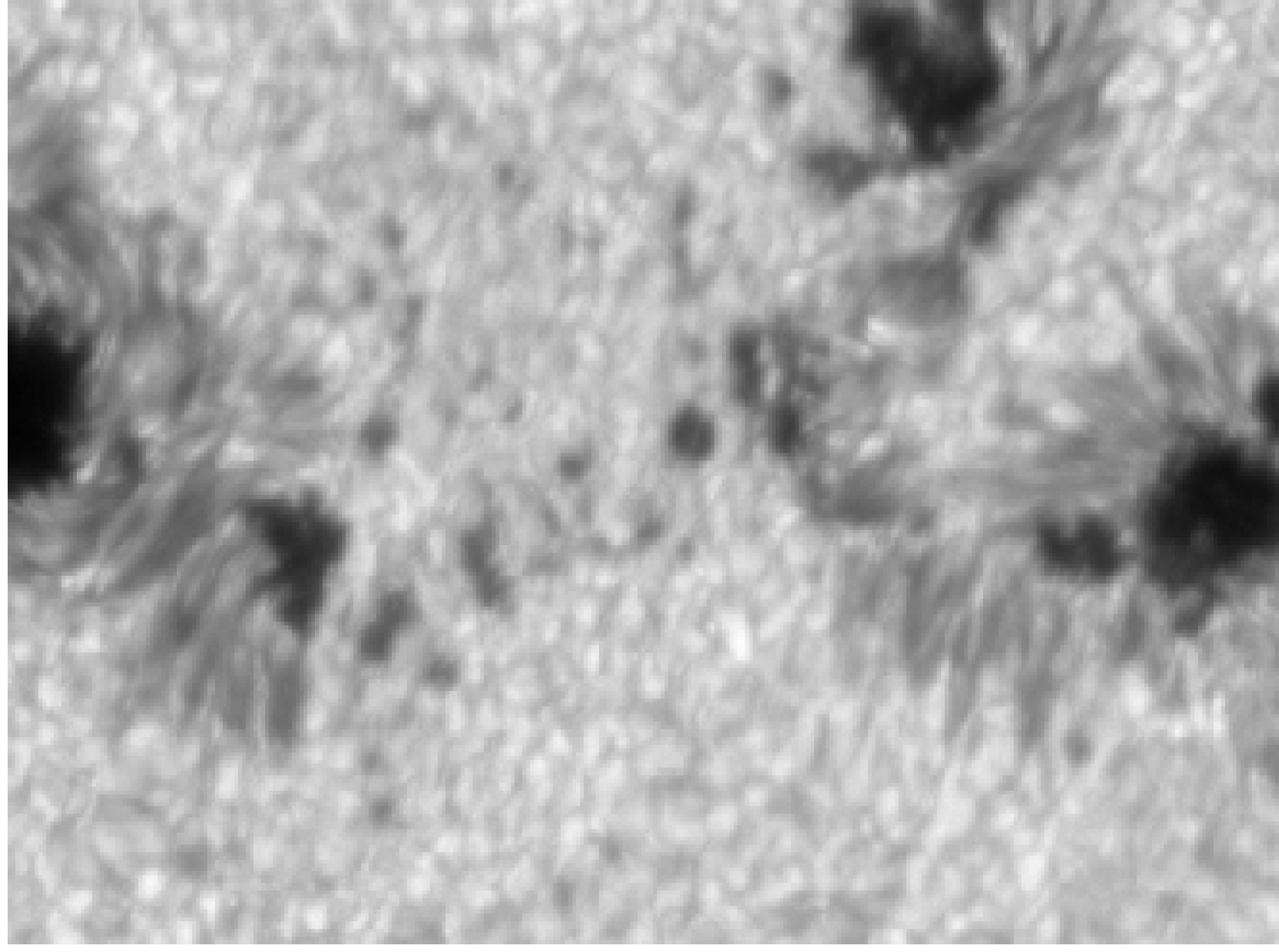} \\
\includegraphics[width=.48\linewidth]{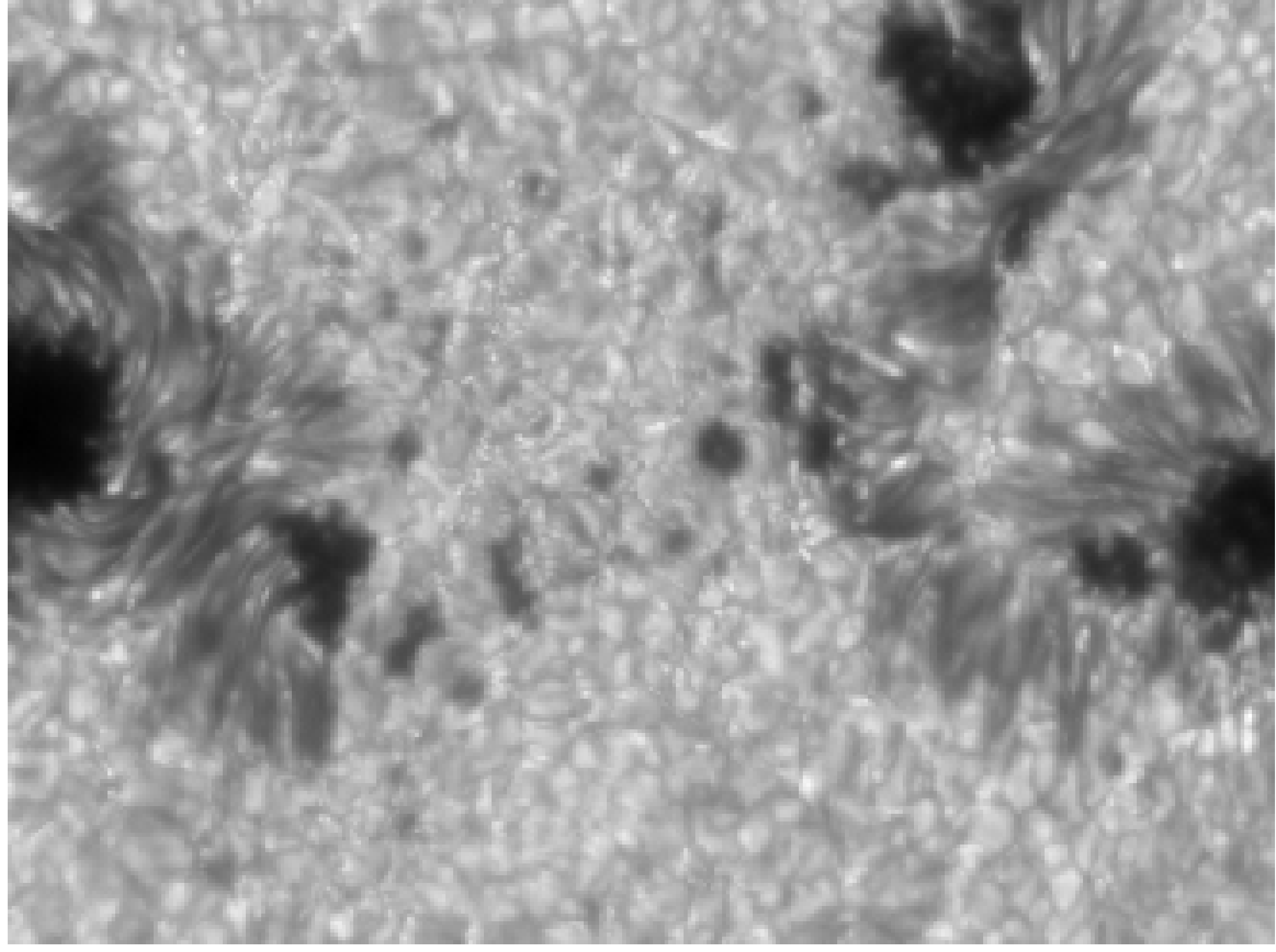} &
\includegraphics[width=.48\linewidth]{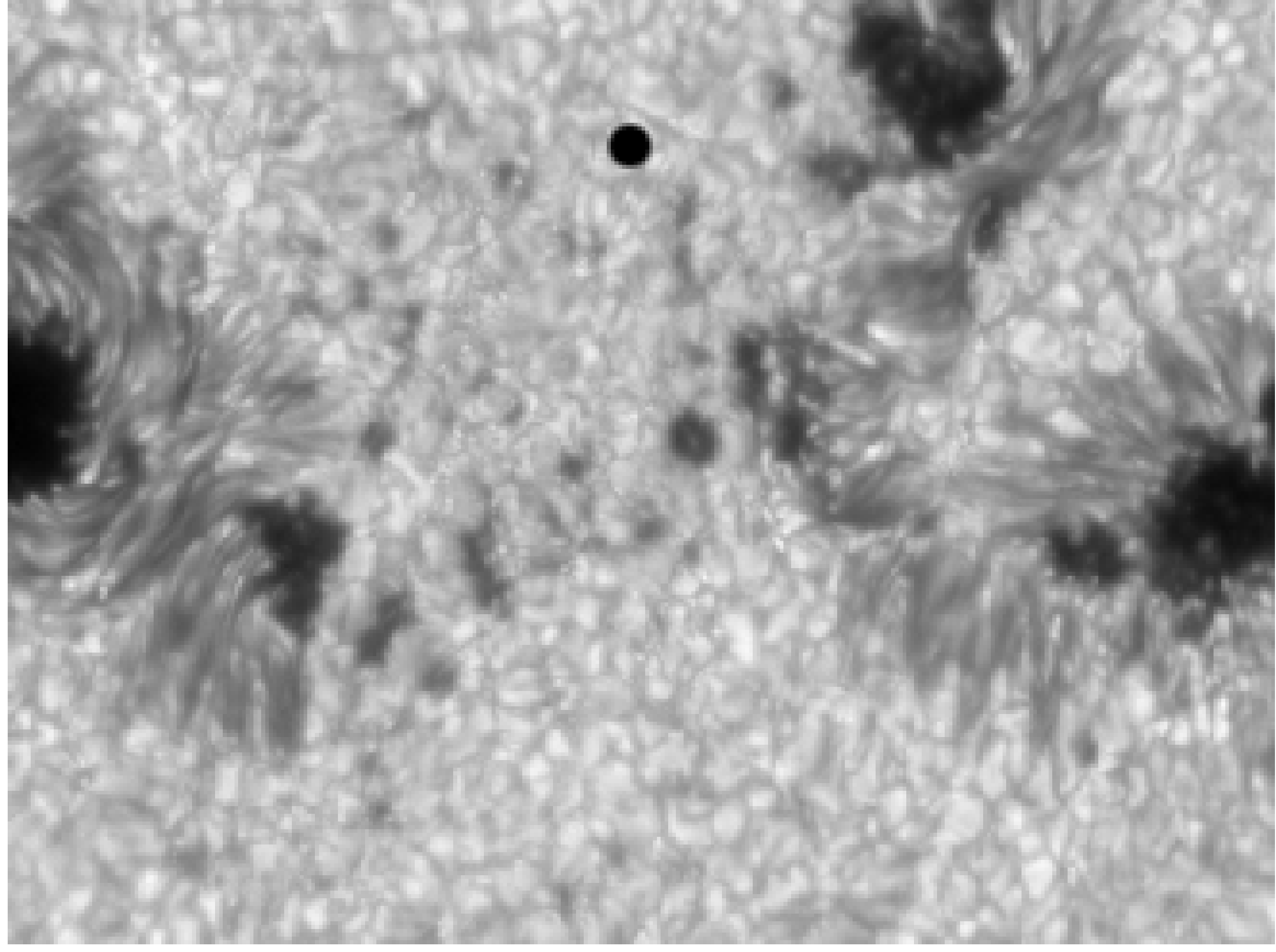} \\
\includegraphics[width=.48\linewidth]{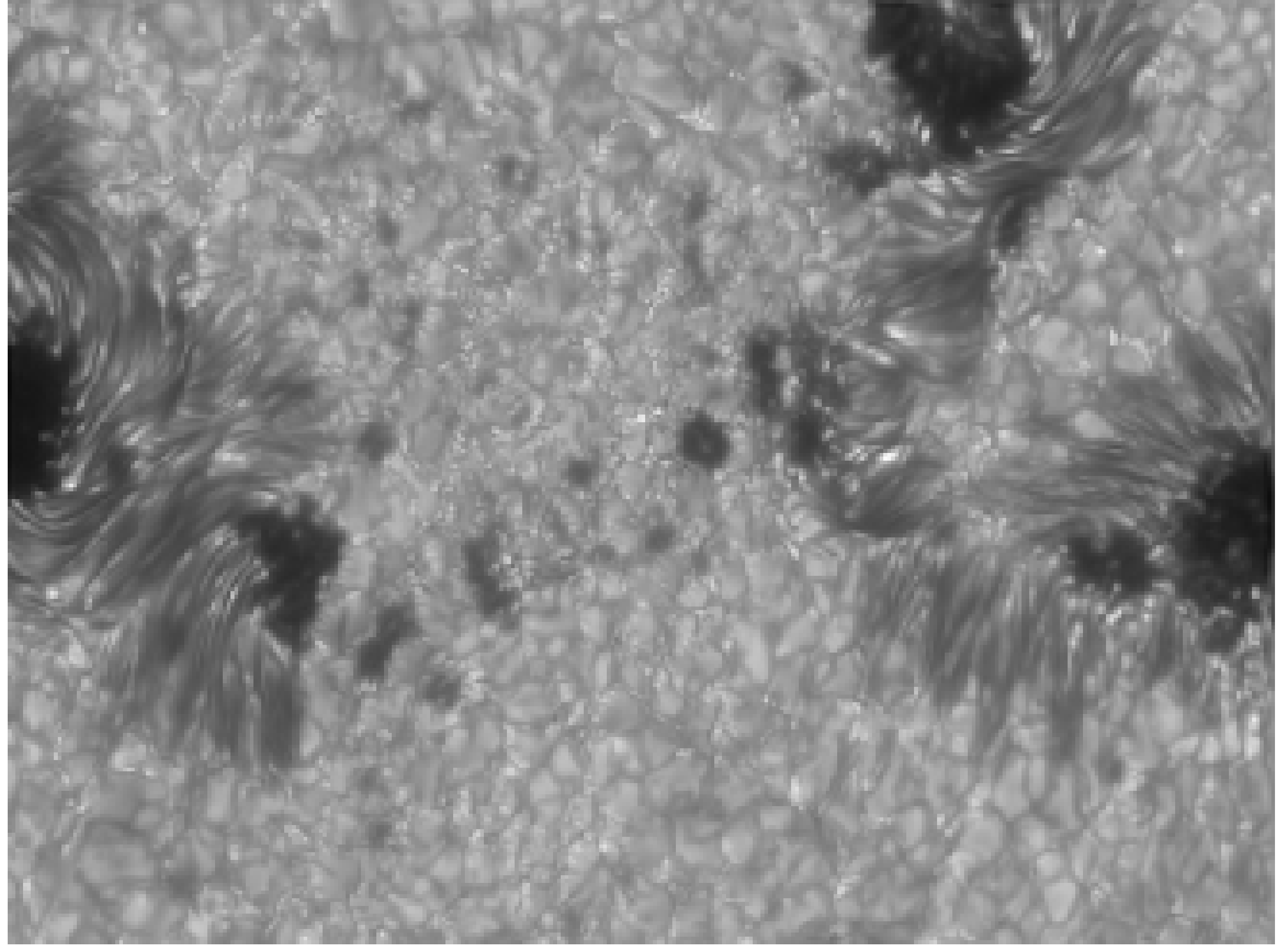}&
\includegraphics[width=.48\linewidth]{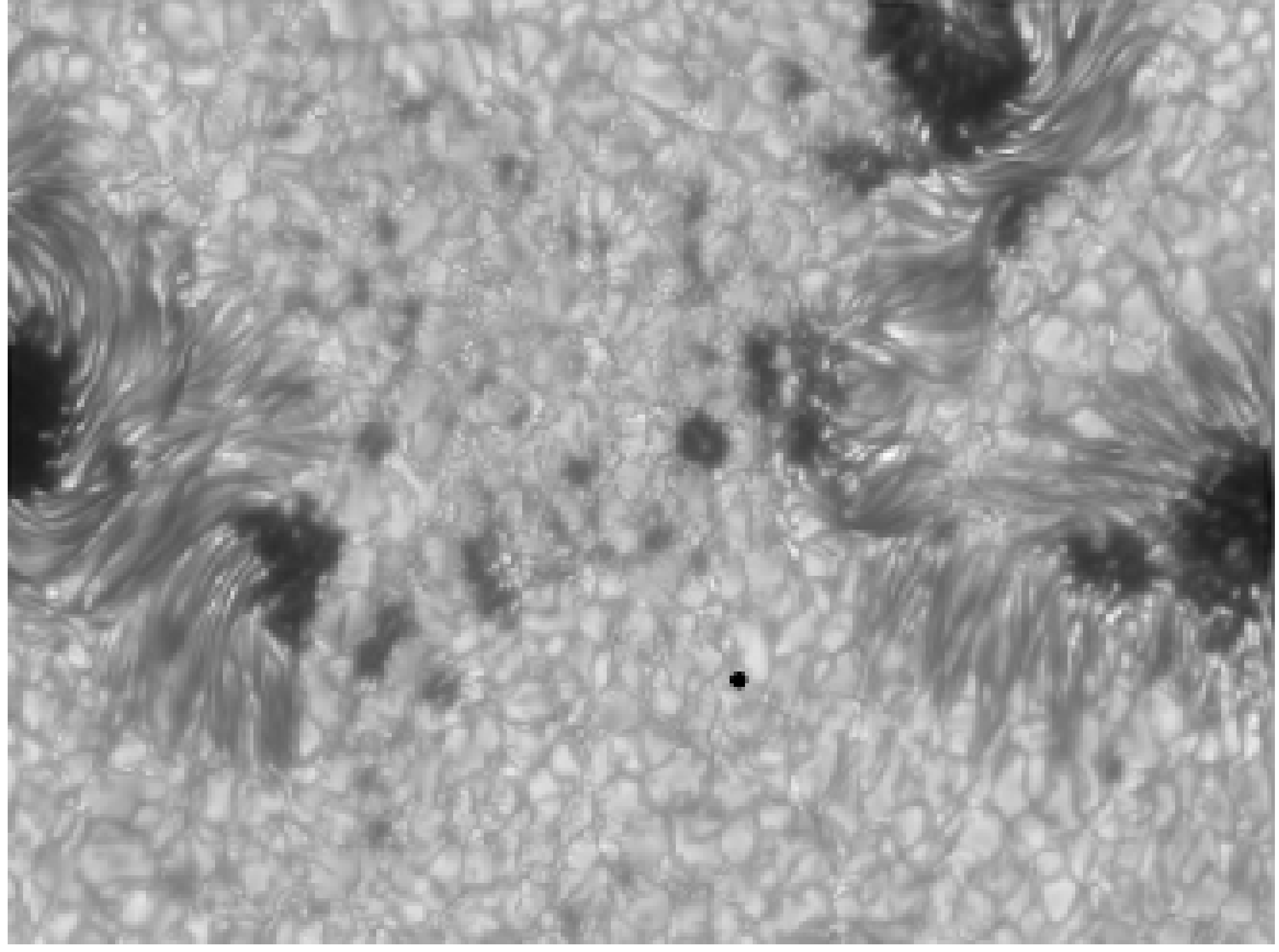}
\end{tabular}
\caption[\sf Example of the final images in G-band and G-cont after applying the restoration process]{\sf Example of the final images in G-band (\emph{left panels}) and G-cont (\emph{right panels}) after applying the restoration process. Displayed are the worst (\emph{upper panels}) and the best (\emph{middle panels}) quality images out of the 18 images in a particular set among the 472 sets. Using the MOMFBD method, two restored images are generated, one for the G-band (\emph{lower left panel}) and another one for G-cont (\emph{lower right panel}). The total FOV in the original images is 63$\times$42 "  whereas for the final ones reduces to \mbox{60$\times$40 ".}}
\label{restauraciones}
\end{figure}

Post-processing for image restoration was performed by employing the Multi-Object Multi-Frame Blind Deconvolution (MOMFBD) described in section~\S1.5. In the blue beam, the observing strategy to apply this technique consisted in taking G-band images and simultaneous G-cont phase diversity image-pairs. Sets of about 18 images per channel (i.e.\ 3 $\times$ 18 images) within a temporal cadence of 10.0517 s, were combined to produce a pair of simultaneous G-band and G-cont restored images. Table~\ref{restaura} shows the information about the images employed in the restoration process.\\

Following this procedure, two time series of reconstructed images were produced: 472 images for each channel, G-band and G-cont, that correspond to 79 min of continuous solar observation. Figure~\ref{restauraciones} displays an example of the restorations pursued, for one out of the 472 image groups processed. The total FOV in the original non-restored images was 63$\times$42 " whereas the final FOV in the final restored images was 60$\times$40 ".\\

\begin{table}
\sffamily
\centering
\begin{tabular}{|l|cc|}\hline
CHANNEL              & IMAGES& Time, UT  \\\hline\hline
G-band              & 221-8818 & 7:47:45 - 9:09:15\\
G-cont (focus)   & 115-8712 & 7:47:45 - 9:09:15\\
G-cont (defocus)& 228-8825 & 7:47:45 - 9:09:15\\\hline
\end{tabular}
\caption[\sf Images information in the restoration process]{\sf Images information in the restoration process. Tabulated in the second column are the number of the initial and final images for each time series in every channel.}
\label{restaura}
\end{table}

\subsection{De-stretching and subsonic filtering}
\label{S:destret}

Once the images were restored, they were, firstly, de-rotated to compensate for diurnal field rotation observed in the focal plane, induced by the alt-azimuthal configuration of the telescope and, secondly, aligned by cross correlation.\\

Additional image processing included the so-called de-stretching. The stretching or distortion effect on the images, commented in section~\S1.1, consist on differential local displacements over the FOV. They are due to differential deviations from the original tilts in the wavefronts coming from different point sources. Atmospheric turbulence is the responsible for this problem. The effect is specially strong during the morning time when the line-of-sight (hereafter LOS) to the Sun is more inclined hence inducing a greater optical thickness in the light path and consequently, a larger atmospheric turbulence contribution.\\

Thus, the stretching is a local effect manifested with different strength at different positions over the FOV and can be corrected at medium- and large-spatial scales by applying local correlation techniques. The local correlation tracking (LCT) technique was originally designed for the removal of distortions in image sequences \citep{november1986} and later on, was used to infer the proper motions of the granulation in a time series of white-light images \citep{november1988}.\\

Due to the strong distortion present in our time series, the de-stretching algorithm was applied repeatedly in four subsequent steps, to completely eliminate residual effects. In every step, the size of the correlation tracking window was modified from greater to smaller sizes to ensure corrections at different spatial scales.\\

\begin{figure}
\begin{center}
\includegraphics[width=.9\linewidth]{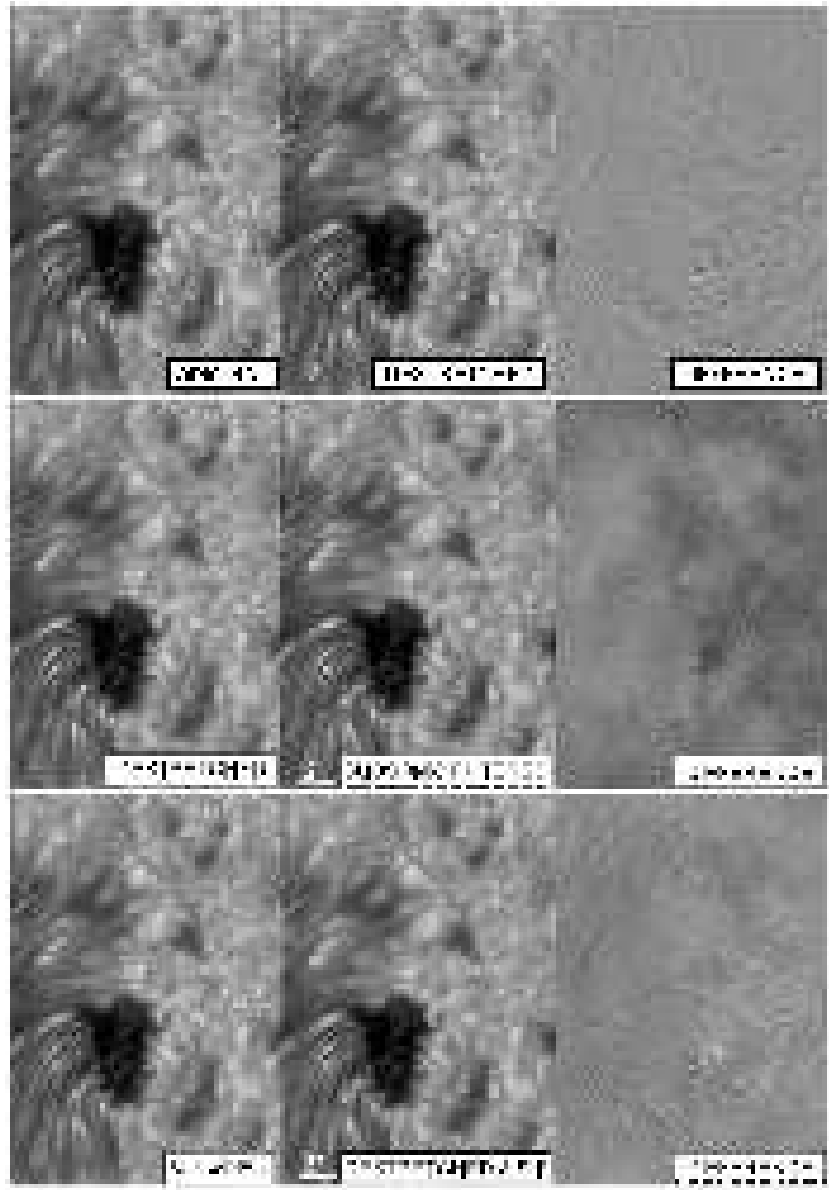}
\caption[\sf Steps in the correction of a time series of images]{\sf Steps in the correction of a time series of images. The correction of \emph{stretching} eliminates the distortion in the images (\emph{upper panel}) caused by atmospheric turbulence. The subsonic filtering eliminates the intensity fluctuations produced by solar high-frequency oscillations and limits the phase-velocity of the solar structures to a maximum of 4 km s$^{-1}$ (\emph{middle panel}). After the described corrections, the time series is ready for its physical analysis (\emph{lower panel}).}
\label{diferencias}
\end{center}
\end{figure}

Subsonic filtering is the last stage in the corrections applied to the images and deals with the cleaning process pursued to eliminate high frequency oscillations, such as the p-modes and residual jitter stemming from de-stretching, in the frequency  $k-w$ domain ~\citep{title1986}. A filter with a phase-velocity threshold of $4$ km s$^{-1}$ was employed with satisfactory results (see Figure~\ref{diferencias}
\emph{center}). The sharp cutoff in the subsonic filter was replaced by a smooth transition from 1 to 0.\\

\begin{figure}
\begin{center}
\includegraphics[width=.8\linewidth]{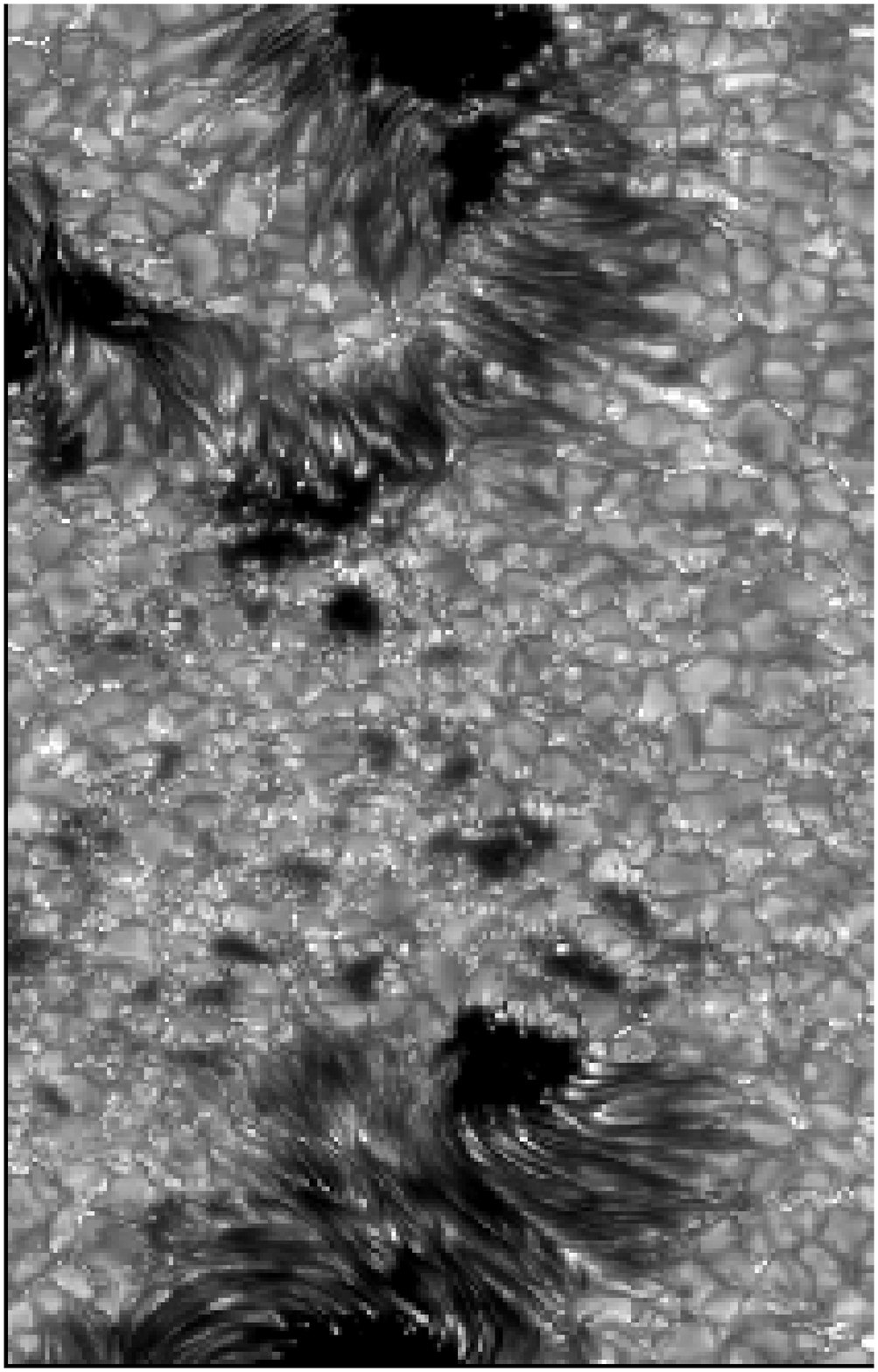}
\caption[\sf G-band image taken at the SST after the restoration process]{\sf G-band best quality image, taken at the SST on 9 July 2005 during the ITP campaign, after the data processing including the MOMFBD restoration.}
\label{gbandbest}
\end{center}
\end{figure}

An apodization in the temporal dimension (8$\%$ of the total length at each end) was also applied to minimize border effects in the filtering process. The final product were two movies (G-band and G-cont)\footnote{\sf See the movies at the web
site: \emph{http://www.iac.es/galeria/svargas/movies.html}} of 428
frames each, spanning over 71 min with a cadence of 10.0517 s, and covering a FOV of 57$\farcs$8 $\times$ 34$\farcs$4. The time series reached the high-resolution level shown in Figure~\ref{diferencias} (\emph{lower central panel}). This quality is pretty much suitable for subsequent scientific analysis. Figure~\ref{gbandbest} shows the superb quality achieved in the best image of the time series.

\vspace{8mm}

\vspace{8mm}

\subsection{Maps of displacements}
\label{S:propermot}

The LCT method has been commonly used for determining velocity fields in time-resolved image data and was first applied by \cite{november1988} to measure proper-motions in solar granulation. Since then, different authors have used LCT techniques at many different spatial scales to study the solar dynamics, such as long-term evolution and flows in supergranulation \citep{shine2000,derosa2000}.\\

The method is based on a simple but powerful idea that works on the principle of finding local concordances between two frames (images). A correlation window selects corresponding sub-areas in both images (the patches to be correlated). The selected sub-areas include those structures we want to track, such as solar granules that evolve in time. The displacement vector is given by the difference of the coordinates of the sub-area centers when the best match between them is found. Mathematically this idea is expressed in terms of a correlation function between two instants.\\

The selection of the sub-areas is performed by means of a window which is usually set to be a Gaussian function with a \emph{Full-Width at Half Maximum} (FWHM) adapted to the desired size for the correlation patch and depending on the spatial-scale of the structures we are about to track.\\

It is well documented in the literature that LCT-techniques in general produce some systematic errors in the determination of displacements. The nature of the interpolation algorithm to fix the position of the correlation maximum with sub-pixel accuracy may play a role in the results. The LCT-method typically underestimate the velocity fields. Tests performed by the authors of the LCT-code we are employing in this work lead to conclude that in the worst cases this underestimation can amount to \mbox{20-30~\%} \citep{yi1992, molownyphd1994}. \\

\begin{figure}
\centering
\begin{tabular}{l}
\includegraphics[width=.7\linewidth]{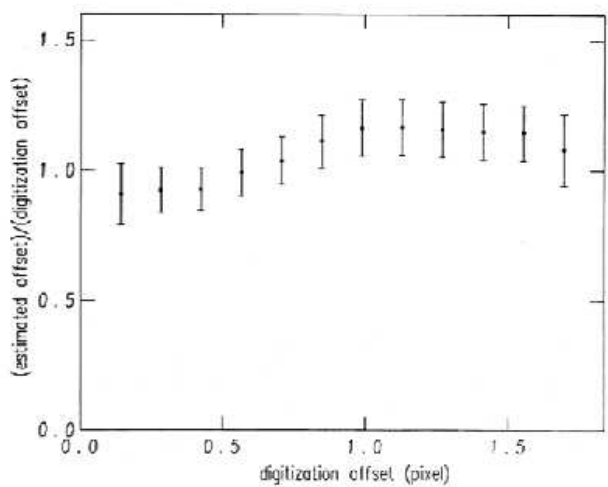}
\end{tabular}
\caption[\sf Precision of the measurements of displacement by LCT using a quadratic interpolation formulae.]{\sf Precision of the measurements of displacement by LCT as shown by \cite{november1988}. A quadratic interpolation formulae has been used to determine the correlation maximum. See the text for more information.}
\label{systerrors}
\end{figure}

Figure~\ref{systerrors}, taken from \cite{november1988}, displays a plot of the precision of the displacement measurements by means of LCT and using a quadratic interpolation formulae to determine the correlation maximum. These authors have used solar granulation images recorded on photographic film support, digitized at 12 positions with increments of 0.1 pixel. Then, they have computed displacement maps by comparison with the unshifted image applying their LCT algorithm. 
Figure~\ref{systerrors} shows the ratio between the computed and the real displacements. The former being the average of displacements over the area of each displacement map with error bars representing $\pm1\sigma$.\\

\subsection{Magnetograms}
\label{S:magnetogram}

In the red beam, the SOUP tunable birefringent filter  working at Fe I 6302\AA~in combination with a modulator composed by liquid crystal retarders, produced longitudinal magnetograms as mentioned above. The SOUP filter recorded images in the two flanks of the line at $\pm$50 m\AA~ from its core. The modulator switched between two states to obtain 10 left and 10 right circular polarization states (LCP and RCP) at each line position. The total time required to record all four-type images was 33 s. A beamsplitter in front of the SOUP filter deflected 10\% of the light to obtain broad-band phase diversity image-pairs, that combined with the narrow-band images of the SOUP allowed MOMFBD reconstructions. The MOMFBD restoration was therefore performed from subsets of 40 SOUP images including all polarization states and line positions, and 40 simultaneously recorded wide-band PD image-pairs, that is, a total of 120 images to produce 5 restored ones: $I_{RCP}^{blue}$, $I_{LCP}^{blue}$, $I_{RCP}^{red}$, $I_{LCP}^{red}$ and $I_{cont}$. Magnetograms (M) and also Dopplergrams (D) can be generated from the saved restored images by means of the following expressions

\begin{equation}
M= \frac{1}{2} \left [ \frac{I_{RCP}^{blue} -  I_{LCP}^{blue}}{I_{RCP}^{blue} + I_{LCP}^{blue}} -
\frac{I_{RCP}^{red} -  I_{LCP}^{red}}{  I_{RCP}^{red} + I_{LCP}^{red}} \right ],
\label{mag}
\end{equation}

\begin{equation}
D= \frac{1}{2} \left [ \frac{I_{RCP}^{blue} -  I_{RCP}^{red}}{  I_{RCP}^{blue} + I_{RCP}^{red}} +
\frac{I_{LCP}^{blue} -  I_{LCP}^{red}}{  I_{LCP}^{blue} + I_{LCP}^{red}} \right ].
\label{dop}
\end{equation}

\vspace{3mm}

\begin{figure}
\hspace{-1cm}\includegraphics[width=1.2\linewidth]{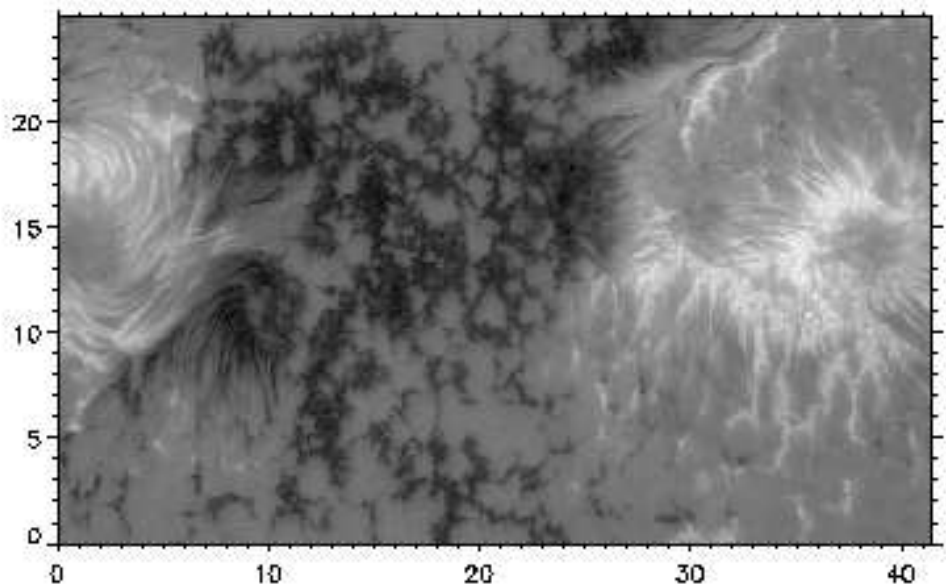}
\caption[\sf Magnetogram of the region under study obtained with the SOUP filter]{\sf Magnetogram corresponding to the region under study obtained with the
SOUP filter in the SST on 9 July 2005. The $\delta$-spot neutral lines in Figure~\ref{F:2} were obtained from this
frame. The coordinates are expressed in Mm.}
\label{F:4}
\end{figure}

The Doppler velocity $v$ is given by: $v = f_{c}D$, where $f_c$ is a constant to convert Doppler signal to Doppler velocities and is estimated to be $\sim 10$ km s$^{-1}$ by comparing the Doppler signal with the mean LOS Doppler velocity of granules ($\sim 1$ km s$^{-1}$). The FOV of the restored SOUP images is $90\arcsec \times 60\arcsec$ (with a pixel size of 0$\farcs$061 pixel$^{-1}$). Figure~\ref{F:4} displays one of the resulting magnetograms having a spatial resolution of about 0$\farcs$2.

\section{Results}
\subsection{General description of proper motions}

The G-band series was employed to study the proper motions of the structures
in the FOV by using a LCT algorithm implemented by \cite{molowny1994}. 
For our analysis, we have preferably chosen a Gaussian tracking window with FWHM 0$\farcs$78 being half of the typical granular size, which is suitable for tracking mainly granules.\\

With this code we obtained two maps of horizontal displacements or proper motions per time step (i.e.\ the horizontal velocity components $v_x$ and $v_y$), which were averaged in time. Averages of horizontal velocity components were performed over 5-min and 71-min intervals.\\

\begin{figure}
\hspace{-1cm}\includegraphics[width=1.2\linewidth]{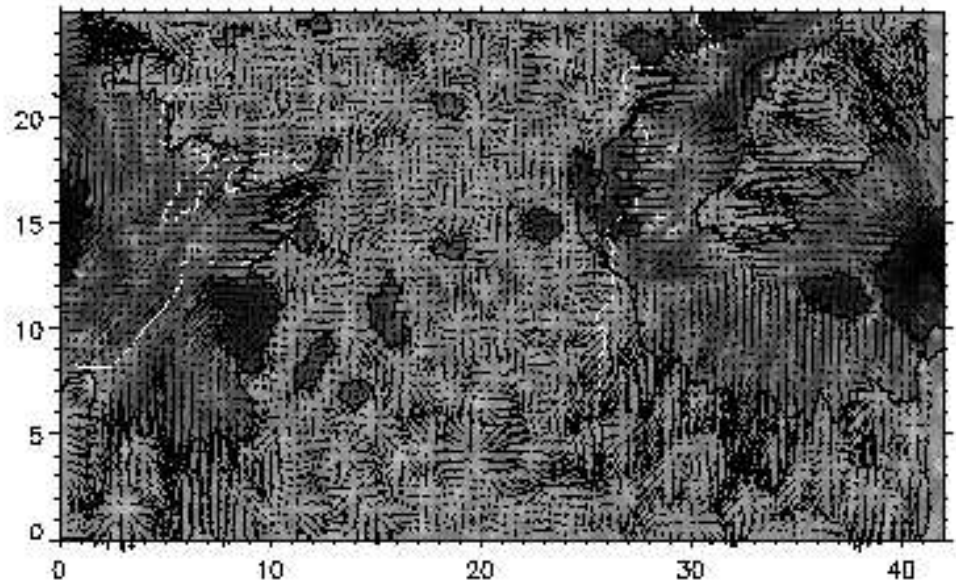} 
\caption[\sf Map of horizontal velocities]{\sf Map of horizontal velocities in the entire FOV (71 min average). The \emph{black contours} outline the borders of umbrae, penumbrae and pores. The \emph{white lines} delineate the LOS neutral lines inferred from a magnetogram. The
coordinates are expressed in Mm. The length of the black bar at coordinates
(0,0) corresponds to 4 km s$^{-1}$. The background represents the average
image of the G-band series.}
\label{F:2}
\end{figure}

Figure~\ref{F:2} shows the resulting flow map averaged over the whole series (71 min).
Moat flows are seen in the spot on the lower left of the FOV but more prominently
in the spot on the right side in the upper and lower penumbral regions (moats flows will be treated in more detail in section~\ref{S:largeflows}). Several centers of diverging horizontal motions are present in the entire FOV.
Some of them surround sunspots (lower left and right sides of the figure) but the most
conspicuous ones are seen in the lesser magnetized area
(lower part of the FOV between the spots),
displaying greater velocities and more symmetrical shapes. 
These velocity structures, related to recurrent expansion and
splitting of granules, are commonly associated with mesogranulation \citep{roudier2003,roudier2004,bonet2005}.\\ 

The \emph{white lines} in Figure~\ref{F:2} delineate the LOS neutral lines inferred from a SOUP magnetogram (see Figure~\ref{F:4}). The regions crossed by these lines
correspond naturally to horizontal fields, similar to those found in penumbrae.
However, we caution that these horizontal fields can have different flows as
those that occur in normal penumbra (such as shear flows, as found by \cite{deng2006} or the supersonic nozzles observed by \cite{pillet1994}.
Close inspection also shows that areas near neutral lines have a filamentary
appearance that is much less distinct than observed in normal penumbrae
extending radially from umbral cores.\\

We measured a mean speed in the granulation outside the moats of $\sim$0.35 km s$^{-1}$. This small value reflects the presence of high magnetic signal (dense population of pores and G-band bright points all over the field), however, it is even smaller than the reported, also for magnetized granulation, by other authors that employ tracking windows of similar widths \citep{berger1998, bonet2005}.\\

\begin{figure}
\includegraphics[height=.48\linewidth]{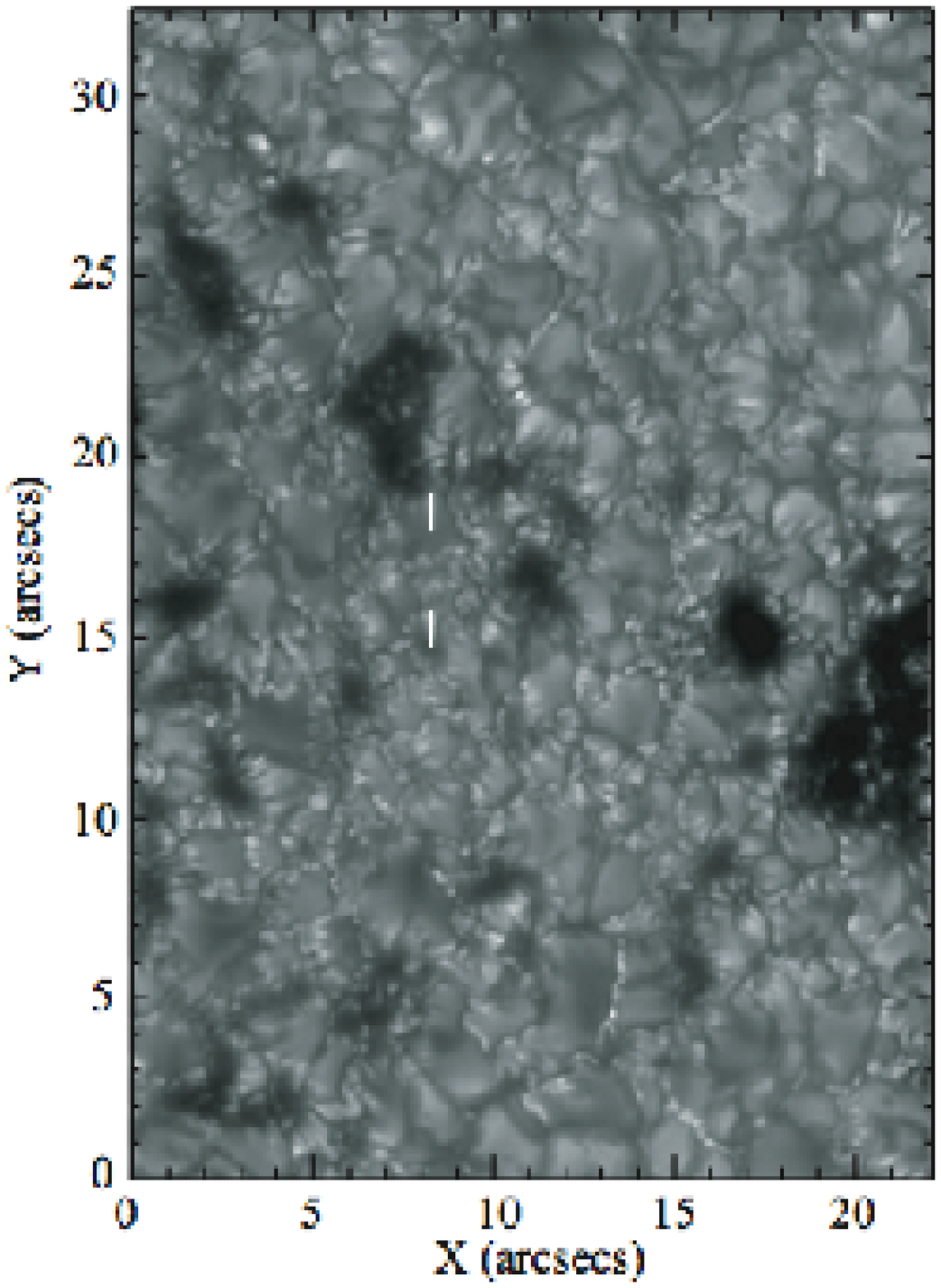}\includegraphics[height=.48\linewidth]{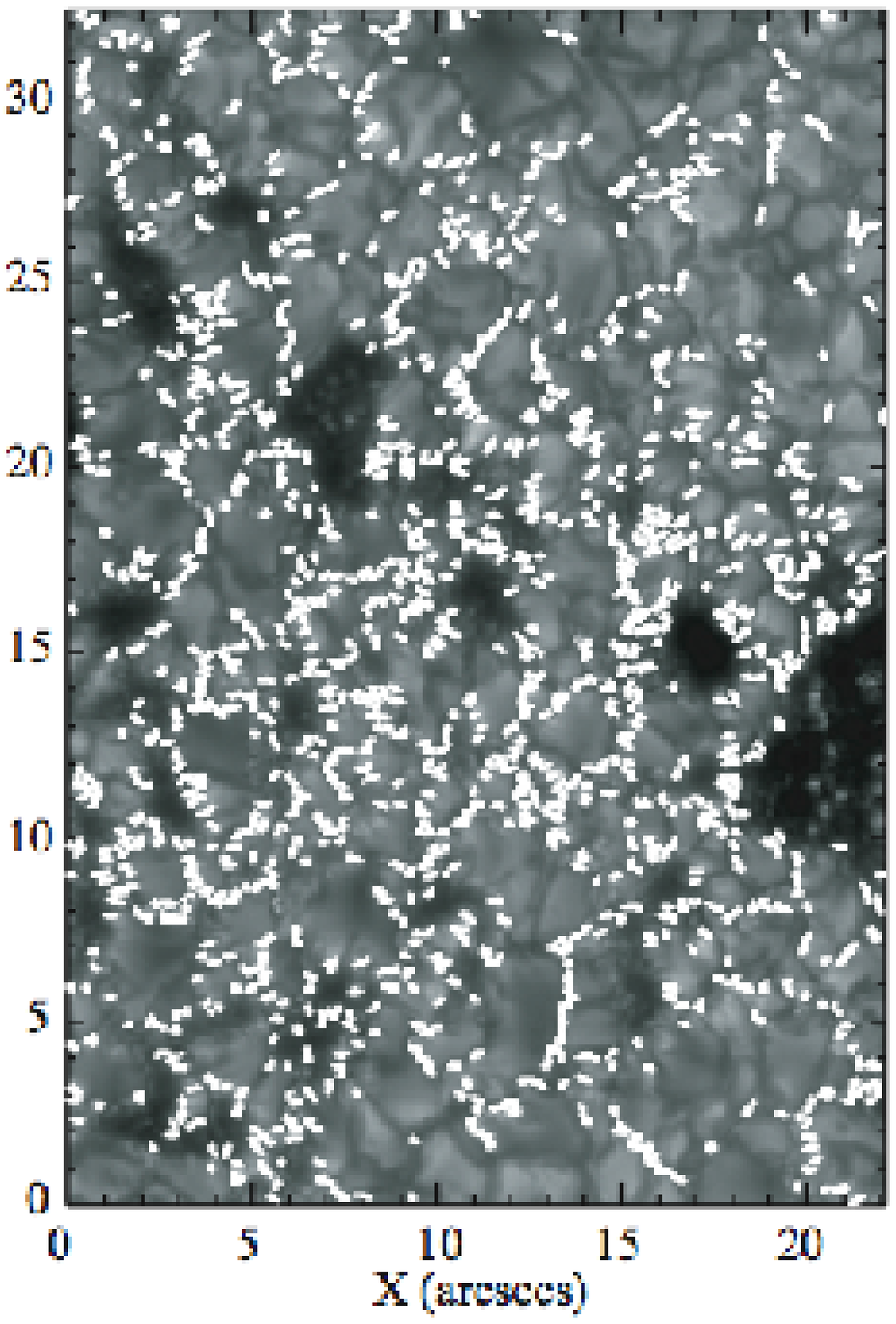}\includegraphics[height=.48\linewidth]{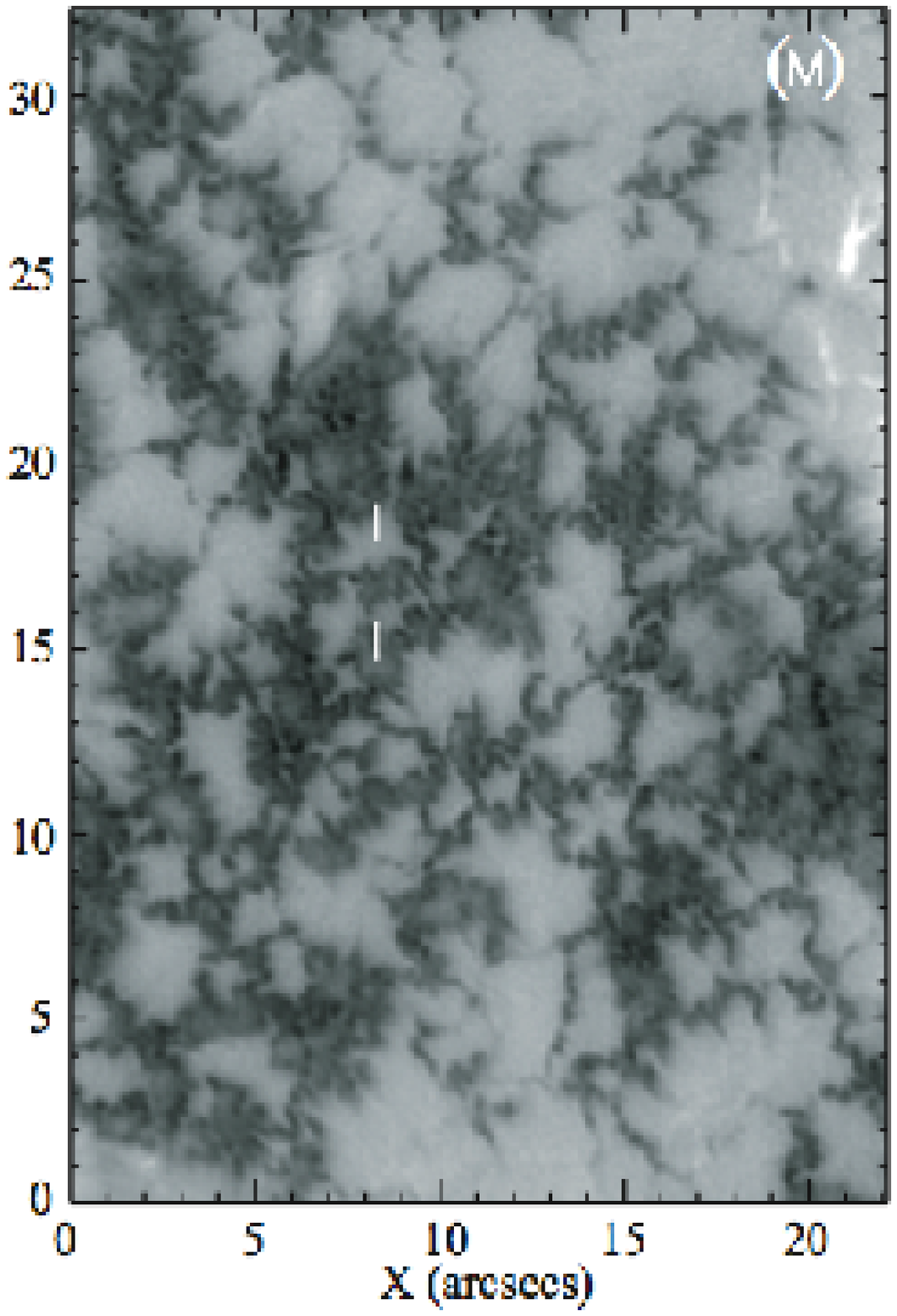} 
\caption[\sf G-band image and magnetogram for a plage region in the central part of the $\delta$-configuration active region.]{\sf \emph{Left}: G-band image displaying the central part of the $\delta$-configuration active region (turned upside-down respect to the one in Figure~\ref{F:2} and scales in arc seconds). \emph{Center}:  G-band bright points are selected and marked in \emph{white}. \emph{Right}: magnetogram (M) of the same region in gray scale ranging from -2795 (black) to 1000 (\emph{white}) Gauss.}
\label{F:ishikawa}
\end{figure}

The region in between the left and right hand side sunspots in the active region under study is indeed a plage region of abnormal granules full of magnetic activity as revealed by \cite{ishikawa2007}. Figure~\ref{F:ishikawa} shows the results from these authors working with the central region of the FOV (in an upside-down position respect to the one shown in Figure~\ref{F:2}). The G-band image shows a large amount of bright structures that are selected and marked in \emph{white} (see Figure~\ref{F:ishikawa} \emph{center}). These G-band bright points are ubiquitously distributed over the whole plage region and point out the intense magnetic activity in this so-called magnetized granulation. The corresponding magnetogram (see Figure~\ref{F:ishikawa} \emph{right}) also reveals the magnetic polarities behavior and intense activity in the same region.\\

\begin{figure}
\hspace{-1cm}\includegraphics[width=1.2\linewidth]{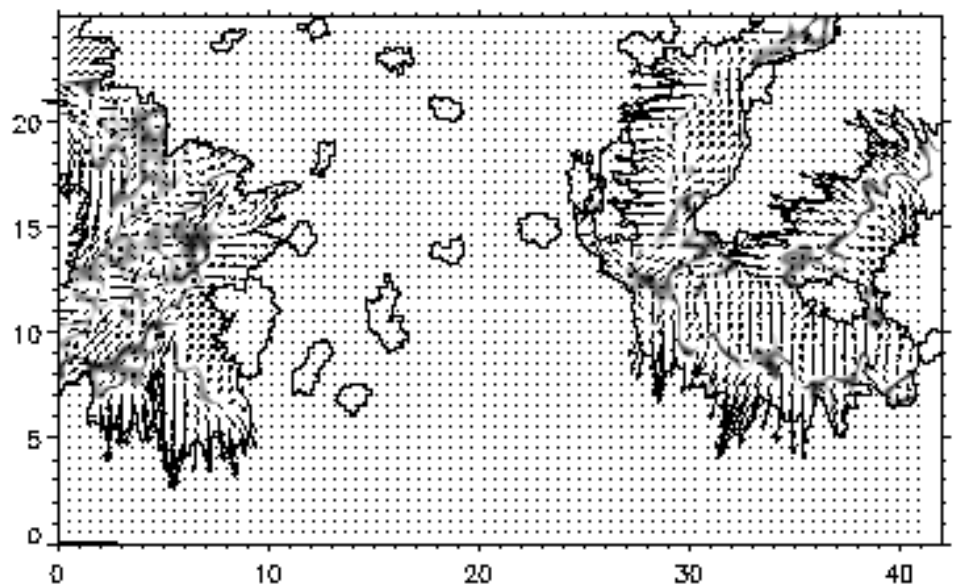}
\caption[\sf Proper motions in penumbrae]{\sf Proper motions inside penumbrae (FWHM= 0$\farcs$78, 71 min average). The figure displays the close-to-zero horizontal speed ($v_h  <$ 0.13 km s$^{-1}$) as gray traces inside \emph{white-colored} penumbrae. These lines delimit the changes in the privileged flow directions: 1) flows towards the umbra in the inner penumbra, and 2) towards the surrounding granulation in the outer penumbra. Umbrae and pores are delimited by \emph{black contours}. The coordinates are expressed in Mm. The length of the black bar at coordinates (0,0) corresponds to $4$ km s$^{-1}$.}
\label{F:linespenumbra}
\end{figure}

Inside penumbrae the proper motions were studied to infer the systematic directions of the flows.  Figure~\ref{F:linespenumbra} shows, with enlarged velocity scale, the proper motions inside penumbra using a tracking window FWHM= 0$\farcs$78 and averaging over 71 minutes. These displacements do not necessarily correspond to real motions of plasma but could reflect displacements of brightness. We are aware that the tracking window used allows us to determine only averaged displacements of the penumbral fine features, however, the description of the proper motions of this features is out of the scope of the present work. Figure~\ref{F:linespenumbra} reveals mean flows at spatial resolution of $\sim$~1\arcsec (mean velocity $\sim$~0.4 km s$^{-1}$), moving toward the umbra in the inner penumbra and toward the surrounding granulation in the outer penumbra. The filamentary like structures in grey, mapping small horizontal velocities ($ <$ 0.13 km s$^{-1}$), outline the frontiers separating both kinds of flow. This clear division of the velocities field in two quite differentiable tendencies has already been mentioned by some authors \citep[see][]{bonet2004} but not in a systematically way inside sunspots in a complex active region as presented in this work.\\

\begin{figure}
\begin{center}
\includegraphics[angle=90,width=.8\linewidth]{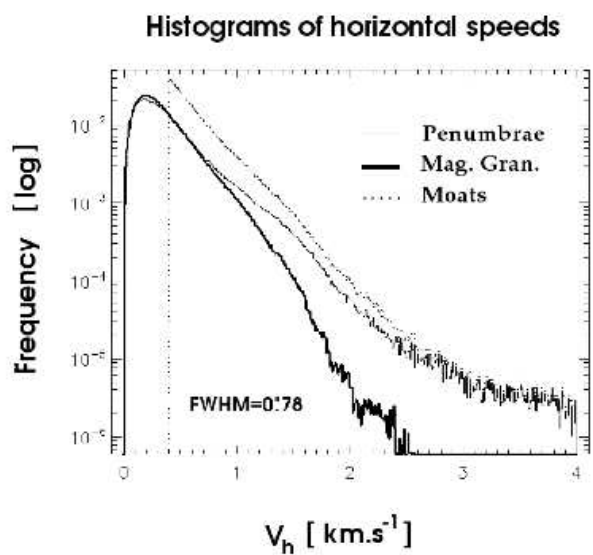}
\caption[\sf Histogram of horizontal speeds for a $\delta$-configuration active region]{\sf Histogram of horizontal speeds $v_h$ (from 14 maps, 5 minutes averaged each) for a $\delta$-configuration active region. A local correlation tracking technique has been employed to derive the horizontal velocity components using a Gaussian  tracking window of FMWH= 0$\farcs$78. Table~4.5 lists the statistics of velocity magnitudes in every region.}
\label{F:histy}
\end{center}
\end{figure}

In order to perform a statistical study of horizontal velocities, a sequence of 14 velocity maps was obtained averaging over 5-min windows along the whole series. As expected, the resulting maps are a little more noisy but with a wider range of variation of the speeds (i.e.\ velocity magnitudes) as compared with the case of averaging over 71 min. Figure~\ref{F:histy} shows the histogram of the speeds as measured in this sequence of maps, for different regions within the FOV: penumbral areas, granulation outside the moats and moat regions (see also Table~\ref{T:velocities} for mean and rms values). Note that the moat histogram in the Figure~\ref{F:histy} shows a cutoff imposed by the lower threshold selected to define moat velocities. \\ 

The statistical analysis of the results (Figure~\ref{F:histy} and Table~4.5) renders a value for the mean speed in the moats of $\sim$ 0.67 km s$^{-1}$, similar to that reported by \cite{bonet2005}, that is almost twice the mean value in the rest of the granulation field. The confinement of such significantly higher speeds in restricted areas close to the sunspots makes moats a sharply differentiated structure in the photospheric granulation.\\

We pursued the same statistical calculations for velocity maps computed with a tracking window with FWHM 0$\farcs$2. The results revealed that when reducing the size of the tracking window, the measured mean and rms speed values increment since we are also including small structures, as bright points, moving faster than the general flows observed in the granulation pattern evolution. In our experiment, for instance, the difference between the mean and rms values is about twofold when going from FWHM 0$\farcs$78 to 0$\farcs$2.

\begin{table}
\sffamily
\begin{center}
    \renewcommand{\arraystretch}{1.2}
     \begin{tabular}[h]{lcccc}
      {\footnotesize REGION} & mean & rms \\\hline 
      Penumbrae & 0.40 & 0.35 \\
      Magnetic  Gran.  & 0.35 & 0.24 \\
      Moats & 0.67 & 0.32  \\
      \hline \\
      \end{tabular}
\caption[\sf Statistics of the horizontal speeds]{\sf Statistics of the horizontal speeds $v_h$ [km s$^{-1}$] for various regions within the FOV of a complex solar active region. The calculations are done using the velocity magnitudes derived from 14 maps (5 min averaged, FWHM 0$\farcs$78) covering 70 minutes of solar evolution.}   
\end{center}
\label{T:velocities}
\end{table}

\subsection{Large-scale horizontal granular flows}
\label{S:largeflows}

\begin{figure}
\hspace{-1cm}\includegraphics[width=1.2\linewidth]{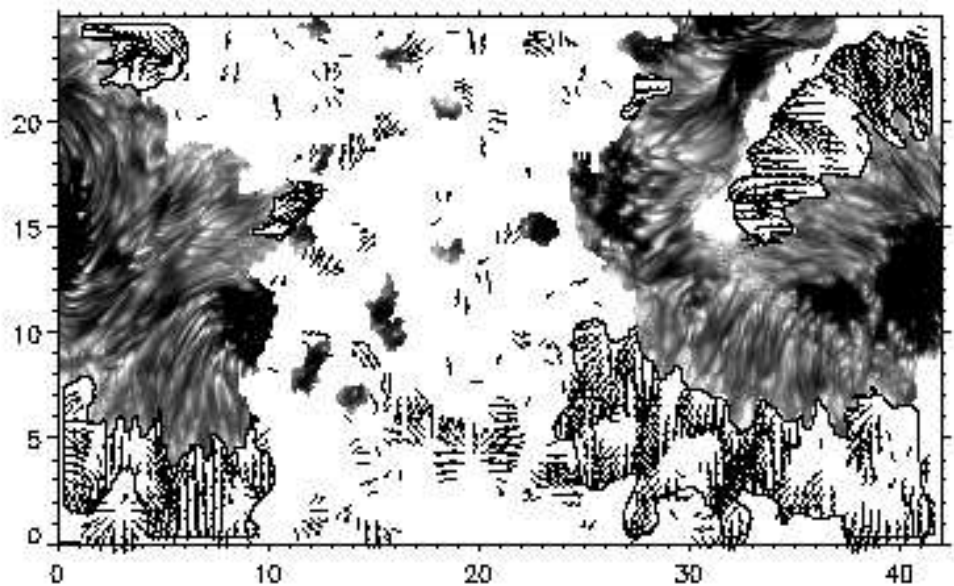}
\caption[\sf Map of the horizontal velocities with magnitude  $>$ 0.4 km s$^{-1}$]{\sf Map of the horizontal velocities with magnitude  $>$ 0.4 km s$^{-1}$ (71 min average). The contrast within penumbrae has been enhanced by removal of a spatial running mean of the original image. Strong radial outflows (moats) are evident surrounding filamentary penumbrae. They are confined by the \emph{black contours}.
These moats are found to be closely associated with the existence of penumbra.
Note that moat region surrounding coordinates (1,5) corresponds to a
penumbra located to the left of the image and not visible in our FOV. The
coordinates are expressed in Mm. The length of the black bar at coordinates
(0,0) corresponds to 4 km s$^{-1}$.}
\label{F:3}
\end{figure}

To study the proper motions surrounding the sunspots and in particular the behavior of the large-scale flows in the FOV, a velocity threshold can be selected to only plot a given range. The map in Figure~\ref{F:3} shows only those velocities in the granulation field with magnitude above 0.4 km s$^{-1}$. Strong radial outflows (moats) are evident surrounding the sunspots and 0.4 km s$^{-1}$ appears to be a characteristic speed value to define the moats. The \emph{black contours} in the figure outline
the area of the moats. It was drawn by hand following visual criteria of
proximity to the sunspot and avoiding a few strong exploding events at the
lower part of the FOV that contaminate the large-scale moat flow.  Empty areas
in the moat region in the velocity maps are a result of the chosen value for the
speed threshold and correspond to the interaction with exploding events
that occur inside the moat.\\ 

From inspection of Figure~\ref{F:3}, it becomes evident that the moats are closely
associated with the presence of a penumbra. The examples in the lower left, lower
right and upper right corners are quite revealing. The right pointed velocity
vectors on the lower left corner correspond to the moat flow of a penumbra that
is not visible in the FOV of this figure, but can be seen in an image covering a wider FOV, of the same active region in Figure~\ref{F:DOT09julio}. Moats are missing in the sunspot sides with no penumbral structure. This is readily seen in the right side of the 
umbral core at coordinates (9,10), but also in the left side of the one
at (25,18). None of the pores in the middle of the FOV is associated with any moat-like
flow. More importantly, moats are also absent in granulation regions
that are located side by side with a penumbra, but in the 
direction perpendicular to that marked by the penumbral filaments.
A clear example is the region around coordinates (9,6-7) in Figure~\ref{F:3}. As a \emph{normal} 
(meaning here far from the neutral line) penumbra, it is a candidate to develop
a moat flow, but none is seen in this region. The moat flow only appears along the 
direction delineated by the penumbral filaments. Moats, thus, appear to
be a natural extension of the flow along the direction of penumbral filaments
and do not exist as outflows in the transverse direction.

\subsection{Neutral lines affecting the flow behavior}
\label{S:neutral-line}

\begin{figure}
\begin{center}
\begin{tabular}{c}
\includegraphics[width=.65\linewidth]{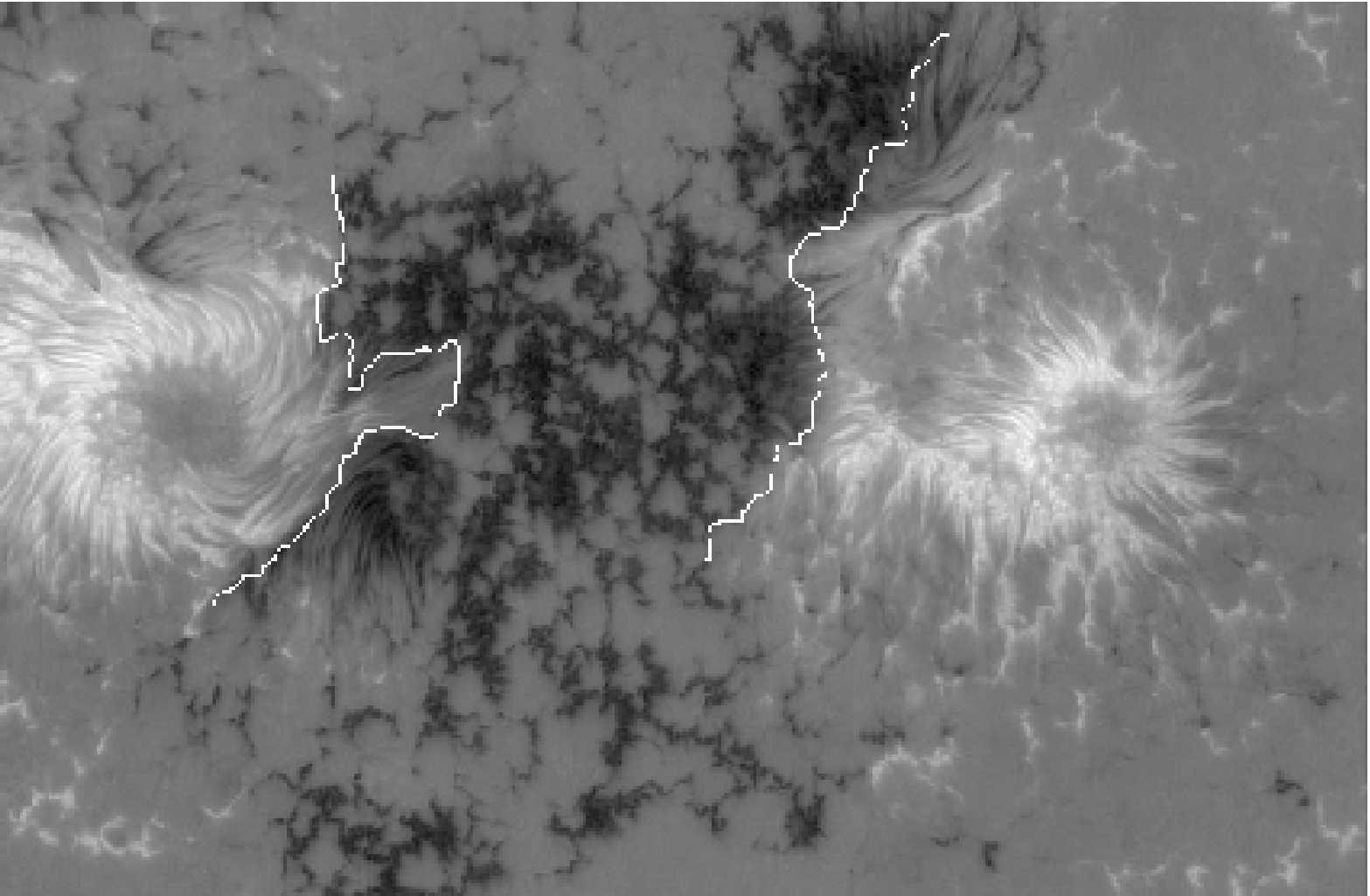} \\
\includegraphics[width=.65\linewidth]{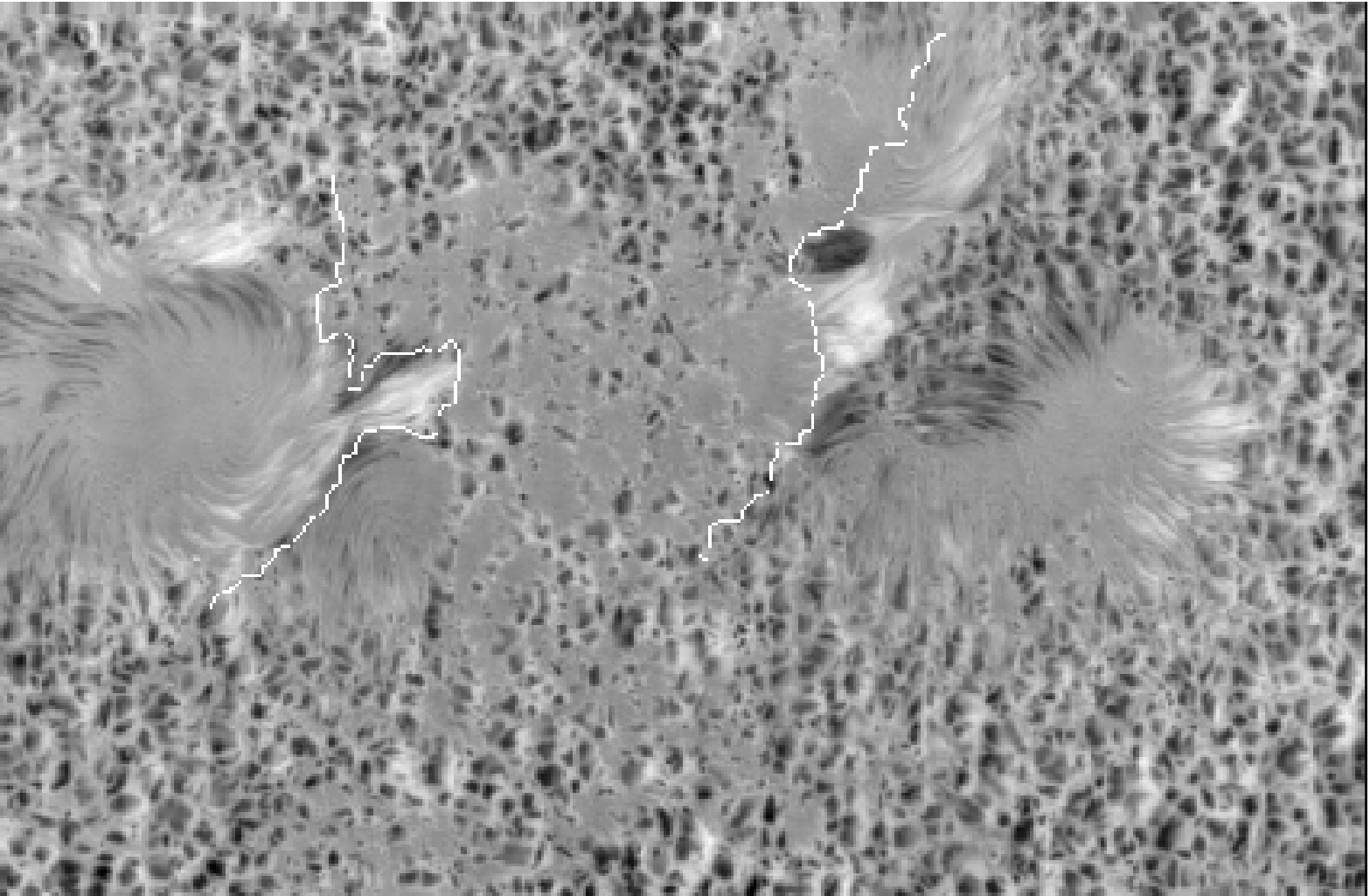}\\
 \includegraphics[width=.65\linewidth]{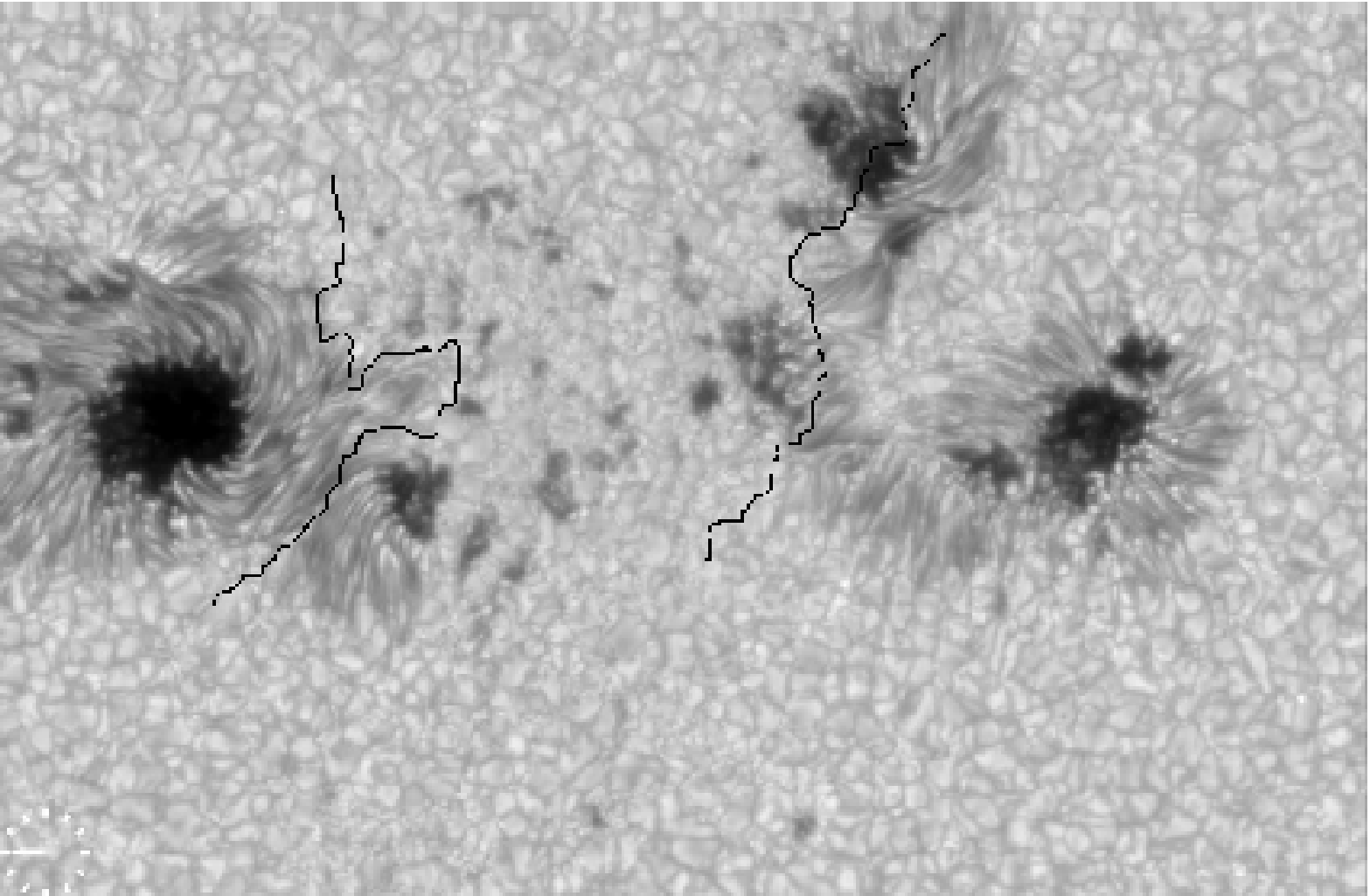}\\
\end{tabular}
\caption[\sf Filtergrams for a $\delta$-configuration sunspot taken by the DOT]{\sf Magnetogram (\emph{upper panel}), Dopplergram (\emph{middle panel}) and white light image (\emph{lower panel}) for the $\delta$-configuration sunspot group, observed by P. S\"utterlin with the DOT on 9 July 2007. Neutral lines are identified in the magnetogram (\emph{white lines}) and overplotted in the Dopplergram (\emph{white lines}) and in the white light image (\emph{black lines}).}
\label{F:DOT09julio}
\end{center}
\end{figure}

As commented in section~\S\ref{S:magnetogram} we produced magnetograms of the FOV by means of the SOUP filter. The high-resolution magnetogram in Figure~\ref{F:4} shows the two polarities. The positive polarities (\emph{in white}) dominate the left and right areas of the FOV and correspond widely to the polarities of the respective sunspots. The central area of the image is dominated by a negative polarity (\emph{in black}).  Two neutral lines can be then identified separating opposite polarities. These neutral lines are also overplotted in the magnetogram, dopplergram and white-light image taken by the DOT during the same observing run but with a wider FOV (see Figure~\ref{F:DOT09julio}).\\

It is seen that when the neutral lines run close to a penumbral border and parallel to it, the expected behavior of the outflows in the granulation region side by side with that penumbra changes somehow and the moats do not appear in those regions.  The penumbral region in Figure~\ref{F:3}, centered at coordinates (5,20), does not show clear evidence of a moat flow (although localized outflows are seen above and below it).  This region corresponds to a strongly sheared neutral line (see Figure~\ref{F:2}) commonly
seen in $\delta$ spots \citep{deng2006}. In the SOUP magnetogram (Figure~\ref{F:4}) the spot centered on the left side of the FOV  (\emph{white polarity}) does not show a radially outward directed penumbra, but penumbral filaments that run parallel to the neutral line. This is a configuration 
commonly referred as sheared. This configuration is also inferred from the 
filaments observed in the G-band frames. We suspect that both conditions, the 
presence of a neutral line and the absence of radially oriented filaments, are
responsible for the absence of a clearly developed moat flow.\\

The nature of neutral lines with sheared configurations in $\delta$-spots
is not well understood. But we stress that if they correspond to relatively 
shallow structures, they would also block the upward propagation of heat and
should be able to generate a moat flow. Such a moat
flow is not observed in this sunspot. 

\subsection{Motions of centers of divergence}
\label{S:diver}

Following  ~\cite{november1988}, \citep[see also][]{marquez2006}, the divergences and the vertical velocities can be calculated from the horizontal velocities, according to the expression

\begin{equation}
v_z(x,y)=\mbox{h}_{\mbox{\scriptsize{m}}} \nabla \cdot \vec v_h (x,y),
\label{vz}
\end{equation}

\noindent where $\mbox{h}_{\mbox{\scriptsize{m}}}=$150 km stands for the scale height of the flux of mass and $\nabla \cdot \vec v_h (x,y)$ the divergence of the horizontal velocity field. \\

\begin{figure}
\centering
\begin{tabular}{cc}
{\bf{\footnotesize \textsc{\sf AVERAGED: 5 MIN}}} ~  {\footnotesize [\sf km s$^{-1}$] } & {\bf{\footnotesize\textsc{\sf AVERAGED: 15 MIN}}} ~   {\footnotesize \sf [km s$^{-1}$]} \\
\begin{tabular}{cccc}
{\footnotesize \sf min} & {\footnotesize \sf max} & {\footnotesize \sf rms} & {\footnotesize \sf mean} \\
{\footnotesize \sf -7.77} &{\footnotesize \sf 12.41} & {\footnotesize \sf 1.78} & {\footnotesize \sf 0.04} \\
\end{tabular}
&  
\begin{tabular}{cccc}
\hspace{-4mm}{\footnotesize \sf min} & {\footnotesize \sf max} & {\footnotesize \sf rms} & {\footnotesize \sf mean} \\
\hspace{-4mm}{\footnotesize \sf -6.30} & {\footnotesize \sf 10.64} &  {\footnotesize \sf 1.46} & {\footnotesize \sf 0.03} \\
\end{tabular}
 \\
\hspace{-4mm}\includegraphics[width=.47\linewidth]{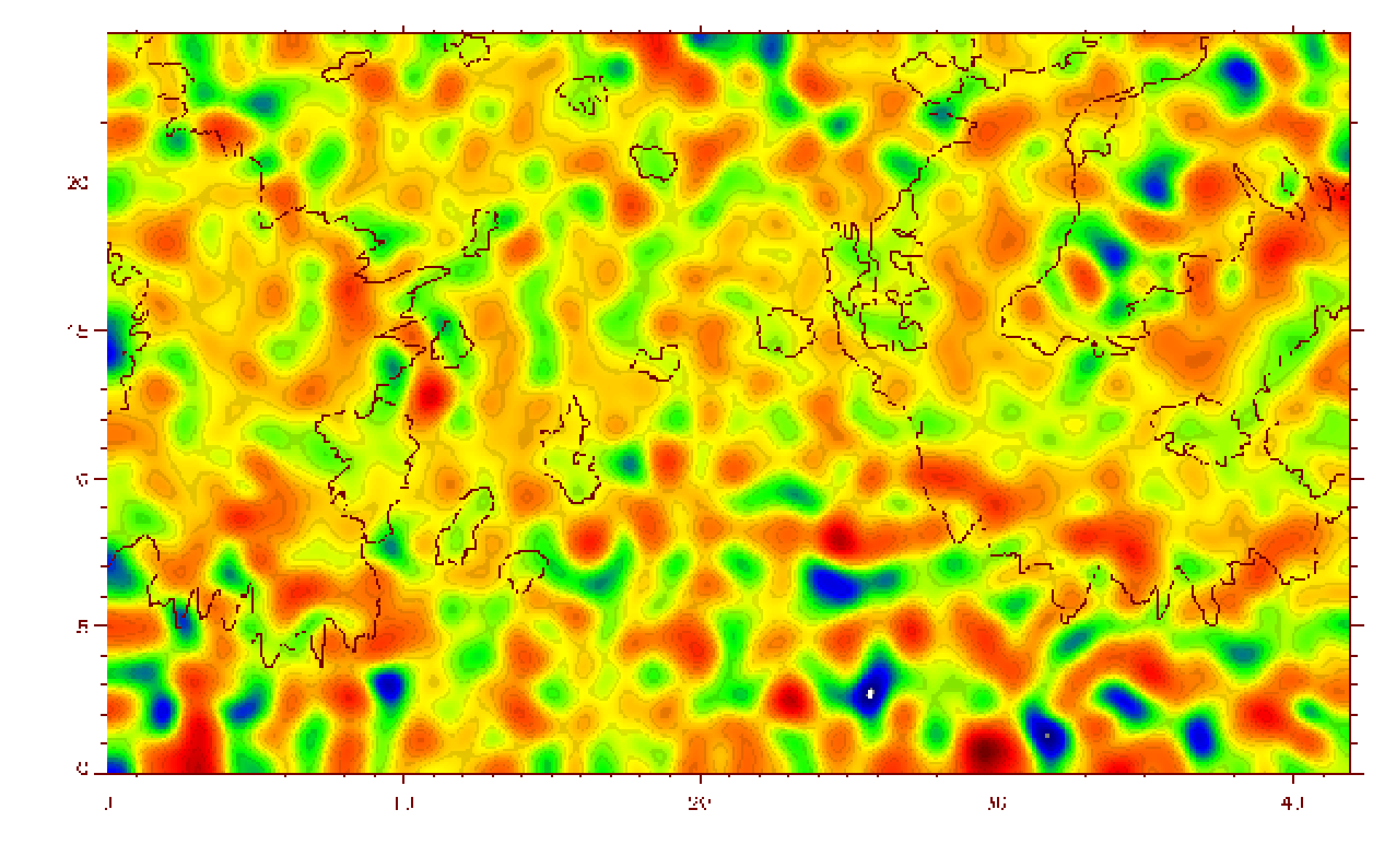}\includegraphics[height=.3\linewidth]{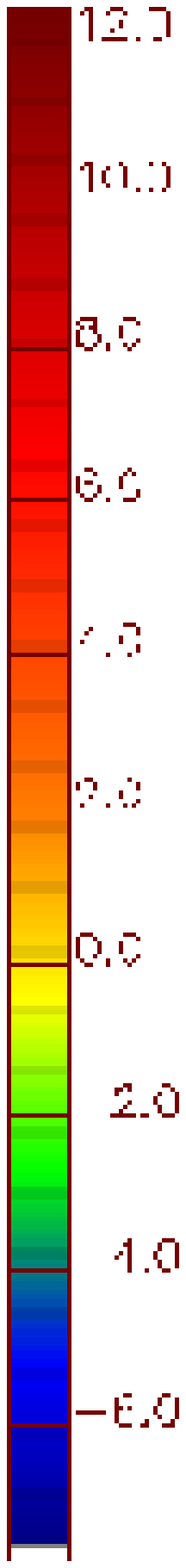} &
\hspace{-4mm}\includegraphics[width=.47\linewidth]{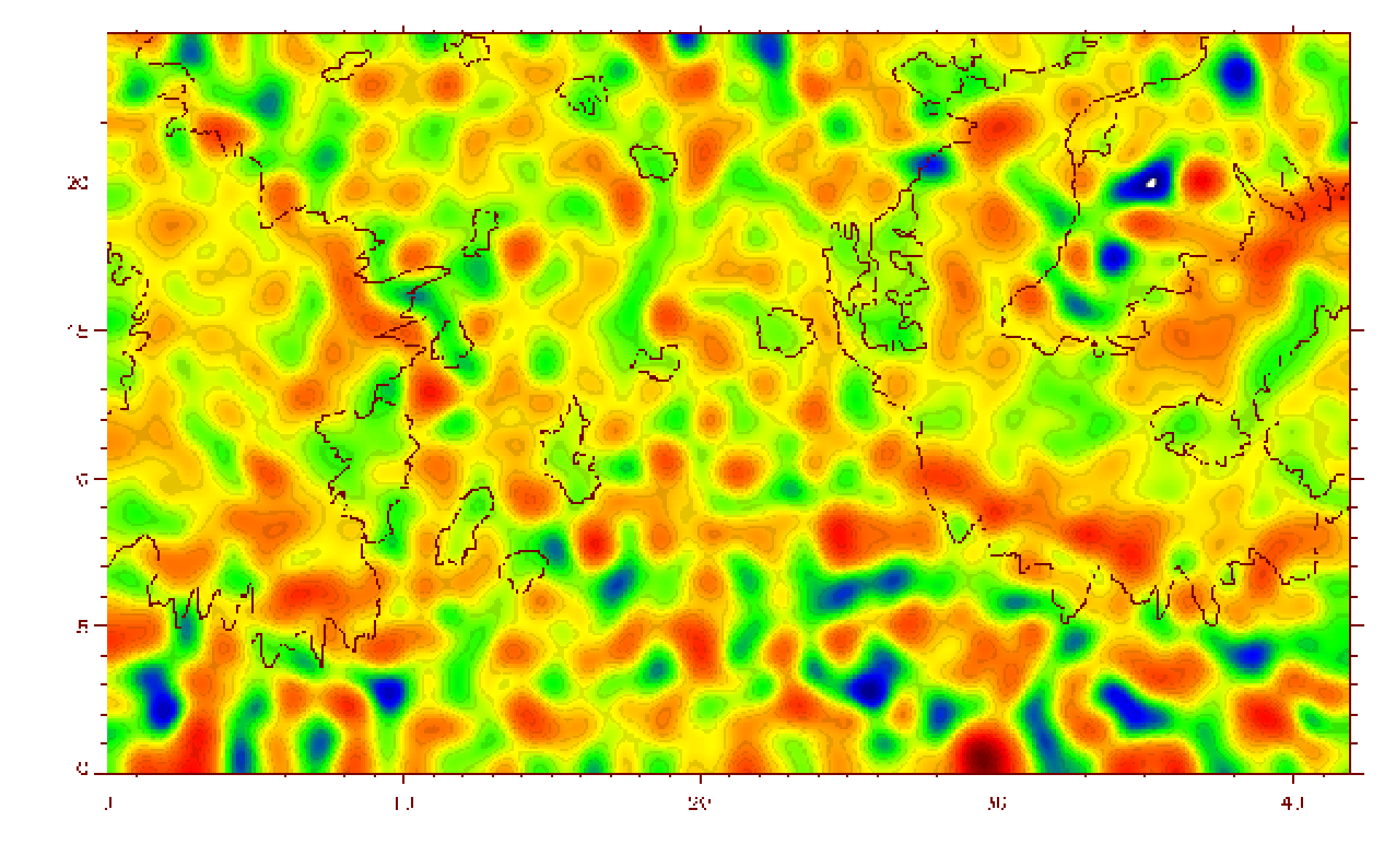}\includegraphics[height=.3\linewidth]{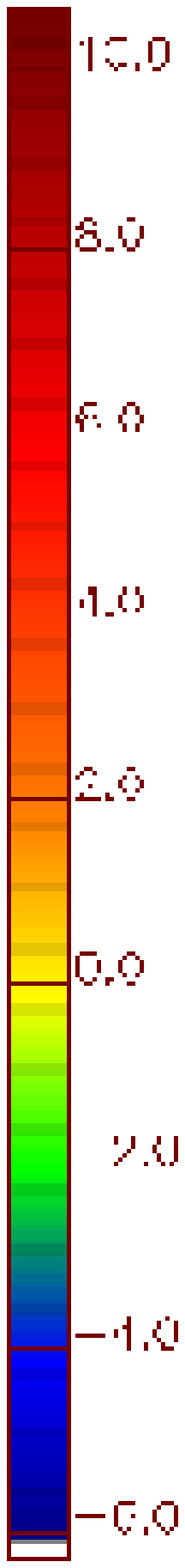}  \\ \\

{\bf {\footnotesize \textsc{\sf AVERAGED: 40 MIN}}}  ~  {\footnotesize\sf [km s$^{-1}$] } & {\bf{\footnotesize\textsc{\sf AVERAGED: 70 MIN }}}   ~ {\footnotesize \sf [km s$^{-1}$]}\\
\begin{tabular}{cccc}
{\footnotesize \sf min} & {\footnotesize \sf max} & {\footnotesize \sf rms} & {\footnotesize \sf mean} \\
{\footnotesize \sf -4.99} & {\footnotesize \sf 7.80} &  {\footnotesize \sf 1.35} & {\footnotesize \sf 0.01} \\
\end{tabular}
&  
\begin{tabular}{cccc}
\hspace{-4mm}{\footnotesize \sf min} & {\footnotesize \sf max} & {\footnotesize \sf rms} & {\footnotesize \sf mean} \\
\hspace{-4mm}{\footnotesize \sf -4.23} & {\footnotesize \sf 6.15} &  {\footnotesize \sf 1.22} & {\footnotesize \sf 0.01} \\
\end{tabular}
 \\
\hspace{-4mm}\includegraphics[width=.47\linewidth]{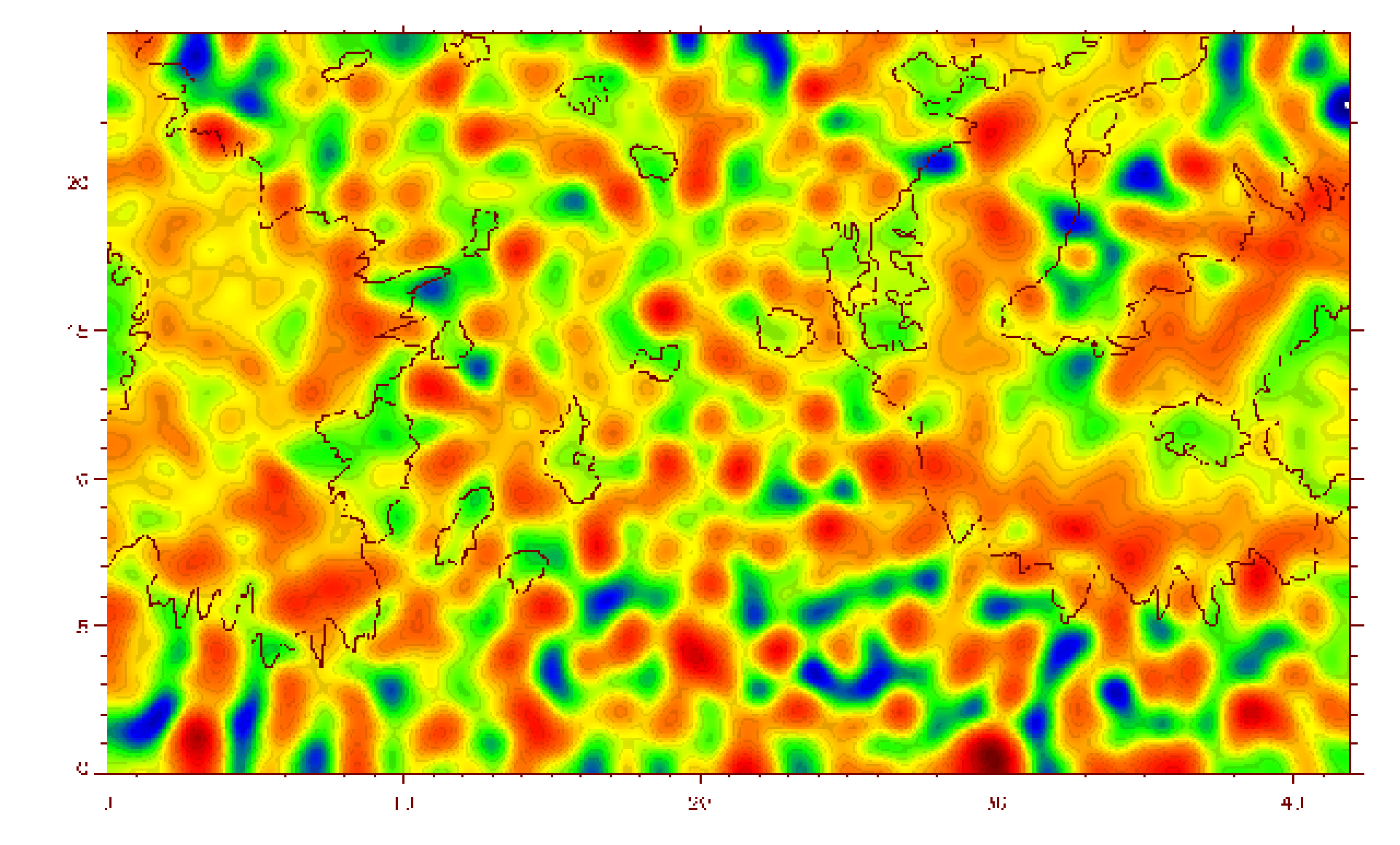}\includegraphics[height=.3\linewidth]{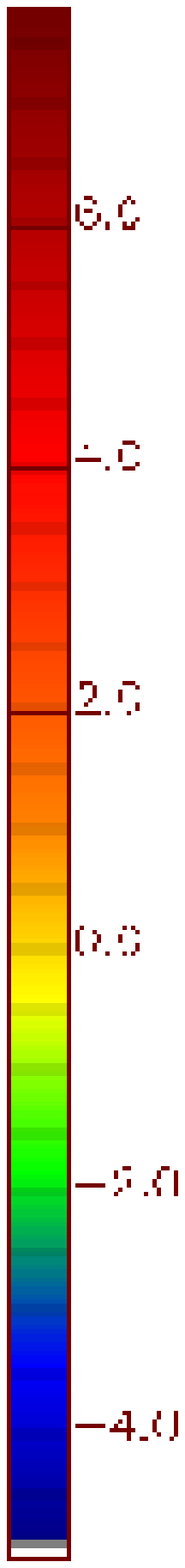}  &
\hspace{-4mm}\includegraphics[width=.47\linewidth]{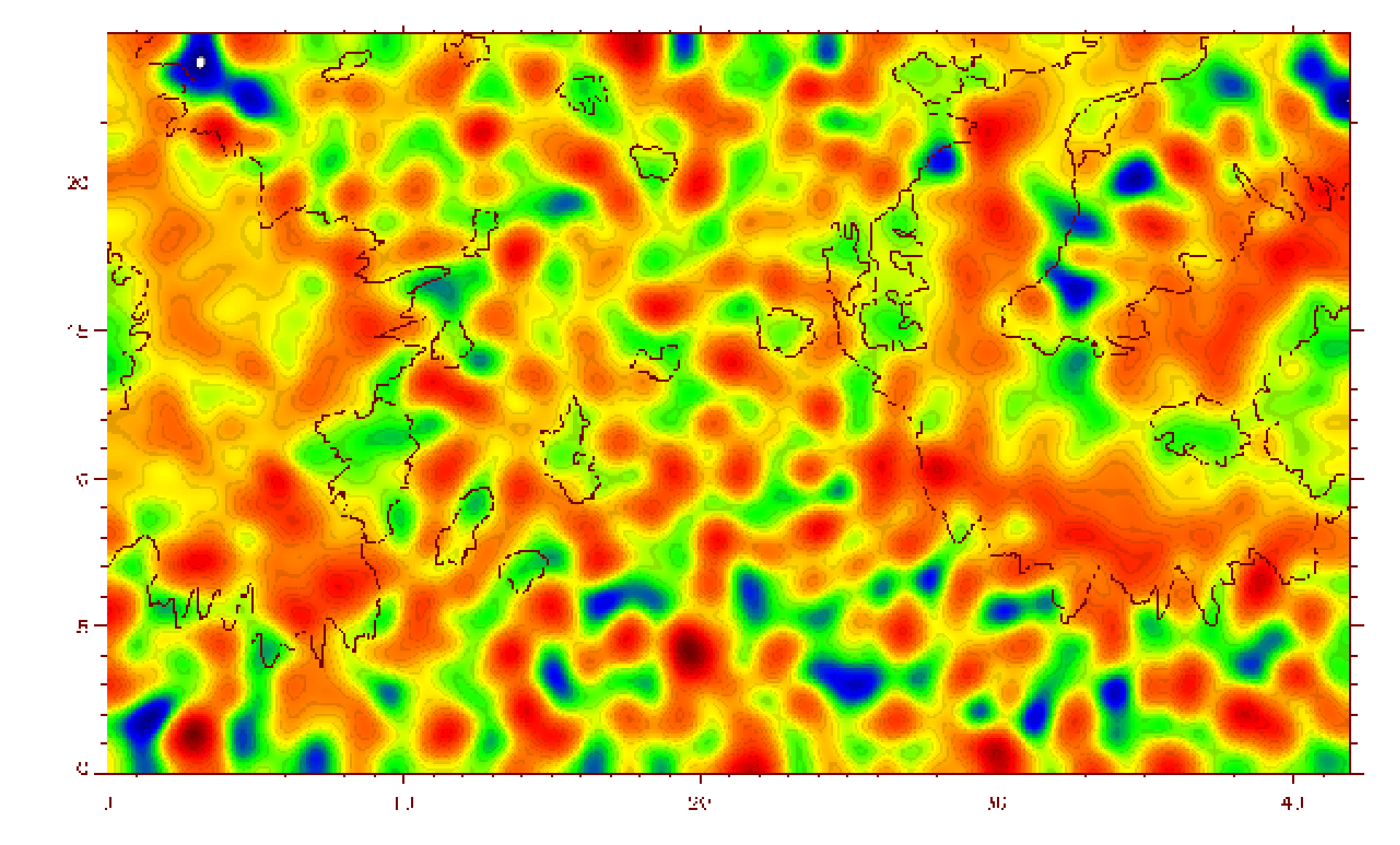}\includegraphics[height=.3\linewidth]{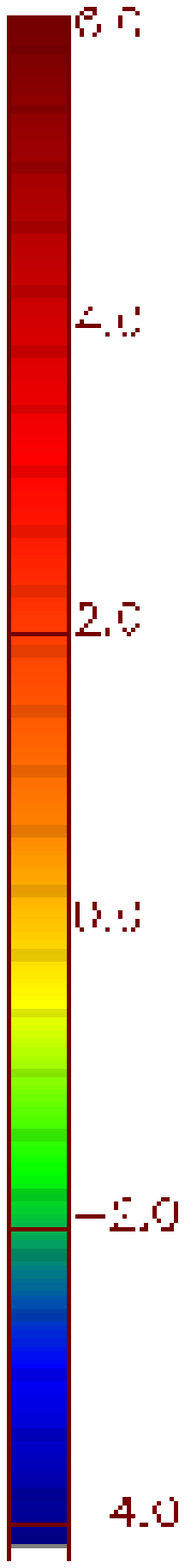}  \\
\end{tabular}
\caption[\sf Maps of vertical velocities for a $\delta$-configuration sunspot]{\sf Maps of vertical velocities averaged over different time windows starting from the beginning of the time series (5, 15, 40 and 70 minutes, FMWH= 0$\farcs$78).  In order to identify general trends in the vertical flows, the maps have been smoothed by convolving with a Gaussian running window of FWHM=1$\farcs$47. \emph{Yellow} represents close-to-zero velocities. Extreme values of divergence can be easily identified: exploding granules with positive divergences (\emph{in dark red}) at coordinates (30,0) and powerful sinks with negative divergence (\emph{in blue}) at the upper right part of the image. Regardless the average applied, from 5 up to 70 minutes, the exploding events and sinks are recurrently seen and clearly defined within the FOV. The statistics of vertical velocities in the FOV is presented for every averaged map. \emph{Black contours} outline umbrae, penumbrae and pores. Note that the color-scales  (the representation in pseudo-color is the opposite to the one used for Doppler shifts) are slightly different for every map since  the values for the minimum and maximum velocity change from one case to the other as shown by the statistics tabulated at the top of each map. The coordinates are expressed in Mm.}
\label{F:divergen}
\end{figure}

For this purpose, we generated four horizontal velocity maps (Gaussian tracking window of FMWH= 0$\farcs$78) averaging over time windows of 5, 15, 40 and 70 min, and starting from the beginning of the time series. From these maps we obtained the divergences and the corresponding vertical velocities. Since we were not interested in describing details but only general trends in the granulation vertical flows, the resulting $v_z(x,y)$ maps were smoothed by convolving with a Gaussian running window of FWHM=1$\farcs$47, so that the identification of coherent structures was facilitated. Figure~\ref{F:divergen} displays the resulting four vertical velocity maps as images representing the velocity magnitudes by using a pseudocolor scale. Structures corresponding to positive divergences (i.e.\ recurrent exploding granules) and sinks with negative divergences are coherently distributed all over the FOV for all the different temporal averages though an evident evolution can be noticed between these averages. The figure shows extreme cases for both positive divergences at coordinates (30,0) \emph{in dark red}, and sinks in the upper right part of the image \emph{in blue}.\\

\begin{figure}
\begin{center}
\includegraphics[width=.6\linewidth]{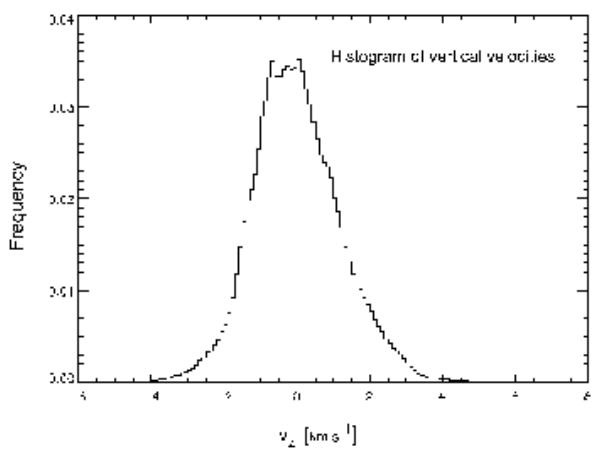}
\caption[\sf Histogram of vertical velocities for a $\delta$-configuration sunspot]{\sf Histogram of vertical velocities in the FOV, calculated from the 5-min averaged velocities maps.}
\label{F:histyvz}
\end{center}
\end{figure}

For the particular case of 5-min horizontal velocity averages we computed 14 of these maps covering the whole time series (70 min), thus deriving 14 vertical velocity maps similar to the one in the upper left corner in Figure~\ref{F:divergen}. Figure~\ref{F:histyvz} shows the histogram of vertical velocities as calculated from these 14 maps, that shows a quasi-symmetric behavior around the mean value ($v \sim$ 0).\\

The divergence structures displace in time dragged by flows at scales larger than the mesogranular ones. In order to follow the evolution of the centers of divergence along the time series, we employed the same 14 vertical velocity maps described in the previous paragraph\footnote{\sf For the purpose we are interested here, the distinction between vertical velocity maps and divergence maps is irrelevant since both quantities only differ by a constant factor $h_m$.}. Then, by applying the LCT again but this time upon the sequence of 14 images of divergence, we can determine how the centers of divergence displace in time. Because of the larger size of the structures we want to track, the correlation tracking window was set to FWHM 3$\arcsec$.\\

\begin{figure}
\hspace{-1cm}\includegraphics[width=1.2\linewidth]{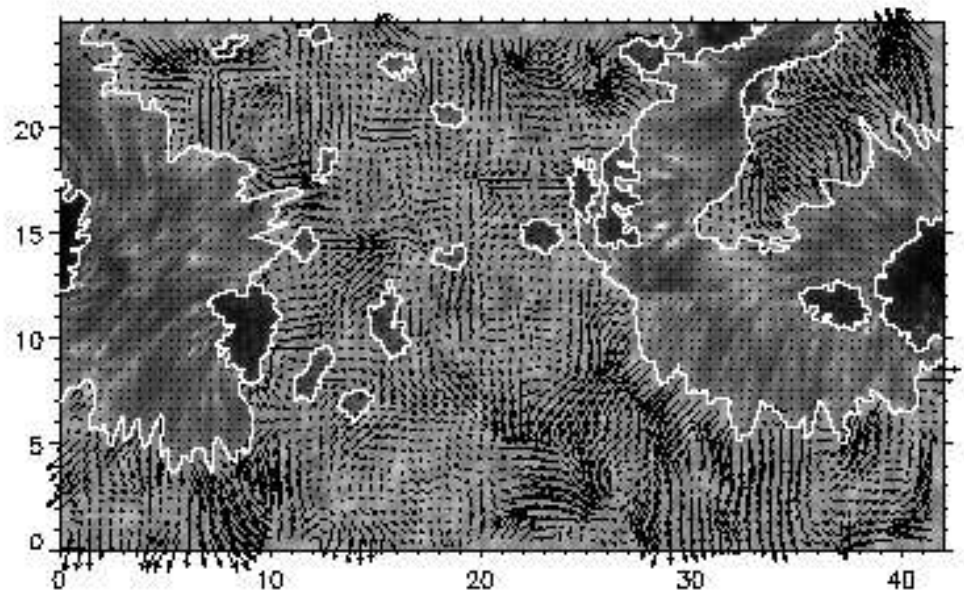}
\caption[\sf Map of velocities of the centers of divergence for a $\delta$-configuration sunspot]{\sf Map of velocities of the centers of divergence, calculated from a sequence of fourteen 5-min averaged maps of divergence structures, along the whole time series (71 min). The velocity magnitudes are in a range from 0 to 1.18 km s$^{-1}$. The length of the black bar at coordinates (0,0) corresponds to 1 km s$^{-1}$.}
\label{F:mapdiv}
\end{figure}

Figure~\ref{F:mapdiv} displays the resulting map corresponding to the averaged velocities of the centers of divergence during the whole time series (70 min). The velocities inside sunspots and pores are set to zero since we are only interested in the granulation proper motions. Figure~\ref{F:mapdiv} reveals the smooth character of the flow transporting the centers of divergence; the flows are more organized and uniform as compared with those in Figure~\ref{F:2} but they still preserve signatures of mesogranular flows. The well-organized moat flows are also clearly defined in the figure as they were in Figure~\ref{F:2}. Note that the scales employed to represent the velocity vectors in Figures~\ref{F:mapdiv} and \ref{F:2} are quite different so that the velocities displayed in the former are much lower than those represented in Figure~\ref{F:2}.\\

\vspace{8mm}

\subsection{Study of convective cells}
\label{S:cells}

\begin{figure}
\centering
\begin{tabular}{rlll}
\hspace{-1cm}\tiny{\sf 0} & \includegraphics[width=.48\linewidth]{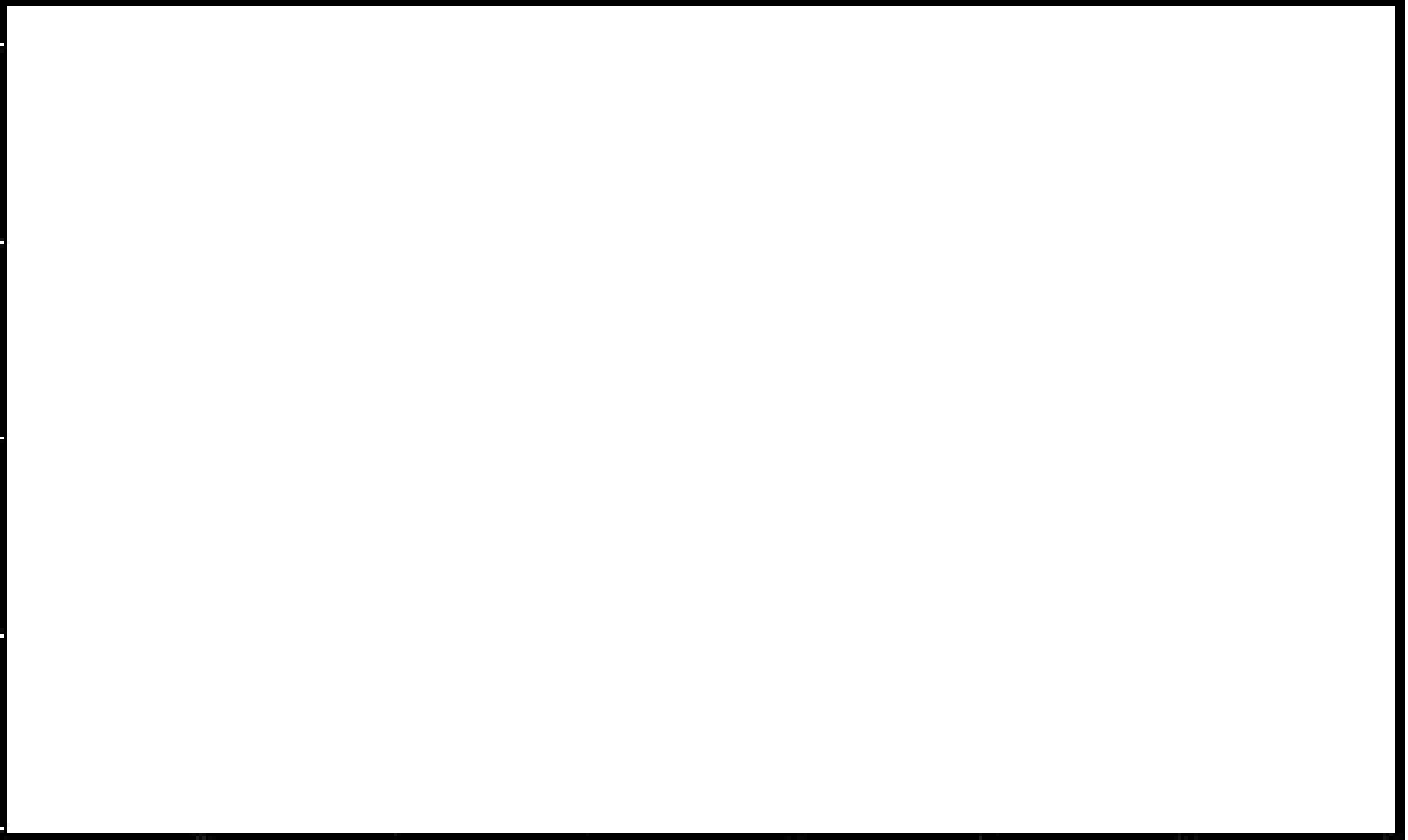} &  \includegraphics[width=.48\linewidth]{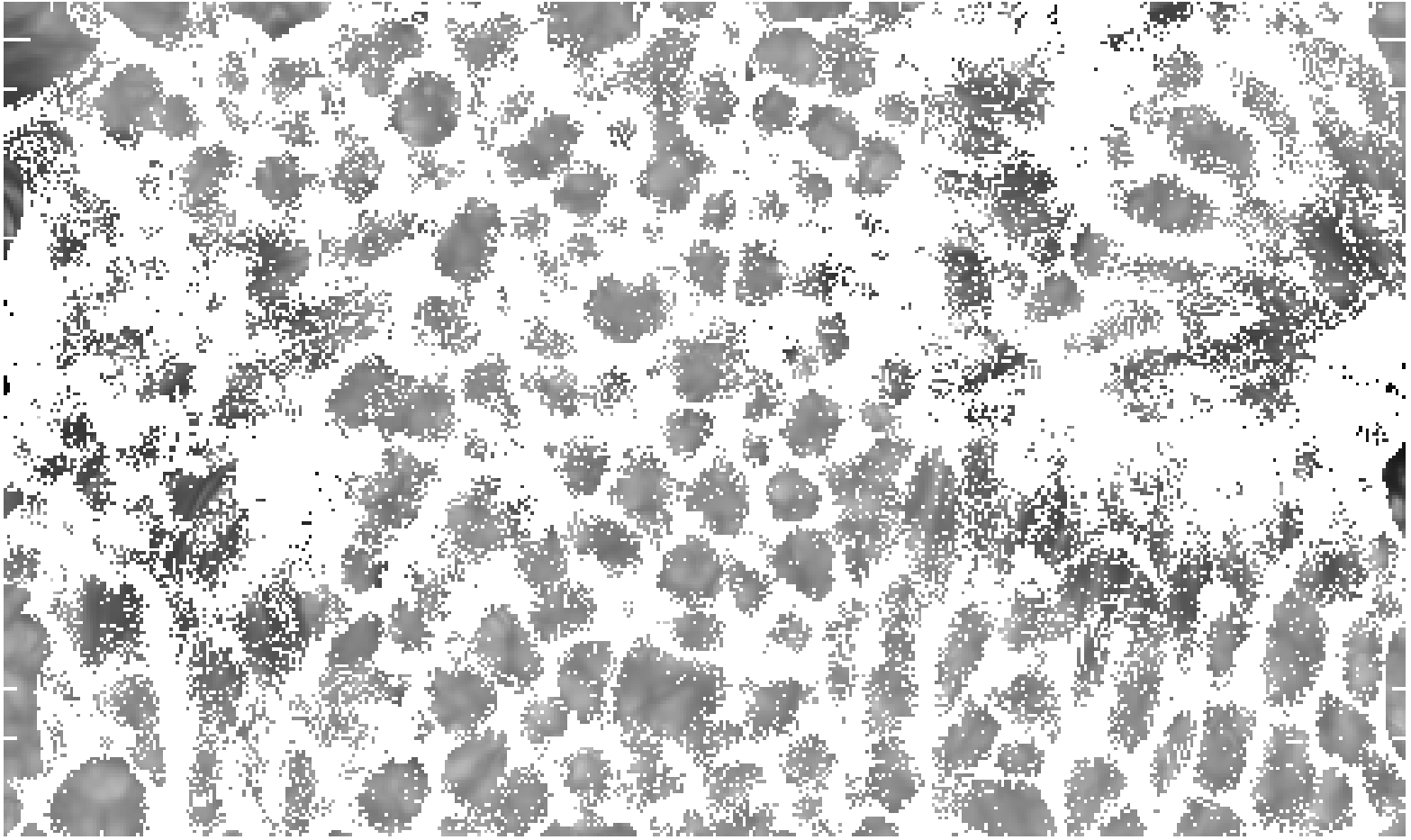} & \tiny{\sf 22} \\
\hspace{-1cm}\tiny{\sf 45} & \includegraphics[width=.48\linewidth]{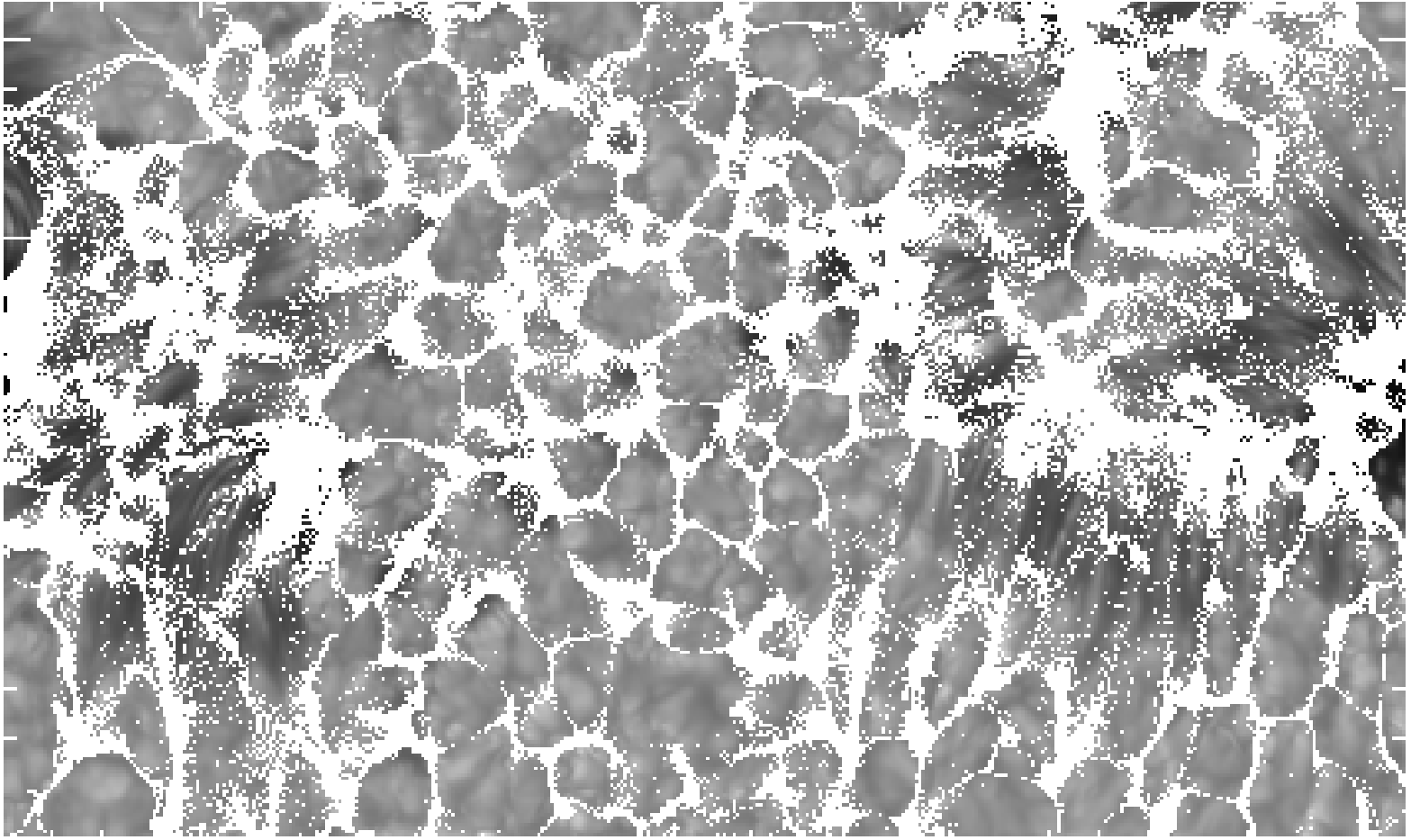} &
\includegraphics[width=.48\linewidth]{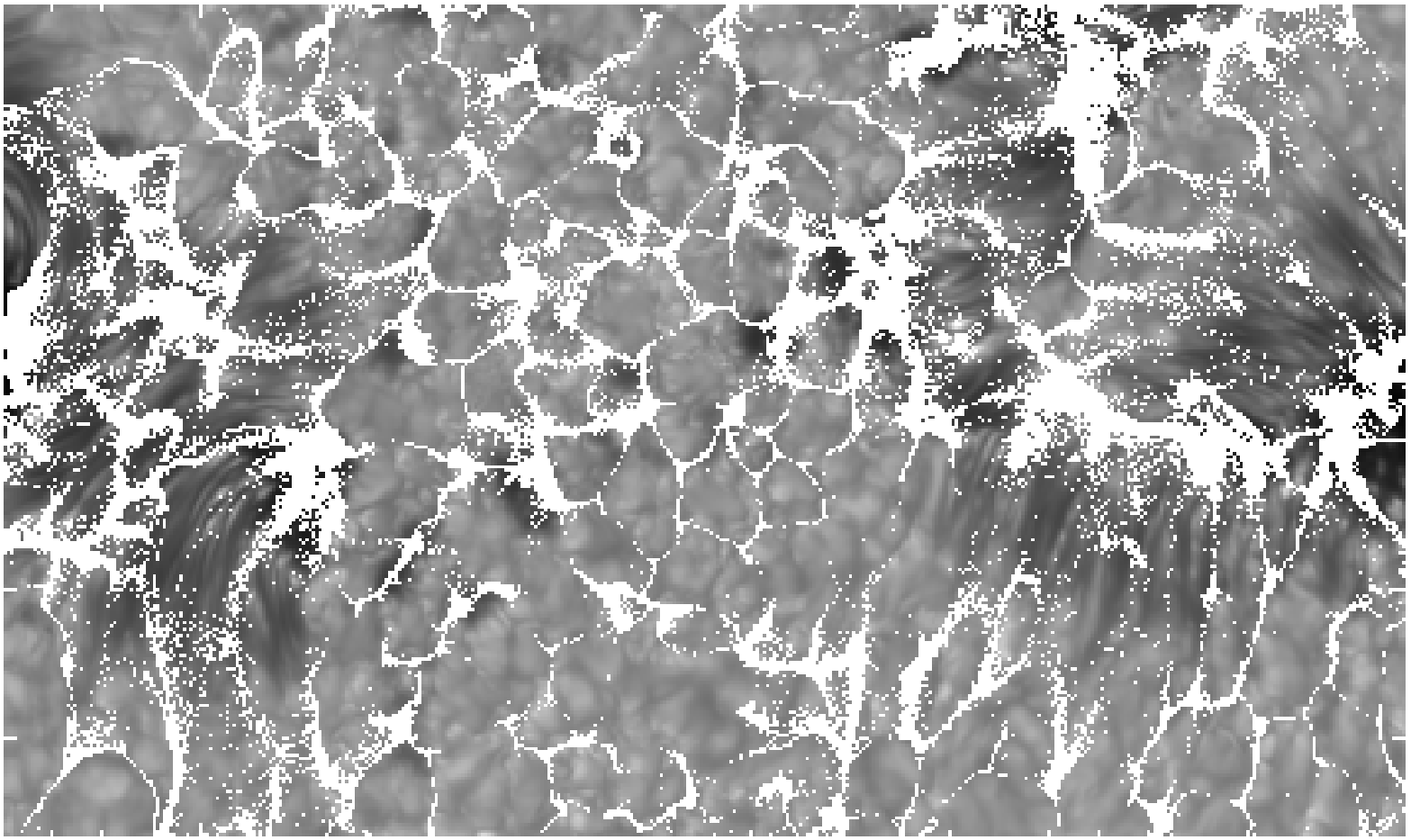} & \tiny{\sf 67} \\
\hspace{-1cm}\tiny{\sf 89} &\includegraphics[width=.48\linewidth]{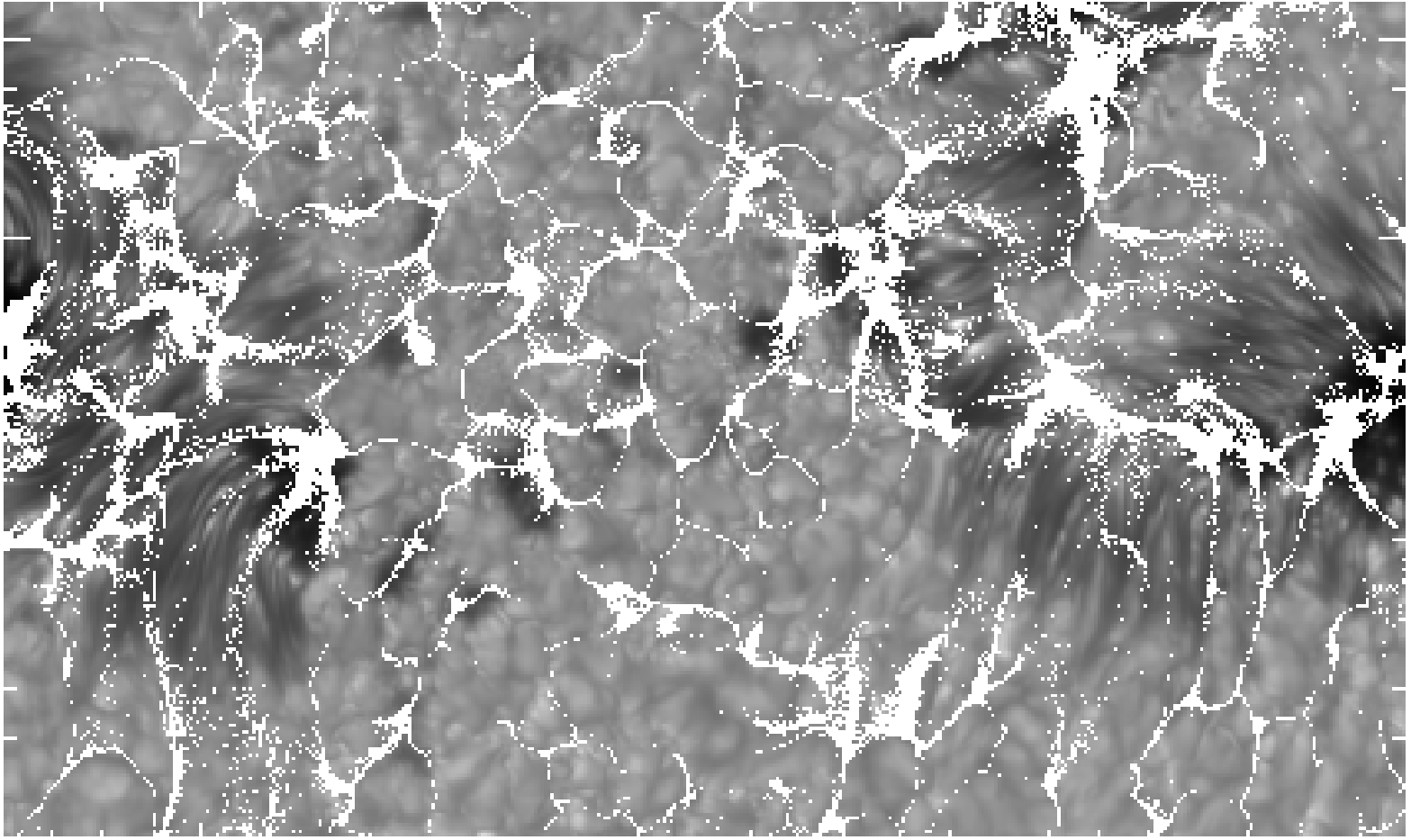} & 
\includegraphics[width=.48\linewidth]{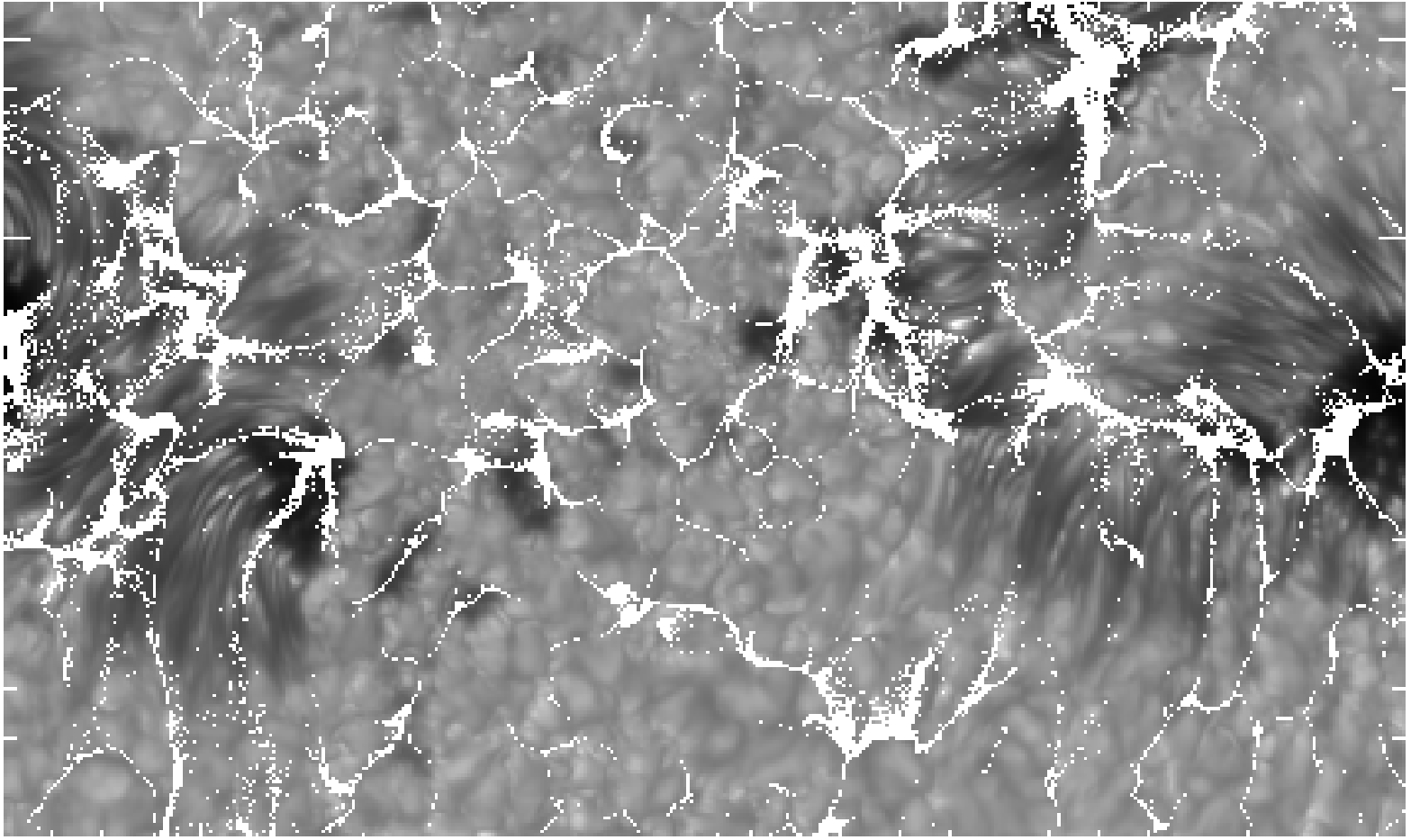} & \tiny{\sf 111} \\
\hspace{-1cm}\tiny{\sf 134} & \includegraphics[width=.48\linewidth]{fig71e.ps} &
\includegraphics[width=.48\linewidth]{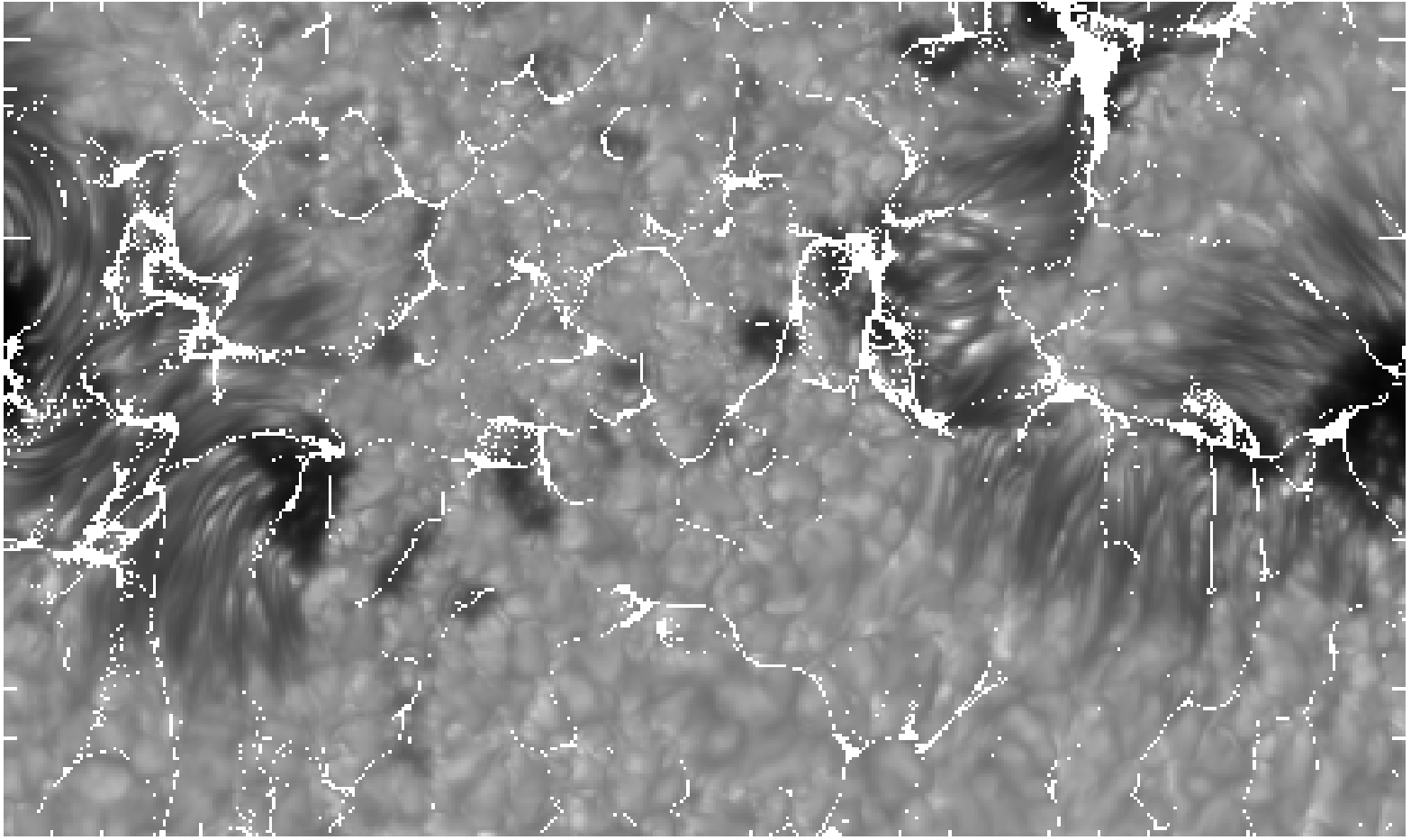} & \tiny{\sf 156}\\
\end{tabular}
\caption[\sf Evolution of artificial passive corks influenced by the averaged velocities map]{\sf Evolution of artificial passive corks moving under the influence of the velocities averaged over the whole time series (71 minutes), displayed in Figure~\ref{F:2}. In one of the \emph{lower corners} of each panel is written down the elapsed time (in minutes) from the beginning. A movie with the animation of the corks is available in the web site: \emph{http://www.iac.es/galeria/svargas/corks.html}}
\label{F:corks}
\end{figure}

Using the map of averaged velocities calculated in section~\S\ref{S:propermot} from the 428 images conforming the time series (Figure~\ref{F:2}), we can study the evolution (displacement) of passive corks homogeneously distributed in the FOV at time equal zero, one per pixel. By applying the mentioned averaged velocity field to the spread corks we can follow the displacement of every single cork in time steps of 10.0517 s. With this procedure, a sequence of images showing the locations of corks in every moment can be generated. Figure~\ref{F:corks} displays the position of corks (in \emph{white}) within the FOV after 0, 22, 45, 67, 89, 111, 134 and 156 minutes respectively. It is important to differentiate between the sequence of 428 solar images conforming the time series and the sequence of images showing the evolution of the corks.\\

This evolution can be extended to times longer than the total duration of the series (71 min). This is done by assuming the same map of velocities affecting the corks, as shown in the last 4 panels of Figure~\ref{F:corks}.\\

Some properties concerning the displacement of corks in the FOV can be inferred from Figure~\ref{F:corks}. Regions having stronger horizontal velocities are quicker emptied and correspond to the large-scale flows (moats) presented in Figure~\ref{F:3}. After 22 minutes there are evidences of mesogranulation and exploding granules affecting the displacements of the corks. Already in minute 45 the mesogranular flows dominate the FOV, these mesogranules appear as elongated structures when located inside the moat flows due to the dragging effect. After 67 minutes when the mean lifetime of mesogranules has been reached or even overtaken, the mesogranular structures tend to disappear, except in the upper central part of the FOV where the intense magnetic field confine and limits the horizontal proper motions of the plasma. A considerable number of corks are distributed along dark penumbral filaments as pointed out by \cite{marquez2006}. In the more quiet granulation regions (\emph{lower central area} of the FOV) and in the last panels of Figure~\ref{F:corks} the remaining bright structures trace a pattern resembling supergranular structures which mean lifetime is superior to the duration of the present sequence.

\noindent

\section{Conclusions and discussion}
\label{S:conclusions}

Time series of G-band and G-cont images of a sunspot group with
$\delta$-configuration (NOAA active region 10786), spanning over 71 min, were
observed and corrected for atmospheric and instrumental degradation. Proper
motions in the entire FOV have been measured in the G-band series by means of
local correlation tracking techniques. In this chapter, we concentrate in the
results obtained in the identified moat flows (or lack thereof) of several
structures within the FOV. The main conclusions from the analysis
can be summarized as follows:\\

1) We have identified inside the penumbra frontiers separating mean longitudinal flows (parallel to the direction of the filaments) at spatial resolution $\sim$1\arcsec ~ (mean velocity about  0.4 km s$^{-1}$), moving toward the umbra in the inner penumbra and toward the surrounding granulation in the outer penumbra.\\

2) We have detected strong (mean speeds of 0.67 km s$^{-1}$ with an rms of 0.32 km s$^{-1}$) outflows streaming from penumbrae radially oriented from an umbral core, the so-called sunspot moats. Our velocity values are comparable to those reported by i.e, \cite{sheeley1972}, \cite{vrabec1974}, \cite{harvey1973}, \cite{hagenaar2005} and \cite{bonet2005}, which describe the outflow velocities in the range 0.5~-~1 km s$^{-1}$. Umbral cores sides with no penumbra do not display moat flows.\\

3) Furthermore, the moats are also absent on penumbral sides \emph{parallel} to
the direction defined by the penumbral filaments. They are not found in 
directions transverse to them. A special case is sheared penumbral configurations
tangential to the umbral core. No moat flow is found there either.\\

This evidence is clearly suggestive of a link between the moat flow and flows aligned with the penumbral filaments. The possible connection with the Evershed flow is inescapable. Although it can be argued that a physical relationship between these two flows is not firmly established in this chapter, we believe the evidence is clear-cut and that a statistical analysis of a larger number of regions, as done here, can establish this connection more solidly. This is one of the goals in the next chapter, where we extend the sample of solar active regions in order to consolidate the main conclusions of the present chapter.\\

\subsection{Implications of the results}
\label{S:implications}

The results in this chapter should be put in the context of the recent findings by \citet{sainz2005}; see also \citet{ravindra2006} mentioned in chapter~\S3. These authors find that the
penumbral filaments extend beyond the photometric sunspot boundary and cross
the region dominated by the moat flow. This region is also the location where the MMF
activity is normally found. Indeed, some of these MMFs are seen by these authors
to start inside the penumbra. All these results point to an origin linking
the moat flow and MMF activity with the well-known Evershed flow. \citet{cabrera2006} also suggest that the Evershed clouds inside the penumbrae propagate to the moat, becoming MMFs once they leave the sunspot.
It is indeed somewhat paradoxical that while the final fate of the Evershed
flow has remained unknown for decades, independent explanations and physical
scenarios have been proposed for the generation of the moat flow (that starts exactly
where the Evershed flow is seen to vanish). \\

Similarly, the results obtained from local helioseismology near sunspots, in particular those
related to the f-mode \citep{gizon2000}, are also quite
relevant to the results presented here. The interpretation of these results
in terms of convective cells surrounding sunspots \citep[e.g.,][]{bovelet2003}
may be different if the Evershed flow turns out to be the major process
that injects mass into the moats surrounding sunspots.\\

%% file: chap5.tex
\chapter{\sf Moats flows surrounding sunspots}\label{cap5}
\dropping[0pt]{2}{T}his chapter is an extended and more detailed version of the paper by \cite{vargas2008} published in the \emph{The Astrophysical Journal (ApJ)}. The analysis is performed employing ground-based observations of solar active regions, and complements the ideas already presented in the previous chapter concerning the plasma motions at photospheric level.

\section{Introduction}
The sunspot moat, as previously commented in this thesis, consists of an organized horizontal radial flow
pattern surrounding the sunspots, which ends quite abruptly at a distance that can be comparable to supergranular sizes or even larger. \\

In the last chapter the moat flow in a
complex active region was studied and described in detail. As a result, some conclusions were established relating these large outflows and certain physical characteristics of the sunspots: a) No outflow was detected in the
granulation next to umbral cores that lack penumbrae, b) Outflows were only
found in the granulation regions adjacent to the penumbrae in the
direction following the penumbral filaments.  c) Granulation regions
located next to penumbral sides parallel to the direction of the
filaments show no moat flow. \\

The aim of the present chapter is to extend the study done in chapter~\S\ref{cap4} to a larger sample of cases, that is, to establish whether this moat-penumbra relation is systematically found in other 
active regions and how the granulation convective pattern surrounding sunspots behaves.\\

In doing so, we have used several time series of images observed at high-spatial resolution. A total of 7
different sunspot series have been processed and analyzed. 
The sample includes sunspots with different penumbral
configurations, varying from well-developed penumbrae to 
rudimentary penumbral morphologies.

\section{Observations and data processing}
\label{S:obs}
The observations were obtained with the \mbox{Swedish 1-m Solar Telescope}
\citep[SST,][]{scharmer2003a} on La Palma between 2003 and 2006.
All observations benefited from the use of the SST adaptive optics
system \citep{SSTAO} that minimized the degrading effects of seeing. 
Image post-processing techniques were applied to increase the
homogeneity in the quality of the time series and to enhance the image
quality over the whole FOV. 
For the 2003 data set we applied Multi-Frame Blind Deconvolution
(MFBD) using the implementation developed by \citet{lofdahl02MFBD}.
The MFBD code was succeeded by the Multi-Object Multi-Frame Blind
Deconvolution code (MOMFBD, \citet{momfbd2005}) which
employs multiple objects and phase-diversity (see section~1.5 for more details on this method).
MOMFBD was applied to the data sets after 2003. 
For all the time series we present in this chapter, the seeing
conditions were generally very good and sometimes excellent, being able to record for more than 40 min in all cases.
A large fraction of the restored images in the different time series
approached the diffraction limit of the telescope. \\

After image restoration, we applied standard techniques to the time
series including correction for the diurnal field rotation, rigid
alignment and de-stretching to correct for seeing-induced image
warping. A p-modes filtering was also applied to the series (phase-velocity threshold of 4 km\,s$^{-1}$).  \\

\begin{figure}
\centering
\begin{tabular}{ccc}
\textsf{S1} & \textsf{S2} & \textsf{S3}  \\
\includegraphics[width=0.3\linewidth]{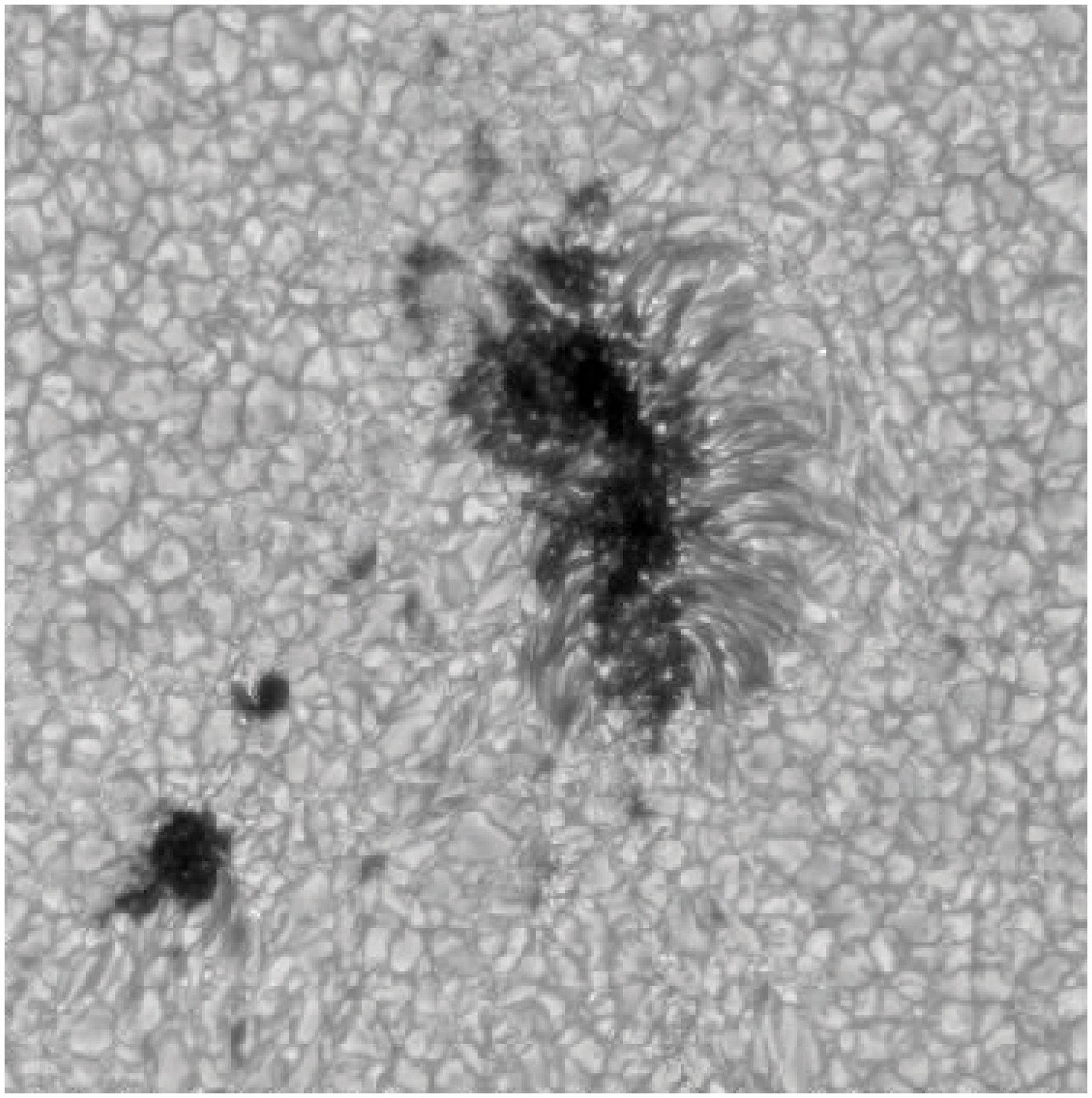} &
\includegraphics[width=0.3\linewidth]{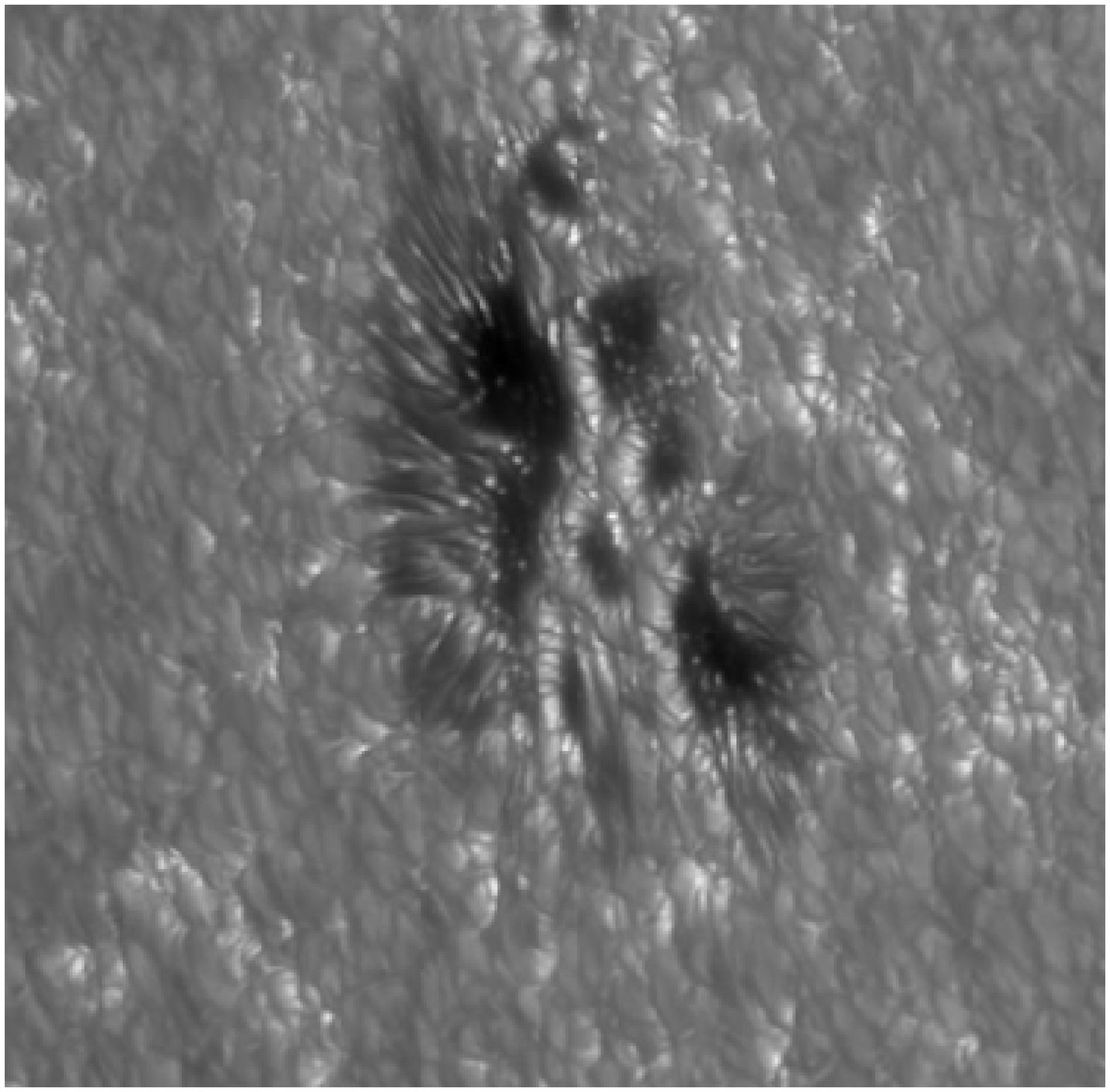} & 
\includegraphics[width=0.3\linewidth]{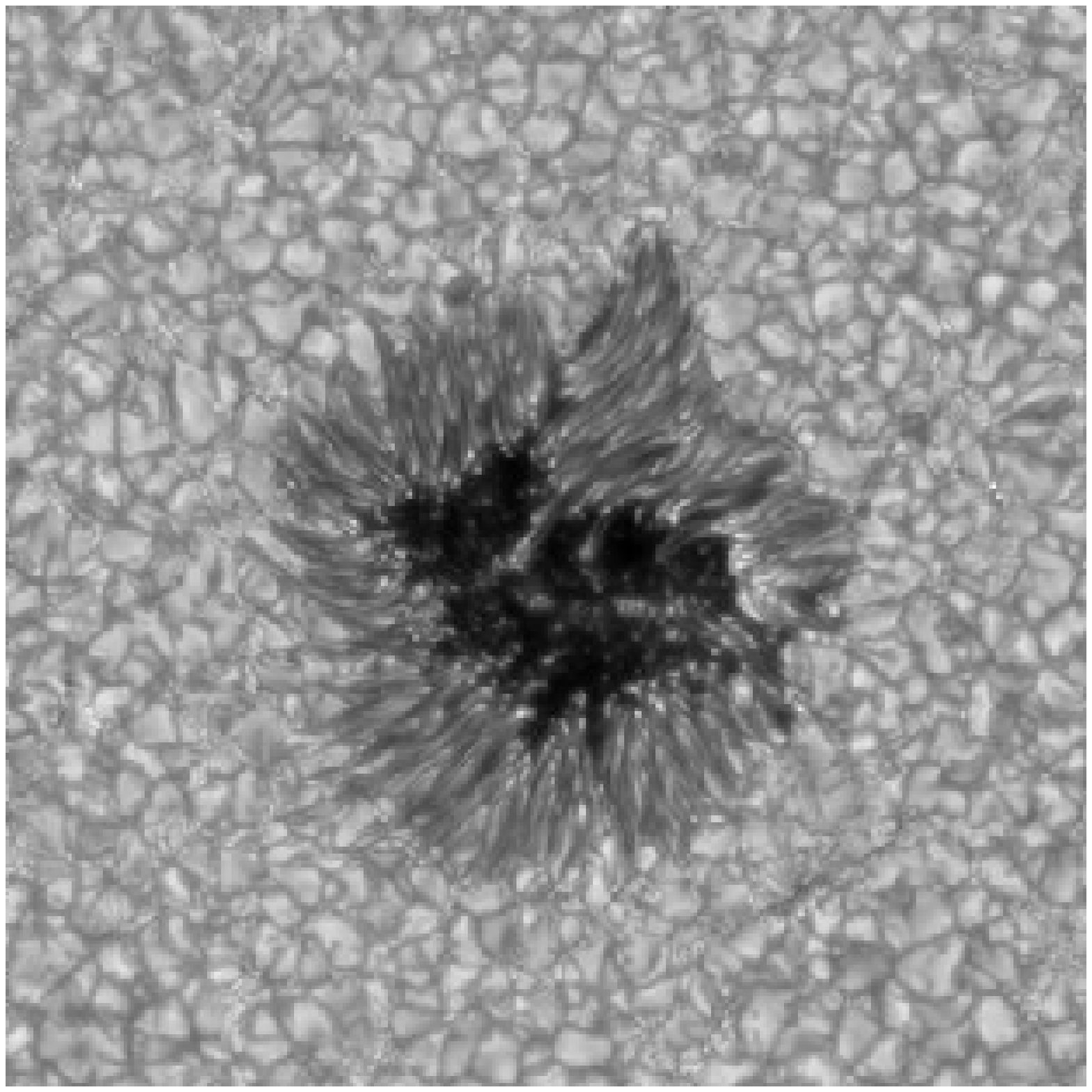} \\\\
\textsf{S4} & \textsf{S5} & \textsf{S6}  \\
\includegraphics[width=0.3\linewidth]{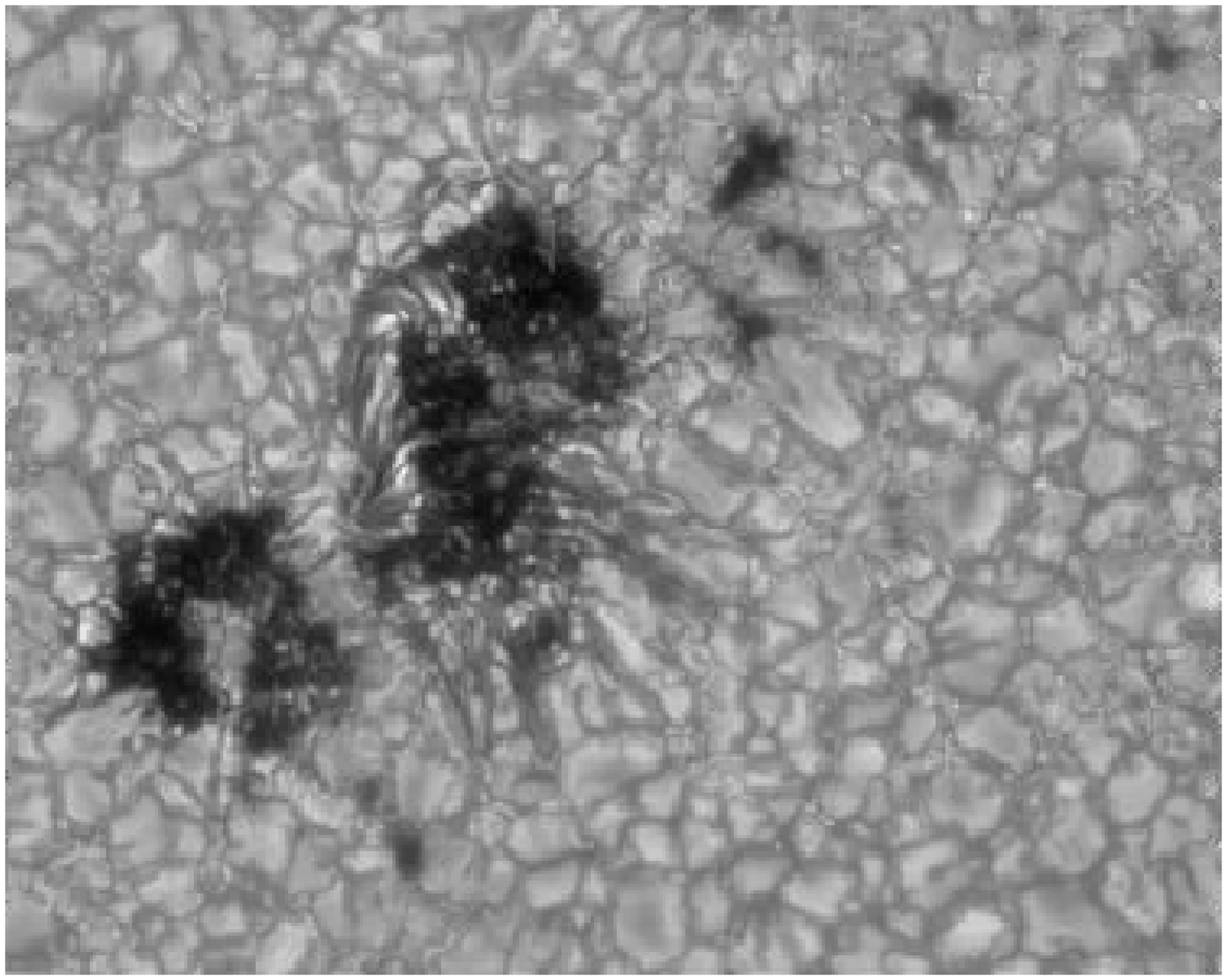} & 
\includegraphics[width=0.3\linewidth]{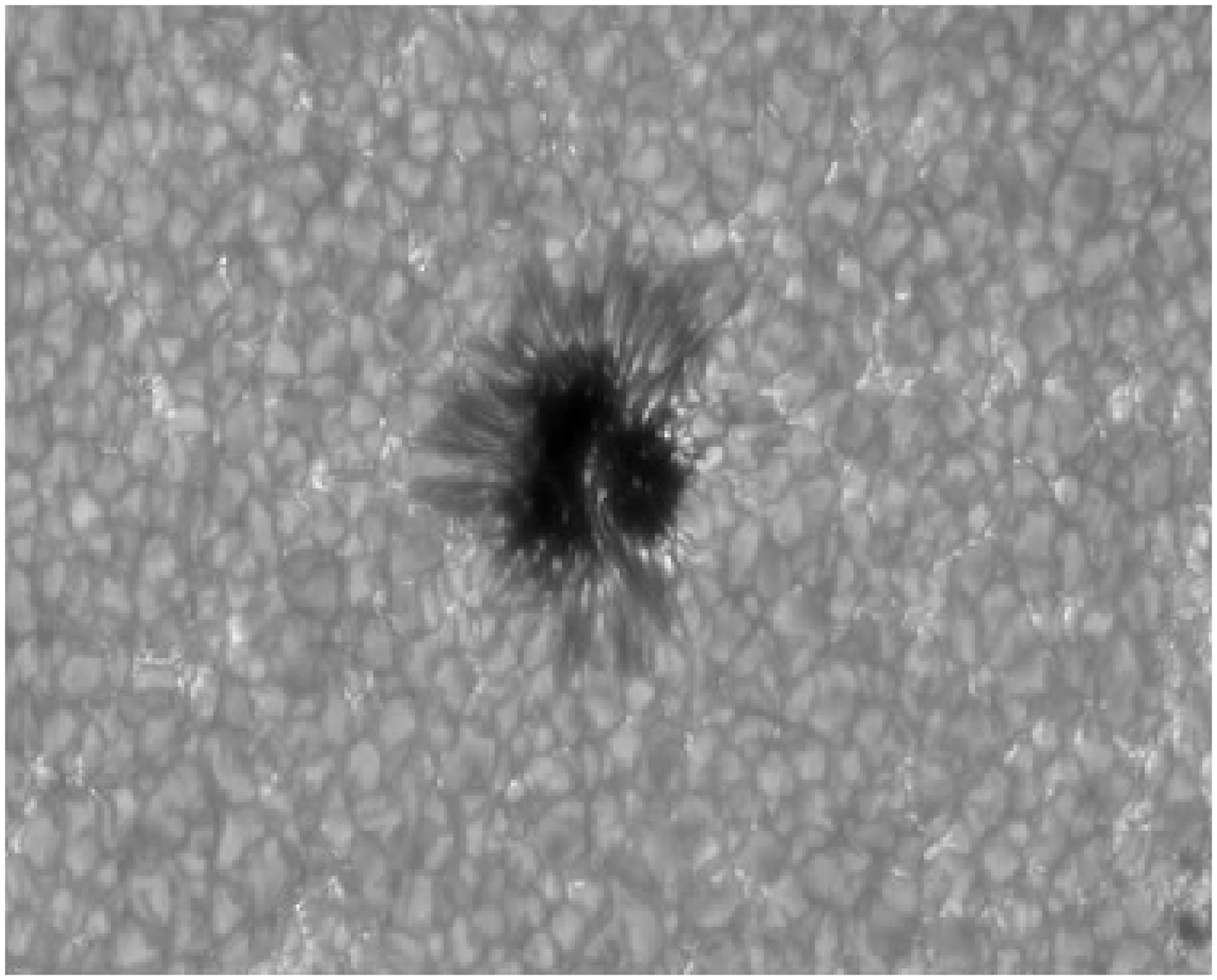} & 
\includegraphics[width=0.3\linewidth]{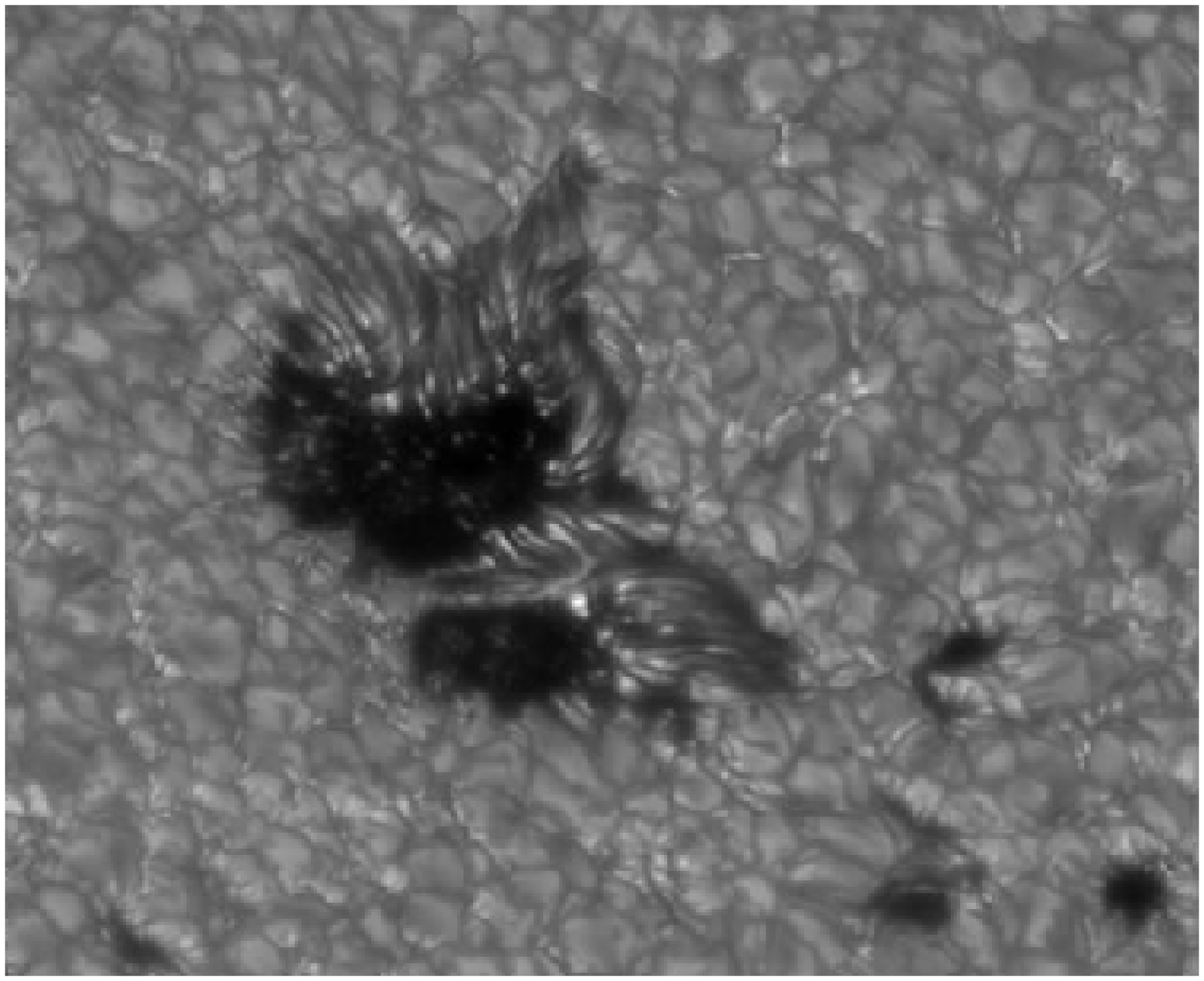}  \\\\
\textsf{S7} & \\
\includegraphics[width=0.3\linewidth]{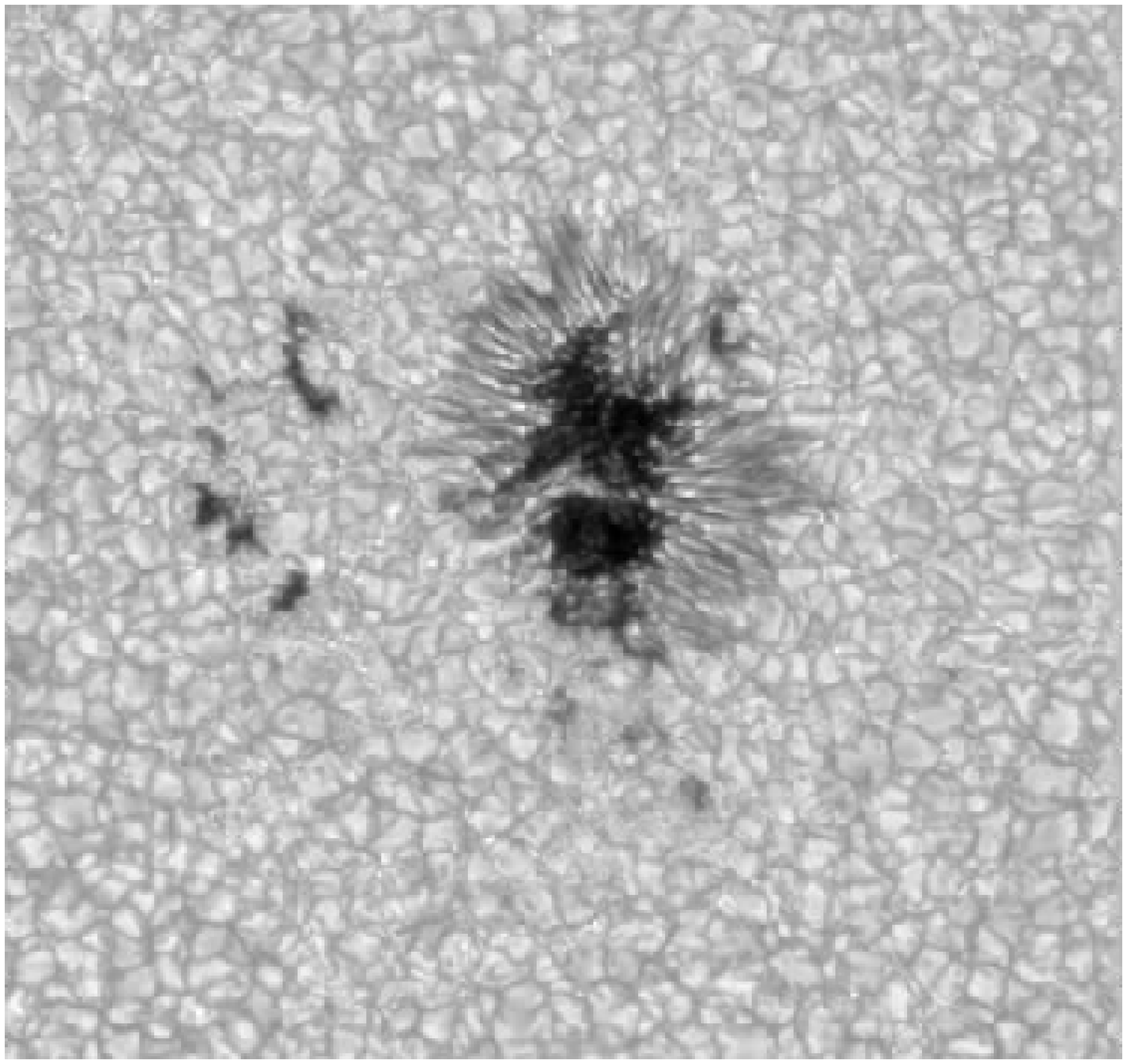} &
\includegraphics[width=0.33\linewidth]{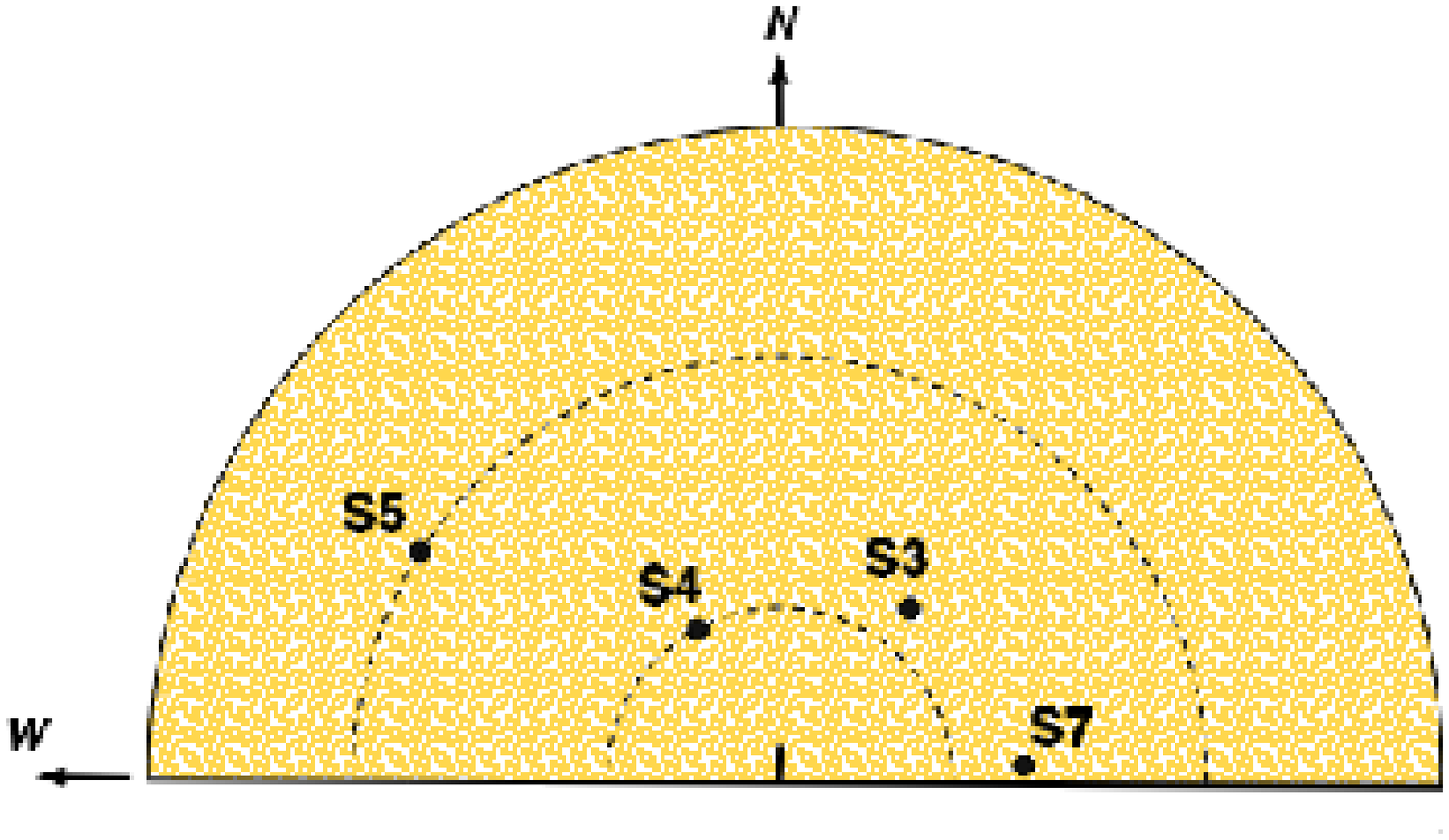} &
\includegraphics[width=0.33\linewidth]{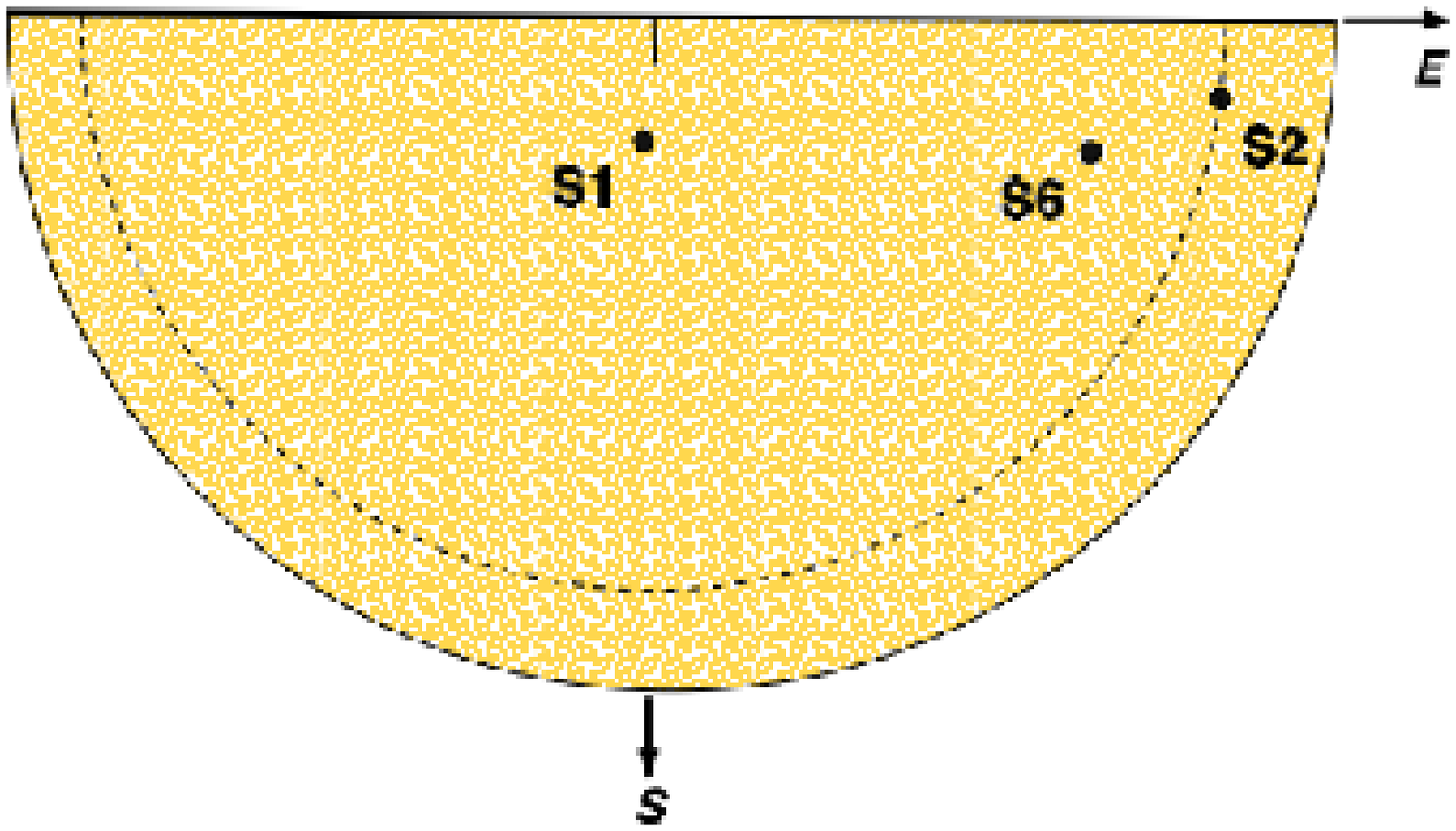}
\end{tabular}
\caption[\sf Solar active regions studied]{\sf Solar active regions, from sunspot S1 to S7, showing the different penumbral configurations analyzed in this chapter. The spatial scale is not the same in all the sunspots. The last two panels show a representation of the solar hemispheres including the positions of the observed sunspots. The \emph{dashed arcs} determine the locations for $\mu =$0.88 ( \emph{the outer one in the northern hemisphere}), $\mu =$0.99  (\emph{the inner in the northern hemisphere}),  and $\mu =$0.55 (\emph{in the southern hemisphere}) respectively.}
\label{sunspots}
\end{figure}

Table~\ref{table1} presents details on the different active region
targets which are also shown in Figure~\ref{sunspots}. The observations were performed in regions at different positions on the solar disc as listed in the columns 4 and 5 in Table~\ref{table1}, including one of them (S2) very close to the limb. Table~\ref{table2} gives some details on the restored time series. The duration of the time series is in all cases longer than 40 minutes so that we can use them to get a reliable average of the displacements of the structures.
Below we provide more detailed information on the different data sets, including the restoration process applied in each case.

\begin{table*}[]
\sffamily
\caption[\sf Sunspots sample]{\sf Sunspots Sample}
\centering
\begin{tabular}{cccccc}
\hline
Name & Active region & Observing date & Heliographic coordinates & $\mu$ \\
     &      NOAA     &&& \\
\hline
S1 & 10440 & 2003, Aug 18 & (S10, W1) & 0.96 \\
S2 & 10608 & 2004, May 10 & (S5, E57) & 0.55 \\
S3 & 10662 & 2004, Aug 20 & (N13, E7) & 0.99 \\
S4 & 10662 & 2004, Aug 21 & (N11, W5) & 0.99 \\
S5 & 10789 & 2005, Jul 13 & (N17, W32) & 0.88 \\
S6 & 10813 & 2005, Oct 04 & (S7, E37) & 0.78 \\
S7 & 10893 & 2006, Jun 10 & (N1, E17) & 0.95 \\
\hline
\end{tabular}
\label{table1}
\end{table*}

\subsection{S1: AR440, 22-Aug-2003}
An interference filter centered at the continuum at 436.4~nm (FWHM
1.1~nm) was used. In the remainder, this filter will be referred to as the
``G-cont'' filter.
Every 25~s, the three highest contrast images were selected and stored to
disk. 
These three images were used for MFBD processing. This sunspot is shown in Figure~\ref{sunspots} (\textsf{S1}).

\subsection{S2: AR608, 10-May-2004}
An interference filter centered at the G-band at 430.5~nm (FWHM
1.2~nm) was used. In the remainder, this filter will be referred to as
``G-band'' filter. 
This observation involved two cameras set-up as a phase-diversity pair:
one in focus and one slightly out of focus. 
20 exposures from each camera, acquired over the course of 7.5
seconds, were used for MOMFBD processing.
During processing of the time series, a few bad quality images,
suffering from too much blurring, were dropped.
The exposure time was 11~ms. This sunspot is shown in Figure~\ref{sunspots} (\textsf{S2}).

\subsection{S3: AR662, 20-Aug-2004}
This time series was observed with the G-cont filter. 
The MOMFBD processing was performed using 10 exposures (acquired during 7 s)
from a phase-diversity pair of cameras. 
For the central part of the FOV, centered on the sunspot, the MOMFBD
processing was extended to involve a G-band phase-diversity pair of
cameras. 
After MOMFBD restoration, the best quality image from 3 subsequent
images was selected for further processing of the time series.
The exposure time was 11.3~ms. This sunspot is shown in Figure~\ref{sunspots} (\textsf{S3}).

\subsection{S4: AR662, 21-Aug-2004}
This time series was processed in a manner similar to S3. After MOMFBD
restorations, the worst 14\% of the images were dropped from the time series. This sunspot is shown in Figure~\ref{sunspots} (\textsf{S4}). 

\subsection{S5: AR789, 13-Jul-2005}
For this time series, MFBD processing was applied to 18 subsequent
G-band images, acquired during 10~s. 
Images that were too blurred and when the AO system was not actively
compensating, were dropped from the MFBD sets. This sunspot is shown in Figure~\ref{sunspots} (\textsf{S5}).

\subsection{S6: AR813, 4-Oct-2005}
This time series was observed with an interference filter centered on
the spectral region between Ca~K and Ca~H at 395.4~nm (FWHM 1.0~nm).
For the MOMFBD processing, 20 exposures (acquired during 8.3~s) from a
phase-diversity pair of cameras were used.
In addition, simultaneous exposures from two other cameras were used
for the restoration: one equipped with a narrow-band filter centered
in the Ca~H wing at 396.5~nm and one equipped with a narrow-band
filter centered on the Ca~H core (396.8~nm).
The exposure time was 11~ms. This sunspot is shown in Figure~\ref{sunspots} (\textsf{S6}).

\subsection{S7: AR893, 10-Jun-2006}
This time series was observed with an interference filter centered at
630.2~nm (FWHM 0.8~nm). This is the pre-filter for the Lockheed SOUP
filter. 
MOMFBD processing was applied to 400 exposures obtained during 19~s. 
The restorations included images from a phase-diversity pair of
cameras with the wide-band filter, and narrow-band SOUP exposures in
the Fe~I~630.2~nm spectral line. This sunspot is shown in Figure~\ref{sunspots} (\textsf{S7}).

\begin{table}[]
\sffamily
\caption[\sf Restored sunspot time series]{\sf Restored sunspots time series}
\centering
\begin{tabular}{cccc}
\hline
Name & N. images & Cadence [s] & Duration [min:s] \\
\hline
S1 .... & 128 & 24.7 & 52:47 \\
S2 .... & 376 & 7.5 & 47:10 \\
S3 .... & 144 & 19.8 & 47:20\\
S4 .... & 344 & 8.0 & 45:57\\
S5 .... & 240 & 10.1 & 40:12 \\
S6 .... & 556 & 8.7 & 80:30 \\
S7 .... & 124 & 19.7 & 40:26\\
\hline
\end{tabular}
\label{table2}
\end{table}

\section{Data analysis and results}
\label{S:analysis}

The high quality, stability and long duration of the restored sunspot series, enable us to study the
dynamics of the plasma around the sunspots. We have computed
proper-motion velocity fields (horizontal velocities) employing the LCT technique described in the previous chapter.
A Gaussian tracking window of FWHM 1$\farcs$0 suitable for mainly tracking granules has been used for all the series. The method produces a sequence of map-pairs each one describing, within a short time interval (tens of milliseconds), the horizontal velocity components, $v_x$ and $v_y$, along the FOV. 
These maps of components have been averaged over 5 and 10 min periods, and also over the respective total duration of every sunspot series (more than 40 min in all cases). From the averaged maps of velocity components, the distribution of velocity magnitudes over the FOV (flow map) has been easily derived.\\

As described in section~\S4.3.4 , it is well documented in the literature \citep{november1988} that LCT-techniques in general produce some systematic errors in the determination of displacements. Their significance depends also on several factors like the width of the tracking window in relation to the size of the elements used as motion tracers, and the presence of large intensity gradients in the images (see section~\S4.3.4 for more details). Even so, this drawback does not change the main conclusions of the present work since we are not interested in fixing absolute values of velocities but rather in the detection of large-scale regularly organized flows around the sunspots in comparison with those typically found in a  normal granulation field. In both cases we use the solar granules as the tracers of motion and the same size for the tracking window so that in case of some bias it will affect to both velocity fields similarly.

\subsection{De-projection of horizontal velocities}
\label{S:proj} 

The sample of sunspots studied in this paper includes a variety of
heliocentric positions on the solar disc as shown in
Table~\ref{table1}  (with $\mu=\cos \theta$, where $\theta$ is the
heliocentric angle). Away from the solar disc center,  the measured proper
motions are in fact projections of the real horizontal velocities in the
sunspot plane onto the plane perpendicular to the LOS. Thus, to evaluate the real horizontal velocities we have de-projected the velocities obtained from the LCT technique to a plane tangent to the solar surface at the center of the sunspot. \\

\begin{figure}
\centering
\begin{tabular}{l}
\includegraphics[width=.7\linewidth]{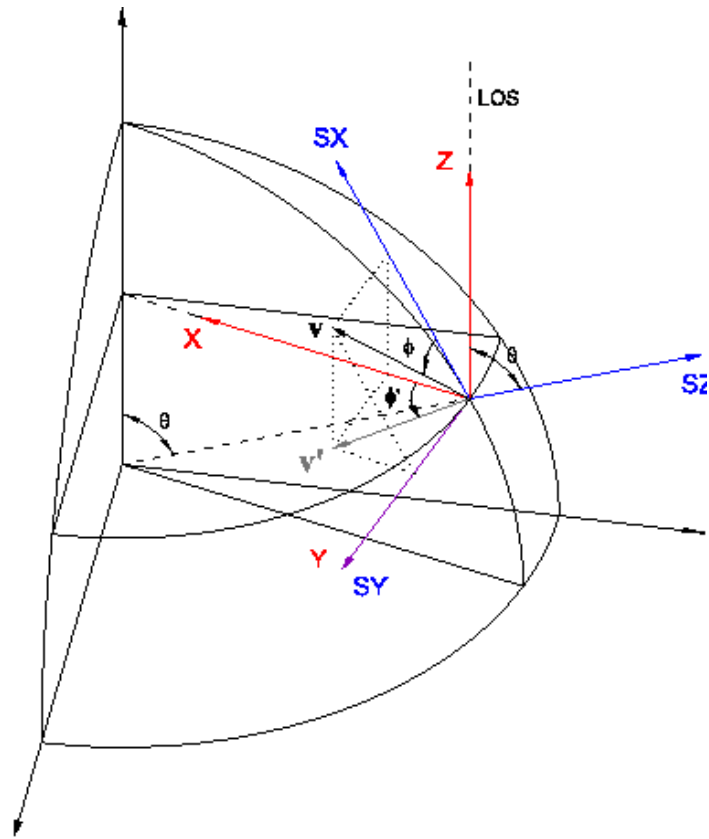}
\end{tabular}
\caption[\sf Sketch showing two orthogonal coordinate systems for the projection analysis of the horizontal velocities]{\sf Sketch showing two orthogonal coordinate systems for the projection analysis of the horizontal velocities: 1) The Sunspot System (SX,SY,SZ) with the Z-axis perpendicular to the solar surface at the sunspot location, and 2) the Observing System (X,Y,Z), where the observations are described, with the Z-axis coinciding with the LOS.}
\label{sketch}
\end{figure}

To that aim, let us consider in Figure~\ref{sketch} the orthogonal
coordinate system SX,SY,SZ (the Sunspot System, SS) with the SZ-axis
perpendicular to the plane tangent to the solar surface at the sunspot
location and the SX-axis tangent to the meridian of the solar sphere
at this location. Also included in the drawing is the coordinate system X,Y,Z (the Observing System, OS) 
with a common origin, the Z-axis coinciding with the LOS and the X-axis pointing to the axis crossing the Sun center in a direction parallel to the LOS. Note that the axis,
X, SX, Z, and SZ are coplanar and SY and Y are collinear. The real
horizontal velocities, $\bf v$($v,\phi)$, are contained in the (SX,
SY) plane whereas we observe their projections {\bf $\bf
  v'$}($v',\phi')$, in the (X,Y) plane.\\

Calculations based on Figure~\ref{sketch} lead to the following relationships (equations
\ref{eq:1} and \ref{eq:2}) between the magnitudes $v$ and $v' $, and
the azimuths $\phi$ and $\phi' $, that will allow us to de-project our
observations.

\begin{equation}
v'~^2=(v\sin \phi)^2 + (v \cos \phi  \cos \theta)^2 = v^2 (\sin^2\phi + \cos^2\phi ~ \cos^2 \theta).
\label{eq:1}
\end{equation}

\begin{equation}
\tan \phi' = \frac{v \sin \phi}{v \cos \phi \cos \theta} = \frac{\tan \phi}{\cos \theta}.
\label{eq:2}
\end{equation}

Since the X- and Y- axes do not necessarily coincide with our
natural reference system (i.e.\ the edges of the CCD frame), the
measurement of the projected azimuths, $\phi '$, requires the
knowledge of the orientation of our FOV with respect to the X-axis 
that points to the solar disc center.

\subsection{Masking moats}
\label{masking}

As mentioned above, for every sunspot series, the map of horizontal velocities averaged over the whole time series is evaluated. In order to coherently detect the moats and compare their statistics (section~\S\ref{S:statist}) in all active regions of our sample, we have de-projected the observed velocity vectors onto the sunspot plane (SX,SY) as described in section~\S\ref{S:proj}.\\

Overlaying the observed images, we construct maps showing in the granulation field only those projected velocity vectors having de-projected magnitudes above a certain threshold. In these maps, extensive organized outflows coming out from the sunspots can easily be identified so that we can outline masks delimiting in a qualitative way the area of the moats for the purpose of statistical calculations and graphical representation.\\
 
We have fixed the mentioned threshold to $\sim$~0.3~km\,s$^{-1}$ as the best compromise to define the limits of the moats. This value should not be understood as an absolute (universal) velocity threshold to define moats in general. As we have mentioned before, the particular method used here to measure the displacements may underestimate the velocity magnitudes. Moreover, outside the moat one can easily find velocities larger than that. The point is that this particular threshold makes in our case more evident the existence of a well organized radial outflows around the sunspots in comparison with the rest of the FOV.  Lower values (even zero) for the threshold do not extend the areas of organized flows around the sunspots but produce maps with a very dense and noisy (exploding granules everywhere) representation of arrows where the outline of the frontiers of the moats results more difficult.\\

\begin{figure}
\centering
\begin{tabular}{cc}
\hspace{-5mm}\textsf{a})\hspace{-5mm}\includegraphics[width=0.9\linewidth]{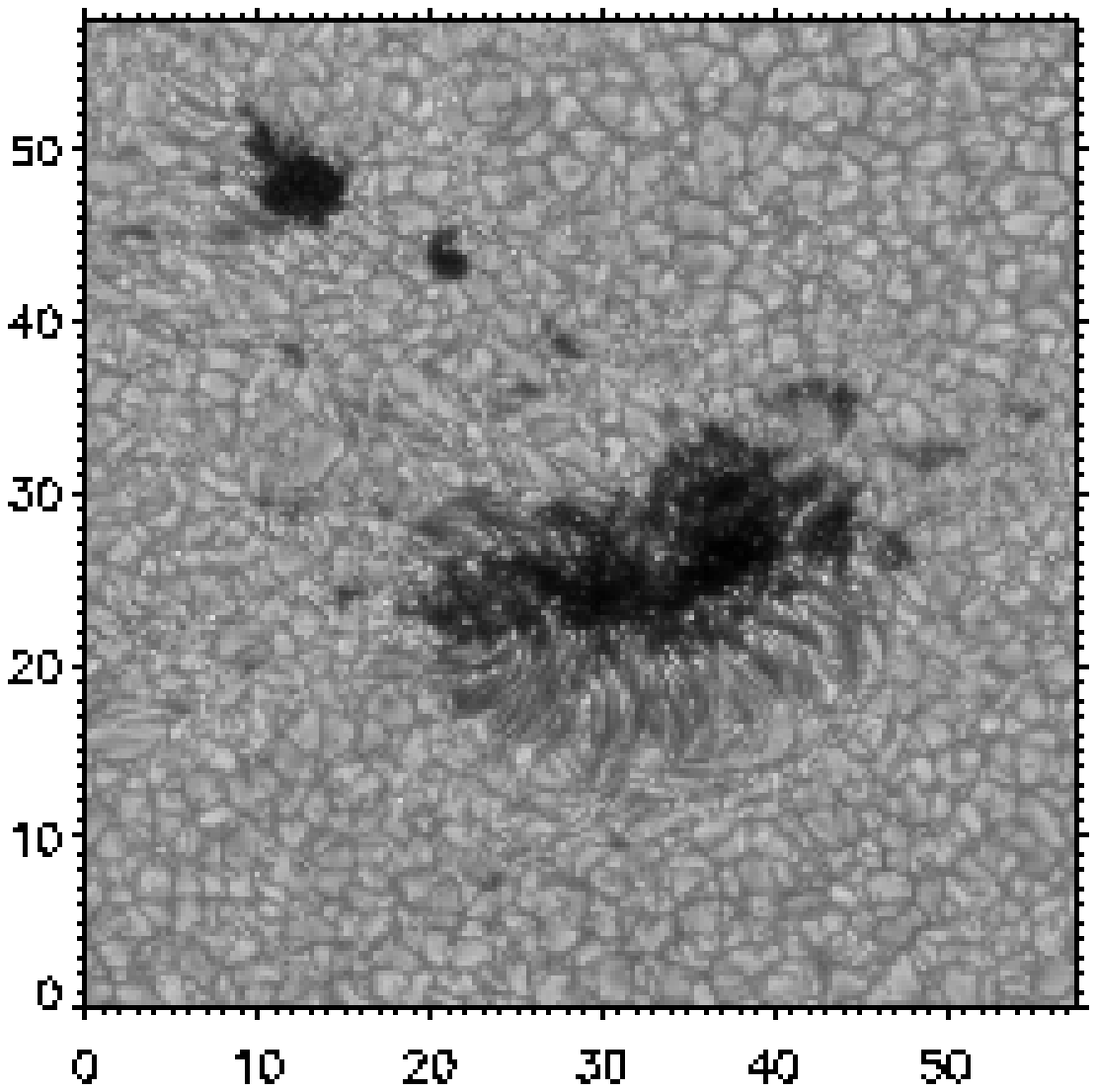} &
\hspace{-55mm}\textsf{b})\hspace{-5mm}\includegraphics[width=0.9\linewidth]{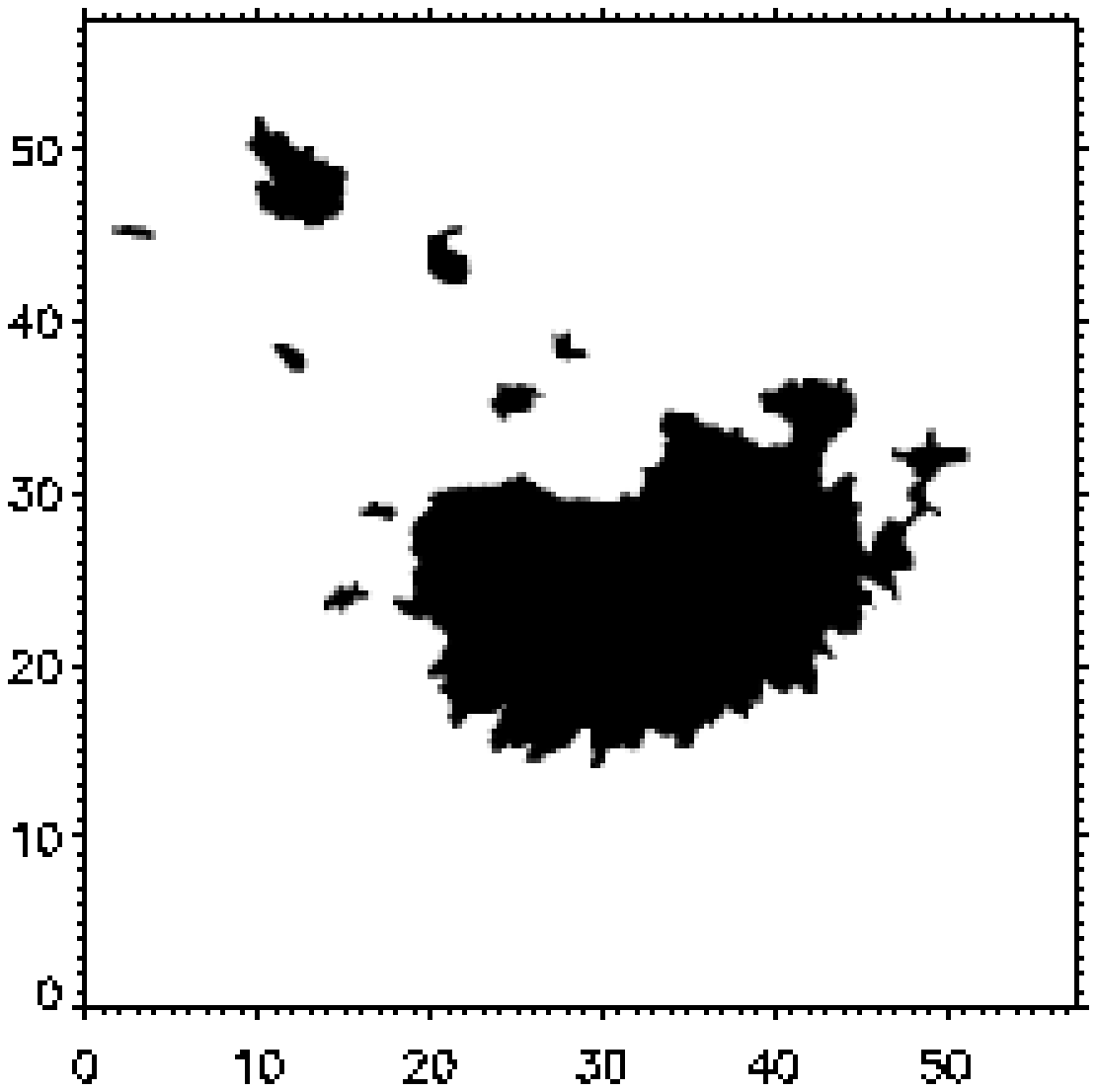} \\
\hspace{-5mm}\textsf{c})\hspace{-5mm}\includegraphics[width=0.9\linewidth]{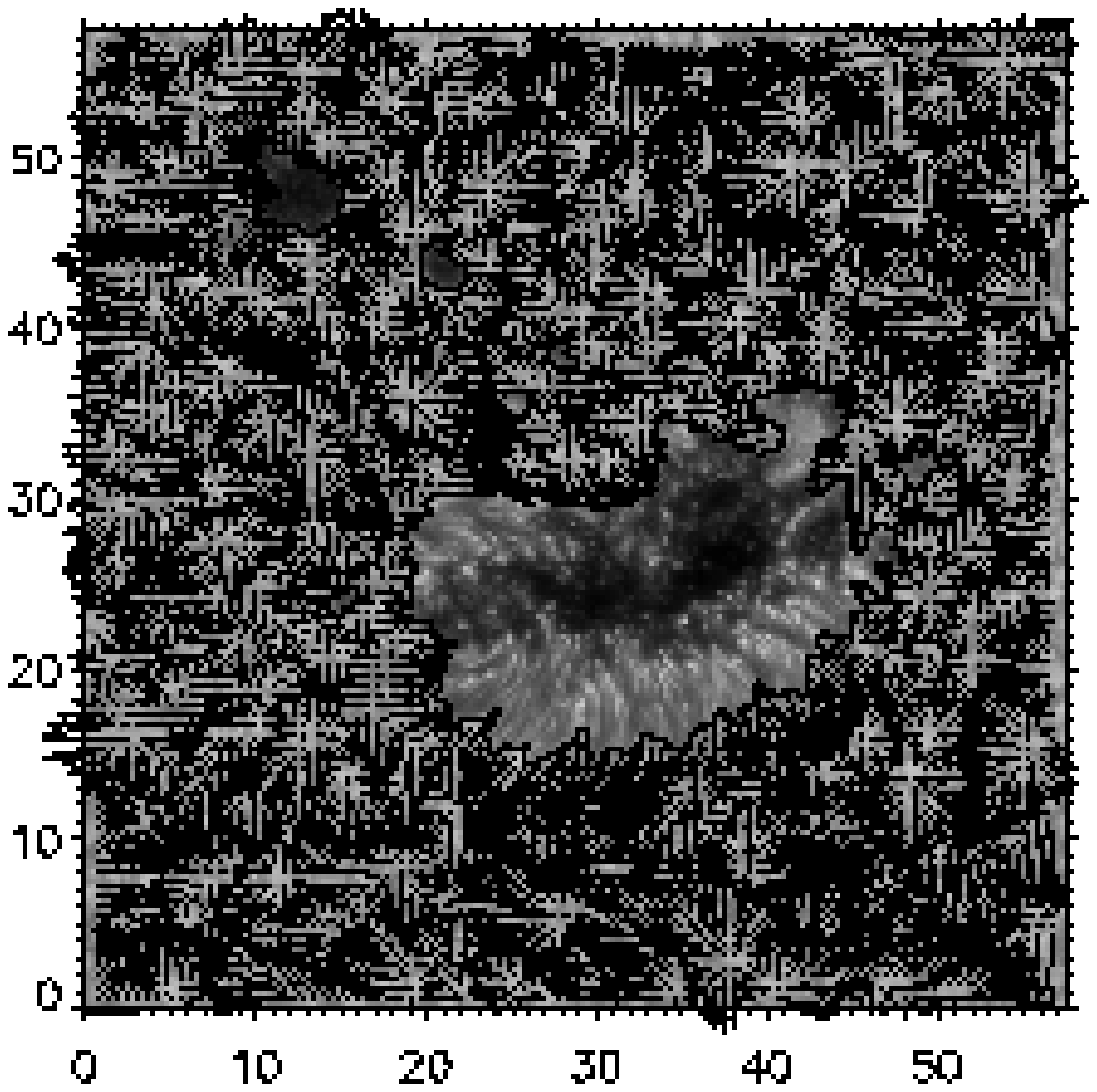} &  
\hspace{-55mm}\textsf{d})\hspace{-5mm}\includegraphics[width=0.9\linewidth]{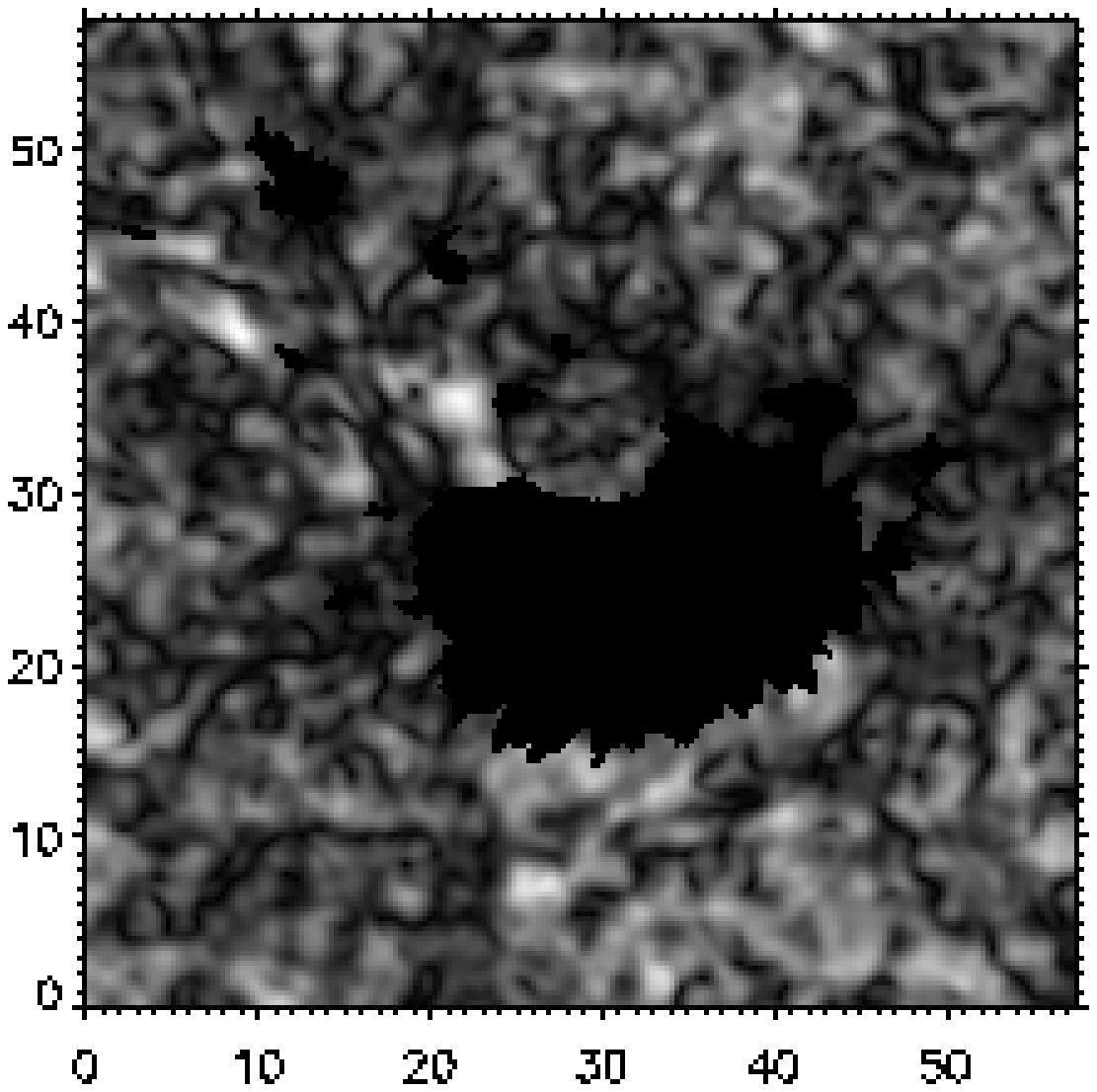} 
\end{tabular}
\caption[\sf Steps of the process applied to generate the flow map with the moat flows]{\sf Steps of the process applied to generate the flow map showing the large horizontal outflows surrounding the sunspots. \textsf{a)} FOV. \textsf{b)} Sunspot mask. \textsf{c)} Flow map. \textsf{d)} De-projected velocities magnitudes. The coordinates are expressed in arc seconds.}
\label{steps}
\end{figure}

\begin{figure}
\centering
\begin{tabular}{cc}
\hspace{-5mm}\textsf{e})\hspace{-5mm}\includegraphics[width=0.9\linewidth]{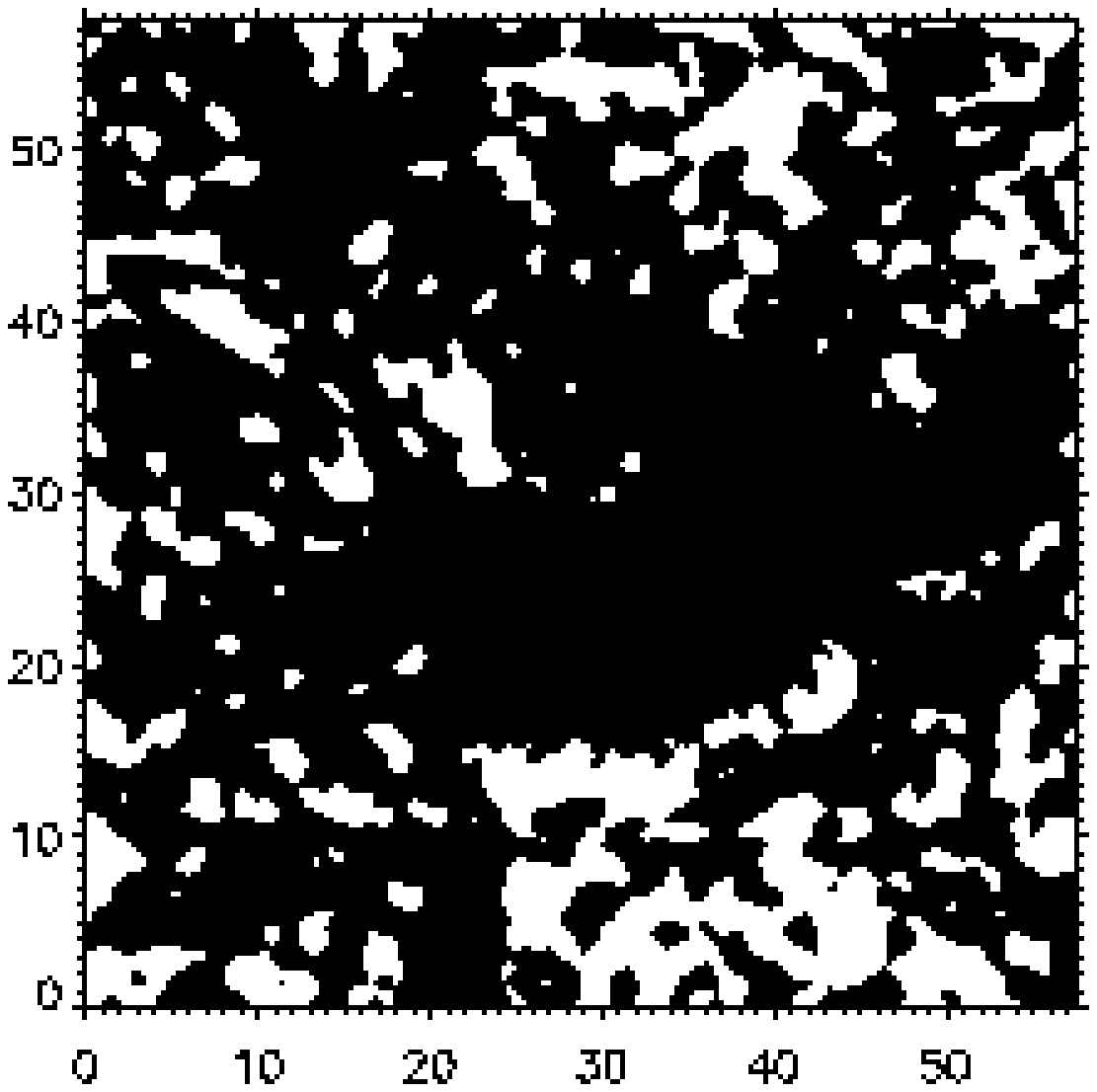} &
\hspace{-55mm}\textsf{f})\hspace{-5mm}\includegraphics[width=0.9\linewidth]{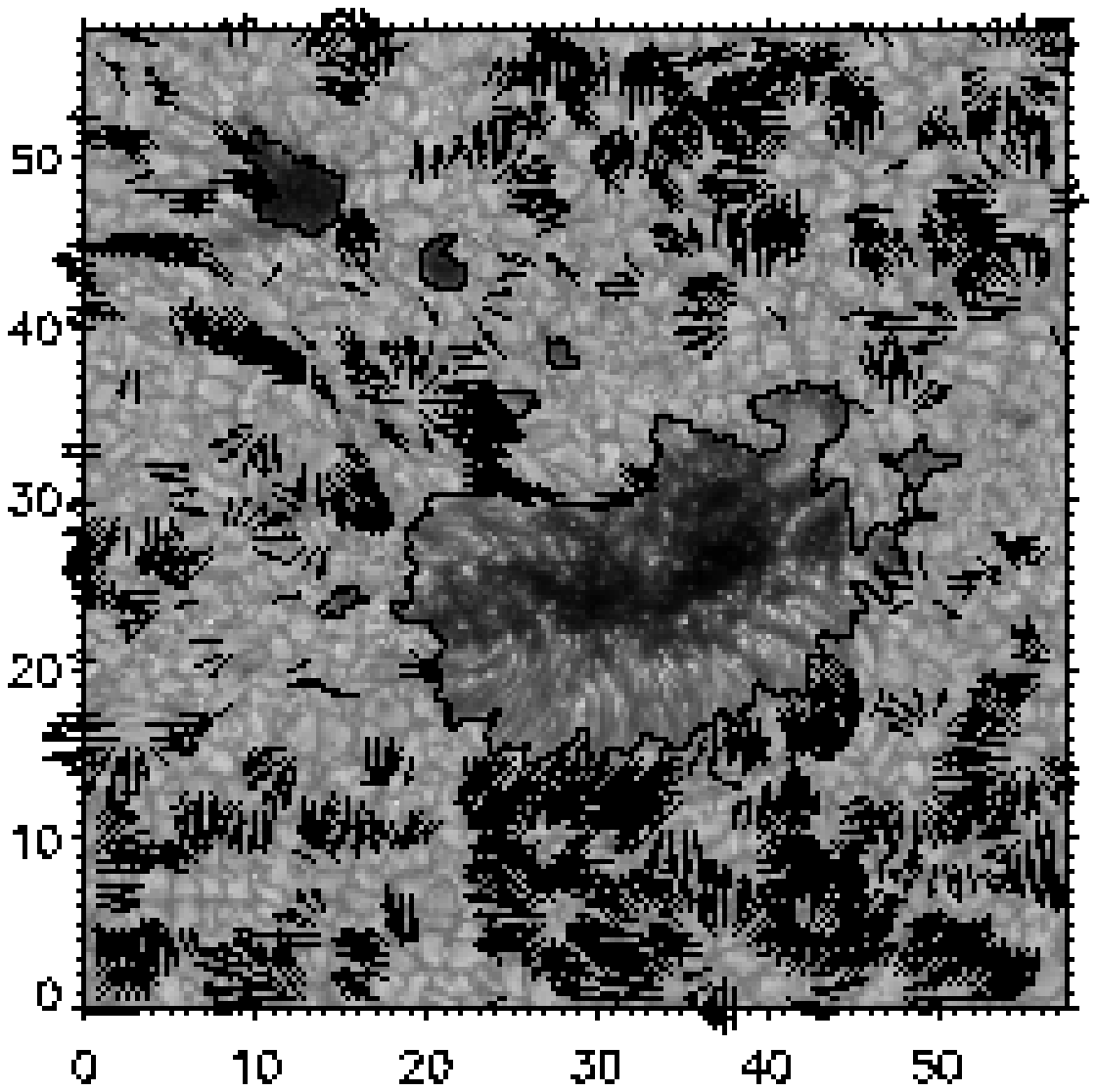}\\
\hspace{-5mm}\textsf{g})\hspace{-5mm}\includegraphics[width=0.9\linewidth]{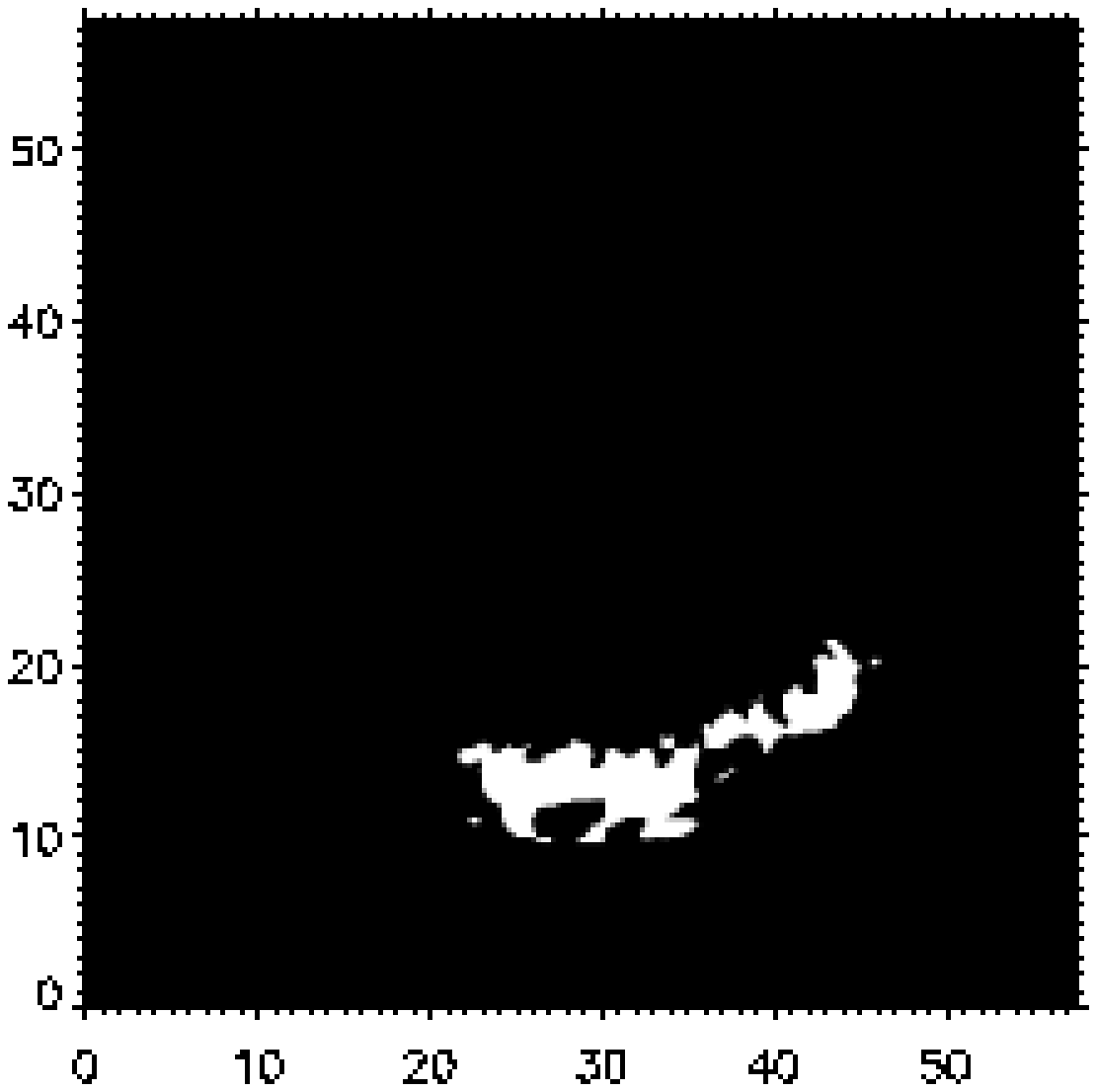} &
\hspace{-55mm}\textsf{h})\hspace{-5mm}\includegraphics[width=0.9\linewidth]{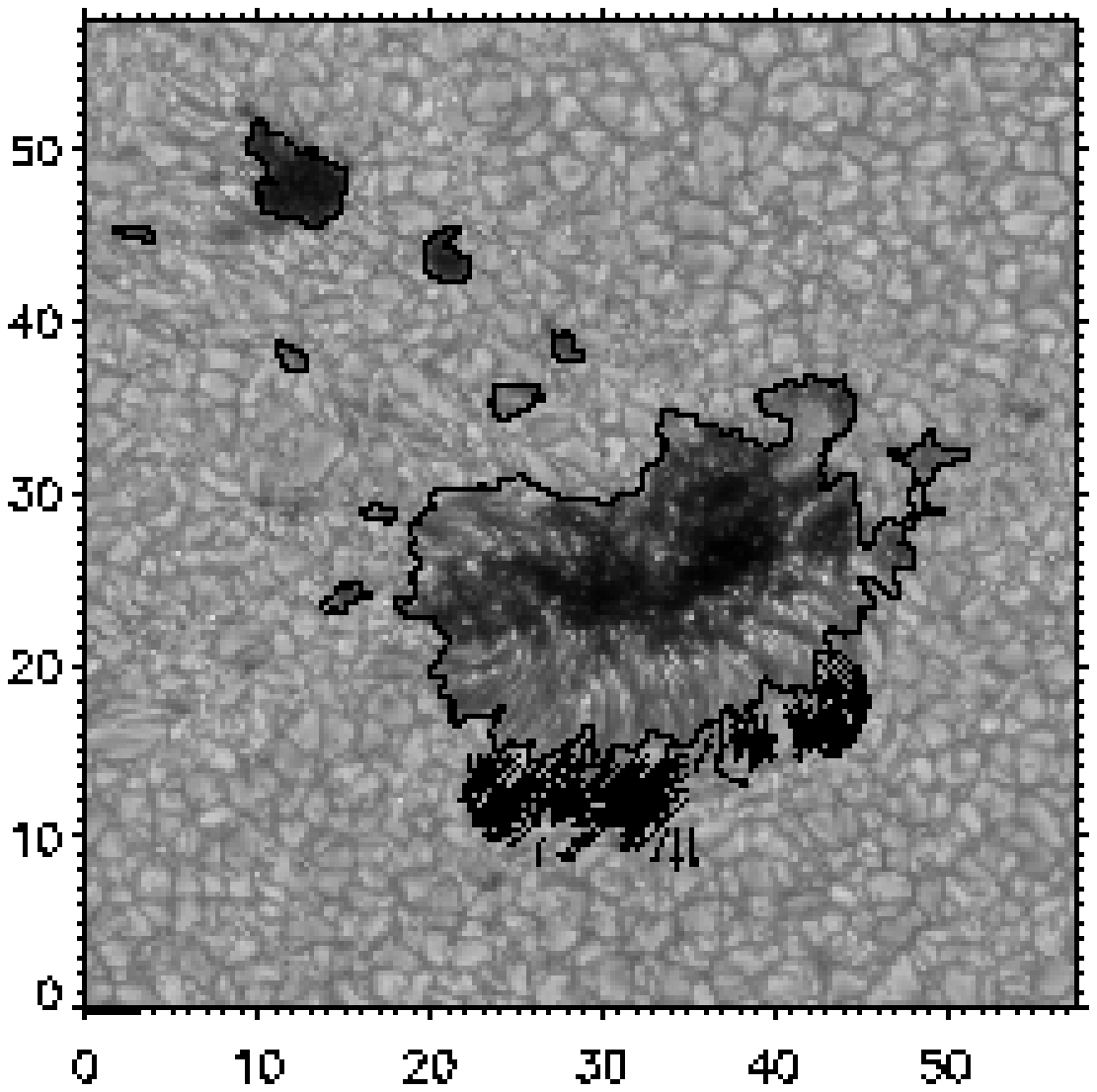}\\\\
\end{tabular}
\flushleft {\footnotesize F}{\scriptsize IGURE} {\footnotesize\ref{steps} (\sf \emph{cont.}) ---  \textsf{e)} \sf Mask of de-projected velocities with magnitudes over a certain threshold. \textsf{f)} \sf Flow map of projected velocities with de-projected magnitudes over the threshold. \textsf{g)} \sf Moat mask. \textsf{h)} \sf Flow map of moat flows. The coordinates are expressed in arc seconds. The lenght of the black bar at coordinates (0,0) in panels f) and h) represents 1.5 km\,s$^{-1}$ for the projected velocities.}
\end{figure}

The whole masking process for one of the time series is illustrated throughout the panels in Figure~\ref{steps} and consist of:

\begin{itemize}
\item \textsf{a}) Selecting the FOV we want to analyze. Hereafter we will use the spatial-scale in arc seconds.
\item \textsf{b}) Creating, by thresholding, a binary mask for the sunspots (including umbra and penumbrae) and pores.
\item \textsf{c}) Computing the proper motions of structures by applying LCT and plotting the flow map for the whole FOV in granulation regions.
\item \textsf{d})  De-projecting the velocities according to the equations \ref{eq:1} y \ref{eq:2}. The de-projected magnitudes are represented in gray-scale. Bright/dark areas correspond to large/small velocity magnitudes. The velocity units are in km\,s$^{-1}$.
\item \textsf{e}) Creating a binary mask for the de-projected velocity magnitudes according to the following rule: pixels with de-projected velocity magnitudes greater than 0.3~km\,s$^{-1}$  are set to "1" and the rest to "0".
\item \textsf{f}) Applying the mask in \textsf{e}) to the flow map in panel \textsf{c}), so that we get only those projected velocities that have a de-proyected magnitude above the threshold of 0.3~km\,s$^{-1}$. Over these new flow map we identified visually the flows coming out from the sunspot that correspond to the moat flows described before in this work. These moat flows are easily recognized surrounding the sunspot as large-scale outflows and hence they can be outlined without difficulty.
\item \textsf{g}) Creating a binary mask of moats from the mask generated in \textsf{e}), with the help of the moat boundaries outlined from the visual inspection in \textsf{f}). 
\item \textsf{h}) Plotting the final flow map showing the moat flows. We apply the moat mask generated in \textsf{g}) onto the flow map of velocities in panel \textsf{c}).  
\end{itemize}

\subsection{Moat flows around the sunspots}
\label{moataround}

Our sunspots sample includes different penumbral configurations as a key factor to establish the moat-penumbra relation in a more robust way. Figures~\ref{S1} to \ref{S6} show within the moats, only those velocity vectors with de-projected magnitudes above 0.3~km\,s$^{-1}$ after applying the whole masking procedure described above. In Figure~\ref{S7} we extend this representation to the entire granulation field showing that large velocities are also present outside the moat. These velocities are generally grouped and associated with exploding granules.\\

First we focus on granulation regions that display moat flows. Close inspection of Figures~\ref{S1} to \ref{S7}, and the upper panel of Figure \ref{S3}  reveals that the velocity vectors in the moats are oriented following the direction of the penumbral filaments. {\it We observe that in all cases the moat flow direction lines up with penumbral filaments that are oriented radially with respect to the sunspot center.} Nevertheless, the completeness of the velocity vectors (density of arrows) in the flow maps depends on the threshold previously imposed, as expected. This can be seen in Figure~\ref{S5} where the left-side penumbra does not seem to be strictly associated with large flows in the
granulation region. However, this penumbral part has in fact an
associated moat flow similar to the other penumbral regions that is
not visible in the representation since the magnitude of the
velocities is slightly lower than the threshold of
0.3~km\,s$^{-1}$. \\

\subsection{Lacking moat flows}
\label{moatlack}

Next we consider granulation regions close to the sunspots that do not display moat flows. All sunspots of our sample, except S3, have parts of the umbral core sides that are free from penumbrae, having direct contact with granulation regions. In all these granulation regions we do not detect systematic large-scale outflows that fulfill our criteria for moat flows. \emph{This supports the conclusion that umbral core boundaries with no penumbra do not display moat flows.}\\

\begin{figure}
\begin{tabular}{l}
\hspace{-1cm}\includegraphics[width=1.7\linewidth]{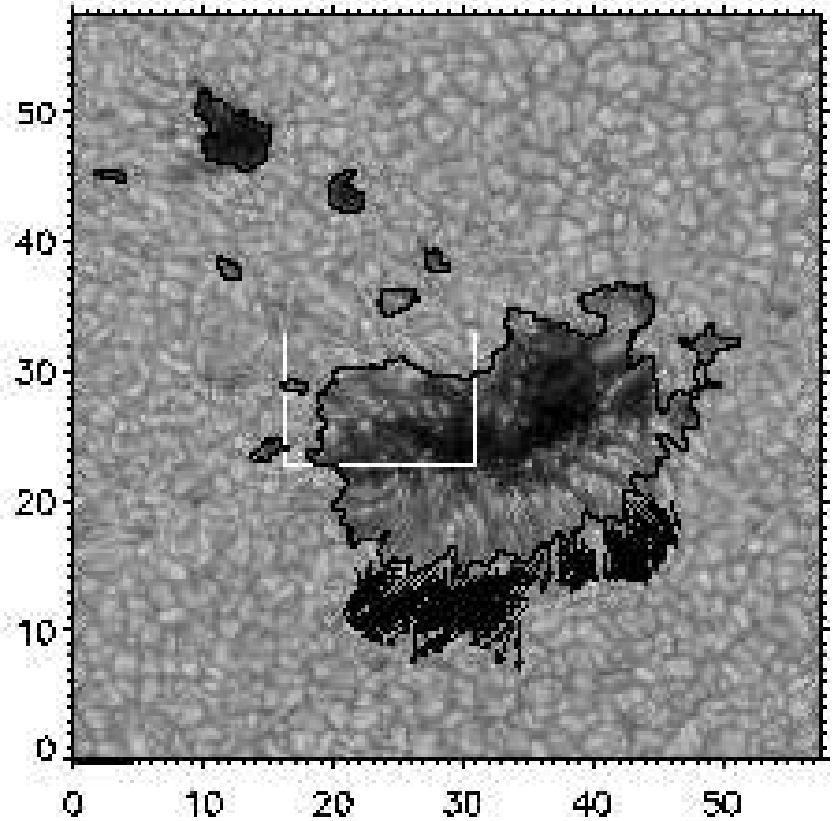}
\end{tabular}
\caption[\sf Map of the horizontal velocities for sunspot S1]{\sf Sunspot {\bf S1}: Map of the horizontal velocities inside the moat with de-projected magnitudes $>$ 0.3~km\,s$^{-1}$. The background corresponds to the average image of the whole series. The sunspot exhibits a well developed penumbra in the lower region. The moat flow is clearly found in the same region following the penumbra filamentary direction and in no other regions surrounding the sunspot. The coordinates are expressed in arc seconds. The black bar at (0,0) represents 1.5 km\,s$^{-1}$ for the projected velocities.  A peculiar region is shown with a \emph{white square}.}
\label{S1}
\end{figure}

\begin{figure}
\begin{tabular}{l}
\hspace{-1cm}\includegraphics[width=1.55\linewidth]{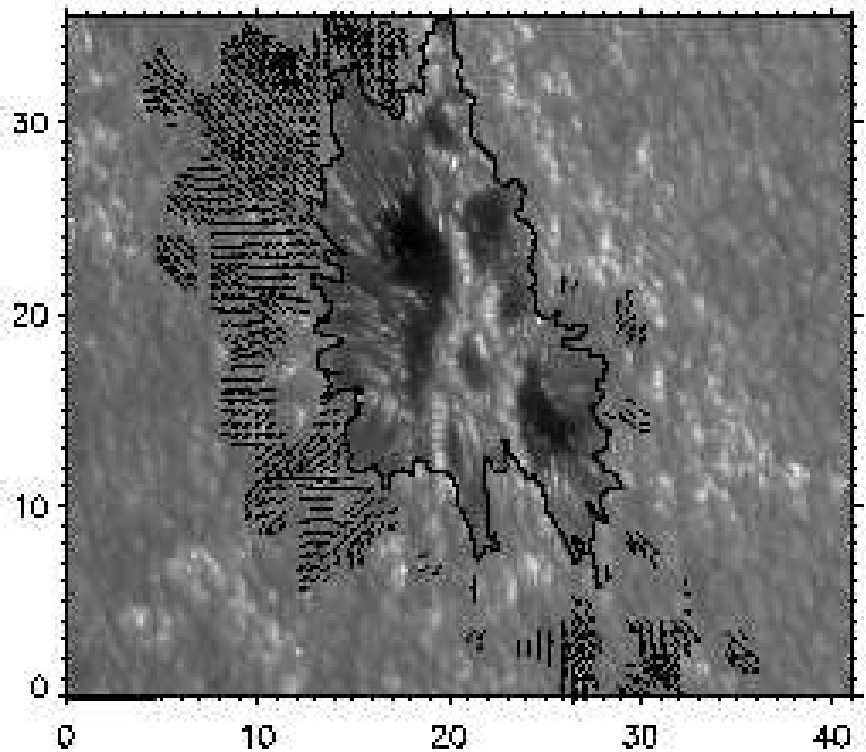}
\end{tabular}
\caption[\sf Map of the horizontal velocities for sunspot S2]{\sf Sunspot {\bf S2}: Map of the horizontal velocities inside the moat with de-projected magnitudes $>$ 0.3~km\,s$^{-1}$. The background corresponds to the average image of the whole series. This sunspot shows different penumbrae shapes all around it, from short to long and from wide to narrow penumbrae. Except for the places where the umbra is adjacent to the surrounding granulation, we identified moat flows all around the sunspot in roughly the same configuration of the penumbra. The black bar at (0,0) represents 1.5 km\,s$^{-1}$ for the projected velocities. The coordinates are expressed in arc seconds.}
\label{S2}
\end{figure}

\begin{figure}
\begin{tabular}{l}
\hspace{-1cm}\includegraphics[width=1.5\linewidth]{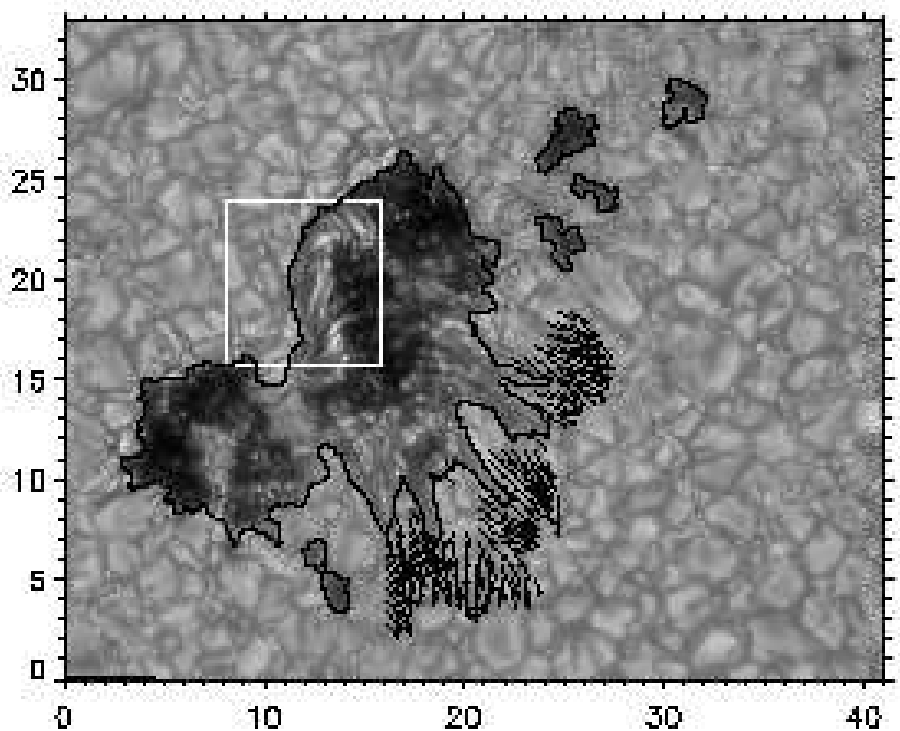}
\end{tabular}
\caption[\sf Map of the horizontal velocities for sunspot S4]{\sf Sunspot {\bf S4}: Map of the horizontal velocities inside the moat with de-projected magnitudes $>$ 0.3~km\,s$^{-1}$. The background corresponds to the average image of the whole series. The figure corresponds to a sunspot with two distinct penumbrae. One tangential to the umbral core and the other one coming out radially from the right part of the umbra. The moat flow continues as a prolongation of the last one. The black bar at (0,0) represents  1.5 km\,s$^{-1}$ for the projected velocities. The coordinates are expressed in arc seconds. A peculiar region is shown with a \emph{white square}.}
\label{S4}
\end{figure}

\begin{figure}
\begin{tabular}{l}
\hspace{-1cm}\includegraphics[width=1.65\linewidth]{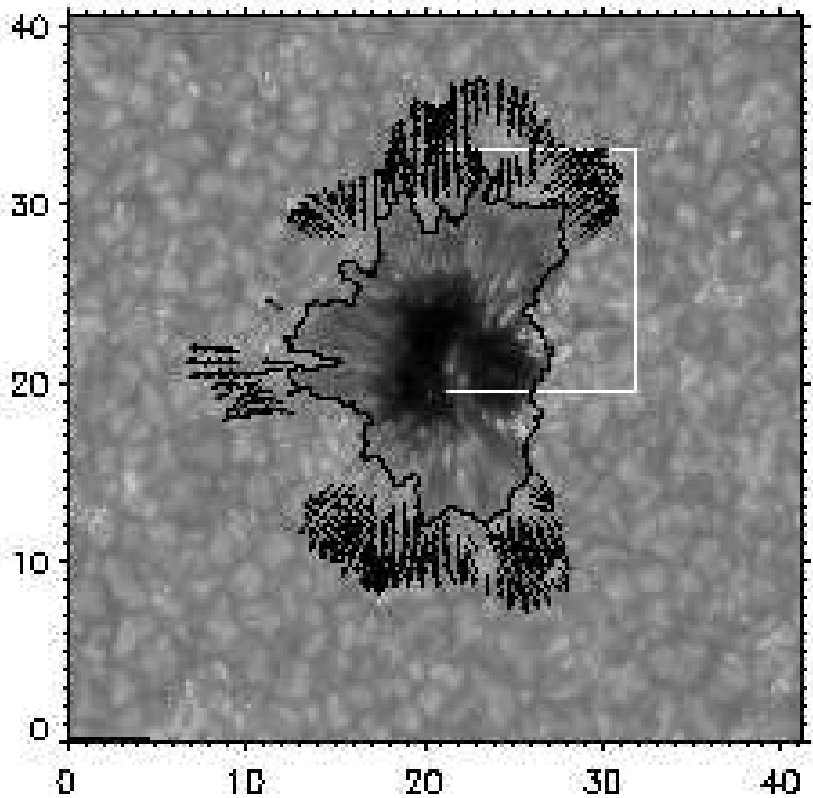}
\end{tabular}
\caption[\sf Map of the horizontal velocities for sunspot S5]{\sf Sunspot {\bf S5}: Map of the horizontal velocities inside the moat with de-projected magnitudes $>$ 0.3~km\,s$^{-1}$. The background corresponds to the average image of the whole series. This sunspot presents very distinct areas with and without penumbrae. The penumbral distributions follow radial directions coming out from the umbra. 
Moat flows are organized radially as the penumbra. The black bar at (0,0) represents 1.5 km\,s$^{-1}$ for the projected velocities. The coordinates are expressed in arc seconds. A peculiar region is shown with a \emph{white square}.}
\label{S5}
\end{figure}

\begin{figure}
\begin{tabular}{l}
\hspace{-1cm}\includegraphics[width=1.5\linewidth]{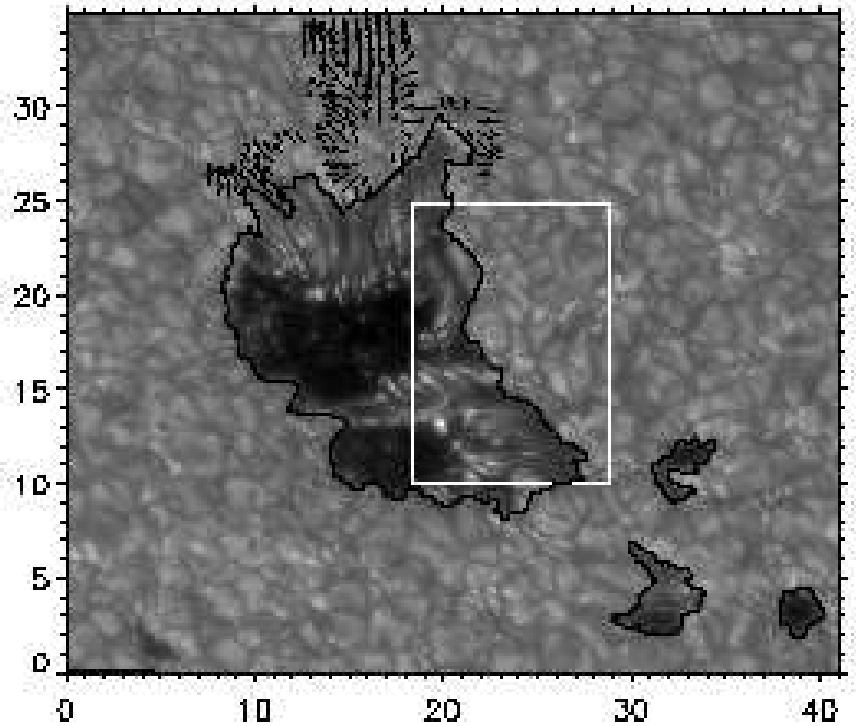}
\end{tabular}
\caption[\sf Map of the horizontal velocities for sunspot S6]{\sf Sunspot {\bf S6}: Map of the horizontal velocities inside the moat with de-projected magnitudes $>$ 0.3~km\,s$^{-1}$. The background corresponds to the average image of the whole series. This sunspot exhibits penumbrae on the right and upper sides. The penumbra on the right is mostly tangential to the sunspot border. In the upper penumbra we identified the moat flow as the prolongation of the penumbral filaments as expected. The black bar at (0,0) represents  1.5 km\,s$^{-1}$ for the projected velocities. The coordinates are expressed in arc seconds.  A peculiar region is shown with a \emph{white square}.}
\label{S6}
\end{figure}

\begin{figure}
\begin{tabular}{l}
\hspace{-1cm}\includegraphics[width=1.7\linewidth]{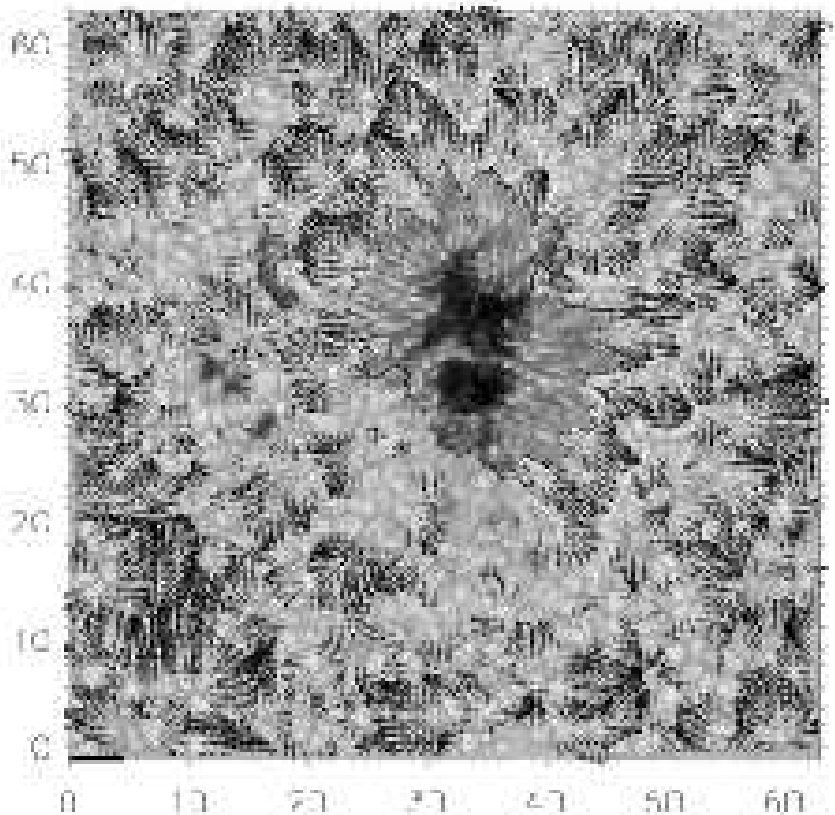} \\
\end{tabular}
\caption[\sf Map of the horizontal velocities for sunspot S7]{\sf Sunspot {\bf S7}: Map of the horizontal velocities surrounding the sunspot with de-projected magnitudes $>$ 0.3~km\,s$^{-1}$. The background corresponds to the average image of the whole series. The sunspot includes a well developed penumbra but also an empty region with no penumbra. Looking at the large flows, they completely disappear in the emptied areas with no prenumbra. The \emph{white contours} outline the moat regions. The black bar at (0,0) represents  1.5 km\,s$^{-1}$ for the projected velocities. The coordinates are expressed in arc seconds}
\label{S7}
\end{figure}

\begin{figure}
\centering
\begin{minipage}{1.1\textwidth}
\begin{tabular}{ll}
\textsf{T}\includegraphics[width=0.57\linewidth]{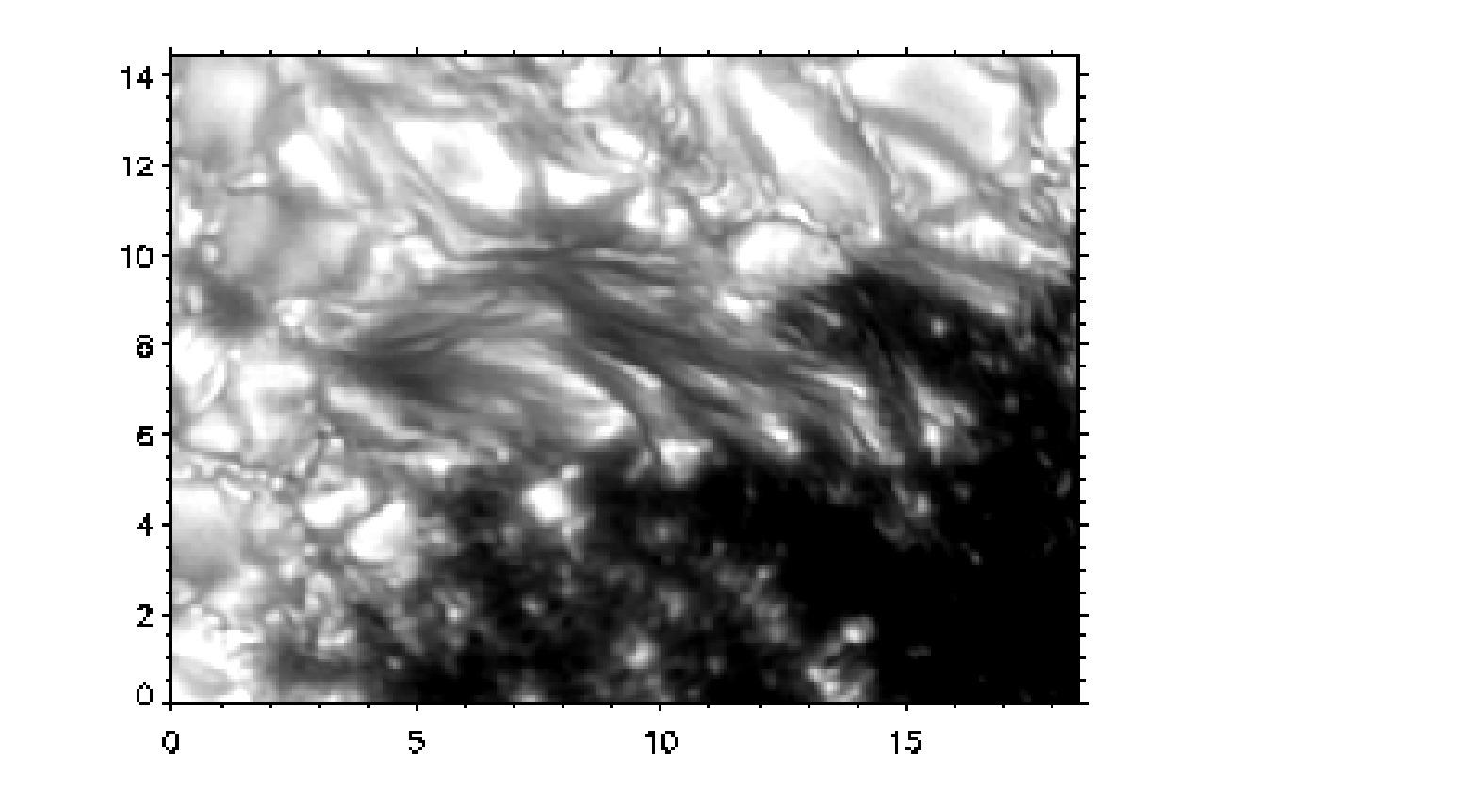} &
\hspace{-2.7cm}\textsf{T}\includegraphics[width=0.65\linewidth]{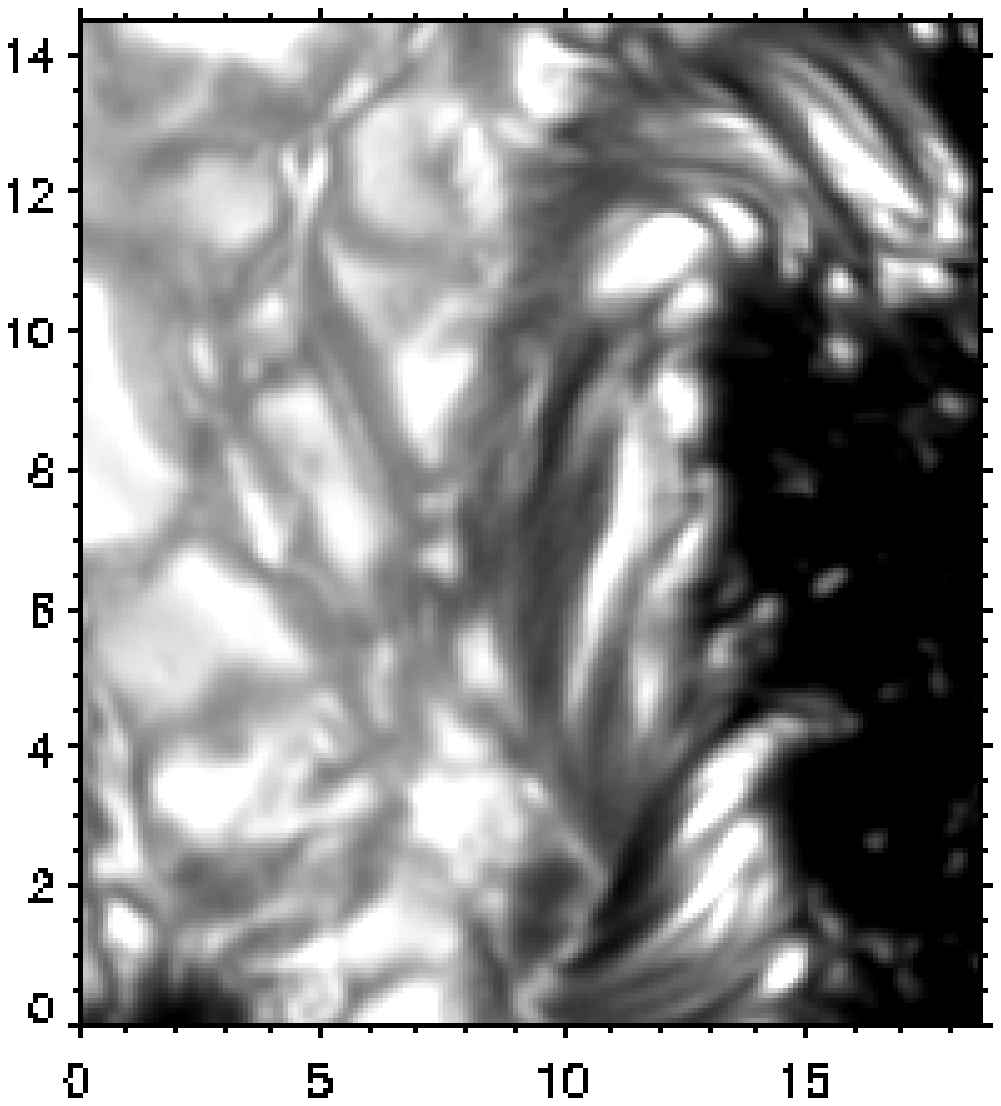}\\
\textsf{T}\includegraphics[width=0.67\linewidth]{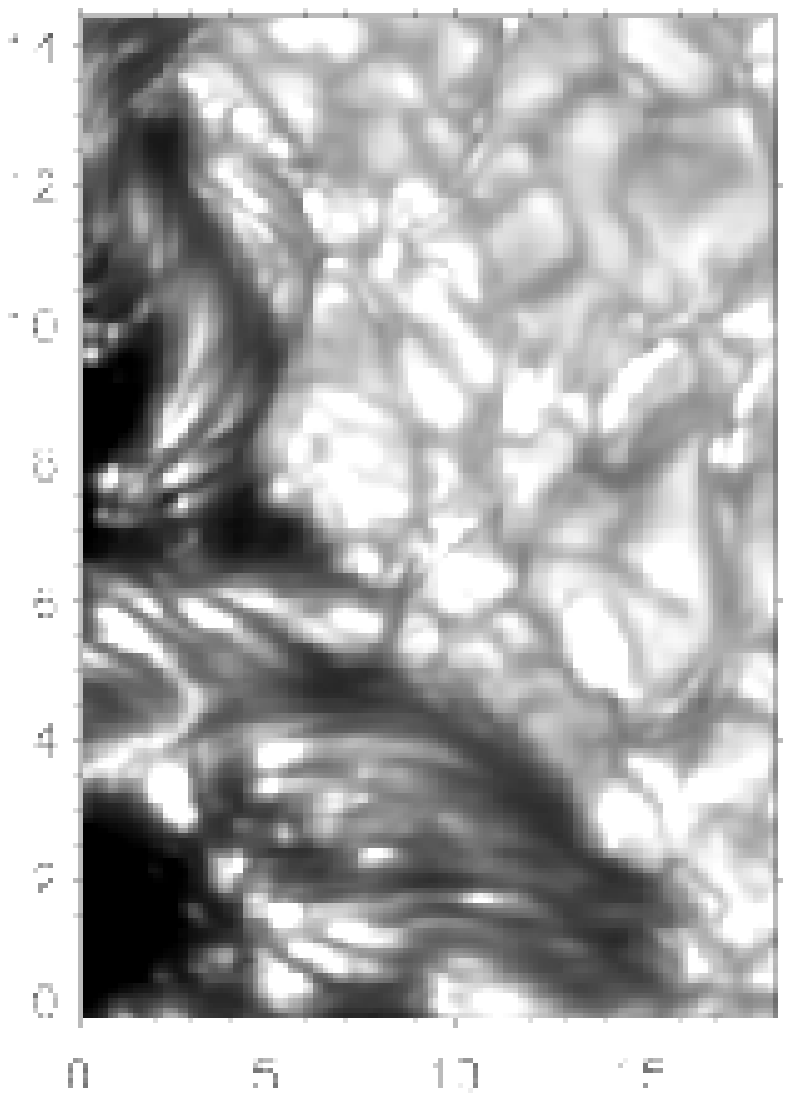} &
\hspace{-2.7cm}\textsf{R}\includegraphics[width=0.67\linewidth]{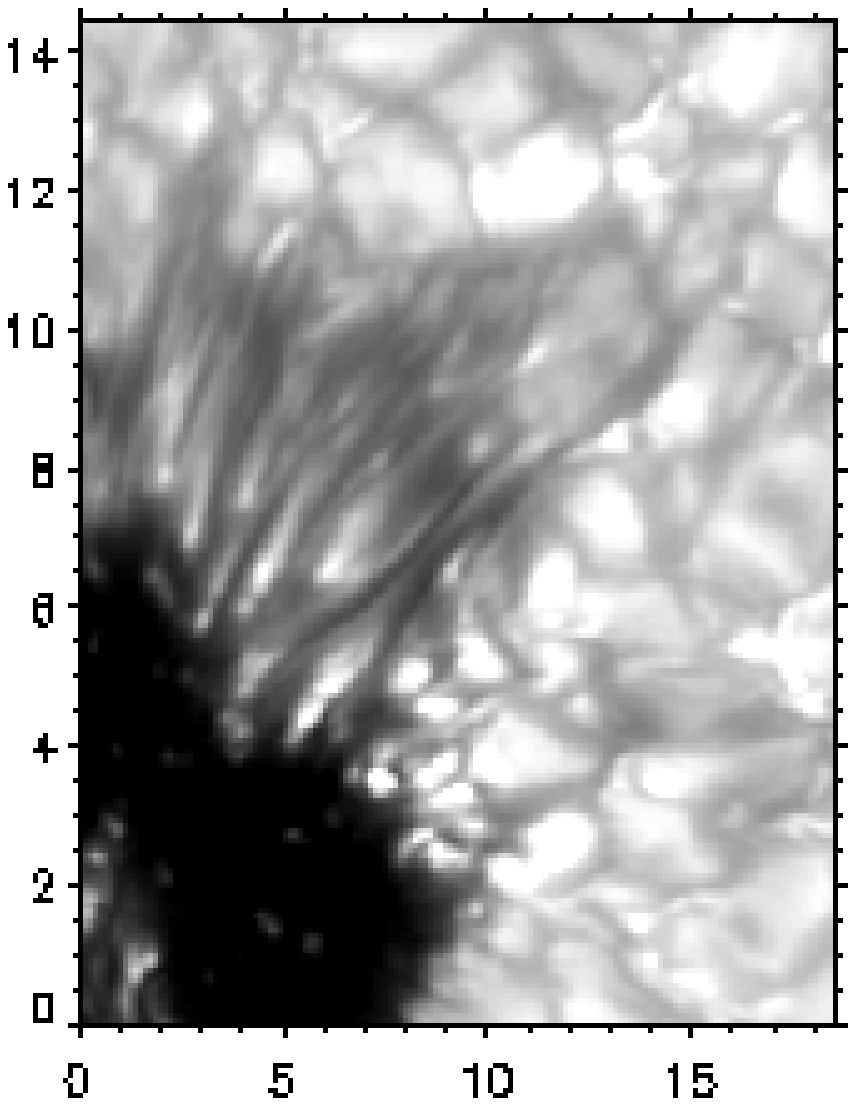}
\end{tabular}
\end{minipage}
\vspace{1cm}
\caption[\sf Close-ups for peculiar regions in sunspots S1, S4, S6 and S5]{\sf From top to bottom and left to right: Close-ups for peculiar regions in sunspots S1, S4, S6 and S5 respectively. The first three regions (\textsf{T}) show a penumbra extending tangential to the sunspot border. The lower right panel (\textsf{R}) shows a radial penumbra. The coordinates are expressed in arc seconds.}
\label{zoom}
\end{figure}

Finally, there are also granulation regions in the vicinity of penumbrae that lack significant moat flows. To study these cases in more detail we mark peculiar regions of interest in Figures~\ref{S1}, \ref{S4}, \ref{S5} and \ref{S6} with rectangular \emph{white boxes}. These regions are represented as close-ups in Figure~\ref{zoom}. In the first three panels (T), the penumbral filaments display significant curvature such that they are not radially oriented with respect to the sunspot center but extend in a direction tangential to the sunspot border (as marked with \emph{black lines} in the overview images Figures~\ref{S1}, \ref{S4} and \ref{S6}). The lower right panel (R) in Figure~\ref{zoom}  shows a penumbra extending radially from the umbra. The surrounding photosphere exhibits moat flows only in the direction of the penumbral filaments (also see Figure~\ref{S5}). \emph{ These four examples lead to the conclusion that moat flows are not found in directions transverse to the penumbral filaments.}\\

In a few cases we found small pores in the vicinity of penumbrae in a
region where we would expect moat flows. Such cases are seen around
coordinates (19,29) in Figure~\ref{S1} and coordinates (29,11) in
Figure~\ref{S6}. For these regions the measured proper motions are less reliable since we have a FWHM=1$\farcs$0 tracking window acting on a region of only a
few arc seconds ($\sim$ 2-3) between the pore and penumbra. Another possibility is that the pores themselves are somehow blocking the large outflows changing the expected behavior. In the case presented in Figure~\ref{S6} the most plausible explanation for the absence of moats is the presence of a neutral line close to coordinates (29,11) as we shall describe in section~\S\ref{S:neutral}.\\

The findings described above suggest a link between the moat flows and the Evershed flows in penumbrae.  We come back to this relation in section~\S\ref{S:dis}.\\

The debate regarding the existence of a moat flow around umbral cores and individual pores is still ongoing. In a work by \cite{deng2007}, these authors found that the dividing line between radial inward and outward proper motions in the inner and outer penumbra, respectively, survived the decay phase, suggesting that the moat flow is still detectable after the penumbra disappeared. These issues will be deeper mentioned in the next chapter contrasting the results presented in this thesis and some more examples in the literature.

\subsection{Neutral lines affecting the flows behavior}
\label{S:neutral}

Only one of the sunspots we study displays a complete regular and well-developed penumbra completely surrounding the umbral core. Figure~\ref{S3} shows the flow map calculated for this active region, plotting the horizontal velocities surrounding the sunspot with de-projected magnitudes $>$ 0.3~km\,s$^{-1}$. Large outflows are not found in a part of the right side of the upper panel of Figure~\ref{S3} (\emph{rectangular box)}.\\

Following the findings of section~\S\ref{moataround}, we would expect to
find a moat flow in this region. The penumbral filaments are oriented
radially from the umbral core which suggest the presence of a moat
flow in the granulation region in the direct vicinity. Even with a
lower velocity threshold, no moat flow can be discerned in this
region. Nevertheless, when comparing with a simultaneous magnetogram (see \emph{lower panel} in Figure~\ref{S3}), we found an inversion in magnetic polarity just
outside the right border of the sunspot: the magnetogram displays
positive polarity (\emph{in white}) for the sunspot but negative polarity
for the small magnetic elements and pore just outside the
penumbra. The reversal of polarity (or neutral line) is confined to
a narrow region that roughly coincides with the sunspot border. The absence of large outflows following this penumbra is suggestively related to the presence of this neutral line that  might somehow be acting as a blocking agent for the moat flow in this region. The position of the neutral line measured by LOS magnetograms is generally influenced by the location of the sunspot ($\theta$ angle). In this case since the sunspot is very close to the disc center we can claim for a reliable determination of the neutral line.\\

\begin{figure}
\centering
\begin{tabular}{l}
\includegraphics[width=1.\linewidth]{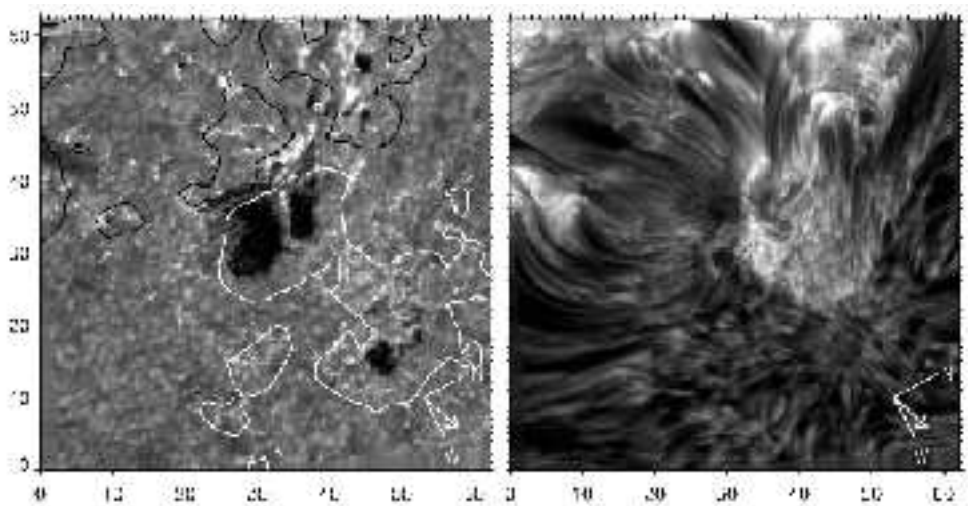}
\end{tabular}
\caption[\sf Images taken in H$\alpha$ wide-band and H$\alpha$ line-center of NOAA AR 10783]{\sf Images taken in H$\alpha$ wide-band (\emph{left}) and H$\alpha$ line-center (\emph{right}) of NOAA AR 10783 taken at the SST on 4-Oct-2005 as displayed by \cite{depontieu2007}. The FOV shown in both images is 90 degrees counterclockwise rotated with respect to the one in Figure~\ref{S6}. Tick marks are in arc seconds. The \emph{white} and \emph{black contours} in the left panel outline the positive and negative magnetic flux from a full-disk MDI magnetogram. The \emph{white arrow} indicates the direction of the disc center.}
\label{fig2depontieu}
\end{figure}

\begin{figure}
\centering
\vspace{-3mm}
\begin{tabular}{l}
\vspace{-6mm}
\hspace{2cm}\includegraphics[width=1.18\linewidth]{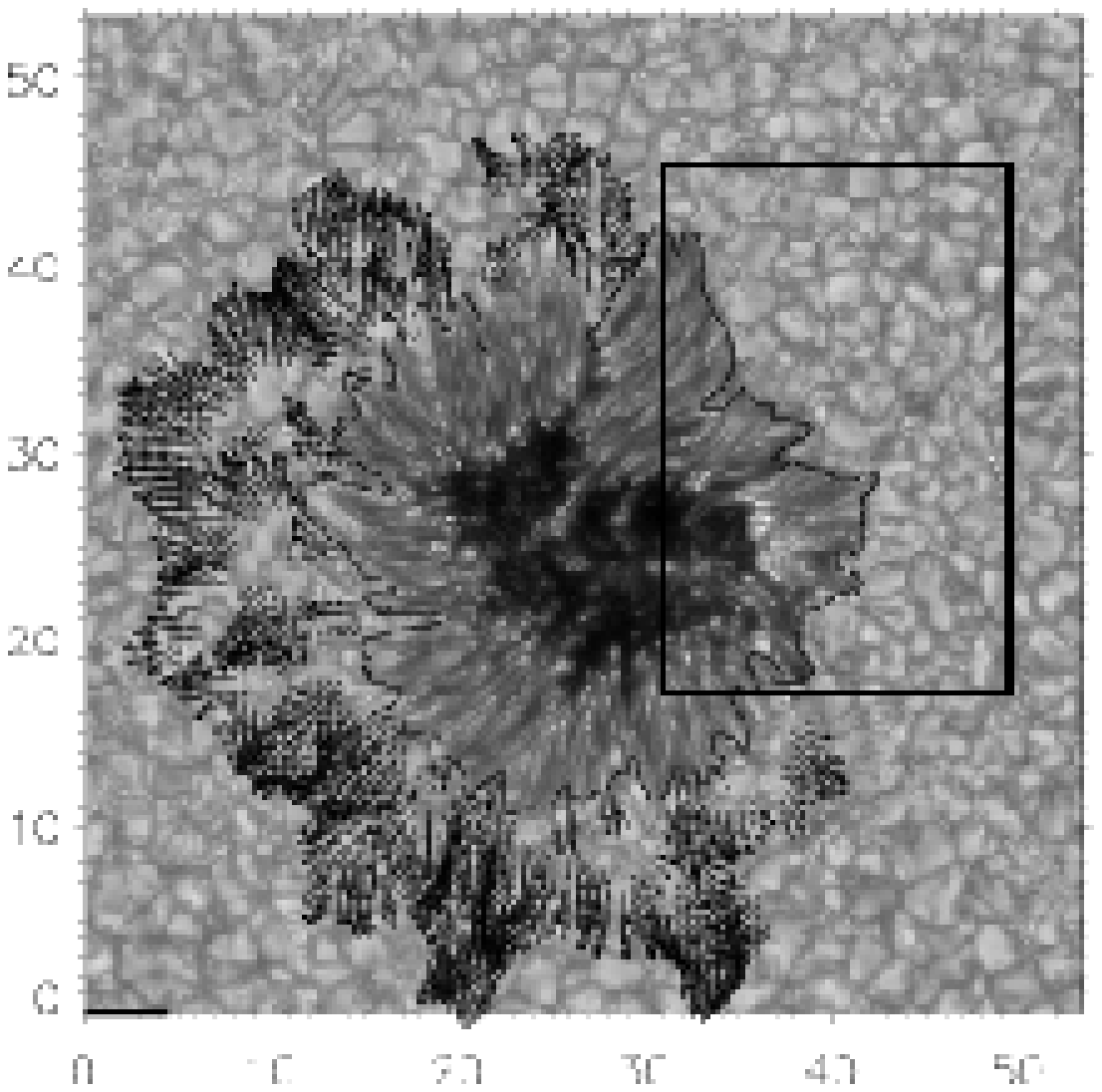}\\\vspace{-6mm}
\hspace{2cm}\includegraphics[width=1.15\linewidth]{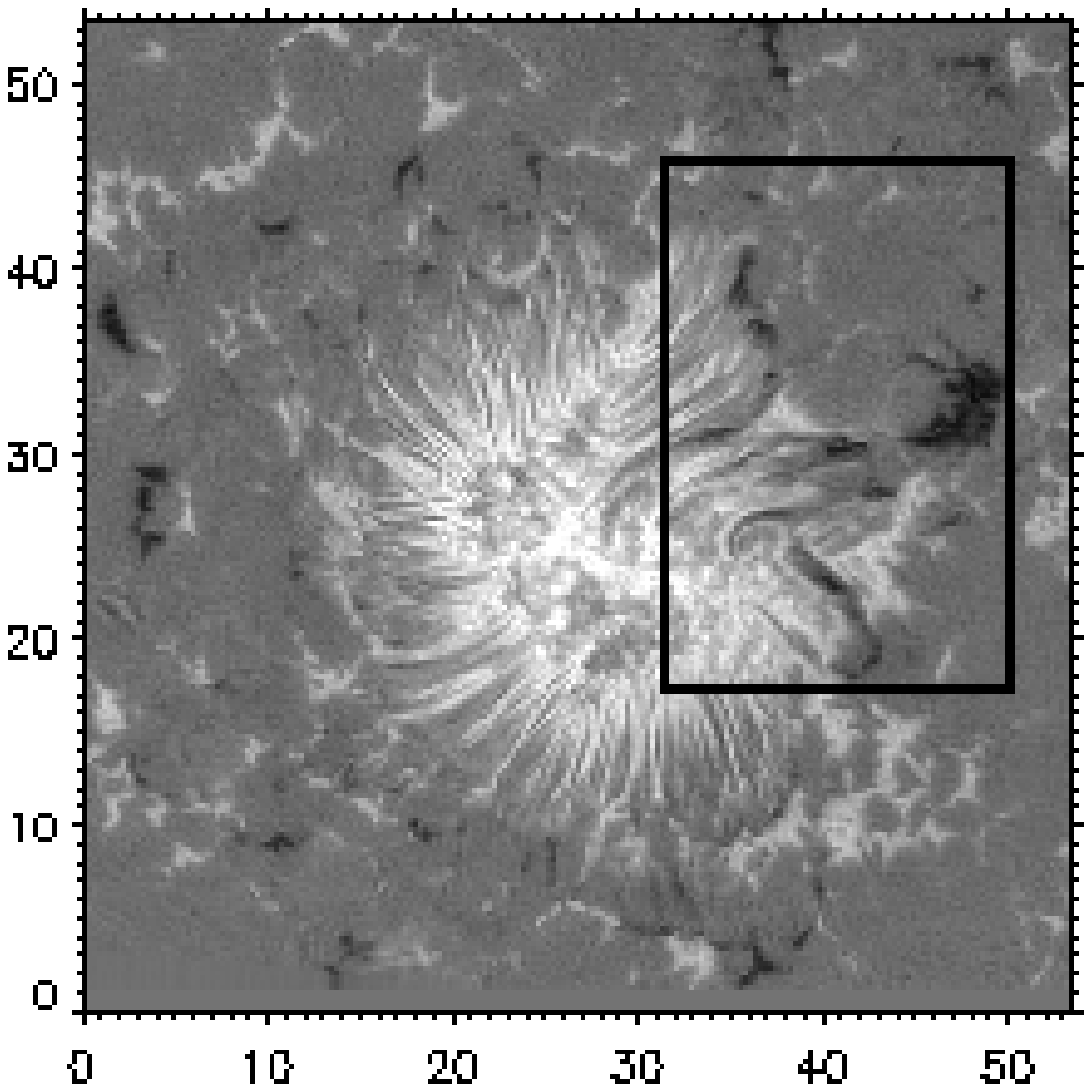} 
\end{tabular}
\caption[\sf Map of the horizontal velocities for sunspot S3]{\sf Sunspot {\bf S3}: Map of the horizontal velocities with de-projected magnitudes $>$ 0.3~km\,s$^{-1}$ (\emph{upper panel}). The sunspot in the figure is surrounded all around by a well-developed penumbra. Large outflows are not found in a certain part of the right-hand side of the spot. When looking at the magnetogram for the same FOV (\emph{lower panel}) we found a neutral line which is  suspiciously acting as a blocking agent for the flow coming out from the spot. The \emph{black box} is a common region in both figures. The coordinates are expressed in arc seconds. The black bar at (0,0) in the flow map represents 1.5 km\,s$^{-1}$ for the projected velocities.}
\label{S3}
\end{figure}

For sunspot in Figure~\ref{S6} there is also a neutral line crossing all along the right-hand border of the sunspot from top to bottom around coordinates (20,28) to (28,11), respectively, as reported by \cite{depontieu2007} and shown here as Figure~\ref{fig2depontieu}. The penumbral filaments extend from the umbra to the right and seems to bend, being forced to follow the direction of the neutral line as approaching to it. The sheared configuration of this penumbra is arranged so that the penumbral filaments end up along directions parallel to the sunspot border. Moat flows are then not found beyond this sort of penumbral configuration as mentioned in section~\S\ref{moatlack}. The same behavior when having neutral lines was presented in section~\S4.4.3 for a complex $\delta$-configuration active region. For that active region we found a strong sheared neutral line crossing a penumbral border in a place where a moat flow was expected to follow the penumbral filaments direction but actually not detected. 
More observations of complex active regions with neutral lines present in the vicinity of penumbrae are needed to firmly establish the relation between the absence of moat flows and magnetic neutral lines.\\

\subsection{Statistics of velocity fields in moats vs.\ quiet granulation}
\label{S:statist}

In this section the statistics of time-averaged horizontal velocity fields in moats is performed. For the sake of comparison, the velocity statistics in quiet granulation areas has also been computed. To that aim, boxes  ($\sim$ 9$\times$9 arcsec$^2$)  in regions of less-magnetized (quiet) granulation and far from the moat flows have been manually selected in every FOV. Table~\ref{table3} summarizes the statistical properties of the de-projected magnitudes of horizontal velocity fields within both moat masks and selected granular boxes. Here we again remark that because of the possible underestimation of the displacements by using LCT, we intend to perform not an absolute but a comparative statistical study between moat regions, where we find radially organized outflows, and quiet granulation regions far from the sunspots.\\

For each sunspot of our sample three velocity magnitude maps result from averaging over different time periods, namely: 5 min, 10 min and the whole duration of the corresponding time series \footnote{\sf Actually, when averaging over 5 min (or 10 min) we obtain more than one map since we then compute maps every 5 min (or 10 min) up to complete the total duration of the series (i.e when having 40 min of duration for a time series, we would finally obtain 8 maps of 5 min or 4 maps of 10 min). This enable us to improve the statistics of the velocity magnitudes and reduce the noise in our calculations.}. Table~\ref{table3} shows that in most cases the mean velocity magnitude in moats is greater than in granulation. In a few cases and only for averages in short time periods (5, 10 min), we find similar values in both or even slightly lower velocities in moats. However, when averaging over a long time period, the difference of mean velocities in moats and granulation is in all cases positive and more conspicuous than for short time period averages (see column 10 in the Table). A similar behavior is obtained for the rms parameter (see column 9 in the Table). The maximum velocity values are also systematically larger in moats than in granulation. \\

The described statistical behavior is expected since for short time averaging periods, of the order of the granulation lifetime (5-16 min, see \citet{hirzberger1999} and references there in), the proper motions of the granulation structures compete in magnitude with the velocity of large-scale flows. However, averaging velocity components over long time periods results in local velocity cancellations in short-lived structures while steady motions at large spatial scales prevail. So moats can be considered as a long-term and large-scale phenomenon where the mean velocity exceeds that of quiet granulation by about 30\%.\\

The significant fluctuations in $\Delta$(rms) and $\Delta$(mean)  (columns 9 and 10 in Table~\ref{table3}) for the velocity fields averaged over the whole time series, deserve a comment and possibly future improved measurements. They could be ascribed to morphological differences in the various cases considered or they could be related to the particular evolutionary state or magnetic field strength in the various sunspots of our sample. But we cannot forget that the de-projection formulas (\ref{eq:1}) and (\ref{eq:2})  include products, divisions and sums of products of trigonometric functions. Accordingly, the accuracy in the measurement of $\theta$ and $\phi '$ is crucial for the determination of $v$. Thus, two sources of inaccuracy are: a) the assumption of constant heliocentric angle $\theta$ all along the sunspot area (this has important impact in regions far from the solar disc center), and b) the determination in our FOV of the direction pointing to the solar disc center which defines the X-axis of the Observing Coordinate System, i.e.\ the angular origin for $\phi '$.\\

\begin{figure*}
\centering
\vspace{-4mm}
\begin{tabular}{l}
\includegraphics[width=.97\linewidth]{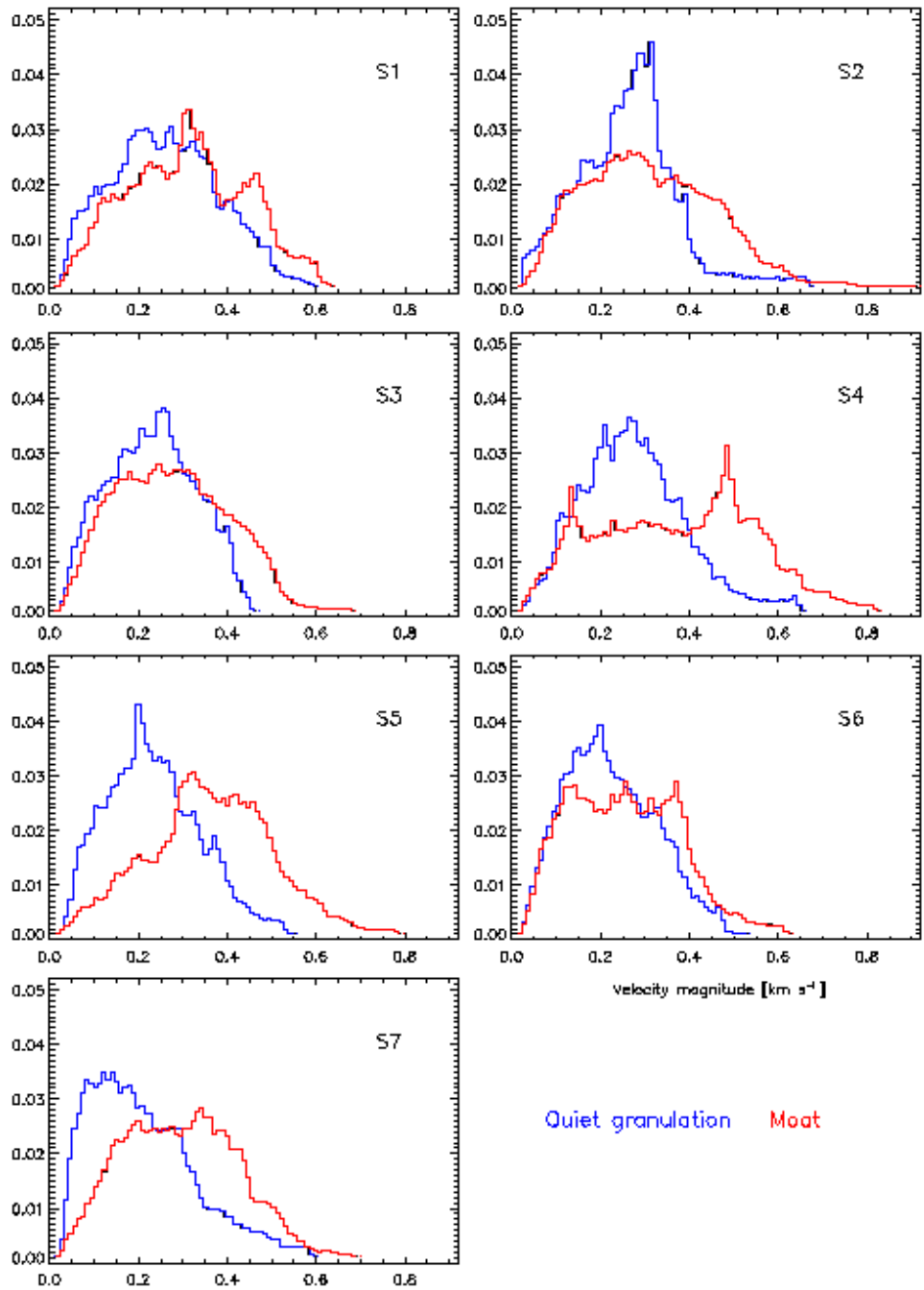} \\
\end{tabular}
\caption[\sf Histograms of the velocity magnitudes for sunspots S1 to S7]{\sf Histograms of the velocity magnitudes averaging over the whole time series (more than 40 minutes in every case) for sunspots S1 to S7. The values for moats (\emph{red line}) and quiet granulation (\emph{blue line}) are plotted. The vertical and horizontal axes represent the percentages and the velocity magnitudes [km\,s$^{-1}$], respectively.}
\label{hist}
\end{figure*}

Figure~\ref{hist} shows the histograms of de-projected velocity magnitudes for averages over the whole time series for sunspots S1 to S7. Red lines correspond to velocities inside the moats and the blue ones to velocities in quiet granulation boxes. In all cases we find the same general trend in both histograms. The \emph{red} histograms are globally shifted toward the right with respect to the \emph{blue} ones.\\

The left wings of the histograms for granulation lie systematically above those of the histograms for moats, and for S1, S2, S3, S4 and S6, both histograms nearly coincide at the farthest left end. This means that within the moats standard velocity values for quiet granulation are still present although with a lower weight. Such is the case for the fragments of exploding granules that additionally are swept by large-scale moat flows. \\

At some point from left to right, both histograms cross each other so that the right wing of the moat histogram surpasses the corresponding quiet granulation wing and extends to larger velocity values ($>$ 0.6~km\,s$^{-1}$). This confirms the predominance of large velocities in moats. The intersection point of both histograms corresponds to  $\sim$ 0.3~km\,s$^{-1}$ in almost all cases (a bit higher in S1, S2 and S4). Interestingly, this value coincides with the threshold empirically selected a priori (section~\S\ref{masking}) to clearly distinguish the frontiers of the moats, which confirms the goodness of the limit-value (0.3~km\,s$^{-1}$) chosen to detect moats in our data set. 

\begin{table}[]
\sffamily
\caption[\sf Statistics of de-projected horizontal velocity magnitudes]{\sf Statistics of de-projected horizontal velocity magnitudes [m s$^{-1}$]}
\begin{center}
\begin{tabular}{ccrrrrrrcc}
\hline\hline\\
 &&\multicolumn{3}{c}{{\footnotesize MOAT}}&\multicolumn{3}{c}{{\footnotesize GRANULATION}}& & \\
 \cline{3-8} \\
 &{\scriptsize DURATION}&&&&&&&{\scriptsize$\Delta$(rms)$^\dag$}&{\scriptsize$\Delta$(mean)$^\dag$}\\
 
{\scriptsize SUNSPOT}&{\scriptsize (minutes)}&{\footnotesize Max}&{\footnotesize rms}&{\footnotesize Mean}&{\footnotesize Max}&{\footnotesize rms}&{\footnotesize Mean}& {\footnotesize (\%)}&{\footnotesize (\%)}\\ \hline\\
 
{\footnotesize S1 ..........} & {\footnotesize5} & {\footnotesize1048} & {\footnotesize174} & {\footnotesize333} & {\footnotesize947} & {\footnotesize157} & {\footnotesize315} & {\footnotesize10.8} & {\footnotesize5.7} \\
 & {\footnotesize10} & {\footnotesize947} & {\footnotesize162} & {\footnotesize315} & {\footnotesize842} & {\footnotesize157} & {\footnotesize314} & {\footnotesize3.2} & {\footnotesize0.3} \\
 & {\footnotesize53} & {\footnotesize639} & {\footnotesize137} & {\footnotesize301} & {\footnotesize596} & {\footnotesize123} & {\footnotesize250} & {\footnotesize11.4} & {\footnotesize20.4}\\ 
 
{\footnotesize S2 ..........} & {\footnotesize5} & {\footnotesize1187} & {\footnotesize183} & {\footnotesize329}  & {\footnotesize898} & {\footnotesize162} & {\footnotesize307} & {\footnotesize13.0} & {\footnotesize7.1}\\
 & {\footnotesize10} & {\footnotesize1278} & {\footnotesize182} & {\footnotesize327}  &  {\footnotesize862} & {\footnotesize155} & {\footnotesize300} & {\footnotesize17.4} & {\footnotesize9.0}\\
 & {\footnotesize47} & {\footnotesize921} & {\footnotesize150} & {\footnotesize303} & {\footnotesize678} & {\footnotesize115} & {\footnotesize249} & {\footnotesize30.4} & {\footnotesize21.7}\\
 
{\footnotesize S3 ..........} & {\footnotesize5} & {\footnotesize1036} & {\footnotesize146} & {\footnotesize290} & {\footnotesize870} &  {\footnotesize147} & {\footnotesize282} & {\footnotesize-0.7}& {\footnotesize2.8}\\
 & {\footnotesize10} &  {\footnotesize868} & {\footnotesize146} & {\footnotesize281} & {\footnotesize671} & {\footnotesize137} &  {\footnotesize289} & {\footnotesize6.6} & {\footnotesize-2.8} \\
 & {\footnotesize47} & {\footnotesize689} & {\footnotesize126} & {\footnotesize265} & {\footnotesize451} & {\footnotesize101} &  {\footnotesize220} & {\footnotesize24.8} & {\footnotesize20.5}\\
 
{\footnotesize S4 ..........} & {\footnotesize5} & {\footnotesize1060} & {\footnotesize198} &  {\footnotesize411} & {\footnotesize863} & {\footnotesize164} & {\footnotesize362} & {\footnotesize20.7} & {\footnotesize13.5} \\
 & {\footnotesize10} & {\footnotesize954} & {\footnotesize180} & {\footnotesize379} & {\footnotesize780} &  {\footnotesize154} &  {\footnotesize348} & {\footnotesize16.9} & {\footnotesize8.9}\\
 & {\footnotesize45} & {\footnotesize834} & {\footnotesize179}& {\footnotesize365} &  {\footnotesize655} &  {\footnotesize118} &  {\footnotesize262} & {\footnotesize51.7} & {\footnotesize39.3}\\
 
{\footnotesize S5 .......... } & {\footnotesize5} & {\footnotesize999} & {\footnotesize178} & {\footnotesize355} & {\footnotesize968} & {\footnotesize172} & {\footnotesize307} & {\footnotesize3.5} & {\footnotesize15.6}\\
 & {\footnotesize10} & {\footnotesize972} & {\footnotesize187} & {\footnotesize372} & {\footnotesize1017} & {\footnotesize159} & {\footnotesize291} & {\footnotesize17.6} & {\footnotesize27.8} \\
 & {\footnotesize40} & {\footnotesize802} & {\footnotesize141} & {\footnotesize355} & {\footnotesize545} & {\footnotesize107} & {\footnotesize220} & {\footnotesize31.8} & {\footnotesize61.4}\\
 
{\footnotesize S6 ..........} & {\footnotesize5} & {\footnotesize699} & {\footnotesize129} & {\footnotesize257} &  {\footnotesize683} &  {\footnotesize130} &  {\footnotesize278}  & {\footnotesize-0.8} & {\footnotesize-7.6}\\
& {\footnotesize10} & {\footnotesize673} & {\footnotesize131} & {\footnotesize248} & {\footnotesize723} & {\footnotesize136} & {\footnotesize277} & {\footnotesize-3.7} & {\footnotesize-10.5} \\
& {\footnotesize40} & {\footnotesize622} & {\footnotesize125} & {\footnotesize243} &  {\footnotesize526} & {\footnotesize104} & {\footnotesize210} & {\footnotesize20.2} & {\footnotesize15.7}\\ 

{\footnotesize S7 ..........}  & {\footnotesize5} & {\footnotesize872} & {\footnotesize158} & {\footnotesize300} & {\footnotesize759} & {\footnotesize142} & {\footnotesize226} & {\footnotesize11.3} & {\footnotesize32.7} \\
& {\footnotesize10} & {\footnotesize785} & {\footnotesize159} & {\footnotesize320} & {\footnotesize828} &  {\footnotesize149} & {\footnotesize227} & {\footnotesize6.7} & {\footnotesize41.0}\\
& {\footnotesize40} & {\footnotesize692} & {\footnotesize131} & {\footnotesize285} & {\footnotesize593} & {\footnotesize121} & {\footnotesize196} & {\footnotesize8.3} & {\footnotesize45.4}\\
\hline
\end{tabular}
\end{center}
\footnotesize{$\dag$ $\Delta$(rms) and  $\Delta$(mean) stand for increments of the rms and mean values in moats with respect to quiet granulation.}
\label{table3}
\end{table}

\newpage
\section{Conclusions and discussion}
\label{S:dis}

Seven time series of sunspots were restored from instrumental and
atmospheric aberrations using MFBD and MOMFBD techniques. The great
quality of the images allowed us to study proper motions of granules
outside the sunspots and measure their time-averaged velocities which
describe the general trends of motions (not the same as the
instantaneous velocities). \\

The results from chapter~\S4 are extended in this chapter to a larger sample of active regions and systematically confirmed the same findings: a) moat flows are oriented following the direction of the penumbral filaments;  b) in granulation regions found adjacent to an irregular penumbral side parallel to the penumbral filaments, moats are absent, or in other words, moats do not develop in the direction transverse to the penumbral filaments.  Note that if the moat flows were originated by the blockage of the heat flux from below by the penumbra, one would expect moat flows directed along, but also, transverse to the direction of the penumbral filaments;  c) umbral core sides with no penumbra do not display moat flows. Moreover we include in our sample a case in which a neutral line extends along a penumbral border where we would expect a moat flow continuation. For this sunspot we do not find any moat flow following the direction of the penumbral filaments after crossing the penumbral border where we see a change in magnetic polarity. The same result was found in section~\S4.4.3 in a penumbral portion of a complex active region crossed by a strong sheared neutral line. \\

All these results indicate a likely connection between the moat flows and flows aligned with penumbral filaments.  In a recent work, \citet{cabrera2006} suggest that the Evershed clouds inside penumbrae propagate to the surrounding moat and then become MMFs after crossing the sunspot border. The MMFs displacements trace very well-defined paths that can actually be clearly seen when averaging magnetograms in time. Some of these MMFs are seen to start inside the penumbra \citep{sainz2005}. We refer the reader to section~\S3.5.2 for more information about these new studies.\\

In this chapter we also complement the results of section~\S4.5 with a statistical analysis describing in a quantitative way the differences between velocity fields in moat flows and in less-magnetized solar granulation in the nearby. In contrast to the case of granulation, moat flows are well-organized, steady and large-scale motions. For averages made over more than 40 minutes, the mean velocity in moats (0.3~km\,s$^{-1}$) exceeds that of quiet granulation ($\sim$ 0.23~km\,s$^{-1}$) by about 30\%, although we obtain a considerable dispersion in the results. Also the rms of the velocity magnitude is greater in moats by a similar percentage. \\

The histograms of velocity magnitudes in the moats are broader than those in granulation (see Figure~\ref{hist}). The histograms of granulation show conspicuous maxima, most of them in a range from 0.2 to 0.3~km\,s$^{-1}$, whereas the histograms of moats present a flatter top. Systematically, the left wing of the granulation histogram lies above that corresponding to the moat. At some point about  0.3~-~0.4~km\,s$^{-1}$, both histograms cross each other and the right wing of the moat histogram extends beyond that of the granulation to larger velocity values ($>$ 0.6~km\,s$^{-1}$). \\

We have studied one case of a sunspot penumbra displaying a neutral line all along a sector of the penumbra (Figure~\ref{S3}). This neutral line is detected at the penumbral boundary but also emphasized by a large opposite polarity concentration nearby. Interestingly, this penumbral sector shows no moat flow. We interpret this evidence as an indication of the Evershed flow being forced to go into deeper sub-photospheric layers at a faster pace than what is normally though to occur in penumbral regions not associated to neutral lines \citep{westendorp1997}. A similar case, but in a sheared neutral line of a $\delta$-spot, was found in section~\S4.4.3.\\

The important questions related to the flows inside and outside sunspots are yet being studied. In the present work we contribute with a new sample of sunspots observed between 2003 and 2006 in six different observing campaigns. The data sets have been selected on basis of the seeing quality (sharpness, homogeneity, duration) and the availability of suitable targets: the presence of spots with some form of irregular penumbra.
In all of our samples we follow the evolution of the sunspots for more than 40 min. Although this only represents a snapshot in the evolution of the sunspots through all their emerging and decaying processes, our sample includes sunspots in different evolutionary stages and penumbral configurations.

%% file: chap6.tex
\chapter{\sf Flow field around solar pores}\label{cap6}
\dropping[0pt]{2}{T}his chapter is devoted to the study of flows around solar pores. After the last two chapters, meant to analyze horizontal flows surrounding different types of sunspots in detail, this one deals specifically with pores. We are interested in observing and analyzing pores because, different from sunspots, they do not have penumbra at all, so that our main conclusions about the relation between moats and penumbrae can be tested in these scenarios as explained in the following sections.

\section{Introduction}
\label{intro}

\begin{figure}
\begin{tabular}{l}
\includegraphics[width=1.\linewidth]{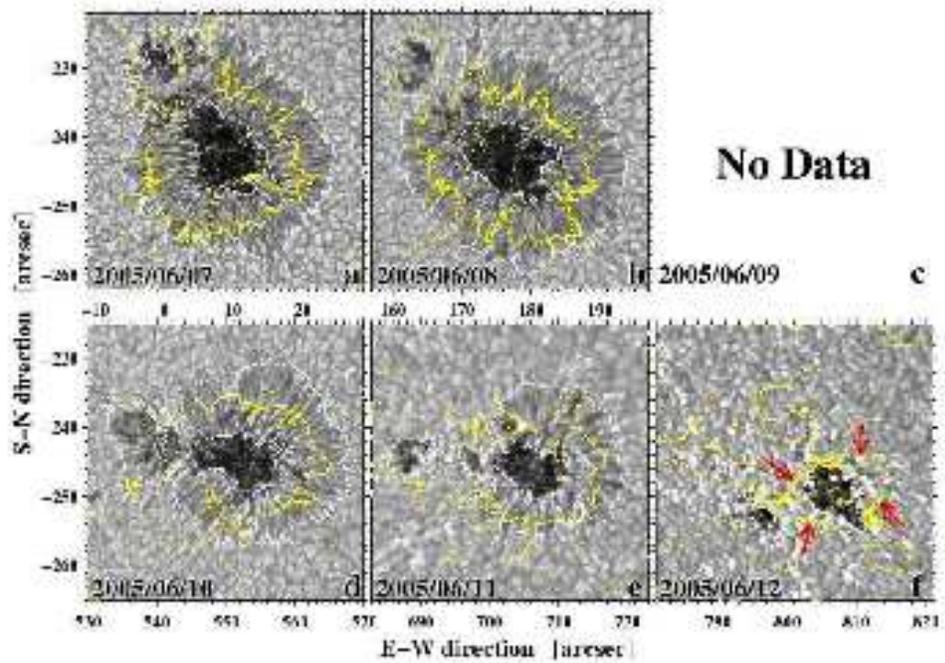}
\end{tabular}
\caption[\sf Sunspot evolution over six days during its decaying  phase]{\sf Sunspot evolution over six days during its decaying  phase taken from  \cite{deng2007}. \emph{White contour lines} mark the umbral and penumbral boundaries. The \emph{yellow corks} trace the dividing line between the inward and the outward proper motions in the penumbra and around the residual pore. The \emph{red arrows} in panel f point to the dividing line around the pore.}
\label{deng}
\end{figure}

Though there is increasing evidence linking the moat flows and the Evershed flow along the penumbral filaments, the debate regarding the existence of a moat flow around umbral cores and individual pores is still undergoing. In a recent work, \cite{deng2007}, found that the dividing line between radial inward and outward proper motions in the inner and outer penumbra, respectively, survived the decay phase, suggesting that the moat flow is still detectable after the penumbra disappeared. Figure~\ref{deng} shows the active region studied by these authors. The sunspot evolution is shown over six days during its decaying phase. \emph{Yellow corks} trace the dividing line between inward and outwards proper motions in the penumbra.\\

\begin{figure}
\begin{tabular}{l}
\hspace{-4mm}\includegraphics[width=.32\linewidth]{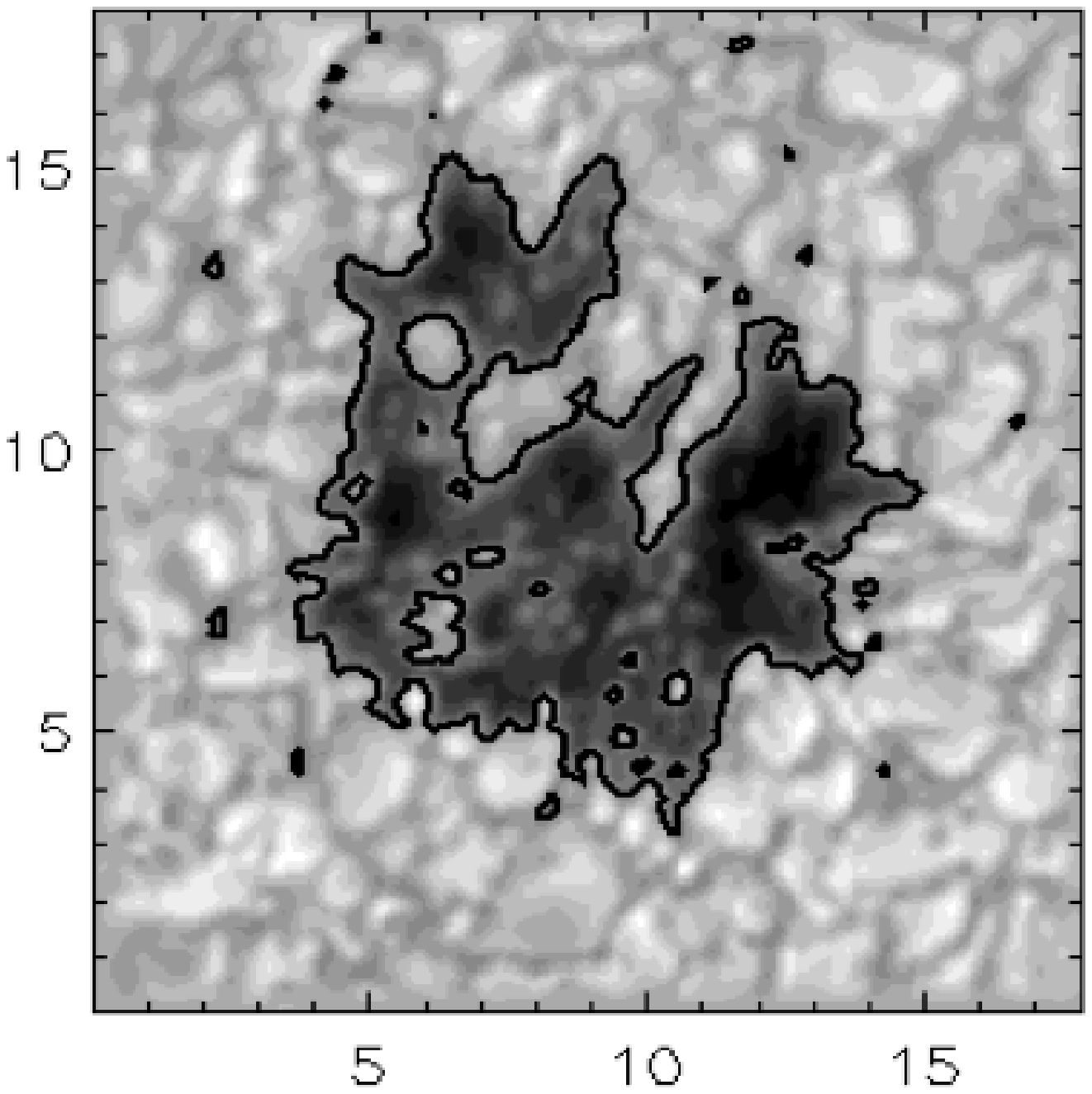}\includegraphics[width=.313\linewidth]{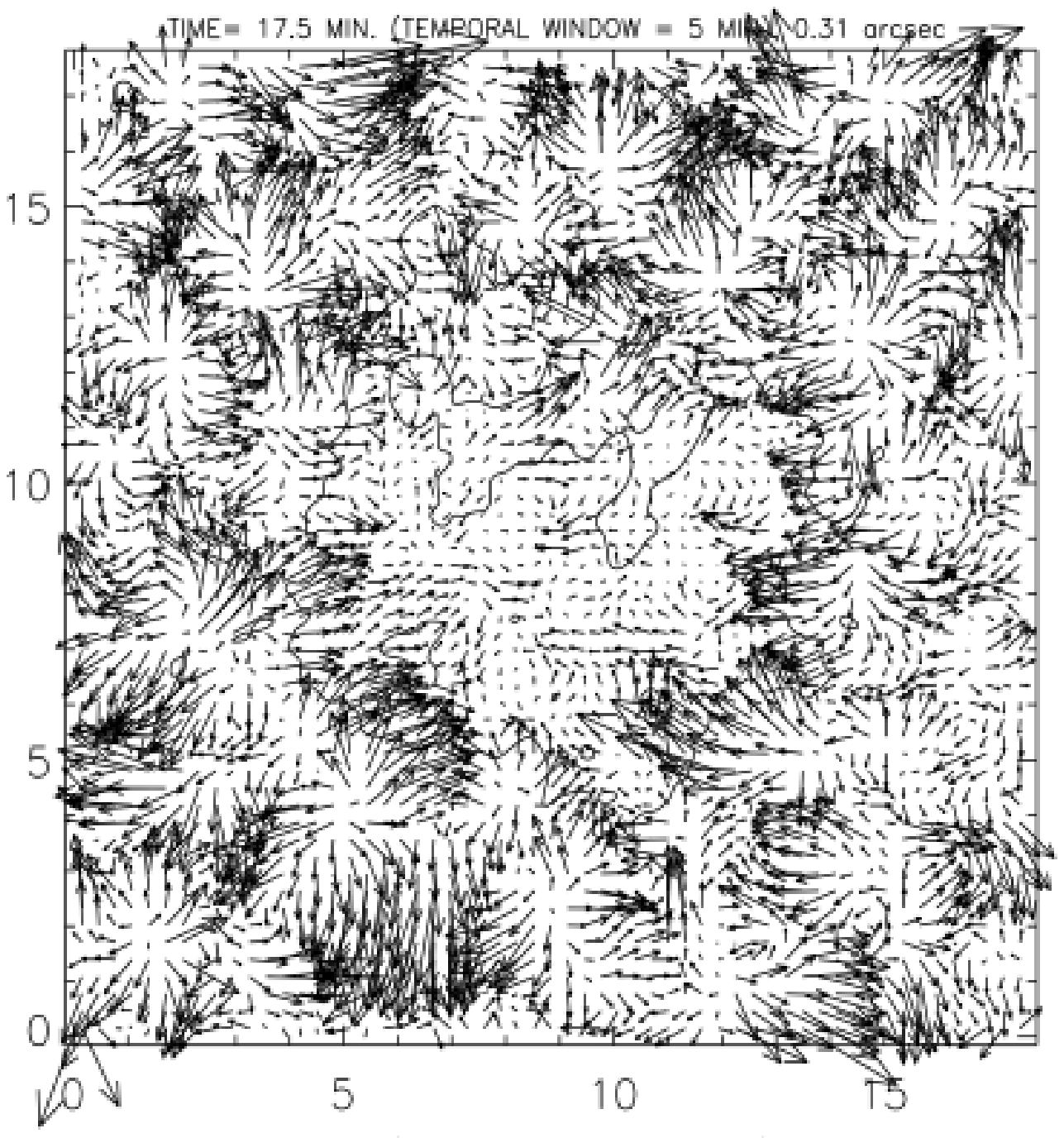}\includegraphics[width=.39\linewidth]{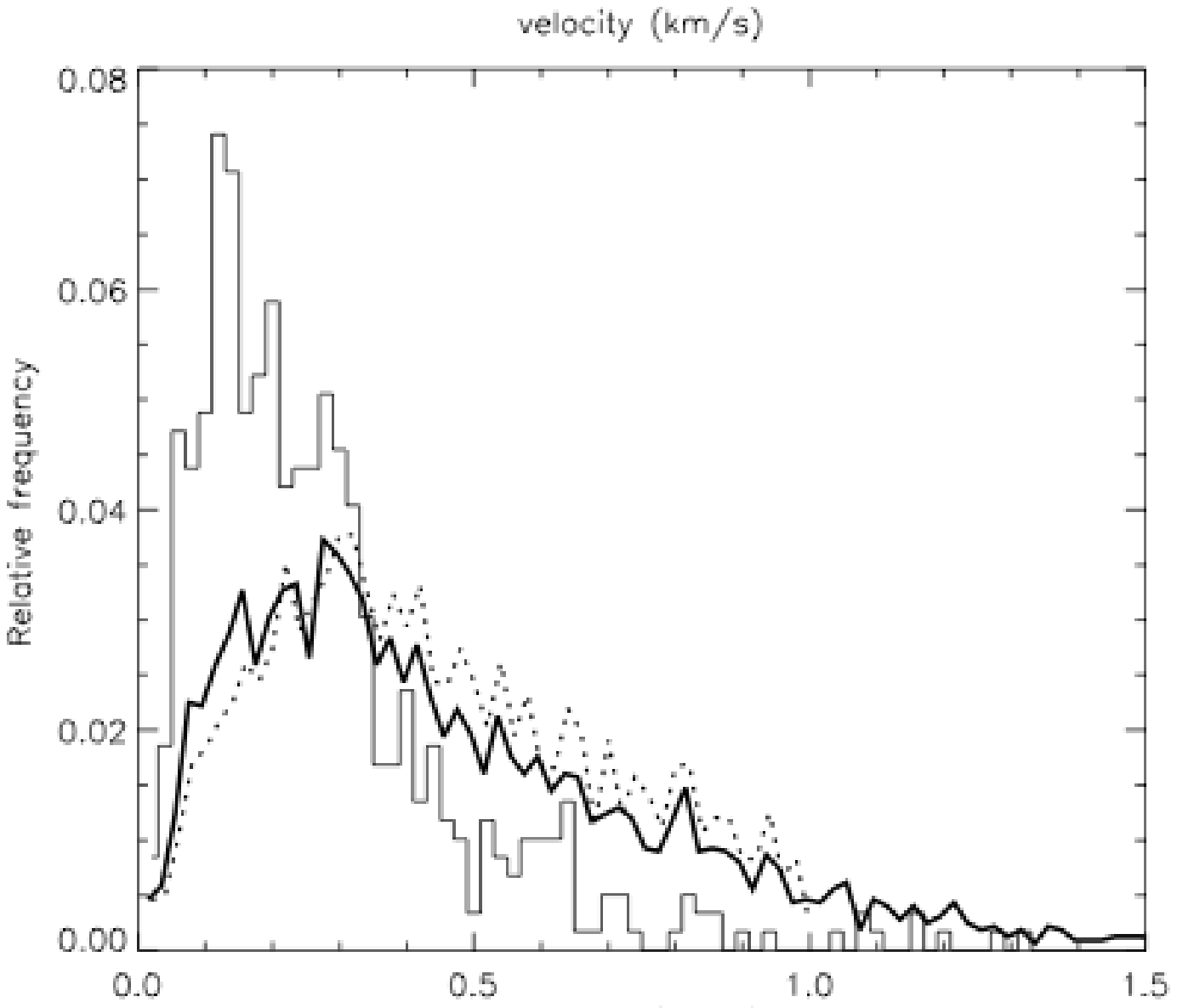}
\end{tabular}
\caption[\sf Horizontal flows observed in a solar pore on 1995 Jun 29.]{\sf Horizontal flows in a solar pore observed on 1995 June 29 and studied by \citet{roudier2002}. \emph{Left}: Image of the pore observed on 1995 Jun 29 at the Swedish Vacuum Solar Telescope. \emph{Center}: Horizontal velocity field with 0$\farcs$3 resolution and averaged over 5 min. \emph{Right}: Histogram of horizontal velocity magnitudes in the whole FOV (\emph{thick solid line}), inside (\emph{dotted line}) and outside the pore (\emph{thin solid line}).  The spatial scales in the first two panels are expressed in arc seconds.}
\label{roudierpores}
\end{figure}

Previous works \citep{sobotka1999, roudier2002, hirzberger2003} have measured horizontal proper motions in and around pores and have observed some penetrating flows at the umbral boundaries and a ringlike structure of positive divergence ("rosettas") around the pores which is related to a continuous activity of exploding granules in the granulation around them. \cite{roudier2002} identified a very clear inflow around pores which corresponds to the penetration of small granules and granular fragments from the photosphere into the pores, pushed by granular motions originated in the divergence centres around them. They conclude that the motions at the periphery of the pore are substantially and continuously influenced by the external plasma flows deposited by the exploding granules. Figure~\ref{roudierpores} shows the pore studied by \cite{roudier2002} and the horizontal velocity field averaged over 5 minutes where the centers of divergence are clearly identified forming the ringlike structure around the pore with mean velocities $\sim$ 0.3 km s$^{-1}$.\\

Pores are interesting to analyze since, as they do not display penumbrae \citep{keil1999}, what we actually observe is the direct interaction between the umbra (with a strong vertical magnetic field that inhibits convection inside it) and the convective plasma in the surrounding photosphere, without any intermediate structure in between. Many observed features such as bright granules moving in the border of a pore \citep{sobotka1999} show the complex exchange taking place between the pore and the granulation surrounding it. \\

In this chapter we aim to get velocity flow fields from high-resolution time series showing the evolution of solar regions having pores. Thus, we shall be able to analyze the horizontal flows around the solar pores under study to figure out from the flow maps how the plasma behaves. The data analyzed are sets taken at both ground-based and space telescopes. The observations were obtained with the Swedish 1-m Solar Telescope and the space telescope HINODE in 2007. In the next two sections the observations and the processing applied to the images are described in some detail. 

\section{Observations and data processing: SST Data}
\label{S:obsSST}

\subsection{Observations}
\label{S:obser}

The campaign at the Swedish 1-m Solar Telescope was carried out during September-October 2007 and corresponds to an important collaboration among several European and Japanese institutions. This was a long campaign (24 days) with joint observations using not only the SST but also three other solar telescopes  (DOT,VTT and THEMIS) at the Canary Islands Observatories in La Palma and Tenerife, respectively. Moreover, and for the very first time, coordinated observations with the space solar telescope HINODE \citep{kosugi2007} were performed as part of the Hinode Operation Program 14.\\

The data from the SST analyzed in this chapter were recorded during a particular observing run on 30 September 2007 and correspond to the active region NOAA 10971. The characteristics of the telescope and some important details concerning the observations are sketched and listed in Figure 4.1 and Table 4.1 respectively. The main target was a region close to the solar disc center ($\mu$=0.98) with some pores of different sizes embedded in a plage region that exhibits an intense magnetic activity.\\

The optical setup consisted in two channels: (\emph{blue} and \emph{red}). The blue channel was devoted to imaging and the red one to feed the SOUP filter. In the blue channel, images in G-band and CaII~H were acquired at a high cadence so that the MFBD restoration technique (see section~\S1.5) could be applied to correct for atmospheric and instrumental aberrations. Table~\ref{SSTblue} enumerates the main parameters of the images in the blue channel.\\

\begin{table}
\sffamily
\centering
\begin{tabular}{|l|cc|}\hline
FILTER & G-band & CaII~H \\\hline\hline
$\lambda$ [\AA] & 4305.6 &  3969 \\
$\Delta\lambda_{fwhm}$ [\AA] & 11.5 & 1.1 \\
Exposure time [ms] & $\sim$10 & $\sim$10\\
CCD (pixels) & 2048 $\times$ 2048  & 2048 $\times$ 2048\\
Field of view ["] & 69 $\times$ 69 &  69 $\times$ 69 \\
Pixel size ["] & 0.034 & 0.034\\\hline 
\end{tabular}
\caption[\sf Parameters of the observation for the 2007 campaign]{\sf Parameters for the blue channel used during the 2007 campaign.}
\label{SSTblue}
\end{table}

Along the red beam the SOUP filter was placed, as mentioned before. Section~\S4.2.1 details the description of the instrument and its operation. A beam splitter positioned in front of the SOUP deflected a fraction of light to obtain simultaneous broad-band PD image-pairs in the continuum near the selected wavelength (FeI 6302 \AA). These images made up an additional \emph{object} for the MOMFBD algorithm to jointly restore both, the broad-band and the narrow-band images (see section~\S1.5). From the restored narrow-band images we computed magnetograms and dopplergrams.\\

Once the observing run was over, only the best quality images (above a certain threshold) in both channels (blue and red) were preserved for further processing and the rest were discarded.\\

\subsection{Data processing}
\label{S:dataproc}

Post-processing of the images included first the standard flatfielding and dark current subtraction, as well as the elimination of hot and dark pixels and spurious borders in the flatfielded images. After these steps, the images were ready to be restored so that we could group them in sets of restoration.

\subsubsection{\bf Blue beam}
In the blue channel, G-band and CaII~H images were independently restored by employing MFBD. For each wavelength, the image sequence was grouped in sets of about 80 frames acquired within time intervals of 10 seconds each. Every set yielded one restored image. So, from $\sim$ 52640 single exposures in G-band and as many in CaII~H (more than 10$^5$ images adding G-band and CaII~H) we obtained a total of 658 restored images per wavelength.\\

Next steps in the data post-processing were: compensation for diurnal field rotation, rigid alignment of the images, correction for distortion and finally subsonic filtering. For more details see section~\S4.3.3 where similar procedures are described for the SST data of the 2005 campaign.\\

\begin{figure}
\centering
\includegraphics[width=.8\linewidth]{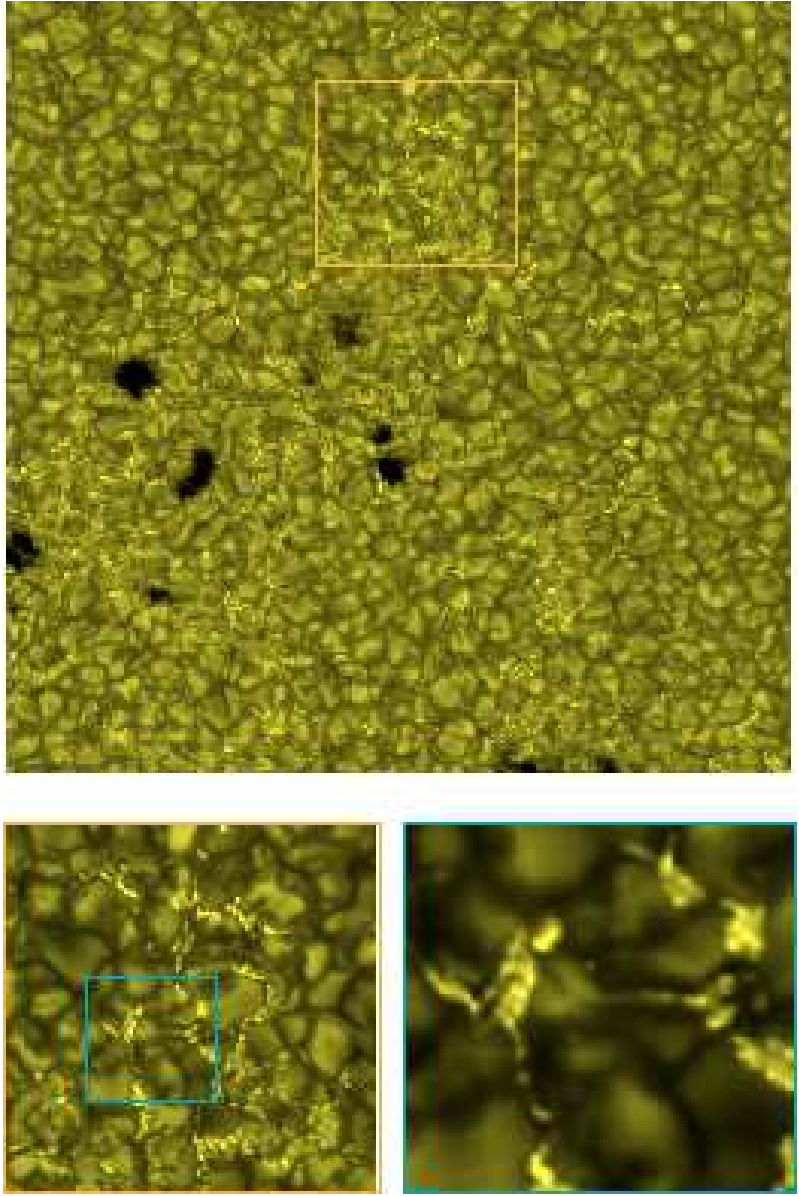} \\
\caption[\sf G-band restored image of the emerging flux region on 30 September 2007]{\sf G-band restored image  (false color) of the emerging active region on 30 September 2007. Small colored boxes are close-ups showing the high-resolution reached after the restoration process that enable us to detail very tiny structures in the bright features placed in the intergranular lanes. The typical size of the granules in the image is $\sim$ 2$\arcsec$.}
\label{images_blue}
\end{figure}

Due to periods of bad seeing in which the quality did not reach the desired top level, some of the images were discarded and we kept only the best and longest consecutive sequence of images. The very final product were 4 time series (2 for G-band, 2 for CaII~H) with the characteristics listed in Table~\ref{poroseries}. The time gap of about 5 min between the two consecutive series for each wavelength was a consequence of a telescope tracking interruption. That is also why the FOV is slightly different in both time series. Most images forming the time series show details near the diffraction limit of the telescope. Figure~\ref{images_blue} displays one of the G-band restored images (in false color) with close-ups enhancing very tiny bright features in the intergranular lanes. \\

\begin{table}
\sffamily
\centering
{ \bf  {\large SST Data }} 
\begin{tabular}{ccccccc}\\
Date 2007  & Serie & Time & Duration & N. images & Cadence & FOV \\
& & {\footnotesize (UT)} & {\footnotesize[min]} & & {\footnotesize[sec]} & {\footnotesize["]}\\\hline\hline
\multirow{2}{*}{30 Sep ..} & 1 & 08:43-09:31 & 48 & 286 & 10 & 64.3$\times$65.0  \\
& 2 & 09:36-09:56 & 20 & 118 & 10 &  64.3$\times$65.0 \\\hline
\end{tabular}
\caption[\sf Time series of solar pores observed with the SST]{\sf Characteristics of the time series of solar pores observed on 30 September 2007 with the SST. For both wavelengths observed (G-band and CaII~H) we have two corresponding time series as listed in the table, with common parameters.}
\label{poroseries}
\end{table}

\subsubsection{\bf Red beam}
\begin{figure}
\centering
\begin{tabular}{cc}
\sf G-band & \sf H$\alpha$ \\
\hspace{-1.5cm}\includegraphics[width=.52\linewidth]{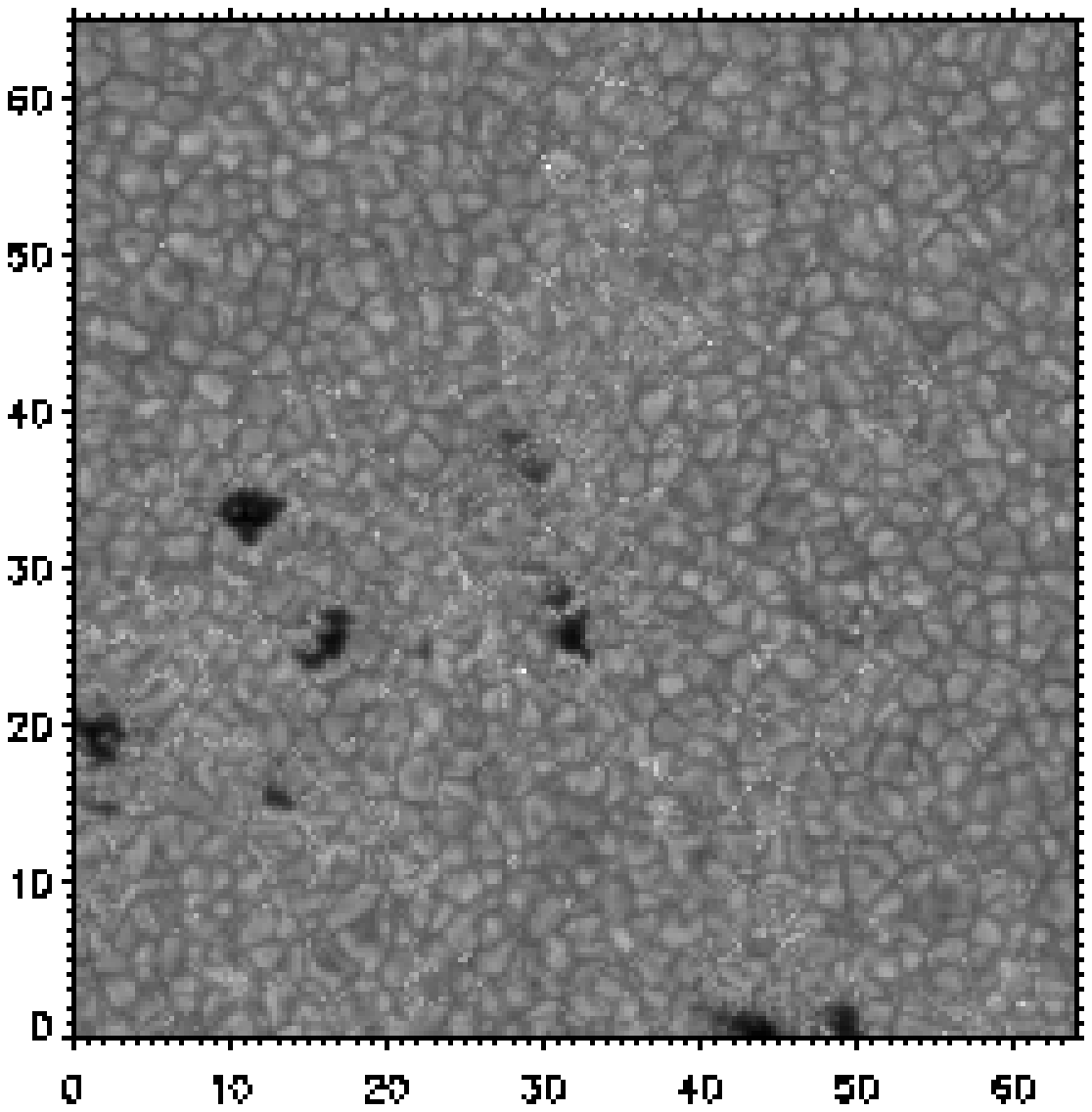} &
\hspace{-1.5cm}\includegraphics[width=.52\linewidth]{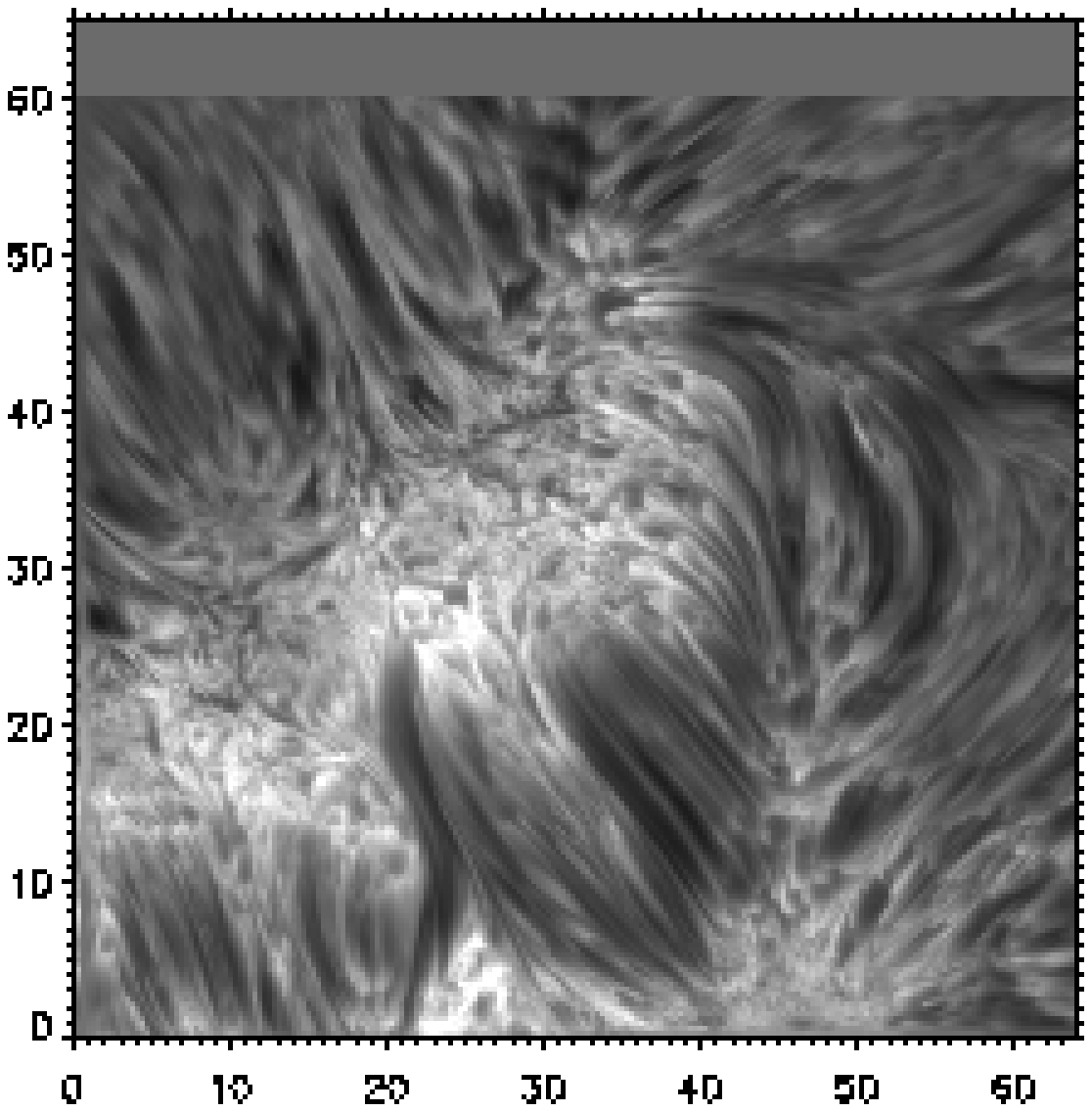}\\
\sf Dopplergram & \sf Magnetogram \\
\hspace{-1.5cm}\includegraphics[width=.52\linewidth]{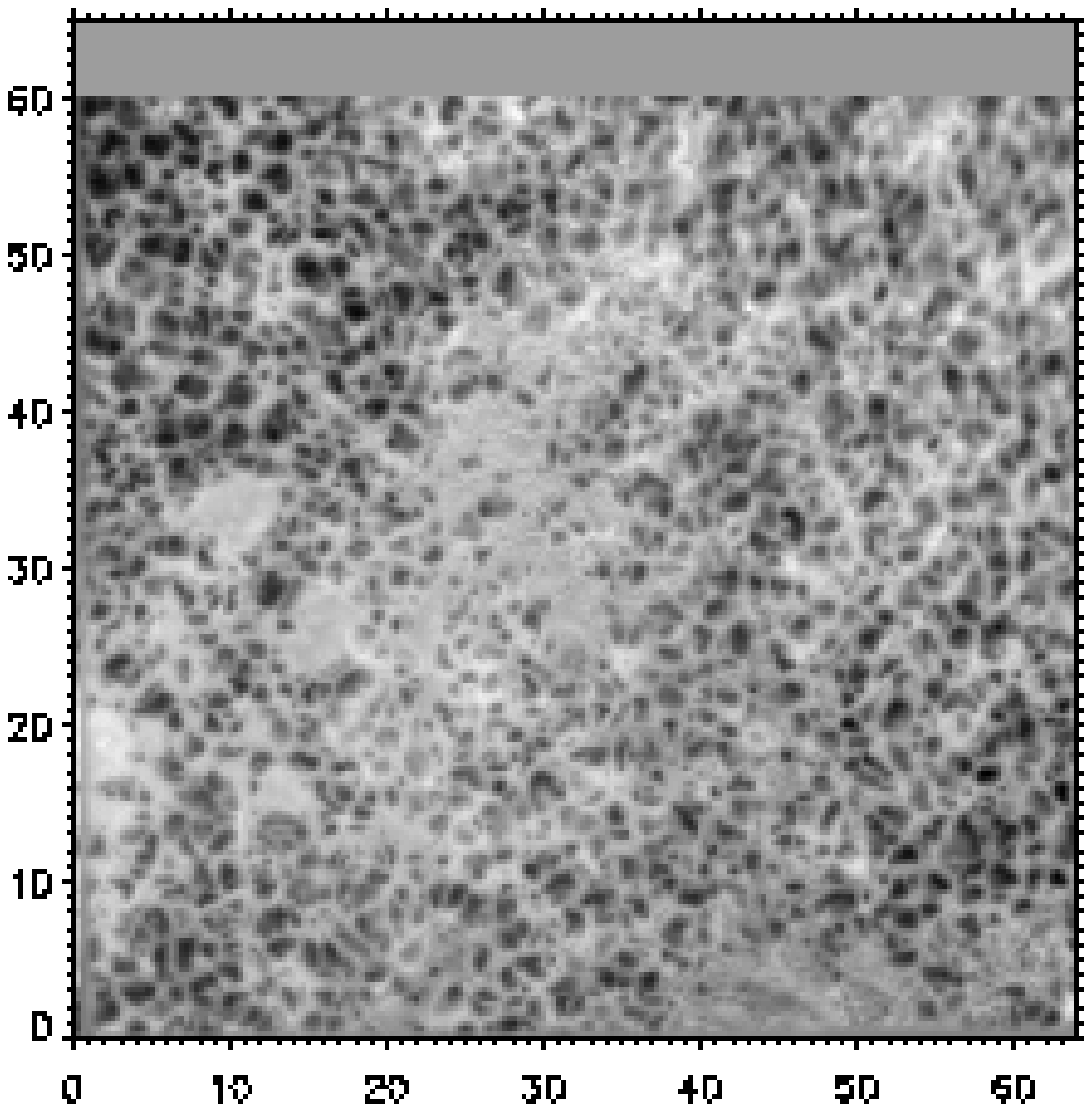} & 
\hspace{-1.5cm}\includegraphics[width=.52\linewidth]{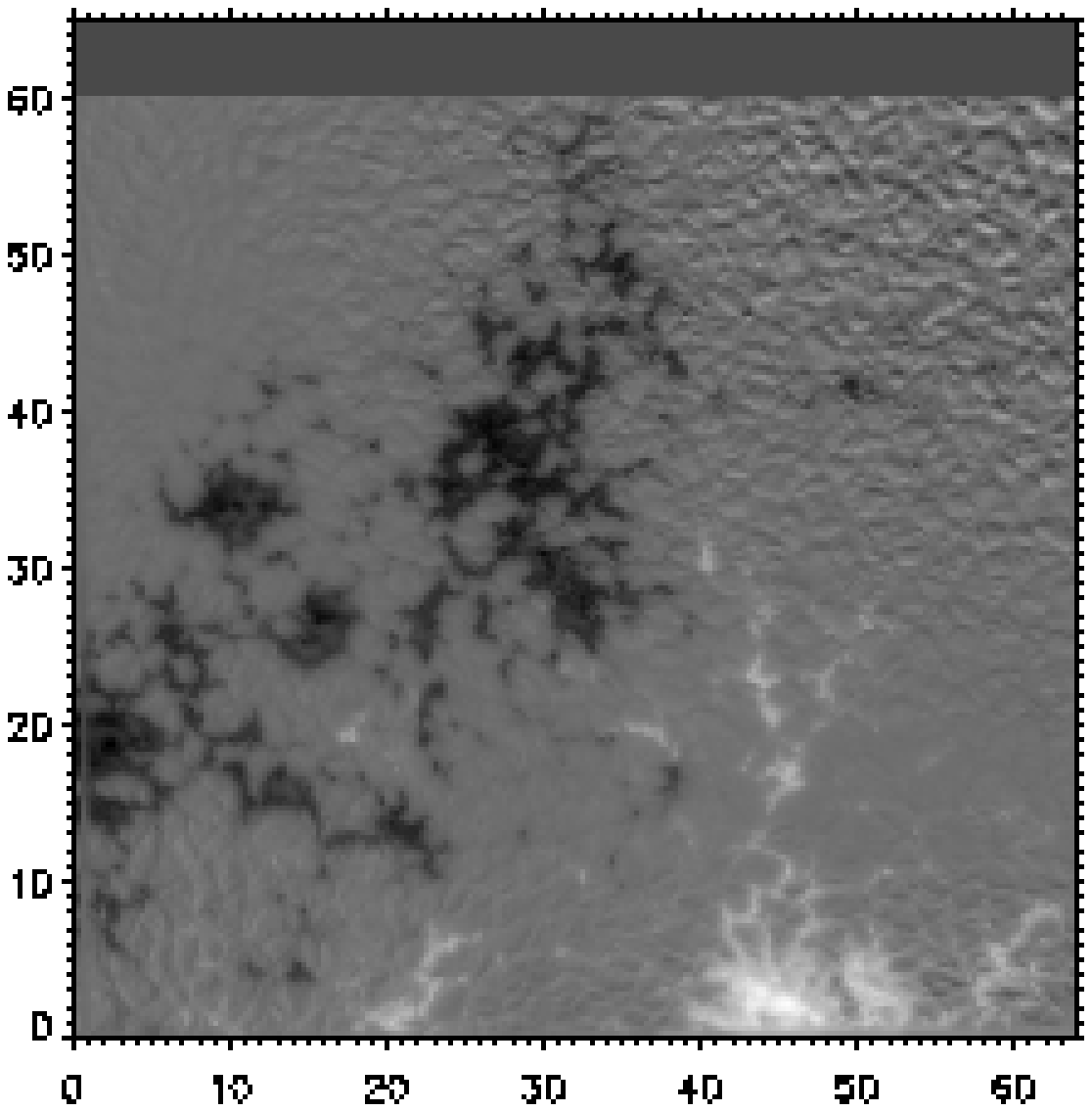} \\

\end{tabular}
\caption[\sf Red channel images of the emerging active region on 30 September 2007]{\sf Co-temporal and co-spatial set of images from the emerging flux region on 30 September 2007. The coordinates are expressed in arc sec.}
\label{images_red}
\end{figure}

In the red beam images were restored with MOMFBD. The procedure followed for image restoration and for obtaining magnetograms and dopplergrams is the same as that applied to the observing material of the 2005 campaign (see section~\S4.3.5). Figure~\ref{images_red} shows one of the co-temporal sets of images of the emerging active region after restorations.

\section{Observations and data processing: HINODE Data}
\label{S:obsHinode}

After launched on the 22$^{nd}$ September 2006 HINODE has become a very useful tool in the solar physics research. It has played an important role since it started pointing to the Sun and managed to observe many solar features in such a high detail by avoiding the distortion effects produced by the Earth's atmosphere. The public archive of HINODE \footnote{\sf See the website: \emph{http://solar-b.nao.ac.jp/hsc\_e/darts\_e.shtml}} is an organized data base where all HINODE observations can be searched and downloaded in such an easy way. \\

We were interested in HINODE observations of solar pores taken with the Solar Optical Telescope (SOT), in order to pursue our study of photospheric flows as explained in section~\S\ref{intro}. Next two sections describe the HINODE data we analyze in the present work. These data extend the sample of cases under study in the present work and moreover give us the possibility to compare with the results stemming from ground-based observations.

\subsection{Data on 1 June 2007}
\label{1jun}

Solar portion including an isolated and round-shaped pore observed by HINODE on 1 June 2007. Images were acquired in G-band with a cadence of 30 seconds covering a FOV of 27$\farcs$85 $\times$ 55$\farcs$70 close to the disk center ($\mu$=0.87). After subsonic filtering we obtained two time series: one from 21:35 to 21:55 UT with 40 images and a second one from 22:26 to 23:33 UT with 134 images. Table~\ref{poroseriesHINODE} (1 Jun) summarizes the parameters of the time series in more detail.\\

The treatment performed with HINODE data does not include any restoration process because of their excellent quality as a consequence of the absence of atmospheric turbulence.
%

\subsection{Data on 30 September 2007}
\label{30sep}

\begin{figure}
\centering
\includegraphics[angle=90,width=.77\linewidth]{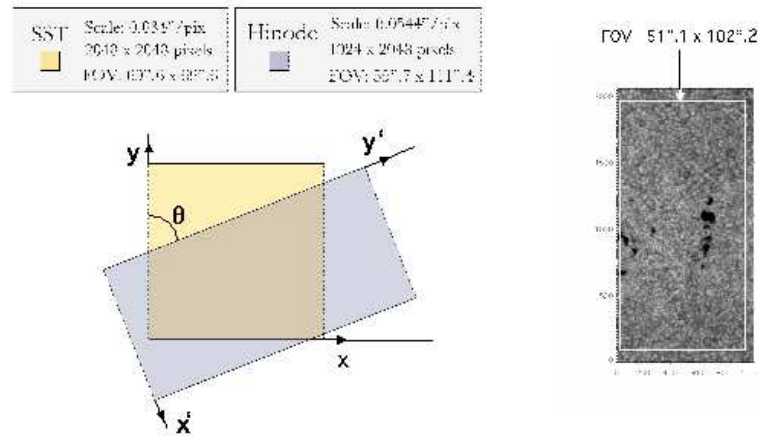}
\vspace{-4mm}
\caption[\sf HINODE data on 30 September 2007]{\sf \emph{Left}: Sketch showing the correspondence between SST and HINODE observed FOVs on 30 September 2007 during a coordinated campaign. Black arrows denote the two orthogonal coordinate systems and $\theta$ stands for the rotation angle between them that amount to $\sim$~65 degrees. \emph{Right}: First image of the series taken with HINODE on 30 September 2007 at 00:14:55 UT in G-band. The original FOV and the final one after data processing are shown for comparison. The spatial coordinates are in pixels.}
\label{FOV_SST_Hinode}
\end{figure}

The data from HINODE on 30 September 2007 corresponds to the coordinated observations described in section~\S\ref{S:obser}.  HINODE observed in G-band the emerging flux region during almost the whole day from 00:14 to 17:59 UT with a few brief interruptions for calibrations. Table~\ref{poroseriesHINODE} (30 Sep) summarizes the parameters of the time series in more detail. Figure~\ref{FOV_SST_Hinode} (\emph{left}) sketches the correspondence between the FOV observed by HINODE (51$\farcs$1 $\times$ 102$\farcs$2) and the one framed by the SST (69$\farcs$6 $\times$ 69$\farcs$6).  A large region of the SST FOV is covered by HINODE observations as shown in the figure.\\

HINODE data presents quite a few misalignments due to tiny tracking flaws and also to every temporal interruption. Hence, we decided to align all the 1030 images (18 hours) at a sub-pixel level. Taking the first image as the reference, next consecutive images were aligned to it by using a routine that made a first alignment approach (in case of large misalignments). In a second iteration, the alignment with the first image was refined by correlating pairs of subsequent images and accumulating their misalignments. Once all images were perfectly aligned, the FOV was clipped to get rid of spurious borders resulting from the alignment process. \\

Next step on the processing is the subsonic filtering of the series. This task resulted a bit tedious not only because of the short temporal interruptions (a total of 7 during the 18 hours) in the series but also because of the big size of the images and the large amount of them. To overcome this drawback we fragmented the total 18 hours series in 8 data cubes of about 130 images each so that consecutive images in every cube were equally spaced. Subsonic filtering was applied independently to every cube. The final FOV (57$\farcs$1 $\times$ 102$\farcs$2) is shown in Figure~\ref{FOV_SST_Hinode} (\emph{right panel}) for the first image at 00:14:55 UT.

\begin{table}
\sffamily
\centering
{ \bf  {\large HINODE Data }} 
\begin{tabular}{ccccccc} \\ 
Date 2007  & Serie & Time & Duration & N. images & Cadence & FOV\\
& & {\footnotesize (UT)} & {\footnotesize[min]} & & {\footnotesize[sec]} & {\footnotesize["]}\\\hline\hline
\multirow{2}{*}{1 Jun ...} & 1 & 21:35-21:55 & 20 & 40 & 30 & 27.9$\times$55.7 \\
& 2 & 22:26-23:33 & 67 & 134 & 30 &  64.26$\times$65.0 \\\\
30 Sep ...& 1 & 00:14-17:59 &  960 & 1030 & 60 & 55.7$\times$111.4  \\\hline
\end{tabular}
\caption[\sf Time series of solar pores observed with HINODE]{\sf Characteristics of the time series of solar pores observed on 1 June and 30 September 2007 with HINODE.}
\label{poroseriesHINODE}
\end{table}

\section{Data analysis and results}
\label{S:res}

\subsection{General description of horizontal proper motions in the FOV}
\label{horizontalpattern}
The G-band series taken from SST and HINODE have been used to analyze the horizontal proper motions of structures in every FOV. Proper motions were measured by using the same LCT-technique employed in previous chapters. In this section we show the maps of horizontal velocities calculated for the different time series using a Gaussian tracking window of FWHM $1.0 \arcsec$ which is roughly half of the typical granular size. \\
 
\begin{figure}
\centering
\vspace{-3mm}
\begin{tabular}{c}
{\footnotesize \sf SST 30.09.2007 / AVERAGE: 08:43-09:31} \\
\vspace{-8mm}\hspace{-2cm}\includegraphics[width=.72\linewidth]{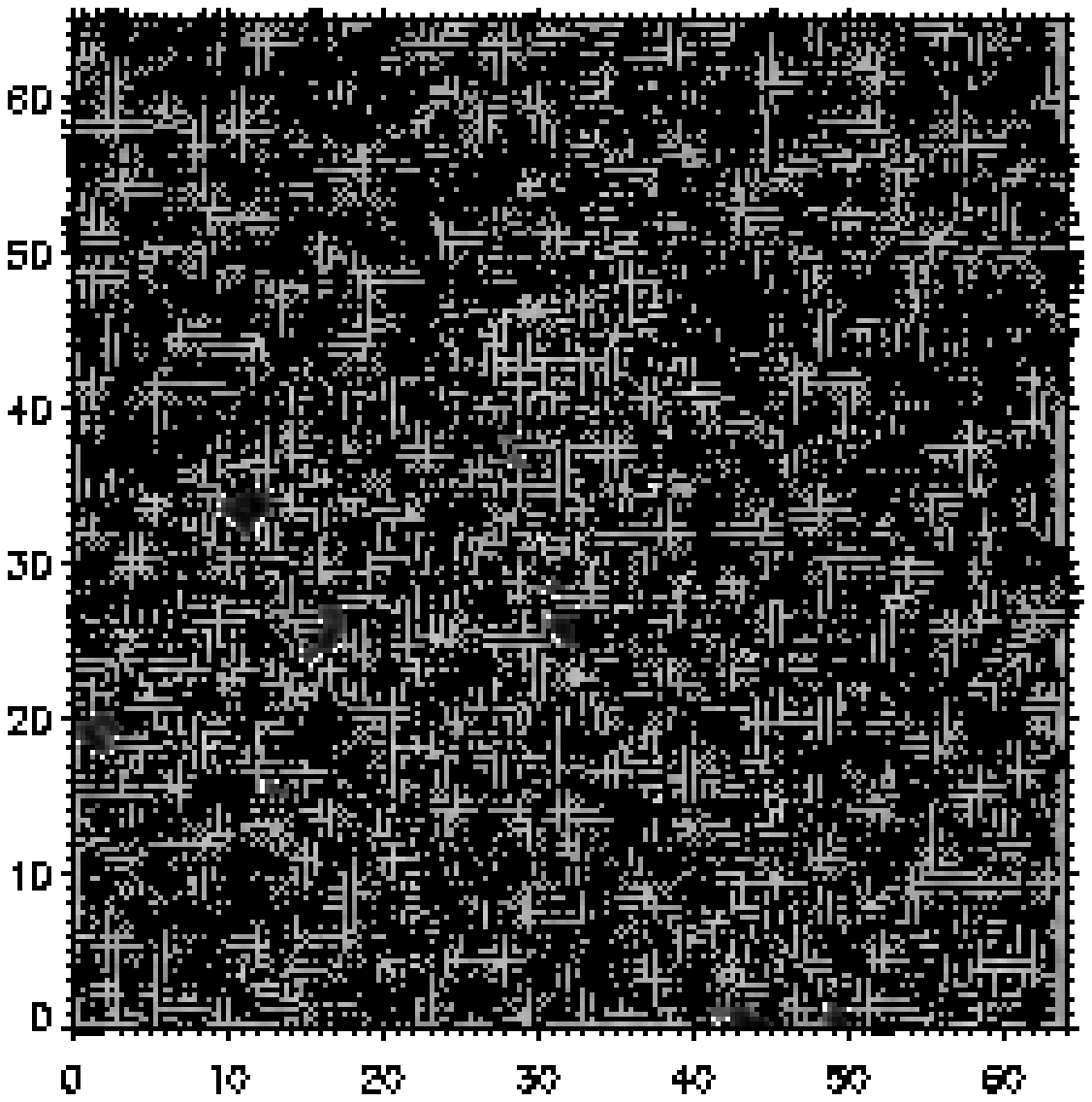} \\
{\footnotesize \sf SST 30.09.2007 / AVERAGE: 09:36-09:56} \\
\hspace{-2cm}\includegraphics[width=.72\linewidth]{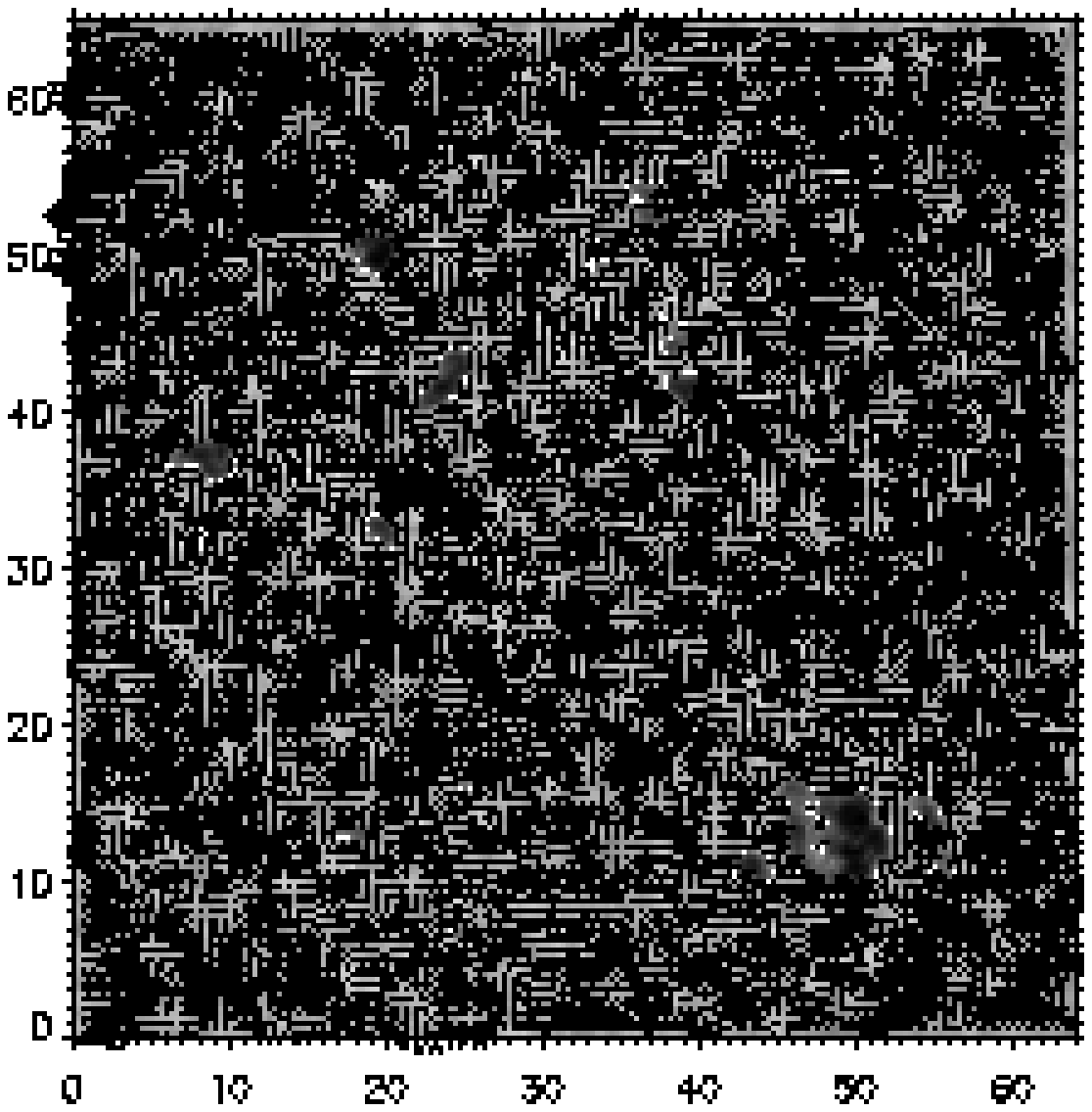}   
\end{tabular}
\vspace{-1.1cm}
\caption[\sf Horizontal flows observed in solar pores regions with SST data]{\sf Map of horizontal velocities (FWHM 1$\farcs$0) from the restored time series taken at the SST on September 30, 2007. The velocities are averaged over 48 minutes (\emph{upper panel}) and 20 minutes (\emph{lower panel}). The \emph{white contours} outline the border of solar pores. The length of the black bar at coordinates (0,0) corresponds to 1.6 km s$^{-1}$. The coordinates are expressed in arc sec. The background represents the average image of the G-band series in every case, respectively.}
\label{porosflowmapSST}
\end{figure}

Figure~\ref{porosflowmapSST} shows the flow maps computed from the two restored time series in the SST. The velocities were averaged over the total duration of each series. The underlying background in the maps is the average image of the respective series. As commented above, the FOV is slightly different in both cases so that, for instance, pore at coordinates (17,26) in the upper map of the figure is located at (24,42) in the lower map.\\

The maps are dominated by flows coming from exploding granules taking place all over the FOV. The map averaging over a longer time period (48 min) is slightly smoother than the other one (20 min), and displays rather lower velocities. Nevertheless, both maps (48 and 20 min average) reproduce in general the same proper motions all over the FOV. The top of the FOV in Figure~\ref{porosflowmapSST} (\emph{upper panel}) shows large flows caused by exploding granular events. These  events are grouped at every upper corner of the FOV forming two structures that fit well the supergranular scale (see also the magnetogram in Figure~\ref{images_red}). A smaller portion of these structures can also be identified in the lower panel of Figure~\ref{porosflowmapSST}. Even though a complete description of the proper motions around the pores in the FOV will be done in section~\S\ref{velpores}, a glance at the figure reveals no evidence of a moat-like pattern around any of the pores. The central part of the FOV where the smaller pores are embedded, exhibits a lower magnitude of horizontal velocities. This behavior is explained by the intense magnetic activity in this part of the FOV as unveiled by the correspondent magnetogram of the zone shown in Figure~\ref{images_red}.\\

\begin{figure}
\centering
\begin{tabular}{c}
{\footnotesize \sf HINODE 01.06.2007 / AVERAGE: 22:26-23:33} \\
\hspace{-3cm}\includegraphics[width=.8\linewidth]{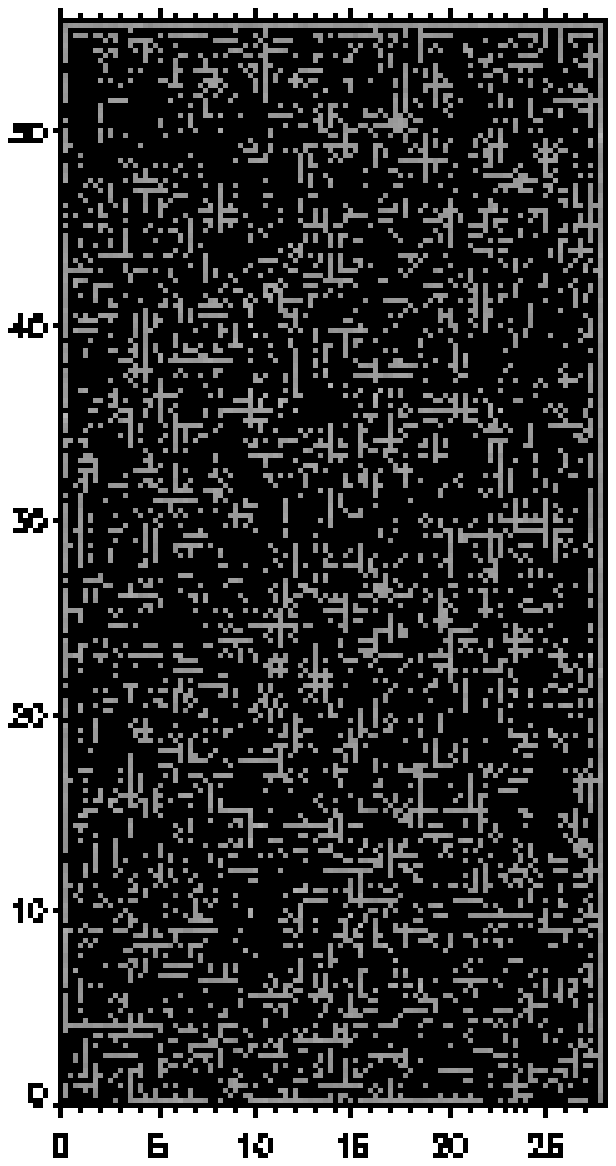} \\
\end{tabular}
\vspace{-6mm}
\caption[\sf Horizontal flows observed in solar pores regions with HINODE data]{\sf Maps of horizontal velocities (FWHM 1$\farcs$0, 67 minutes average) for the processed time series taken by HINODE on June 1, 2007. The length of the black bar at coordinates (0,0) corresponds to 2.3 km s$^{-1}$. The coordinates are expressed in arc sec. The background represents the average image of the G-band series.}
\label{porosflowmapHinode}
\end{figure}

Concerning HINODE data, the computed map of horizontal velocities for the time series on 1 June 2007 is displayed in Figure~\ref{porosflowmapHinode}. The flow map shown in this figure is calculated for the longest time series (67 minutes, see Table~\ref{poroseriesHINODE}). The isolated solar pore is immerse in a granular region displaying several and recurrent large exploding granular events all around it. The connection of the centers of the exploding events outlines a round-shaped contour at a distance from the pore border similar to the pore radius. The periphery of the pore is hereby influenced by plasma flows being deposited by exploding granular events. Velocity magnitudes in the outer part of the outlined contour are clearly larger than those in the inner part between the contour and the pore border.


\subsection{Averaging horizontal flows in different time intervals}
\label{averages}

\begin{figure}
\centering
\vspace{-3mm}
\begin{tabular}{c}
{\footnotesize \sf SST 30.09.2007 / AVERAGE: 5 min} \\
\vspace{-8mm}\hspace{-1.5cm}\includegraphics[width=.72\linewidth]{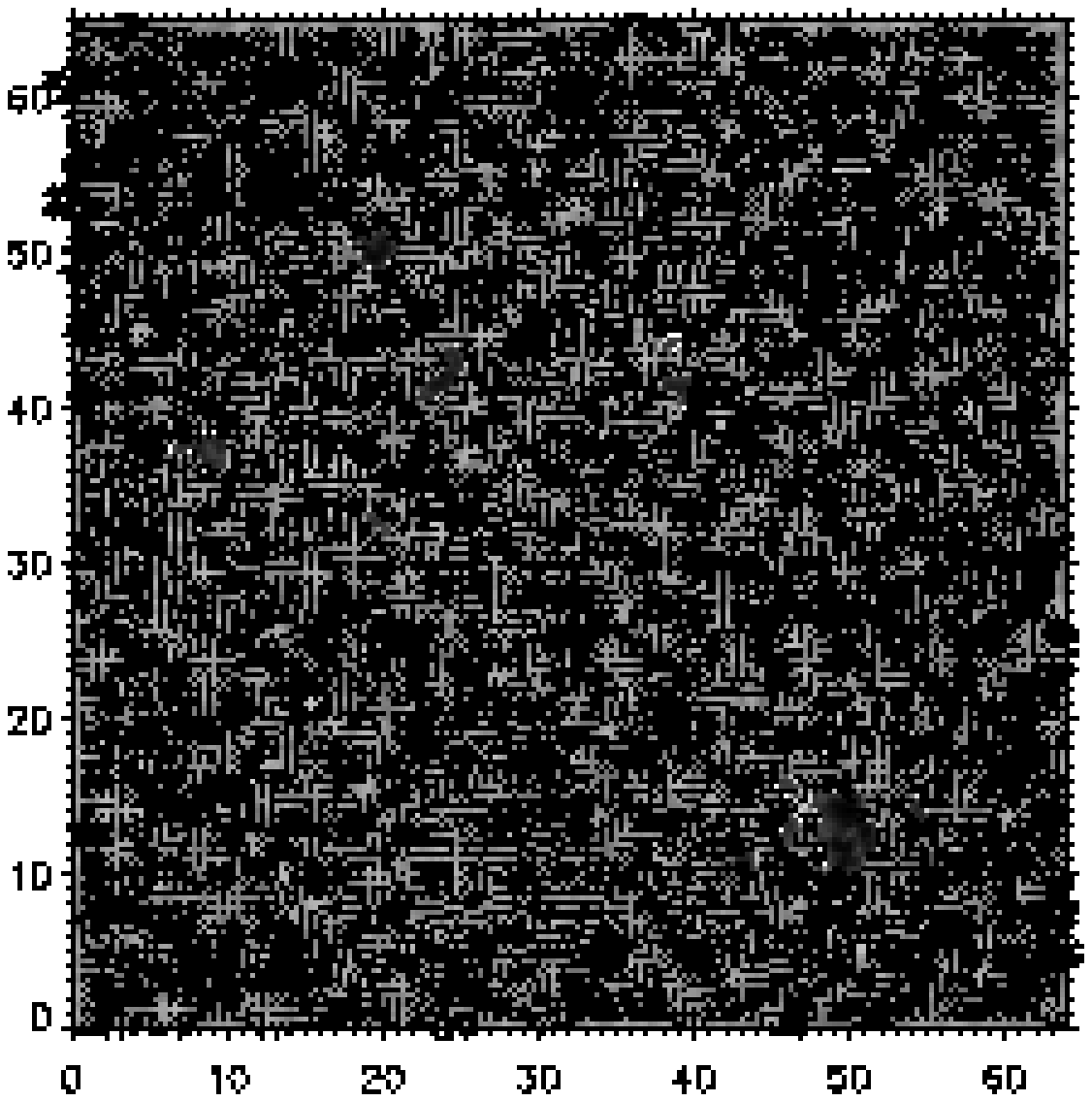} \\ 
{\footnotesize \sf SST 30.09.2007 / AVERAGE: 10 min} \\
\hspace{-1.5cm}\includegraphics[width=.72\linewidth]{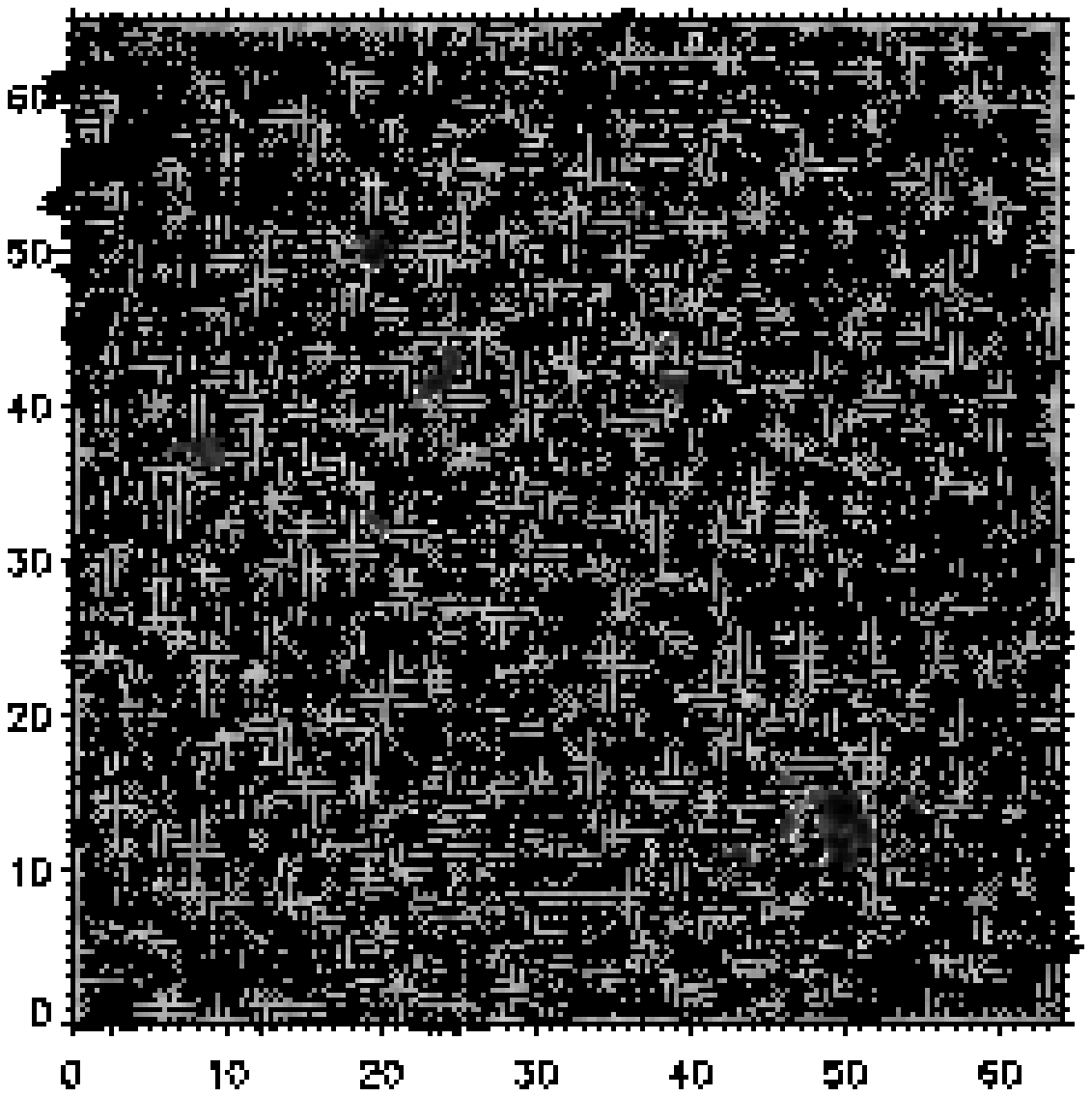} 
\end{tabular}
\vspace{-1.1cm}
\caption[\sf Horizontal flows observed in solar pores regions with SST data averaged over different time intervals]{\sf Maps of horizontal velocities (FWHM 1$\farcs$0) for the restored time series taken at the SST on September 30, 2007, averaged over different time intervals: 5 minutes (\emph{upper panel}) and 10 minutes (\emph{lower panel}). The \emph{white contours} outline the border of solar pores. The length of the black bar at coordinates (0,0) corresponds to 1.6 km s$^{-1}$. The coordinates are expressed in arc sec. The background represents the average image of the G-band series in every case, respectively.}
\label{porosflowmapSSTinterv1}
\end{figure}

\begin{figure}
\centering
\vspace{-3mm}
\begin{tabular}{c}
{\footnotesize \sf SST 30.09.2007 / AVERAGE: 15 min} \\
\vspace{-8mm}\hspace{-1.5cm}\includegraphics[width=.72\linewidth]{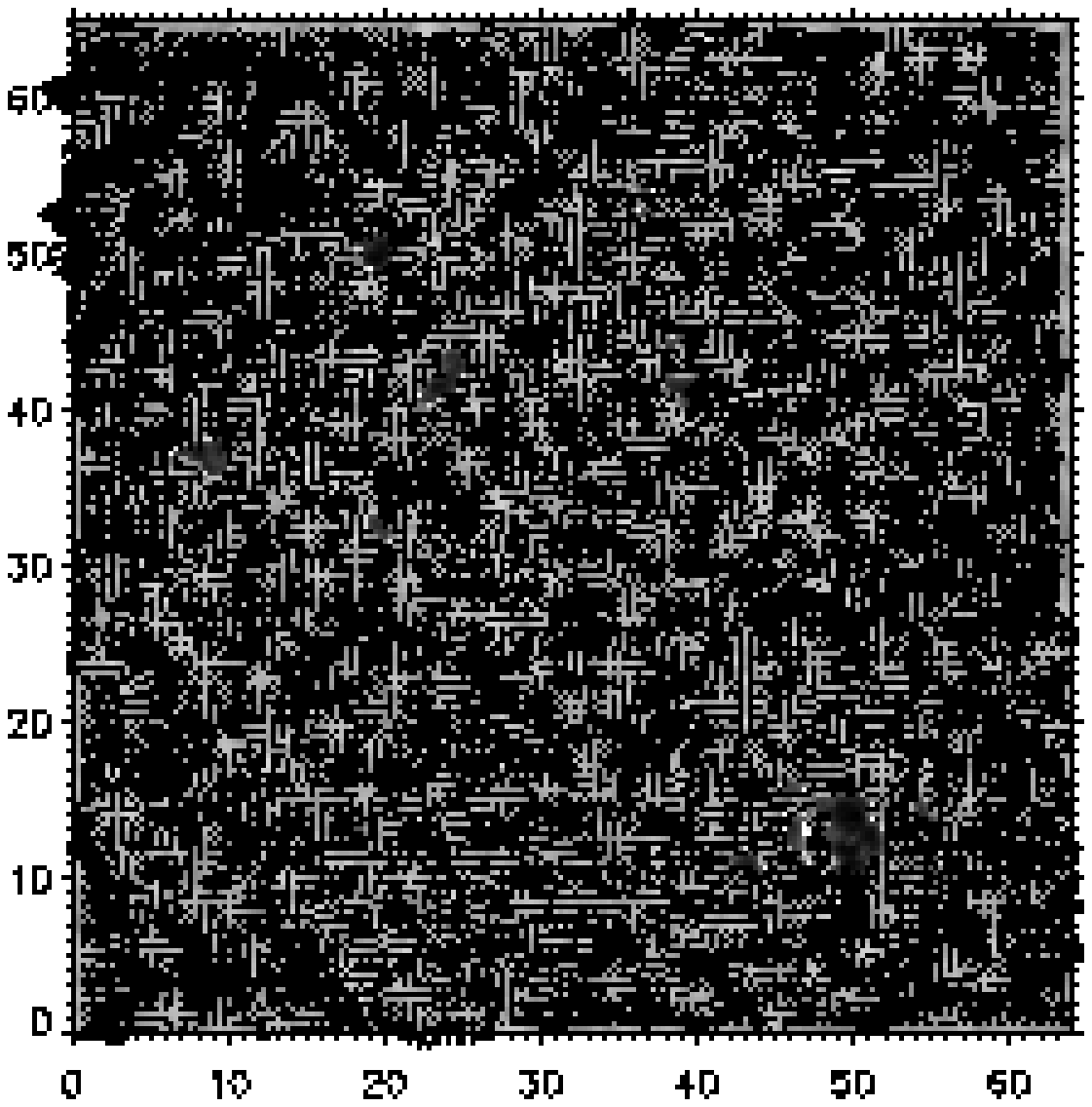} \\
{\footnotesize \sf SST 30.09.2007 / AVERAGE: 20 min} \\
\hspace{-1.5cm}\includegraphics[width=.72\linewidth]{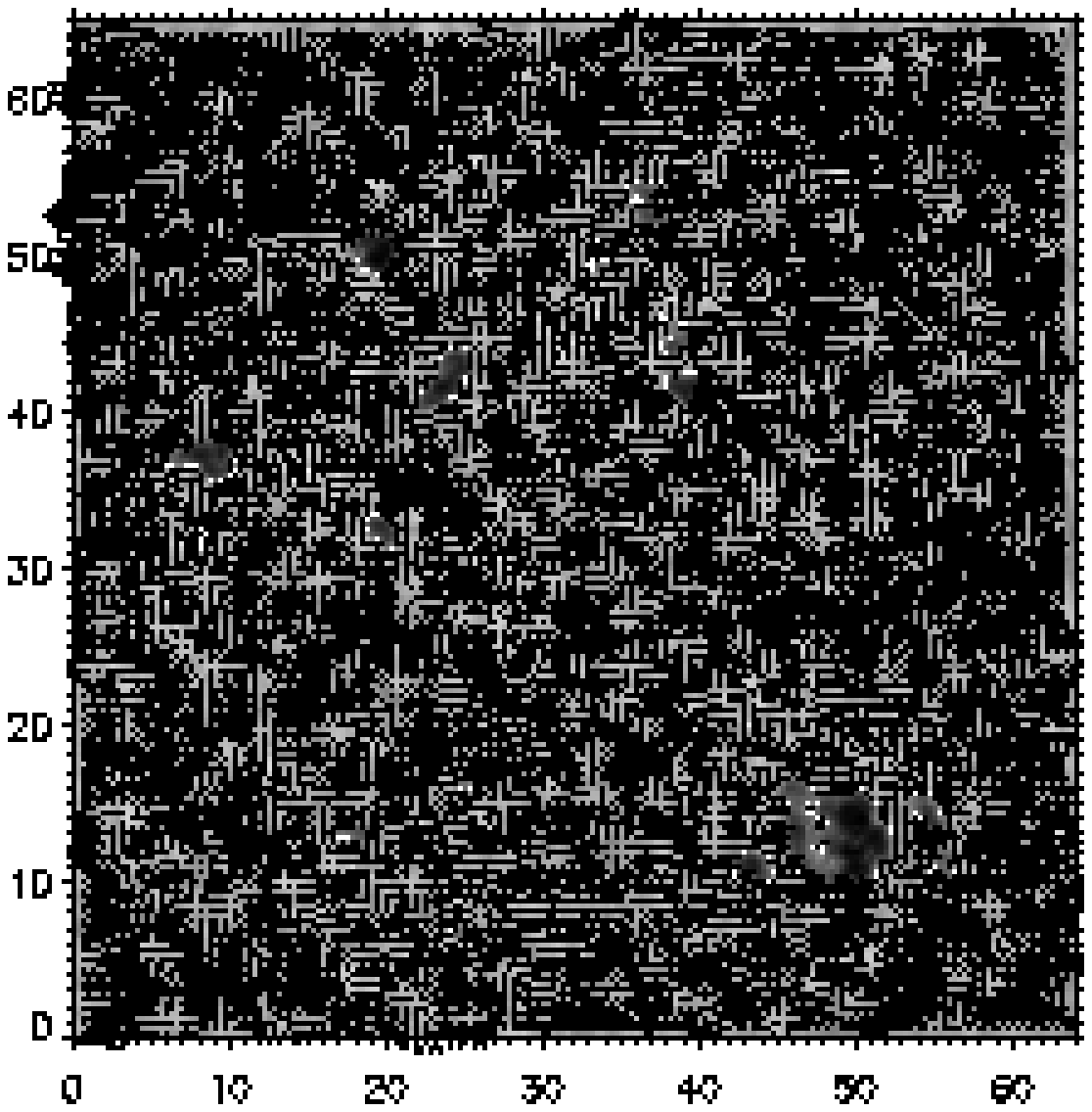}
\end{tabular}
\vspace{-1.1cm}
\caption[\sf Horizontal flows observed in solar pores regions with SST data averaged over different time intervals]{\sf Maps of horizontal velocities (FWHM 1$\farcs$0) for the restored time series taken at the SST on September 30, 2007, averaged over different time intervals: 15 minutes (\emph{upper panel}) and 20 minutes (\emph{lower panel}). The \emph{white contours} outline the border of solar pores. The length of the black bar at coordinates (0,0) corresponds to 1.6 km s$^{-1}$. The coordinates are expressed in arc sec. The background represents the average image of the G-band series in every case respectively.}
\label{porosflowmapSSTinterv2}
\end{figure}

Here we study the influence of different time averages of velocity maps in a time series of solar pores. To that aim we consider averaging intervals of 5, 10, 15 and 20 min in the best quality G-band time series recorded at the SST (the second series in Table~\ref{poroseries}). Figures~\ref{porosflowmapSSTinterv1} and \ref{porosflowmapSSTinterv2} show the resulting mean velocity map for each case. As expected, the maps are smoother for longer averaging periods. Note that regardless we average over 5 or 20 min intervals, we find essentially the same general trends and exploding granular patterns all over the FOV, meaning that the averaging periods we are employing are smaller than or about the lifetime of the observed structures, and that even from a short time sequence we get a reliable picture of the plasma dynamics in the FOV. It is also worth mentioning that we do not recognize any moat-like flow around the pores for such different time averages.\\

\begin{figure}
\centering
\hspace{-1.cm}\includegraphics[width=.8\linewidth]{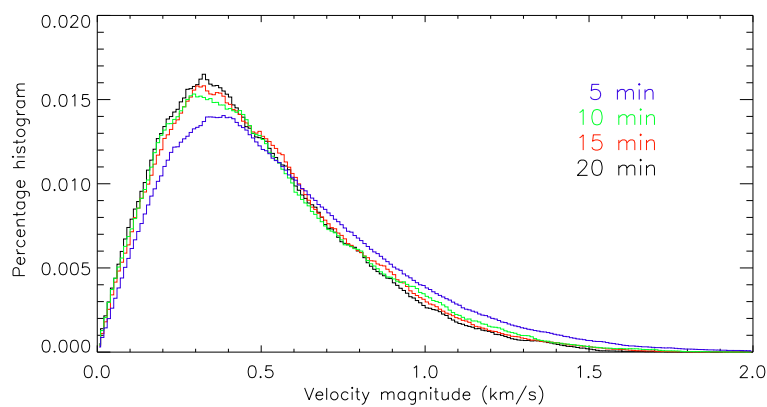}
\caption[\sf Histogram of horizontal velocity magnitudes for different averaging time intervals]{\sf Histogram of horizontal velocity magnitudes for different time intervals. A local correlation tracking technique has been used to derive the horizontal velocities (FWHM=1$\farcs$0).}
\label{histogram_averages}
\end{figure}

The statistics of the velocity magnitudes for every time average has been computed and the resulting histograms are shown in Figure~\ref{histogram_averages}. Although the distribution of velocity magnitudes is very similar in the four cases considered, one can easily detect that the histograms shift to the left as long as the averaging periods increase. Thus, the largest velocity magnitudes range from 1.83 to 1.97 km s$^{-1}$ and the mean values from 0.48 to 0.50 km s$^{-1}$. In both cases, smaller/larger values correspond to longer/shorter time averages.

\subsection{Long-term evolution of the velocity field}
\label{evol_map}
In order to investigate how the evolution of the emerging region affects the velocity in the FOV for long periods of time, we have used the HINODE time series lasting for several hours. HINODE data on 30 September 2007 corresponds to 18 hours of almost continuous solar observation as reported in section~\S\ref{30sep}. The stability and long duration of the series is ideal for the analysis of proper motions in the emerging flux region. In particular we are interested in the evolution of the maps of horizontal velocities averaged over every hour. We have computed 14 of this maps starting from the first hour of observation 00:14 UT up to the time 14:00 UT; during this long period of time we have the region of interest in our FOV. Afterwards, because of failures in the telescope tracking and a stop for calibration purposes, the FOV changes significantly and the main pore within the region moves further away to the border of the FOV.\\

\begin{figure}
\centering
\vspace{-4mm}
\begin{tabular}{cccc}
\hspace{0.5cm}{\scriptsize \sf HINODE 30.09.2007 / 00:14-00:59 UT} &  \hspace{0.7cm}{\scriptsize \sf HINODE 30.09.2007 / 01:00-01:59 UT} \\
\hspace{-1cm}\includegraphics[width=.43\linewidth]{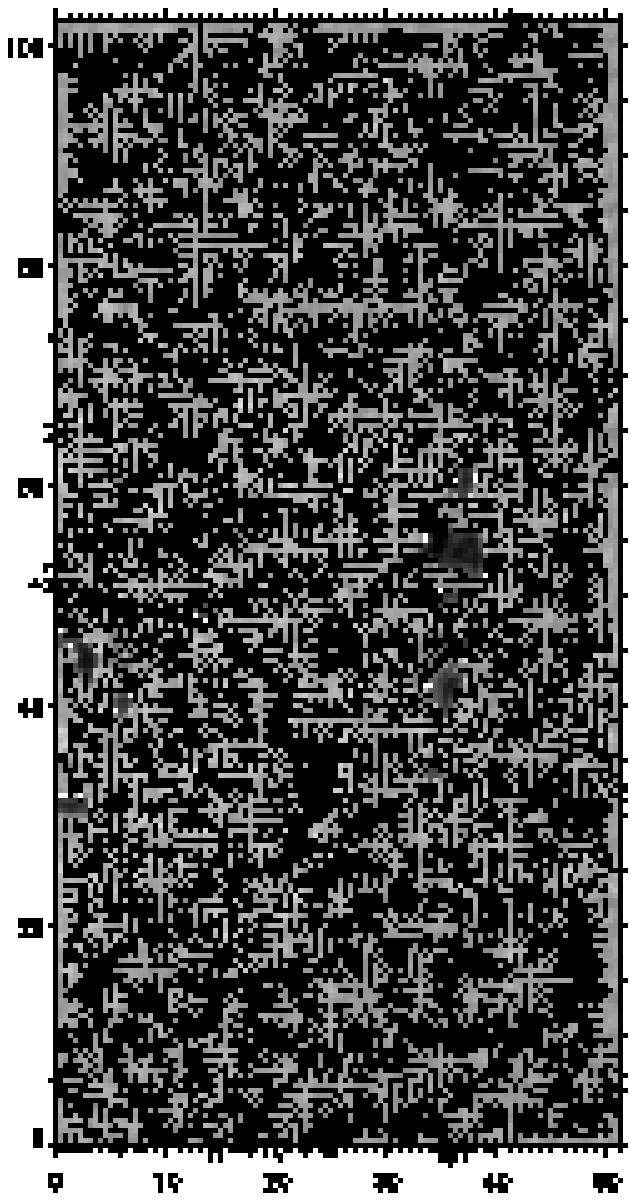} & 
\vspace{-7mm}\hspace{-1cm}\includegraphics[width=.43\linewidth]{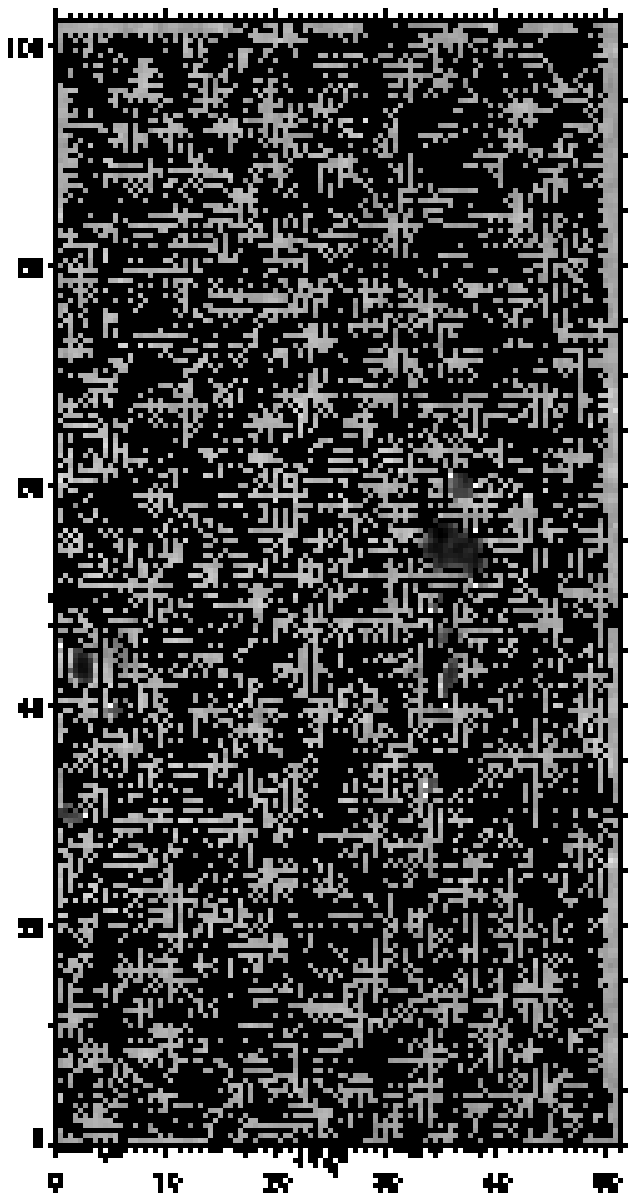} \\
\hspace{0.5cm}{\scriptsize \sf HINODE 30.09.2007 / 02:00-02:59 UT} &  \hspace{0.7cm}{\scriptsize \sf HINODE 30.09.2007 / 03:00-03:59 UT} \\
\hspace{-1cm}\includegraphics[width=.43\linewidth]{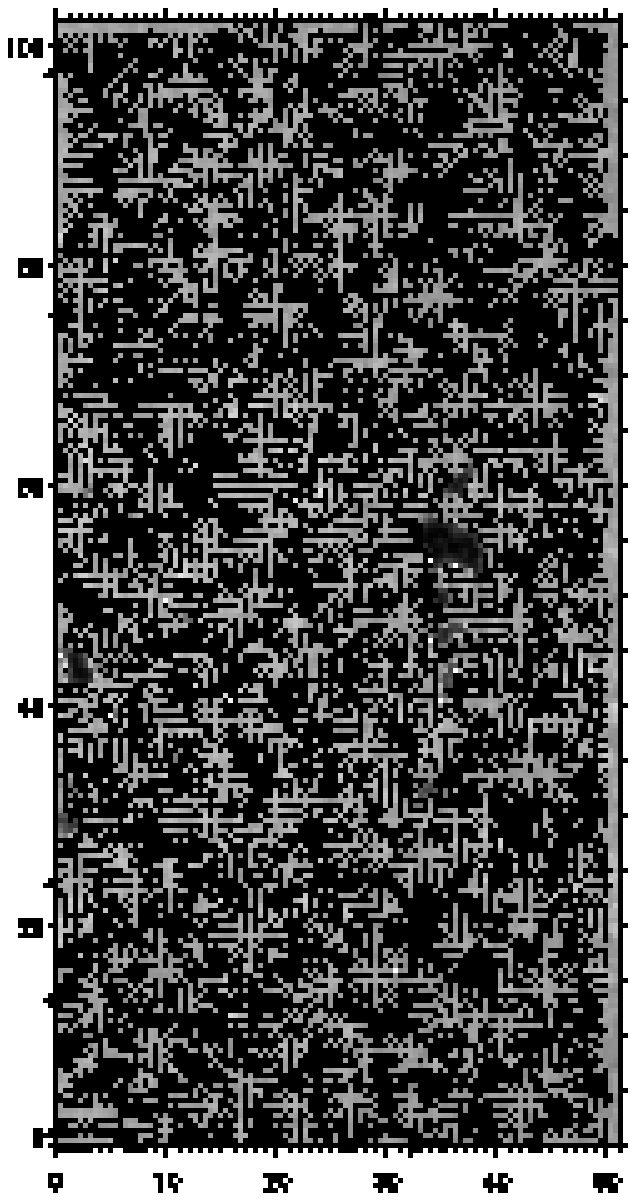} &
\hspace{-1cm}\includegraphics[width=.43\linewidth]{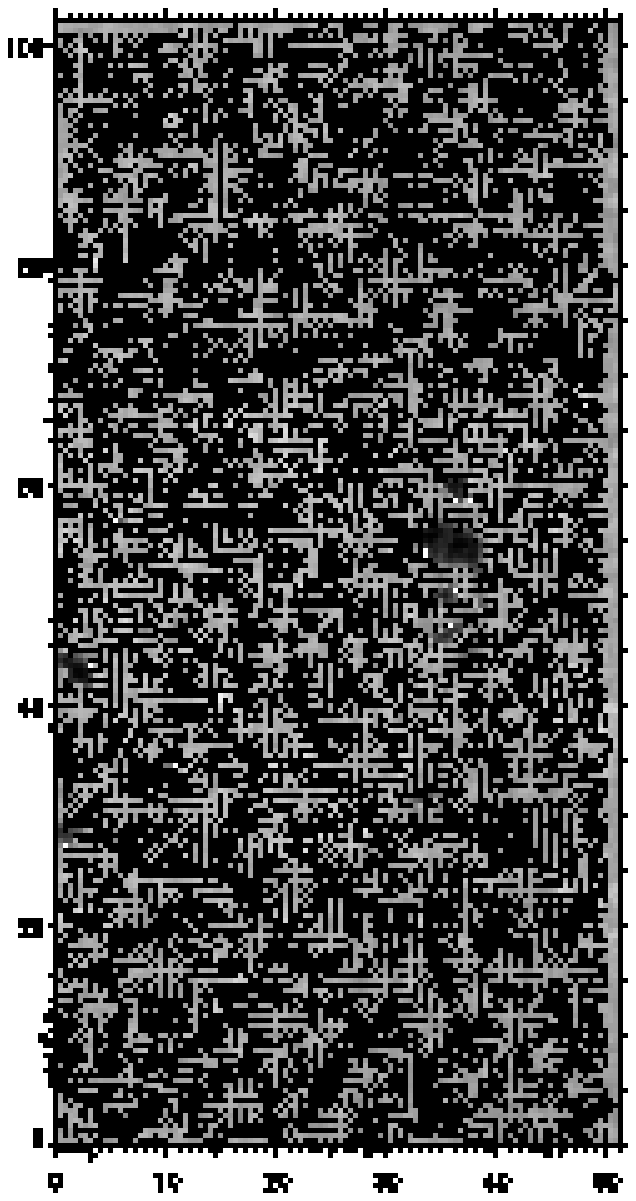} 
\end{tabular}
\vspace{-0.9cm}
\caption[\sf Horizontal flows observed in solar pores regions with HINODE data on 30 Sep 2007.]{\sf Horizontal flows observed in solar pores regions with HINODE data on 30 Sep 2007. The length of the black bar at coordinates (0,0) corresponds to 2.3 km s$^{-1}$. The coordinate are expressed in arc sec in all the maps hereafter.}
\label{porosflowmapHINODEall_a}
\end{figure}

\begin{figure}
\centering
\vspace{-4mm}
\begin{tabular}{cc}
\hspace{0.5cm}{\scriptsize \sf HINODE 30.09.2007 / 04:00-04:59 UT} &  \hspace{0.7cm}{\scriptsize \sf HINODE 30.09.2007 / 05:00-05:59 UT}\\
\hspace{-1cm}\includegraphics[width=.43\linewidth]{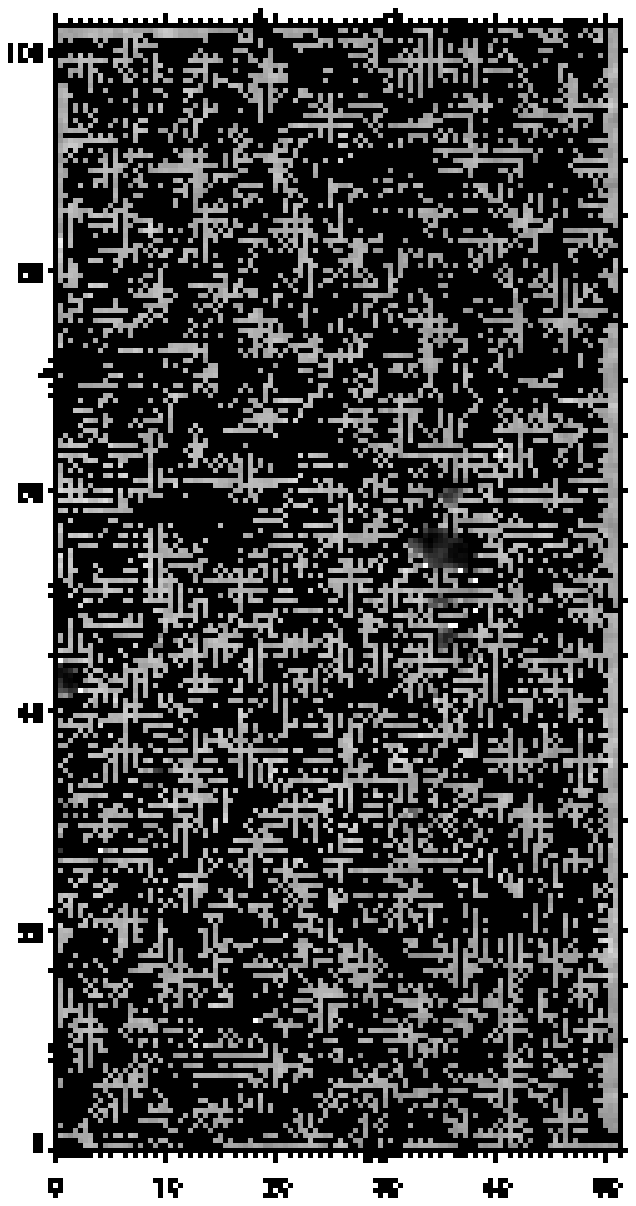} & 
\vspace{-7mm}\hspace{-1cm}\includegraphics[width=.43\linewidth]{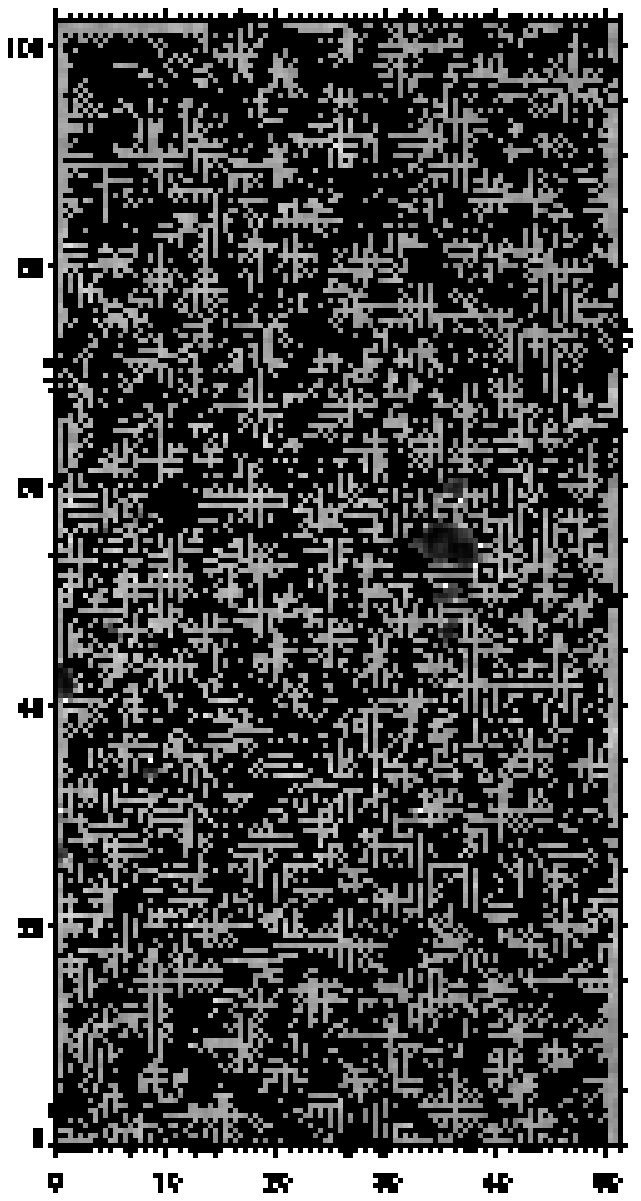} \\
\hspace{0.5cm}{\scriptsize \sf HINODE 30.09.2007 / 06:15-06:59 UT} &  \hspace{0.7cm}{\scriptsize \sf HINODE 30.09.2007 / 07:00-07:59 UT} \\
\hspace{-1cm}\includegraphics[width=.43\linewidth]{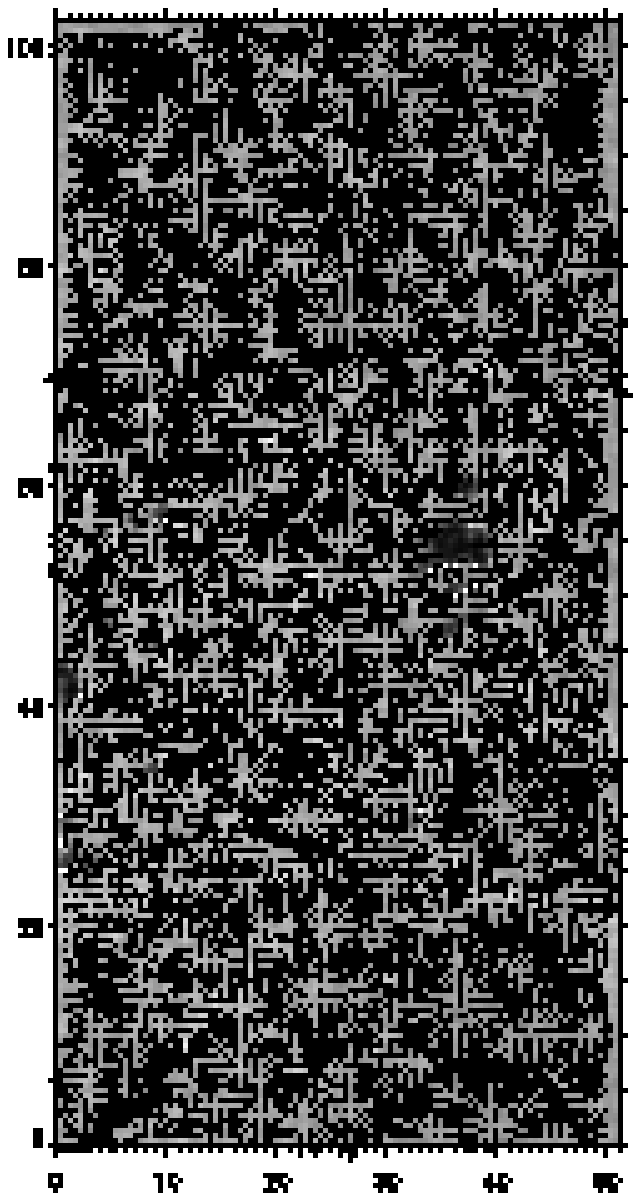} &
\hspace{-1cm}\includegraphics[width=.43\linewidth]{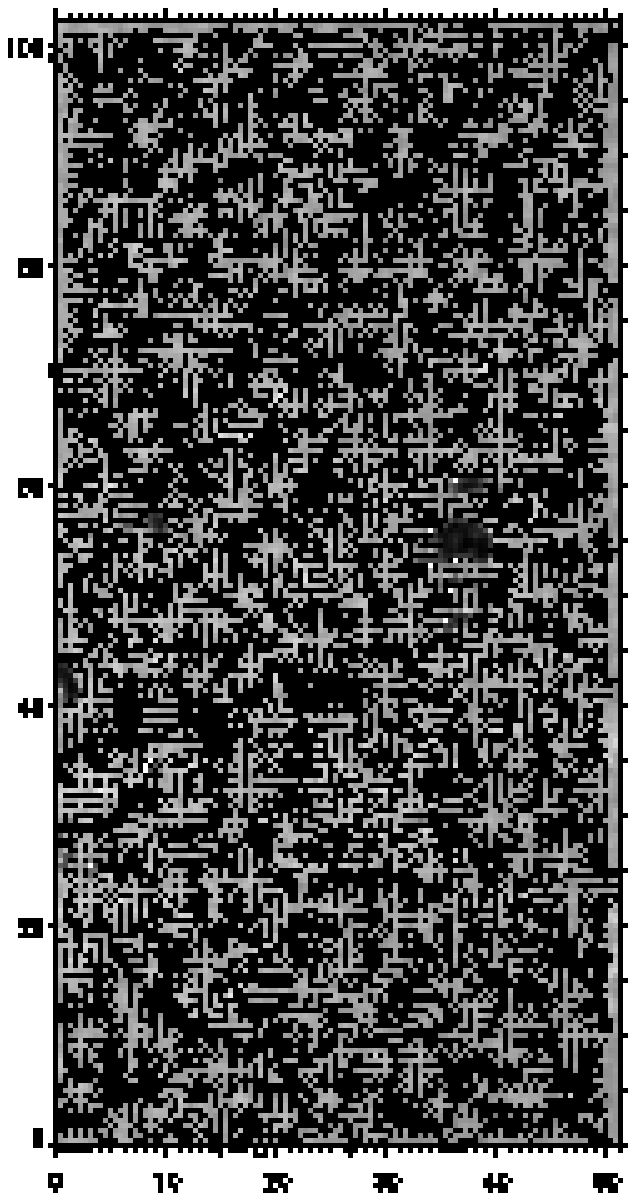} 
\end{tabular}
\vspace{-0.9cm}
\caption[\sf Horizontal flows observed in solar pores regions with HINODE data on 30 Sep 2007.]{\sf Horizontal flows observed in solar pores regions with HINODE data on 30 Sep 2007 (\emph{cont}).}
\label{porosflowmapHINODEall_b}
\end{figure}

\begin{figure}
\centering
\vspace{-4mm}
\begin{tabular}{cc}
\hspace{0.5cm}{\scriptsize \sf HINODE 30.09.2007 / 08:00-08:59 UT} &  \hspace{0.7cm}{\scriptsize \sf HINODE 30.09.2007 / 09:00-09:59 UT} \\
\hspace{-1cm}\includegraphics[width=.43\linewidth]{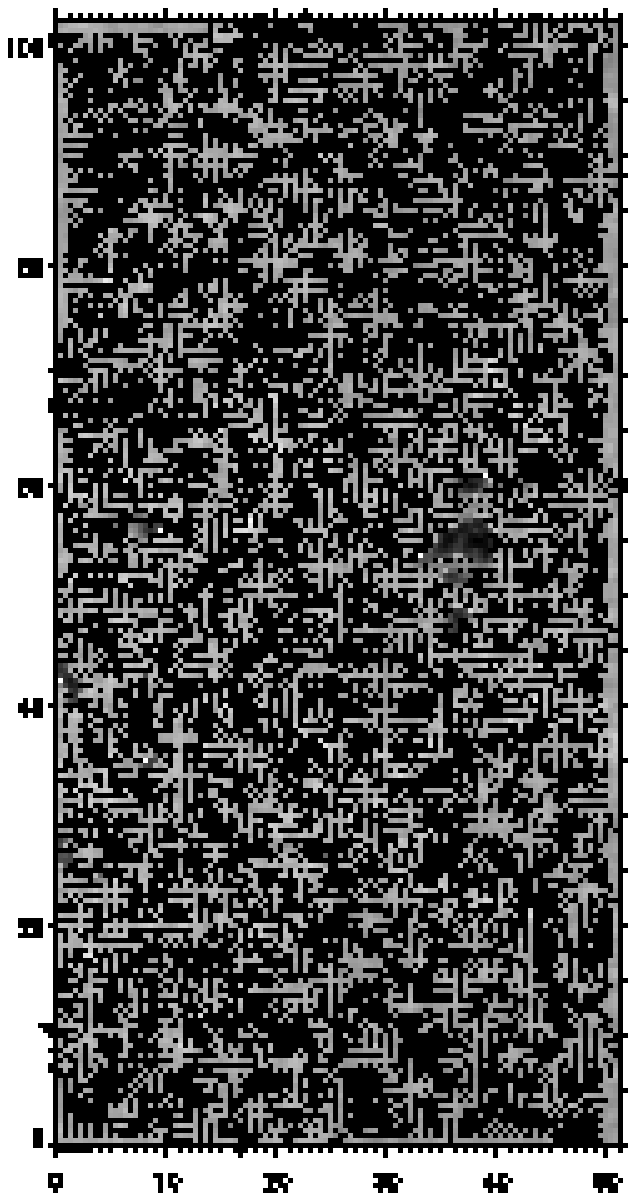} & 
\vspace{-7mm}\hspace{-1cm}\includegraphics[width=.43\linewidth]{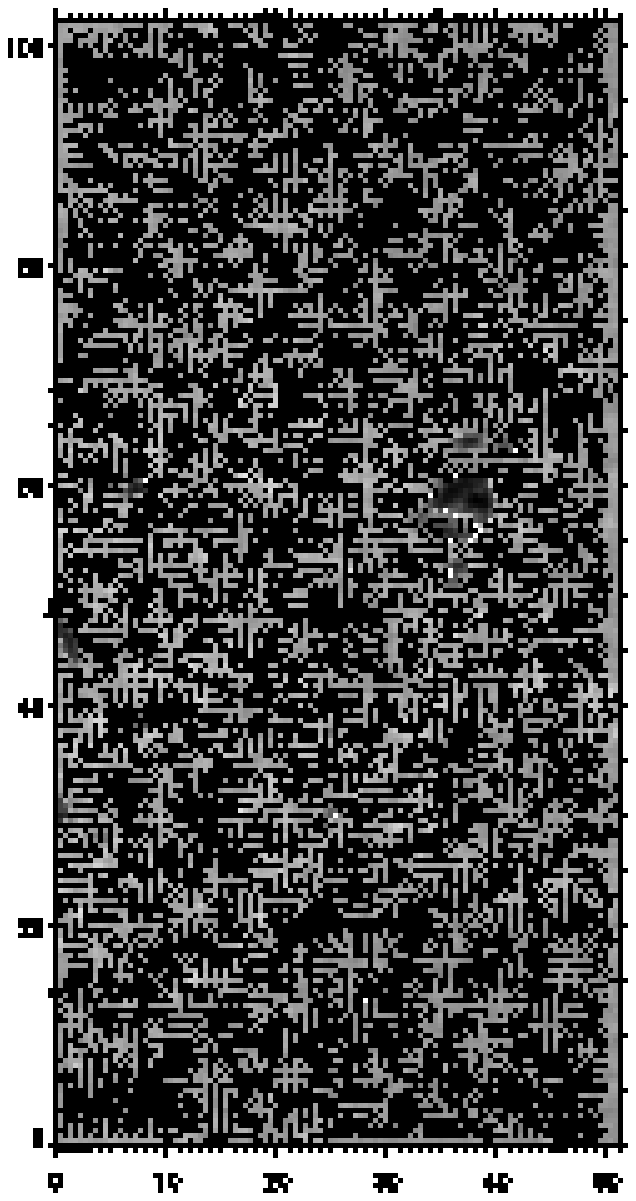} \\
\hspace{0.5cm}{\scriptsize \sf HINODE 30.09.2007 / 10:00-10:59 UT} &  \hspace{0.7cm}{\scriptsize \sf HINODE 30.09.2007 / 11:00-11:59 UT} \\
\hspace{-1cm}\includegraphics[width=.43\linewidth]{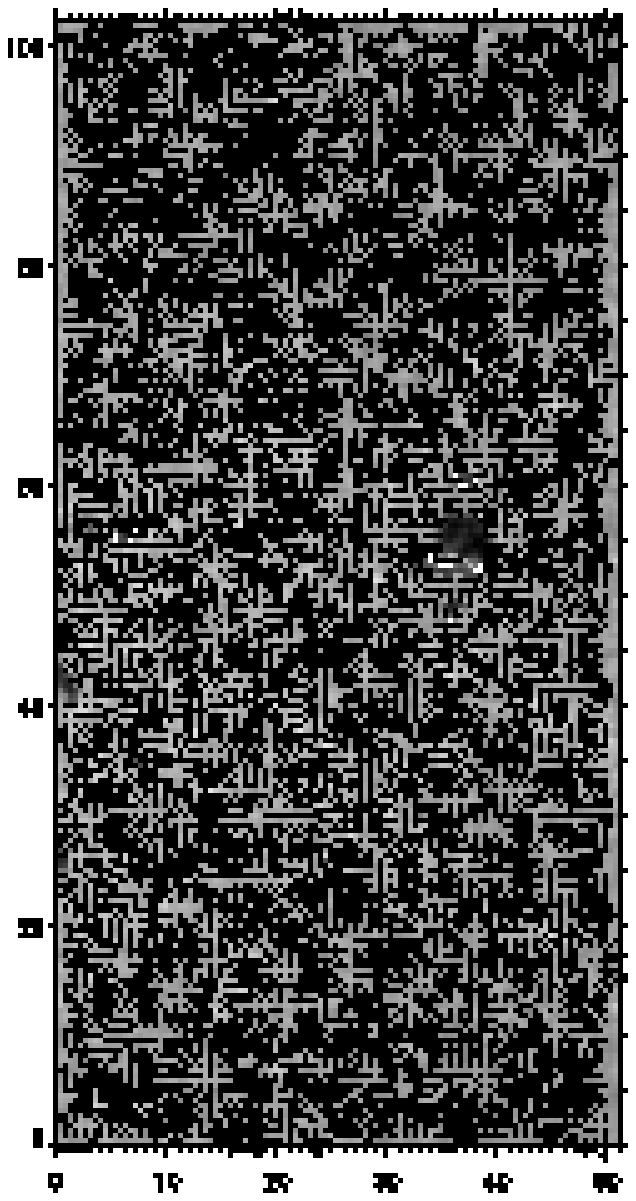} &
\hspace{-1cm}\includegraphics[width=.43\linewidth]{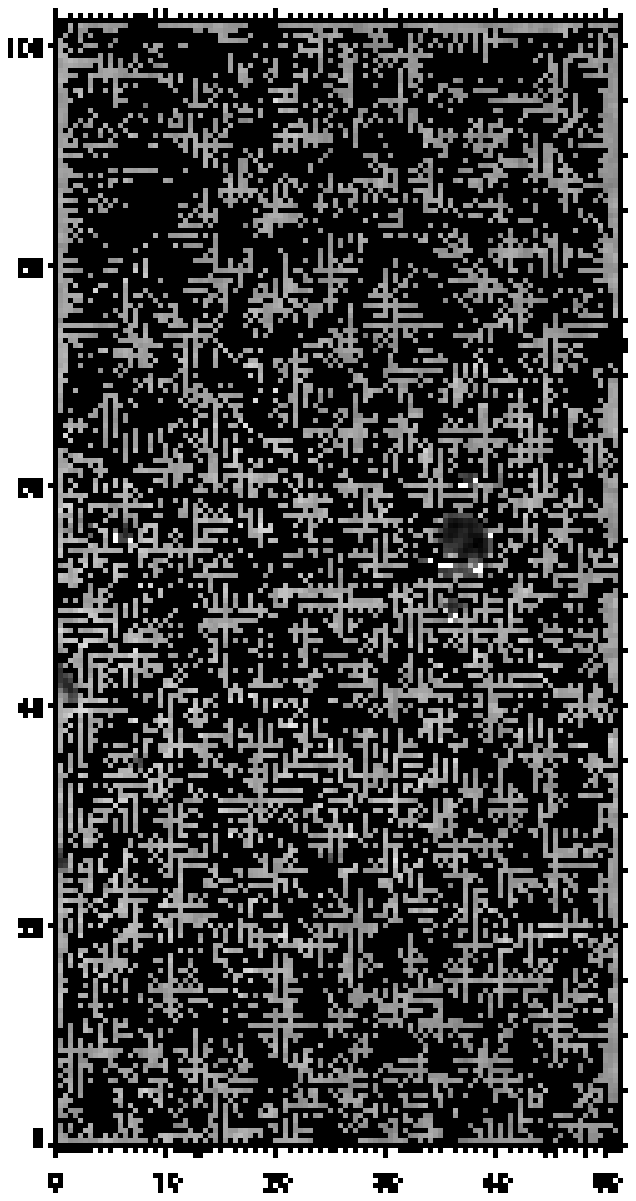} 
\end{tabular}
\vspace{-0.9cm}
\caption[\sf Horizontal flows observed in solar pores regions with HINODE data on 30 Sep 2007.]{\sf Horizontal flows observed in solar pores regions with HINODE data on 30 Sep 2007 (\emph{cont}).}
\label{porosflowmapHINODEall_c}
\end{figure}

\begin{figure}
\centering
\vspace{-1mm}
\begin{tabular}{cc}
\hspace{0.5cm}{\scriptsize \sf HINODE 30.09.2007 / 12:00-12:59 UT} &  \hspace{0.7cm}{\scriptsize \sf HINODE 30.09.2007 / 13:00-13:59 UT} \\
\hspace{-1cm}\includegraphics[width=.43\linewidth]{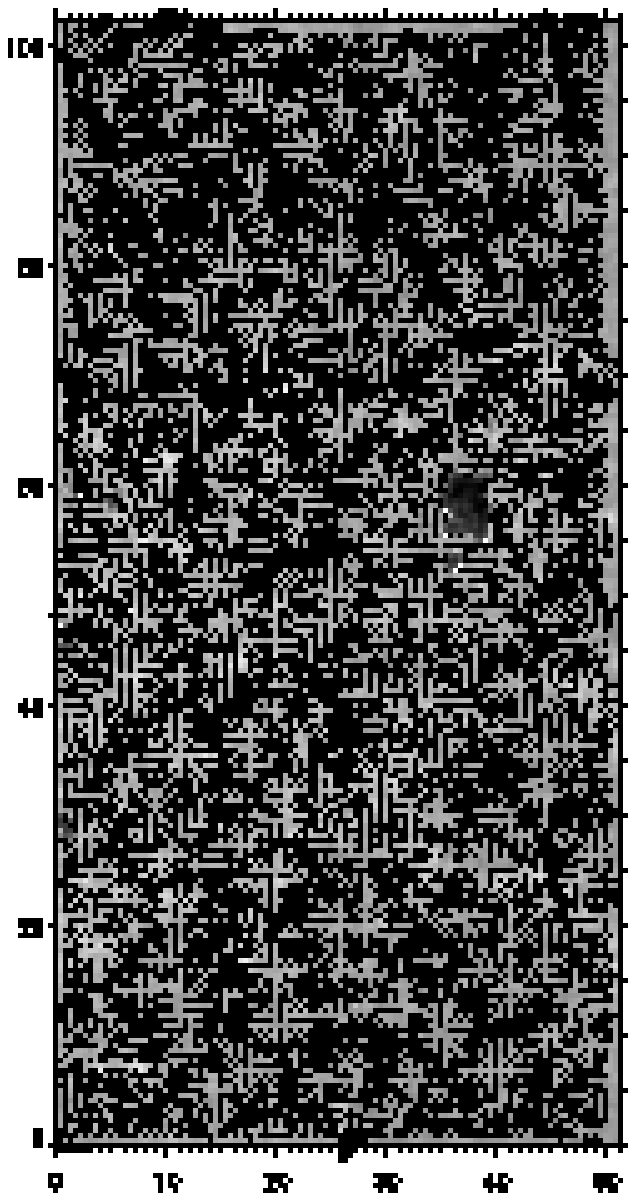} & 
\vspace{-4mm}\hspace{-1cm}\includegraphics[width=.43\linewidth]{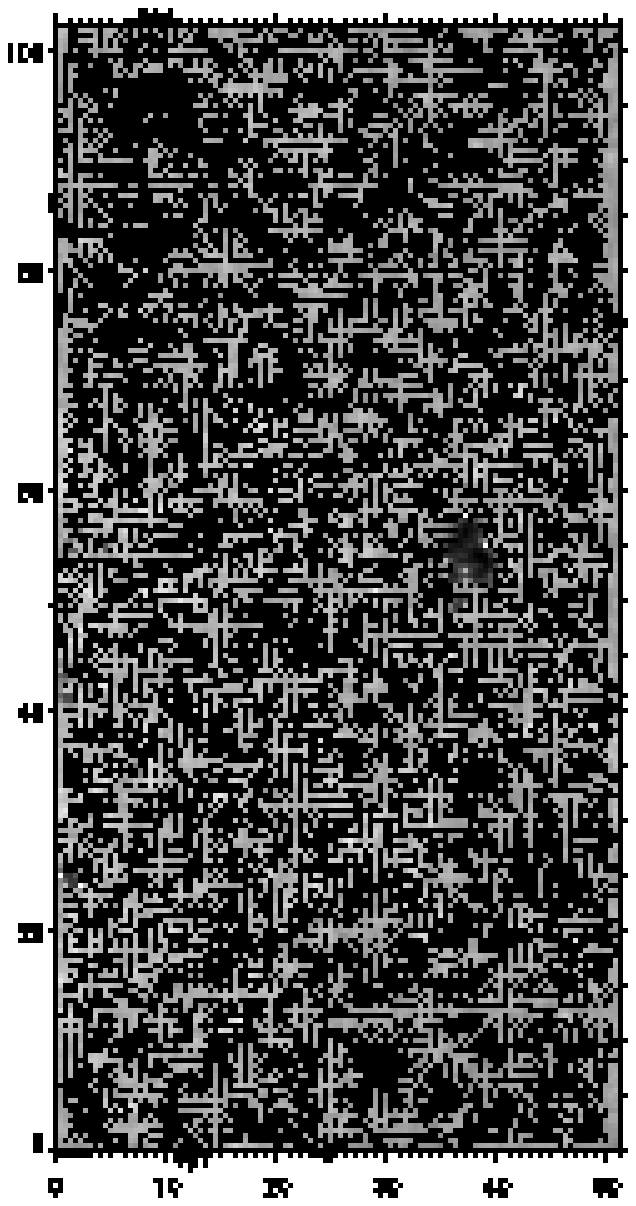} \\
\end{tabular}
\caption[\sf Horizontal flows observed in solar pores regions with HINODE data on 30 Sep 2007.]{\sf Horizontal flows observed in solar pores regions with HINODE data on 30 Sep 2007 (\emph{cont}).}
\label{porosflowmapHINODEall_d}
\end{figure}

Images are grouped then in sets of one hour and every set is processed independently resulting in a map of horizontal velocities averaged over the total duration of the set. The majority of the sets have a duration of 60 minutes except for four of them with 46,45,45 and 55 min respectively. These four sets will be understood hereafter as one-hour sets for description purposes although the average to calculate the map of velocities is performed over the actual duration of every set, respectively. The 14 maps of horizontal velocities are computed using a Gaussian tracking window of FWHM 1$\farcs$0. Figures~\ref{porosflowmapHINODEall_a}-\ref{porosflowmapHINODEall_d} show all the flow maps with the indication of the time period they correspond to, and the background image represents the averaged image of every 1-hour set.\\

The main pore is located in the spatial position (37,55) in the first map (Figures~\ref{porosflowmapHINODEall_a}) and surrounded by some smaller pores which form altogether a sort of vertical and elongated structure in the figure. The collection of these pores is evolving in time and some of them start merging and disappearing. The final picture of the region displayed in Figure~\ref{porosflowmapHINODEall_d} shows the isolated main pore with only a very tiny companion.\\

We do not identify any signal of moat-like flow around the pores in any of the evolutionary stages shown in the maps sequence but continuous activity caused by exploding granules. Centers of divergence are systematically identified, some of them very close to the pore border. Proper motions displaying inward components are more common around the pores and any outward regular and large-scale flow, as corresponding to a moat-flow, is found. These results are firmly validated because of the long duration of the sample and reinforce the previous analysis pursued for the only 1-hour SST data.

\subsection{Distribution of horizontal speeds in the FOV}

\begin{figure}
\centering
\vspace{-1mm}
\begin{tabular}{c}
{\footnotesize \sf SST 30.09.2007 / VELOCITY RANGE: $< $0.3 km s$^{-1}$} \\
\vspace{-4mm}\hspace{-1.5cm}\includegraphics[width=.72\linewidth]{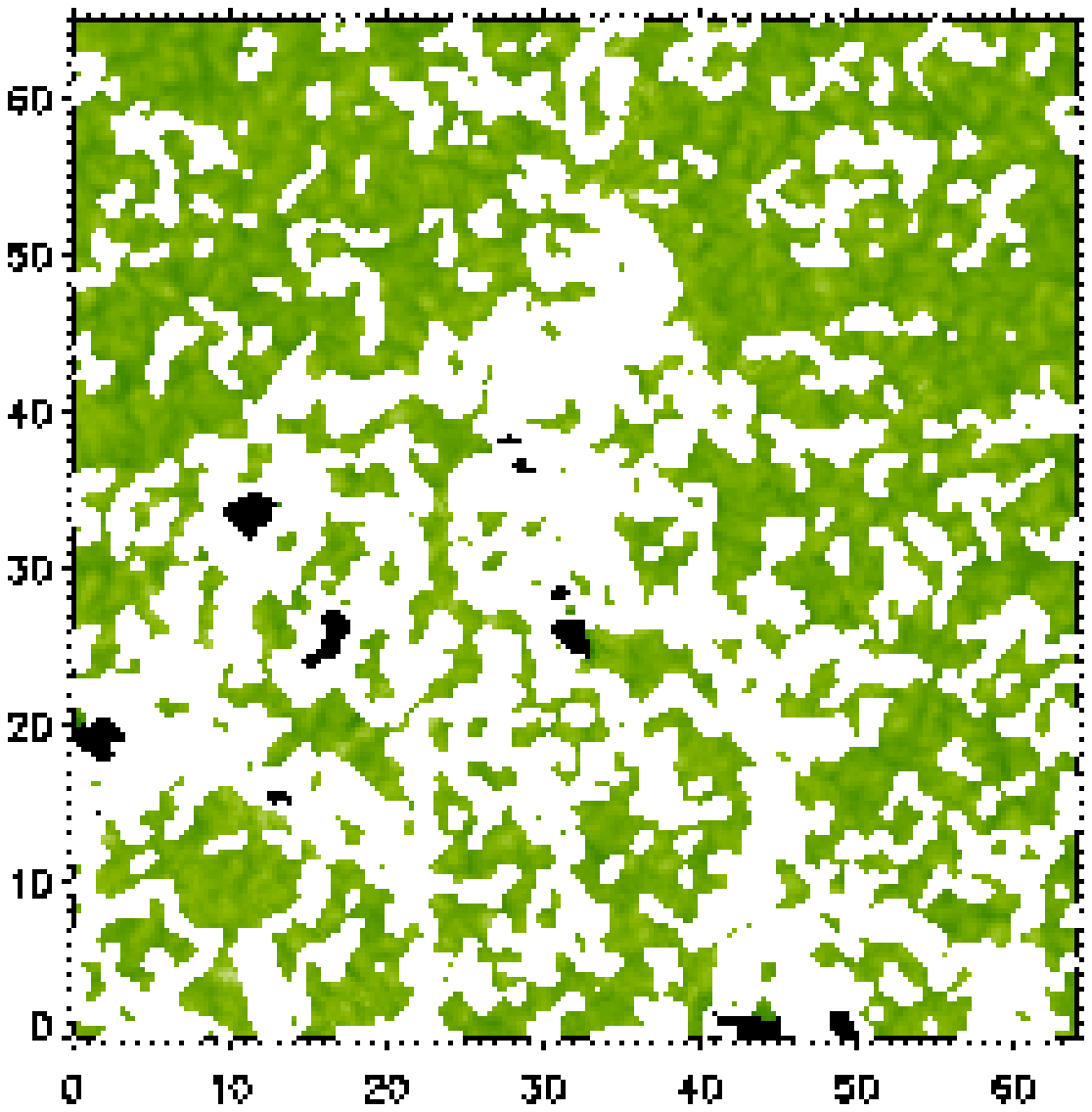} \\ 
{\footnotesize \sf SST 30.09.2007 / VELOCITY RANGE: 0.3-0.8 km s$^{-1}$} \\
\hspace{-1.5cm}\includegraphics[width=.72\linewidth]{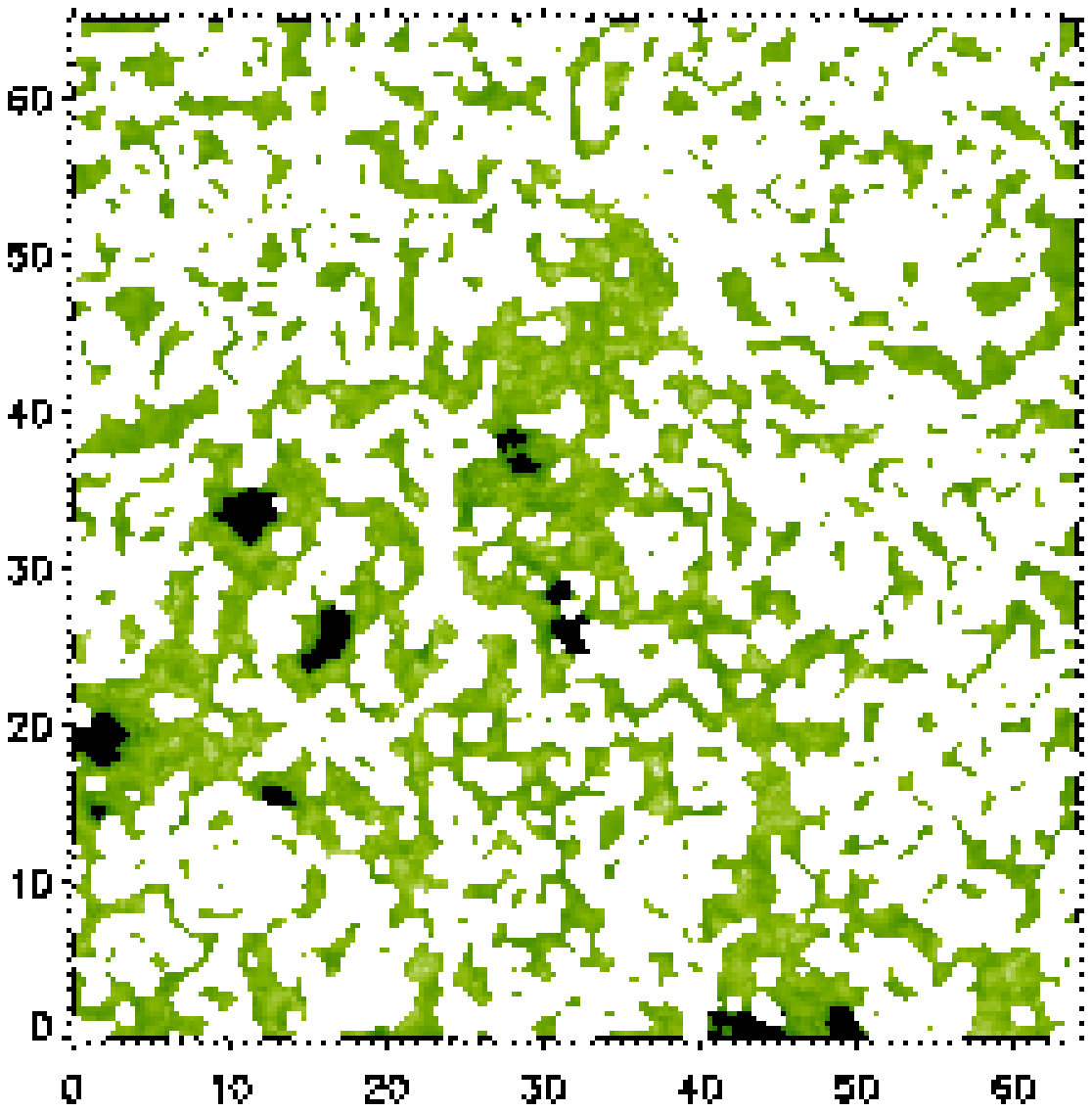} 
\end{tabular}
\vspace{-1cm}
\caption[\sf Locations of the areas for different velocity ranges.]{\sf Location of the areas (\emph{in white}) where the magnitude of horizontal velocities is lower than 0.3 km s$^{-1}$ (\emph{upper panel}) and between 0.3 - 0.8 km s$^{-1}$(\emph{lower panel}). The coordinates are expressed in arc seconds.}
\label{vranges1}
\end{figure}

\begin{figure}
\centering
\vspace{-1mm}
\begin{tabular}{c}
{\footnotesize \sf SST 30.09.2007 / VELOCITY RANGE: $>$ 0.8 km s$^{-1}$} \\
\vspace{-4mm}\hspace{-1.5cm}\includegraphics[width=.72\linewidth]{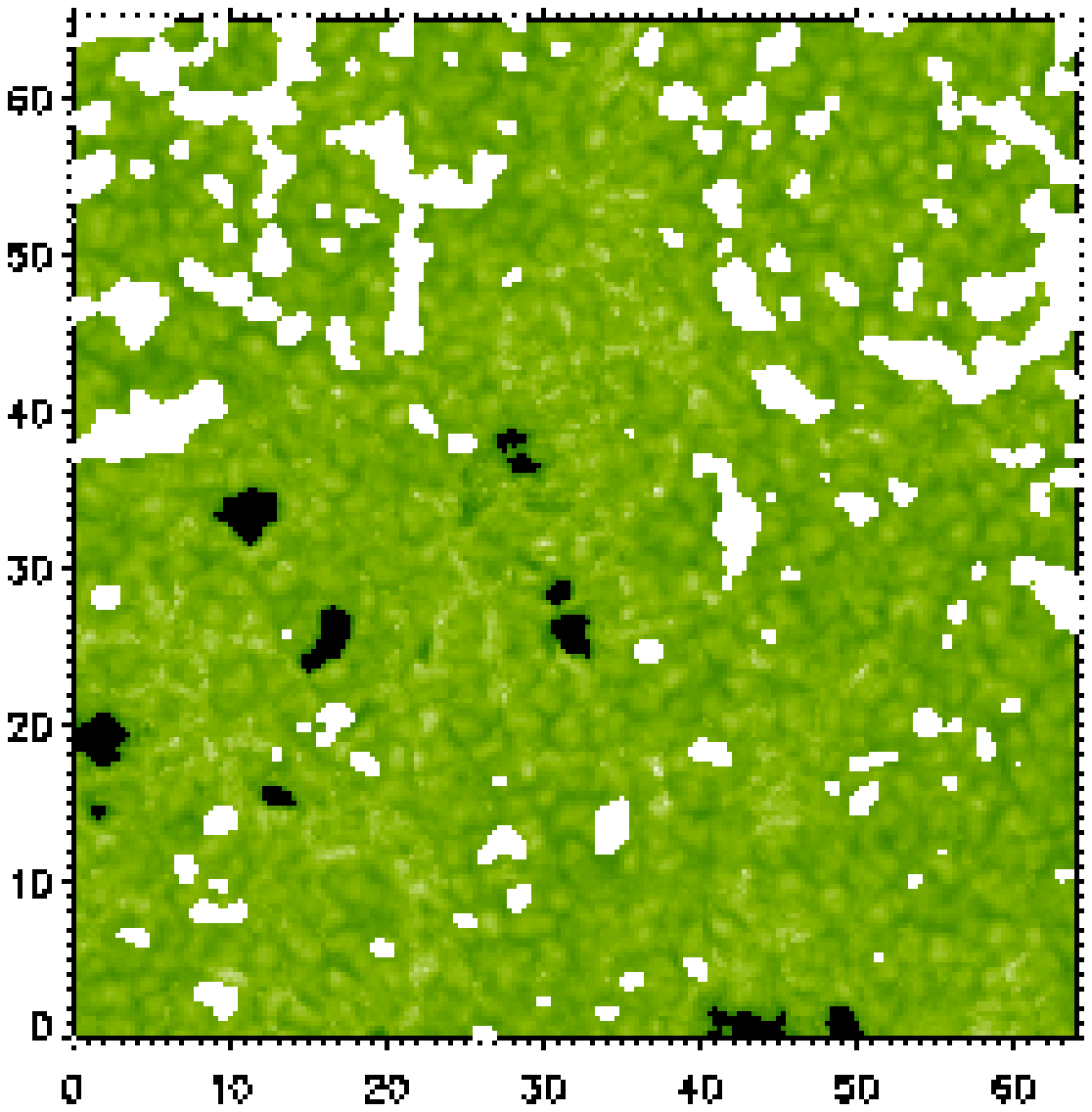} \\
{\footnotesize \sf SST 30.09.2007 / VEL. RANGE: $<$ 0.2 km s$^{-1}$ / BACKGR: H$\alpha$ }\\
\hspace{-1.5cm}\includegraphics[width=.72\linewidth]{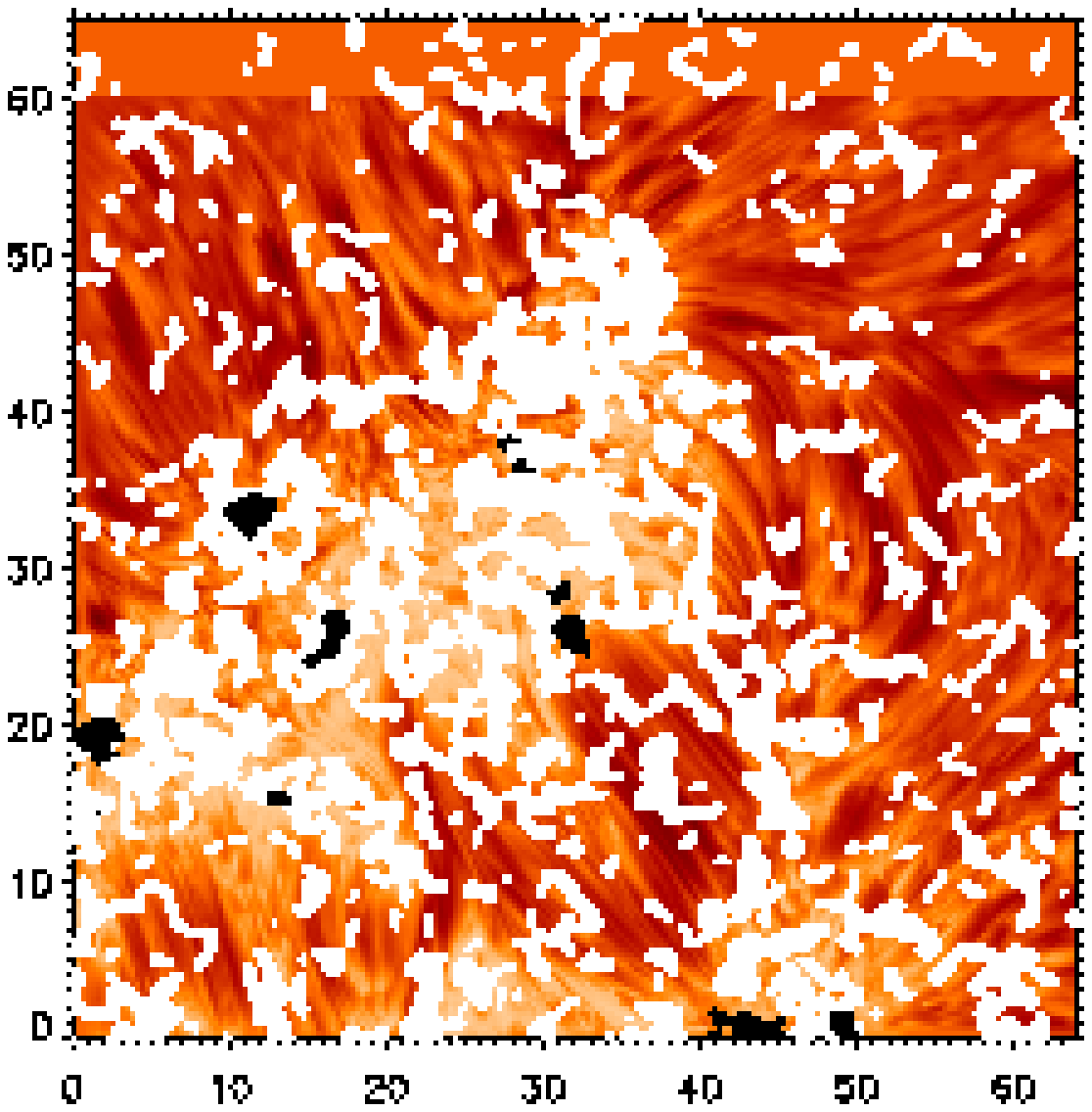}
\end{tabular}
\vspace{-1cm}
\caption[\sf Locations of the areas for different velocity ranges.]{\sf Location of the areas (\emph{in white}) where the magnitude of horizontal velocities is greater than 0.8 km s$^{-1}$(\emph{upper panel}). Location of small velocities overplotted (areas in \emph{white}) on a co-temporal H$\alpha$ restored image (\emph{lower panel}). The coordinates are expressed in arc seconds.}
\label{vranges2}
\end{figure}

The global speed distribution in a granulation field can be easily visualized by means of false-color maps showing the areas including velocity magnitudes within certain ranges. Figures~\ref{vranges1} and \ref{vranges2} show the location (\emph{white areas}) of velocity magnitudes in three different ranges in km s$^{-1}$, for the SST data (the first series in Table~\ref{poroseries}):

\begin{itemize}
\item Low velocity magnitudes: $<$ 0.3.
\item Medium velocity magnitudes: 0.3~-~0.8.
\item Large velocity magnitudes: $>$ 0.8.
\end{itemize}

Small speeds are mainly grouped in the central part of the FOV where an intense magnetic activity is detected as evidenced by the high concentration of G-band bright points and faculae present in this region (see Figures~\ref{images_blue} and \ref{images_red}). Around pores, the velocity magnitudes mainly correspond to the lower range \mbox{($<$ 0.3 km s$^{-1}$)} so that they are surrounded by white areas in Figure~\ref{vranges1} (\emph{upper panel}). \\

The areas mapping medium velocity magnitudes are regularly spread out all over the FOV except in the proximity of pores (Figure~\ref{vranges1} \emph{lower panel}). The shape of empty (\emph{non white}) areas around the pores almost reproduce the pores shape. \\

Large velocities ($>$ 0.8 km s$^{-1}$) are not homogeneously distributed in the FOV but mainly located in the two upper corners of Figure~\ref{vranges2} (\emph{upper panel}) where the granulation is less-magnetized (see magnetogram in Figure~\ref{images_red}). These zones show large velocity flows that might reflect the presence of supergranular cells as commented in section~\S\ref{horizontalpattern}.\\

Figure~\ref{vranges2} (\emph{lower panel}) shows the mask of areas (\emph{in white}) corresponding to very low horizontal velocities ($<$ 0.2 km s$^{-1}$) plotted over co-spatial and temporal H$\alpha$ (\emph{false color}) image. The correspondence between the bright structures in H$\alpha$ and the regions of close-to-zero horizontal velocities is clearly seen. In the filamentary stripe-like structures in H$\alpha$ connecting the two polarities (see magnetogram in Figure~\ref{images_red}), there is a high density of low velocities (\emph{white areas}) whereas this density shrinks when going to the upper part of the FOV in Figure~\ref{vranges2} (\emph{lower panel}).
 


\subsection{Velocity distribution around solar pores}
\label{velpores}

\begin{figure}
\centering
\includegraphics[angle=90,width=.9\linewidth]{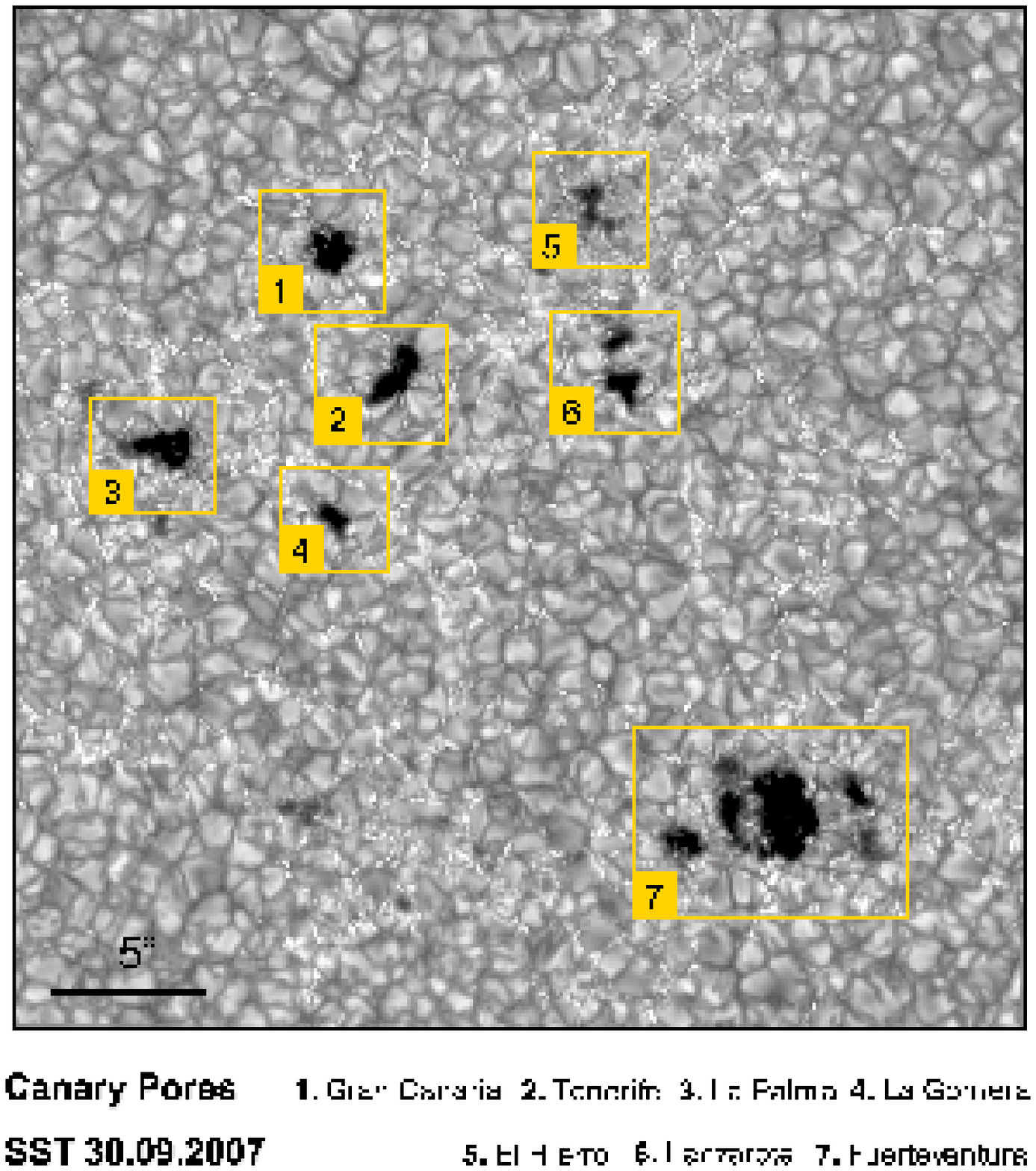}
\caption[\sf Emerging active region observed with the SST on 30 September 2007 ]{\sf Emerging active region observed with the SST on 30 September 2007. The FOV include several pores (hereafter \emph{Canary Pores}) which for the sake of simplicity have been labeled as shown in the figure to facilitate their identification. The coordinates are expressed in arc seconds.}
\label{canarypores}
\end{figure}

From the computed velocity fields we can do detailed analysis of the velocity distribution around the solar pores. To pursue this study we will use the region observed with the SST were we have a collection of pores to work with. As we have two time series for this region (see Table~\ref{poroseries}), we will employ the first one covering a longest time period, except for one of the pores which is out of the FOV. For this pore we will use the second time series. Figure~\ref{canarypores} shows the FOV including all the pores under study which are labeled for easy identification hereafter.\\  

Figure~6.18 illustrates the projection of the velocities into radial and transversal components as a convenient way to compute inward and outward motions. The figure plots two points in the granulation surrounding a solar pore (\emph{red dots}) with their corresponding velocity vectors ${\bf v}$. The pore is centered at coordinates $(x_c,y_c)$ with respect to the orthogonal coordinate system X,Y placed in the lower left corner of the FOV. Vector ${\bf r}$ is the position vector of a given point with respect to the pore center. The pore center is located at its gravity-center calculated by weighting the position of every point inside the pore with the inverse of its respective intensity. Velocity vectors in every point of the granulation surrounding the pore are projected into radial $v_r$ and transversal $v_t$ components. 

\begin{figure}[h]
 \hfill
\begin{minipage}[h]{.60\textwidth}
\hspace{-9mm}\includegraphics[angle=90,width=1.15\linewidth]{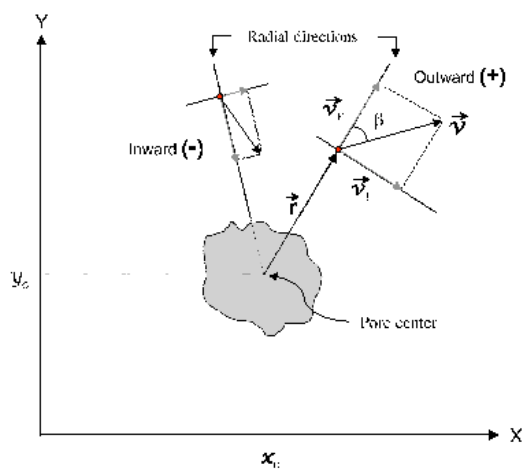}      
  \end{minipage}
  \hfill
\begin{minipage}[h]{.30\textwidth}    
  \caption[\sf Skecth showing the projection applied to the velocities around a solar pore.]{\sf Sketch showing the projection applied to the velocities around a solar pore centered at ($x_c,y_c$) respect to the orthogonal coordinate system  X, Y. The figure shows the velocity vectors ${\bf v}$ for two points in the granulation region around the pore (\emph{red dots}). Projection of velocities is done along both radial and transversal directions (\emph{dotted lines}) obtaining the radial $v_r$ and transversal $v_t$ velocity components, respectively.}
  \end{minipage}
\label{radialvel_sketch}
\end{figure}
\vspace{3mm}

\begin{figure}
\centering
\vspace{-4mm}
\begin{tabular}{cc}
{\footnotesize \textsf{a})}\includegraphics[width=.4\linewidth]{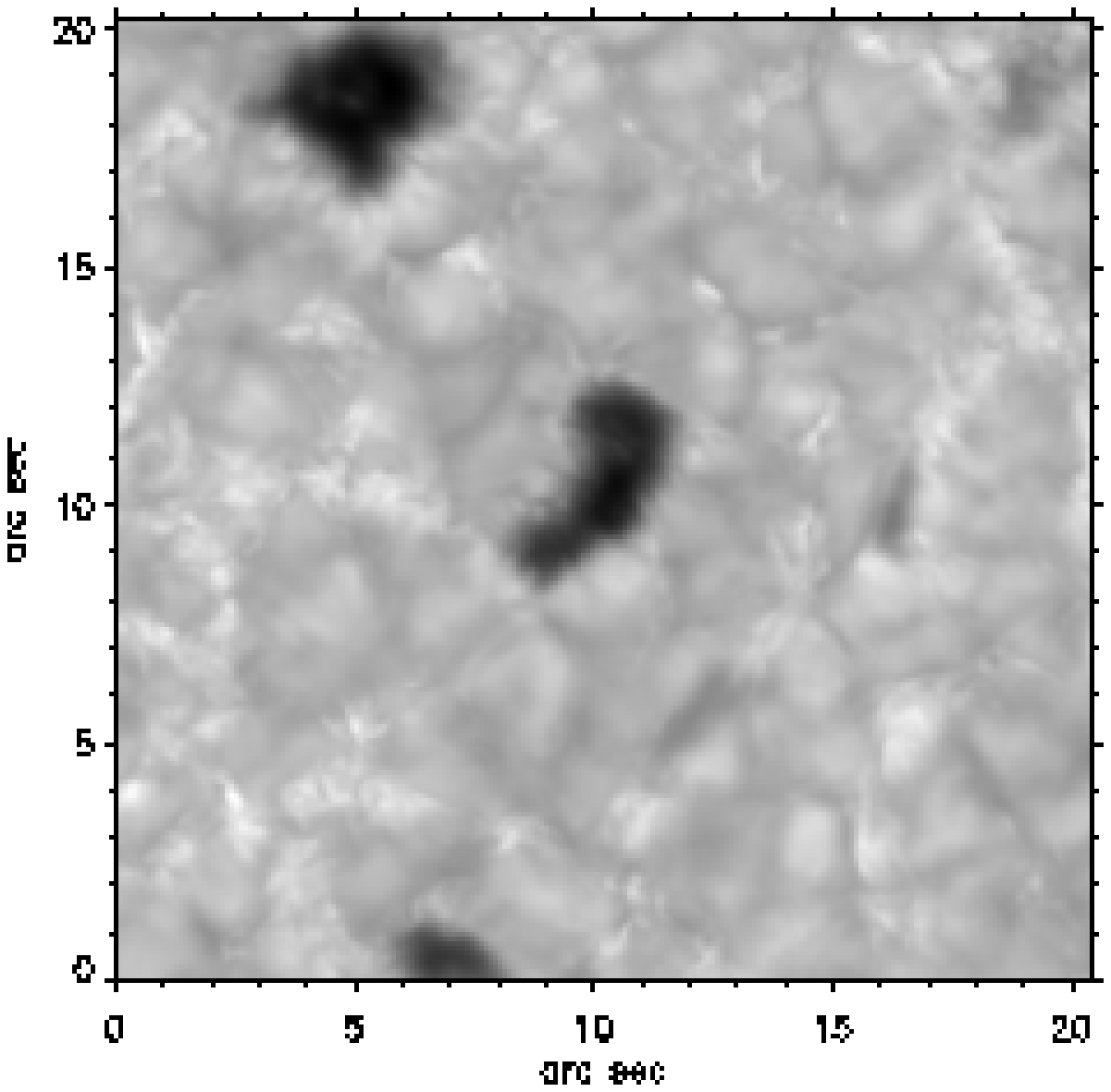} &
{\footnotesize \textsf{b})}\includegraphics[width=.4\linewidth]{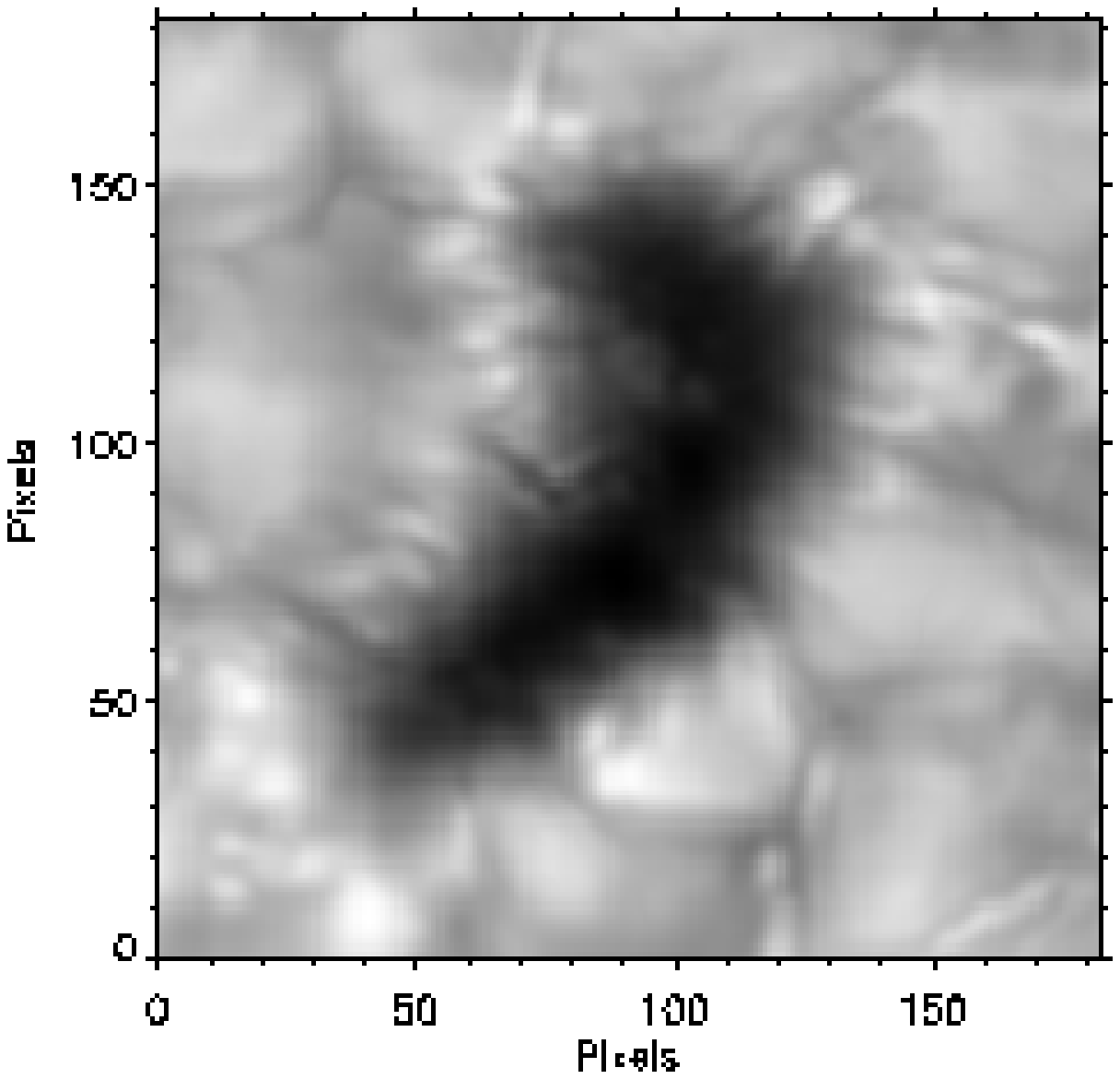} \\
\hspace{0.5cm}{\footnotesize \textsf{c})}\includegraphics[width=.4\linewidth]{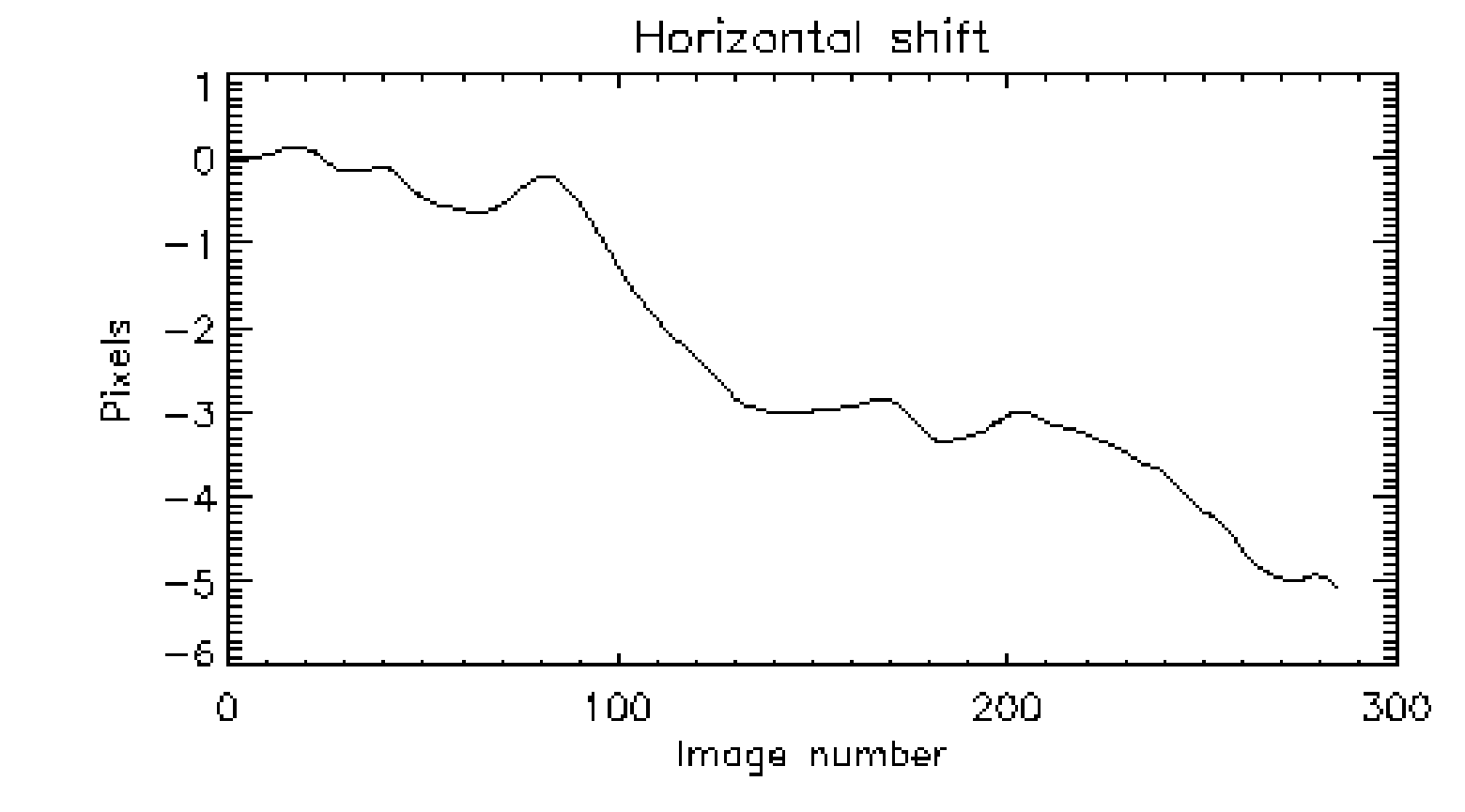} &
\hspace{0.5cm}{\footnotesize \textsf{d})}\includegraphics[width=.4\linewidth]{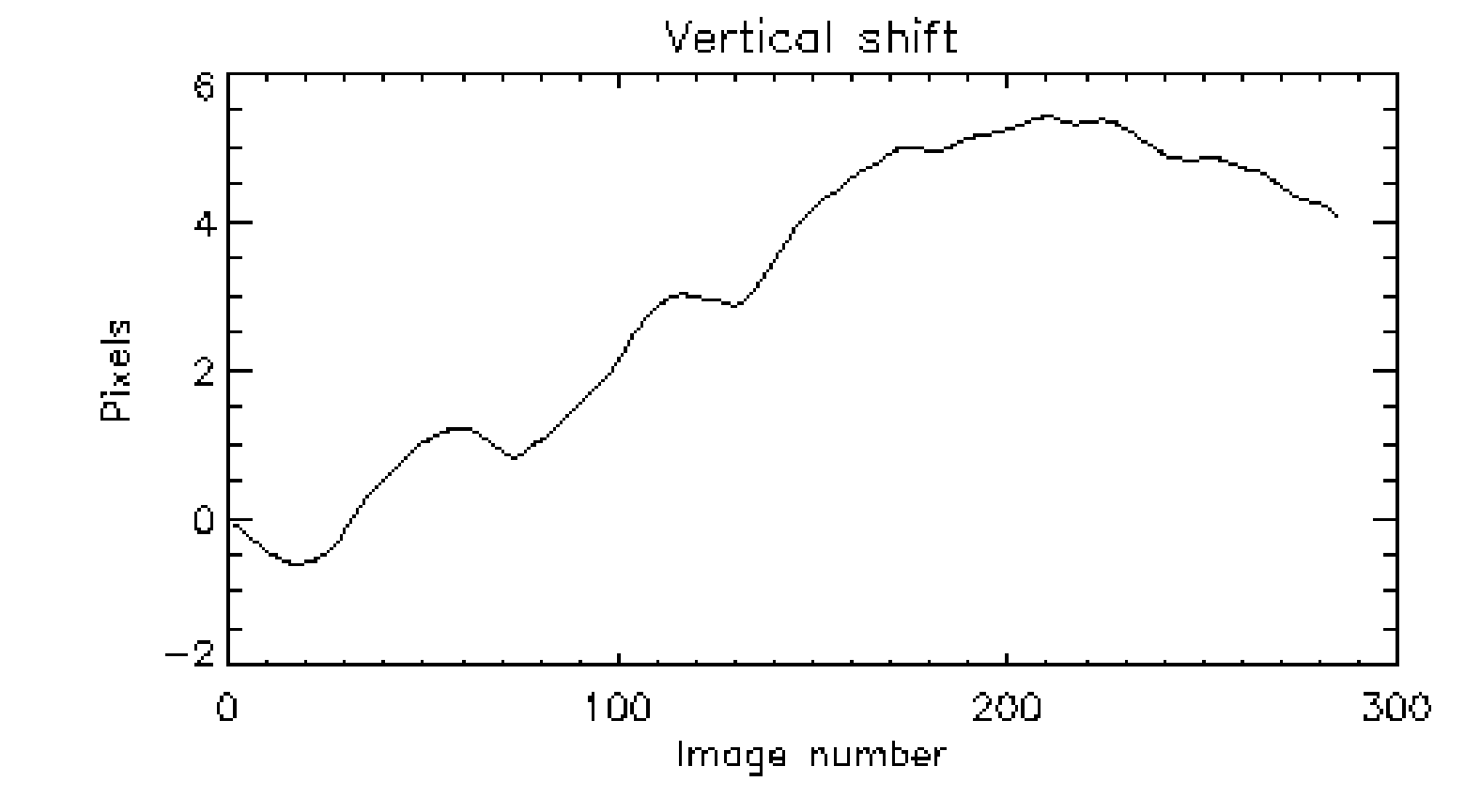} \\
{\footnotesize \textsf{e})}\includegraphics[width=.4\linewidth]{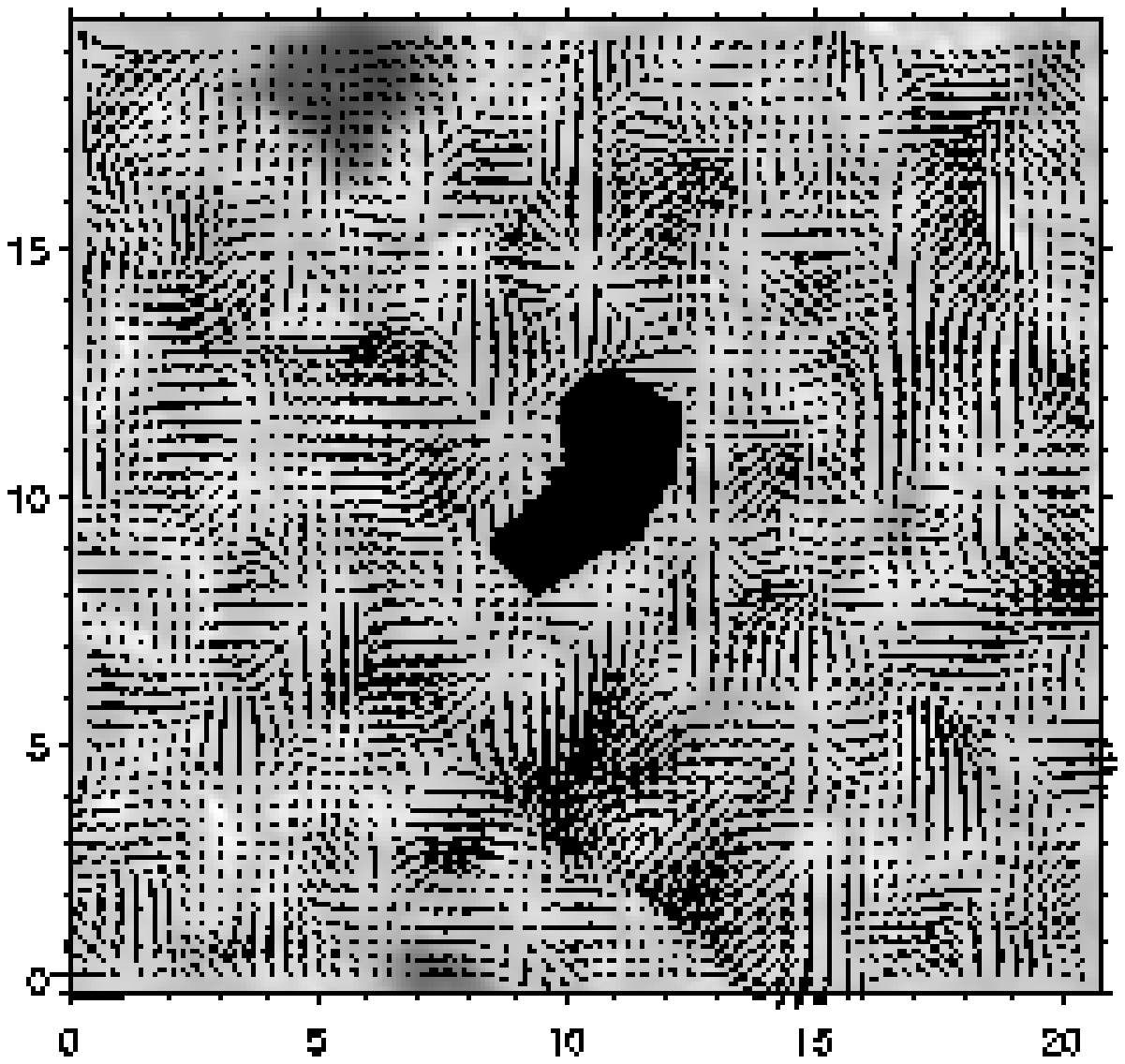} &
{\footnotesize \textsf{f})}\includegraphics[width=.4\linewidth]{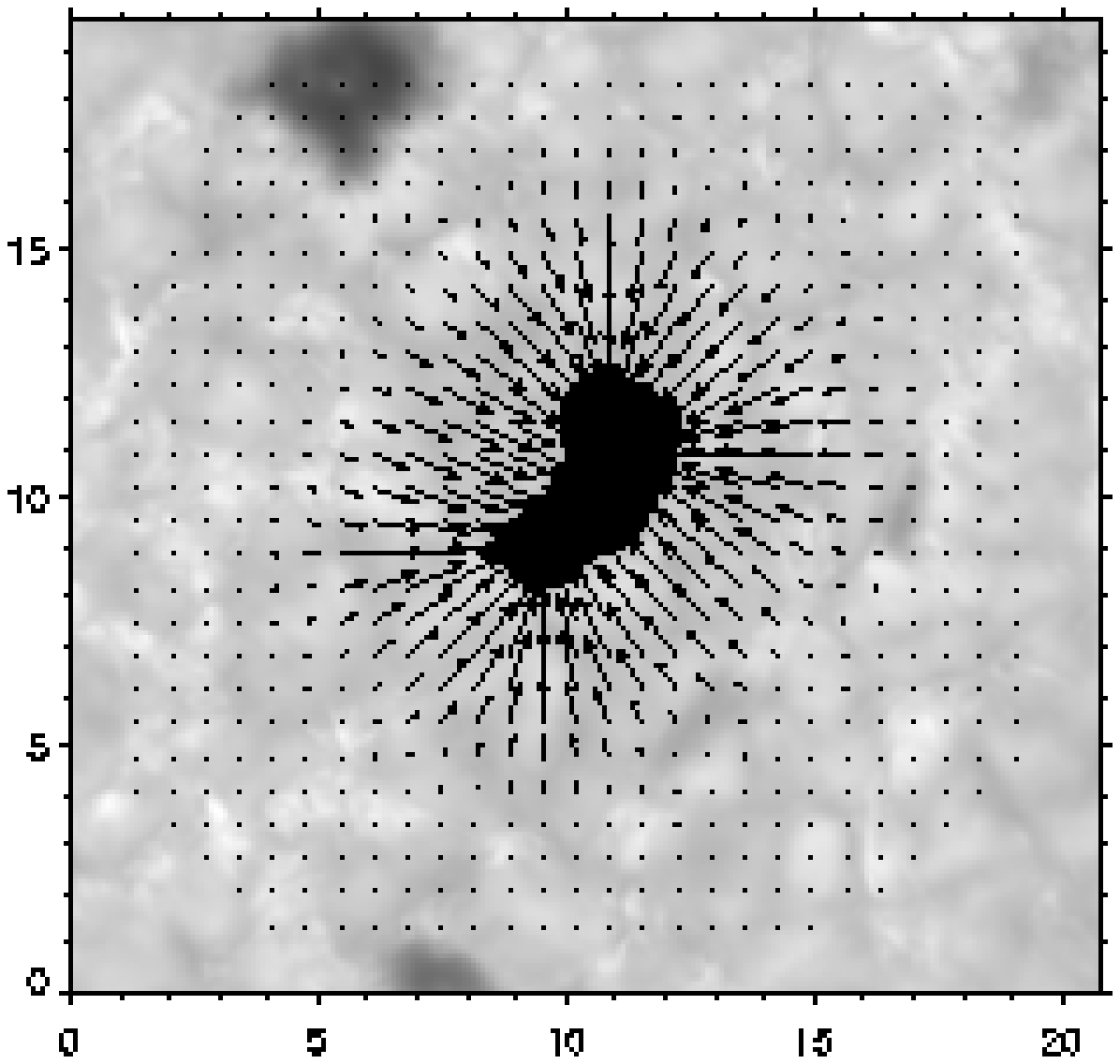} \\
\end{tabular}
\caption[\sf Steps of the process applied to generate a map describing the direction of the  velocities in the vicinity of solar pores.]{\sf Steps of the process applied to generate a map describing the direction of the velocities in the vicinity of solar pores. \textsf{a}) FOV. \textsf{b}) Correlation box. \textsf{c}) Shifts to correct for vertical misalignment \textsf{d})  Shifts to correct for horizontal misalignment. \textsf{e}) Flow map. \textsf{f}) Perpendicular directions to the pore border; arrows point to the inward (negative) directions.}
\label{radialvel_procedure1}
\end{figure}

By using the pore named as \emph{TENERIFE} in Figure~\ref{canarypores} we will illustrate the procedure followed to calculate velocity directions, throughout the panels in Figure~\ref{radialvel_procedure1}. In the following itemization, we will describe the steps corresponding to each panel in the figure:

\begin{itemize}
\item \textsf{a}) Selection of the FOV that includes the pore under study.
\item \textsf{b}) Selection of a correlation box. Since active regions in general, and pores in particular, exhibit their own displacement (due to differential rotation and intrinsic motions) while embedded in the granulation pattern, we align the time series respect to an area framing the pore (correlation box) so that we make sure we are measuring plasma motions with respect to the pore.
\item \textsf{c}) Horizontal shifts for alignment correction as computed by cross-correlation.
\item \textsf{d}) Vertical shifts for alignment correction as computed by cross-correlation.
\item \textsf{e}) Map of horizontal velocities as computed from LCT.
\item \textsf{f}) Calculation of radial directions for short distances. The "radial directions" that we use as the reference to project the velocities are defined as perpendicular to the pore border. These perpendicular directions are calculated from the gradients of intensity in a smoothed pore-mask image. This way, we can also manage with non round-shaped pores. The limit for short distances is defined by thresholding the intensity gradients. The threshold depends on the pore shape and size.
\item \textsf{g}) Calculation of radial directions for long distances. At large distances we consider all pores as round-shaped structures (as in Figure~6.18) and the radial directions are defined by the position vector ${\bf r}$ of a given point with respect to the pore center. 
\item \textsf{h}) Determination of inward and outward motions. According to Figure~6.18, inward/outward motions correspond to $\cos\beta$ negative/positive, where $\beta$ is the angle formed by $\bf v$ and the positive radial direction (outward) at each point of the FOV. The value of $\cos\beta$ is mapped in gray scale in panel \textsf{h)}, ranging from 1 for purely radial outward velocities (\emph{in black}) to -1 for purely inward velocities (\emph{in white}).
\item \textsf{i}) Creation of a binary mask. A binary mask is created from the gray-scaled map in \textsf{h}). Areas in \emph{black} and  \emph{white} correspond to velocities with positive (outward) and negative (inward) radial components, respectively.
\item \textsf{j}) Selection of a final clipped FOV.
\end{itemize}
%

\begin{figure}
\centering
\vspace{-4mm}
\begin{tabular}{cc}
{\footnotesize \textsf{g})}\includegraphics[width=.41\linewidth]{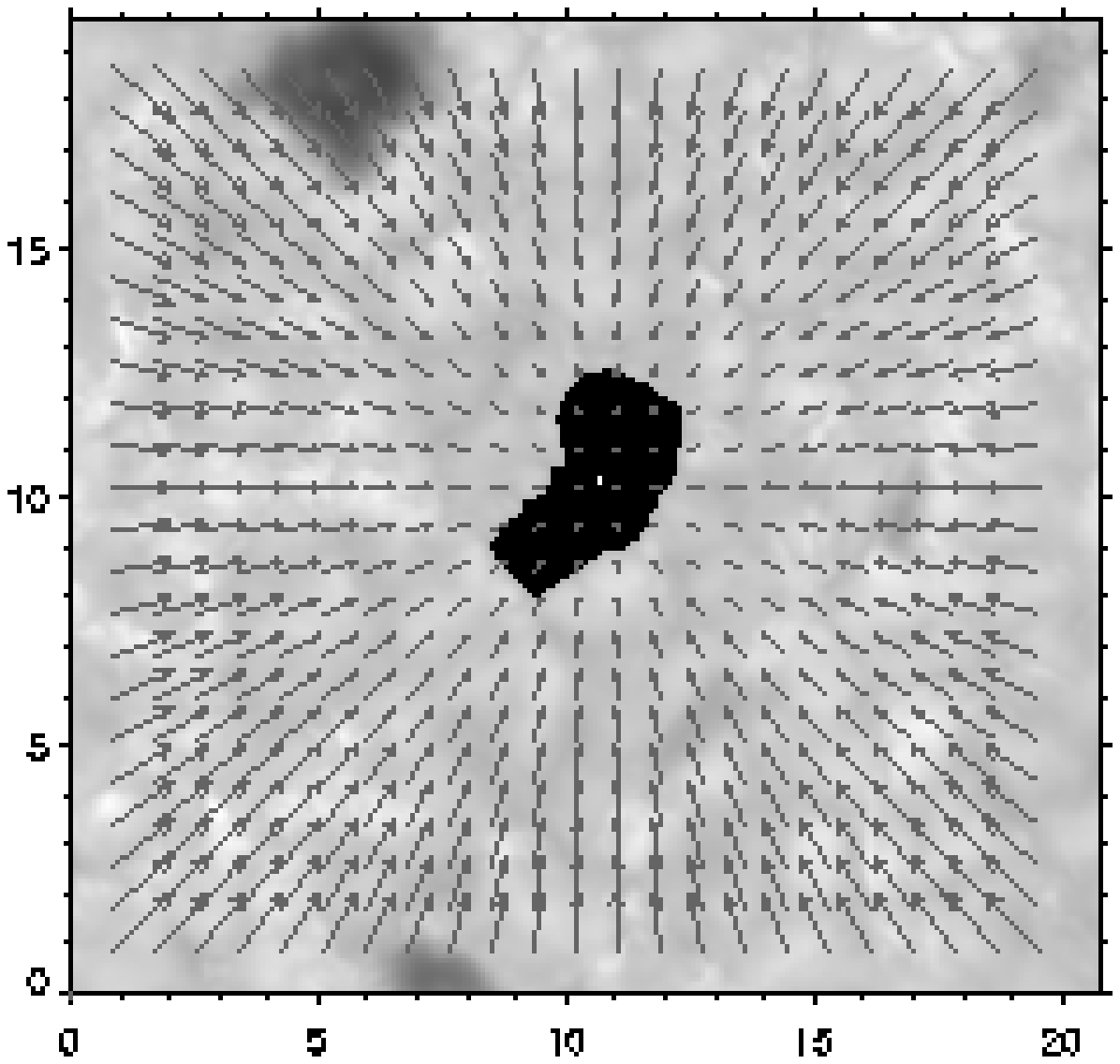} &
{\footnotesize \textsf{h})}\includegraphics[width=.4\linewidth]{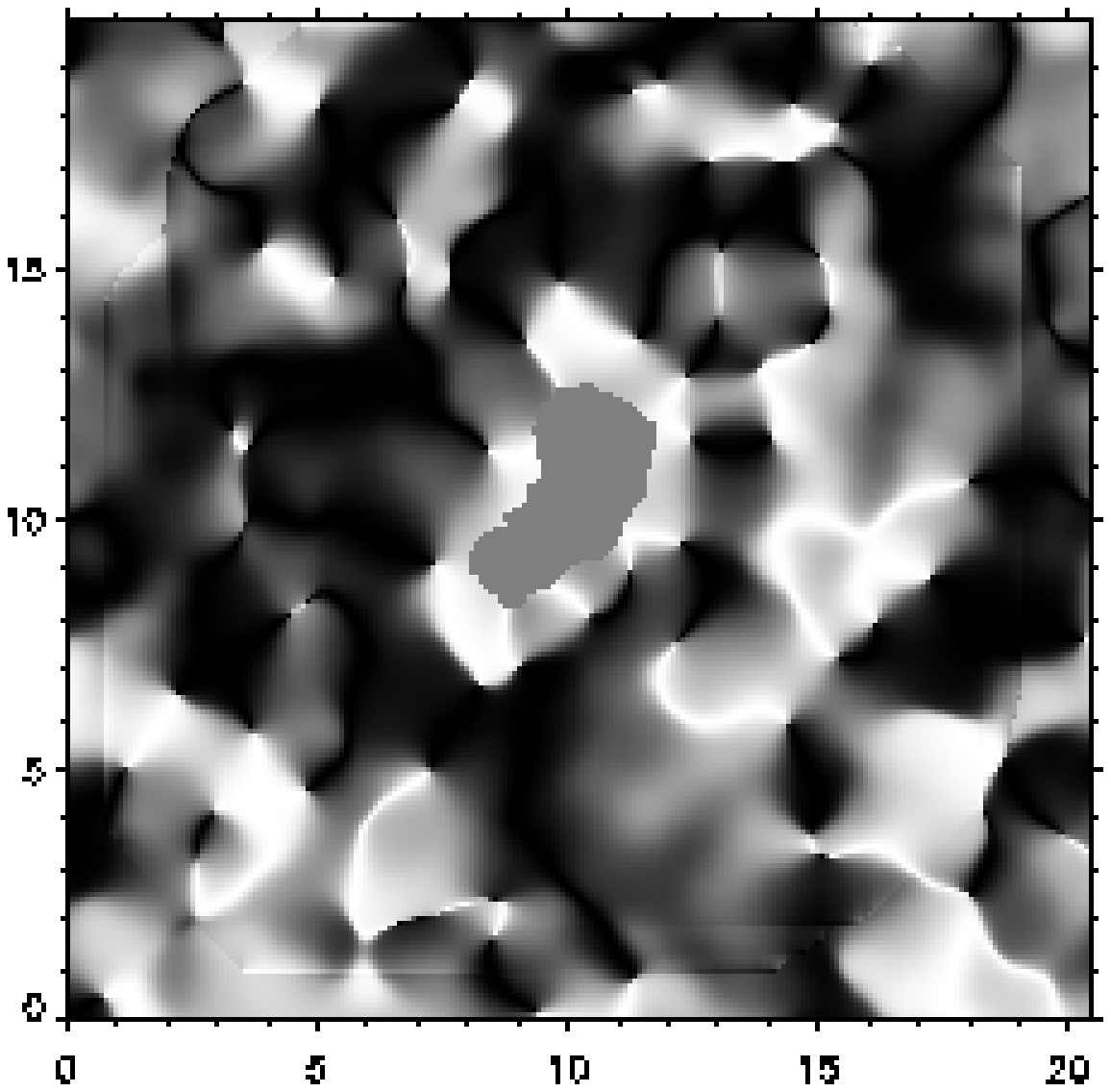} \\
{\footnotesize \textsf{i})}\includegraphics[width=.41\linewidth]{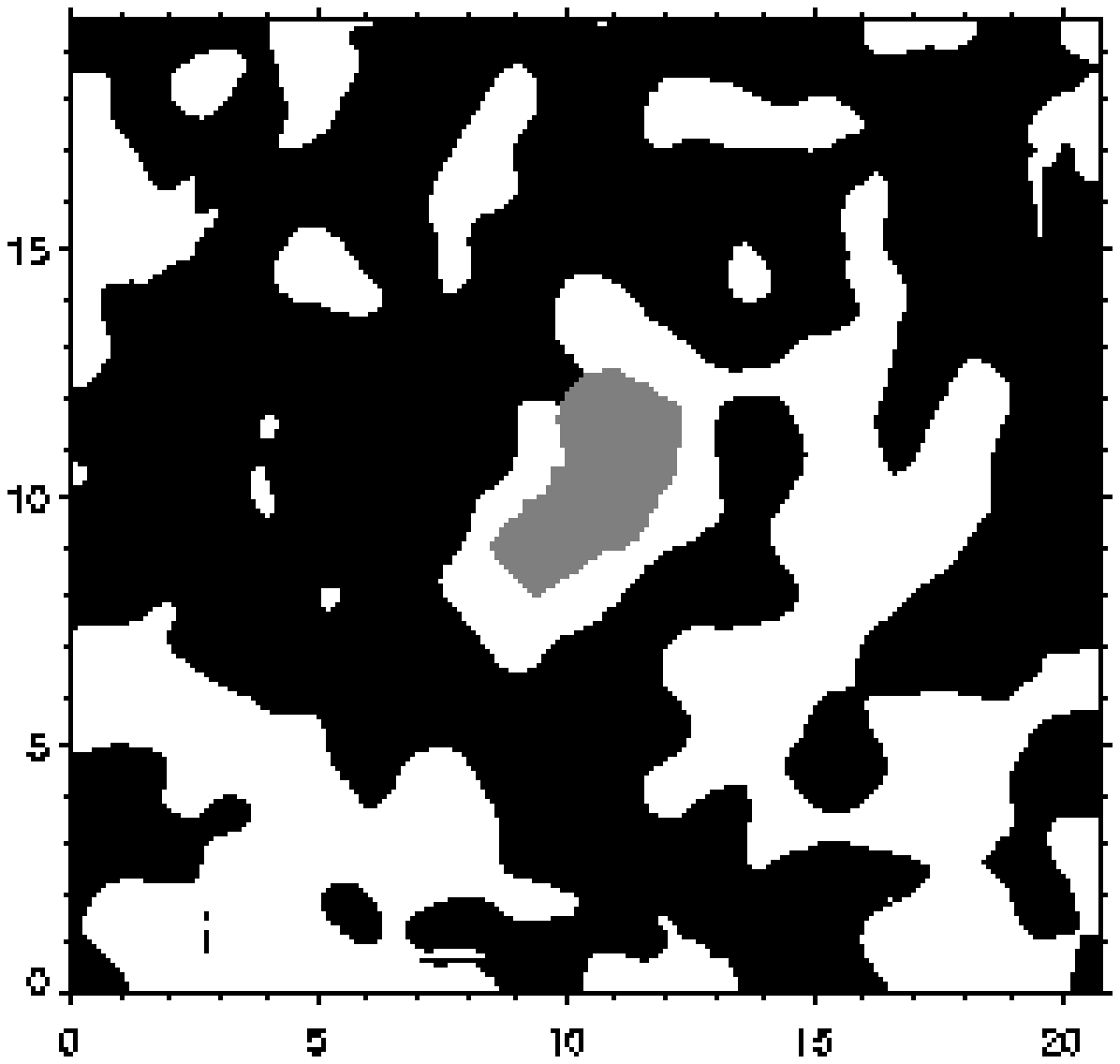} &
{\footnotesize \textsf{j})}\includegraphics[width=.4\linewidth]{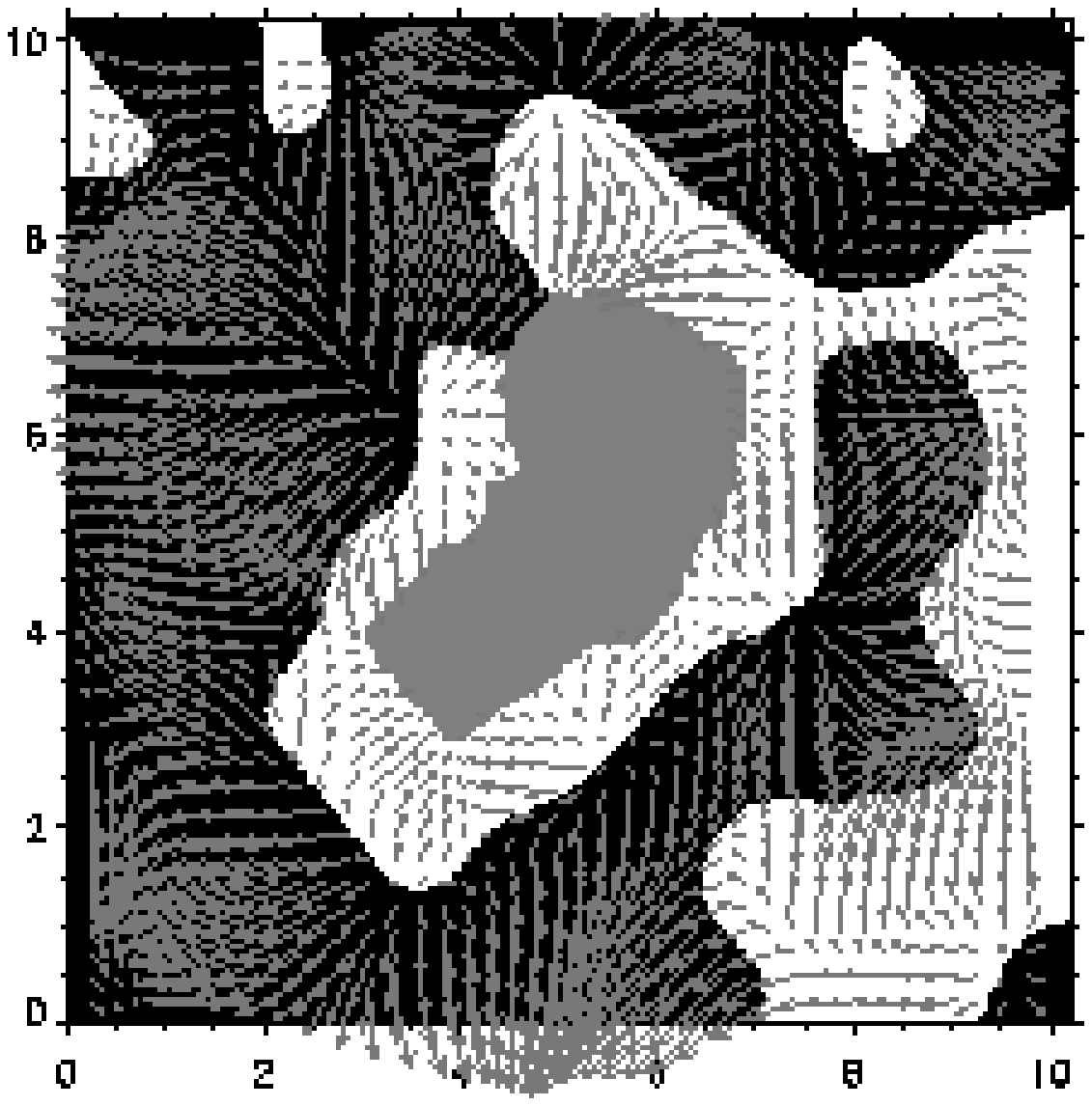} \\
\end{tabular}
\flushleft {\footnotesize F}{\scriptsize IGURE} {\footnotesize\ref{radialvel_procedure1} (\sf \emph{cont.}) --- \textsf{g}) \sf Radial directions; arrows point to the inward (negative) directions and the (\emph{white dot}) represents the pore gravity center. \textsf{h}) \sf Map of $\cos\beta$. \textsf{i}) \sf Mask of inward (\emph{white}) and outward (\emph{black}) radial velocities. \textsf{j}) \sf Map of velocities masked with \textsf{i}).}
\end{figure}

\begin{figure}
\centering
{\footnotesize \sf SST 30.09.2007 - PROPER MOTIONS ANALYSIS AROUND \emph{GRAN CANARIA} PORE}\\
\vspace{1cm}
\begin{tabular}{cc}
\hspace{-1cm}\includegraphics[width=.45\linewidth]{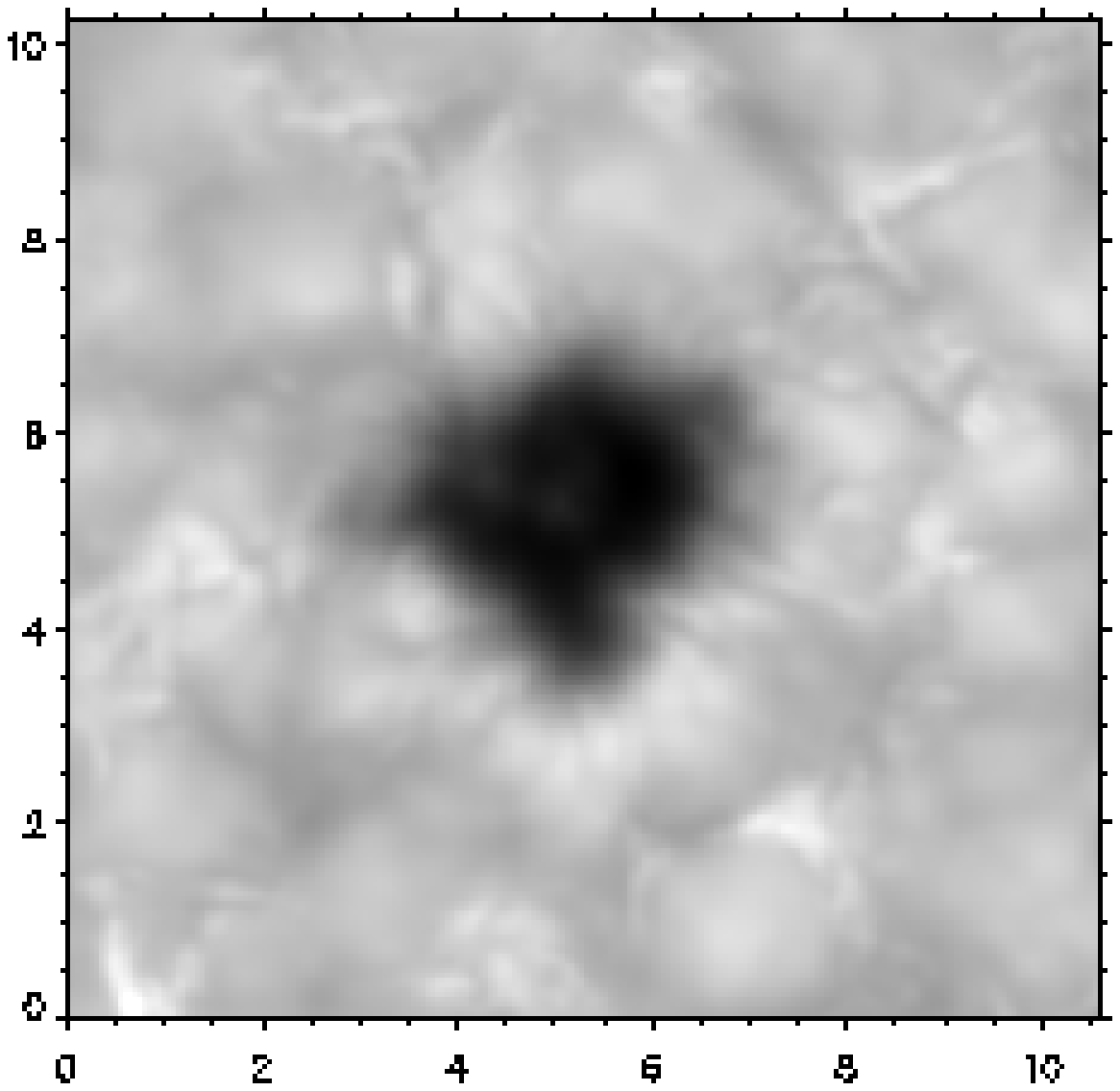} &
\hspace{-0.8cm}\includegraphics[width=.45\linewidth]{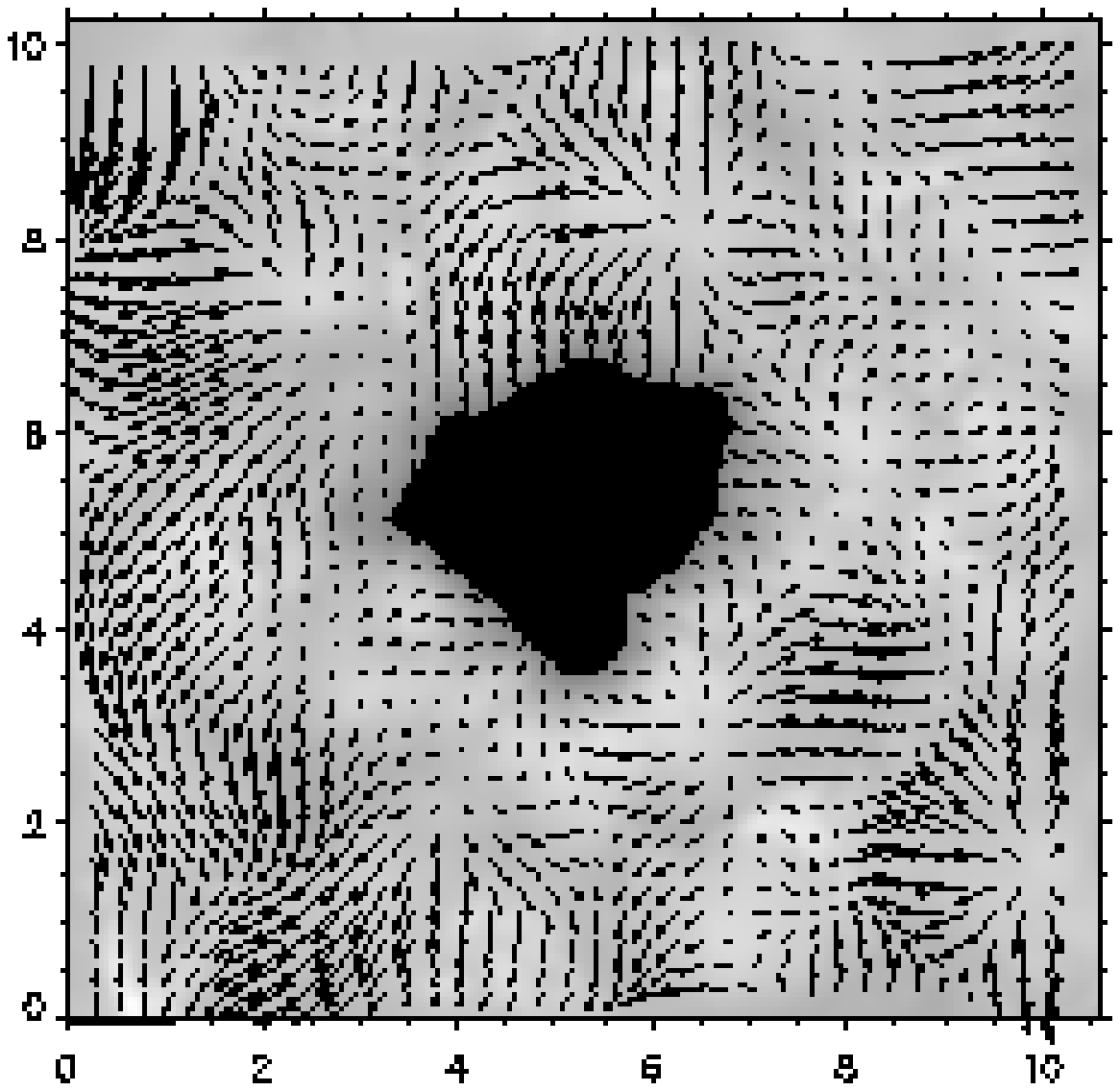} \\
\hspace{-1cm}\includegraphics[width=.45\linewidth]{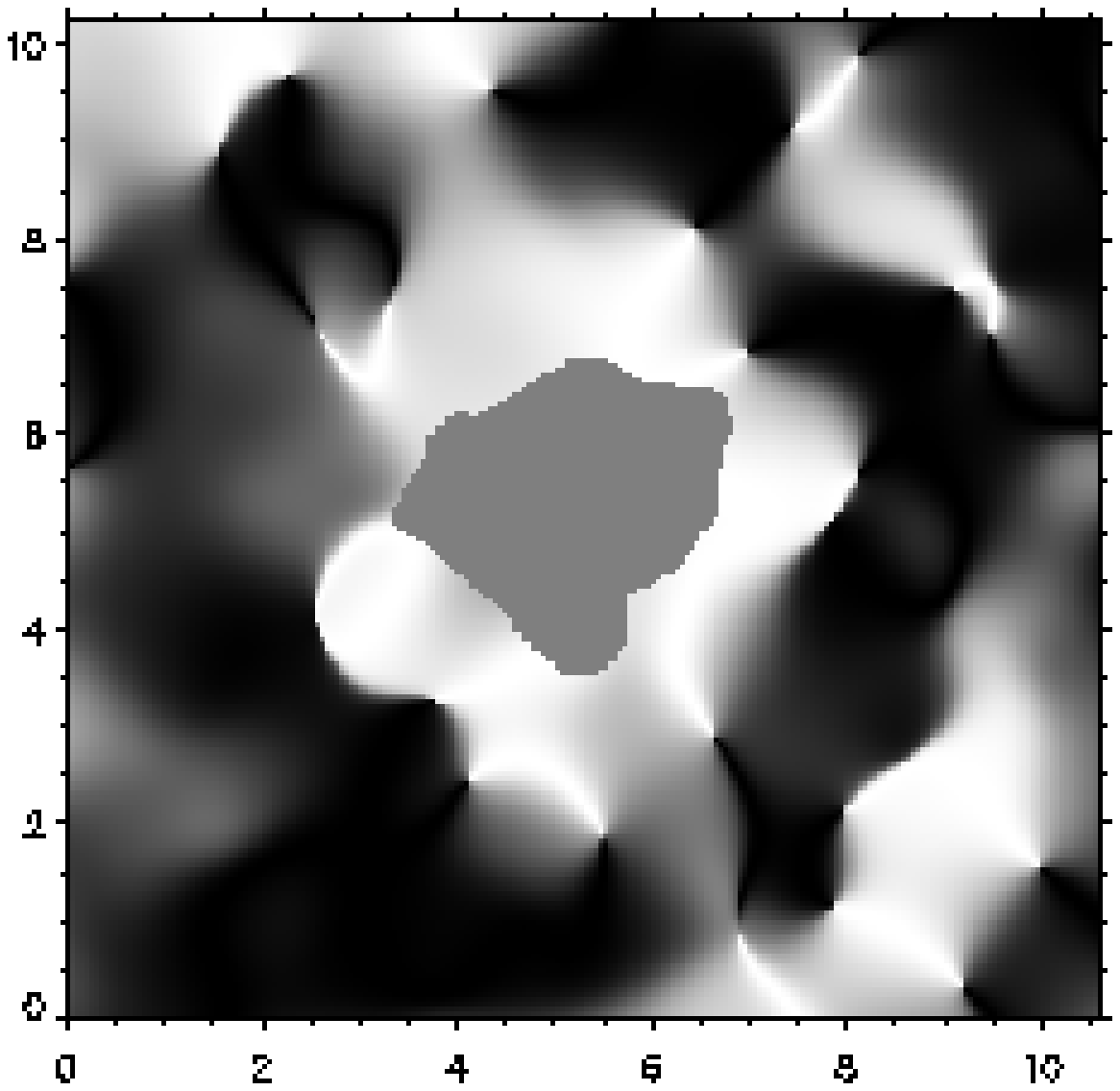} &
\hspace{-0.8cm}\includegraphics[width=.45\linewidth]{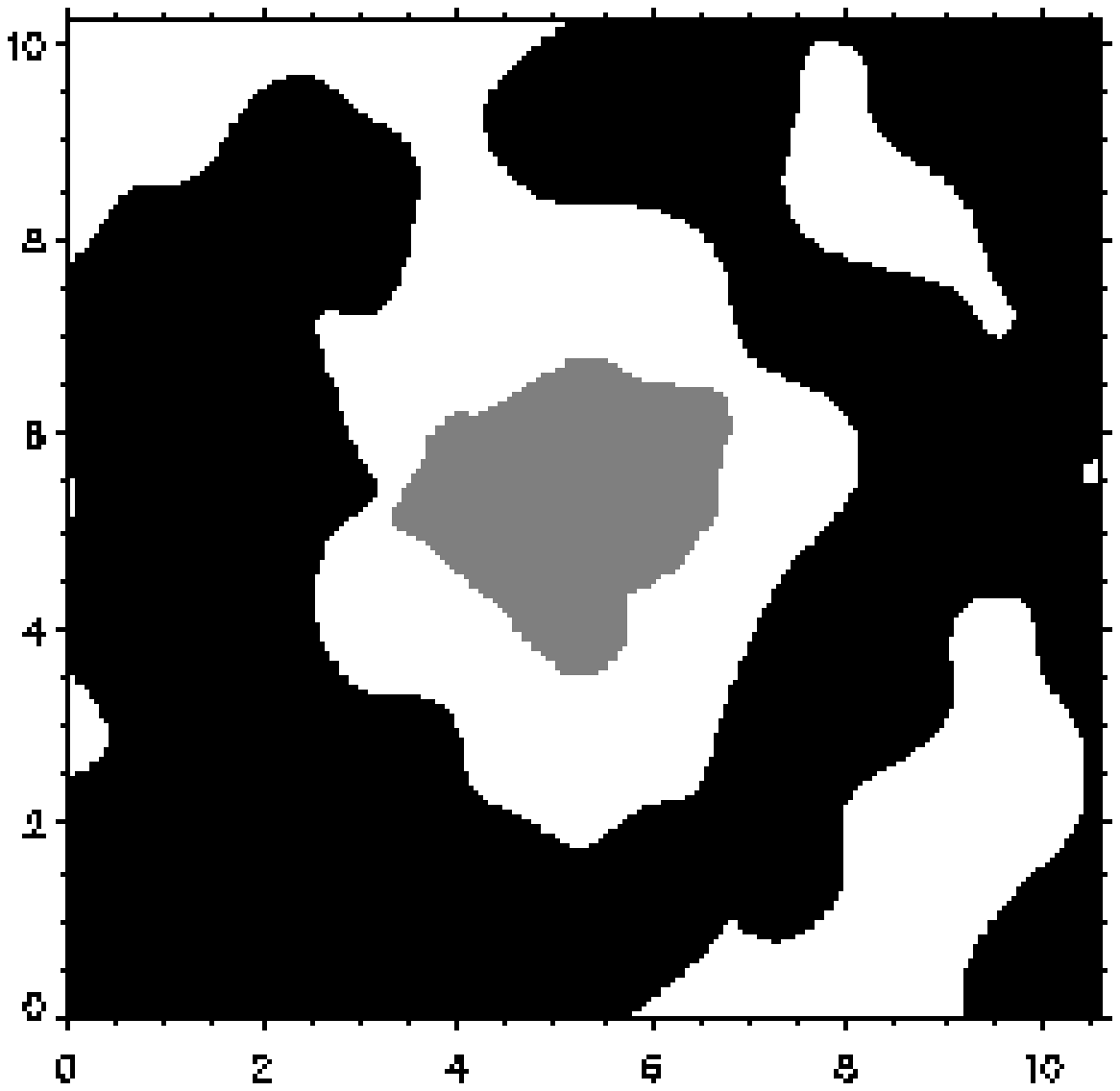} 
\end{tabular}
\caption[\sf Plot of the inward and outward motions around \emph{GRAN CANARIA} pore.]{\sf Discrimination between inward and outward motions surrounding \emph{GRAN CANARIA} pore. Black and white areas in the last panel correspond to velocities displaying outward and inward radial components, respectively. The length of the black bar at coordinates (0,0) in the upper right panel corresponds to 1.6 km s$^{-1}$. The spatial units are in arc sec.}
\label{radialvel_pore1}
\end{figure}

\begin{figure}
\centering
{\footnotesize \sf SST 30.09.2007 - PROPER MOTIONS ANALYSIS AROUND \emph{TENERIFE} PORE}\\
\vspace{1cm}
\begin{tabular}{cc}
\hspace{-1cm}\includegraphics[width=.45\linewidth]{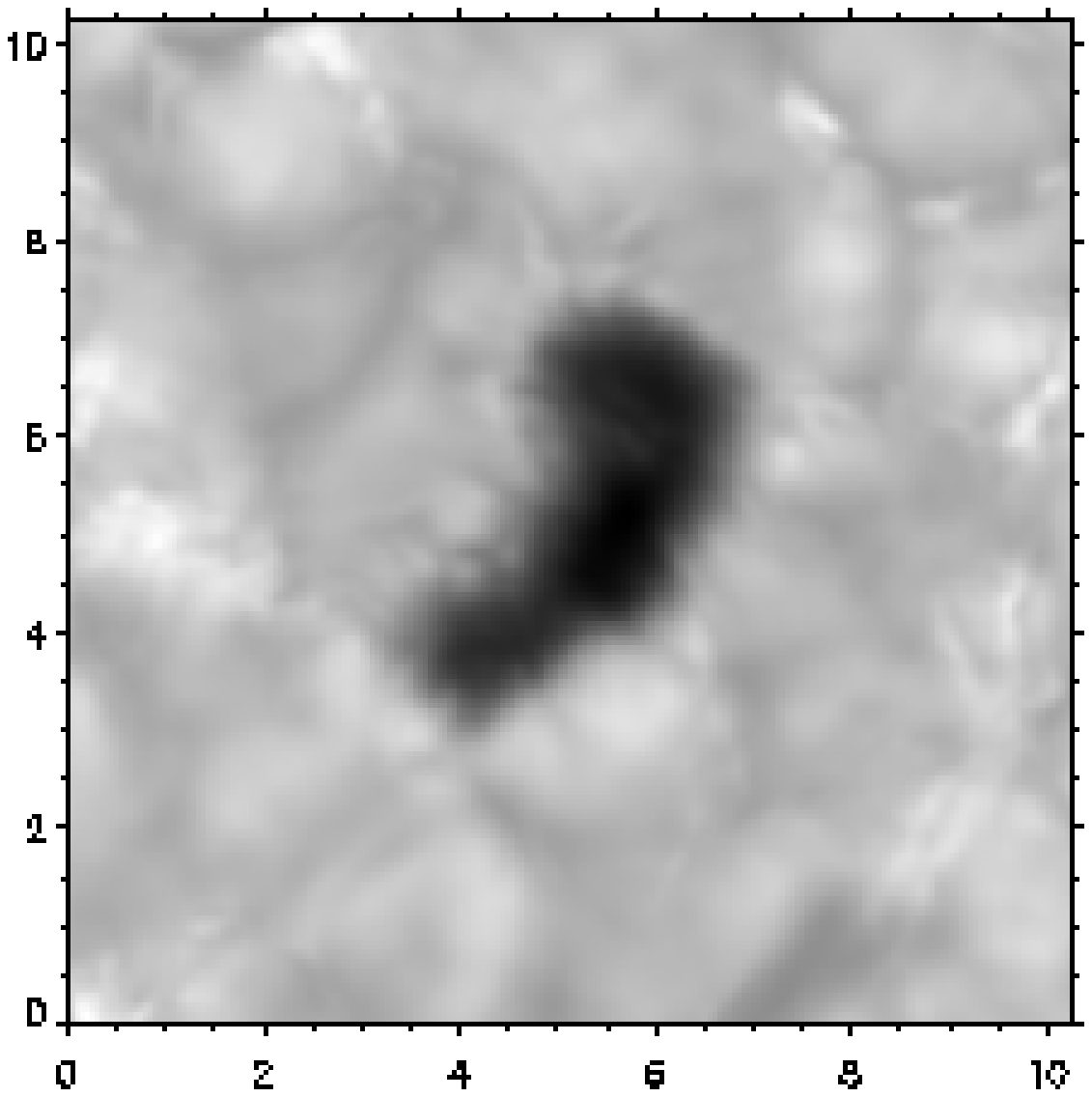} &
\hspace{-0.8cm}\includegraphics[width=.45\linewidth]{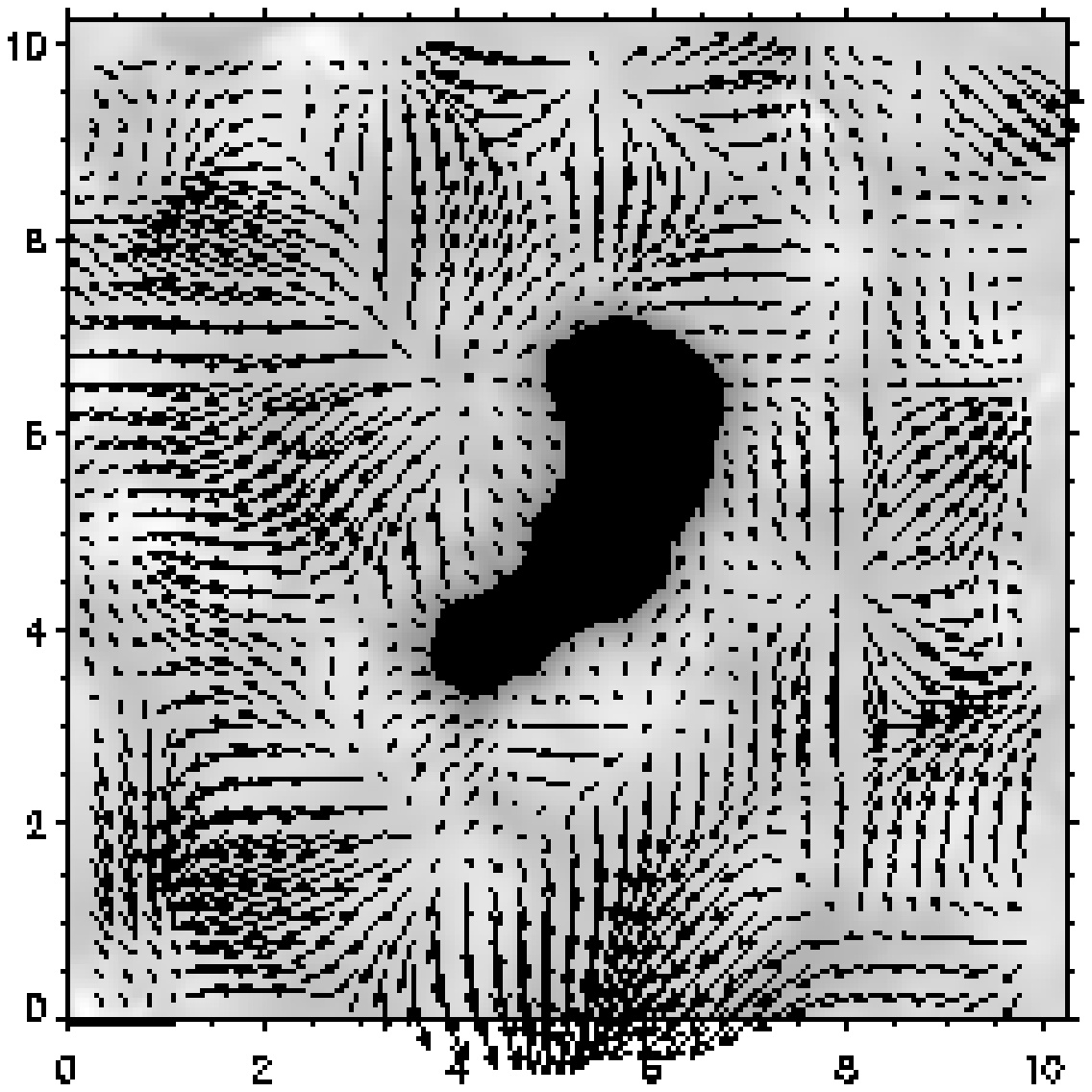} \\
\hspace{-1cm}\includegraphics[width=.45\linewidth]{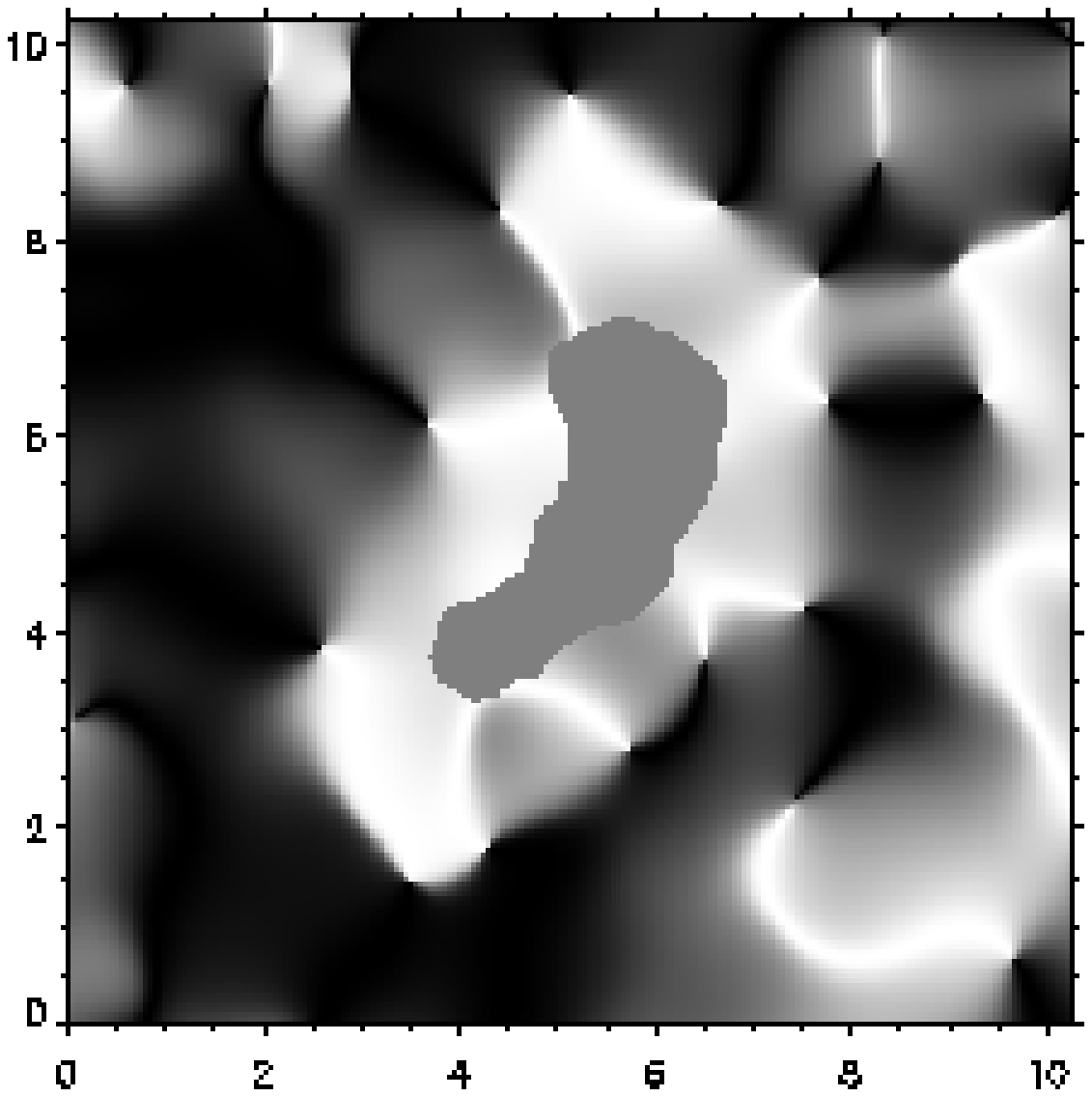} &
\hspace{-0.8cm}\includegraphics[width=.45\linewidth]{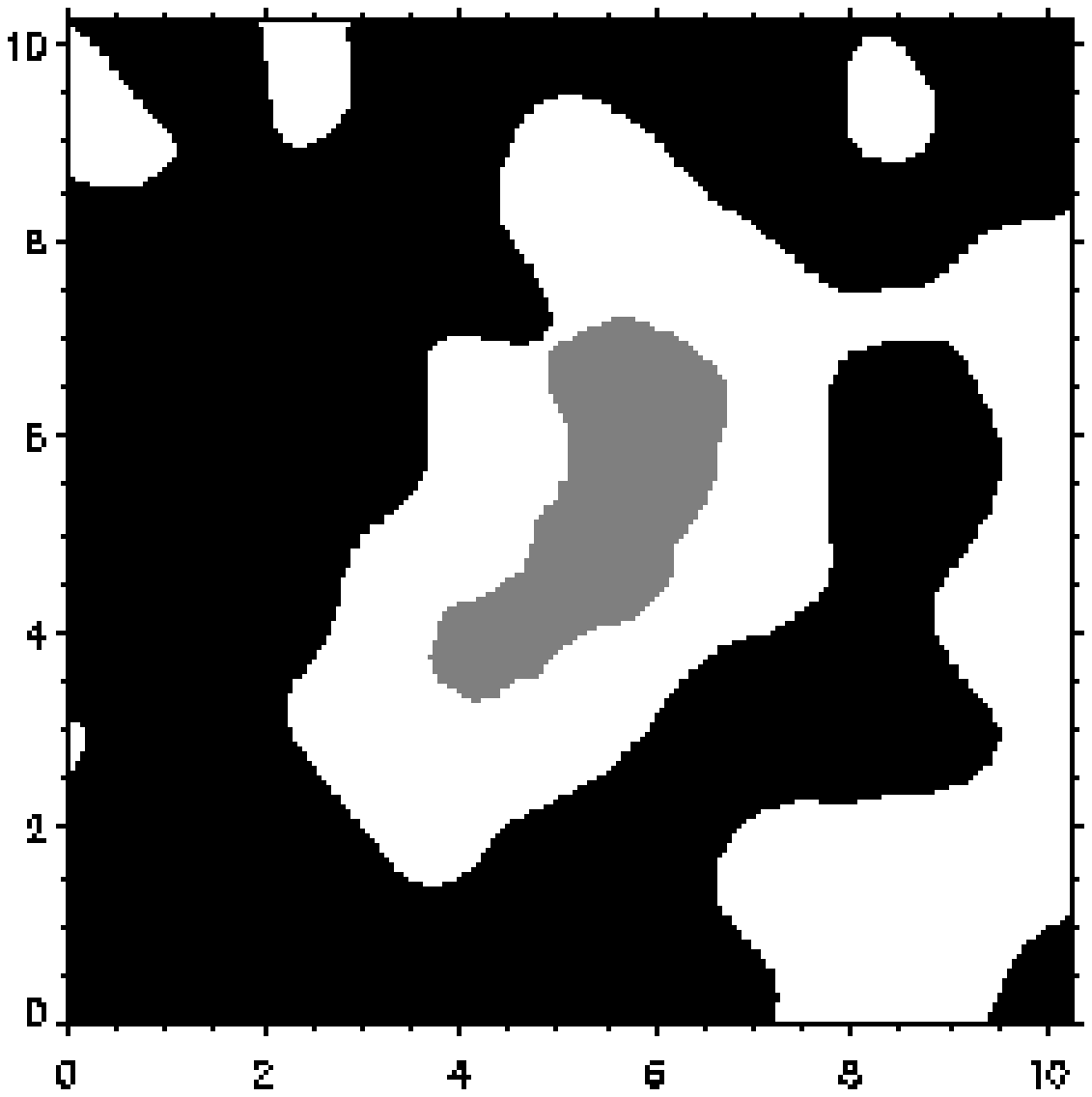} 
\end{tabular}
\caption[\sf Plot of the inward and outward motions around \emph{TENERIFE} pore.]{\sf Discrimination between  inward and outward motions surrounding \emph{TENERIFE} pore. Black and white areas in the last panel correspond to velocities displaying outward and inward radial components, respectively. The length of the black bar at coordinates (0,0) in the upper right panel corresponds to 1.6 km s$^{-1}$. The spatial units are in arc sec.}
\label{radialvel_pore2}
\end{figure}

\begin{figure}
\centering
{\footnotesize \sf SST 30.09.2007 - PROPER MOTIONS ANALYSIS AROUND \emph{LA PALMA} PORE}\\
\vspace{1cm}
\begin{tabular}{cc}
\hspace{-1cm}\includegraphics[width=.45\linewidth]{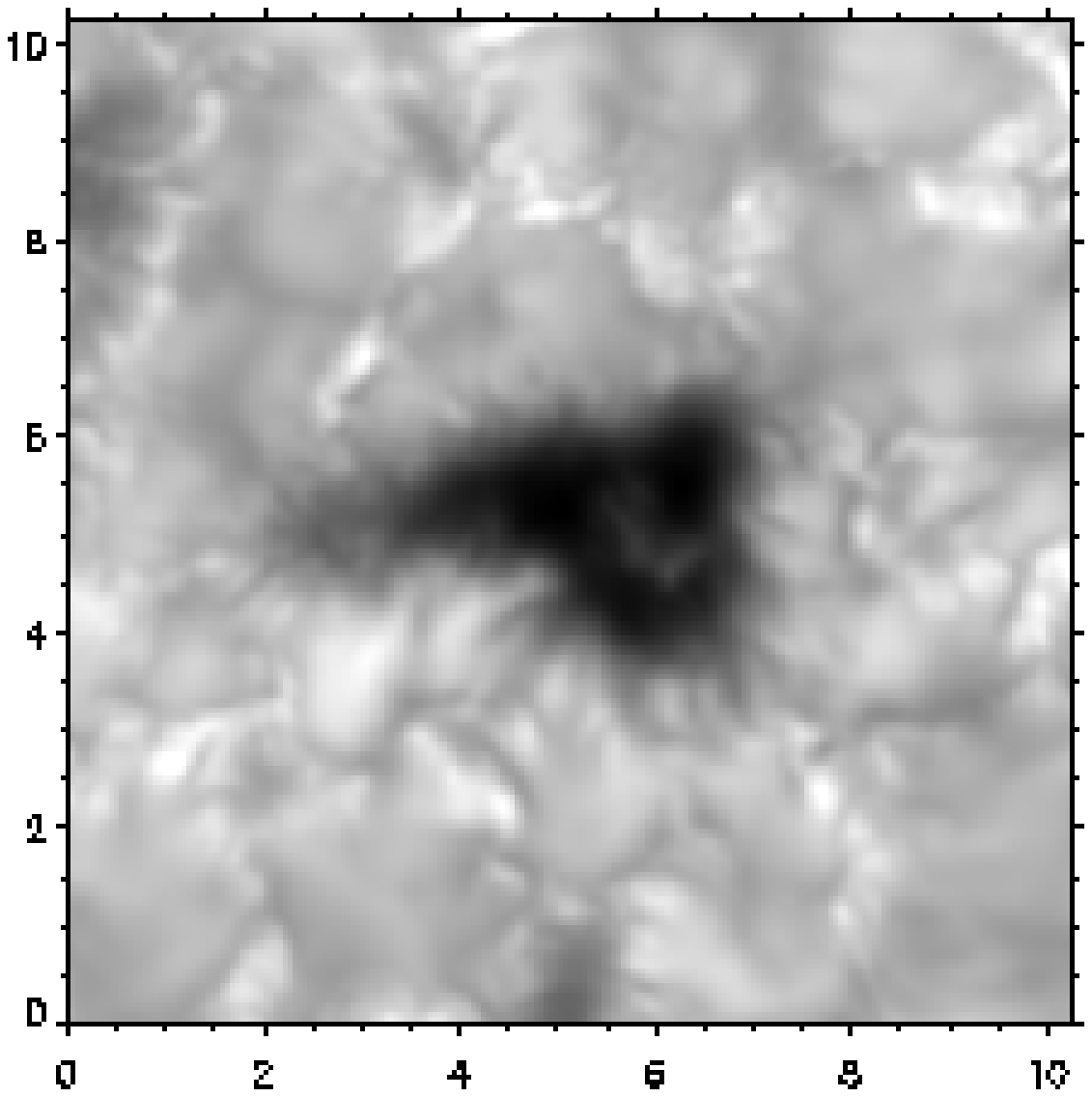} &
\hspace{-0.8cm}\includegraphics[width=.45\linewidth]{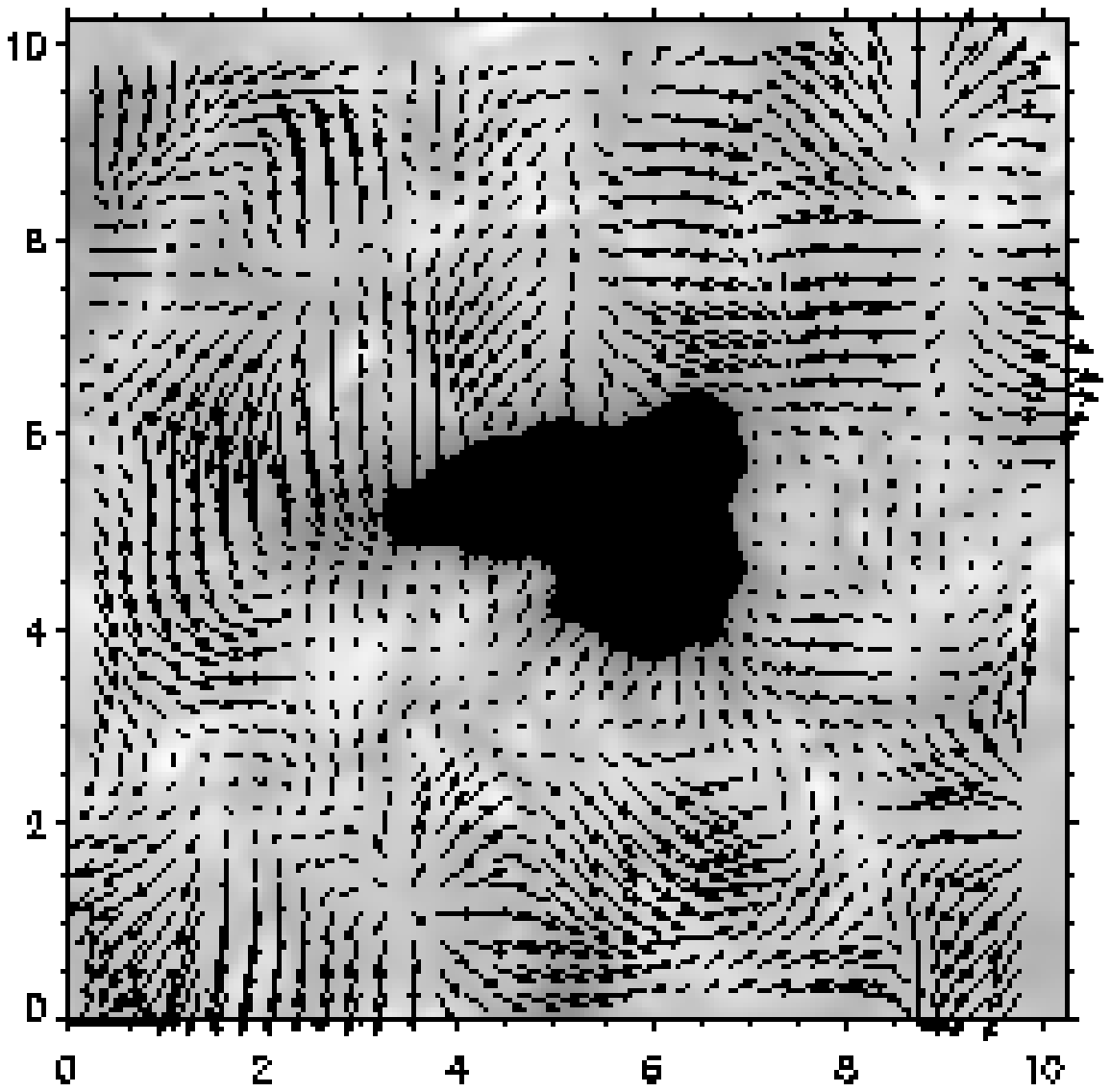} \\
\hspace{-1cm}\includegraphics[width=.45\linewidth]{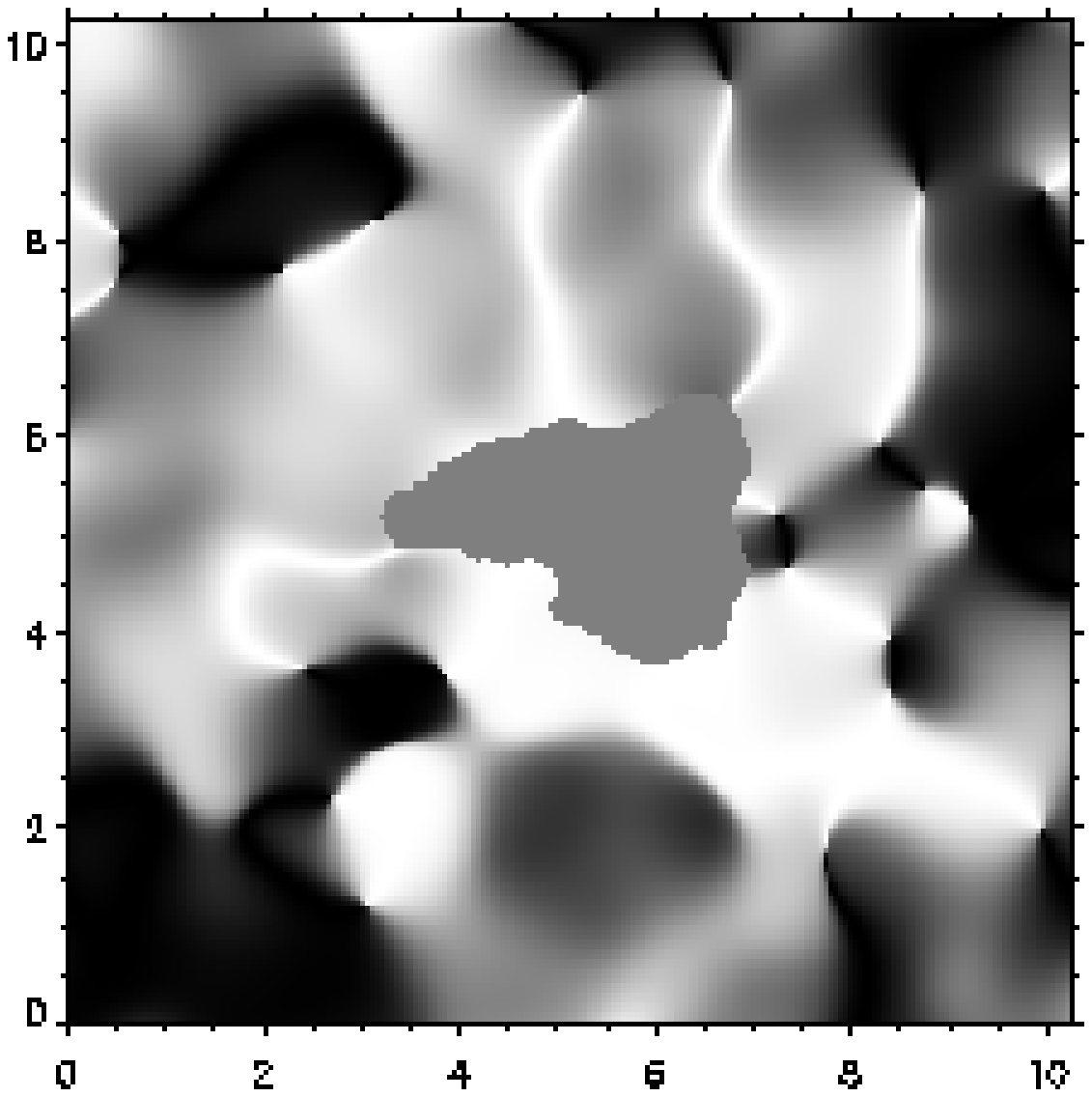} &
\hspace{-0.8cm}\includegraphics[width=.45\linewidth]{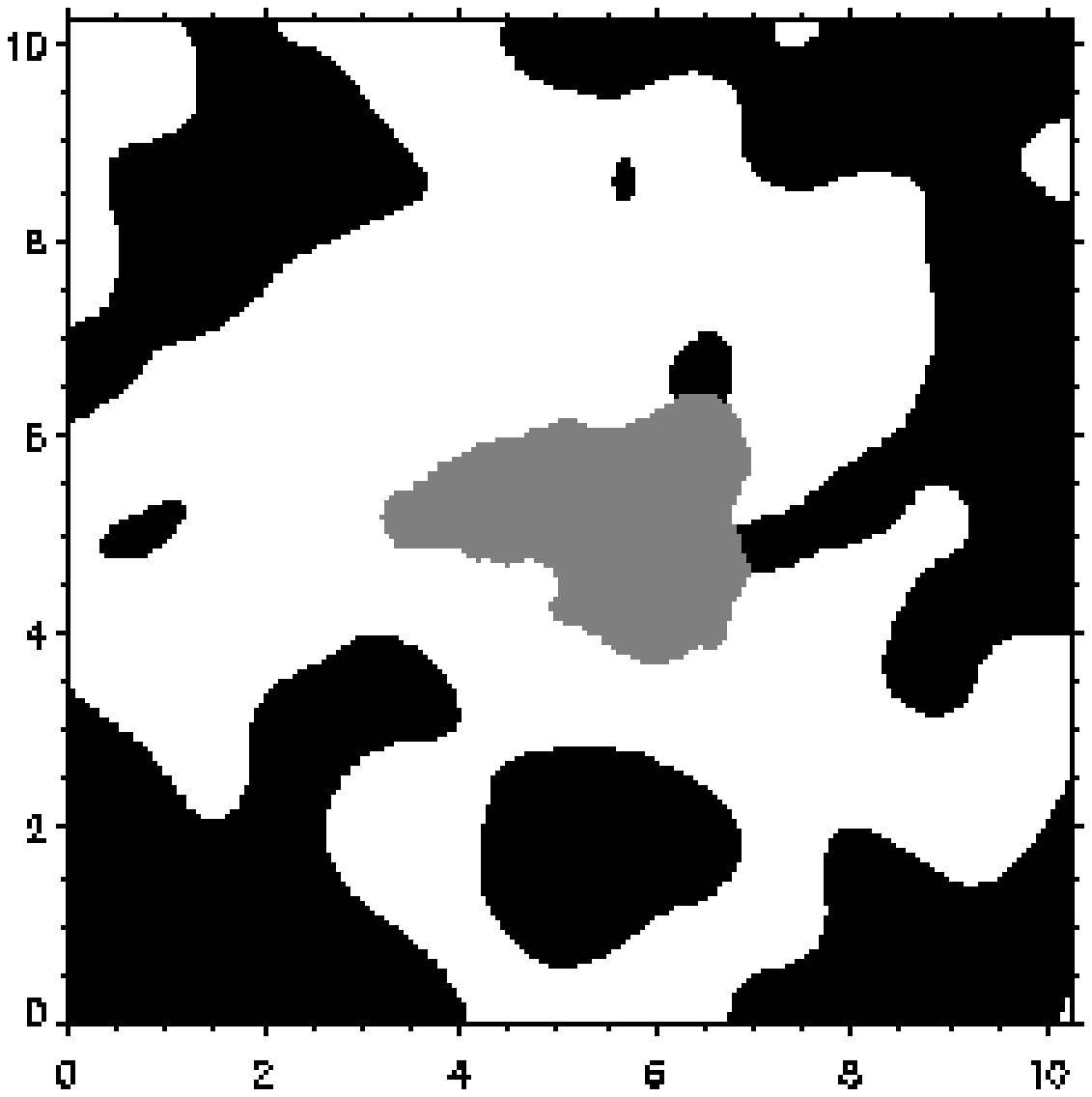} 
\end{tabular}
\caption[\sf Plot of the inward and outward motions around \emph{LA PALMA} pore .]{\sf Discrimination between  inward and outward motions surrounding \emph{LA PALMA} pore. Black and white areas in the last panel correspond to velocities displaying outward and inward radial components, respectively. The length of the black bar at coordinates (0,0) in the upper right panel corresponds to 1.6 km s$^{-1}$. The spatial units are in arc sec.}
\label{radialvel_pore3}
\end{figure}

\begin{figure}
\centering
{\footnotesize \sf SST 30.09.2007 - PROPER MOTIONS ANALYSIS AROUND \emph{LA GOMERA} PORE}\\
\vspace{1cm}
\begin{tabular}{cc}
\hspace{-1cm}\includegraphics[width=.45\linewidth]{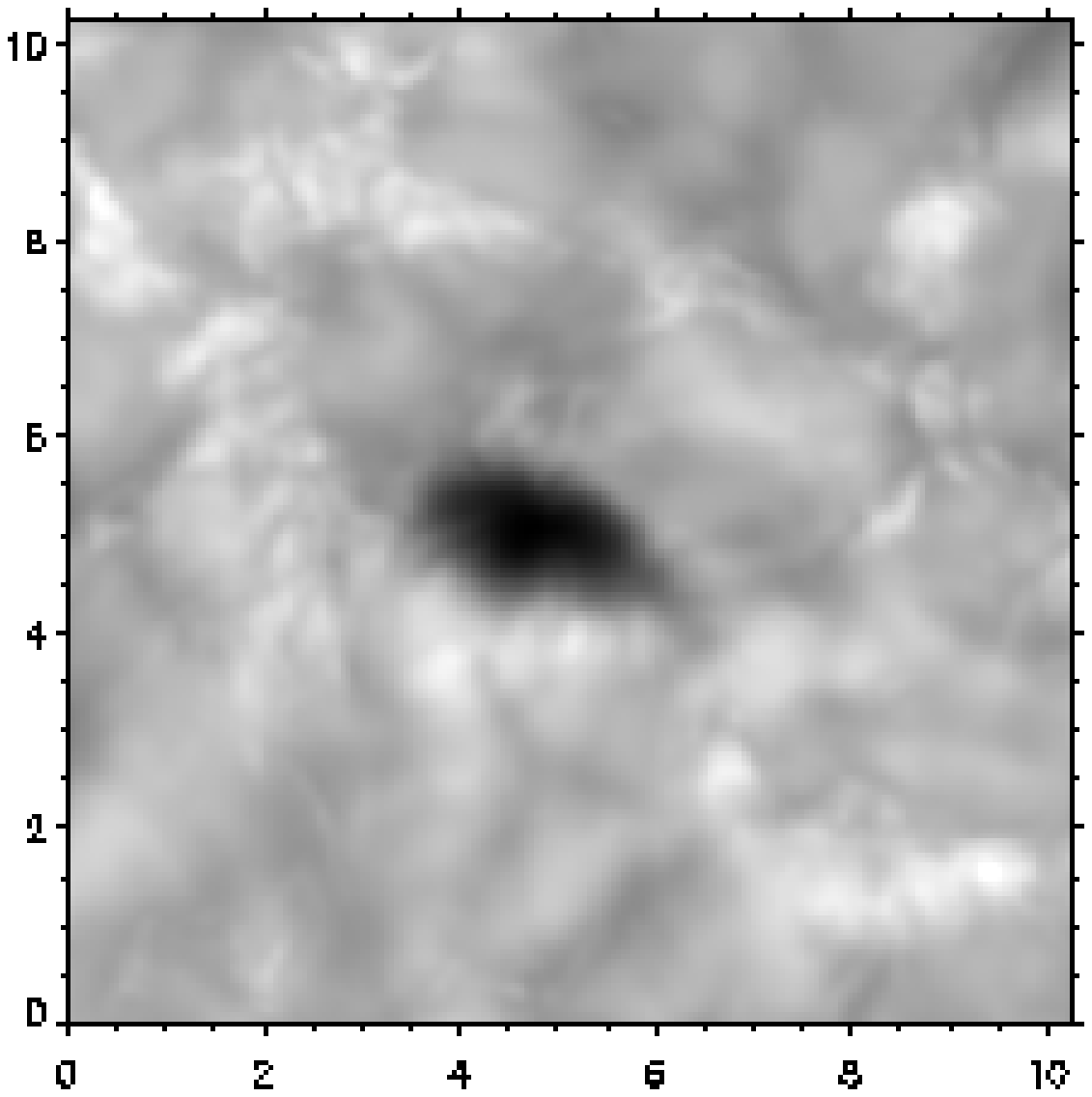} &
\hspace{-0.8cm}\includegraphics[width=.45\linewidth]{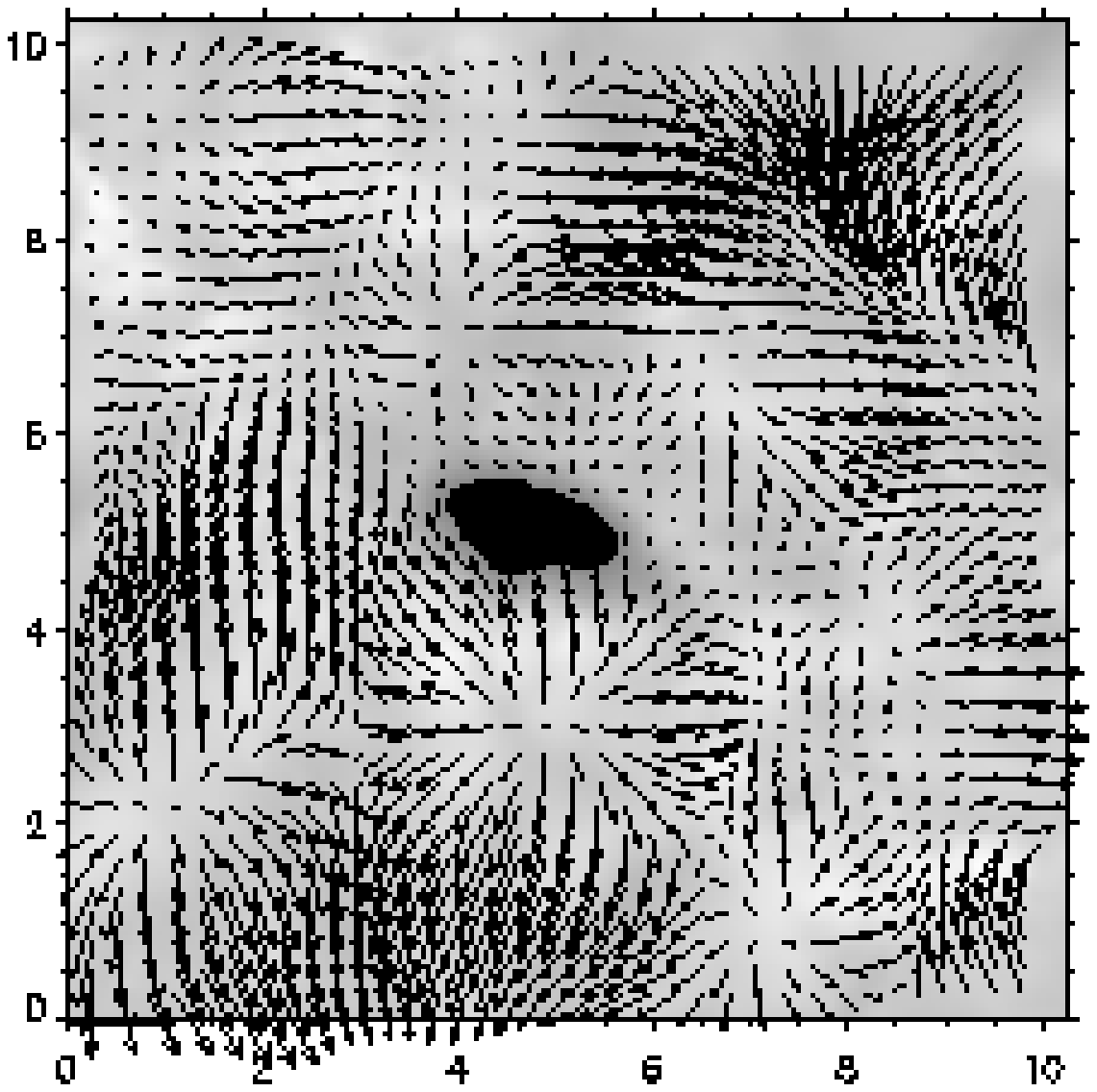} \\
\hspace{-1cm}\includegraphics[width=.45\linewidth]{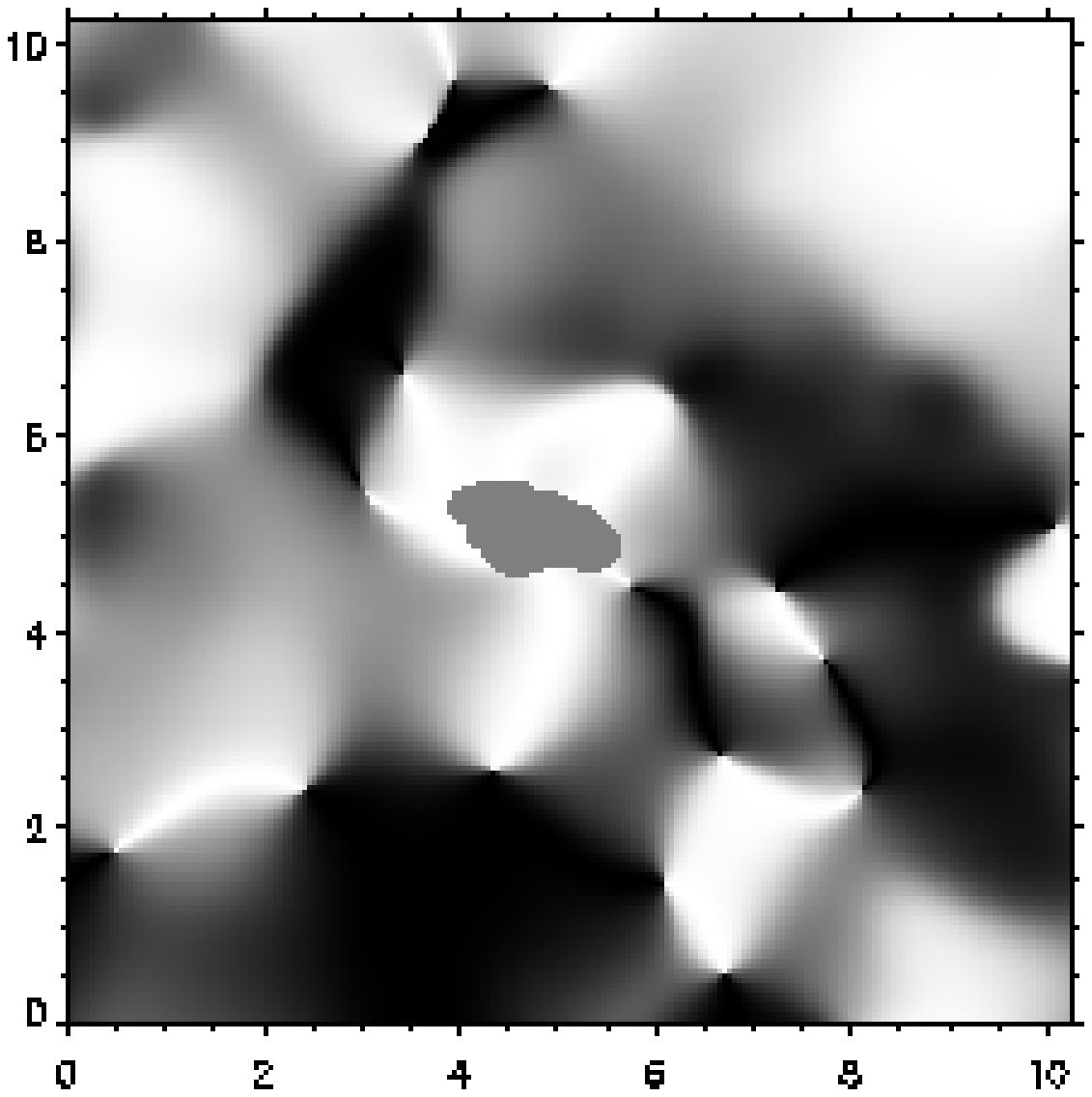} &
\hspace{-0.8cm}\includegraphics[width=.45\linewidth]{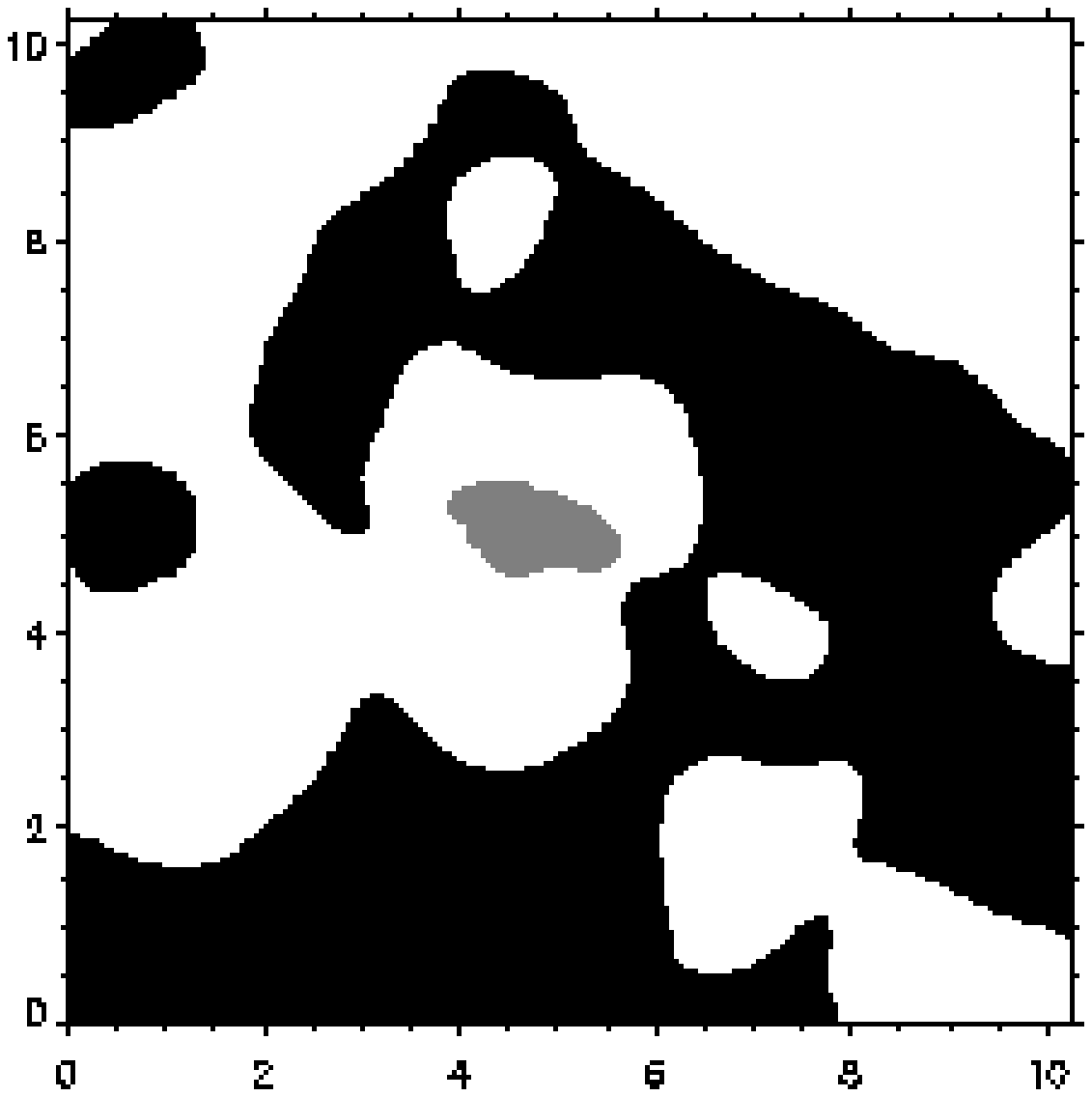} 
\end{tabular}
\caption[\sf Plot of the inward and outward motions around \emph{LA GOMERA} pore.]{\sf Discrimination between  inward and outward motions surrounding \emph{LA GOMERA} pore. Black and white areas in the last panel  correspond to velocities displaying outward and inward radial components, respectively. The length of the black bar at coordinates (0,0) in the upper right panel corresponds to 1.6 km s$^{-1}$. The spatial units are in arc sec.}
\label{radialvel_pore4}
\end{figure}

\begin{figure}
\centering
{\footnotesize \sf SST 30.09.2007 - PROPER MOTIONS ANALYSIS AROUND \emph{EL HIERRO} PORE}\\
\vspace{1cm}
\begin{tabular}{cc}
\hspace{-1cm}\includegraphics[width=.45\linewidth]{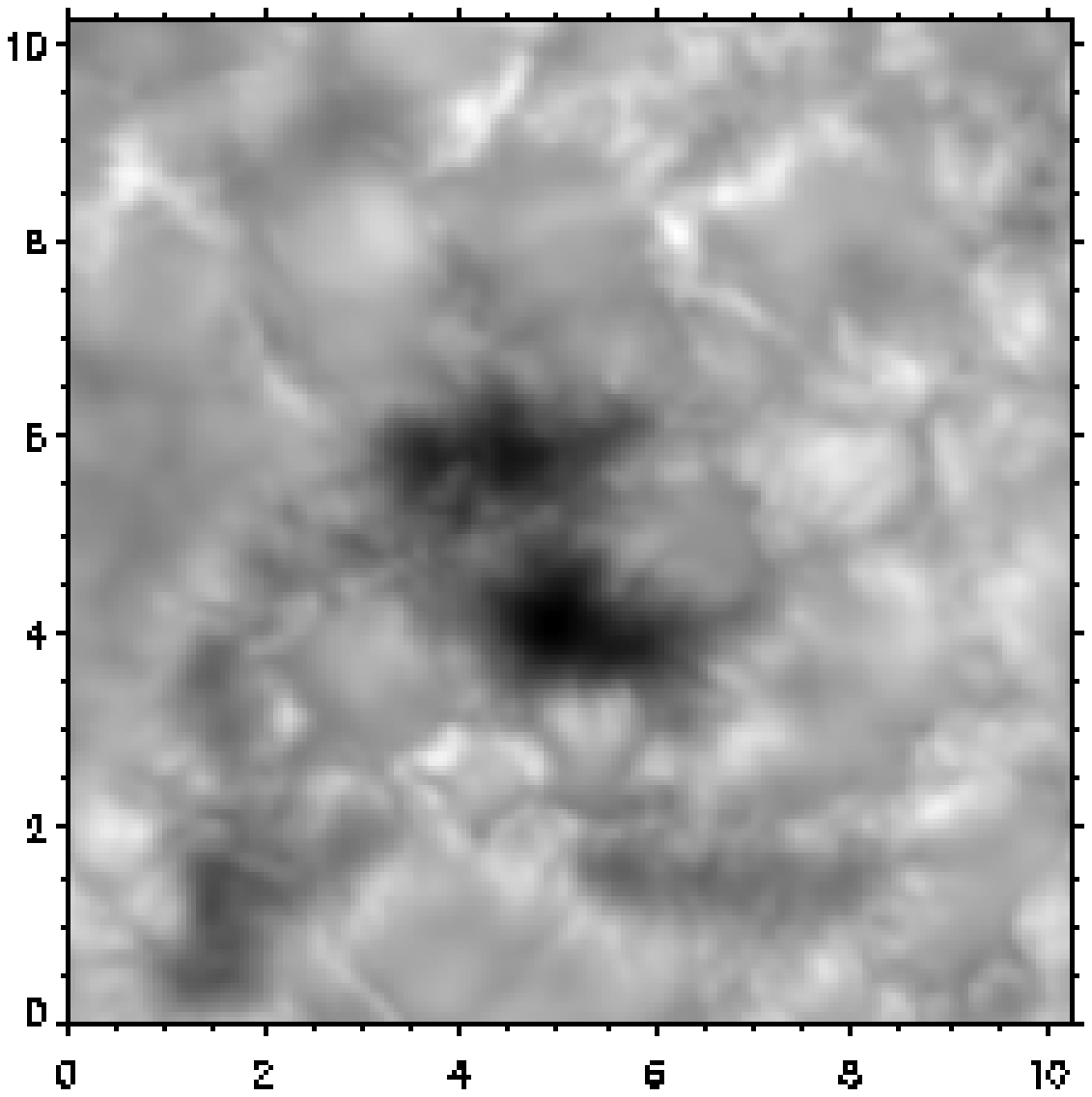} &
\hspace{-0.8cm}\includegraphics[width=.45\linewidth]{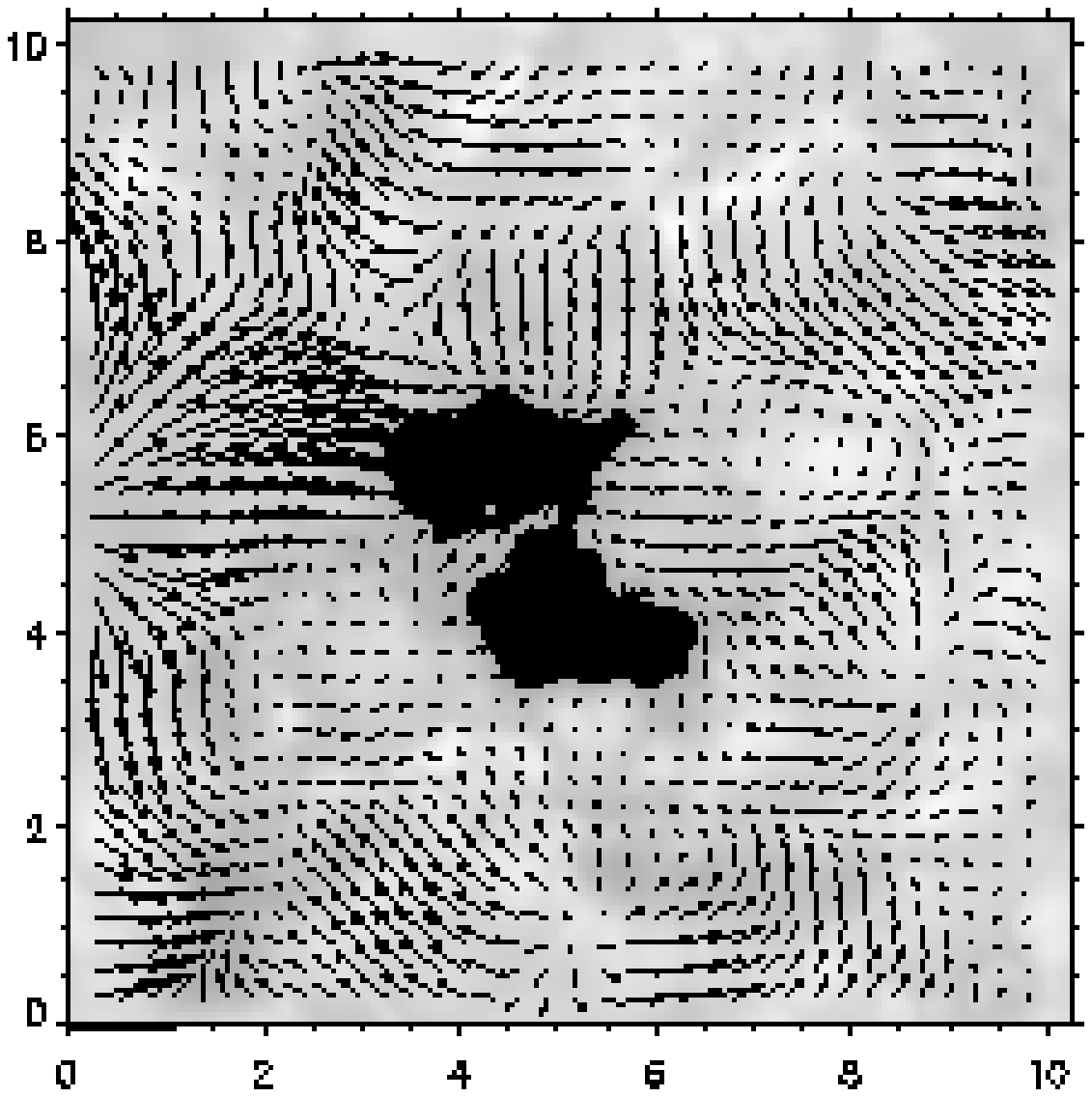} \\
\hspace{-1cm}\includegraphics[width=.45\linewidth]{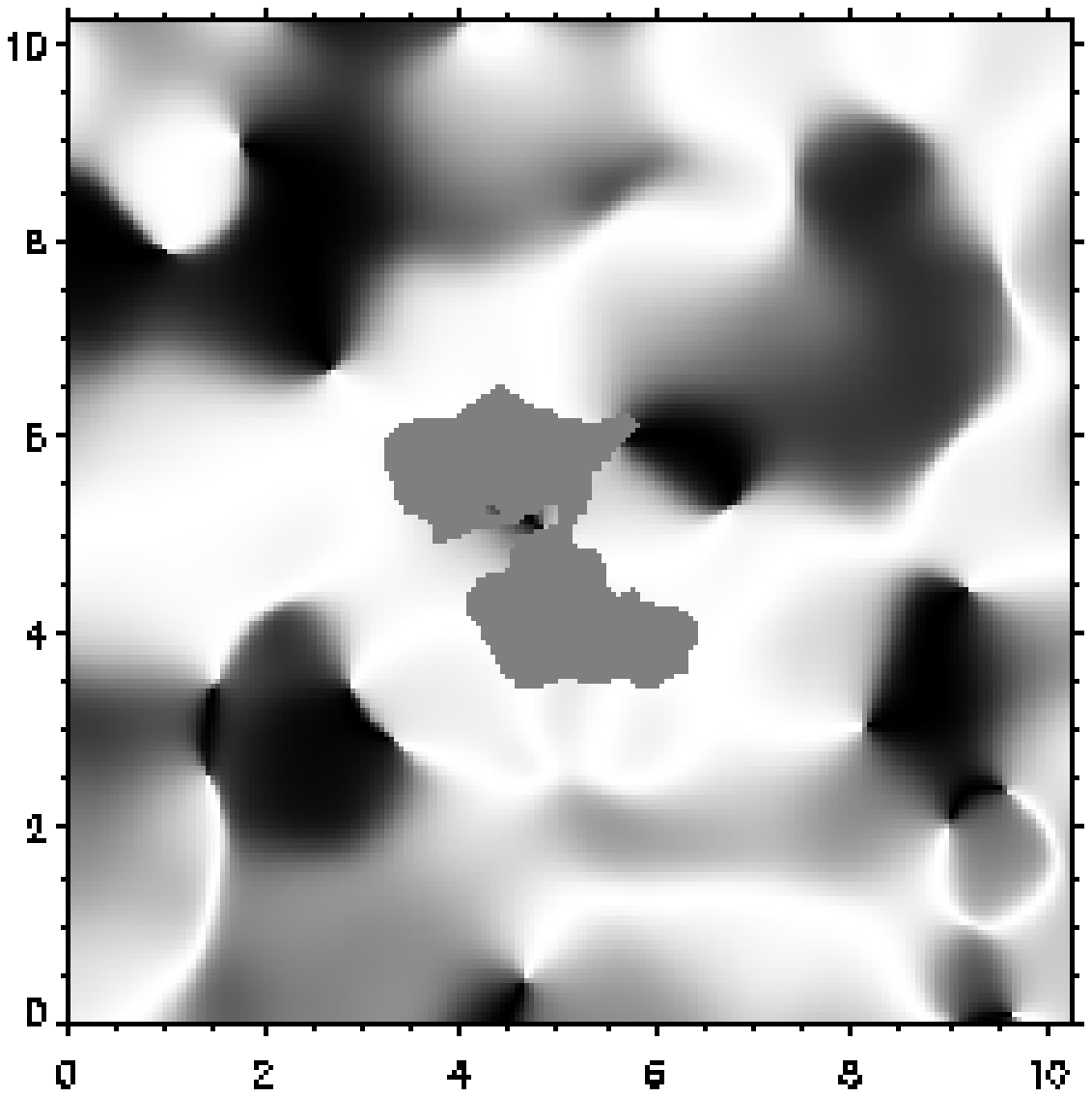} &
\hspace{-0.8cm}\includegraphics[width=.45\linewidth]{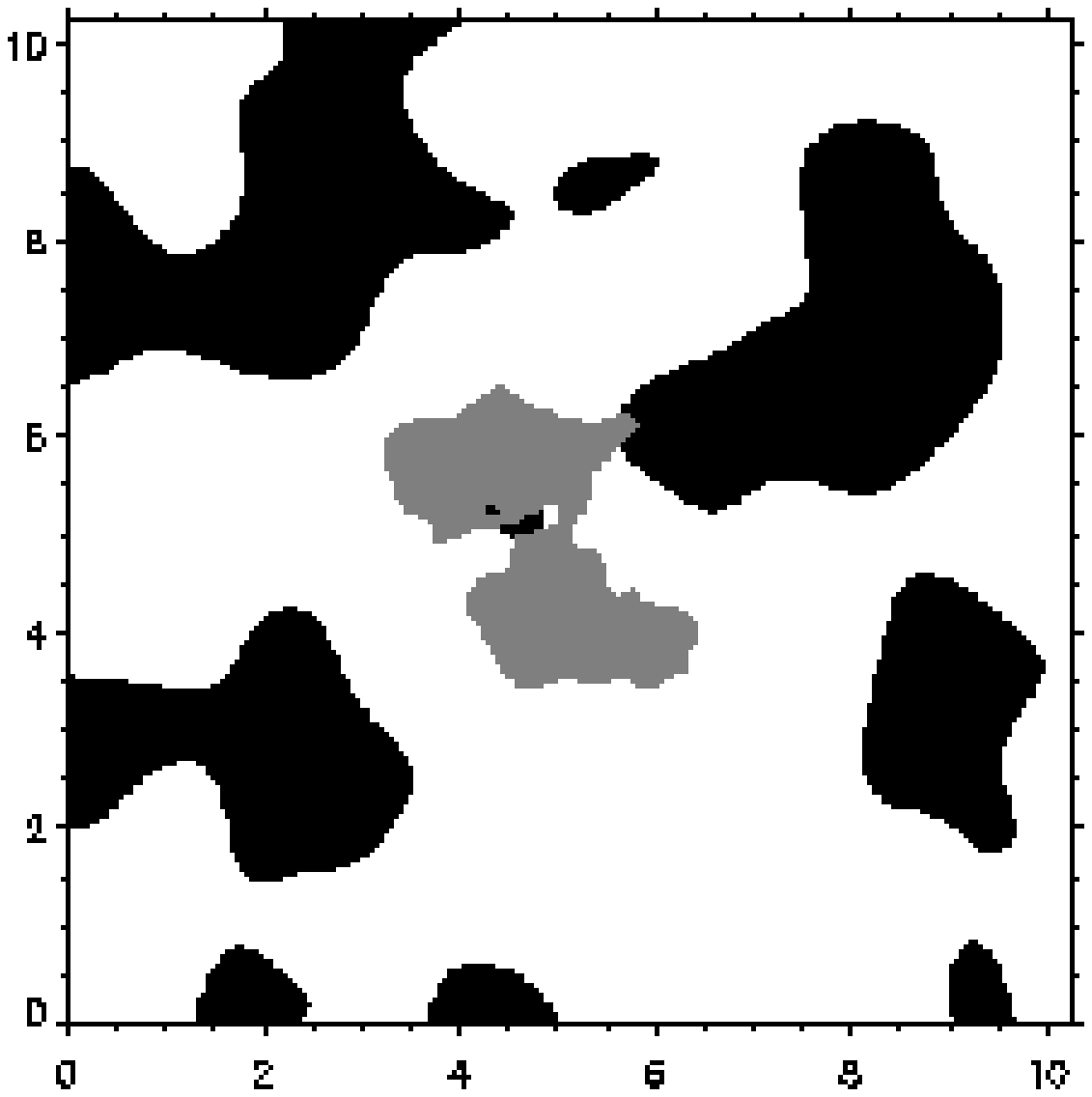} 
\end{tabular}
\caption[\sf Plot of the inward and outward motions around \emph{EL HIERRO} pore.]{\sf Discrimination between inward and outward motions surrounding \emph{EL HIERRO} pore. Black and white areas in the last panel correspond to velocities displaying outward and inward radial components, respectively. The length of the black bar at coordinates (0,0) in the upper right panel corresponds to 1.6 km s$^{-1}$. The spatial units are in arc sec.}
\label{radialvel_pore5}
\end{figure}

\begin{figure}
\centering
{\footnotesize \sf SST 30.09.2007 - PROPER MOTIONS ANALYSIS AROUND \emph{LANZAROTE} PORE}\\
\vspace{1cm}
\begin{tabular}{cc}
\hspace{-1cm}\includegraphics[width=.45\linewidth]{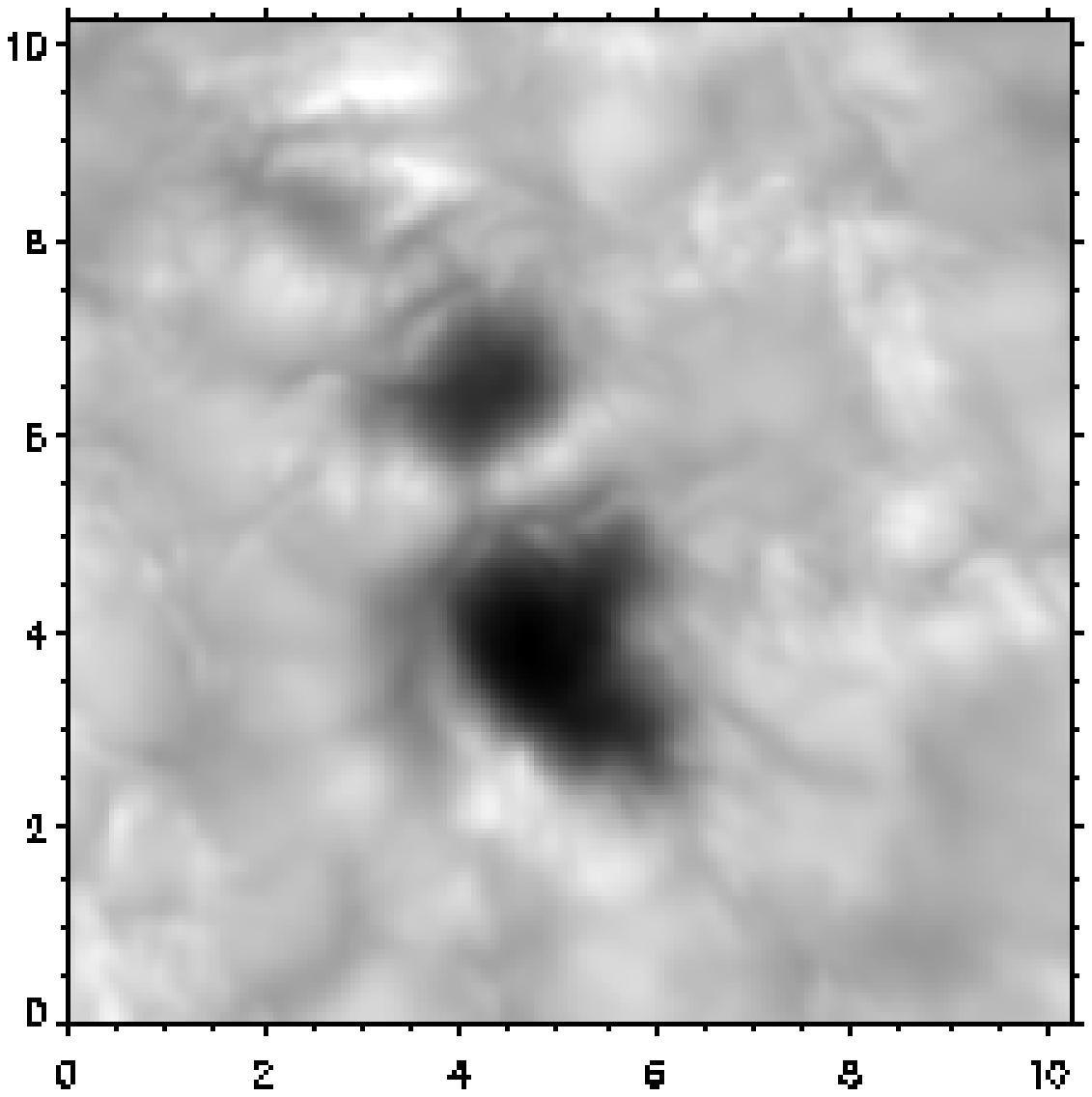} &
\hspace{-0.8cm}\includegraphics[width=.45\linewidth]{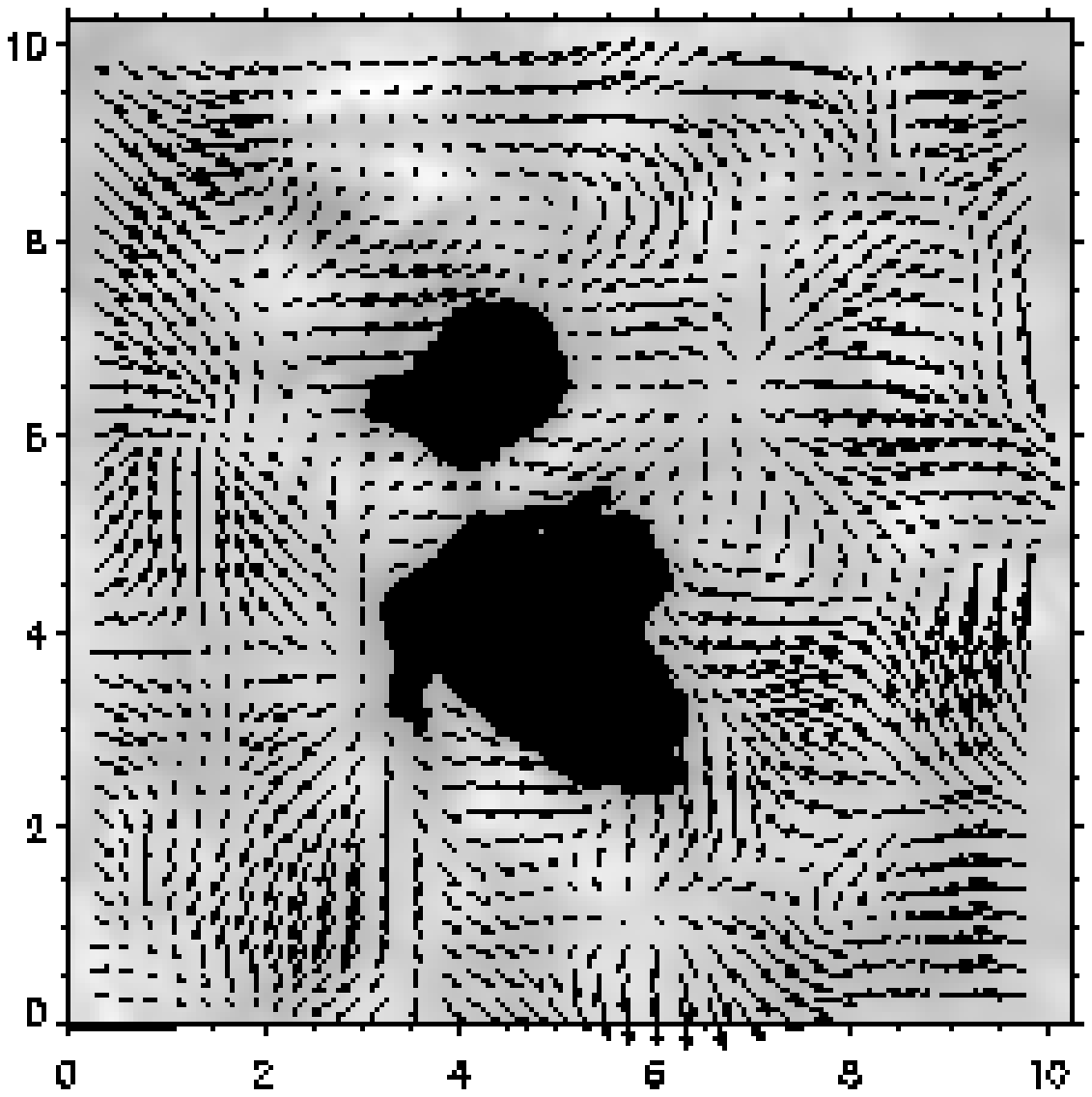} \\
\hspace{-1cm}\includegraphics[width=.45\linewidth]{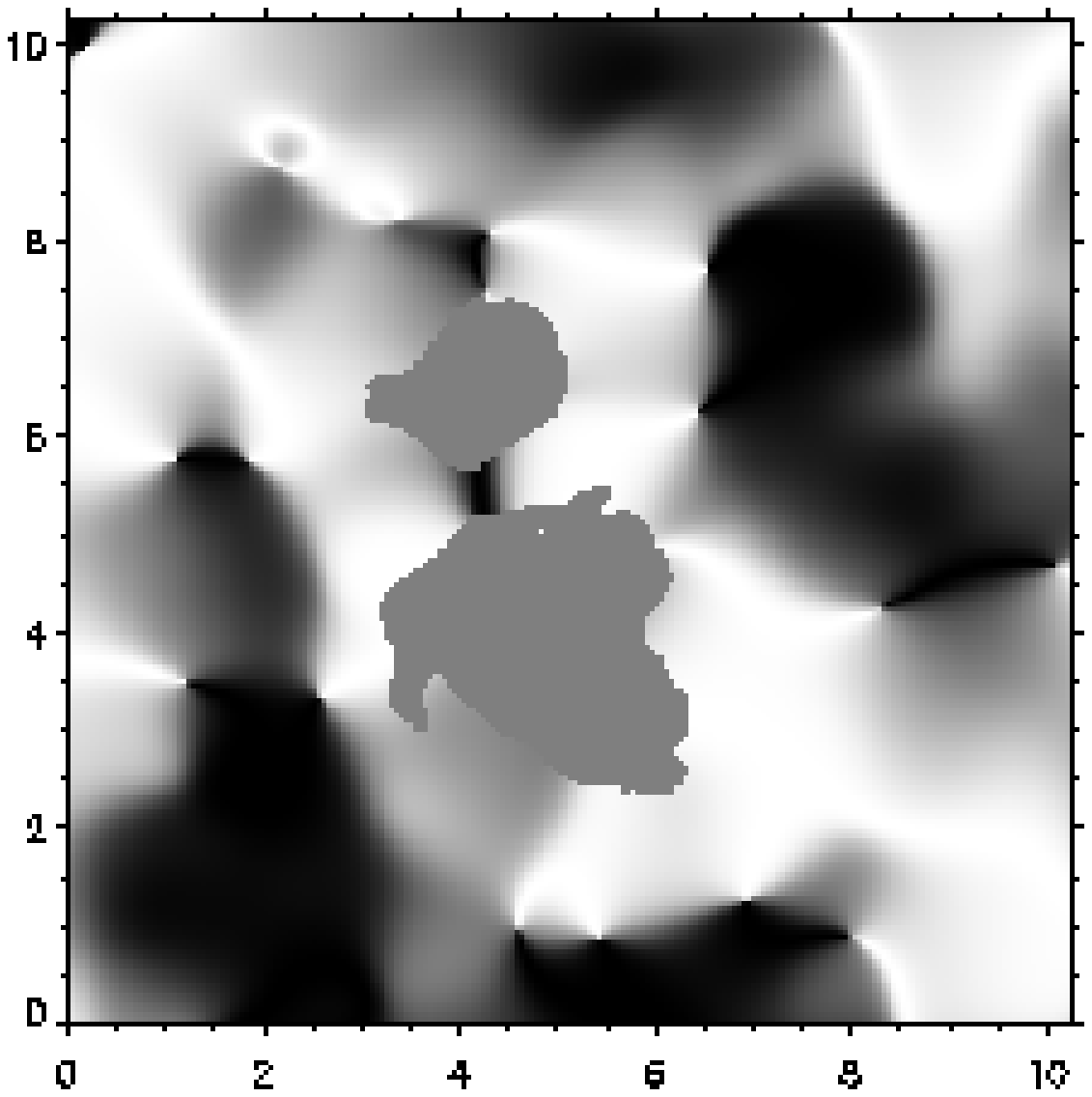} &
\hspace{-0.8cm}\includegraphics[width=.45\linewidth]{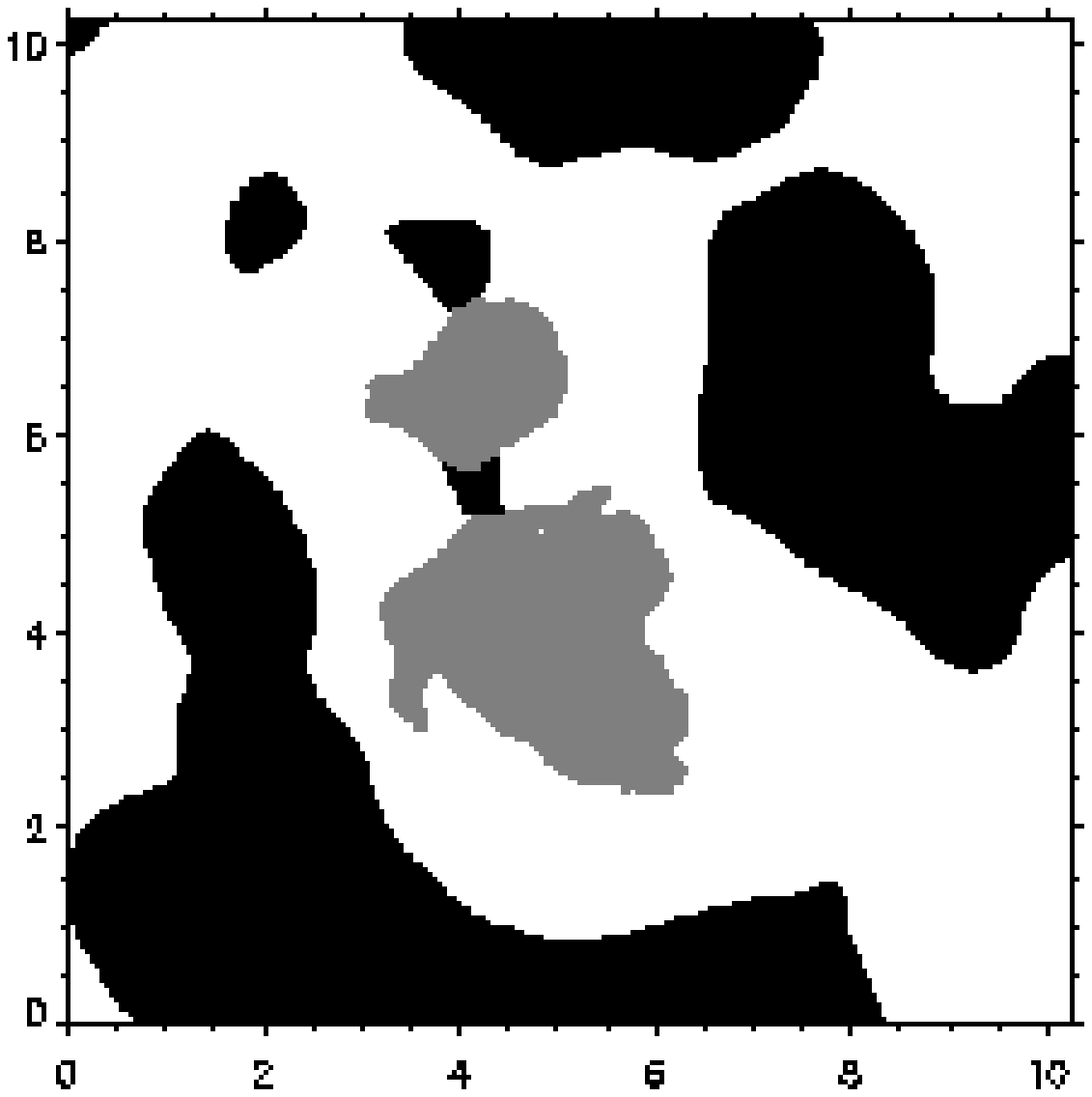} 
\end{tabular}
\caption[\sf Plot of the inward and outward motions around \emph{LANZAROTE} pore.]{\sf Discrimination between  inward and outward motions surrounding \emph{LANZAROTE} pore. Black and white areas in the last panel correspond to velocities displaying outward and inward radial components, respectively. The length of the black bar at coordinates (0,0) in the upper right panel corresponds to 1.6 km s$^{-1}$. The spatial units are in arc sec.}
\label{radialvel_pore6}
\end{figure}

\begin{figure}
\centering
{\footnotesize \sf SST 30.09.2007 - PROPER MOTIONS ANALYSIS AROUND \emph{FUERTEVENTURA} PORE}\\
\vspace{1cm}
\begin{tabular}{cc}
\hspace{-1cm}\includegraphics[width=.45\linewidth]{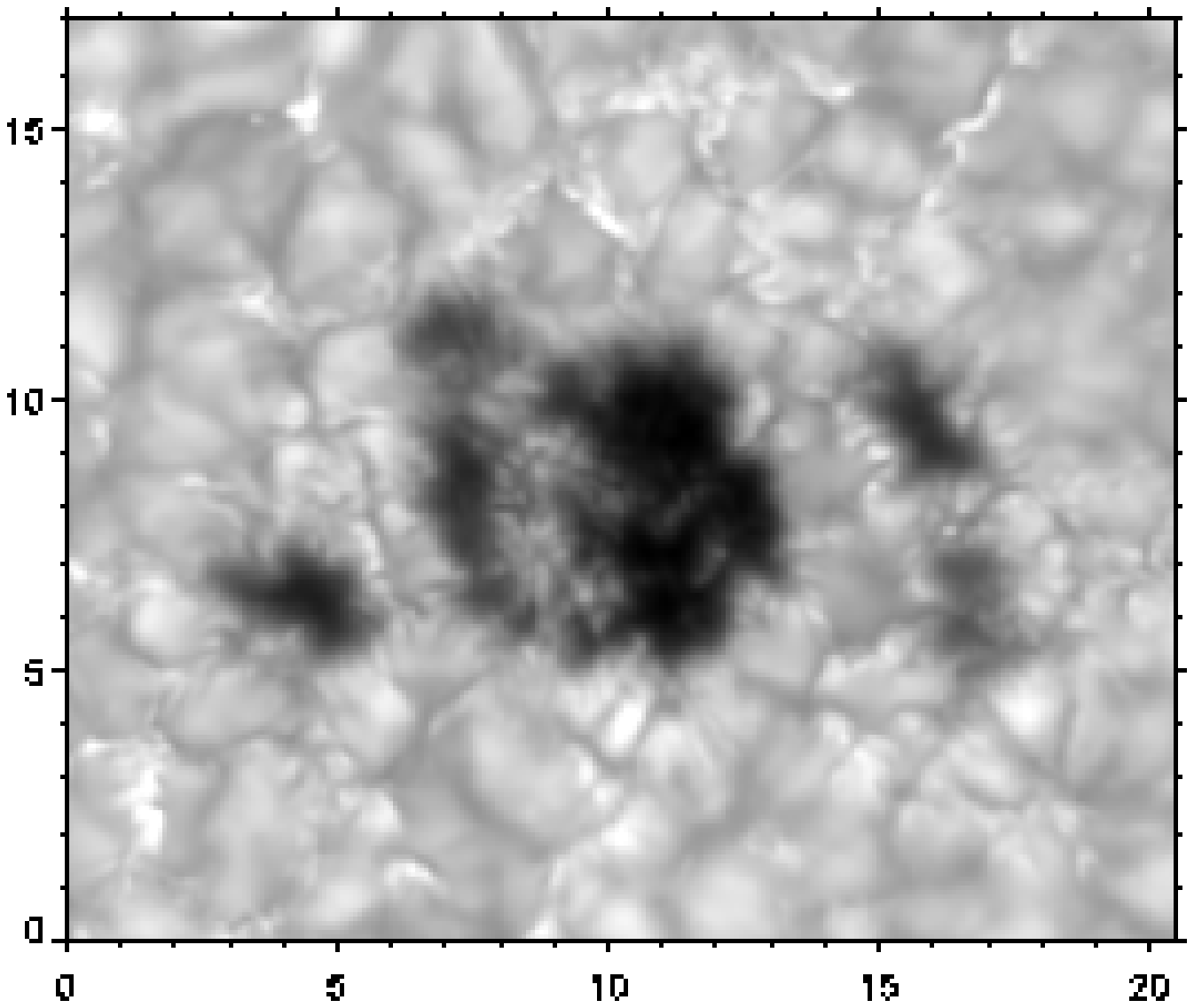} &
\hspace{-0.6cm}\includegraphics[width=.45\linewidth]{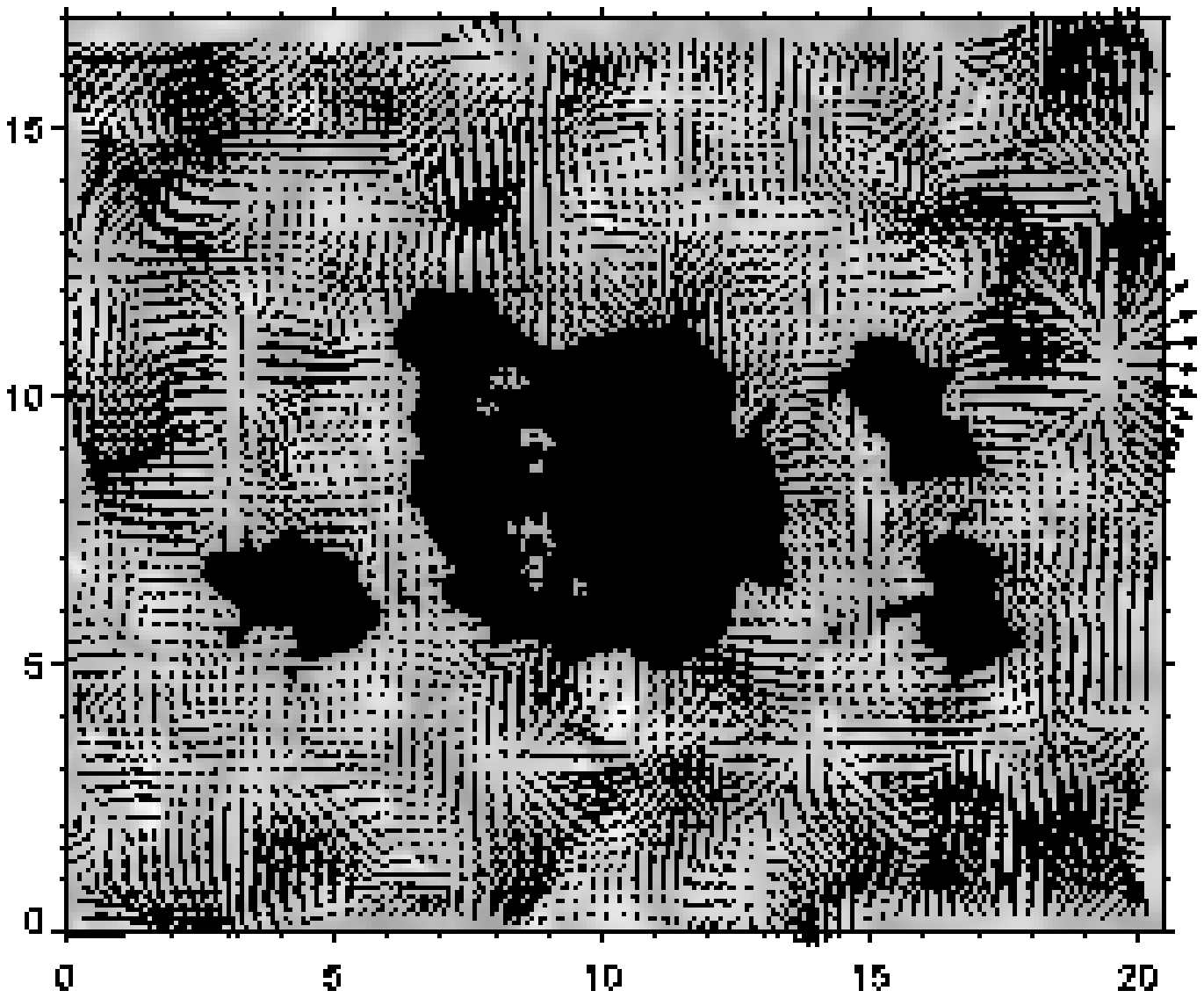} \\
\hspace{-1cm}\includegraphics[width=.45\linewidth]{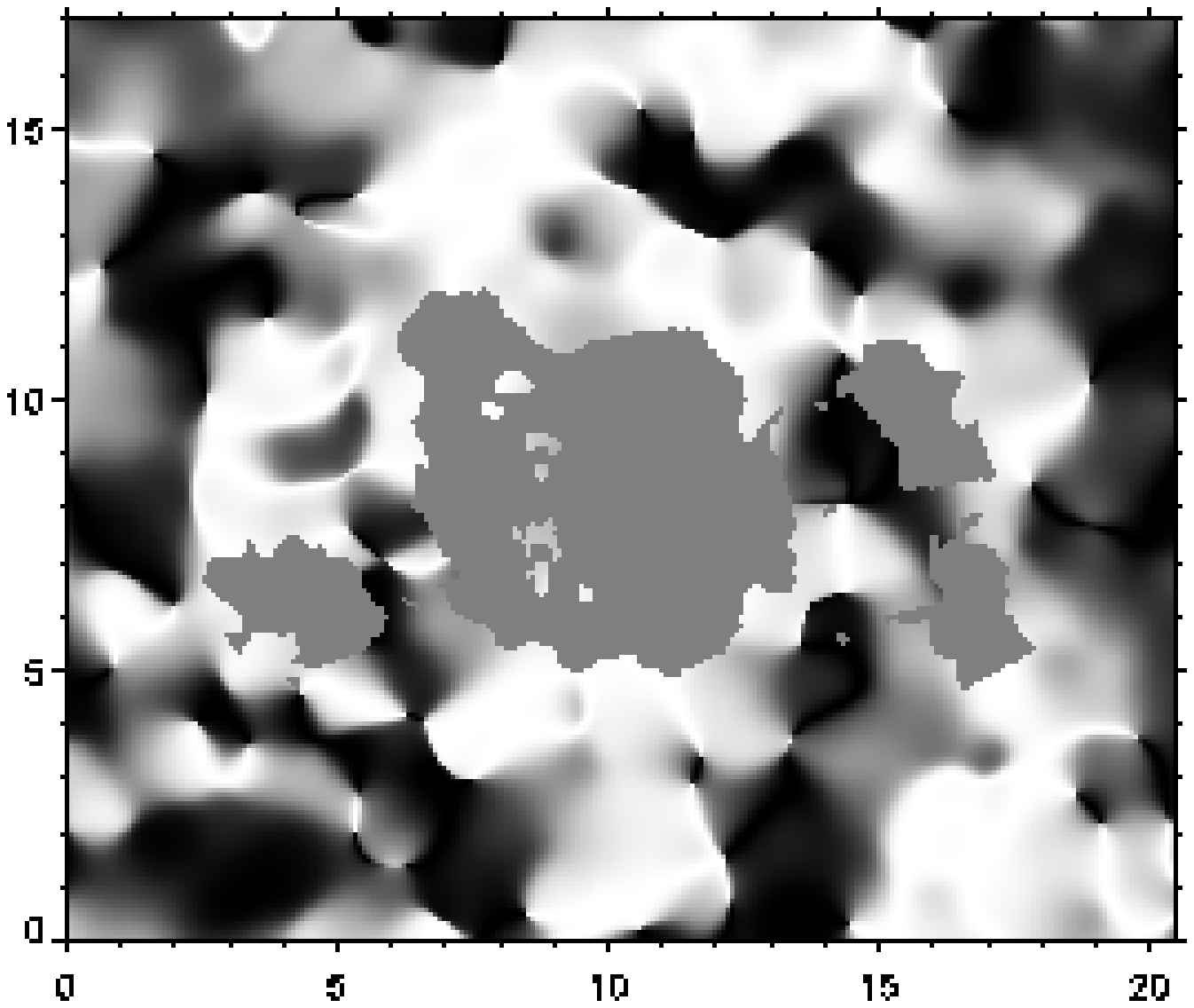} &
\hspace{-0.6cm}\includegraphics[width=.45\linewidth]{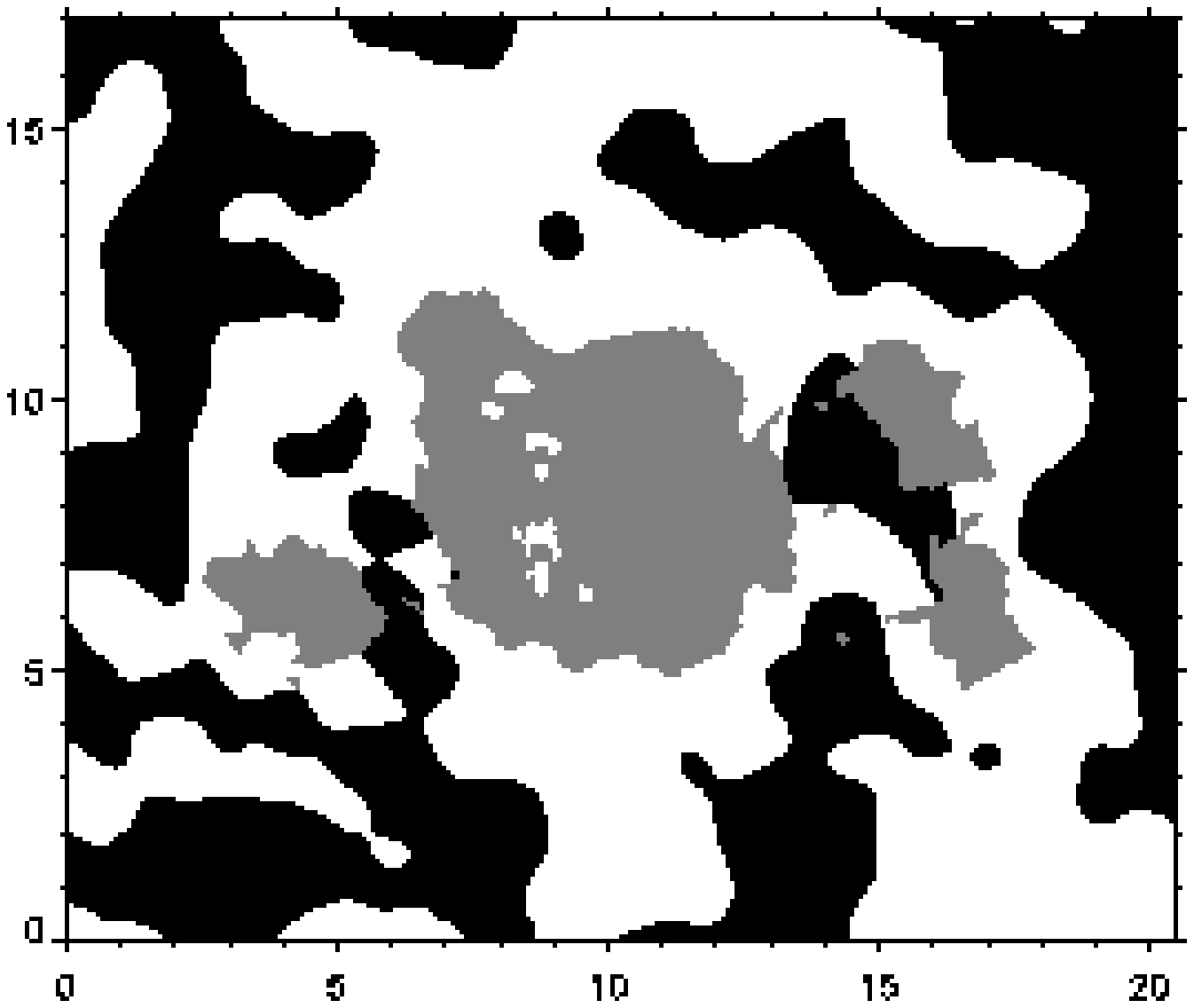} 
\end{tabular}
\caption[\sf Plot of the inward and outward motions around \emph{FUERTEVENTURA} pore.]{\sf Discrimination between inward and outward motions surrounding \emph{FUERTEVENTURA} pore. Black and white areas in the last panel correspond to velocities displaying outward and inward radial components, respectively. The length of the black bar at coordinates (0,0) in the upper right panel corresponds to 1.6 km s$^{-1}$. The spatial units are in arc sec.}
\label{radialvel_pore7}
\end{figure}

Once the method to describe the velocity directions has been set, we apply it to our sample of 7 different pores shown in Figure~\ref{canarypores}. The results are displayed in Figures~\ref{radialvel_pore1} to \ref{radialvel_pore7} where we have: the averaged FOV around the pore, the velocity field, the gray-scale representation of $\cos\beta$, and the final mask of velocities with inward (\emph{white}) and outward (\emph{black}) radial components. The analysis of all cases establish that the flows display a clear preference for inward directions around the pores. This fact is systematically found in all examples. The more regular-shaped pores (Figures~\ref{radialvel_pore1} and \ref{radialvel_pore2}) are surrounded by an also regular annular-like area with inward velocity components, having a mean width similar to the center-to-border distance in the pore. The dividing line between inward and outward motions is connecting the centers of divergence.\\

\begin{figure}
\centering
{\footnotesize \sf SST 30.09.2007 - VELOCITY MAGNITUDES} \\
{\footnotesize \sf OF PROPER MOTIONS AROUND \emph{GRAN CANARIA} PORE}\\
\vspace{0.6cm}
\begin{tabular}{l}
\hspace{1cm}\includegraphics[width=.45\linewidth]{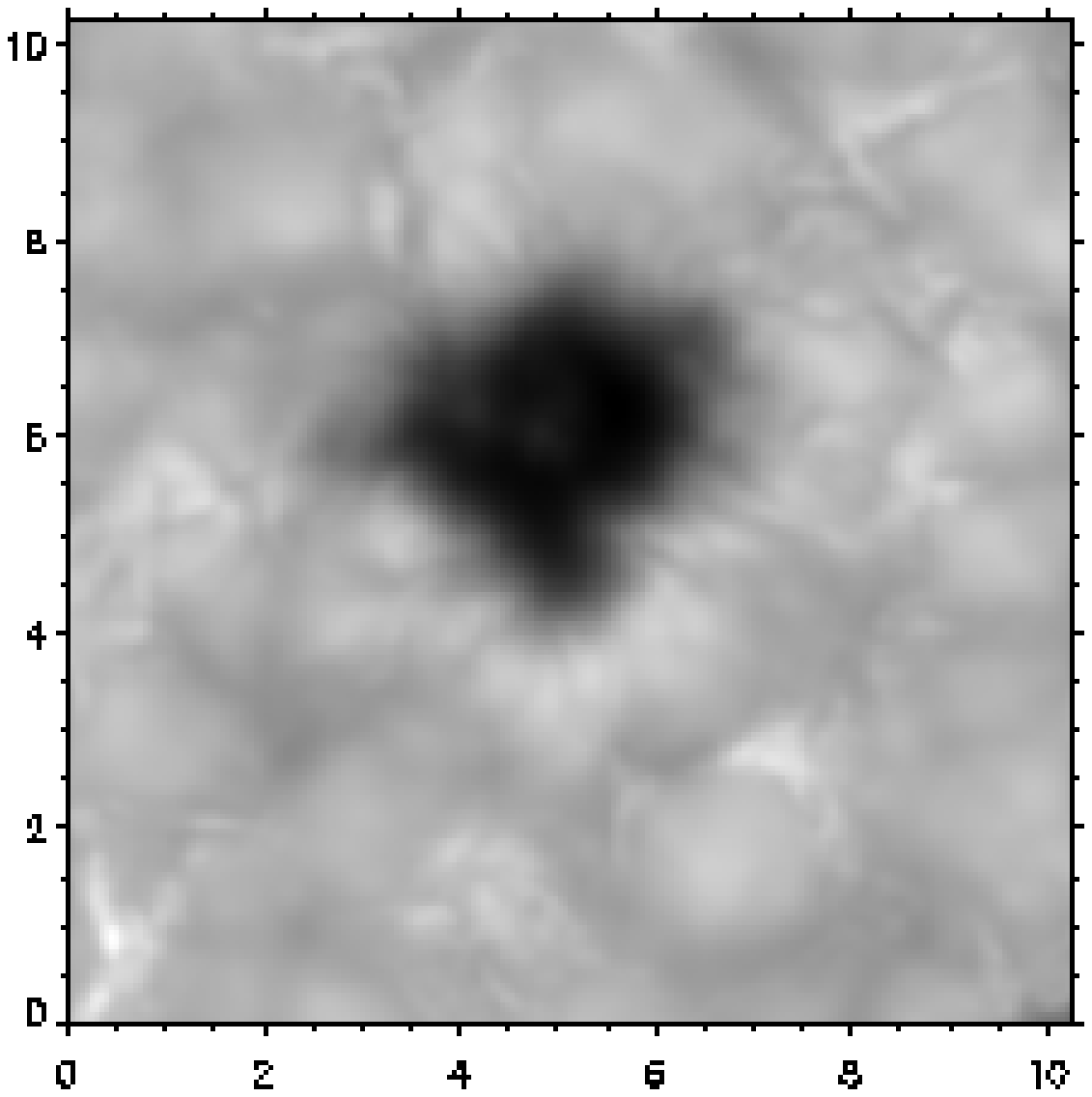} \\
\hspace{1cm}\includegraphics[width=.45\linewidth]{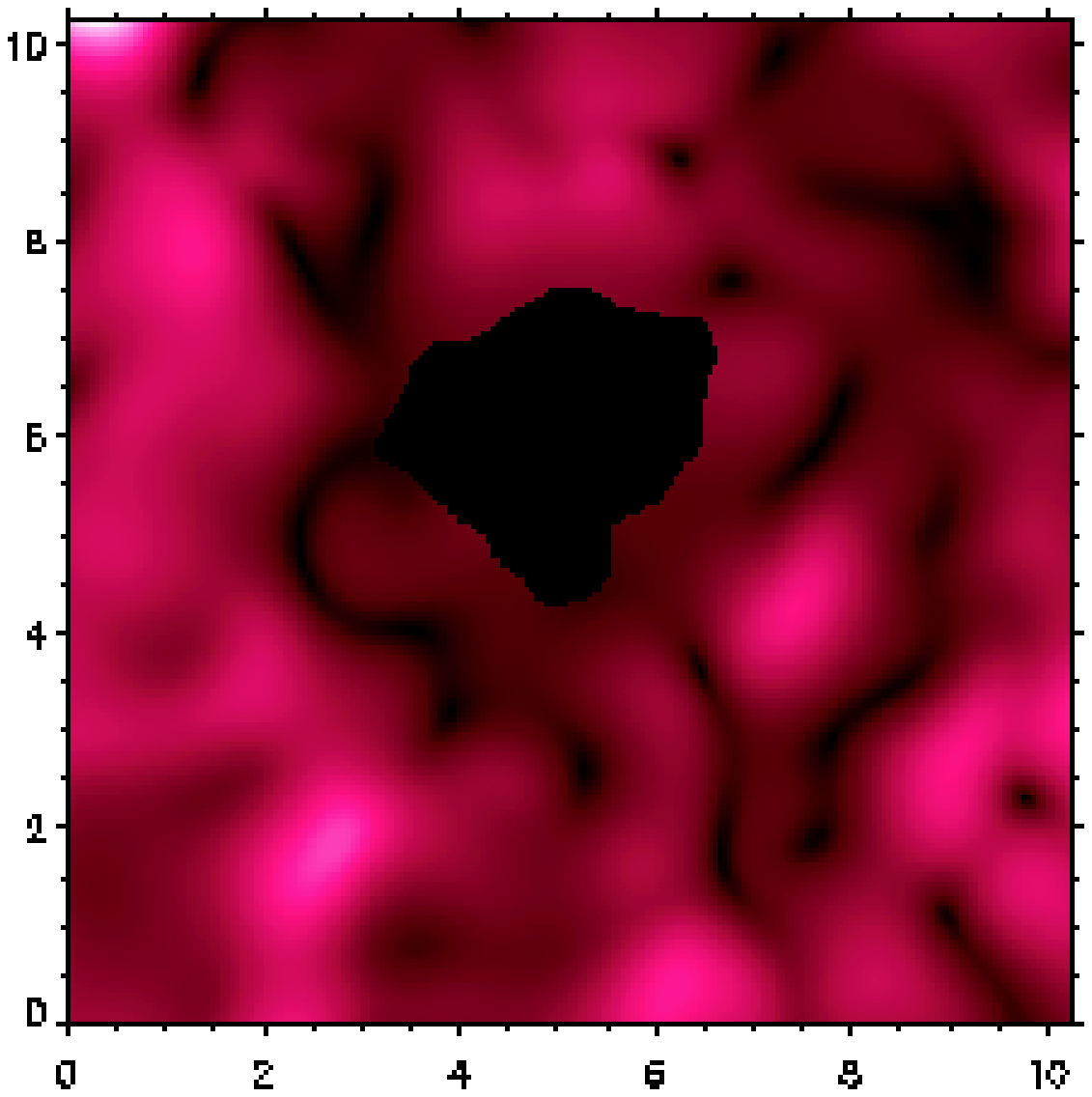}\includegraphics[width=.19\linewidth]{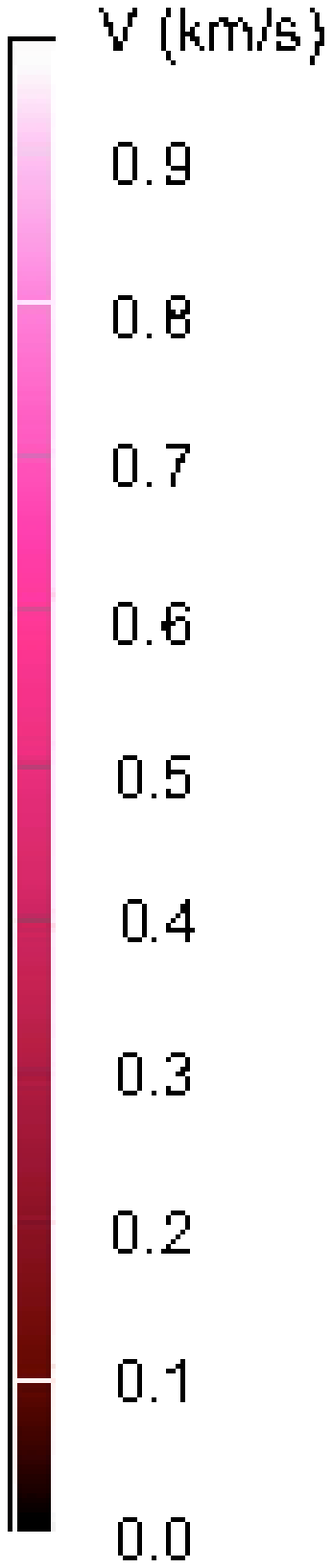} \\
\hspace{1cm}\includegraphics[width=.55\linewidth]{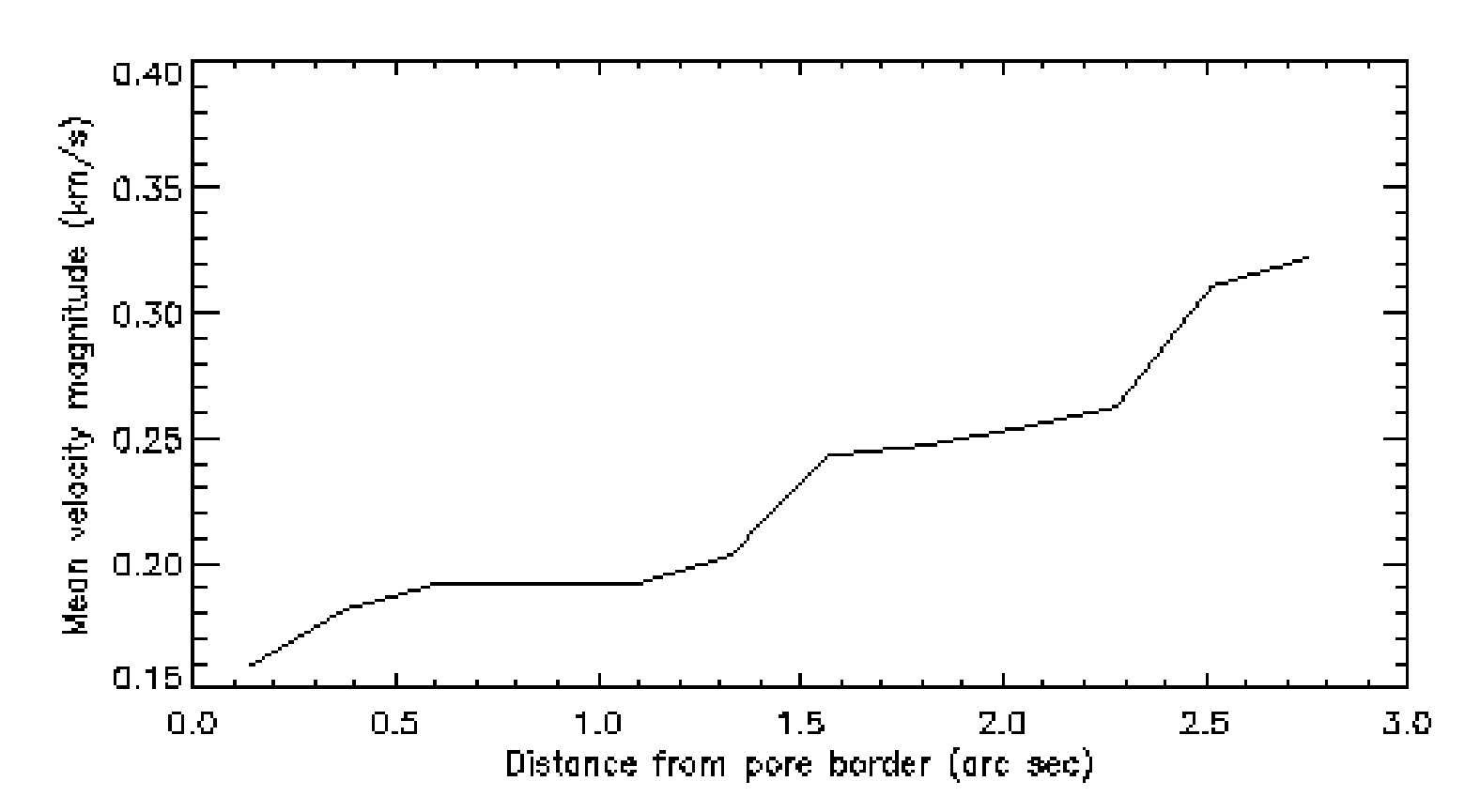} 
\end{tabular}
\caption[\sf Analysis of the velocity magnitudes of proper motions around \emph{GRAN CANARIA} pore]{\sf Analysis of the velocity magnitudes of proper motions around \emph{GRAN CANARIA} pore. \emph{Upper panel}: Averaged FOV in G-band. \emph{Central panel}: Magnitude of horizontal velocities in false-color representation. \emph{Lower panel}: Plot of the mean velocity magnitude versus the distance from the pore border. The spatial units are in arc sec.}
\label{magvel_pore1}
\end{figure}

\begin{figure}
\centering
{\footnotesize \sf SST 30.09.2007 - VELOCITY MAGNITUDES}\\
{\footnotesize \sf  OF PROPER MOTIONS AROUND \emph{TENERIFE} PORE}\\
\vspace{0.6cm}
\begin{tabular}{l}
\vspace{-3mm}
\hspace{0.85cm}\includegraphics[width=.45\linewidth]{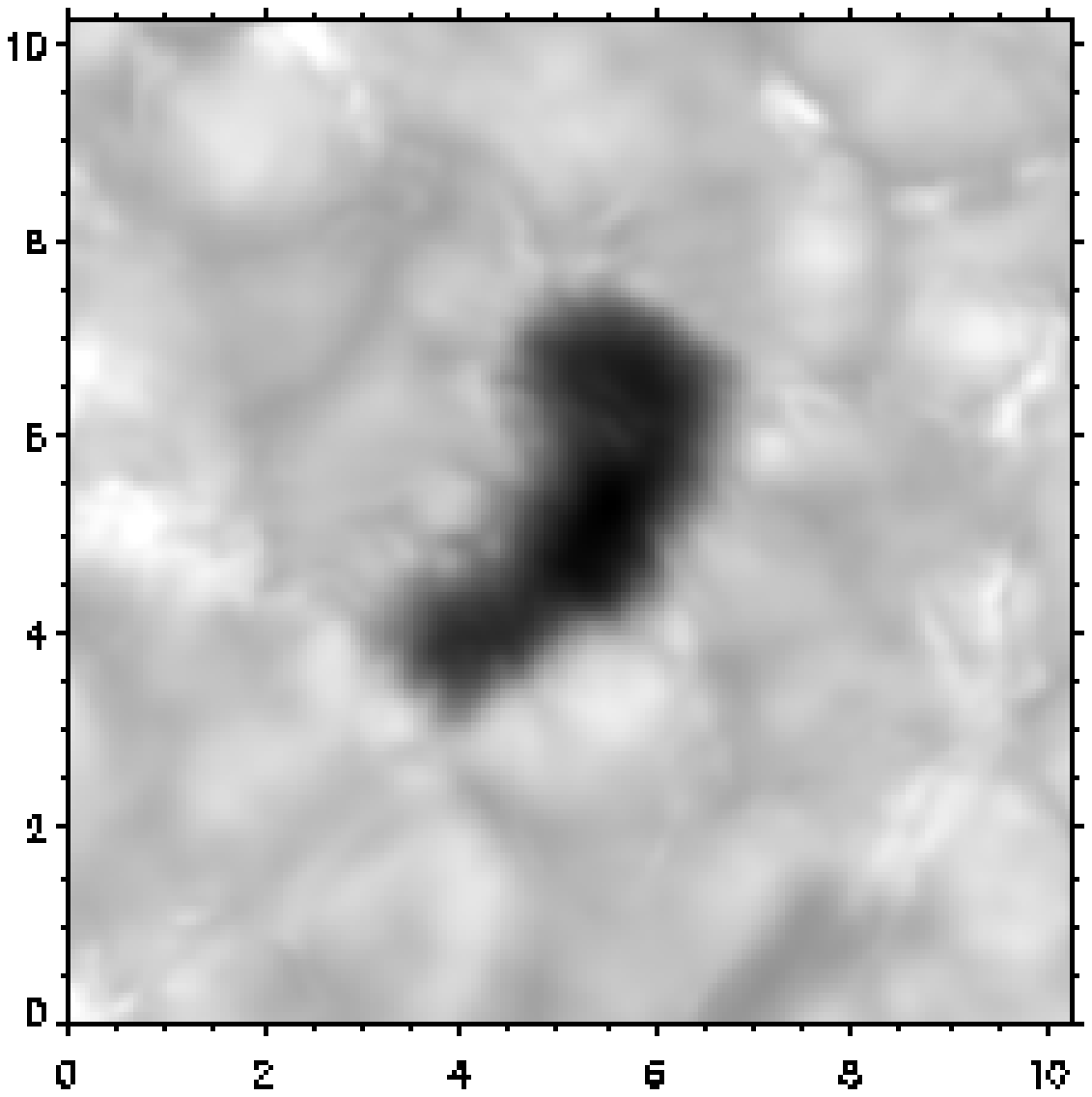} \\
\vspace{-3mm}\hspace{1cm}\includegraphics[width=.45\linewidth]{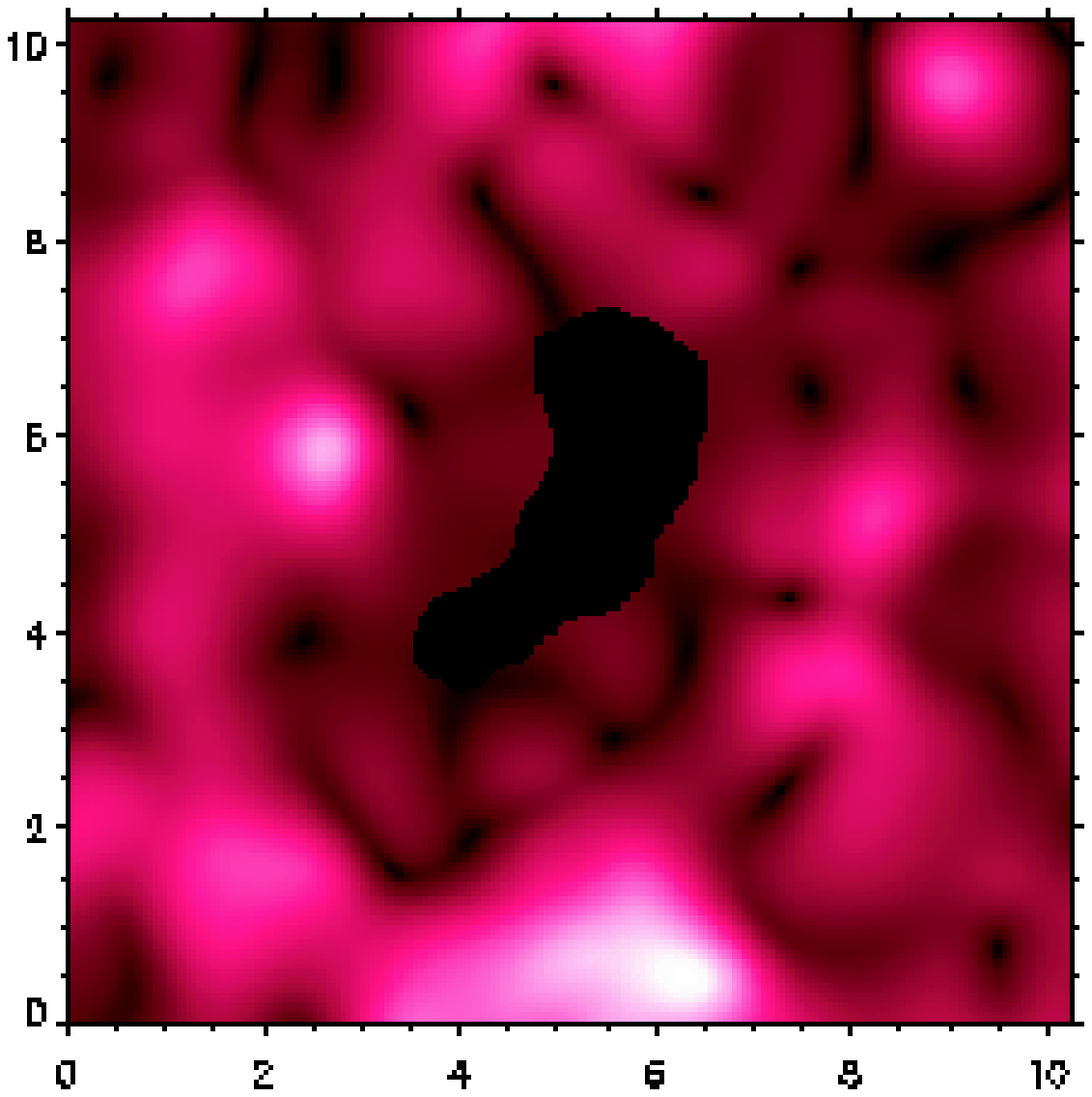}\includegraphics[width=.19\linewidth]{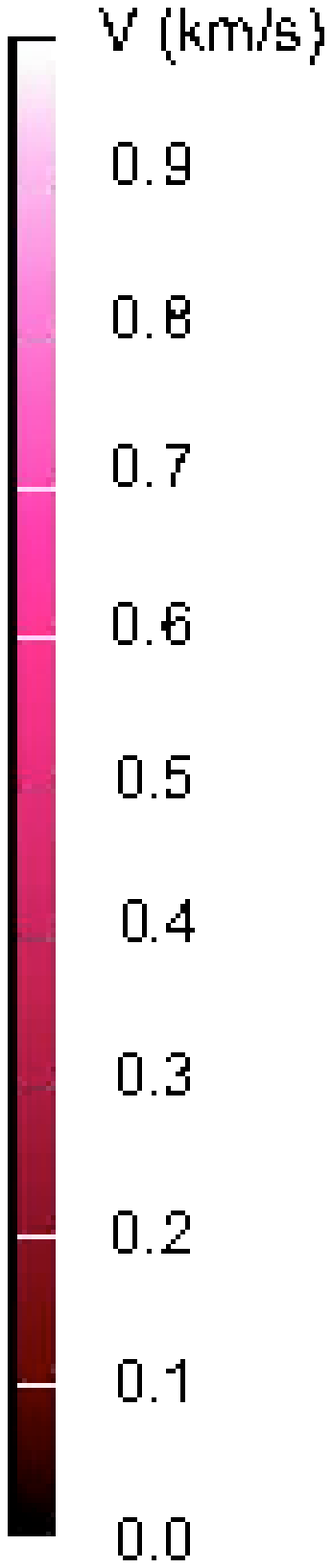} \\ \\
\hspace{1cm}\includegraphics[width=.55\linewidth]{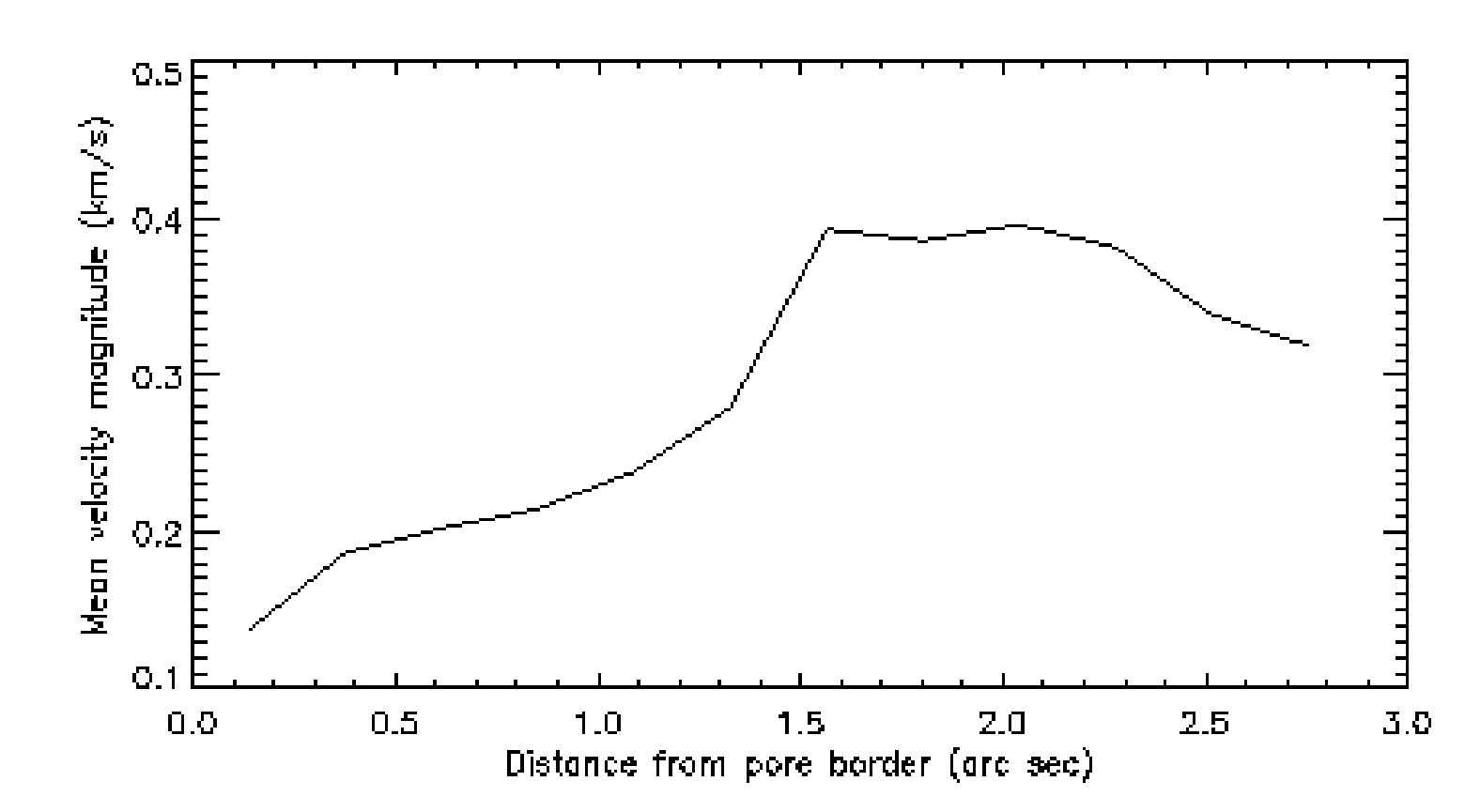} 
\end{tabular}
\caption[\sf Analysis of the velocity magnitudes of proper motions around \emph{TENERIFE} pore]{\sf Analysis of the velocity magnitudes of proper motions around \emph{TENERIFE} pore. \emph{Upper panel}: Averaged FOV in G-band. \emph{Central panel}: Magnitude of horizontal velocities in false-color representation. \emph{Lower panel}: Plot of the mean velocity magnitude versus the distance from the pore border. The spatial units are in arc sec.}
\label{magvel_pore2}
\end{figure}

Figures~\ref{magvel_pore1} and \ref{magvel_pore2} show for the two regular-shaped pores, an average image (48 min) of the time series (\emph{upper panels}) and a color-scaled image representing the horizontal velocity magnitudes (\emph{middle panels}). Both pores exhibit a sort of proto-penumbral structure though not very emphasized. Signals of proto-penumbra appearance are also visible in other pores of our sample (e.g.\ see Figure~\ref{radialvel_pore3}). \\

The FOV in Figures~\ref{magvel_pore1} and \ref{magvel_pore2} is dominated by flows with velocity magnitudes up to 1 km s$^{-1}$. The centers of divergence are also visible and clearly identified in these images as \emph{black} structures around the pores. Another important distinctive feature that comes out from these images is that the highest magnitudes of horizontal velocities are placed beyond these \emph{black} structures away from the pore.\\
 
In order to study how speeds of proper motions around the pore are changing, we calculate the mean velocity magnitude in a sort of ribbons (stripes) around the pore. A total of 12 adjacent ribbons ($\sim$ 0$\farcs$24 wide), referred to as ring-like structures surrounding the pore, are used for this analysis. Figures~\ref{magvel_pore1} and \ref{magvel_pore2} (\emph{lower panels}) plot the relation between speeds of proper motions versus distance to the pore border. The velocity magnitudes increment as a function of this distance: 1) the mean value of speeds increments rapidly at distances ranging from $\sim$ 0$\farcs$15 to 0$\farcs$4; 2) in the range 0$\farcs$4~-~1$\farcs$0 the variation curve is flatter; 3) in the range 1$\farcs$0~-~1$\farcs$6 we again obtain very sharp increments. Up to this distance from the pore border, we observe the same behavior in both pores. Nevertheless, the mean speed values at 1$\farcs$6 are quite different (250 vs.\ 390 m s$^{-1}$). At further distances ($>$ 1$\farcs$6) the trend in the mean velocity magnitudes differs substantially. We must have in mind that the flows around the pore can also be affected by the intrinsic characteristics of every single pore and by the contribution from other sources in the neighbourhood, e.g.\ pores in the vicinity.

\section{Conclusions and discussion}
\label{S:conclusions}

We obtained time series for solar regions displaying pores from both ground-based and space observations. We then analyzed the horizontal displacements of the plasma around solar pores by means of LCT. This is the first time we have tested with HINODE data, our standard reduction routines for LCT and p-modes filtering. The long duration, stability and high-resolution of the data achieved by HINODE enable us to get extremely useful time series to study dynamical properties of the horizontal flows in the photosphere in periods much longer than those we are used to have in ground-based observations which are restricted by the changing seeing conditions.\\

The LCT technique applied to the series allowed us to track the proper motions of structures in the regions nearby solar pores. We conclude that the flows calculated from different solar observations are coherent among each other and show the determinant and overall influence of exploding events in the granulation around the pores and in the whole FOV for every single set. Motions towards the pores in their nearest vicinity are the dominant characteristic we claim to observe systematically.

The motions at the periphery of the pores are basically influenced by the external plasma flows deposited by the exploding events, as suggested by other authors in previous works  \citep{sobotka1999, roudier2002}.\\

We interpret the dividing line between radial inward and outward motions, found by \cite{deng2007} outside the residual pore (see panel \emph{f} in Figure~\ref{deng}), as corresponding to the location of the centres of divergence of the exploding events around the pore. The outward motions these authors describe, which are not in the immediate surroundings of the pore but separated by the annular inward motion, would then correspond not to moat flows but to the outward flows originated in the regular mesh of divergence centers around the pore.

%% file: chap7.tex
\chapter{\sf Conclusions}\label{cap7}
\dropping[0pt]{2}{T}he first part of this work deals with the problem of image degradation caused by atmospheric and instrumental aberrations. By using image restoration techniques we can correct the images and bring them near the diffraction limit of the telescope. One of the aims at working with these novel techniques is the correction of the instrumental aberrations that will be present in the SUNRISE mission, in particular in the images produced by the IMaX instrument, a magnetograph designed and built by a Spanish consortium leaded by the Instituto de Astrof\'isica de Canarias.\\

Nowadays, the use of numerical restoration techniques is determinant and compulsory to obtain high-resolution images, and stands as a key tool for recent and future solar physics advances and developments. In that sense, the present work represents an effort to have high-quality observations in IMaX by defining the method to calibrate the in-flight instrumental aberrations. The method is based on the Phase Diversity (PD) inversion algorithm.\\

The robustness of the method has been tested by numerical experiments simulating different aberration components applied to synthetic solar images. Degradation followed by PD-inversion has allowed to compare the goodness of the match between the input values for degradation and those resulting from the inversion. The sources of aberration have been modelled and added in subsequent experiments. The repercussion of every new added ingredient in the final result from the inversion has been evaluated. The PD-code does not accurately reproduce the shape of the wavefront errors but provides suitable OTFs for subsequent satisfactory restorations. The results of the inversions can be considered reliable enough as to validate the method proposed for the calibration of errors in the images of IMaX, at least for aberrations lying in the range: \emph{rms-ripple} $\le$ 2/60 and \emph{rms-LOA} $\le$ 1/4 waves.\\

The second part of the work is devoted to the observation and analysis of solar active regions. In particular we have studied the dynamical behaviour of the plasma around sunspots and solar pores at a photospheric level. We analyzed data from ground-based (SST) and space (HINODE) telescopes. Ground-based  data are firstly corrected for atmospheric and instrumental aberrations with the methods described in the first part of the work, generating in this way time series of images displaying the solar evolution in active regions.\\
 
Local correlation tracking techniques have been employed to infer proper motion of structures within the FOV. We are mainly interested in the photospheric horizontal flows around sunspots and pores. A sample including various sunspots with different penumbral configurations, and solar pores was analyzed.\\

The main findings concerning the existence of moats surrounding sunspots depending on the penumbral configurations exhibit by them are:

\begin{itemize}
\item Moat flows are oriented following the direction of the penumbral filaments.
\item In granulation regions found adjacent to penumbral sides parallel to the penumbral filaments, moats are absent, or in other words, moats do not develop in the direction transverse to the penumbral filaments. 
\item Umbral core sides with no penumbra do not display moat flows.
\item Around solar pores there are not signs of moat flows. In all the cases studied we have found evidence of exploding granules activity in the pore's surroundings.   
\end{itemize}

Moreover we include in our study two examples in which a neutral line extends along a penumbral border where we would expect a moat flow continuation. For these cases we do not find any moat flow following the direction of the penumbral filaments after crossing the penumbral border where we see a change in magnetic polarity. The presence of neutral lines appears to be significantly affecting the behaviour of the plasma flows, preventing the appearance of moats.\\\

\section{Future prospects}
More upcoming data from the HINODE telescope as well as from new facilities and missions like SUNRISE will provide long-time series of active regions with a constant image quality and enough spatial-resolution to establish firm confirmation of the evidences and findings presented in this thesis. The addition of simultaneous Doppler and magnetogram data to the continuum intensity (or G-band) data sets
will enhance our understanding of the link between the Evershed flow inside penumbrae and the moat flows around sunspots. Needless to say, the study of those stages were penumbrae are just formed or destroyed becomes of extreme importance to validate our findings. Similarly, the results that can be expected in the coming years from local helioseismology, in particular those
related to the f-mode \citep{gizon2000}, describing the flow patterns
in the deeper layers near sunspots will be crucial for the establishment of a clear link between these two well-known flow patterns, Evershed and moat, that has not been appreciated yet.\\
   
The interpretation of these results in terms of convective cells surrounding sunspots \citep[e.g.,][]{bovelet2003} may be different if the Evershed flow turns out to be the major process that injects mass into the moats surrounding sunspots.\\

More observations of complex active regions with neutral lines present in the vicinity of penumbrae are needed to firmly establish the relation between the absence of moat flows and magnetic neutral lines.\\

We believe that one of the more relevant results put forward in this work suggesting that the moat flow is the continuation outside the spots of the Evershed flow, deserves adequate attention in the future from both an observational and a theoretical point of view.\\

%% file: conclusiones.tex
\chapter*{\sf Conclusiones}
\dropping[0pt]{2}{E}n la primera parte de este trabajo de tesis se aborda el problema de degradaci\'on de imagen causada por aberraciones atmosf\' ericas e instrumentales. Gracias a las nuevas t\' ecnicas de reconstrucci\'on de im\'agenes podemos corregir estos efectos perturbadores y alcanzar un alto grado de resoluci\' on cercano al l\'imite de resoluci\'on del telescopio.\\

Uno de los principales objetivos que se presiguen en esta tesis al trabajar con estas innovadoras t\' ecnicas es el de aplicarlas en la correcci\' on de las aberraciones instrumentales en IMaX. Este instrumento es un magnet\'ografo solar que forma parte de la misi\' on SUNRISE (un telescopio solar de un metro de apertura que volar\'a en un globo aerost\' atico y recoger\' a im\' agenes solares continuamente durante 15 dias). IMaX est\'a desarrollado enteramente por un consorcio de instituciones espa\~nolas, siendo el Instituto de Astrof\'isica de Canarias la instituci\'on que lidera el proyecto. \\

Actualmente las t\'ecnicas de restauraci\'on de im\'agenes se han convertido en herramientas indispensables para el avance en el estudio de la F\'isica Solar. En este sentido, el presente trabajo representa un esfuerzo para tener observaciones de alta calidad en IMaX definiendo el m\'etodo para calibrar las aberraciones instrumentales en vuelo. Este m\'etodo se basa en el algoritmo de inversi\'on de Diversidad de Fase (DF).\\

Se ha validado la robustez del m\'etodo de calibraci\'on de aberraciones mediante experimentos num\'ericos, simulando diferentes componentes de aberraci\'on sobre imagenes solares sint\'eticas. La degradaci\'on de las im\'agenes seguida de su inversi\'on con DF, ha sido el proceso utilizado para comparar la bondad del ajuste entre los valores de entrada al degradar y los obtenidos despues de la inversi\'on. Las diversas fuentes de aberraci\'on se han modelado y a\~nadido en sucesivos experimentos. Hemos evaluado la repercusi\'on que tiene cada ingrediente a\~nadido, sobre el resultado final de la inversi\'on. El c\'odigo de DF no reproduce en detalle la forma de los errores del frente de onda sino que proporciona OTFs adecuadas para producir restauraciones satisfactorias. Los resultados de las inversiones pueden considerarse lo suficientemente fiables como para validar el m\'etodo propuesto para la calibraci\'on de errores en las im\'agenes de IMaX, al menos para aberraciones dentro del rango: \emph{rms-ripple} $\le$ 2/60 and \emph{rms-LOA} $\le$ 1/4 waves.\\

La segunda parte de la tesis esta dirigida al estudio y an\' alisis de regiones solares activas. En particular se ha estudiado el comportamiento y din\' amica del plasma a nivel fotosf\'erico alrededor de manchas y poros solares. Para ello se han utilizado observaciones hechas en Tierra (SST) y del telescopio espacial (HINODE) recientemente lanzado al espacio. A trav\'es de series temporales se han estudiado los movimientos propios de estructuras en la zona granular que circunda manchas y poros, considerando una amplia muestra de ambos.\\

Los principales hallazgos que surgen de este trabajo de tesis son:

\begin{itemize}
\item Los flujos de gran escala aldededor de las manchas solares (llamados flujos de foso) se orientan siguiendo la direcci\'on de los filamentos penumbrales.
\item En las regiones de granulaci\' on adjacentes a penumbras en donde los filamentos son paralelos al borde penumbral, no se encuentran signos de flujos de \emph{foso}. 
\item Las regiones de granulaci\'on adjacentes a nucleos umbrales (es decir sin penumbra), no muestran indicios de presencia de flujos de \emph{foso}. Es decir, los flujos de foso no se producen en direcci\'on transversal a los filamentos penumbrales.
\item Alrededor de los poros analizados no se encuentran signos de flujos de \emph{foso}. En todos los casos se han encontrado evidencias de flujos provenientes de gr\'anulos explosivos alrededor de estos poros.
\end{itemize}

Sumado a esto, se han analizado tambi\'en un par de ejemplos en donde hay presencia de l\'ineas neutras que se extienden longitudinalmente a lo largo de bordes penumbrales en donde se esperar\'ia observar flujos de foso. Para estos casos sin embargo no se ha detectado este tipo de flujos. La presencia de l\'ineas neutras parece afectar de forma notable el comportamiento de los flujos de plasma impidiendo la aparici\'on de flujos de foso.\\\

\section*{\sf \bf Futuros estudios}

Datos provenientes del telescopio HINODE y de nuevos instrumentos y misiones como SUNRISE, permitir\'an tener muchos m\'as ejemplos de series temporales estables, de alta resoluci\'on espacial y temporal, sobre las cuales se podr\'a estudiar la din\' amica de las estructuras en las zonas activas para verificar los hallazgos presentados en este trabajo. El an\'alisis de series de magnetogramas y dopplergramas simult\'aneas a series en banda G, enriquecer\'an indudablemente nuestro entendimiento de la relaci\'on entre el flujo Evershed dentro de la penumbra y el flujo de foso alrededor de las manchas solares.\\
 
Sin duda el estudio de manchas en diferentes estados evolutivos donde la penumbra ha sido recientemente formada o esta en fase de disoluci\'on, es de vital importancia para validar los resultados y conclusiones que se exponen en el presente trabajo. \\

De igual forma aquellos resultados aportados por nuevos estudios de heliosismologia local, en particular los relacionados con los modos-f \citep{gizon2000} tendientes a describir el patr\'on de flujos a diferentes profundidades bajo las manchas solares y sus regiones circundantes, ser\'an cruciales para establecer  una clara relaci\' on entre estos dos flujos: dentro y fuera de la penumbra. La interpretaci\'on los flujos de foso en t\'erminos de celdas convectivas de gran escala alrededor de las manchas \citep[ver][]{bovelet2003} puede cambiar radicalmente si el flujo Evershed resultara ser el principal proceso que inyecta masa en las regiones \emph{moat} alrededor de las manchas solares. \\

Creemos que uno de los resultados mas relevantes que se defienden en este trabajo y que supone que el flujo de foso es una continuaci\'on, fuera de la penumbra, del flujo Evershed, merece especial atenci\'on en el futuro tanto desde el punto de vista observacional como del te\'orico.

%% file: appenA.tex
\chapter{\sf Zernike polynomials}
\dropping[0pt]{2}{T}he wavefront aberration $\phi(r,\theta)$ in the pupil of an optical system is often parametrized by using a convenient set of basis functions. Because of their simple analytical forms \emph{circle polynomials of Zernike} $Z_j(\rho,\theta)$ \citep{zernike1934} defined in a unit circle  $(0 \le \rho \le1)$ are often used as basis functions to expand $\phi(r,\theta)$, as follows:

\begin{equation}
\phi(r, \theta) = \sum \alpha_j Z_j (\rho,\theta)
\label{wavefront}
\end{equation}

\noindent where:\\

\noindent
$(r,\theta)$ are the polar coordinates centered on the pupil.\\
$\rho=r/R$ is the normalized radial distance ($R$ is the radius of the pupil).\\
$\alpha_j$ are the coefficients of the terms in the expansion.\\

For the sake of simplicity, e.g. the orthogonality properties, we use the Zernike polynomials as modified by \cite{noll1976}. Table~\ref{zernikes} list the first Zernike terms up to 6$^{th}$.\\

\begin{figure}
\begin{center}
\vspace{-1cm}
\includegraphics[angle=-180,width=1.\linewidth]{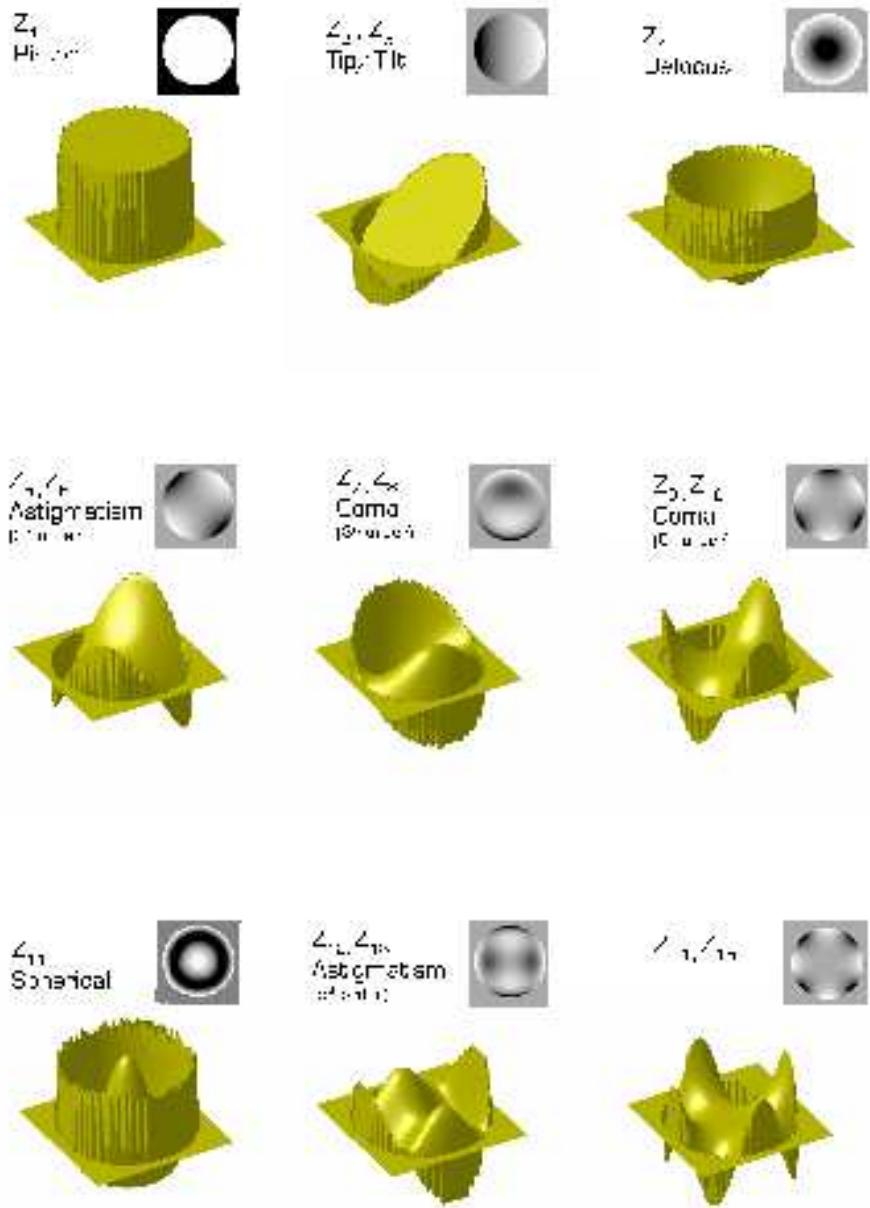}
\caption[\sf Classical Seidel's aberrations]{\sf Classical Seidel's aberrations corresponding to the first few Zernike polynomials in Table~\ref{zernikes}. In the panels labeled with two Z's the second one differs only by a simple azimuthal rotation.}
\label{seidel}
\end{center}
\end{figure}

Figure~\ref{seidel} shows the first few Zernike polynomials which correspond to the classical Seidel's aberrations\footnote{\sf Philipp Ludwig von Seidel (1821-1896) was a German mathematician. In 1857, von Seidel decomposed the first order monochromatic aberrations into five constituent aberrations. They are now commonly referred to as the five Seidel Aberrations.}.

\newpage
\thispagestyle{empty}

\begin{table}
\sffamily
\centering
\caption[\sf Zernike polynomials]{\sf Zernike polynomials \citep[as formulated by][]{noll1976}.}
\begin{tabular}{llcc}\\\hline\\
& {\footnotesize ~~~ZERNIKE MODE} & {\footnotesize RADIAL} & {\footnotesize AZIMUTHAL}\\
& & {\footnotesize DEGREE} & {\footnotesize FREQUENCY}\\
& & n & m\\\hline\\
$Z_1$ & = 1 & 0 & 0\\
$Z_2$ & = $2\rho \cos{\theta}$ & 1 & 1\\
$Z_3$ & = $2\rho \sin{\theta}$ & 1 & 1 \\
$Z_4$ & = $\sqrt{3}(2\rho^2-1)$ & 2 & 0\\
$Z_5$ & = $\sqrt{6} \rho^2 \sin{2\theta}$ & 2 & 2\\
$Z_6$ & = $\sqrt{6} \rho^2 \cos{2\theta}$& 2 & 2\\
$Z_7$ & = $\sqrt{8}(3\rho^3 - 2\rho ) \sin{\theta}$& 3 & 1\\
$Z_8$ & = $\sqrt{8}(3\rho^3 - 2\rho ) \cos{\theta}$& 3 & 1\\
$Z_9$ & = $\sqrt{8}\rho^3 \sin{3\theta}$& 3 & 3\\
$Z_{10}$ & = $\sqrt{8}\rho^3 \cos{3\theta}$& 3 & 3\\
$Z_{11}$ & = $\sqrt{5}(6\rho^4 - 6\rho^2 + 1)$ & 4 & 0\\
$Z_{12}$ & = $\sqrt{10}(4\rho^4 - 3\rho^2 ) \cos{2\theta}$ & 4 & 2\\
$Z_{13}$ & = $\sqrt{10}(4\rho^4 - 3\rho^2 ) \sin{2\theta}$& 4 & 2\\
$Z_{14}$ & = $\sqrt{10}\rho^4\cos{4\theta}$& 4 & 4\\
$Z_{15}$ & = $\sqrt{10}\rho^4\sin{4\theta}$& 4 & 4\\
$Z_{16}$ & = $\sqrt{12}(10\rho^5 - 12\rho^3 + 3\rho) \cos{\theta}$ & 5 & 1\\
$Z_{17}$ & = $\sqrt{12}(10\rho^5 - 12\rho^3 + 3\rho) \sin{\theta}$& 5 & 1\\
$Z_{18}$ & = $\sqrt{12}(5\rho^5 - 4\rho^3) \cos{3\theta}$& 5 & 3\\
$Z_{19}$ & = $\sqrt{12}(5\rho^5 - 4\rho^3) \sin{3\theta}$& 5 & 3\\
$Z_{20}$ & = $\sqrt{12}\rho^5\cos{5\theta}$& 5 & 5 \\
$Z_{21}$ & = $\sqrt{12}\rho^5\sin{5\theta}$& 5 & 5 \\
$Z_{22}$ & = $\sqrt{7}(20\rho^6 - 30\rho^4 + 12\rho^2 -1)$& 6 & 0 \\
$Z_{23}$ & = $\sqrt{14}(15\rho^6 - 20\rho^4 + 6\rho^2) \sin{2\theta}$& 6 & 2 \\
$Z_{24}$ & = $\sqrt{14}(15\rho^6 - 20\rho^4 + 6\rho^2) \cos{2\theta}$ & 6 & 2\\
$Z_{25}$ & = $\sqrt{14}(6\rho^6 - 5\rho^4) \sin{4\theta}$& 6 & 4\\
$Z_{26}$ & = $\sqrt{14}(6\rho^6 - 5\rho^4) \cos{4\theta}$& 6 & 4 \\
$Z_{27}$ & = $\sqrt{14} \rho^6 \sin{6\theta}$& 6 & 6 \\
$Z_{28}$ & = $\sqrt{14} \rho^6 \cos{6\theta}$& 6 & 6\\\\\hline
\end{tabular}
\label{zernikes}
\end{table}

\thispagestyle{empty}

\thispagestyle{empty}

%% file: appenB.tex
\chapter{\sf Glossary of acronyms}
\dropping[0pt]{2}{T}he list of acronyms used in the thesis are:\\\\

\noindent  {\textsc{\sf {\large INSTITUTES, ORGANISATIONS AND GROUPS}} --------------------}\\
\begin{tabular}{ll}\\
\textsc{\sf CSIRO} & Commonwealth Scientific and Industrial Research Organisation\\
\textsc{\sf GACE} & Grupo de Astronom\'ia y Ciencias del Espacio.\\
\textsc{\sf IAA} & Instituto de Astrof\'isica de Andaluc\' ia\\
\textsc{\sf IAC} & Instituto de Astrof\'isica de Canarias\\
\textsc{\sf INTA} & Instituto Nacional de T\' ecnica Aeroespacial\\
\textsc{\sf ISP} & Institute of Solar Physics - Royal Swedish Academy of Sciences\\
\textsc{\sf LMSAL} & Loockhead Martin Solar Astrophysical Laboratory\\
\textsc{\sf NOAA} & National Oceanic and Atmospheric Administration\\
\textsc{\sf SPARG} & Solar Physics and upper-Atmosphere Research Group\\
\textsc{\sf ULL} & University of La Laguna
\end{tabular}
\vspace{7mm}

\noindent {\textsc{\sf {\large TELESCOPES AND INSTRUMENTATION}} -----------------------------}\\\\
\noindent \textsc{\sf {\large OBSERVATORIES}}\\
\begin{tabular}{ll}\\
\textsc{\sf ENO} & European Northern Observatory\\
\textsc{\sf ORM} & Roque de los Muchachos Observatory\\
\textsc{\sf OT} & Teide Observatory
\end{tabular}
\vspace{4mm}

\newpage
\thispagestyle{empty}
\vspace{5mm}
\noindent \textsc{\sf {\large INSTRUMENTATION}} \\
\begin{tabular}{ll}\\
\textsc{\sf AO} & Adaptive Optics \\
\textsc{\sf AOM} & Adaptive Optics Mirror\\
\textsc{\sf BS} & Beamsplitter\\
\textsc{\sf CAM} & Camera optics\\
\textsc{\sf CCD} & Charge-Coupled Device\\ 
\textsc{\sf COLL} & Collimator \\
\textsc{\sf CT} & Correlation Tracker\\
\textsc{\sf CWS} & Correlation Tracker and Wavefront Sensor\\
\textsc{\sf E} & Etalon\\
\textsc{\sf FP} & Fabry-Perot\\
\textsc{\sf I} & Iris\\
\textsc{\sf IF} & Interference Filter\\
\textsc{\sf IMaX} & Imaging Magnetograph Experiment\\
\textsc{\sf ISLiD} & Imaging Stabilisation and Light Distribution\\
\textsc{\sf LCVR} & Liquid Cristal Variable Retarders\\
\textsc{\sf M1,M2,M3} & Folding Mirrors\\
\textsc{\sf MDI} & Michelson Doppler Image \\
\textsc{\sf MO} & Microscope Objective\\
\textsc{\sf RB} & Reflecting Ball\\
\textsc{\sf RE} & Reference Plate\\
\textsc{\sf SOT} & Solar Optical Telescope\\
\textsc{\sf SF} & Spatial Filter\\
\textsc{\sf SOUP} & Solar Optical Universal Polarimeter \\
\textsc{\sf SUFI} & SUNRISE Filter Imager\\
\textsc{\sf SUPOS} & Polarimetric Spectrograph\\
\textsc{\sf WFE} & Wavefront Error\\
\textsc{\sf WFS} & Wavefront Sensor\\
\textsc{\sf XRT} & X-rays Telescope
\end{tabular}

\newpage
\thispagestyle{empty}

\noindent \textsc{\sf {\large GROUND-BASED TELESCOPES}}\\
\begin{tabular}{ll}\\
\textsc{\sf DOT} & Dutch Open Telescope\\
\textsc{\sf SST} & Solar Swedish Tower\\
\textsc{\sf THEMIS} & Heliographic Telescope for the Study of \\
\textsc{ }& the Magnetism and Instabilities on the Sun\\
\textsc{\sf VTT} & Vacuum Tower Telescope
\end{tabular}

\vspace{4mm}
\noindent \textsc{\sf {\large SPACE TELESCOPES}}\\
\begin{tabular}{ll}\\
\textsc{\sf HINODE} & Solar-B Telescope\\
\textsc{\sf SOHO} & Solar Heliospheric Observatory\\
\textsc{\sf TRACE} & Transition Region and Coronal Explorer
\end{tabular}
\vspace{7mm}

\noindent {\textsc{\sf {\large GENERAL DEFINITIONS}}  ------------------------------------------------}\\
\begin{tabular}{ll}\\
\textsc{\sf FOV} & Field-of-view\\
\textsc{\sf IDL} & Interactive Data Language\\
\textsc{\sf ITP} & International Time Program\\
\textsc{\sf LCT} & Local Correlation Tracking\\
\textsc{\sf LDB} & Long Duration Balloon\\
\textsc{\sf MTF} & Modulation Transfer Function\\
\textsc{\sf PD} & Phase Diversity\\
\textsc{\sf PSF} & Point Spread Function\\
\textsc{\sf PV} & Peak-to-Valley\\
\textsc{\sf RMS} & Root Mean Square\\
\textsc{\sf RSS} & Root Sum Square\\
\textsc{\sf SNR} & Signal-to-Noise Ratio
\end{tabular}
\vspace{7mm}

\noindent {\textsc{\sf {\large SOLAR DEFINITIONS}} -----------------------------------------------------}\\
\begin{tabular}{ll}\\
\textsc{\sf BP} & Bright Points\\
\textsc{\sf CME} & Coronal Mass Ejections\\
\textsc{\sf G-cont} & G continuum\\
\textsc{\sf LB} & Light Bridge\\
\textsc{\sf MMF} & Moving Magnetic Features\\
\textsc{\sf PG} & Penumbral Grains\\
\textsc{\sf UD} & Umbral Dots
\end{tabular}

%% file: appenC.tex
\chapter{\sf Scientific contributions}
\dropping[0pt]{2}{T}he list of scientific contributions on the topic presented in this work is given below:\\\\

{\bf  {\large Conference Contributions:}}\\

[1] {\bf S. Vargas Dom\'{\i}nguez}, L.Rouppe van der Voort, J. A. Bonet, V. Mart\'{\i}nez Pillet, Y. Katsukawa \emph{Moat flow as the continuation of the Evershed Flow}. Initial Result from Hinode, 2007. \\

[2] {\bf S. Vargas Dom\'{\i}nguez}, J.A. Bonet Navarro, V. Mart\'{\i}nez Pillet. \emph{Evidence of an association between the presence of penumbra and strong radial outflows in sunspots}. Proc. Second Solar Orbiter Workshop, 2006.\\

[3]  D. Orozco Su\'arez, L.R. Bellot Rubio, {\bf S. Vargas Dom\'{\i}nguez}, J.A. Bonet Navarro, V. Mart\'{\i}nez Pillet. \emph{From simulations and IMaX observations to the VIM design}. Proc. Second Solar Orbiter Workshop, 2006.\\

{\bf {\large Refereed Publications:}}\\

[1] {\bf S. Vargas Dom\'{\i}nguez}, J. A. Bonet, V. Mart\'{\i}nez Pillet. \emph{Characterization of horizontal flows around solar pores from high-resolution time series}. Astrophysical Journal (in preparation).\\

[2] R.J. Rutten, P. S\"utterlin, {\bf S. Vargas Dom\' {\i}nguez}. \emph{DOT tomography of the solar atmosphere. VIII. Nature and visibility of pseudo Ellerman bombs}. Astronomy \& Astrophysics (in preparation).\\

[3] {\bf S. Vargas Dom\' {\i}nguez}, L.Rouppe van der Voort,  J. A. Bonet, V. Mart\' {\i}nez Pillet, M. Van Noort, Y. Katsukawa. \emph{Moat flow in the vicinity of sunspots for various penumbral configurations}. 2008,  Astrophysical Journal, 679, 900\\

[4] R. Ishikawa, S. Tsuneta, Y. Kitakoshi, Y. Katsukawa, J.A. Bonet, { \bf S. Vargas Dom\'{\i}nguez}, L.H.M.Rouppe van der Voort, Y. Sakamoto and T.Ebisuzaki.  \emph{Relationships between magnetic foot points and G-band bright structures}. 2007, Astronomy and Astrophysics, 472, 911\\

[5] {\bf S. Vargas Dom\'{\i}nguez}, J. A. Bonet, V. Mart\'{\i}nez Pillet, Y. Katsukawa, Y. Kitakoshi, L.Rouppe van der Voort.  \emph{On the moat - penumbra relation}.2007, Astrophysical Journal, 660L, 165\\

%% file: agra.tex
\chapter*{\sf Agradecimientos}
\dropping[0pt]{2}{D}espu\'es de todos estos a\~nos en el IAC, de todas las experiencias y los momentos que he compartido con mucha gente, es casi imposible mencionar una a una las personas que han contribuido de una u otra forma para tratar de alcanzar mis metas y superar mis desafios a nivel personal y profesional, en definitiva para tratar de ser feliz. Entre estas personas, mis supervisores son sin duda unas a las cuales no solo debo, sino a las que quiero agradecer en primer lugar. Estoy muy agradecido a Jos\'e Antonio Bonet y Valent\'{\i}n Mart\'inez Pillet por haber compartido conmigo toda su invaluable experiencia, por hacerme sentir y contarme como uno m\'as en el equipo, en pocas palabras agradecerles por ser los directores de tesis que cualquier estudiante desear\'ia tener. Dentro del grupo tambi\'en quiero agradecer especialmente a Alberto Sainz por su colaboraci\'on y a In\'es M\'arquez por su labor de \'arbitro interno de esta tesis. Ram\' on Garc\' ia L\'opez ha sido mi tutor dentro del programa de doctorado y deseo agradecerle sus gestiones y ayuda al comienzo del mismo.

Este trabajo de tesis ha sido parcialmente financiado por el Ministerio de Ciencia e Innovaci\'on de Espa\~na a trav\'es de los proyectos AYA2007-63881, ESP2003-07735-C04 y a la beca de Formaci\'on de Personal Investigador (FPI) que me ha sido otorgada para el proyecto IMaX en el Instituto de Astrof\'isica de Canarias. Financial support by the European Commission through the SOLAIRE Network (MTRN-CT-2006-035484) is gratefuly acknowledged.\\

A todos mis compa\~neros del corral\'on y a todos los que han pasado por aqu\'i dejando buenos recuerdos, especialmente Ingrid, Nieves, Caro, Leonel, Humberto, ....  Pero dentro del IAC tampoco debo olvidar a todas las personas que con sus peque\~nos y grandes detalles hacen que todo funcione como una engrasada maquinaria para facilitarnos nuestra labor diaria: las secretarias de ense\~nanza y de investigaci\'on, el personal del CAU, personal administrativo, de recepci\'on, de viajes, de la OTRI, el servicio de limpieza y seguridad. \\

I am very thankful to the Institute of Solar Physics of the Royal Swedish Academy of Siences in Stockholm, in particular to G\"oran Scharmer for the invitation to join the team for a short stay. Also to Mats L\"ofdahl, Dan Kisselman, Michiel Van Noort and the PhD students Jaime, Vasco and Javier for welcoming me and made my stay in Sweden a very enjoyable one.\\

Al Grupo de Astronom\'ia y Ciencias del Espacio de la Universidad de Valencia, a Vicente y Jos\'e Luis les agradezco su buena energ\'{\i}a, animo y apoyo brindado, a Iballa y Judith su valiosa colaboraci\'on, especialmente en estos \'ultimos meses. Al Instituto de T\'ecnica Aeroespacial (INTA) que conforma tambi\'en el equipo de trabajo de IMaX, y en especial a Carmen Pastor y Luis Miguel Gonz\'alez, les agradezco su disposici\'on y por suministrarnos tablas llenas de datos, imprescindibles para la realizaci\'on de la primera parte de la tesis. \\

Many thanks to all the different collaborators, specially, David Orozco, Luc Rouppe van der Voort and Yukio Katsukawa for providing important data I employed in this thesis, and to the  International Time Program and Hinode Operation Plan 14 for the observing campaigns I collaborated with. The european referees Thierry Roudier and Johann Hirzberger have done and excellent report on this work that enabled it to be presented for the \emph{Doctor Europeus} mention.\\

A mis amigos Tom\'as, Jorge y Guille que dejo en Canarias y que espero volver a ver con sus esposas y familias numerosas en un futuro no muy lejano. Y a los que conoc\'i en este encantador archipi\'elago y que s\'e que nunca dejar\'an de ser parte importante de mis vivencias, a Rene, Pablo, Alejandro.  Estas islas me dejan recuerdos muy buenos, el f\' utbol (PlayStation !!) de los lunes con los colegas, las doradas, las chuletadas, los paseitos, el Teide ..... etc, etc, espero volver y volver\'e. Quiero mencionar tambi\'en a Juanlu y recordarlo con alegr\'ia, la misma que \'el irradiaba.\\

Ha sido sin duda un buen trayecto recorrido en este camino desde que sal\'i de casa hace casi 8 a\~nos, y en \'el, algunas personas especiales compartieron sue\~nos y luchas. A mis amigos dudle\~nos Juan Felipe, H\'ector, Pablo Mateo, Santiago, Manuel, Dario y Gilberto que comenzaron conmigo esta aventura, agradezco el saber que cuento con ellos pese a las distancias y que siempre ser\'a un placer reunirnos como en aquellas \'epocas de Lingwood Hall y Firs Street en Dudley Dudley. Steve was also an important part in the English experience, thanks mate for all your kindness and support, and all the best for ASTRAL.\\

Una nueva vida sin duda se presenta ante mis ojos y eso se lo debo a Laura, quien a lo largo de estos \'ultimos a\~nos ha sido la persona que con su energ\'ia, comprensi\'on y amor ha sabido llenar los espacios vac\'ios que ten\'ia y que espero lo siga haciendo por muchos, muchos m\'as. Gracias a ella y a su inigualable familia, a sus padres Jes\'us y Maria Jos\'e, a su hermano Carlos, a los yayos Rafa, Encarna y J\'ulia a los t\'ios y primos, he sentido un cari\~no excepcional dificilmente igualable. Ahora ir\'e all\'i a reunirme con ellos y a seguir disfrutando de unas buenas paellas domingueras, uhhmmmm \~nam \~nam.\\

A mi familia en Colombia que ha sido un apoyo incondicional en toda mi vida, a mi abuelita, a mis t\'ios, t\'ias, primas y a cada uno de los que se ha alegrado con mi victorias y sentido tristeza con mis penas; ocupar\'an siempre un amplio espacio en mi coraz\'on.\\

Y el final, lo dedico al m\'as importante apoyo en mi vida, mis padres y mi hermano. Tendr\'ia que escribir un cap\'itulo entero o quiz\'as una y hasta varias tesis para agradecerles con palabras todo su amor y su dedicaci\'on. Los quiero inmensamente y cada d\'ia los llevo en mis pensamientos sabiendo que todo este esfuerzo es por y para ellos.\\

A Dios, a la vida y al Universo por hacerme tan feliz y darme siempre lo que he querido.

\vspace{8mm}
\hspace{8cm}\begin{minipage}{.4\textwidth}
Tenerife, Estocolmo, Valencia \\
Santiago Vargas Dom\'{\i}nguez\\
2008
\end{minipage}